\documentclass[mathleft
]{an}
\usepackage{graphicx}
\usepackage{times}
\usepackage{lscape,supertabular}       % inserted by author
\begin{document}

% The following seven commands are intended for editorial usage and should be ignored by
% the author(s).
\Pagespan{789}{}% Document's page range. 
% If second parameter is left empty, the last page is computed automatically.
\Yearpublication{2012}%
\Yearsubmission{2012}%
\Month{11}%   
\Volume{999}%  
\Issue{88}% 
% \DOI{This.is/not.aDOI}% 

\title{The stellar content of the young open cluster Trumpler 37}

\author{R. Errmann\inst{1}\fnmsep\thanks{Corresponding author:
  \email{ronny.errmann@uni-jena.de}\newline}
\and R. Neuh\"auser\inst{1}  %ralph.neuhaeuser@uni-jena.de
\and L. Marschall\inst{2}   %marschal@gettysburg.edu
\and G. Torres\inst{3}     %gtorres@cfa.harvard.edu
\and M. Mugrauer\inst{1}     %markus@astro.uni-jena.de
\and W.P. Chen\inst{4}         %wchen@astro.ncu.edu.tw
\and S.C.-L. Hu\inst{4,5}         %m989001@astro.ncu.edu.tw
\and C. Briceno\inst{6}      %briceno@cida.ve
\and R. Chini\inst{7,8}        %chini@astro.ruhr-uni-bochum.de
\and {\L}. Bukowiecki\inst{9}  %lukasz.bukowiecki@astri.uni.torun.pl
\and D.P. Dimitrov\inst{10}        %dinko@astro.bas.bg
\and D. Kjurkchieva\inst{11}  %kyurk@shu-bg.net
\and E.L.N. Jensen\inst{12}      %ejensen1@swarthmore.edu
\and D.H. Cohen\inst{12}      %dcohen1@swarthmore.edu
\and Z.-Y. Wu\inst{13} %zywu@nao.cas.cn
\and T. Pribulla\inst{14}       %pribulla@ta3.sk
\and M. Va\v{n}ko\inst{14}        %vanko@ta3.sk
\and V. Krushevska\inst{15}       %vkrush@mao.kiev.ua
\and J. Budaj\inst{14}          %budaj@ta3.sk
\and Y. Oasa\inst{16}     %yummy@mail.saitama-u.ac.jp
\and A.K. Pandey\inst{17}         %anil_pandey@yahoo.com
\and M. Fernandez\inst{18}   %matilde@iaa.es
\and A. Kellerer\inst{19}    %kellerer@bbso.njit.edu, a.n.c.kellerer@durham.ac.uk
\and C. Marka\inst{1} %p3macl@uni-jena.de
%Example 
%for footnote, note the usage of the \texttt{fnmsep}
%command as separator between institute number and footnote mark} 
}
\titlerunning{Stellar content of Trumpler 37}
\authorrunning{R. Errmann et al.}
\institute{
Astrophysikalisches Institut und Universit{\"a}ts-Sternwarte Jena, Schillerg{\"a}{\ss}chen 2-3, 
D-07745 Jena, Germany
\and %2
Gettysburg College Observatory, Department of Physics, 300 North Washington St., Gettysburg, PA 17325, USA
\and
Harvard-Smithsonian Center for Astrophysics, 60 Garden St., Mail Stop 20, Cambridge MA 02138, USA
%\and %\and F.M. Walter
%Department of Physics and Astronomy, Stony Brook University, Stony Brook, NY 11794-3800, USA
\and
Graduate Institute of Astronomy, National Central University, Jhongli City, Taoyuan County 32001, Taiwan (R.O.C.)
\and %5
Taipei Astronomical Museum, 363 Jihe Rd., Shilin, Taipei 11160, Taiwan
\and
Centro de Investigaciones de Astronomia, Apartado Postal 264, Merida 5101, Venezuela
\and
Astronomisches Institut, Ruhr-Universit\"at Bochum, Universit\"atsstr. 150, D-44801 Bochum, Germany
\and
Instituto de Astronom\'{i}a, Universidad Cat\'{o}lica del Norte, Antofagasta, Chile
\and
Toru\'n Centre for Astronomy, Nicolaus Copernicus University, Gagarina 11, PL–87-100 Toru\'n, Poland
\and
Institute of Astronomy and NAO, Bulg. Acad. Sci., 72 Tsarigradsko Chaussee Blvd., 1784 Sofia, Bulgaria
\and %Diana
Shumen University, 115 Universitetska str., 9700 Shumen, Bulgaria
\and
Dept. of Physics and Astronomy, Swarthmore College, Swarthmore, PA 19081-1390, USA
\and
Key Laboratory of Optical Astronomy, NAO, Chinese Academy of Sciences, 20A Datun Road, Beijing 100012, China
\and
Astronomical Institute, Slovak Academy of Sciences, 059 60, Tatransk\'a Lomnica, Slovakia
\and
Main Astronomical Observatory of National Academy of Sciences of Ukraine, 27 Akademika Zabolotnoho St., 03680 Kyiv, Ukraine
%\and %\and H. Takahashi
%Institute of Astronomy, University of Tokyo, 2-21-1 Osawa, Mitaka, Tokyo 181-0015, Japan
\and %17
Dept. of Astronomy and Earth Science, Saitama University, 255 Shimo-Okubo, Sakura, Saitama 338-8570, Japan
\and %18
Aryabhatta Research Institute of Observational Science, Manora Peak, Naini Tal, 263 129, Uttarakhand, India
%\and %\and H. Harutyunyan %\and T. Movsessian %\and E.H. Nikogossian
%Byurakan Astrophysical Observatory, 378433 Byurakan, Armenia 
\and
Instituto de Astrofisica de Andalucia, CSIC, Apdo. 3004, 18080 Granada, Spain
\and
Department of Physics, Durham University, South Road, Durham DH1 3LE, United Kingdom
}

\received{}
\accepted{}
\publonline{later}

 % next with --

\abstract{%  
With an apparent cluster diameter of $1.5^{\circ}$ and an age of $\sim 4\,$Myr, Trumpler\,37 is an ideal target for photometric monitoring of young stars as well as for the search of planetary transits, eclipsing binaries and other sources of variability. The YETI consortium has monitored Trumpler\, 37 throughout 2010 and 2011 to obtain a comprehensive view of variable phenomena in this region. In this first paper we present the cluster properties and membership determination as derived from an extensive investigation of the literature. We also compared the coordinate list to some YETI images.
For 1872 stars we found literature data. Among them 774 have high probability of being member and 125 a medium probability. Based on infrared data we re-calculate a cluster extinction of $0.9-1.2$\,mag. We can confirm the age and distance to be $3-5$\,Myr and $\sim870$\,pc. Stellar masses are determined from theoretical models and the mass function is fitted with a power-law index of $\alpha=1.90$ ($0.1-0.4\,M_\odot$) and $\alpha=1.12$ ($1-10\,M_\odot$).
}

\maketitle

\section{Introduction: Trumpler\,37}
Trumpler 37 is a young open cluster in the Cepheus OB2 region. Based on optical spectroscopy and photometry, and main sequence fitting, Contreras et al. (\cite{con02}) derived a distance of about 870\,pc. The latest age estimation yields \linebreak$\sim4$\,Myr (Kun, Kiss \& Balog \cite{kkb08}, Sicilia-Aguilar \cite{sic05}), using also optical spectroscopy and photometry for comparison to theoretical isochrones. Thereby the average extinction was measured to be $A_{\mathrm{V}}=1.56\pm0.55$\,mag. Mercer et al. (\cite{mer09}) found an average extinction in the central region of  $A_{\mathrm{V}}\sim1.3$\,mag.
Several studies were devoted to distinguish between members and foreground or background stars. 

The first classification as a cluster was done by Trumpler (\cite{tru30}), who used the brightness and spectral types of the stars to derive their distance moduli. This resulted in a cluster distance of 670 to 890\,pc. Similar work was done by Simonson (\cite{sim68}), and Garrison \& Kormendy (\cite{gar76}), who both obtained a distance of 1000\,pc. The stars in young clusters are expected to display common space velocities which surpass their random movements. Therefore, studying the kinematics of a stellar aggregate allows calculating the membership probability. Marschall \& van Altena (\cite{mar87}) measured  the proper motions while Sicilia-Aguilar et al. (\cite{sic06-2}) determined the radial velocities to infer members of Trumpler\,37.

Young clusters offer additional membership tracers\linebreak which use particular signs of star formation to discriminate young stellar objects from older field stars. A prominent property of young stars is their photometric variability due to spots or accretion. Gieseking (\cite{gie76}), Sicilia-Aguilar et al. (\cite{sic04}) and Morales-Calder{\'o}n et al. (\cite{mor09}) applied this technique to Trumpler\,37. The youth of stars and therefore high membership probability can also be derived from lithium absorption (Sicilia-Aguilar et al. \cite{sic04,sic05}), because most of the primordial lithium is depleted after a few Myr (e.g. Piau \& Turck-Chi{\`e}ze \cite{pia02}). A useful tracer for disk accretion is H$\alpha$ emission. 
This behavior was employed by Kun (\cite{kun86}), and Kun \& Pasztor (\cite{kun90}) to find cluster members. Infrared excess in the spectral energy distribution is a hint for circumstellar disks and therefore another indicator for youth (Sicilia-Aguilar et al. \cite{sic06-1}). The 
variability of young stars which are still embedded in a dark cloud can be studied in the infrared (Morales-Calder{\'o}n \cite{mor09}). Likewise, they show enhanced X-ray emission due to higher activity. This was used by Mercer et al. (\cite{mer09}) to investigate membership in Trumpler\,37.

The YETI (\textit{Young Exoplanet Transit Initiative}) consortium was established, to monitor young clusters like Trumpler\,37 in a continuous way (see Neuh{\"a}user et al. \cite{neu11}). The consortium consists of $0.4$ to $2$\,m sized telescopes, which are located at different longitudes all over the world. Data from the Jena 90/60\,cm Schmidt telescope, from the Xinglong 90\,cm telescope, and from the Rozhen 60\,cm telescopes were used for this work. The big fields of view of Jena ($53^\prime$x$53^\prime$) and Xinglong ($94^\prime$x$94^\prime$) covered the largest areas of Trumpler\,37, while the 2x2 mosaic of the Rozhen 60\,cm telescope provides a good resolution ($0.53^{\prime\prime}$/px).

The main motivation for this paper is to present a comprehensive view
of the properties of Trumpler 37, by collating information scattered
throughout the literature, compiling the most complete list available
of stars in the field of the cluster from various existing astrometric
and photometric sources, assessing their individual membership using a
suite of kinematic and astrophysical criteria, and deriving additional
properties of the cluster including the mass distribution. This will
provide the basic framework for extensive variability studies of
members of Trumpler 37 that are currently underway within the YETI
project.

\section{Data collection}

We combined the data from several publications and data\-bases: Marschall \& van Altena (\cite{mar87}), Contrer\-as et al. \linebreak (\cite{con02}), Sicilia-Aguilar et al. (\cite{sic04,sic05,sic06-1,sic06-2}), \linebreak Mercer et al. (\cite{mer09}), Morales-Calder{\'o}n (\cite{mor09}) and  the \linebreak WEBDA\footnote{http://www.univie.ac.at/webda/} database. The WEBDA catalog contains the data on Trumpler\,37 from publications before the year 2000. One has to take into account the fact that all observations and publications deal with biased samples, either because only brighter stars are included or because they include only a selected subset of the stars, e.g. late-type stars or those showing photometric variability over limited time scales, band passes, and/or magnitude ranges.

Additional information was added from the Two-Micron All Sky Survey Point Source Catalog (2MASS PSC, Skrutskie et al. \cite{skr06}). If two 2MASS sources were located next to the star, the literature data were connected to both of the 2MASS sources, resulting in two entries for them. Probably, the other data, like the optical brightness, of the two close stars are unresolved in this cases. Furthermore, the proper motion catalogs UCAC3 (Zacharias et al. \cite{zac10}), and \mbox{PPMXL} (Roeser, Demleitner \& Schilbach \cite{roe10}) were attached to the 2MASS positions. PPMXL was more complete than UCAC3.
%Separation $\xi$ between different catalog positions were calculated with $$\xi=\arccos\left(\sin\delta_1\sin\delta_2+\cos\delta_1\cos\delta_2\cos(\alpha_2-\alpha_1)\right),$$ with right ascension $\alpha_i$ and declination $\delta_i$.

\begin{figure}
 \includegraphics[width=8.3cm]{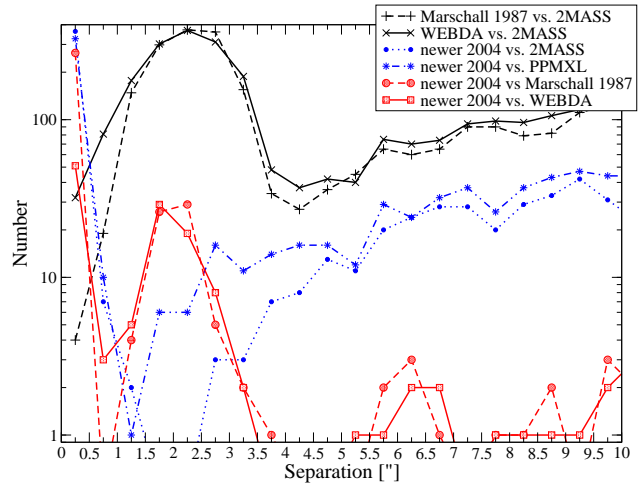}  % Distance means separation
 \caption{The measured minimal separation between stars of two catalogs, resulting in bimodal distributions to find the maximum allowable separation. The comparison of the coordinates from literature after 2004 with WEBDA and Marschall \& van Altena (\cite{mar87}) (red filled circles and squares) results in an optimized search radius of $\le 4.5^{\prime\prime}$. This is also the best search radius, when comparing the later positions to 2MASS (black + and x). For the catalogs after 2004 the coordinates are more accurate, which results in optimal search radii of $\le 1.5^{\prime\prime}$ (blue dots and stars).}
 \label{disthist}
\end{figure}

We used the J2000 coordinates as given in the literature for the cross correlation. The B1950 coordinates were transformed in J2000 for WEBDA and Marschall \& van Altena (\cite{mar87}), using the proper motion from the latter publication, if available.

The cross correlation of coordinates from two different catalogs results in bimodal distributions (see Fig.~\ref{disthist} for some examples) in the separation (as minimum between two peaks). We therefore used two criteria for identifying stars in different tables: a $4.5^{\prime\prime}$ search radius was used if at least one of the coordinate sets was created before 2004; otherwise a $1.5^{\prime\prime}$ search radius was used for comparison of more recently obtained pairs of coordinate measurements. A search radius of $2.0^{\prime\prime}$ was used for adding 2MASS, \mbox{PPMXL} and UCAC3 to the catalogs after 2004. Additionally, the identifiers are used to check the correct entries, if available. We also compared brightness measurements from different data sources to check the plausibility of our cross-identifications. But the latter method was limited due to different photometric sources and variability of the stars.%Contreras 2002 also 4.5

We compared the coordinate lists also with data taken with the telescopes in the YETI network. The used images were taken on 2009 July 25 with the Jena telescope, and on 2010 August 5 and 6 with the Xinglong and Rozhen telescopes, respectively. The images were reduced in a standard way (bias, dark and flat-field correction) and were astrometrically calibrated, using the program \textit{GAIA\footnote{http://star-www.dur.ac.uk/~pdraper/gaia/gaia.html}}. No galaxies were found during the inspection of the images.

We found and corrected a few problems, which are all marked in Table~\ref{supertab}:
\begin{itemize}
 \item Some stars with the same position got two entries: in the WEBDA database and in one case in the Sicilia-Aguilar et al. publications. %IDs 12-44 and 54-1547
These data were combined into one entry in our table and marked with footnotes.
 \item The positions were in some cases imprecise. This was the case for the declination as given by Contreras et al. (\cite{con02}) and for WEBDA entries. The discrepancy in the first one was up to $10^{\prime\prime}$ using the identifiers and magnitudes.
 \item The star position in the finding chart and the catalog position did not match in some cases for the stars from Marschall \& van Altena (\cite{mar87}). In the most cases we were able to fix the catalog entry. In our list they are marked with ``new coordinates''. If no star is visible at the new position it is marked with ``no star''. 
 \item Additionally, when we plotted the stars from Marschall \& van Altena (\cite{mar87}) in the YETI images, we found for some stars no or only a very faint star in our observation. We annotated these `missing' stars in our table as ``no/faint star'' or ``no star''. If these stars are true objects, detected by Marschall \& van Altena (\cite{mar87}) on the plates from both 1937 and 1973, they disappeared by becoming much fainter or being very variable. Objects detected on only one epoch (either 1937 or 1973) could also be very fast moving objects, in which case they probably are foreground objects, not cluster members.
\end{itemize}

The literature data are given in Table~\ref{supertab} and partly in Table~\ref{supermem} (Columns 2 to 10).

\begin{landscape}
 \begin{@twocolumnfalse}
  \begin{table*}
   \caption{Literature data for stars in Trumpler 37}
   \scriptsize
   \begin{tabular}{l @{\hspace{1.5mm}} l @{\hspace{1.5mm}}l @{\hspace{1.5mm}} l @{\hspace{1.5mm}} l @{\hspace{1.5mm}} l @{\hspace{1.5mm}} l @{\hspace{1.5mm}} l @{\hspace{1.5mm}} l @{\hspace{1.5mm}} l@{\hspace{1.5mm}} l @{\hspace{1.5mm}} l@{\hspace{1.5mm}}l@{\hspace{1.5mm}}l @{\hspace{1.5mm}} l @{\hspace{0.3mm}} l @{\hspace{1.5mm}} l @{\hspace{1.5mm}}  l@{\hspace{1.5mm}}l @{\hspace{1.5mm}} l@{\hspace{1.5mm}}l @{\hspace{1.5mm}} l@{\hspace{1.5mm}} l@{\hspace{1.5mm}} l}                                                                                                                                                                                                                                                                               
 No. & RA          & Dec        & MVA  & WEB-      &  SHB-         &$U$    &$B$    & $V$   &$R$    &$I$    &$J$   &$H$   &$K$   & SpT      & Class &$A_{\mathrm{V}}$&$\mu_\alpha$  & $\mu_\delta$ & $\mu_\alpha$ & $\mu_\delta$ & $\mu_\alpha$ & $\mu_\delta$ &  Comments \\
     & \multicolumn{2}{c}{J2000}&      & DA        &  2004         &       &       &       &       &       &      &      &      &          &                        &     & \multicolumn{2}{l}{PPMXL}   & \multicolumn{2}{l}{UCAC3}   & \multicolumn{2}{l}{MVA [j]} &  \\
     & hh:mm:ss.ss & dd:mm:ss.s &      &           &               &  mag  &  mag  &  mag  &  mag  &  mag  &  mag & mag  & mag  &          &                        & mag & mas/yr       & mas/yr       & mas/yr       & mas/yr       & mas/yr       & mas/yr       &  \\
   \hline
%No. & RA          & Dec                  & MVA  & WEBDA                     &  SHB                          & U               &  B              & V               &   R   &  I              & J           & H           & K                     & SpT                & Class           & A_V           & $\mu_\alpha$ & $\mu_\delta$ & $\mu_\alpha$ & $\mu_\delta$ &mu_a & mu_b&    Comments \\
1    & 21:36:46.57 &  57:11:25.4$^{\rm r}$& 2    & 3002                      &                               &                 &                 &   14.7$^{\rm j}$&       &                 &11.532\,(24) &10.787\,(28) & 10.583\,(20)$^{\rm r}$&                    &                 &               &-3.2\,(4.1)   &-2.4\,(4.1)   &-11.3\,(6.8)  &0.1\,(6.8)    &     &     &                                                                                      \\
2    & 21:36:44.78 &  57:11:53.0$^{\rm r}$& 3    & 3003                      &                               &                 &                 &   13.9$^{\rm j}$&       &                 &12.647\,(27) &12.547\,(37) & 12.452\,(29)$^{\rm r}$&                    &                 &               &-10.4\,(4.1)  &2.2\,(4.1)    &-30.9\,(6.8)  &12.1\,(6.8)   &-0.17&0.05 &                                                                                      \\
3    & 21:36:42.64 &  57:13:01.0$^{\rm r}$& 4    & 3004                      &                               &                 &                 &   13.3$^{\rm j}$&       &                 &11.880\,(27) &11.530\,()   &   11.445\,()$^{\rm r}$&                    &                 &               &              &              &              &              &3.15 &-4.9 &                                                                                      \\
4    & 21:36:20.30 &  57:12:55.9$^{\rm r}$& 5    & 3005                      &                               &                 &                 &   13.6$^{\rm j}$&       &                 &11.562\,(26) &11.227\,(28) & 11.101\,(20)$^{\rm r}$&                    &                 &               &-8.2\,(4.1)   &1.9\,(4.1)    &-11.9\,(6.8)  &6.6\,(6.8)    &0.13 &-0.23&                                                                                      \\
5    & 21:36:29.82 &  57:12:48.0$^{\rm j}$& 6    & 3006                      &                               &                 &                 &   14.6$^{\rm j}$&       &                 &             &             &                       &                    &                 &               &              &              &              &              &     &     &no/faint star                                                               \\
6    & 21:36:41.77 &  57:13:40.8$^{\rm r}$& 7    & 3007                      &                               &                 &                 &   13.9$^{\rm j}$&       &                 &10.779\,(26) &10.051\,(27) & 9.849 \,(23)$^{\rm r}$&                    &                 &               &-0.3\,(5.1)   &5.3\,(5.1)    &-8.2\,(7.1)   &41.1\,(7.2)   &0.01 &0.75 &                                                                                      \\
7    & 21:36:40.66 &  57:13:39.2$^{\rm r}$& 8    & 3008                      &                               &                 &                 &     15$^{\rm j}$&       &                 &12.718\,(28) &12.333\,(41) & 12.227\,(29)$^{\rm r}$&                    &                 &               &-19.4\,(4.1)  &-6.3\,(4.1)   &-70.6\,(7.1)  &-5.5\,(7.1)   &     &     &                                                                                      \\
8    & 21:36:46.12 &  57:12:53.3$^{\rm r}$& 9    & 3009                      &                               &                 &                 &   14.8$^{\rm j}$&       &                 &12.685\,(26) &12.395\,(30) & 12.256\,(25)$^{\rm r}$&                    &                 &               &-3.9\,(4.1)   &-10.2\,(4.1)  &-6.6\,(6.8)   &-29.3\,(6.8)  &0.1  &-0.13&                                                                                      \\
9    & 21:36:47.04 &  57:13:01.7$^{\rm r}$& 10   & 3010                      &                               &                 &                 &   14.5$^{\rm j}$&       &                 &12.718\,(26) &12.458\,(31) & 12.361\,(25)$^{\rm r}$&                    &                 &               &2.7\,(4.1)    &7.3\,(4.1)    &5.2\,(7.6)    &13.9\,(7.6)   &-0.12&0.26 &                                                                                      \\
10   & 21:36:50.76 &  57:12:41.4$^{\rm r}$& 11   & 3011                      &                               &                 &                 &   14.9$^{\rm j}$&       &                 &12.423\,(31) &12.017\,(32) & 11.878\,(25)$^{\rm r}$&                    &                 &               &-13\,(4.1)    &6.5\,(4.1)    &-23.2\,(6.9)  &28.7\,(7)     &     &     &                                                                                      \\
11   & 21:36:27.84 &  57:14:05.7$^{\rm r}$& 12   & 3012                      &                               &                 &                 &   14.9$^{\rm j}$&       &                 &11.658\,(24) &11.039\,(27) & 10.860\,(21)$^{\rm r}$&                    &                 &               &-37.1\,(4.1)  &-55.6\,(4.1)  &-32.8\,(6.8)  &-50.7\,(6.8)  &     &     &                                                                                      \\
12   & 21:36:32.90 &  57:14:20.1$^{\rm r}$& 13   & 3013                      &                               &                 &                 &   13.6$^{\rm j}$&       &                 &12.133\,(28) &11.785\,(33) & 11.716\,(28)$^{\rm r}$&                    &                 &               &-8.9\,(4.1)   &-18.9\,(4.1)  &1.6\,(7)      &-59.8\,(7)    &0.65 &-0.37&                                                                                      \\
13   & 21:36:32.90 &  57:14:52.2$^{\rm r}$& 14   & 3014                      &                               &                 &                 &   13.4$^{\rm j}$&       &                 &11.662\,(26) &11.384\,(28) & 11.279\,(23)$^{\rm r}$&                    &                 &               &-6\,(4.1)     &0.4\,(4.1)    &-10.8\,(6.8)  &17.5\,(6.8)   &0.22 &-0.28&                                                                                      \\
14   & 21:36:55.07 &  57:15:23.6$^{\rm r}$& 15   & 3015                      &                               &                 &                 &   13.8$^{\rm j}$&       &                 &12.213\,(22) &11.835\,(28) & 11.769\,(21)$^{\rm r}$&                    &                 &               &-0.1\,(4.1)   &11.4\,(4.1)   &-1.3\,(6.8)   &11.3\,(6.8)   &-0.85&1.14 &                                                                                      \\
15   & 21:36:55.96 &  57:13:39.7$^{\rm r}$& 16   & 3016                      &                               &                 &                 &   14.8$^{\rm j}$&       &                 &10.627\,(24) &9.679 \,(28) & 9.406 \,(21)$^{\rm r}$&                    &                 &               &-3.6\,(5.1)   &-2.3\,(5.1)   &-13.7\,(6.9)  &1.6\,(6.9)    &0.32 &0.17 &                                                                                      \\
16   & 21:36:58.46 &  57:13:46.0$^{\rm r}$& 17   & 3017                      &                               &                 &                 &     15$^{\rm j}$&       &                 &12.959\,(22) &12.565\,(28) & 12.481\,(25)$^{\rm r}$&                    &                 &               &-3.4\,(4.1)   &9.5\,(4.1)    &0\,(6.8)      &7.7\,(6.8)    &     &     &                                                                                      \\
17   & 21:36:46.85 &  57:17:11.5$^{\rm r}$& 18   & 3018                      &                               &                 &                 &     15$^{\rm j}$&       &                 &12.827\,(27) &12.484\,(33) & 12.370\,(28)$^{\rm r}$&                    &                 &               &-2.6\,(4.1)   &3\,(4.1)      &3.6\,(6.8)    &9.8\,(7)      &     &     &                                                                                      \\
18   & 21:36:35.45 &  57:17:33.0$^{\rm r}$& 19   & 3019                      &                               &                 &                 &   14.3$^{\rm j}$&       &                 &12.236\,(24) &11.890\,(28) & 11.817\,(24)$^{\rm r}$&                    &                 &               &-10.1\,(4.1)  &2.5\,(4.1)    &-14.5\,(7.4)  &8.7\,(7.4)    &0.74 &0.04 &                                                                                      \\
19   & 21:36:30.67 &  57:19:25.5$^{\rm r}$& 20   & 3020                      &                               &                 &                 &   12.6$^{\rm j}$&       &                 &11.613\,(24) &11.496\,(31) & 11.367\,(23)$^{\rm r}$&                    &                 &               &-5.8\,(4.1)   &0.2\,(4.1)    &-5.1\,(2.3)   &-5.4\,(2.1)   &0.05 &-0.43&                                                                                      \\
20   & 21:36:41.27 &  57:18:43.6$^{\rm r}$& 22   & 3022                      &                               &                 &  16.13$^{\rm l}$&  14.53$^{\rm i}$& 13.63 &  12.75$^{\rm i}$&11.284\,(24) &10.614\,(28) & 10.357\,(21)$^{\rm r}$&                    &                 &               &-6.3\,(4.1)   &-1.7\,(4.1)   &-18.3\,(6.8)  &0.3\,(6.8)    &0.39 &-0.01&                                                                                      \\
21   & 21:36:46.24 &  57:18:47.6$^{\rm r}$& 23   & 3023                      &                               &  15.66$^{\rm l}$&   15.3$^{\rm h}$&  14.42$^{\rm h}$& 13.9  &   13.4$^{\rm i}$&12.685\,(26) &12.411\,(31) & 12.302\,(26)$^{\rm r}$&        F6$^{\rm h}$&                 & 1.31$^{\rm h}$&-7.4\,(4.1)   &-0.5\,(4.1)   &-0.9\,(6.8)   &6.8\,(6.8)    &0.25 &0.06 &Dec [h] imprec.                                                               \\
22   & 21:36:50.20 &  57:19:07.2$^{\rm r}$& 24   & 3024                      &                               &                 &                 &   14.4$^{\rm j}$&       &                 &8.750 \,(27) &7.551 \,(42) & 7.140 \,(21)$^{\rm r}$&                    &                 &               &-4.6\,(5.1)   &0.9\,(5.1)    &-18.3\,(6.7)  &5\,(6.7)      &     &     &                                                                                      \\
23   & 21:36:50.49 &  57:18:15.0$^{\rm r}$& 25   & 3025                      &                               &                 &  14.55$^{\rm l}$&  12.42$^{\rm l}$&       &                 &8.164 \,(23) &7.239 \,(34) & 6.914 \,(31)$^{\rm r}$&                    &                 &               &-4.7\,(13.8)  &2.8\,(13.8)   &-1.8\,(7.8)   &0.8\,(7.8)    &-0.32&0.21 &                                                                                      \\
24   & 21:37:00.18 &  57:18:27.1$^{\rm r}$& 26   & 445                       &                               &                 &                 &   11.8$^{\rm j}$&       &                 &11.102\,(21) &10.989\,(28) & 10.930\,(21)$^{\rm r}$&        B8$^{\rm q}$&                 &               &-6.5\,(13.3)  &3.4\,(13.3)   &-4\,(1.2)     &-1.3\,(1.1)   &-0.35&0.32 &                                                                                      \\
25   & 21:36:55.01 &  57:19:43.2$^{\rm r}$& 27   & 3027                      &                               &                 &                 &   14.7$^{\rm j}$&       &                 &12.806\,(24) &12.394\,(32) & 12.264\,(24)$^{\rm r}$&                    &                 &               &14.6\,(4.1)   &-2.7\,(4.1)   &10.2\,(6.8)   &-0.8\,(6.8)   &     &     &                                                                                      \\
26   & 21:37:01.56 &  57:19:47.3$^{\rm r}$& 28   & 3028                      &                               &                 &                 &   13.8$^{\rm j}$&       &                 &11.048\,(22) &10.404\,(28) & 10.179\,(21)$^{\rm r}$&                    &                 &               &5.1\,(4.1)    &9.7\,(4.1)    &14.7\,(6.8)   &15.4\,(6.8)   &-1.43&1.09 &                                                                                      \\
27   & 21:37:08.60 &  57:18:03.3$^{\rm r}$& 29   & 3029                      &                               &                 &                 &   13.7$^{\rm j}$&       &                 &12.380\,(21) &12.181\,(27) & 12.113\,(24)$^{\rm r}$&                    &                 &               &-3.5\,(4.1)   &3.5\,(4.1)    &20\,(6.8)     &-8.4\,(6.8)   &0.12 &0.14 &                                                                                      \\
28   & 21:37:10.54 &  57:18:39.9$^{\rm r}$& 30   & 3030                      &                               &                 &                 &   14.7$^{\rm j}$&       &                 &11.413\,(22) &10.619\,(27) & 10.465\,(20)$^{\rm r}$&                    &                 &               &-5.6\,(4.1)   &1\,(4.1)      &-7.6\,(6.8)   &-6.1\,(6.8)   &-0.22&0.33 &                                                                                      \\
29   & 21:36:34.30 &  57:20:53.6$^{\rm r}$& 31   & 3031                      &                               &                 &                 &   14.8$^{\rm j}$&       &                 &10.745\,(22) &9.736 \,(29) & 9.435 \,(21)$^{\rm r}$&                    &                 &               &-12\,(5.1)    &-1.1\,(5.1)   &-13.3\,(6.8)  &3\,(6.9)      &     &     &                                                                                      \\
  \end{tabular} 

\small
plus 1862 stars in electronic table (at the end of the document: Table~\ref{supertabfull}).        

\footnotesize
Remarks: The superscript letters behind the values indicate the source for the value:                                     \newline
{\bf$[$a$]$}~Morales-Calder{\'o}n et al. (\cite{mor09}); {\bf$[$b$]$}~Mercer et al. (\cite{mer09}); {\bf$[$c$]$}~Sicilia-Aguilar et al. (\cite{sic06-2}); {\bf$[$d$]$}~Sicilia-Aguilar et al. (\cite{sic06-1}); {\bf$[$e$]$}~Sicilia-Aguilar et al. (\cite{sic05}); {\bf$[$f$]$}~Sicilia-Aguilar et al. (\cite{sic04}); {\bf$[$g$]$}~WEBDA (consists of Sicilia-Aguilar  et al. (\cite{sic04}) and Morbidelli et al. (\cite{morb97}); {\bf$[$h$]$}~Contreras et al. (\cite{con02}) (used for photometry Marschall, Karshner \& Comins (\cite{mar90})); {\bf$[$i$]$}~Marschall et al. (\cite{mar90}); {\bf$[$j$]$}~Marschall \& van Altena (\cite{mar87}) ($V$ magnitudes from fitting instrumental magnitudes to photometry from Garrison \& Kormendy (\cite{gar76}) and de Lichtbuer (\cite{lic82})); {\bf$[$k$]$}~Kun (\cite{kun86}); 
{\bf$[$l$]$}~WEBDA (consists of Marschall et al. (\cite{mar90}), Garrison \& Kormendy (\cite{gar76}), Simonson (\cite{sim68}) and other publications for few stars); {\bf$[$m$]$}~WEBDA (coordinate source); 
{\bf$[$n$]$}~WEBDA (consists of Marschall \& van Altena (\cite{mar87}) and internal WEBDA information); {\bf$[$o$]$}~WEBDA (consists of 6 publications for 7 stars); {\bf$[$p$]$}~WEBDA (consists of Garrison \& Kormendy (\cite{gar76}) and other publications for few stars); {\bf$[$q$]$}~WEBDA (consists of Alkansis (\cite{alk58}), Contreras et al. (\cite{con02}), Sicilia-Aguilar et al. (\cite{sic04}), Balazs et al. (\cite{bal96}) and other publication for few stars); {\bf$[$r$]$}~2MASS (Skrutskie et al. \cite{skr06}). The different WEBDA tables were compiled from different literature, the main publications are given in brackets\newline
MVA, WEBDA and SHB-2004 are star numbers in papers {$[$j$]$}; {$[$l$]$}-{$[$q$]$}; and {$[$c$]$}-{$[$f$]$}, {$[$h$]$}, respectively.
If data from different literature are available, the more recent one is given. Please note, that the $V$ magnitude was measured from photographic plate, photoelectrical or with CCD, making comparison difficult. The source for $R$ and $I$ magnitude is the same (given after $I$) and the source for $J$, $H$ and $K$ magnitude is the same (given after $K$). Errors in \textit{JHK}-photometry are given only, if the 2MASS quality flag is ``A'', ``B'', ``C'' or ``D'', otherwise an empty parenthesis indicates uncertainties in the 2MASS photometry.        \newline
{\bf Comments}: If two stars were located close to each other ($<5^{\prime\prime}$), the stars were marked with ``near \#''.  ``no star'' or ``no/faint star'' means we were not able to find the star from Marschall \& van Altena (\cite{mar87}) in our images (see also the text).
``new coordinates'' means, we changed the coordinates from Marschall \& van Altena (\cite{mar87}) to match the position that was given in their finding chart (see also text).
In cases of infrared data (Sicilia-Aguilar et al. \cite{sic06-1}), we were not able to see some stars in our optical images, resulting in comments ``no opt. cp.'' or ``very faint opt. cp.'' (opt. cp. standing for optical counterpart).
Because Sicilia-Aguilar et al. (\cite{sic04}) used the earlier compilation of the 2MASS catalog (Cutrie et al. \cite{cut03}) some stars get the comment ``\textit{JHK} in {$[$f$]$} different''.
In case of two not distinguishable 2MASS sources near the star, the entry was duplicated in the consecutive row, so both sources were connected. The comment ``2x{$[$r$]$}'' was added and the fainter one marked. Probably, the other data from the literature, like optical brightness, is not resolved in this case.
In Marschall \& van Altena (\cite{mar87}) and the WEBDA database stars outside all YETI telescope fields of view (FoV) are marked with ``outFoV''.
In some cases stars with the same names (and properties) differ in the coordinates in different catalogs. The more reliable coordinate was used and in the comments ``Dec {$[$h$]$} imprec.'' or ``{$[$m$]$} imprec.'' was attached, meaning that problems occurred in Contreras et al. \cite{con02} or the WEBDA database. In some entries the WEBDA entries were even wrong, resulting in ``WEBDA wrong''. \newline
Spectroscopic binaries were marked with ``SB1'' or SB2'' as given in Sicilia-Aguilar et al. (\cite{sic06-2}).
 \label{supertab}                                                                                                                                                                                                                                                                                                                                           
\end{table*}                                                                                                                                                                                                                                                                                                                                          
\end{@twocolumnfalse}                                                                                                                                                                                                                                                                                                                                          
\end{landscape}

\begin{table*}
 \caption{Literature data and membership probabilities for stars in Trumpler 37}
  \footnotesize
    %                                No                  PM                  RV               &  EW(Li)           &  EW(Li)           &  EW(H$\alpha$)    &  EW(H$\alpha$)    &  $\dot M$        &   L_x,c            &  TTS              &   Li       &           H$\alpha$       &   RV         &        $\dot M$      &     X-ray        &       IR-Excess      &     Varia-              PM                  OWN aV              Mass
   
  \begin{tabular}{@{\hspace{0.8mm}}l @{\hspace{1.6mm}} l @{\hspace{1.6mm}} l @{\hspace{1.6mm}} l @{\hspace{1.6mm}} l @{\hspace{1.6mm}} l @{\hspace{1.6mm}} l @{\hspace{1.6mm}} l @{\hspace{1.6mm}} l @{\hspace{1.6mm}} l | @{\hspace{1.0mm}} l @{\hspace{1.6mm}} l @{\hspace{1.6mm}} l @{\hspace{1.6mm}} l @{\hspace{1.6mm}} l  @{\hspace{1.6mm}} l  @{\hspace{1.6mm}} l @{\hspace{1.6mm}} l | @{\hspace{1.0mm}} l @{\hspace{1.6mm}} l @{\hspace{0.8mm}} } 
 No. &  RV  & PM        &EW(Li)&EW(Li)&EW(H$\alpha$)&EW(H$\alpha$)& $\dot M$    &$L_{\mathrm{X,c}}$& TTS & Li    &H$\alpha$&   RV  &$\dot M$&X-ray&IR ex-&Varia- & PM  &$A_{\mathrm{V}}$ & Mass          \\
     &      & [j]       & max  & min  & max         & min         &$10^{-8}$    &$10^{30}$         &     &       &         &       &       &      &cess  &bility &     &(\textit{JHK})   & (models)      \\
     & km/s & \%        & \AA  & \AA  & \AA         & \AA         &M$_\odot$/yr & erg/s            &     &       &         &       &[c]    &[b]   &[d]   &[a,e,f]& [j] &mag              & M$_\odot$    \\
   \hline
  % No  &    RV                    & PM &EW(Li          &EW(Li          &EW(Ha          &EW(Ha)          & $\dot M$   &L_x,c  & TTS           &Li     & Halpha & RV    & Mdot  & X-ray & IR-E  & Var & PM &own Av&MassMod                                           \\
   2    &                          & 93 &               &               &               &                &            &       &               &       &        &       &       &       &       &     & h  &      &                                                  \\
   3    &                          & 0  &               &               &               &                &            &       &               &       &        &       &       &       &       &     & l  &      &                                                  \\
   4    &                          & 92 &               &               &               &                &            &       &               &       &        &       &       &       &       &     & h  &      &                                                  \\
   6    &                          & 9  &               &               &               &                &            &       &               &       &        &       &       &       &       &     & l  &      &                                                  \\
   8    &                          & 94 &               &               &               &                &            &       &               &       &        &       &       &       &       &     & h  &      &                                                  \\
   9    &                          & 90 &               &               &               &                &            &       &               &       &        &       &       &       &       &     & h  &      &                                                  \\
   12   &                          & 37 &               &               &               &                &            &       &               &       &        &       &       &       &       &     & l  &      &                                                  \\
   13   &                          & 89 &               &               &               &                &            &       &               &       &        &       &       &       &       &     & h  &      &                                                  \\
   14   &                          & 0  &               &               &               &                &            &       &               &       &        &       &       &       &       &     & l  &      &                                                  \\
   15   &                          & 79 &               &               &               &                &            &       &               &       &        &       &       &       &       &     & h  &      &                                                  \\
   18   &                          & 14 &               &               &               &                &            &       &               &       &        &       &       &       &       &     & l  &      &                                                  \\
   19   &                          & 78 &               &               &               &                &            &       &               &       &        &       &       &       &       &     & h  &      &                                                  \\    \vspace{-2mm}
   20   &            44.9$^{\rm h}$& 83 &               &               &               &                &            &       &               &       &        &l      &       &       &       &     & h  &      &                                                  \\    \hspace{3mm}\vdots  &                          &    &               &               &               &                &            &       &               &       &      &       &       &       &       &     &    &      &                                                  \\
   1410 &                          &    &  0.3$^{\rm c}$&               &   -7$^{\rm c}$&  -7.3$^{\rm e}$&           0&       &    w$^{\rm c}$&h      & l      &       &l      &       &l      &     &    & 1.04 &0.2                                               \\
   1411 &                          &    &  0.5$^{\rm c}$&               & -4.8$^{\rm e}$&    -5$^{\rm c}$&           0&       &    w$^{\rm c}$&h      & l      &       &l      &       &l      &h    &    &      &                                                  \\
   1412 &           -14.6$^{\rm c}$&    &  0.5$^{\rm c}$&               & -1.8$^{\rm e}$&    -2$^{\rm c}$&        0.13&       &    w$^{\rm c}$&h      & l      &h      &h      &       &l      &h    &    &      &                                                  \\
   1413 &                          &    &               &               &   -3$^{\rm f}$&                &            &       &               &       & l      &       &       &       &       &     &    & 0.49 &0.15                                              \\
   1414 &           -42.8$^{\rm c}$&    &  0.3$^{\rm c}$&               &   -5$^{\rm c}$&                &           0&       & w(c)$^{\rm c}$&h      & h      &l      &l      &       &l      &l    &    &      &                                                  \\
   1415 &                          &    &    1$^{\rm c}$&               &  -13$^{\rm c}$& -13.4$^{\rm e}$&       0.12:&       & w(c)$^{\rm c}$&h      & h      &       &h      &       &l      &l    &    &      &                                                  \\
   1416 &           -17.2$^{\rm c}$&    &  0.4$^{\rm c}$&               &   -5$^{\rm c}$&                &         1.6&       &    c$^{\rm c}$&h      & h      &h      &h      &       &h      &h    &    & 2.65 &0.1                                               \\
   1417 &           -19.9$^{\rm c}$&    &  0.5$^{\rm c}$&  0.3$^{\rm f}$&  -43$^{\rm c}$&   -63$^{\rm f}$&    0.97-2.5&       &    c$^{\rm c}$&h      & h      &m      &h      &       &h      &h    &    & 1.18 &0.1                                               \\
   1418 &                          &    &  0.7$^{\rm f}$&  0.5$^{\rm c}$&   -4$^{\rm f}$&   -10$^{\rm c}$&         1.1&       &    c$^{\rm c}$&h      & h      &       &h      &       &h      &h    &    & 2.64 &0.1                                               \\
   1419 &           -15.4$^{\rm c}$&    &  0.5$^{\rm c}$&               &  -28$^{\rm c}$&   -33$^{\rm c}$&   16.2-13.2&       &    c$^{\rm c}$&h      & h      &h      &h      &       &h      &h    &    & 1.93 &0.1                                               \\
   1420 &                          &    &  0.5$^{\rm e}$&               &   -8$^{\rm e}$&                &            &       &    w$^{\rm e}$&h      & l      &       &       &       &l      &     &    & 1.12 &0.1                                               \\
   1421 &            -9.9$^{\rm c}$&    &  0.4$^{\rm c}$&               &  -18$^{\rm c}$&   -23$^{\rm c}$&         0.8&       &    c$^{\rm c}$&h      & h      &m      &h      &       &h      &h    &    & 1.11 &0.1                                               \\
   1422 &                          &    &               &               &-80.8$^{\rm e}$&                &            &       &    c$^{\rm e}$&       & h      &       &       &       &h      &h    &    & 1.53 &0.1                                               \\
   1423 &                          &    &  0.4$^{\rm c}$&               & -3.9$^{\rm e}$&    -4$^{\rm c}$&           0&       &    w$^{\rm c}$&h      & l      &       &l      &       &l      &l    &    &      &                                                  \\
   1424 &                          &    &  0.3$^{\rm e}$&               & -7.2$^{\rm e}$&                &            &       &    w$^{\rm e}$&h      & l      &       &       &       &l      &     &    &      &                                                  \\
   1425 &                          &    &  1.3$^{\rm f}$&  0.3$^{\rm c}$&  -23$^{\rm c}$&   -37$^{\rm c}$&      $<$0.1&       &    c$^{\rm c}$&h      & h      &       &m      &       &h      &l    &    & 2.21 &0.1                                               \\
   1426 &           -68.2$^{\rm c}$&    &               &               &   -9$^{\rm c}$&                &            &       &    w$^{\rm c}$&       & h      &l      &       &       &l      &h    &    &      &                                                  \\
   1427 &           -18.4$^{\rm c}$&    &  0.6$^{\rm c}$&               &   -4$^{\rm c}$&  -4.5$^{\rm e}$&           0&       &    w$^{\rm c}$&h      & l      &h      &l      &       &h      &h    &    & 0.39 &0.2                                               \\
   1428 &           -16.5$^{\rm c}$&    &  0.2$^{\rm c}$&               &  -20$^{\rm c}$&   -23$^{\rm c}$&            &       &    c$^{\rm c}$&m      & h      &h      &       &       &h      &l    &    & 1.52 &0.1                                               \\
   1429 &           -15.1$^{\rm c}$&    &  0.6$^{\rm c}$&               & -3.8$^{\rm e}$&    -4$^{\rm c}$&        0.06&       &    w$^{\rm c}$&h      & l      &h      &h      &       &l      &l    &    &      &                                                  \\
   1430 &                          &    &  0.8$^{\rm c}$&               &  -11$^{\rm c}$&                &           0&       & w(c)$^{\rm c}$&h      & h      &       &l      &       &l      &l    &    &      &                                                  \\
   1431 &                          &    &  0.7$^{\rm c}$&               &   -4$^{\rm c}$&    -8$^{\rm f}$&           0&       &    w$^{\rm c}$&h      & l      &       &l      &       &l      &h    &    & 0.02 &0.1                                               \\
   1432 &           -15.8$^{\rm c}$&    &  0.6$^{\rm f}$&  0.5$^{\rm c}$&   -2$^{\rm c}$&   -17$^{\rm c}$&    0.81-3.3&       &    c$^{\rm c}$&h      & h      &h      &h      &       &h      &h    &    & 3.15 &0.1                                               \\
   1433 &                          &    &  0.7$^{\rm c}$&               &  -17$^{\rm c}$&                &           0&       & w(c)$^{\rm c}$&h      & h      &       &l      &       &l      &h    &    &      &                                                  \\
   1434 &           -15.6$^{\rm c}$&    &  0.5$^{\rm c}$&               & -1.5$^{\rm e}$&    -2$^{\rm c}$&      $<$0.1&       &    w$^{\rm c}$&h      & l      &h      &m      &       &l      &h    &    &      &                                                  \\
   1435 &                          &    &               &               &  -13$^{\rm c}$&                &           0&       &w(c:)$^{\rm c}$&       & h      &       &l      &       &l      &l    &    & 0.11 &0.1                                               \\
   1436 &           -13.4$^{\rm c}$&    &  0.9$^{\rm f}$&  0.6$^{\rm c}$&  -13$^{\rm c}$&   -30$^{\rm c}$&        0.88&       &    c$^{\rm c}$&h      & h      &h      &h      &       &h      &h    &    & 0.69 &0.1                                               \\
   1437 &           -25.2$^{\rm c}$&    &               &               &               &                &            &       &    w$^{\rm c}$&       &        &m      &       &       &       &     &    &      &                                                  \\
   1438 &           -15.8$^{\rm c}$&    &  0.6$^{\rm e}$&               &  -10$^{\rm e}$&                &           0&       &    w$^{\rm c}$&h      & h      &h      &l      &       &l      &l    &    &      &                                                  \\
   1439 &           -15.7$^{\rm c}$&    &  0.6$^{\rm f}$&  0.4$^{\rm c}$&  -33$^{\rm c}$&   -37$^{\rm f}$&        0.21&       &    c$^{\rm c}$&h      & h      &h      &h      &       &h      &     &    & 0.24 &0.1                                               \\
   1440 &           -19.1$^{\rm c}$&    &  0.4$^{\rm c}$&               &   -2$^{\rm c}$&    -7$^{\rm f}$&           0&       &    w$^{\rm c}$&h      & l      &m      &l      &       &l      &l    &    &      &                                                  \\
   1441 &           -16.9$^{\rm c}$&    &  0.4$^{\rm c}$&               &   -8$^{\rm c}$& -11.3$^{\rm e}$&      $<$0.1&       &    c$^{\rm c}$&h      & h      &h      &m      &       &h      &h    &    & 0.77 &0.1                                               \\
   1442 &                          &    &  0.7$^{\rm c}$&               & -4.8$^{\rm e}$&    -5$^{\rm c}$&           0&       &    w$^{\rm c}$&h      & l      &       &l      &       &l      &l    &    & 1.13 &0.2                                               \\
   1443 &          -117.9$^{\rm c}$&    &               &               &   -4$^{\rm e}$&                &            &       &  w w$^{\rm c}$&       & h      &l      &       &       &l      &l    &    & 1.01 &0.2                                               \\
  \end{tabular}

\small
plus 1421 more stars in electronic table (at the end of the document: Table~\ref{supermemfull}).          %1496 stars with data in table

\footnotesize
Remarks: The literature sources and numbering are the same as in Table~\ref{supertab}, empty lines were omitted.
  The proper motion (PM) membership probability as it is given in {$[$j$]$}.
  If the literature gives more than one value for Li or H$\alpha$ equivalent width, the minimal and maximal values are given, otherwise the value is written in the maximum columns. 
  The mass accretion $\dot M$ is only from {$[$c$]$}, the corrected X-ray luminosity only from {$[$b$]$}.
  Column TTS indicates a classical (c) or a weak (w) T Tauri star. If an additional T Tauri state follows in parentheses, the classification differs between low and high resolution spectra (see source literature for more details), colons indicate uncertainty.    \newline
  The next to last column gives the re-calculated extinction as decribed in the text. %An anoted value of ``neg'' indicates negative extinction.
  The last column contains the masses determinded by the models by Siess et al. (\cite{sie00}) from the infrared color-magnitude diagram (Fig.~\ref{CMD-IR_ownAV-iso}).       \newline
  \textbf{The membership prediction:} h, m and l stand for high, medium and low membership probability, as a result of the following criteria:                                               \newline
  $\cdot$ Lithium absorption: see Table~\ref{tabli}. \newline
  $\cdot$ H$\alpha$ emission: if spectral type earlier than K0 and EW(H$\alpha$)$<0 \rightarrow$ h, if spectral type later than K0 we follow White \& Basri (\cite{whi03}) to distingish between h and l. \newline
  $\cdot$ radial velocity (RV): if within $1\sigma$ (3.6\,km/s) around $-15\,\rm{km/s} \rightarrow$ h, if within $3\sigma \rightarrow$ m, otherwise l. \newline
  $\cdot$ Accretion: if $\dot M>0.05\cdot10^{-8}$M$_\odot$/yr $\rightarrow$ h, if $\dot M>0\cdot10^{-8}$M$_\odot$/yr $\rightarrow$ m, if $\dot M=0\cdot10^{-8}$M$_\odot$/yr $\rightarrow$ l. \newline
  $\cdot$ X-ray: {$[$b$]$} analyzed only bright X-ray sources with corrected luminosity $L_{x,c}>0.75\cdot10^{30}$\,erg/s, so all $\rightarrow$ h.\newline
  $\cdot$ Infrared excess: if excess visible in SEDs from Sicilia-Aguilar et al. (\cite{sic06-1}), then h, otherwise l. \newline
  $\cdot$ Variability: if marked as ``$V$'' or ``\textit{RI}'' in the source literature $\rightarrow$ h, if ``$I$'' $\rightarrow$ m, if marked as ``N'' or ``No'' $\rightarrow$ l.  \newline
  $\cdot$ Proper motion: if $p\ge75$\% $\rightarrow$ h, if $p\ge50$\% $\rightarrow$ m, otherwise l.                    %\newline
  \label{supermem}
 \end{table*}

\section{Membership determination}

We established a three-level scale of probabilities of membership in Trumpler\,37: high (h), medium (m), and low (l) based on the following data from the literature: lithium absorption (from Sicilia-Aguilar et al. \cite{sic06-2, sic05, sic04}), H$\alpha$ emission (from  Sicilia-Aguilar et al. \cite{sic06-2, sic05, sic04}), radial velocity (RV) (from Sicilia-Aguilar et al. \cite{sic06-2}, Contreras et al. \cite{con02}), mass accretion on the star (from Sicilia-Aguilar et al. \cite{sic06-2}), X-ray luminosity (Mercer et al. \cite{mer09}) and variability (Morales-Calder{\'o}n et al. \cite{mor09}, Sicilia-Agui\-lar et al. \cite{sic05, sic04}). The following listing describes our considerations for membership determination in detail.

\begin{itemize}
 
\item Due to different temperatures, depth of convection \linebreak zones, rotation,accretion history, and close companions, the primordial lithium in the atmospheres of stars has different life times. We followed Fig.~6 from Neuh\"auser (\cite{neu97}) in our criteria for the equivalent widths (EW) as listed in Table~\ref{tabli}.

\item Accretion disks typically last for about 10 Myr (e.g. Jayawardhana et al. \cite{jay06}). Depending on its temperature the circumstellar dust emits from infrared (IR) to mm wavelengths. We have used the infrared excess\linebreak emission for constraining further membership probability. Consequently, whenever the spectral energy distribution given by Sicilia-Aguilar et al. (\cite{sic06-1}) displays infrared excess, we assigned a high membership probability for the corresponding star; otherwise, if IR data are available, but there is no apparent excess, we assigned a low probability; if no IR data are available, we did not assign a membership probability. % (e.g. TW Hya of the TW Hya association with an age of 5 to 12 Myr, Jayawardhana et al. \cite{jay06})

\item Accreting young stars show strong H$\alpha$ emission, well above the values expected from purely chromospheric activity in K and M type field dwarfs. If a star showed significant H$\alpha$ emission, above the locus for field dwarfs (White \& Basri \cite{whi03}), we assumed its likely a CTTS, and assigned it a high membership probability. If H$\alpha$ emission is weaker, we assigned low membership probability.

\item Young stars often exhibit dramatic changes in their \linebreak brightness, e.g. due to spots or accretion. We used the measured variability from time series analysis \linebreak per\-formed in the infrared (Morales-Calder{\'o}n et al. \linebreak \cite{mor09}) and in $R$ and $I$ band (Sicilia-Aguilar et al. \cite{sic04,sic05}). If significant variability is indicated both in $R$ and $I$, we assigned it high membership probability; if it is only variable in $R$, then we assigned medium membership probability; and otherwise we assigned it low membership probability.

\item Clusters and T associations are also moving groups \linebreak which allow us to use radial velocity (RV) and proper motion (PM) for membership analysis: if the RV is with\-in $1\,\sigma$ of the mean value, then we assigned high membership probability; if the value is between 1 and $3\,\sigma$ from the mean, we assigned it medium membership \linebreak probability; if it is more than 3 $\sigma$ from the mean, we assigned it low membership probability. The radial velocity determined by Sicilia-Aguilar et al. (\cite{sic06-2}) is $-15\pm3.6$\,km/s, using high resolution spectra ($R\sim 34000$) for the known cluster members. The cluster distance of 870\,pc, however, implies small proper motion of the stars, so that it is difficult to exclude background stars. We adopt proper-motion membership probabilities based on the work of Marschall \& van Altena \linebreak (\cite{mar87}).

\item Strong X-ray detection also implies high membership probability. The X-ray observation by Mercer et al. \linebreak (\cite{mer09}) extracted only bright sources, so all those stars got high membership probability. %but they donated some stars as N, if below 100\,Myr isochrone in $V$ vs $V$-$I$ CMD
\end{itemize}

\begin{table}
\caption{Membership probability using the lithium absorption}
\label{tabli}
\begin{tabular}{rccccc}
Spectral type earlier& \multicolumn{3}{c}{EW(Li) [\AA] for}\\ 
than or equal to  & h & m & l \\
\hline
G3     & $>0.15$  & 0.15-0.05 & $<0.05$   \\
G8     & $>0.2$   &  0.2-0.1  & $<0.1$    \\
%K3    & $>0.3$   &  0.3-0.2  & $<0.2$    \\
K7     & $>0.3$   &  0.3-0.2  & $<0.2$    \\
M4     & $>0.2$   &  0.2-0.1  & $<0.1$    \\
M9     & $>0.15$  & 0.15-0.1  & $<0.1$    \\
\end{tabular}
\end{table}

% Marschall:
% RA:
% a               = 1921.12          +/- 50.97        (2.653%)
% m               = -4.69004         +/- 0.1152       (2.457%)
% s               = 3.77494          +/- 0.1165       (3.086%)
% Dec:
% a               = 1968.71          +/- 52.25        (2.654%)
% m               = -2.18353         +/- 0.1346       (6.165%)
% s               = 4.39103          +/- 0.1345       (3.063%)
% Li+Ha:
% RA:
% a               = 531.074          +/- 22.41        (4.219%)
% m               = -3.29901         +/- 0.2028       (6.148%)
% s               = 4.16705          +/- 0.2034       (4.882%)
% Dec:
% a               = 551.252          +/- 29.69        (5.385%)
% m               = -5.6221          +/- 0.2834       (5.042%)
% s               = 4.55835          +/- 0.2835       (6.22%)

\section{Analysis}

\begin{figure}
 \includegraphics[width=8.3cm]{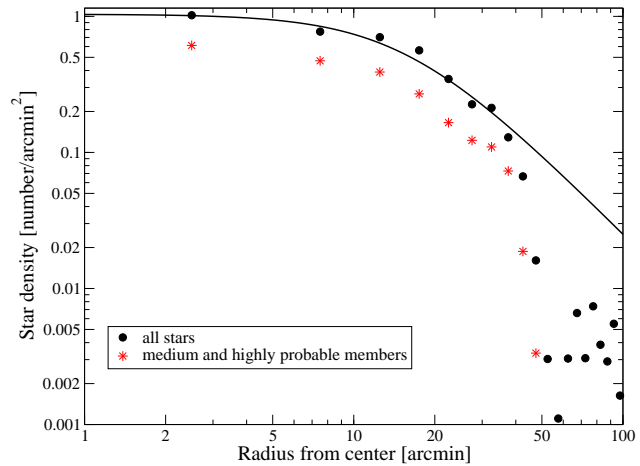}
 \caption{The radial surface density with center at 21:39:06, 57:30:00.}
 \label{stardens}
\end{figure}
%f(x)=s/(1+(x/r)**2); fit f(x) "King_profile_density-radius" using 1:2 via s,r ; plot "King_profile_density-radius",f(x)
%s               = 1.03707          +/- 0.06014      (5.799%)
%r               = 15.7225          +/- 1.32         (8.396%)

The radial surface density of Trumpler\,37 is shown in Fig.~\ref{stardens}. The medium and highly probable members are distributed uniformly. We fitted the King model of form $$\sigma=\frac{\sigma_0}{1+(r/r_c)^2}$$ with the parameters core density $\sigma_0= 1.037\pm\linebreak0.060$\,stars/arcmin$^2$ and core radius $r_c=15.7\pm1.3$\,arcmin.

\begin{figure*}
 \includegraphics[width=8.3cm]{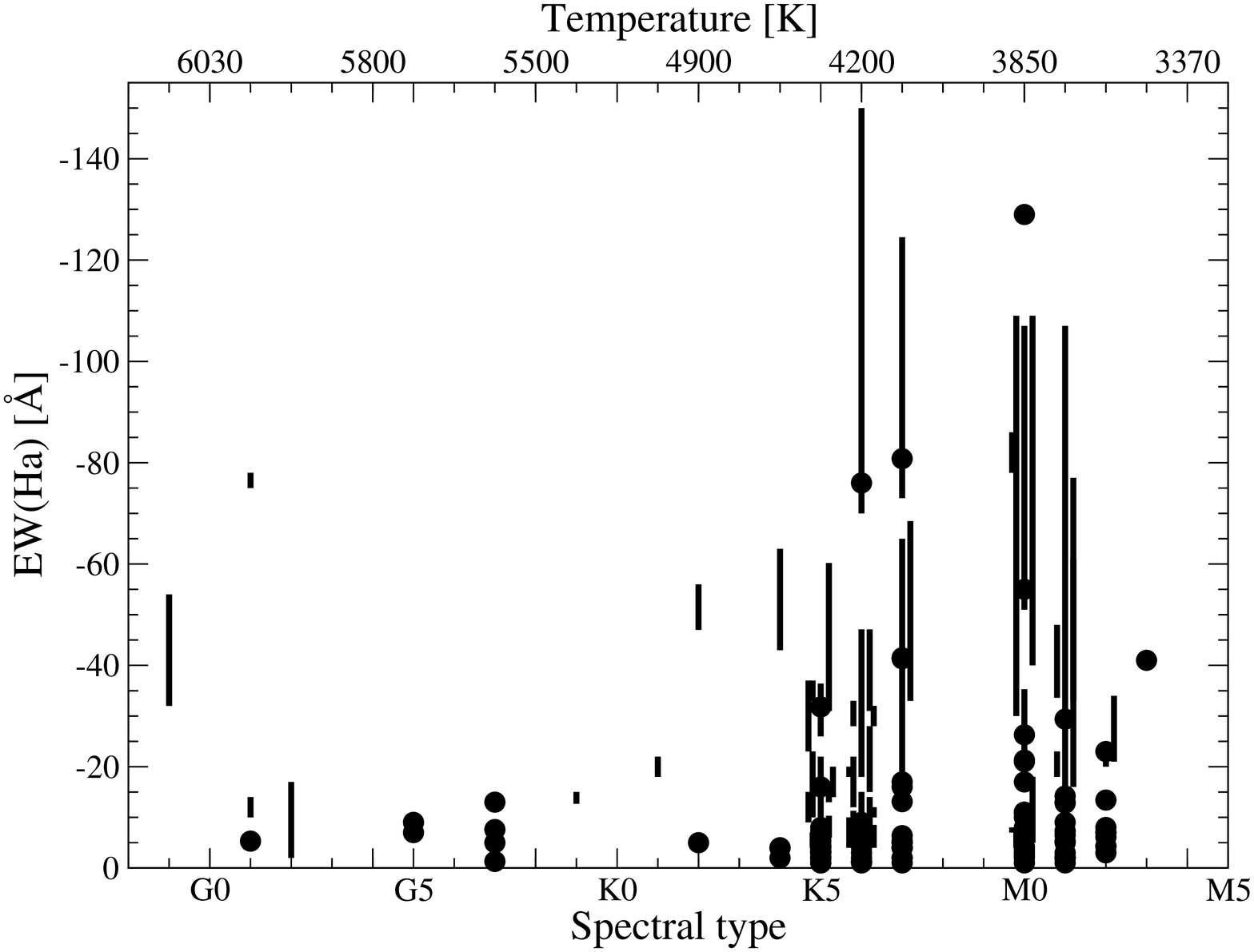}
 \includegraphics[width=8.3cm]{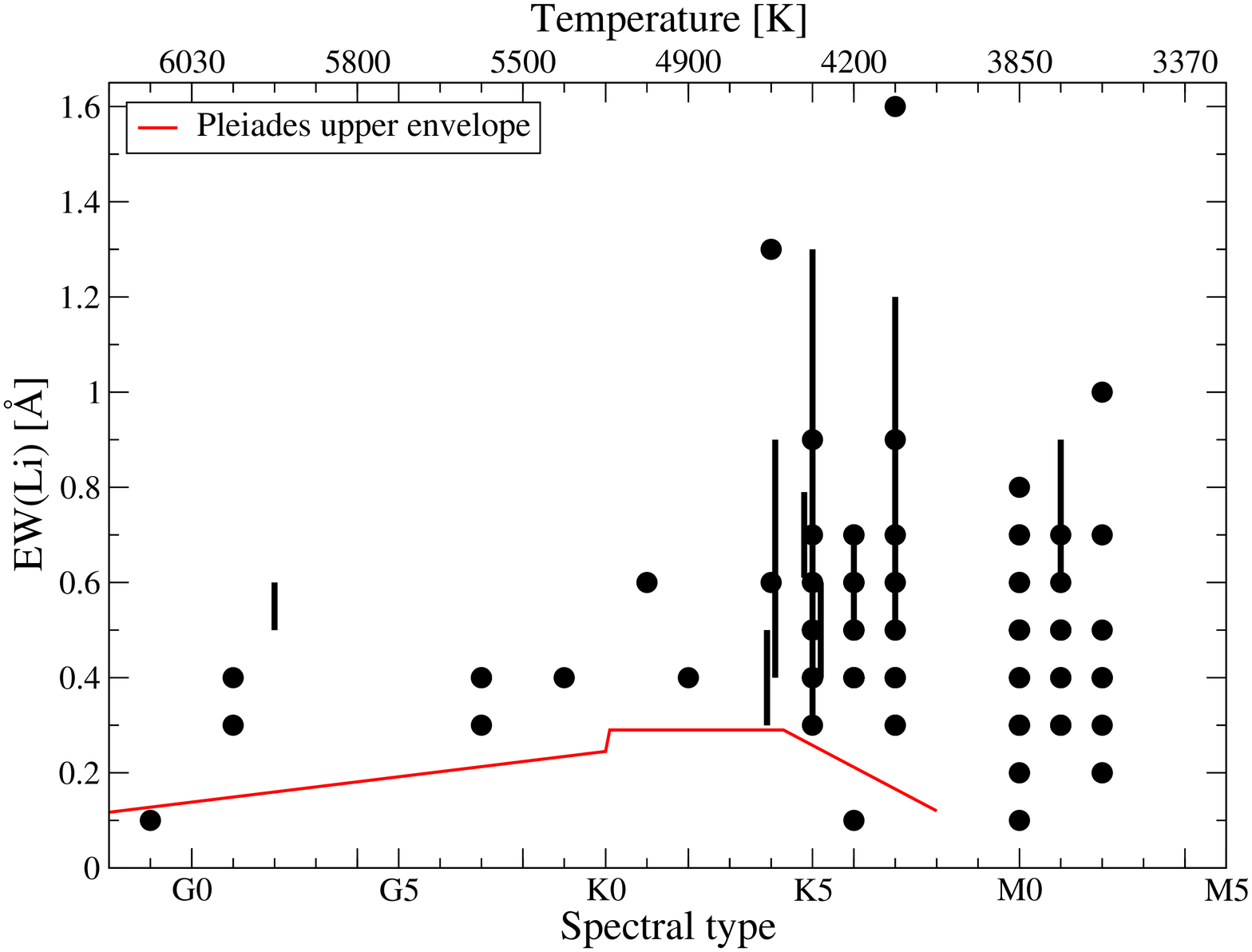}
 \caption{The equivalent widths (EW) of H$\alpha$ and lithium, depending on spectral type. Both values have been taken from the literature. If different EW are available, the range is given, otherwise only a dot. For lithium, the upper envelope of the Pleiades values (Soderblom et al. \cite{sod93}) is also given.}
 \label{EWvsSpT}
\end{figure*}

The equivalent widths for lithium and H$\alpha$ are plotted against the spectral type in Fig.~\ref{EWvsSpT}. All values are used as given in Table~\ref{supermem}.

\begin{figure*}
 \includegraphics[width=8.3cm]{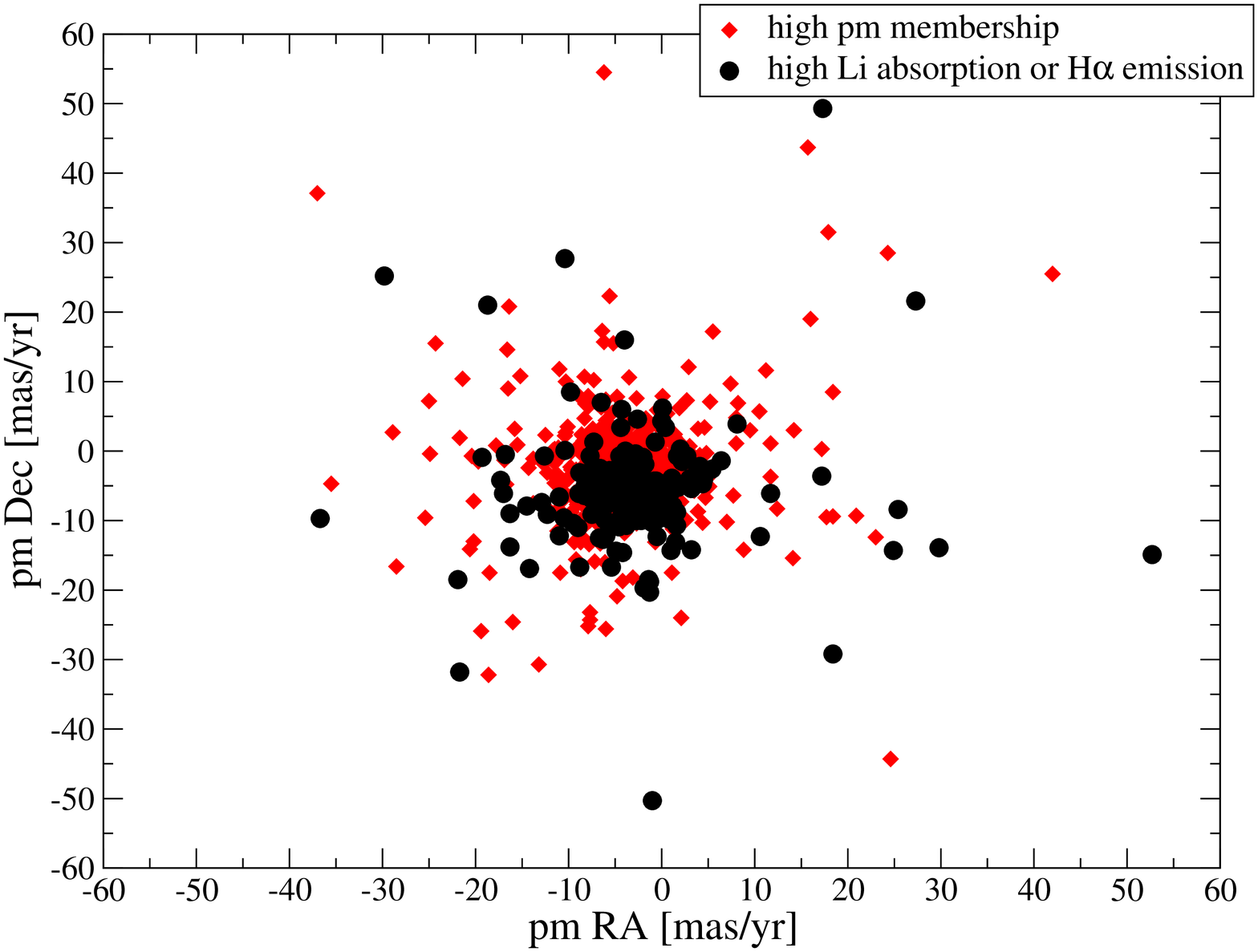}
 \includegraphics[width=8.3cm]{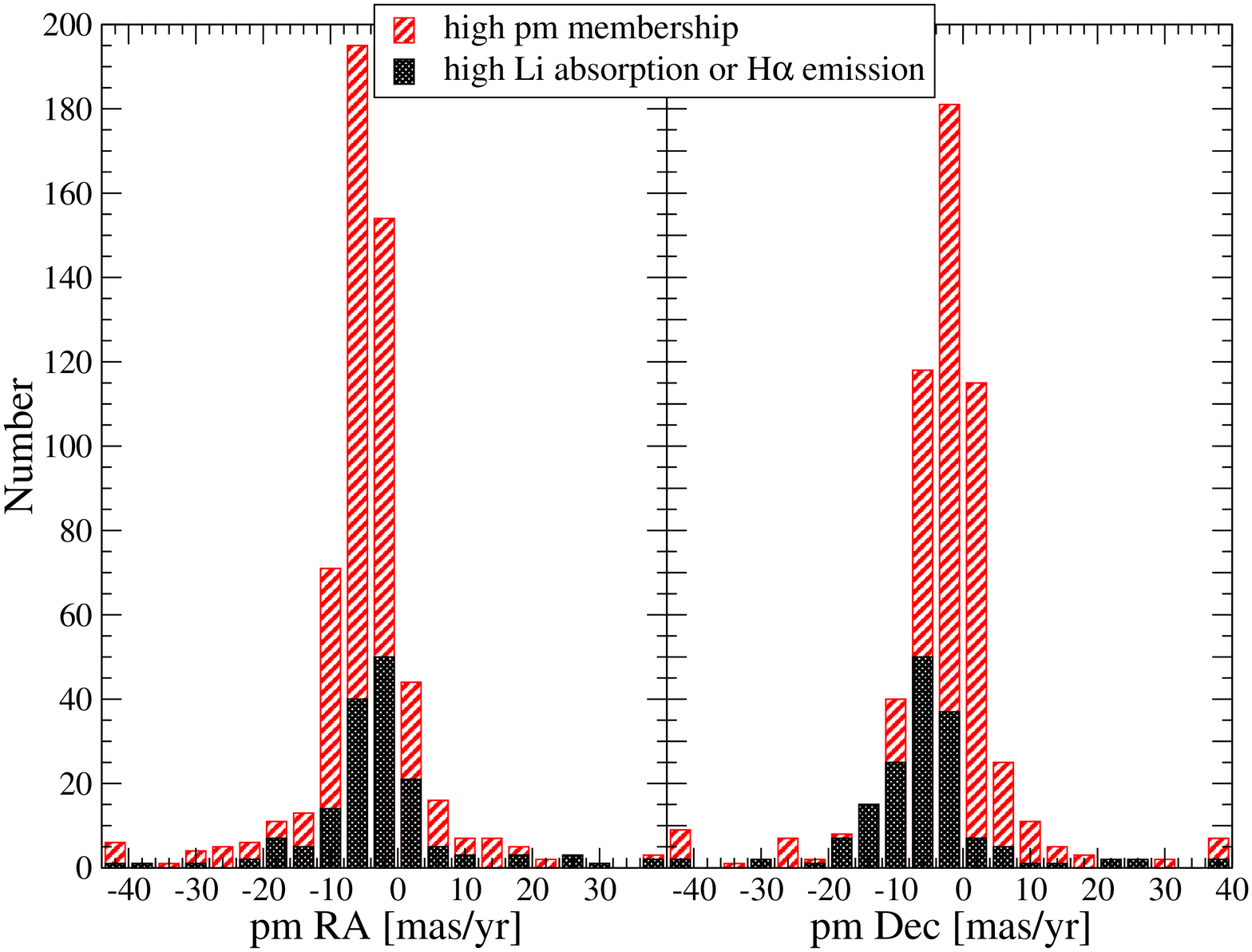}
 \caption{The PPMXL proper motion as 2 dimensional distribution and in the histogram form. The black circles and the black histograms correspond to the high probable member stars from lithium absorption or H$\alpha$ emission, the red diamonds and the red, shaded histograms to the high probable member stars from Marschall \& van Altena \cite{mar87}. Stars outside $\pm$40\,km/s are binned together.}%The distributions are similar spread.}
 \label{PMppmxl}
\end{figure*}

The distribution of the PPMXL proper motion of two subsamples of Trumpler\,37 is shown in Fig.~\ref{PMppmxl}: the proper motion analysis from Marschall \& van Altena (\cite{mar87}) investigated brighter stars, while the search for lithium absorption and H$\alpha$ emission is much more sensitive to the late-type spectral types and therefore to fainter stars. 
The distributions in Fig.~\ref{PMppmxl} are similar. Table~\ref{tabPMdistri} gives the parameters of their fitting with Gaussian %$$N(\mathrm{RV})\propto \frac{1}{\sqrt{2\pi} \sigma} e^{\frac{-(\mathrm{RV}-m)^2}{2\sigma^2}}$$ 
(mean $m$ for the center and width $\sigma$ of the histogram).

\begin{table}
% \centering%%%
\caption{Fit of the proper motion distribution with a Gaussian (mean $m$ and width $\sigma$) for two different star samples (Li: stars with lithium absorption, PM: stars with high membership probability from Marschall \& van Altena (\cite{mar87})).}
\label{tabPMdistri}
\begin{tabular}{r|cc|cc}%\hline
 & \multicolumn{2}{c}{RA} & \multicolumn{2}{c}{DEC}   \\
 & mean $m$ & width $\sigma$ & mean $m$ & width $\sigma$ \\ 
Sample & [mas/yr] & [mas/yr] & [mas/yr] & [mas/yr]  \\
\hline
Li \& H$\alpha$ & -3.30 (20) & 4.17 (20) & -5.62 (28) & 4.56 (28) \\
PM      & -4.69 (12) & 3.77 (12) & -2.18 (13) & 4.39 (13) \\
\end{tabular}
\end{table}

\begin{figure}
 \includegraphics[width=8.3cm]{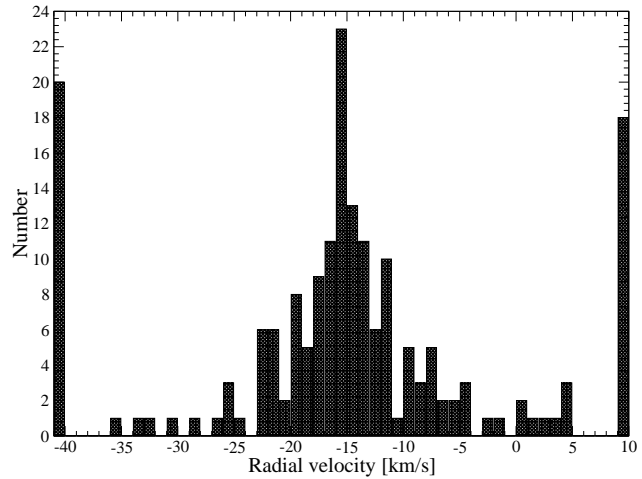}
 \caption{The radial velocity distribution for all stars in Table~\ref{supermem}. Values outside -40\,km/s and 9\,km/s are binned together, resulting in the strong signals at the edges.}
 \label{RVhist}
\end{figure}

The radial velocity distribution of all stars is plotted in Fig.~\ref{RVhist}. For this purpose data from Sicilia-Aguilar et al. (\cite{sic06-2}) and Contreras et al. (\cite{con02}) were used. We fitted the radial velocity distribution by Gaussian with parameters: center at $−15.3$\,km/s and width of $3.6$\,km/s. They are almost the same as those obtained by Sicilia-Aguilar et al. (\cite{sic06-2}): $-15.0$\,km/s and $3.6$\,km/s.

\begin{figure}
 \includegraphics[width=8.3cm]{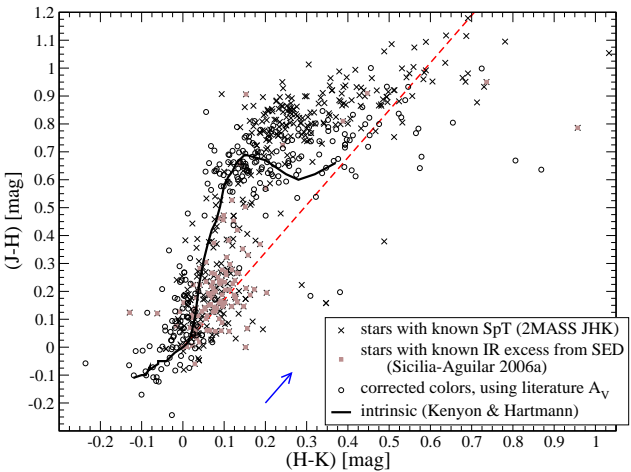}
 \caption{The color-color-diagram from the 2MASS magnitudes. % This figure represents ($J$-$H$) versus ($H$-$K$) for the x and 
  Additionally, stars with known infrared excess are marked with grey squares. The blue arrow shows the reddening vector of 1\,mag (Rieke \& Lebofsky \cite{rie85}). The intrinsic colors of the main sequence are from Kenyon \& Hartmann (\cite{ken95}). Stars to the lower right of the diagonal (dashed, red line) have circumstellar excess. The open circles show the correction with the literature extinction.}
 \label{new_AV}
\end{figure}

\begin{figure}
 \includegraphics[width=8.3cm]{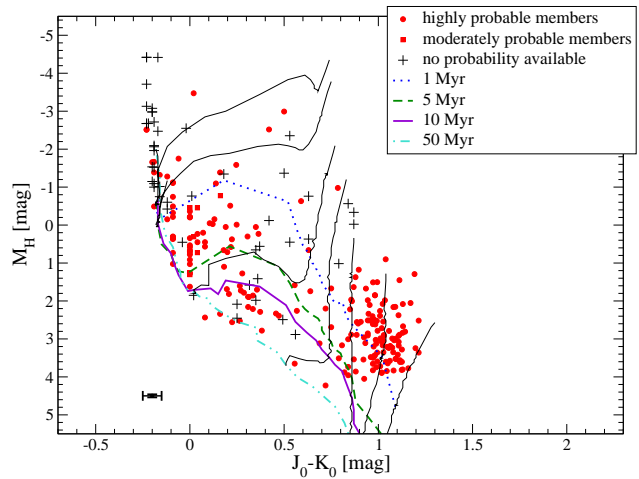}
 \caption{Dereddened infrared color-magnitude diagram: The 2MASS magnitudes were adjusted for the excess and our fitted extinction (using the interstellar extinction law from Rieke \& Lebofsky \cite{rie85}). The absolute $H$ brightness was calculated for the cluster distance of 870\,pc. Additional model data from Siess et al. (\cite{sie00}) are included: the 1, 5, 10, and 50\,Myr isochrones and the evolutionary tracks for 0.1, 0.2, 0.5, 1, 2, 5, and 7\,M$_\odot$. The mean error is shown in the lower left.}
 \label{CMD-IR_ownAV-iso}
\end{figure}

For the stars with known spectral types, we re-calculated the extinction by means of the infrared color-color diagram (Fig.~\ref{new_AV}), using 2MASS \textit{JHK}. We corrected the excess from circumstellar dust. 
We used the law from Rieke \& Lebofsky (\cite{rie85}) and intrinsic colors from Kenyon \& Hartmann (\cite{ken95}). For 103 out of 423 stars we got  unphysical extinctions values ($A_{\mathrm{V}}<0$). Stars with $A_{\mathrm{V}}<0$ or lying left of the main sequence could be variable stars. %
Using a $3\,\sigma$ threshold, 80 stars show infrared excess.
The median and average extinctions of the cluster members are correspondingly 0.9 and 1.2\,mag.  The extinctions are listed in Table~\ref{supermem}.
The open circles in Fig.~\ref{new_AV} show the correction done only with the literature extinction,  resulting in big deviations \linebreak from the intrinsic colors.

The color-magnitude diagram (Fig.~\ref{CMD-IR_ownAV-iso}) was created with the derived extinction. The 2MASS photometry was corrected for distance, excess and extinction, meaning all stars with known spectral type and re-calculated extinction are plotted. Assuming the previously derived distance of 870\,pc, our results are consistent with an age younger than 10\,Myr. Only 12\% of the stars lie below the 5\,Myr isochrone.

The masses of these young stars were determined using the corrected 2MASS magnitudes and the theoretical tracks from Siess, Dufour \& Forestini (\cite{sie00}). The masses are listed in the last column of Table~\ref{supermem}. 

\begin{figure}
 \includegraphics[width=8.3cm]{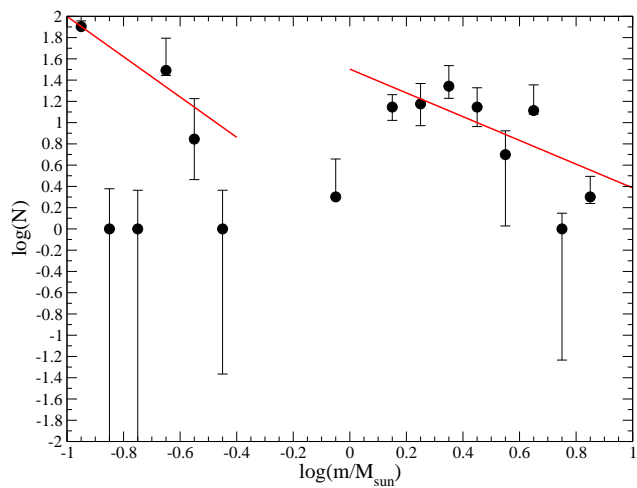}
 \caption{Initial mass function (IMF) of the Trumpler\,37 high and medium member stars with known spectral type in the literature. The points represent bins in the masses from Table~\ref{supermem}, with a width of 0.1 on a log scale. The IMF was fitted with the power-law index $\alpha=1.12\pm0.37$ in the range of $0.0<\log(m/M_\odot)<1.0$ and with $\alpha=1.90\pm0.44$ in the range of $-1.0<\log(m/M_\odot)<-0.4$.}
 \label{IMF}
\end{figure}

We could plot an initial mass function of these masses (Fig.~\ref{IMF}). We fitted the power-law index $\alpha$ from equation $\mathrm{d}N=k\cdot m^{-\alpha}\mathrm{d}m$, with constant $k$, following the typical zoning with changes of $\alpha$ at $0.08$ and $0.5\,M_\odot$. We skipped the obviously incomplete mass regime of $0.5-0.8\,M_\odot$ for the fit and therefore also no continuity was applied. 
We found, comparing to Kroupa (\cite{kro07}), a higher value of $\alpha=1.90\pm0.44$ ($0.1-0.4\,M_\odot$) and an unusual low value of $\alpha=1.12\pm0.37$ ($1-10\,M_\odot$).
It indicates that our sample may not be complete at the intermediate mass regime.

\section{Results}
We found data for 1872 different stars which were studied in the context of Trumpler\,37; membership was investigated for 1402 stars. Of these, 774 have a high membership probability in terms of at least one criterion; 125 stars have a medium, and 503 stars a low probability of being member of Trumpler\,37.
We re-calculated the extinction. Our color-magnitude diagram is consistent with the best values for the age in the literature of 3-5\,Myr and distance of $\sim870$\,pc.
The unusual power-law index demands a search for more cluster members.

In the upcoming papers we will present planetary transit candidates and other results from variability studies. In future work we will also try to improve the knowledge of the cluster properties by homogeneous photometric and spectroscopic analysis including narrow-band photometry.

\acknowledgements
This publication makes use of data products from the Two Micron All Sky Survey, which is a joint project of the University of Massachusetts and the Infrared Processing and Analysis Center/California Institute of Technology, funded by the National Aeronautics and Space Administration and the National Science Foundation. This research has made use of the WEBDA database, operated at the Institute for Astronomy of the University of Vienna. This research has made use of the VizieR catalog access tool, CDS, Strasbourg, France.

RN and RE would like to thank DFG for support in the Priority Programme SPP 1385 on the {\em First ten Million years of the Solar System} in project NE 515 / 34-1.
RE also thanks the Abbe-School of Photonics for support.
We would like to acknowledge financial support from the Thuringian government (B 515-07010) for the STK CCD camera used in this project.
TP, MV and JB thank for the support to the projects APVV-0158-11 and VEGA 2/0094/11.
MM acknowledges DFG for support in program MU2695/13-1. The research was supported partly by funds of projects DO~02-85 and DDVU 02/40-2010 of the Bulgarian Scientific Foundation. E.L.N.J. and D.H.C. gratefully acknowledge the support of the National Science Foundation's PREST program, which helped to establish the Peter van de Kamp Observatory through grant AST-0721386, and of the Provost's Office of Swarthmore College for their support maintaining and operating the observatory.

\newpage%%%%%%%%%%%%%%%%%%%%%%%%%%%%%%%%%%%%%%%%%%%%%%%%%%%%%%

\appendix
\section{Full tables}

\begin{landscape}
\begin{@twocolumnfalse}
%\caption{Literature data for stars in Trumpler 37}
  %\small
  %\footnotesize
  \scriptsize
 \tablefirsthead{\multicolumn{10}{l}{\bf Table~\ref{supertabfull} Literature data for stars in Trumpler 37}\\
 No. & RA          & Dec        & MVA  & WEB-      &  SHB-         &$U$    &$B$    & $V$   &$R$    &$I$    &$J$   &$H$   &$K$   & SpT      & Class &$A_{\mathrm{V}}$&$\mu_\alpha$  & $\mu_\delta$ & $\mu_\alpha$ & $\mu_\delta$ & $\mu_\alpha$ & $\mu_\delta$ &  Comments \\
     & \multicolumn{2}{c}{J2000}&      & DA        &  2004         &       &       &       &       &       &      &      &      &          &                        &     & \multicolumn{2}{l}{PPMXL}   & \multicolumn{2}{l}{UCAC3}   & \multicolumn{2}{l}{MVA [j]} &  \\
     & hh:mm:ss.ss & dd:mm:ss.s &      &           &               &  mag  &  mag  &  mag  &  mag  &  mag  &  mag & mag  & mag  &          &                        & mag & mas/yr       & mas/yr       & mas/yr       & mas/yr       & mas/yr       & mas/yr       &  \\
   \hline}
 \tablehead{\multicolumn{10}{l}{\bf Table~\ref{supertabfull} Literature data for stars in Trumpler 37 -- continued}\\
 No. & RA          & Dec        & MVA  & WEB-      &  SHB-         &$U$    &$B$    & $V$   &$R$    &$I$    &$J$   &$H$   &$K$   & SpT      & Class &$A_{\mathrm{V}}$&$\mu_\alpha$  & $\mu_\delta$ & $\mu_\alpha$ & $\mu_\delta$ & $\mu_\alpha$ & $\mu_\delta$ &  Comments \\
     & \multicolumn{2}{c}{J2000}&      & DA        &  2004         &       &       &       &       &       &      &      &      &          &                        &     & \multicolumn{2}{l}{PPMXL}   & \multicolumn{2}{l}{UCAC3}   & \multicolumn{2}{l}{MVA [j]} &  \\
     & hh:mm:ss.ss & dd:mm:ss.s &      &           &               &  mag  &  mag  &  mag  &  mag  &  mag  &  mag & mag  & mag  &          &                        & mag & mas/yr       & mas/yr       & mas/yr       & mas/yr       & mas/yr       & mas/yr       &  \\
   \hline}
 \bottomcaption{\small Literature data for stars in Trumpler 37                                               \newline
Remarks: The superscript letters behind the values indicate the source for the value:                                     \newline
{\bf$[$a$]$}~Morales-Calder{\'o}n et al. (\cite{mor09}); {\bf$[$b$]$}~Mercer et al. (\cite{mer09}); {\bf$[$c$]$}~Sicilia-Aguilar et al. (\cite{sic06-2}); {\bf$[$d$]$}~Sicilia-Aguilar et al. (\cite{sic06-1}); {\bf$[$e$]$}~Sicilia-Aguilar et al. (\cite{sic05}); {\bf$[$f$]$}~Sicilia-Aguilar et al. (\cite{sic04}); {\bf$[$g$]$}~WEBDA (consists of Sicilia-Aguilar  et al. (\cite{sic04}) and Morbidelli et al. (\cite{morb97}); {\bf$[$h$]$}~Contreras et al. (\cite{con02}) (used for photometry Marschall et al. (\cite{mar90})); {\bf$[$i$]$}~Marschall et al. (\cite{mar90}); {\bf$[$j$]$}~Marschall \& van Altena (\cite{mar87}) ($V$ magnitudes from fitting instrumental magnitudes to photometry from Garrison \& Kormendy (\cite{gar76}) and de Lichtbuer (\cite{lic82})); {\bf$[$k$]$}~Kun (\cite{kun86}); 
{\bf$[$l$]$}~WEBDA (consists of Marschall et al. (\cite{mar90}), Garrison \& Kormendy (\cite{gar76}), Simonson (\cite{sim68}) and other publications for few stars); {\bf$[$m$]$}~WEBDA (coordinate source); 
{\bf$[$n$]$}~WEBDA (consists of Marschall \& van Altena (\cite{mar87}) and internal WEBDA information); {\bf$[$o$]$}~WEBDA (consists of 6 publications for 7 stars); {\bf$[$p$]$}~WEBDA (consists of Garrison \& Kormendy (\cite{gar76}) and other publications for few stars); {\bf$[$q$]$}~WEBDA (consists of Alkansis (\cite{alk58}), Contreras et al. (\cite{con02}), Sicilia-Aguilar et al. (\cite{sic04}), Balazs et al. (\cite{bal96}) and other publication for few stars); {\bf$[$r$]$}~2MASS (Skrutskie et al. \cite{skr06}). The different WEBDA tables were compiled from different literature, the main publications are given in brackets\newline
MVA, WEBDA and SHB-2004 are star numbers in papers {$[$j$]$}; {$[$l$]$}-{$[$q$]$}; and {$[$c$]$}-{$[$f$]$}, {$[$h$]$}, respectively.
If data from different literature are available, the more recent one is given. Please note, that the $V$ magnitude was measured from photographic plate, photoelectrical or with CCD, making comparison difficult. The source for $R$ and $I$ magnitude is the same (given after $I$) and the source for $J$, $H$ and $K$ magnitude is the same (given after $K$). Errors in \textit{JHK}-photometry are given only, if the 2MASS quality flag is ``A'', ``B'', ``C'' or ``D'', otherwise an empty parenthesis indicates uncertainties in the 2MASS photometry.        \newline
{\bf Comments}: If two stars were located close to each other ($<5^{\prime\prime}$), the stars were marked with ``near \#''.  ``no star'' or ``no/faint star'' means we were not able to find the star from Marschall \& van Altena (\cite{mar87}) in our images (see also the text).
``new coordinates'' means, we changed the coordinates from Marschall \& van Altena (\cite{mar87}) to match the position that was given in their finding chart (see also text).
In cases of infrared data (Sicilia-Aguilar et al. \cite{sic06-1}), we were not able to see some stars in our optical images, resulting in comments ``no opt. cp.'' or ``very faint opt. cp.'' (opt. cp. standing for optical counterpart).
Because Sicilia-Aguilar et al. (\cite{sic04}) used the earlier compilation of the 2MASS catalog (Cutrie et al. \cite{cut03}) some stars get the comment ``\textit{JHK} in {$[$f$]$} different''.
In case of two not distinguishable 2MASS sources near the star, the entry was duplicated in the consecutive row, so both sources were connected. The comment ``2x{$[$r$]$}'' was added and the fainter one marked. Probably, the other data from the literature, like optical brightness, is not resolved in this case.
In Marschall \& van Altena (\cite{mar87}) and the WEBDA database stars outside all YETI telescope fields of view (FoV) are marked with ``outFoV''.
In some cases stars with the same names (and properties) differ in the coordinates in different catalogs. The more reliable coordinate was used and in the comments ``Dec {$[$h$]$} imprec.'' or ``{$[$m$]$} imprec.'' was attached, meaning that problems occurred in Contreras et al. \cite{con02} or the WEBDA database. In some entries the WEBDA entries were even wrong, resulting in ``WEBDA wrong''. \newline
Spectroscopic binaries were marked with ``SB1'' or SB2'' as given in Sicilia-Aguilar et al. (\cite{sic06-2}).
 }
   \begin{mpsupertabular}{l @{\hspace{1.5mm}} l @{\hspace{1.5mm}}l @{\hspace{1.5mm}} l @{\hspace{1.5mm}} l @{\hspace{1.5mm}} l @{\hspace{1.5mm}} l @{\hspace{1.5mm}} l @{\hspace{1.5mm}} l @{\hspace{1.5mm}} l@{\hspace{1.5mm}} l @{\hspace{1.5mm}} l@{\hspace{1.5mm}}l@{\hspace{1.5mm}}l @{\hspace{1.5mm}} l @{\hspace{0.3mm}} l @{\hspace{1.5mm}} l @{\hspace{1.5mm}}  l@{\hspace{1.5mm}}l @{\hspace{1.5mm}} l@{\hspace{1.5mm}}l @{\hspace{1.5mm}} l@{\hspace{1.5mm}} l@{\hspace{1.5mm}} l}                                                                                                                                                                                                                                                                               
%No. & RA          & Dec                  & MVA  & WEBDA                     &  SHB                          & U               &  B              & V               &   R   &  I              & J           & H           & K                     & SpT                & Class           & A_V           & $\mu_\alpha$ & $\mu_\delta$ & $\mu_\alpha$ & $\mu_\delta$ &mu_a & mu_b&    Comments \\
1    & 21:36:46.57 &  57:11:25.4$^{\rm r}$& 2    & 3002                      &                               &                 &                 &   14.7$^{\rm j}$&       &                 &11.532\,(24) &10.787\,(28) & 10.583\,(20)$^{\rm r}$&                    &                 &               &-3.2\,(4.1)   &-2.4\,(4.1)   &-11.3\,(6.8)  &0.1\,(6.8)    &     &     &                                                                                      \\
2    & 21:36:44.78 &  57:11:53.0$^{\rm r}$& 3    & 3003                      &                               &                 &                 &   13.9$^{\rm j}$&       &                 &12.647\,(27) &12.547\,(37) & 12.452\,(29)$^{\rm r}$&                    &                 &               &-10.4\,(4.1)  &2.2\,(4.1)    &-30.9\,(6.8)  &12.1\,(6.8)   &-0.17&0.05 &                                                                                      \\
3    & 21:36:42.64 &  57:13:01.0$^{\rm r}$& 4    & 3004                      &                               &                 &                 &   13.3$^{\rm j}$&       &                 &11.880\,(27) &11.530\,()   &   11.445\,()$^{\rm r}$&                    &                 &               &              &              &              &              &3.15 &-4.9 &                                                                                      \\
4    & 21:36:20.30 &  57:12:55.9$^{\rm r}$& 5    & 3005                      &                               &                 &                 &   13.6$^{\rm j}$&       &                 &11.562\,(26) &11.227\,(28) & 11.101\,(20)$^{\rm r}$&                    &                 &               &-8.2\,(4.1)   &1.9\,(4.1)    &-11.9\,(6.8)  &6.6\,(6.8)    &0.13 &-0.23&                                                                                      \\
5    & 21:36:29.82 &  57:12:48.0$^{\rm j}$& 6    & 3006                      &                               &                 &                 &   14.6$^{\rm j}$&       &                 &             &             &                       &                    &                 &               &              &              &              &              &     &     &no/faint star                                                               \\
6    & 21:36:41.77 &  57:13:40.8$^{\rm r}$& 7    & 3007                      &                               &                 &                 &   13.9$^{\rm j}$&       &                 &10.779\,(26) &10.051\,(27) & 9.849 \,(23)$^{\rm r}$&                    &                 &               &-0.3\,(5.1)   &5.3\,(5.1)    &-8.2\,(7.1)   &41.1\,(7.2)   &0.01 &0.75 &                                                                                      \\
7    & 21:36:40.66 &  57:13:39.2$^{\rm r}$& 8    & 3008                      &                               &                 &                 &     15$^{\rm j}$&       &                 &12.718\,(28) &12.333\,(41) & 12.227\,(29)$^{\rm r}$&                    &                 &               &-19.4\,(4.1)  &-6.3\,(4.1)   &-70.6\,(7.1)  &-5.5\,(7.1)   &     &     &                                                                                      \\
8    & 21:36:46.12 &  57:12:53.3$^{\rm r}$& 9    & 3009                      &                               &                 &                 &   14.8$^{\rm j}$&       &                 &12.685\,(26) &12.395\,(30) & 12.256\,(25)$^{\rm r}$&                    &                 &               &-3.9\,(4.1)   &-10.2\,(4.1)  &-6.6\,(6.8)   &-29.3\,(6.8)  &0.1  &-0.13&                                                                                      \\
9    & 21:36:47.04 &  57:13:01.7$^{\rm r}$& 10   & 3010                      &                               &                 &                 &   14.5$^{\rm j}$&       &                 &12.718\,(26) &12.458\,(31) & 12.361\,(25)$^{\rm r}$&                    &                 &               &2.7\,(4.1)    &7.3\,(4.1)    &5.2\,(7.6)    &13.9\,(7.6)   &-0.12&0.26 &                                                                                      \\
10   & 21:36:50.76 &  57:12:41.4$^{\rm r}$& 11   & 3011                      &                               &                 &                 &   14.9$^{\rm j}$&       &                 &12.423\,(31) &12.017\,(32) & 11.878\,(25)$^{\rm r}$&                    &                 &               &-13\,(4.1)    &6.5\,(4.1)    &-23.2\,(6.9)  &28.7\,(7)     &     &     &                                                                                      \\
11   & 21:36:27.84 &  57:14:05.7$^{\rm r}$& 12   & 3012                      &                               &                 &                 &   14.9$^{\rm j}$&       &                 &11.658\,(24) &11.039\,(27) & 10.860\,(21)$^{\rm r}$&                    &                 &               &-37.1\,(4.1)  &-55.6\,(4.1)  &-32.8\,(6.8)  &-50.7\,(6.8)  &     &     &                                                                                      \\
12   & 21:36:32.90 &  57:14:20.1$^{\rm r}$& 13   & 3013                      &                               &                 &                 &   13.6$^{\rm j}$&       &                 &12.133\,(28) &11.785\,(33) & 11.716\,(28)$^{\rm r}$&                    &                 &               &-8.9\,(4.1)   &-18.9\,(4.1)  &1.6\,(7)      &-59.8\,(7)    &0.65 &-0.37&                                                                                      \\
13   & 21:36:32.90 &  57:14:52.2$^{\rm r}$& 14   & 3014                      &                               &                 &                 &   13.4$^{\rm j}$&       &                 &11.662\,(26) &11.384\,(28) & 11.279\,(23)$^{\rm r}$&                    &                 &               &-6\,(4.1)     &0.4\,(4.1)    &-10.8\,(6.8)  &17.5\,(6.8)   &0.22 &-0.28&                                                                                      \\
14   & 21:36:55.07 &  57:15:23.6$^{\rm r}$& 15   & 3015                      &                               &                 &                 &   13.8$^{\rm j}$&       &                 &12.213\,(22) &11.835\,(28) & 11.769\,(21)$^{\rm r}$&                    &                 &               &-0.1\,(4.1)   &11.4\,(4.1)   &-1.3\,(6.8)   &11.3\,(6.8)   &-0.85&1.14 &                                                                                      \\
15   & 21:36:55.96 &  57:13:39.7$^{\rm r}$& 16   & 3016                      &                               &                 &                 &   14.8$^{\rm j}$&       &                 &10.627\,(24) &9.679 \,(28) & 9.406 \,(21)$^{\rm r}$&                    &                 &               &-3.6\,(5.1)   &-2.3\,(5.1)   &-13.7\,(6.9)  &1.6\,(6.9)    &0.32 &0.17 &                                                                                      \\
16   & 21:36:58.46 &  57:13:46.0$^{\rm r}$& 17   & 3017                      &                               &                 &                 &     15$^{\rm j}$&       &                 &12.959\,(22) &12.565\,(28) & 12.481\,(25)$^{\rm r}$&                    &                 &               &-3.4\,(4.1)   &9.5\,(4.1)    &0\,(6.8)      &7.7\,(6.8)    &     &     &                                                                                      \\
17   & 21:36:46.85 &  57:17:11.5$^{\rm r}$& 18   & 3018                      &                               &                 &                 &     15$^{\rm j}$&       &                 &12.827\,(27) &12.484\,(33) & 12.370\,(28)$^{\rm r}$&                    &                 &               &-2.6\,(4.1)   &3\,(4.1)      &3.6\,(6.8)    &9.8\,(7)      &     &     &                                                                                      \\
18   & 21:36:35.45 &  57:17:33.0$^{\rm r}$& 19   & 3019                      &                               &                 &                 &   14.3$^{\rm j}$&       &                 &12.236\,(24) &11.890\,(28) & 11.817\,(24)$^{\rm r}$&                    &                 &               &-10.1\,(4.1)  &2.5\,(4.1)    &-14.5\,(7.4)  &8.7\,(7.4)    &0.74 &0.04 &                                                                                      \\
19   & 21:36:30.67 &  57:19:25.5$^{\rm r}$& 20   & 3020                      &                               &                 &                 &   12.6$^{\rm j}$&       &                 &11.613\,(24) &11.496\,(31) & 11.367\,(23)$^{\rm r}$&                    &                 &               &-5.8\,(4.1)   &0.2\,(4.1)    &-5.1\,(2.3)   &-5.4\,(2.1)   &0.05 &-0.43&                                                                                      \\
20   & 21:36:41.27 &  57:18:43.6$^{\rm r}$& 22   & 3022                      &                               &                 &  16.13$^{\rm l}$&  14.53$^{\rm i}$& 13.63 &  12.75$^{\rm i}$&11.284\,(24) &10.614\,(28) & 10.357\,(21)$^{\rm r}$&                    &                 &               &-6.3\,(4.1)   &-1.7\,(4.1)   &-18.3\,(6.8)  &0.3\,(6.8)    &0.39 &-0.01&                                                                                      \\
21   & 21:36:46.24 &  57:18:47.6$^{\rm r}$& 23   & 3023                      &                               &  15.66$^{\rm l}$&   15.3$^{\rm h}$&  14.42$^{\rm h}$& 13.9  &   13.4$^{\rm i}$&12.685\,(26) &12.411\,(31) & 12.302\,(26)$^{\rm r}$&        F6$^{\rm h}$&                 & 1.31$^{\rm h}$&-7.4\,(4.1)   &-0.5\,(4.1)   &-0.9\,(6.8)   &6.8\,(6.8)    &0.25 &0.06 &Dec [h] imprec.                                                               \\
22   & 21:36:50.20 &  57:19:07.2$^{\rm r}$& 24   & 3024                      &                               &                 &                 &   14.4$^{\rm j}$&       &                 &8.750 \,(27) &7.551 \,(42) & 7.140 \,(21)$^{\rm r}$&                    &                 &               &-4.6\,(5.1)   &0.9\,(5.1)    &-18.3\,(6.7)  &5\,(6.7)      &     &     &                                                                                      \\
23   & 21:36:50.49 &  57:18:15.0$^{\rm r}$& 25   & 3025                      &                               &                 &  14.55$^{\rm l}$&  12.42$^{\rm l}$&       &                 &8.164 \,(23) &7.239 \,(34) & 6.914 \,(31)$^{\rm r}$&                    &                 &               &-4.7\,(13.8)  &2.8\,(13.8)   &-1.8\,(7.8)   &0.8\,(7.8)    &-0.32&0.21 &                                                                                      \\
24   & 21:37:00.18 &  57:18:27.1$^{\rm r}$& 26   & 445                       &                               &                 &                 &   11.8$^{\rm j}$&       &                 &11.102\,(21) &10.989\,(28) & 10.930\,(21)$^{\rm r}$&        B8$^{\rm q}$&                 &               &-6.5\,(13.3)  &3.4\,(13.3)   &-4\,(1.2)     &-1.3\,(1.1)   &-0.35&0.32 &                                                                                      \\
25   & 21:36:55.01 &  57:19:43.2$^{\rm r}$& 27   & 3027                      &                               &                 &                 &   14.7$^{\rm j}$&       &                 &12.806\,(24) &12.394\,(32) & 12.264\,(24)$^{\rm r}$&                    &                 &               &14.6\,(4.1)   &-2.7\,(4.1)   &10.2\,(6.8)   &-0.8\,(6.8)   &     &     &                                                                                      \\
26   & 21:37:01.56 &  57:19:47.3$^{\rm r}$& 28   & 3028                      &                               &                 &                 &   13.8$^{\rm j}$&       &                 &11.048\,(22) &10.404\,(28) & 10.179\,(21)$^{\rm r}$&                    &                 &               &5.1\,(4.1)    &9.7\,(4.1)    &14.7\,(6.8)   &15.4\,(6.8)   &-1.43&1.09 &                                                                                      \\
27   & 21:37:08.60 &  57:18:03.3$^{\rm r}$& 29   & 3029                      &                               &                 &                 &   13.7$^{\rm j}$&       &                 &12.380\,(21) &12.181\,(27) & 12.113\,(24)$^{\rm r}$&                    &                 &               &-3.5\,(4.1)   &3.5\,(4.1)    &20\,(6.8)     &-8.4\,(6.8)   &0.12 &0.14 &                                                                                      \\
28   & 21:37:10.54 &  57:18:39.9$^{\rm r}$& 30   & 3030                      &                               &                 &                 &   14.7$^{\rm j}$&       &                 &11.413\,(22) &10.619\,(27) & 10.465\,(20)$^{\rm r}$&                    &                 &               &-5.6\,(4.1)   &1\,(4.1)      &-7.6\,(6.8)   &-6.1\,(6.8)   &-0.22&0.33 &                                                                                      \\
29   & 21:36:34.30 &  57:20:53.6$^{\rm r}$& 31   & 3031                      &                               &                 &                 &   14.8$^{\rm j}$&       &                 &10.745\,(22) &9.736 \,(29) & 9.435 \,(21)$^{\rm r}$&                    &                 &               &-12\,(5.1)    &-1.1\,(5.1)   &-13.3\,(6.8)  &3\,(6.9)      &     &     &                                                                                      \\
30   & 21:36:14.23 &  57:21:30.9$^{\rm r}$& 32   & 3032                      &                               &                 &  12.87$^{\rm f}$&  12.24$^{\rm e}$&       &                 &10.388\,(32) &10.033\,(37) & 9.623 \,(26)$^{\rm r}$&                    &                 &  1.6$^{\rm e}$&-13.8\,(5.7)  &-7.5\,(5.2)   &-4.2\,(14.8)  &-4.3\,(9.2)   &-0.17&0.36 &near 31, [h] imprec.                                           \\     
31   & 21:36:14.62 &  57:21:34.7$^{\rm r}$& 33   & 3033                      &                               &                 &                 &   14.3$^{\rm j}$&       &                 &11.857\,(28) &11.366\,(39) & 11.174\,(25)$^{\rm r}$&                    &                 &               &              &              &              &              &0.1  &0.35 &near 30                                                           \\
32   & 21:36:28.65 &  57:22:53.7$^{\rm r}$& 34   & 3034                      &                               &                 &                 &   14.3$^{\rm j}$&       &                 &12.502\,(32) &12.196\,(39) & 12.119\,(29)$^{\rm r}$&                    &                 &               &12.4\,(4.1)   &18.1\,(4.1)   &47\,(6.7)     &33.3\,(6.8)   &-0.01&0.44 &                                                                                      \\
33   & 21:36:37.70 &  57:23:23.0$^{\rm r}$& 35   & 3035                      &                               &                 &                 &   15.1$^{\rm j}$&       &                 &10.643\,(22) &9.660 \,(31) & 9.288 \,(21)$^{\rm r}$&                    &                 &               &-9.5\,(5.1)   &-4.2\,(5.1)   &-12.9\,(6.8)  &3.8\,(6.8)    &     &     &                                                                                      \\
34   & 21:36:41.91 &  57:23:30.1$^{\rm r}$& 36   & 3036                      &                               &                 &                 &   14.9$^{\rm j}$&       &                 &12.759\,(22) &12.417\,(32) & 12.267\,(24)$^{\rm r}$&                    &                 &               &-15.1\,(4.1)  &-11.8\,(4.1)  &-20.6\,(6.9)  &-4.2\,(6.8)   &     &     &                                                                                      \\
35   & 21:36:52.65 &  57:22:53.9$^{\rm r}$& 37   & 3037                      &                               &                 &                 &   15.1$^{\rm j}$&       &                 &13.502\,(26) &13.233\,(36) & 13.061\,(34)$^{\rm r}$&                    &                 &               &-2.7\,(4.1)   &0.2\,(4.1)    &-4.9\,(7)     &3.8\,(6.9)    &     &     &                                                                                      \\
36   & 21:36:58.51 &  57:23:25.8$^{\rm r}$& 38   & 3038\footnote{also 4603}  & 11-1209                       &  17.70$^{\rm f}$&                 &  15.41$^{\rm e}$& 14.52 &  13.59$^{\rm e}$&12.200\,(24) &11.433\,(33) & 11.122\,(23)$^{\rm r}$&        K6$^{\rm c}$&                 &  0.7$^{\rm e}$&-8.8\,(4.1)   &-3.1\,(4.1)   &-16.9\,(6.8)  &27.6\,(6.9)   &     &     &2 [m] combined                                                              \\
37   & 21:37:02.28 &  57:22:39.0$^{\rm r}$& 40   & 3040                      &                               &                 &                 &   14.9$^{\rm j}$&       &                 &12.546\,(23) &12.038\,(30) & 11.907\,(24)$^{\rm r}$&                    &                 &               &-6.3\,(4.1)   &8.7\,(4.1)    &-6.3\,(6.8)   &12.9\,(6.8)   &     &     &                                                                                      \\
38   & 21:37:11.93 &  57:22:57.3$^{\rm r}$& 41   & 3041                      &                               &                 &                 &   14.8$^{\rm j}$&       &                 &12.985\,(26) &12.620\,(32) & 12.558\,(29)$^{\rm r}$&                    &                 &               &-1.8\,(4.1)   &3.6\,(4.1)    &-6.2\,(6.8)   &4.1\,(6.8)    &     &     &                                                                                      \\
39   & 21:36:22.34 &  57:25:46.2$^{\rm r}$& 43   & 3043                      &                               &                 &                 &   14.3$^{\rm j}$&       &                 &12.303\,(22) &12.027\,(31) & 11.853\,(23)$^{\rm r}$&                    &                 &               &-5.7\,(4.1)   &1.1\,(4.1)    &-2.5\,(6.7)   &12.3\,(6.7)   &-0.64&0.45 &                                                                                      \\
40   & 21:36:26.61 &  57:26:11.7$^{\rm r}$& 44   & 3044                      &                               &                 &                 &   14.5$^{\rm j}$&       &                 &12.731\,(22) &12.389\,(28) & 12.254\,(24)$^{\rm r}$&                    &                 &               &19.6\,(4.1)   &-28.3\,(4.1)  &8.6\,(6.8)    &-15.4\,(6.7)  &     &     &                                                                                      \\
41   & 21:36:39.36 &  57:26:02.5$^{\rm r}$& 45   & 439                       &                               &                 &                 &   12.6$^{\rm j}$&       &                 &10.866\,(24) &10.689\,(31) & 10.618\,(21)$^{\rm r}$&        A0$^{\rm q}$&                 &               &-1.7\,(11.4)  &-5.8\,(11.4)  &-6.8\,(1.4)   &-4.3\,(1.4)   &-0.1 &0.01 &                                                                                      \\
42   & 21:36:49.05 &  57:25:18.4$^{\rm r}$& 49   & 3049                      &                               &                 &                 &   12.7$^{\rm j}$&       &                 &11.686\,(24) &11.445\,(33) & 11.326\,(23)$^{\rm r}$&                    &                 &               &-8.6\,(4.1)   &-1.3\,(4.1)   &-6.3\,(7)     &15.1\,(7)     &0.29 &-0.59&                                                                                      \\
43   & 21:36:48.79 &  57:24:40.8$^{\rm r}$& 50   & 3050                      &                               &                 &                 &   14.8$^{\rm j}$&       &                 &12.925\,(27) &12.619\,(31) & 12.506\,(28)$^{\rm r}$&                    &                 &               &-1\,(4.1)     &1.9\,(4.1)    &1.9\,(6.8)    &12.7\,(6.8)   &     &     &                                                                                      \\
44   & 21:36:54.09 &  57:25:11.0$^{\rm r}$& 51   & 3051                      &                               &                 &                 &   14.9$^{\rm j}$&       &                 &11.632\,(22) &10.904\,(31) & 10.714\,(21)$^{\rm r}$&                    &                 &               &-1.4\,(4.1)   &0.3\,(4.1)    &-1.2\,(6.8)   &12.2\,(6.8)   &     &     &                                                                                      \\
45   & 21:36:55.66 &  57:24:32.8$^{\rm r}$& 52   & 3052                      &                               &                 &                 &   14.2$^{\rm j}$&       &                 &12.415\,(24) &12.116\,(28) & 12.034\,(24)$^{\rm r}$&                    &                 &               &-2.8\,(4.1)   &4.8\,(4.1)    &2.5\,(6.8)    &11.5\,(6.8)   &-0.62&0.91 &                                                                                      \\
46   & 21:37:02.23 &  57:24:51.1$^{\rm r}$& 53   & 3053                      &                               &                 &                 &   13.2$^{\rm j}$&       &                 &10.406\,(23) &9.672 \,(30) & 9.490 \,(23)$^{\rm r}$&                    &                 &               &0.9\,(5.4)    &-1.6\,(5.4)   &4.1\,(6.9)    &10.1\,(6.9)   &0.09 &-0.43&                                                                                      \\
47   & 21:36:14.20 &  57:27:38.0$^{\rm r}$& 54   & 3054                      &                               &                 &                 &   13.5$^{\rm j}$&       &                 &10.088\,(22) &9.226 \,(28) & 8.960 \,(20)$^{\rm r}$&                    &                 &               &-5.1\,(5.3)   &-3.9\,(5.3)   &              &              &-0.28&0.34 &[m] imprec.                                                          \\
48   & 21:36:46.81 &  57:27:54.4$^{\rm r}$& 57   & 3057                      &                               &                 &                 &   14.4$^{\rm j}$&       &                 &12.218\,(26) &11.756\,(32) & 11.582\,(25)$^{\rm r}$&                    &                 &               &-7.8\,(4)     &4.1\,(4)      &-43.8\,(7.4)  &36.4\,(7.5)   &     &     &                                                                                      \\
49   & 21:36:53.95 &  57:27:58.8$^{\rm r}$& 58   & 3058                      &                               &                 &                 &   14.7$^{\rm j}$&       &                 &12.042\,(28) &11.483\,(39) & 11.326\,(23)$^{\rm r}$&                    &                 &               &-7.4\,(4)     &5.8\,(4)      &-12.5\,(7.3)  &21.7\,(7.3)   &-0.42&0.23 &                                                                                      \\
50   & 21:36:41.04 &  57:30:08.3$^{\rm r}$& 59   & 441                       &                               &   9.43$^{\rm l}$&   9.84$^{\rm l}$&    9.5$^{\rm l}$&       &                 &8.723 \,(24) &8.696 \,(42) & 8.627 \,(20)$^{\rm r}$&    B2.5-3$^{\rm p}$&   IV-V$^{\rm p}$&               &-0.3\,(1.3)   &-5\,(1.2)     &-4\,(0.8)     &-4.9\,(0.6)   &-0.7 &-0.07&                                                                                      \\
51   & 21:36:50.72 &  57:31:10.7$^{\rm r}$& 60   & 3060                      &                               &  16.20$^{\rm l}$&  15.12$^{\rm l}$&  13.41$^{\rm i}$& 12.35 &  11.28$^{\rm i}$&9.639 \,(21) &8.908 \,(44) & 8.589 \,(21)$^{\rm r}$&                    &                 &               &-1.7\,(4.9)   &2.2\,(4.9)    &              &              &-0.01&-0.12&same star                                                                             \\
52   & 21:36:18.17 &  57:32:40.4$^{\rm r}$& 61   & 3061                      &                               &                 &                 &   13.6$^{\rm j}$&       &                 &12.019\,(22) &11.778\,(28) & 11.688\,(23)$^{\rm r}$&                    &                 &               &-0.1\,(4)     &-2.1\,(4)     &-9.5\,(7)     &4.2\,(6.8)    &-0.35&0.17 &                                                                                      \\
53   & 21:36:36.91 &  57:34:06.0$^{\rm r}$& 63   & 440                       &                               &  11.26$^{\rm l}$&  11.51$^{\rm f}$&  11.03$^{\rm e}$& 10.73 &  10.38$^{\rm i}$&9.887 \,(22) &9.745 \,(26) & 9.692 \,(20)$^{\rm r}$&        B4$^{\rm e}$&                 &    2$^{\rm e}$&-0.9\,(2)     &0.4\,(2)      &-4.1\,(0.7)   &-2.2\,(0.9)   &-0.06&-0.01&                                                                                      \\
54   & 21:36:23.12 &  57:35:02.0$^{\rm r}$& 65   & 3065                      &                               &                 &                 &   13.8$^{\rm j}$&       &                 &11.863\,(24) &11.515\,(29) & 11.389\,(22)$^{\rm r}$&                    &                 &               &-6.4\,(3.8)   &-5.3\,(3.8)   &-15.2\,(6.7)  &-4.6\,(6.8)   &0.5  &-0.4 &                                                                                      \\
55   & 21:36:15.34 &  57:35:28.3$^{\rm r}$& 66   & 3066                      &                               &                 &                 &   10.5$^{\rm j}$&       &                 &9.468 \,(23) &9.171 \,(26) & 9.124 \,(22)$^{\rm r}$&                    &                 &               &8.2\,(2)      &20.6\,(2)     &4.9\,(0.7)    &13.6\,(1.7)   &-0.79&1.25 &                                                                                      \\
56   & 21:36:25.53 &  57:36:00.7$^{\rm r}$& 67   & 436                       &                               &                 &                 &   10.9$^{\rm j}$&       &                 &9.199 \,(21) &8.583 \,(26) & 8.418 \,(22)$^{\rm r}$&        A0$^{\rm q}$&                 &               &9.2\,(2)      &8.8\,(2)      &4.9\,(1.2)    &-0.3\,(0.8)   &-0.98&0.04 &                                                                                      \\
57   & 21:36:26.53 &  57:37:00.7$^{\rm r}$& 68   & 437                       &                               &                 &                 &   10.7$^{\rm j}$&       &                 &8.325 \,(20) &7.698 \,(29) & 7.513 \,(15)$^{\rm r}$&        G8$^{\rm q}$&                 &               &-11.4\,(10.7) &14\,(10.7)    &-5.8\,(1.3)   &6\,(1.2)      &0.33 &0.56 &                                                                                      \\
58   & 21:36:37.54 &  57:35:24.9$^{\rm r}$& 69   & 3069                      &                               &                 &                 &   14.8$^{\rm j}$&       &                 &12.522\,(23) &12.132\,(28) & 11.993\,(25)$^{\rm r}$&                    &                 &               &36\,(7.9)     &-69.1\,(7.9)  &16.8\,(6.9)   &-13\,(6.9)    &     &     &                                                                                      \\
59   & 21:36:48.70 &  57:35:31.4$^{\rm r}$& 70   & 3070\footnote{also 5068}  &                               &  14.20$^{\rm l}$&  14.03$^{\rm l}$&  13.44$^{\rm l}$&       &                 &12.183\,(23) &12.047\,(28) & 11.993\,(25)$^{\rm r}$&        F0$^{\rm q}$&                 &               &11.2\,(11.4)  &11.6\,(11.4)  &8.6\,(6.8)    &23.5\,(6.8)   &-0.13&0.14 &                                                                                      \\
60   & 21:36:47.86 &  57:35:02.1$^{\rm r}$& 71   & 442                       &                               &   8.33$^{\rm l}$&   8.91$^{\rm l}$&   8.64$^{\rm l}$&       &                 &8.037 \,(19) &8.097 \,(51) & 8.068 \,(27)$^{\rm r}$&        B2$^{\rm p}$&   IV-V$^{\rm p}$&               &-2.8\,(1.3)   &-3.3\,(1.3)   &-4.7\,(0.8)   &-4.2\,(0.8)   &-0.17&-0.27&                                                                                      \\
61   & 21:36:55.91 &  57:33:43.6$^{\rm r}$& 72   & 3072                      &                               &                 &                 &     15$^{\rm j}$&       &                 &12.550\,(35) &12.146\,(41) & 11.961\,(30)$^{\rm r}$&                    &                 &               &              &              &              &              &     &     &                                                                                      \\
62   & 21:36:55.42 &  57:33:49.0$^{\rm r}$& 73   & 3073                      &                               &                 &                 &   15.1$^{\rm j}$&       &                 &11.528\,(36) &10.798\,(40) & 10.530\,(28)$^{\rm r}$&                    &                 &               &-1.4\,(5.3)   &-2.5\,(5.3)   &              &              &     &     &                                                                                      \\
63   & 21:36:59.76 &  57:34:23.7$^{\rm r}$& 74   & 3074                      &                               &                 &                 &   12.5$^{\rm j}$&       &                 &11.120\,(22) &10.841\,(28) & 10.757\,(21)$^{\rm r}$&                    &                 &               &-13.4\,(4)    &-2.4\,(4)     &-9.1\,(2.6)   &-7\,(1.1)     &0.52 &-0.58&                                                                                      \\
64   & 21:37:02.32 &  57:35:36.7$^{\rm r}$& 75   & 3075                      &                               &                 &                 &   13.8$^{\rm j}$&       &                 &11.881\,(22) &11.586\,(27) & 11.468\,(23)$^{\rm r}$&                    &                 &               &-13.6\,(3.8)  &-7.3\,(3.8)   &-7.8\,(6.8)   &3.4\,(6.9)    &0.5  &-0.52&                                                                                      \\
65   & 21:36:47.91 &  57:36:36.4$^{\rm r}$& 77   & 3077                      &                               &                 &                 &     15$^{\rm j}$&       &                 &13.002\,(69) &12.665\,()   &   12.553\,()$^{\rm r}$&                    &                 &               &-8.9\,(3.8)   &-4.7\,(3.8)   &-0.6\,(7.5)   &-8.2\,(7.5)   &     &     &                                                                                      \\
66   & 21:36:57.24 &  57:37:14.0$^{\rm r}$& 78   & 3078                      &                               &                 &                 &   13.8$^{\rm j}$&       &                 &16.525\,(132)&15.508\,()   &   14.958\,()$^{\rm r}$&                    &                 &               &-0.8\,(5.7)   &-17.2\,(5.7)  &              &              &     &     &no star                                                                          \\
67   & 21:36:28.71 &  57:37:20.1$^{\rm r}$& 79   & 3079                      &                               &                 &                 &     15$^{\rm j}$&       &                 &12.981\,(21) &12.578\,(25) & 12.492\,(22)$^{\rm r}$&                    &                 &               &4.9\,(3.8)    &-3.9\,(3.8)   &7.6\,(7.4)    &-11\,(7.5)    &     &     &                                                                                      \\
68   & 21:36:31.19 &  57:37:47.8$^{\rm r}$& 80   & 3080                      &                               &                 &                 &   13.6$^{\rm j}$&       &                 &10.608\,(21) &9.857 \,(26) & 9.657 \,(22)$^{\rm r}$&                    &                 &               &-3\,(4.7)     &-4.2\,(4.7)   &-4.5\,(7.4)   &-5.1\,(7.4)   &0.33 &-0.23&                                                                                      \\
69   & 21:36:23.83 &  57:38:05.4$^{\rm r}$& 81   & 435                       &                               &  12.16$^{\rm l}$&     12$^{\rm f}$&  11.51$^{\rm e}$&       &                 &10.427\,(24) &10.325\,(31) & 10.269\,(23)$^{\rm r}$&        A0$^{\rm e}$&                 &  1.5$^{\rm e}$&3\,(2.7)      &-3.2\,(2.7)   &-3.5\,(1.4)   &-2.9\,(2.3)   &-0.23&-0.11&                                                                                      \\
70   & 21:36:56.12 &  57:37:32.8$^{\rm j}$& 82   & 3082                      &                               &                 &                 &   15.2$^{\rm j}$&       &                 &             &             &                       &                    &                 &               &              &              &              &              &     &     &no/faint star                        \\
71   & 21:36:53.24 &  57:38:05.6$^{\rm r}$& 83   & 3083                      &                               &                 &                 &   13.9$^{\rm j}$&       &                 &12.343\,(23) &12.133\,(31) & 12.022\,(23)$^{\rm r}$&                    &                 &               &-11.1\,(4)    &0.5\,(4)      &-18.9\,(7.4)  &-22.8\,(7.4)  &-0.02&0.05 &                                                                                      \\
72   & 21:36:30.04 &  57:39:09.9$^{\rm r}$& 84   & 3084                      &                               &                 &                 &   13.4$^{\rm j}$&       &                 &11.044\,(23) &10.440\,(29) & 10.322\,(22)$^{\rm r}$&                    &                 &               &-3.4\,(4)     &-2.8\,(4)     &-2.9\,(7.4)   &-9.3\,(7.4)   &0.1  &-0.18&                                                                                      \\
73   & 21:36:25.52 &  57:39:22.4$^{\rm r}$& 85   & 3085                      &                               &                 &                 &   13.9$^{\rm j}$&       &                 &10.935\,(23) &10.241\,(26) & 10.063\,(22)$^{\rm r}$&                    &                 &               &-11.7\,(4)    &-4.7\,(4)     &-14.7\,(7.3)  &-7\,(7.3)     &0.97 &-0.2 &                                                                                      \\
74   & 21:36:27.37 &  57:39:34.5$^{\rm r}$& 86   & 438                       &                               &  13.32$^{\rm l}$&  12.89$^{\rm h}$&  12.26$^{\rm h}$&       &                 &10.963\,(24) &10.763\,(28) & 10.684\,(22)$^{\rm r}$&        A7$^{\rm h}$&                 & 1.37$^{\rm h}$&-5.9\,(3.8)   &-1.1\,(3.8)   &-8.6\,(1)     &-4.3\,(1.1)   &0.25 &-0.34&                                                                                      \\
75   & 21:36:15.97 &  57:39:26.9$^{\rm r}$& 87   & 3087                      &                               &                 &                 &   13.6$^{\rm j}$&       &                 &12.158\,(23) &11.964\,(31) & 11.802\,(22)$^{\rm r}$&                    &                 &               &-11.5\,(4)    &-1.7\,(4)     &-8.6\,(7.4)   &-14.5\,(7.5)  &0.21 &-0.09&                                                                                      \\
76   & 21:36:16.00 &  57:40:16.3$^{\rm r}$& 88   & 3088                      &                               &                 &                 &   13.7$^{\rm j}$&       &                 &12.177\,(28) &12.023\,(29) & 11.889\,(26)$^{\rm r}$&                    &                 &               &2.9\,(3.8)    &12.1\,(3.8)   &50.6\,(7.1)   &50.5\,(7.1)   &0.22 &-0.03&                                                                                      \\
77   & 21:36:47.90 &  57:40:32.2$^{\rm r}$& 89   & 3089                      &                               &                 &  15.81$^{\rm h}$&  14.66$^{\rm h}$&       &                 &12.794\,(26) &12.417\,(36) & 12.354\,(23)$^{\rm r}$&        F4$^{\rm h}$&                 & 2.37$^{\rm h}$&-1.6\,(3.8)   &1\,(3.8)      &-5.3\,(7.9)   &12.5\,(7.4)   &-0.7 &0.34 &                                                                                      \\
78   & 21:36:48.26 &  57:39:18.5$^{\rm r}$& 90   & 443                       &                               &                 &                 &   10.4$^{\rm j}$&       &                 &8.318 \,(18) &7.818 \,(36) & 7.664 \,(17)$^{\rm r}$&                    &                 &               &-5\,(2)       &-7.7\,(2)     &-5.7\,(0.9)   &-6.1\,(1.6)   &0.16 &-0.5 &                                                                                      \\
79   & 21:36:43.56 &  57:41:13.4$^{\rm r}$& 91   & 3091                      &                               &                 &                 &   14.7$^{\rm j}$&       &                 &12.565\,(24) &12.104\,(28) & 12.043\,(23)$^{\rm r}$&                    &                 &               &1.2\,(3.8)    &-2.5\,(3.8)   &1.8\,(7.4)    &-7.6\,(7.3)   &     &     &                                                                                      \\
80   & 21:36:36.10 &  57:41:52.0$^{\rm r}$& 92   & 3092                      &                               &                 &                 &   14.9$^{\rm j}$&       &                 &12.610\,(24) &12.062\,(31) & 11.906\,(23)$^{\rm r}$&                    &                 &               &-1.3\,(3.8)   &1.6\,(3.8)    &-1.5\,(7.3)   &1.9\,(7.3)    &     &     &                                                                                      \\
81   & 21:36:34.96 &  57:42:02.6$^{\rm r}$& 93   & 3093                      &                               &                 &                 &   14.1$^{\rm j}$&       &                 &11.885\,(23) &11.373\,(28) & 11.274\,(22)$^{\rm r}$&                    &                 &               &7.8\,(4)      &18.5\,(4)     &15.6\,(7.3)   &11.5\,(7.4)   &-1.01&2.1  &                                                                                      \\
82   & 21:36:27.89 &  57:42:08.5$^{\rm r}$& 94   & 3094                      &                               &                 &                 &   14.9$^{\rm j}$&       &                 &12.222\,(24) &11.413\,(28) & 11.203\,(23)$^{\rm r}$&                    &                 &               &-8\,(4.1)     &-2.4\,(4.1)   &-7.9\,(7.8)   &-11.2\,(7.9)  &     &     &                                                                                      \\
83   & 21:36:27.18 &  57:43:42.1$^{\rm r}$& 95   & 3095                      &                               &  14.11$^{\rm l}$&  14.22$^{\rm h}$&   13.5$^{\rm h}$&       &                 &11.782\,(23) &11.576\,(29) & 11.463\,(23)$^{\rm r}$&        B2$^{\rm h}$&                 & 2.93$^{\rm h}$&8\,(4)        &4.8\,(4)      &62.3\,(7.3)   &32.1\,(7.3)   &-0.26&0.02 &                                                                                      \\
84   & 21:36:32.96 &  57:43:36.9$^{\rm r}$& 96   & 3096                      &                               &                 &                 &   14.8$^{\rm j}$&       &                 &12.597\,(24) &12.238\,(29) & 12.125\,(26)$^{\rm r}$&                    &                 &               &-7.4\,(3.8)   &1.5\,(3.8)    &-30.2\,(7.4)  &20.8\,(7.4)   &-0.12&0.52 &                                                                                      \\
85   & 21:36:34.85 &  57:43:18.5$^{\rm r}$& 97   & 3097                      &                               &                 &                 &   12.7$^{\rm j}$&       &                 &9.848 \,(23) &9.248 \,(28) & 9.046 \,(22)$^{\rm r}$&                    &                 &               &-51.8\,(4.7)  &60.9\,(4.7)   &-49.2\,(7.4)  &65.2\,(7.3)   &4.14 &7.32 &[m] imprec.                                                          \\
86   & 21:36:50.55 &  57:44:42.1$^{\rm j}$& 98   & 3098                      &                               &                 &                 &   14.5$^{\rm j}$&       &                 &             &             &                       &                    &                 &               &              &              &              &              &     &     &no star                                                                          \\
87   & 21:36:56.24 &  57:44:10.4$^{\rm r}$& 99   & 3099                      &                               &                 &                 &   15.3$^{\rm j}$&       &                 &12.697\,(24) &11.908\,(29) & 11.691\,(23)$^{\rm r}$&                    &                 &               &-0.8\,(3.8)   &-6.3\,(3.8)   &-12.5\,(7.7)  &-11.5\,(7.9)  &     &     &                                                                                      \\
88   & 21:36:55.35 &  57:43:52.9$^{\rm r}$& 100  & 3100                      &                               &                 &                 &   13.4$^{\rm j}$&       &                 &11.912\,(23) &11.439\,(28) & 11.389\,(22)$^{\rm r}$&                    &                 &               &2.1\,(3.8)    &10.1\,(3.8)   &-1.3\,(7.5)   &13.1\,(7.5)   &-0.95&1.57 &                                                                                      \\
89   & 21:36:51.36 &  57:41:45.5$^{\rm r}$& 101  & 444                       &                               &                 &                 &   10.9$^{\rm j}$&       &                 &9.780 \,(23) &9.400 \,(28) & 9.330 \,(20)$^{\rm r}$&        G2$^{\rm q}$&                 &               &-30.1\,(2)    &-27.7\,(2)    &-34.4\,(0.7)  &-24.5\,(1.6)  &2.95 &-2.33&                                                                                      \\
90   & 21:36:52.77 &  57:41:17.1$^{\rm r}$& 102  & 3102                      &                               &                 &                 &   14.8$^{\rm j}$&       &                 &12.816\,(26) &12.464\,(31) & 12.374\,(28)$^{\rm r}$&                    &                 &               &-3.7\,(3.8)   &-7.5\,(3.8)   &-0.8\,(7.1)   &-10.6\,(6.9)  &     &     &                                                                                      \\
91   & 21:36:59.13 &  57:42:46.0$^{\rm r}$& 103  & 3103                      &                               &                 &                 &   14.1$^{\rm j}$&       &                 &11.681\,(24) &11.057\,(28) & 10.932\,(23)$^{\rm r}$&                    &                 &               &8.6\,(3.8)    &-16.7\,(3.8)  &4.4\,(7.4)    &-3.4\,(7.4)   &-0.87&-1.05&                                                                                      \\
92   & 21:37:04.46 &  57:44:14.7$^{\rm r}$& 104  & 3104                      &                               &                 &                 &   14.4$^{\rm j}$&       &                 &12.394\,(26) &12.078\,(30) & 11.985\,(24)$^{\rm r}$&                    &                 &               &-2.9\,(3.8)   &1.3\,(3.8)    &-8\,(7.7)     &-3.4\,(7.5)   &-0.39&0.32 &                                                                                      \\
93   & 21:37:03.69 &  57:11:41.2$^{\rm r}$& 105  & 3105                      &                               &                 &                 &   14.9$^{\rm j}$&       &                 &12.500\,(27) &12.014\,(30) & 11.897\,(25)$^{\rm r}$&                    &                 &               &9.4\,(4.1)    &-0.9\,(4.1)   &26.7\,(6.8)   &-3.8\,(6.8)   &     &     &                                                                                      \\
94   & 21:37:05.99 &  57:12:19.7$^{\rm r}$& 106  & 162                       &                               &                 &                 &   10.2$^{\rm j}$&       &                 &9.160 \,(22) &8.804 \,(27) & 8.771 \,(20)$^{\rm r}$&        F8$^{\rm q}$&                 &               &19.1\,(2)     &6.1\,(2)      &              &              &-3.02&0.36 &                                                                                      \\
95   & 21:37:11.65 &  57:12:56.3$^{\rm r}$& 108  & 3108                      &                               &  14.42$^{\rm l}$&   13.1$^{\rm h}$&  13.03$^{\rm h}$&       &                 &11.906\,(22) &11.636\,(28) & 11.532\,(23)$^{\rm r}$&        A5$^{\rm h}$&                 &  1.5$^{\rm h}$&-4.8\,(4.1)   &7.8\,(4.1)    &-14.9\,(6.8)  &17.1\,(6.8)   &-0.17&0.12 &Dec [h] imprec.                                                               \\
96   & 21:37:38.85 &  57:11:28.1$^{\rm r}$& 110  & 3110                      &                               &                 &                 &   13.9$^{\rm j}$&       &                 &10.584\,(22) &9.767 \,(27) & 9.582 \,(20)$^{\rm r}$&                    &                 &               &-7.8\,(5.1)   &-2.8\,(5.1)   &-18.1\,(6.9)  &10.1\,(6.9)   &0.61 &-0.51&                                                                                      \\
97   & 21:37:45.47 &  57:12:04.8$^{\rm r}$& 111  & 3111                      &                               &                 &                 &   14.3$^{\rm j}$&       &                 &8.875 \,(27) &7.732 \,(31) & 7.343 \,(26)$^{\rm r}$&                    &                 &               &-0.7\,(5.1)   &0.5\,(5.1)    &4\,(6.8)      &-8.4\,(6.9)   &0.07 &0.25 &                                                                                      \\
98   & 21:37:42.52 &  57:12:13.1$^{\rm r}$& 112  & 3112                      &                               &                 &                 &   14.9$^{\rm j}$&       &                 &13.101\,(22) &12.791\,(30) & 12.680\,(26)$^{\rm r}$&                    &                 &               &-0.9\,(4.1)   &6.6\,(4.1)    &-11.2\,(6.8)  &-2.7\,(6.8)   &     &     &                                                                                      \\
99   & 21:37:23.13 &  57:13:23.3$^{\rm r}$& 113  & 3113                      &                               &                 &                 &   14.3$^{\rm j}$&       &                 &12.304\,(22) &11.731\,(26) & 11.631\,(24)$^{\rm r}$&                    &                 &               &              &              &              &              &3.68 &1.31 &                                                                                      \\
100  & 21:37:28.51 &  57:13:49.9$^{\rm r}$& 114  & 3114                      &                               &                 &                 &   12.5$^{\rm j}$&       &                 &9.449 \,(21) &8.549 \,(26) & 8.364 \,(20)$^{\rm r}$&                    &                 &               &0.7\,(5.1)    &-9.2\,(5.1)   &17.3\,(7.8)   &-41.6\,(7.8)  &-0.12&0.15 &                                                                                      \\
101  & 21:37:37.57 &  57:13:54.8$^{\rm r}$& 115  & 3115                      &                               &                 &                 &   14.8$^{\rm j}$&       &                 &10.783\,(32) &9.397 \,(116)& 8.902 \,(86)$^{\rm r}$&                    &                 &               &0.2\,(5.7)    &-3\,(5.7)     &              &              &     &     &                                                                                      \\
102  & 21:37:38.20 &  57:13:54.6$^{\rm r}$& 116  & 3116                      &                               &                 &                 &   14.5$^{\rm j}$&       &                 &9.301 \,(27) &8.128 \,(29) & 7.735 \,(29)$^{\rm r}$&                    &                 &               &128.9\,(18.7) &-1.6\,(18.7)  &              &              &-0.29&0.13 &                                                                                      \\
103  & 21:37:12.17 &  57:15:08.7$^{\rm r}$& 117  & 3117                      &                               &                 &                 &   14.7$^{\rm j}$&       &                 &12.735\,(26) &12.363\,(28) & 12.326\,(25)$^{\rm r}$&                    &                 &               &-3.3\,(4.1)   &-1.1\,(4.1)   &-4.7\,(6.8)   &-7.2\,(6.8)   &0.11 &0.2  &                                                                                      \\
104  & 21:37:20.16 &  57:14:36.8$^{\rm r}$& 118  & 3118                      &                               &                 &                 &   14.2$^{\rm j}$&       &                 &11.042\,(22) &10.406\,(26) & 10.221\,(21)$^{\rm r}$&                    &                 &               &-0.3\,(4.1)   &4\,(4.1)      &22.9\,(6.9)   &22.7\,(6.9)   &-0.03&-0.18&                                                                                      \\
105  & 21:37:39.08 &  57:14:42.5$^{\rm r}$& 119  & 3119                      &                               &                 &                 &   14.9$^{\rm j}$&       &                 &13.038\,(26) &12.726\,(32) & 12.619\,(31)$^{\rm r}$&                    &                 &               &3\,(4.1)      &0.7\,(4.1)    &8.1\,(6.9)    &11.9\,(6.9)   &-0.17&0.29 &                                                                                      \\
106  & 21:37:20.28 &  57:16:00.7$^{\rm r}$& 120  & 3120                      &                               &                 &                 &   14.4$^{\rm j}$&       &                 &11.265\,(22) &10.547\,(27) & 10.329\,(20)$^{\rm r}$&                    &                 &               &-2.6\,(4.1)   &-1.9\,(4.1)   &-7.8\,(6.9)   &5.3\,(6.9)    &0.51 &-0.42&                                                                                      \\
107  & 21:37:39.83 &  57:15:12.2$^{\rm r}$& 122  & 3122                      &                               &                 &                 &   14.4$^{\rm j}$&       &                 &12.559\,(24) &12.104\,(27) & 12.065\,(25)$^{\rm r}$&                    &                 &               &-1.3\,(4.1)   &-1.4\,(4.1)   &-4.2\,(6.8)   &2.7\,(6.8)    &0.09 &-0.49&                                                                                      \\
108  & 21:37:43.64 &  57:15:44.2$^{\rm r}$& 123  & 3123                      &                               &                 &                 &   14.8$^{\rm j}$&       &                 &12.511\,(27) &12.026\,(31) & 11.902\,(28)$^{\rm r}$&                    &                 &               &-3.8\,(4.1)   &9.9\,(4.1)    &-4.1\,(6.8)   &23.6\,(6.8)   &-0.39&0.38 &                                                                                      \\
109  & 21:37:58.37 &  57:12:44.6$^{\rm r}$& 124  & 3124                      &                               &                 &                 &   13.8$^{\rm j}$&       &                 &12.178\,(29) &11.781\,(33) & 11.655\,(21)$^{\rm r}$&                    &                 &               &5.3\,(4)      &-6\,(4)       &10.3\,(6.8)   &-6.6\,(6.8)   &-1.19&-0.13&                                                                                      \\
110  & 21:37:47.82 &  57:16:29.6$^{\rm r}$& 125  & 3125                      &                               &                 &                 &   14.5$^{\rm j}$&       &                 &10.643\,(22) &9.754 \,(27) & 9.529 \,(20)$^{\rm r}$&                    &                 &               &-4.4\,(5.1)   &2.1\,(5.1)    &-5.5\,(6.8)   &8.9\,(6.8)    &-0.1 &0.18 &                                                                                      \\
111  & 21:37:47.04 &  57:16:11.5$^{\rm r}$& 126  & 3126                      &                               &                 &                 &     15$^{\rm j}$&       &                 &11.117\,(22) &10.134\,(27) & 9.908 \,(20)$^{\rm r}$&                    &                 &               &-4.1\,(5.1)   &-3.4\,(5.1)   &-3.4\,(6.8)   &6\,(6.8)      &-0.78&-0.58&                                                                                      \\
112  & 21:37:54.79 &  57:16:02.6$^{\rm r}$& 127  & 3127                      &                               &                 &                 &   14.9$^{\rm j}$&       &                 &12.793\,(27) &12.301\,(33) & 12.198\,(23)$^{\rm r}$&                    &                 &               &-7\,(4)       &0.3\,(4)      &-9.5\,(6.8)   &-0.3\,(6.8)   &     &     &                                                                                      \\
113  & 21:37:52.81 &  57:17:14.6$^{\rm r}$& 129  & 3129                      &                               &                 &                 &   14.7$^{\rm j}$&       &                 &12.741\,(27) &12.277\,(32) & 12.143\,(21)$^{\rm r}$&                    &                 &               &6.3\,(4)      &2.7\,(4)      &-3\,(6.8)     &9.6\,(6.8)    &-2.19&0.52 &                                                                                      \\
114  & 21:37:33.23 &  57:17:12.6$^{\rm j}$& 131  & 3131                      &                               &                 &                 &   13.6$^{\rm j}$&       &                 &             &             &                       &                    &                 &               &              &              &              &              &     &     &no star                                                                          \\
115  & 21:37:35.20 &  57:17:44.7$^{\rm r}$& 132  & 3132                      &                               &                 &                 &   13.8$^{\rm j}$&       &                 &10.969\,(22) &10.276\,(26) & 10.110\,(20)$^{\rm r}$&                    &                 &               &-2.5\,(4.1)   &-4.3\,(4.1)   &-2.6\,(6.9)   &-1.3\,(6.9)   &-0.31&-0.61&                                                                                      \\
116  & 21:37:30.86 &  57:18:33.8$^{\rm r}$& 133  & 3133                      &                               &                 &                 &   12.4$^{\rm j}$&       &                 &11.191\,(24) &10.898\,(28) & 10.794\,(23)$^{\rm r}$&                    &                 &               &-13.2\,(4.1)  &4.1\,(4.1)    &-9\,(1.2)     &-12.3\,(4.5)  &0.49 &-1.07&                                                                                      \\
117  & 21:37:14.31 &  57:20:21.6$^{\rm r}$& 134  & 3134                      &                               &                 &                 &   13.3$^{\rm j}$&       &                 &11.903\,(24) &11.630\,(27) & 11.523\,(24)$^{\rm r}$&                    &                 &               &-0.1\,(4.1)   &-0.1\,(4.1)   &5.3\,(7)      &-11.1\,(7)    &-0.52&0.28 &                                                                                      \\
118  & 21:37:19.41 &  57:20:56.3$^{\rm j}$& 135  & 3135                      &                               &                 &                 &   13.7$^{\rm j}$&       &                 &             &             &                       &                    &                 &               &-1.7\,(8.6)   &90.6\,(8.6)   &              &              &     &     &no star                                                                          \\
119  & 21:37:19.13 &  57:20:26.0$^{\rm r}$& 136  & 3136                      &                               &                 &                 &   14.8$^{\rm j}$&       &                 &13.074\,(26) &12.811\,(28) & 12.655\,(30)$^{\rm r}$&                    &                 &               &-7.4\,(4.1)   &5.2\,(4.1)    &-21.3\,(6.8)  &9.3\,(6.8)    &0.15 &-0.39&                                                                                      \\
120  & 21:37:18.07 &  57:20:01.1$^{\rm r}$& 137  & 3137                      &                               &                 &                 &   14.8$^{\rm j}$&       &                 &11.119\,(22) &10.188\,(27) & 9.968 \,(18)$^{\rm r}$&                    &                 &               &-2\,(5.1)     &2.2\,(5.1)    &0.6\,(6.8)    &10.6\,(6.8)   &     &     &                                                                                      \\
121  & 21:37:24.84 &  57:21:12.0$^{\rm r}$& 138  & 3138                      &                               &                 &                 &     15$^{\rm j}$&       &                 &12.387\,(24) &11.705\,(28) & 11.553\,(24)$^{\rm r}$&                    &                 &               &-4.4\,(4.1)   &-1.2\,(4.1)   &-0.8\,(6.8)   &-9.5\,(6.8)   &     &     &                                                                                      \\
122  & 21:37:25.65 &  57:20:19.2$^{\rm r}$& 139  & 3139                      &                               &                 &                 &   14.7$^{\rm j}$&       &                 &12.742\,(22) &12.376\,(32) & 12.310\,(26)$^{\rm r}$&                    &                 &               &-4.6\,(4.1)   &2.1\,(4.1)    &-8.4\,(6.8)   &5.2\,(6.8)    &0.53 &0.1  &                                                                                      \\
123  & 21:37:34.21 &  57:19:33.1$^{\rm r}$& 140  & 3140                      &                               &                 &                 &   14.8$^{\rm j}$&       &                 &12.947\,(22) &12.639\,(28) & 12.498\,(26)$^{\rm r}$&                    &                 &               &-3.6\,(4.1)   &3.2\,(4.1)    &2.3\,(6.8)    &1.4\,(6.8)    &     &     &                                                                                      \\
124  & 21:37:46.97 &  57:19:06.4$^{\rm r}$& 141  & 3141                      &                               &                 &                 &   14.8$^{\rm j}$&       &                 &12.204\,(24) &11.594\,(27) & 11.476\,(23)$^{\rm r}$&                    &                 &               &7.1\,(4.1)    &1.8\,(4.1)    &-0.6\,(6.8)   &7.2\,(6.7)    &     &     &                                                                                      \\
125  & 21:37:49.67 &  57:18:07.7$^{\rm r}$& 143  & 3143                      &                               &                 &                 &   15.1$^{\rm j}$&       &                 &13.259\,(45) &12.774\,(49) & 12.645\,(47)$^{\rm r}$&                    &                 &               &-15.2\,(5.4)  &-4.6\,(5.4)   &-25.2\,(6.8)  &12.6\,(6.8)   &     &     &                                                                                      \\
126  & 21:37:24.25 &  57:23:08.0$^{\rm r}$& 144  & 451                       &                               &  13.07$^{\rm l}$&  12.81$^{\rm h}$&  12.19$^{\rm h}$&       &                 &10.903\,(23) &10.692\,(30) & 10.618\,(21)$^{\rm r}$&        F0$^{\rm h}$&                 & 0.97$^{\rm h}$&-6.8\,(2.7)   &-8.6\,(2.7)   &-7.6\,(0.7)   &-6.1\,(1.1)   &0.34 &-0.15&Dec [h] wrong                                                                   \\
127  & 21:37:33.54 &  57:22:26.2$^{\rm r}$& 145  & 3145                      &                               &                 &                 &   14.9$^{\rm j}$&       &                 &12.986\,(26) &12.564\,(34) & 12.501\,(29)$^{\rm r}$&                    &                 &               &-18.5\,(4.1)  &-4.7\,(4.1)   &-23.1\,(6.8)  &3.4\,(6.9)    &     &     &                                                                                      \\
128  & 21:37:41.29 &  57:21:25.9$^{\rm r}$& 146  & 3146                      &                               &                 &                 &   14.3$^{\rm j}$&       &                 &12.416\,(24) &12.035\,(27) & 11.967\,(25)$^{\rm r}$&                    &                 &               &-8.2\,(4.1)   &0.9\,(4.1)    &-6.7\,(6.7)   &6.8\,(6.7)    &0.21 &-0.16&                                                                                      \\
129  & 21:37:42.95 &  57:21:03.9$^{\rm r}$& 147  & 453                       &                               &  11.82$^{\rm l}$&  11.08$^{\rm h}$&  11.04$^{\rm h}$&       &                 &10.418\,(24) &10.279\,(27) & 10.248\,(21)$^{\rm r}$&        A0$^{\rm h}$&                 & 1.28$^{\rm h}$&-0.6\,(1.7)   &0.4\,(1.7)    &-5.8\,(2.2)   &-4.5\,(3.4)   &0.19 &-0.03&Dec [h] imprec.                                                               \\
130  & 21:37:54.80 &  57:21:06.1$^{\rm r}$& 148  & 3148                      &                               &  13.74$^{\rm l}$&  12.68$^{\rm l}$&  11.42$^{\rm l}$&       &                 &9.083 \,(27) &8.520 \,(38) & 8.364 \,(24)$^{\rm r}$&                    &                 &               &-5.1\,(2)     &-9\,(2)       &-2.9\,(0.8)   &-0.2\,(1.3)   &-0.04&0.29 &                                                                                      \\
131  & 21:37:47.77 &  57:21:47.2$^{\rm r}$& 149  & 3149                      &                               &                 &  15.43$^{\rm l}$&   13.9$^{\rm l}$&       &                 &10.745\,(26) &10.061\,(28) & 9.880 \,(20)$^{\rm r}$&                    &                 &               &-12.7\,(5.1)  &-1.8\,(5.1)   &1.4\,(7)      &-45.9\,(7)    &0.2  &0.15 &                                                                                      \\
132  & 21:37:51.87 &  57:22:23.5$^{\rm r}$& 150  & 3150                      &                               &  14.38$^{\rm l}$&  14.35$^{\rm h}$&  13.65$^{\rm h}$&       &                 &12.299\,(31) &12.113\,(36) & 12.063\,(27)$^{\rm r}$&        B4$^{\rm h}$&                 & 2.68$^{\rm h}$&-2.6\,(4)     &6\,(4)        &10.4\,(6.9)   &30.2\,(6.8)   &0.13 &-0.03&                                                                                      \\
133  & 21:37:54.85 &  57:22:16.1$^{\rm r}$& 151  & 3151                      &                               &                 &                 &   11.2$^{\rm j}$&       &                 &8.980 \,(27) &8.314 \,(55) & 8.189 \,(23)$^{\rm r}$&                    &                 &               &0.4\,(2.8)    &-10.8\,(2.7)  &-7\,(1.7)     &-3.1\,(1.3)   &0.49 &0.04 &                                                                                      \\
134  & 21:37:55.15 &  57:23:01.0$^{\rm r}$& 152  & 3152                      &                               &                 &                 &   11.9$^{\rm j}$&       &                 &10.614\,(26) &10.254\,(31) & 10.202\,(20)$^{\rm r}$&                    &                 &               &20.6\,(2.7)   &3.9\,(2.7)    &18.1\,(1.3)   &3\,(1.7)      &-2.42&0.79 &                                                                                      \\
135  & 21:37:41.24 &  57:23:09.0$^{\rm r}$& 153  & 454                       &                               &  12.91$^{\rm l}$&  12.07$^{\rm h}$&  12.03$^{\rm h}$&       &                 &11.518\,(25) &11.391\,(30) & 11.399\,(24)$^{\rm r}$&        A0$^{\rm h}$&                 & 1.15$^{\rm h}$&-1.4\,(2.7)   &-6.1\,(2.7)   &-2.8\,(0.9)   &-6.4\,(1.5)   &-0.11&0.01 &Dec [h] imprec.                                                               \\
136  & 21:37:24.52 &  57:24:19.2$^{\rm r}$& 154  & 3154                      &                               &                 &                 &   12.6$^{\rm j}$&       &                 &10.009\,(26) &9.299 \,(30) & 9.082 \,(21)$^{\rm r}$&                    &                 &               &-54.5\,(41.8) &-33.5\,(41.8) &-160.7\,(14.1)&-17.2\,(13.6) &-0.46&-0.33&                                                                                      \\
137  & 21:37:25.58 &  57:24:23.7$^{\rm r}$& 155  & 3155                      &                               &                 &                 &   13.9$^{\rm j}$&       &                 &10.967\,()   &9.908 \,()   &   9.611 \,()$^{\rm r}$&                    &                 &               &              &              &              &              &0.68 &-0.59&                                                                                      \\
138  & 21:37:43.25 &  57:23:03.8$^{\rm r}$& 156  & 3156                      &                               &                 &                 &     15$^{\rm j}$&       &                 &11.609\,(25) &10.801\,(27) & 10.566\,(23)$^{\rm r}$&                    &                 &               &-7.5\,(4.1)   &-0.3\,(4.1)   &-3.6\,(6.9)   &5.5\,(6.8)    &     &     &                                                                                      \\
139  & 21:37:35.65 &  57:23:57.8$^{\rm r}$& 157  & 3157                      &                               &                 &                 &   14.7$^{\rm j}$&       &                 &10.680\,(23) &9.733 \,(27) & 9.468 \,(21)$^{\rm r}$&                    &                 &               &-5.1\,(5.1)   &-3.2\,(5.1)   &-7.1\,(6.8)   &6.7\,(6.8)    &0.18 &0.15 &                                                                                      \\
140  & 21:37:41.88 &  57:24:29.2$^{\rm r}$& 158  & 3158                      &                               &                 &                 &   15.1$^{\rm j}$&       &                 &13.361\,(27) &13.109\,(34) & 12.969\,(36)$^{\rm r}$&                    &                 &               &-5.9\,(4.1)   &-1.1\,(4.1)   &-2.4\,(6.9)   &15.3\,(6.9)   &     &     &                                                                                      \\
141  & 21:37:41.69 &  57:24:51.3$^{\rm r}$& 159  & 3159                      &                               &                 &                 &   14.7$^{\rm j}$&       &                 &10.394\,(25) &9.376 \,(31) & 9.080 \,(21)$^{\rm r}$&                    &                 &               &-26.4\,(18.5) &0.7\,(18.5)   &              &              &     &     &                                                                                      \\
142  & 21:37:10.94 &  57:24:31.6$^{\rm r}$& 160  & 3160                      &                               &                 &                 &   15.1$^{\rm j}$&       &                 &13.035\,(29) &12.648\,(37) & 12.559\,(38)$^{\rm r}$&                    &                 &               &-3.3\,(4.1)   &-3.3\,(4.1)   &14.3\,(6.9)   &14.1\,(6.9)   &     &     &                                                                                      \\
143  & 21:37:12.85 &  57:25:05.3$^{\rm r}$& 161  & 3161                      &                               &                 &                 &   14.9$^{\rm j}$&       &                 &12.895\,(26) &12.535\,(32) & 12.458\,(29)$^{\rm r}$&                    &                 &               &-0.7\,(4.1)   &-2.4\,(4.1)   &4.4\,(6.8)    &5.9\,(6.8)    &     &     &                                                                                      \\
144  & 21:37:10.70 &  57:26:10.3$^{\rm r}$& 162  & 3162                      &                               &                 &                 &   12.7$^{\rm j}$&       &                 &10.655\,(25) &10.143\,(31) & 9.976 \,(24)$^{\rm r}$&                    &                 &               &-8.5\,(5.1)   &0.5\,(5.1)    &-49.6\,(7.2)  &21.9\,(7.2)   &-0.12&-0.19&                                                                                      \\
145  & 21:37:14.81 &  57:25:51.5$^{\rm r}$& 163  & 3163                      &                               &                 &                 &   13.3$^{\rm j}$&       &                 &12.007\,(29) &11.573\,(32) & 11.579\,(29)$^{\rm r}$&                    &                 &               &-9.7\,(4.1)   &-2.7\,(4.1)   &-45\,(6.9)    &8.6\,(6.9)    &0.21 &-0.27&                                                                                      \\
146  & 21:37:20.09 &  57:25:14.1$^{\rm r}$& 164  & 449                       &                               &  13.26$^{\rm l}$&  12.97$^{\rm f}$&  12.44$^{\rm e}$&       &                 &11.169\,(26) &10.983\,(30) & 10.918\,(23)$^{\rm r}$&        A3$^{\rm e}$&                 &  1.4$^{\rm e}$&-4\,(2.7)     &-11.8\,(2.7)  &-7.1\,(0.8)   &-5.7\,(1.1)   &-0.06&0.3  &                                                                                      \\
147  & 21:37:29.36 &  57:25:18.0$^{\rm r}$& 165  & 3165                      &                               &                 &                 &   14.4$^{\rm j}$&       &                 &11.167\,(26) &10.434\,(28) & 10.234\,(25)$^{\rm r}$&                    &                 &               &-3.3\,(4.1)   &0.1\,(4.1)    &-8.2\,(6.8)   &6.1\,(6.8)    &-0.04&0.28 &                                                                                      \\
148  & 21:37:19.62 &  57:26:14.0$^{\rm r}$& 166  & 447                       &                               &                 &  10.56$^{\rm h}$&  10.24$^{\rm h}$&       &                 &9.080 \,(25) &8.954 \,(28) & 8.884 \,(20)$^{\rm r}$&        A3$^{\rm h}$&                 & 0.72$^{\rm h}$&-1.8\,(1.4)   &-2.2\,(1.4)   &-1.5\,(0.6)   &-3.2\,(0.7)   &-0.58&0.1  &                                                                                      \\
149  & 21:37:11.56 &  57:27:10.8$^{\rm r}$& 167  & 3167                      &                               &                 &                 &   14.2$^{\rm j}$&       &                 &12.481\,(26) &12.167\,(32) & 12.083\,(28)$^{\rm r}$&                    &                 &               &1.5\,(4.1)    &0.8\,(4.1)    &8.6\,(6.8)    &15.5\,(6.8)   &-0.76&0.26 &                                                                                      \\
150  & 21:37:21.02 &  57:28:38.2$^{\rm r}$& 168  & 3168                      &                               &                 &                 &   14.4$^{\rm j}$&       &                 &12.726\,()   &12.553\,(38) & 12.430\,(36)$^{\rm r}$&                    &                 &               &-8.4\,(4.1)   &-3.2\,(4.1)   &-28.5\,(6.8)  &-7.8\,(6.8)   &-0.21&-0.05&                                                                                      \\
151  & 21:37:23.95 &  57:28:16.9$^{\rm r}$& 169  & 3169                      &                               &  15.36$^{\rm l}$&  14.74$^{\rm f}$&  14.02$^{\rm e}$&       &                 &12.296\,(23) &12.033\,(27) & 11.956\,(24)$^{\rm r}$&        A4$^{\rm e}$&                 &  1.9$^{\rm e}$&1.4\,(10.9)   &2\,(10.9)     &31.3\,(7.3)   &19.8\,(7.3)   &-0.32&0.2  &                                                                                      \\
152  & 21:37:27.05 &  57:27:06.5$^{\rm r}$& 170  & 3170                      &                               &                 &                 &   12.9$^{\rm j}$&       &                 &9.095 \,(23) &8.167 \,(33) & 7.841 \,(20)$^{\rm r}$&                    &                 &               &-6.9\,(5.1)   &-2.6\,(5.1)   &-23.3\,(7.2)  &-16.2\,(7.2)  &-0.21&0.2  &                                                                                      \\
153  & 21:37:35.31 &  57:27:52.4$^{\rm r}$& 171  & 3171                      &                               &                 &                 &   14.8$^{\rm j}$&       &                 &12.553\,(25) &12.182\,(30) & 12.035\,(26)$^{\rm r}$&                    &                 &               &-11.1\,(3.9)  &-7\,(3.9)     &-9.1\,(6.8)   &5\,(6.8)      &     &     &                                                                                      \\
154  & 21:37:52.32 &  57:26:53.2$^{\rm r}$& 172  & 3172                      &                               &                 &                 &   12.9$^{\rm j}$&       &                 &14.143\,(74) &13.678\,(83) & 13.606\,(58)$^{\rm r}$&                    &                 &               &              &              &-57.1\,(7.3)  &-3.3\,(7.2)   &-0.03&-0.09&                                                                                      \\
155  & 21:37:51.76 &  57:26:52.6$^{\rm r}$& 172  & 3172                      &                               &                 &                 &   12.9$^{\rm j}$&       &                 &10.962\,(29) &10.422\,(35) & 10.294\,(22)$^{\rm r}$&                    &                 &               &-20.4\,(4)    &-0.7\,(4)     &-57.1\,(7.3)  &-3.3\,(7.2)   &-0.03&-0.09&                                                                                      \\
156  & 21:37:53.94 &  57:26:42.3$^{\rm r}$& 173  & 3173                      &                               &                 &                 &   14.5$^{\rm j}$&       &                 &11.326\,(29) &10.614\,(32) & 10.342\,(22)$^{\rm r}$&                    &                 &               &13.3\,(18.1)  &6.6\,(18.1)   &-25.5\,(6.8)  &8\,(6.8)      &-0.12&0.06 &                                                                                      \\
157  & 21:37:54.46 &  57:26:33.3$^{\rm r}$& 174  & 3174                      &                               &                 &                 &   14.9$^{\rm j}$&       &                 &13.026\,(31) &12.489\,(33) & 12.360\,(27)$^{\rm r}$&                    &                 &               &7.9\,(4)      &-9.3\,(4)     &-15.8\,(6.8)  &3.1\,(6.7)    &     &     &                                                                                      \\
158  & 21:37:47.70 &  57:26:17.0$^{\rm j}$& 175  & 3175                      &                               &                 &                 &   15.1$^{\rm j}$&       &                 &             &             &                       &                    &                 &               &              &              &              &              &     &     &no star \footnote[26]{[j] No. exists 2 times on plate}                                    \\
159  & 21:37:47.06 &  57:26:04.6$^{\rm r}$& 176  & 3176                      &                               &                 &                 &   15.2$^{\rm j}$&       &                 &12.544\,(27) &11.850\,(28) & 11.671\,(26)$^{\rm r}$&                    &                 &               &1.1\,(4.1)    &-14.6\,(4.1)  &15.5\,(6.9)   &-31.4\,(6.8)  &     &     &                                                                                      \\
160  & 21:37:41.81 &  57:29:34.0$^{\rm r}$& 177  & 3177                      &                               &                 &                 &   12.6$^{\rm j}$&       &                 &11.274\,(23) &10.990\,(28) & 10.937\,(23)$^{\rm r}$&                    &                 &               &-4.6\,(4.1)   &1\,(4.1)      &-7.8\,(7.1)   &1.6\,(7.1)    &-0.31&0.28 &                                                                                      \\
161  & 21:37:35.43 &  57:30:25.6$^{\rm j}$& 179  & 3179                      &                               &                 &                 &   14.7$^{\rm j}$&       &                 &             &             &                       &                    &                 &               &              &              &              &              &     &     &no star                                                                          \\
162  & 21:37:10.23 &  57:30:30.2$^{\rm r}$& 180  & 3180                      &                               &                 &                 &   14.9$^{\rm j}$&       &                 &12.506\,(27) &12.084\,(33) & 11.959\,(29)$^{\rm r}$&                    &                 &               &6.9\,(3.8)    &9.1\,(3.8)    &              &              &     &     &                                                                                      \\
163  & 21:37:18.41 &  57:31:20.7$^{\rm r}$& 181  & 448                       &                               &  12.67$^{\rm l}$&  12.42$^{\rm f}$&  11.84$^{\rm e}$&       &                 &10.385\,(23) &10.119\,(27) & 10.049\,(23)$^{\rm r}$&        A0$^{\rm e}$&                 &  1.8$^{\rm e}$&-2.7\,(2)     &7.6\,(2)      &-3.9\,(0.6)   &-1.4\,(1.5)   &-0.2 &0.12 &[h] wrong                                               \\    
164  & 21:37:19.57 &  57:30:48.9$^{\rm r}$& 182  & 450                       &                               &  12.43$^{\rm l}$&   12.4$^{\rm f}$&  11.93$^{\rm e}$&       &                 &10.892\,(23) &10.786\,(27) & 10.763\,(20)$^{\rm r}$&        B8$^{\rm e}$&                 &  1.8$^{\rm e}$&-5.6\,(2)     &-0.7\,(2)     &-3.5\,(1.4)   &-5.4\,(1.1)   &-0.19&-0.14&                                                                                      \\
165  & 21:37:23.60 &  57:30:57.7$^{\rm r}$& 183  & 3183                      &                               &                 &                 &   14.8$^{\rm j}$&       &                 &13.125\,(23) &12.894\,(27) & 12.791\,(30)$^{\rm r}$&                    &                 &               &-3.5\,(3.8)   &-2.1\,(3.8)   &-4\,(7.3)     &-4.8\,(7.3)   &     &     &                                                                                      \\
166  & 21:37:19.75 &  57:33:10.7$^{\rm r}$& 184  & 3184                      &                               &                 &                 &   14.9$^{\rm j}$&       &                 &10.554\,(23) &9.485 \,(27) & 9.210 \,(21)$^{\rm r}$&                    &                 &               &1.9\,(4.7)    &-6.6\,(4.7)   &-23.1\,(7.1)  &10.7\,(7)     &0.06 &-0.19&                                                                                      \\
167  & 21:37:09.08 &  57:32:18.3$^{\rm r}$& 185  & 3185                      &                               &                 &                 &   13.3$^{\rm j}$&       &                 &10.306\,(23) &9.537 \,(26) & 9.350 \,(20)$^{\rm r}$&                    &                 &               &-6.3\,(4.9)   &-10.5\,(4.9)  &              &              &0.09 &-0.85&                                                                                      \\
168  & 21:37:08.37 &  57:33:50.9$^{\rm r}$& 186  & 446                       &                               &                 &   9.73$^{\rm h}$&   9.42$^{\rm h}$&       &                 &8.848 \,(23) &8.753 \,(26) & 8.735 \,(21)$^{\rm r}$&        A8$^{\rm h}$&                 & 0.25$^{\rm h}$&6.7\,(1.2)    &-16.9\,(1.2)  &6.6\,(0.6)    &-14.3\,(0.7)  &-1.34&-1.4 &                                                                                      \\
169  & 21:37:32.23 &  57:31:57.7$^{\rm j}$& 188  & 3188                      &                               &                 &                 &   14.8$^{\rm j}$&       &                 &             &             &                       &                    &                 &               &              &              &              &              &     &     &no star                                                                          \\
170  & 21:37:47.16 &  57:31:39.3$^{\rm r}$& 189  & 3189                      &                               &                 &                 &   14.6$^{\rm j}$&       &                 &12.305\,(27) &11.958\,(34) & 11.826\,(28)$^{\rm r}$&                    &                 &               &13.1\,(3.9)   &1.2\,(3.9)    &52.3\,(7.4)   &6.5\,(7.4)    &-0.29&-0.33&                                                                                      \\
171  & 21:37:15.79 &  57:35:00.6$^{\rm r}$& 190  & 3190                      &                               &                 &                 &   14.9$^{\rm j}$&       &                 &10.703\,(22) &9.648 \,(26) & 9.394 \,(21)$^{\rm r}$&                    &                 &               &-5.2\,(4.7)   &-4\,(4.7)     &-0.2\,(6.9)   &7.7\,(7)      &     &     &                                                                                      \\
172  & 21:37:25.95 &  57:34:33.0$^{\rm r}$& 192  & 3192                      &                               &                 &                 &   15.1$^{\rm j}$&       &                 &13.015\,(25) &12.541\,(28) & 12.443\,(26)$^{\rm r}$&                    &                 &               &-5.1\,(3.9)   &4\,(3.9)      &1.3\,(7.1)    &19.6\,(7.1)   &     &     &                                                                                      \\
173  & 21:37:24.42 &  57:35:08.0$^{\rm r}$& 193  & 3193                      &                               &                 &                 &   13.3$^{\rm j}$&       &                 &11.468\,(22) &11.079\,(27) & 11.003\,(21)$^{\rm r}$&                    &                 &               &-13.7\,(3.8)  &-3.1\,(3.8)   &-11.9\,(7)    &10.9\,(6.9)   &0.82 &0.01 &                                                                                      \\
174  & 21:37:40.94 &  57:33:37.3$^{\rm r}$& 194  & 452                       &                               &   8.18$^{\rm l}$&   8.58$^{\rm f}$&   8.24$^{\rm e}$&       &                 &7.697 \,(19) &7.647 \,(31) & 7.611 \,(21)$^{\rm r}$&     B3 B5$^{\rm e}$&      V$^{\rm p}$&1.7 1$^{\rm e}$&-6.5\,(1.9)   &-2.8\,(2.2)   &              &              &-0.39&-0.23&                                                                                      \\
175  & 21:37:41.85 &  57:33:38.4$^{\rm r}$& 195  & 3195                      &                               &                 &                 &   11.9$^{\rm j}$&       &                 &9.578 \,(27) &9.446 \,(42) & 9.459 \,(21)$^{\rm r}$&                    &                 &               &3.2\,(1.7)    &-0.4\,(1.7)   &              &              &-0.27&-0.14&                                                                                      \\
176  & 21:37:51.34 &  57:32:18.7$^{\rm r}$& 196  & 3196                      &                               &                 &                 &   14.8$^{\rm j}$&       &                 &11.463\,(26) &10.768\,(32) & 10.533\,(20)$^{\rm r}$&                    &                 &               &-6.7\,(3.8)   &-6.2\,(3.8)   &-8.8\,(6.8)   &9.6\,(6.9)    &     &     &                                                                                      \\
177  & 21:37:58.59 &  57:31:34.6$^{\rm r}$& 197  & 3197                      &                               &                 &                 &   14.1$^{\rm j}$&       &                 &12.278\,(38) &11.853\,(47) & 11.737\,(35)$^{\rm r}$&                    &                 &               &-4.7\,(5.3)   &-8.3\,(5.3)   &              &              &0.58 &-0.31&near 178                                                                          \\
178  & 21:37:59.13 &  57:31:35.5$^{\rm r}$& 198  & 3198                      &                               &                 &                 &   14.8$^{\rm j}$&       &                 &11.597\,()   &10.849\,()   &   10.642\,()$^{\rm r}$&                    &                 &               &              &              &              &              &     &     &near 177                                                                          \\
179  & 21:37:57.12 &  57:32:33.7$^{\rm r}$& 199  & 3199                      &                               &                 &                 &   15.1$^{\rm j}$&       &                 &11.397\,(31) &10.468\,(37) & 10.162\,(22)$^{\rm r}$&                    &                 &               &-10.4\,(3.9)  &-8.5\,(3.9)   &-18\,(7)      &-41\,(6.9)    &     &     &                                                                                      \\
180  & 21:37:56.68 &  57:32:50.5$^{\rm r}$& 200  & 3200                      &                               &                 &                 &   14.5$^{\rm j}$&       &                 &11.238\,(26) &10.426\,(31) & 10.202\,(20)$^{\rm r}$&                    &                 &               &-6.5\,(3.9)   &2.4\,(3.9)    &-2.4\,(6.8)   &9.2\,(6.8)    &-0.17&0.27 &                                                                                      \\
181  & 21:38:00.35 &  57:32:32.7$^{\rm r}$& 201  & 3201                      &                               &                 &                 &     15$^{\rm j}$&       &                 &11.415\,(27) &10.471\,(32) & 10.184\,(22)$^{\rm r}$&                    &                 &               &-10.4\,(3.9)  &-0.5\,(3.9)   &-20.3\,(6.8)  &17.4\,(6.8)   &     &     &                                                                                      \\
182  & 21:38:00.98 &  57:33:00.1$^{\rm r}$& 202  & 3202                      &                               &                 &                 &   14.9$^{\rm j}$&       &                 &10.763\,(27) &9.789 \,(30) & 9.491 \,(20)$^{\rm r}$&                    &                 &               &-2\,(4.7)     &-5.8\,(4.7)   &-1.8\,(6.3)   &-12.6\,(6.2)  &     &     &                                                                                      \\
183  & 21:37:47.24 &  57:34:54.2$^{\rm r}$& 203  & 3203                      &                               &  15.27$^{\rm l}$&  14.66$^{\rm l}$&  13.55$^{\rm l}$&       &                 &11.287\,(22) &10.915\,(26) & 10.783\,(21)$^{\rm r}$&                    &                 &               &-9.8\,(3.9)   &-3.5\,(3.9)   &-6.2\,(7)     &6.8\,(7)      &0.19 &-0.14&                                                                                      \\
184  & 21:37:26.01 &  57:36:01.6$^{\rm r}$& 204  & 3204                      &                               &                 &  13.21$^{\rm f}$&   12.7$^{\rm e}$&       &                 &11.395\,(22) &11.266\,(26) & 11.176\,(20)$^{\rm r}$&        A1$^{\rm e}$&                 &  1.5$^{\rm e}$&-1.1\,(3.9)   &-5.1\,(3.9)   &-0.7\,(0.7)   &-1.9\,(1.3)   &-0.48&0.04 &same star                                                                             \\
185  & 21:37:18.38 &  57:36:30.8$^{\rm r}$& 205  & 3205                      &                               &                 &                 &   15.1$^{\rm j}$&       &                 &11.975\,(23) &11.280\,(26) & 11.133\,(23)$^{\rm r}$&                    &                 &               &-4.6\,(3.8)   &-7.3\,(3.8)   &-3.7\,(7.2)   &-2.7\,(7.1)   &     &     &                                                                                      \\
186  & 21:37:17.42 &  57:36:49.3$^{\rm r}$& 206  & 3206                      &                               &                 &                 &   15.1$^{\rm j}$&       &                 &12.988\,(26) &12.385\,(28) & 12.311\,(29)$^{\rm r}$&                    &                 &               &-1.2\,(3.8)   &-29.2\,(3.8)  &0.4\,(7.9)    &-26.8\,(7.5)  &     &     &                                                                                      \\
187  & 21:37:32.25 &  57:36:31.6$^{\rm r}$& 207  & 3207                      &                               &                 &                 &   15.2$^{\rm j}$&       &                 &13.242\,(26) &12.877\,(33) & 12.733\,(34)$^{\rm r}$&                    &                 &               &-6.8\,(3.8)   &-12.9\,(3.8)  &-11\,(7.9)    &-13.6\,(7.7)  &     &     &                                                                                      \\
188  & 21:37:23.33 &  57:37:33.3$^{\rm r}$& 208  & 3208                      &                               &                 &                 &   14.8$^{\rm j}$&       &                 &11.723\,(22) &11.040\,(26) & 10.861\,(20)$^{\rm r}$&                    &                 &               &1.5\,(3.8)    &-2.9\,(3.8)   &8.7\,(7.4)    &10.9\,(7.4)   &-0.07&0.38 &                                                                                      \\
189  & 21:37:42.92 &  57:36:31.4$^{\rm r}$& 209  & 3209                      &                               &                 &                 &   12.4$^{\rm j}$&       &                 &10.571\,(22) &10.096\,(27) & 9.837 \,(21)$^{\rm r}$&                    &                 &               &-7.7\,(2.7)   &-23.2\,(2.7)  &-4.1\,(1.3)   &-3.6\,(2)     &-0.02&-0.14&                                                                                      \\
190  & 21:37:42.80 &  57:37:06.5$^{\rm r}$& 210  & 3210                      &                               &                 &                 &     12$^{\rm j}$&       &                 &9.820 \,(23) &9.179 \,(26) & 9.063 \,(18)$^{\rm r}$&                    &                 &               &-7.7\,(2.7)   &-4.9\,(2.7)   &-12.5\,(0.7)  &-4.4\,(2.1)   &0.47 &-0.2 &                                                                                      \\
191  & 21:37:36.30 &  57:37:23.6$^{\rm r}$& 211  & 3211                      &                               &                 &                 &     15$^{\rm j}$&       &                 &11.518\,(22) &10.750\,(27) & 10.535\,(21)$^{\rm r}$&                    &                 &               &-1.4\,(3.8)   &-2.2\,(3.8)   &              &              &     &     &                                                                                      \\
192  & 21:37:47.56 &  57:37:06.6$^{\rm r}$& 212  & 3212                      &                               &                 &                 &   14.4$^{\rm j}$&       &                 &12.463\,(23) &12.052\,(30) & 11.980\,(23)$^{\rm r}$&                    &                 &               &13.2\,(3.9)   &-2\,(3.9)     &21.6\,(7.5)   &-1.5\,(7.4)   &-2.34&0.37 &                                                                                      \\
193  & 21:38:09.43 &  57:35:26.4$^{\rm r}$& 213  & 3213                      &                               &                 &                 &   15.2$^{\rm j}$&       &                 &13.315\,(27) &12.878\,(30) & 12.741\,(22)$^{\rm r}$&                    &                 &               &-1.5\,(3.8)   &-2.6\,(3.8)   &-1.5\,(7.8)   &-3.8\,(7.6)   &     &     &                                                                                      \\
194  & 21:38:11.07 &  57:35:09.5$^{\rm r}$& 214  & 3214                      &                               &                 &                 &   15.1$^{\rm j}$&       &                 &10.244\,(26) &9.201 \,(29) & 8.830 \,(20)$^{\rm r}$&                    &                 &               &-1.6\,(4.8)   &-4.8\,(4.8)   &3.8\,(6.8)    &2.4\,(7)      &     &     &                                                                                      \\
195  & 21:38:01.59 &  57:36:52.6$^{\rm r}$& 217  & 455                       &                               &  11.86$^{\rm l}$&  11.86$^{\rm l}$&  11.12$^{\rm l}$&       &                 &9.366 \,(26) &9.044 \,(31) & 8.938 \,(20)$^{\rm r}$&        B5$^{\rm p}$&  Ib-II$^{\rm p}$&               &-5.8\,(1.7)   &-3\,(1.7)     &-4\,(0.6)     &-1.9\,(0.7)   &-0.03&0.05 &                                                                                      \\
196  & 21:37:57.24 &  57:37:24.0$^{\rm r}$& 218  & 3218                      &                               &                 &                 &   15.1$^{\rm j}$&       &                 &13.344\,(27) &12.999\,(29) & 12.867\,(20)$^{\rm r}$&                    &                 &               &-1.6\,(3.8)   &-1.9\,(3.8)   &-0.6\,(7.7)   &-8.2\,(7.4)   &     &     &                                                                                      \\
197  & 21:38:03.93 &  57:36:35.4$^{\rm r}$& 219  & 3219                      &                               &                 &                 &   12.7$^{\rm j}$&       &                 &10.178\,(26) &9.492 \,(31) & 9.316 \,(20)$^{\rm r}$&                    &                 &               &6.1\,(4.7)    &-3\,(4.7)     &-1.8\,(7.4)   &-7.7\,(7.4)   &-0.46&0.02 &                                                                                      \\
198  & 21:38:10.87 &  57:36:03.8$^{\rm r}$& 220  & 3220                      &                               &                 &                 &   13.9$^{\rm j}$&       &                 &12.270\,(27) &11.922\,(31) & 11.852\,(24)$^{\rm r}$&                    &                 &               &-3.3\,(3.9)   &-2.2\,(3.9)   &0.9\,(6.8)    &7\,(6.8)      &-0.39&0.14 &                                                                                      \\
199  & 21:38:18.55 &  57:35:40.1$^{\rm r}$& 221  & 3221                      &                               &                 &  14.23$^{\rm l}$&  12.94$^{\rm l}$&       &                 &10.088\,(27) &9.438 \,(29) & 9.266 \,(20)$^{\rm r}$&                    &                 &               &-6\,(5)       &-4.9\,(5)     &-6.6\,(7.4)   &-7.4\,(7.3)   &-0.05&-0.26&                                                                                      \\
200  & 21:37:56.50 &  57:38:08.8$^{\rm r}$& 222  & 3222                      &                               &                 &                 &   15.1$^{\rm j}$&       &                 &13.217\,(27) &12.839\,(31) & 12.718\,(24)$^{\rm r}$&                    &                 &               &-4.1\,(3.8)   &0\,(3.8)      &-3.4\,(7.4)   &-1.3\,(7.4)   &     &     &                                                                                      \\
201  & 21:38:00.06 &  57:38:07.0$^{\rm r}$& 223  & 3223                      &                               &                 &                 &   14.7$^{\rm j}$&       &                 &12.894\,(27) &12.515\,(35) & 12.417\,(18)$^{\rm r}$&                    &                 &               &              &              &              &              &-1.34&0.23 &                                                                                      \\
202  & 21:38:00.96 &  57:38:06.9$^{\rm r}$& 224  & 3224                      &                               &  13.92$^{\rm l}$&  13.58$^{\rm f}$&  13.05$^{\rm e}$&       &                 &11.856\,(27) &11.696\,(32) & 11.594\,(21)$^{\rm r}$&        A7$^{\rm e}$&                 &  1.1$^{\rm e}$&-4.8\,(4.4)   &1.2\,(4.4)    &-4.9\,(1.6)   &-3.7\,(0.8)   &0.09 &-0.23&                                                                                      \\
203  & 21:37:57.99 &  57:38:51.3$^{\rm r}$& 225  & 5073                      &                               &                 &                 &   14.7$^{\rm j}$&       &                 &11.471\,(24) &10.774\,(30) & 10.559\,(19)$^{\rm r}$&        K7$^{\rm q}$&                 &               &-0.2\,(3.9)   &-4.5\,(3.9)   &5.3\,(7.3)    &-16.1\,(7.4)  &     &     &                                                                                      \\
204  & 21:38:10.61 &  57:37:59.3$^{\rm r}$& 226  & 3226                      &                               &                 &                 &   14.4$^{\rm j}$&       &                 &12.599\,(27) &12.304\,(33) & 12.154\,(24)$^{\rm r}$&                    &                 &               &6.4\,(3.8)    &-4.6\,(3.8)   &5.3\,(7.3)    &-4.5\,(7.3)   &-0.82&-0.59&                                                                                      \\
205  & 21:38:03.50 &  57:38:43.7$^{\rm r}$& 227  & 3227                      &                               &                 &                 &   14.5$^{\rm j}$&       &                 &11.740\,(26) &11.096\,(30) & 10.936\,(19)$^{\rm r}$&                    &                 &               &21\,(5.1)     &26\,(5.1)     &9.6\,(7.3)    &43.7\,(7.4)   &     &     &new coordinates                                                                       \\     
206  & 21:37:10.09 &  57:40:30.0$^{\rm r}$& 229  & 3229                      &                               &  14.11$^{\rm l}$&  13.07$^{\rm h}$&  13.01$^{\rm h}$&       &                 &11.762\,(23) &11.539\,(24) & 11.514\,(21)$^{\rm r}$&        B8$^{\rm h}$&                 & 2.22$^{\rm h}$&-7.2\,(3.8)   &-6.1\,(3.8)   &-9.7\,(1.1)   &-1.5\,(1.7)   &0.13 &-0.36&                                                                                      \\
207  & 21:37:18.80 &  57:39:52.5$^{\rm r}$& 230  & 3230                      &                               &                 &                 &   14.8$^{\rm j}$&       &                 &11.460\,(25) &10.702\,(26) & 10.518\,(21)$^{\rm r}$&                    &                 &               &0.3\,(3.8)    &-4.3\,(3.8)   &-3.5\,(7.4)   &-0.9\,(7.4)   &     &     &                                                                                      \\
208  & 21:37:32.60 &  57:39:06.4$^{\rm r}$& 231  & 3231                      &                               &                 &                 &   15.1$^{\rm j}$&       &                 &11.964\,(23) &11.261\,(28) & 11.069\,(23)$^{\rm r}$&                    &                 &               &6\,(3.8)      &-13.1\,(3.8)  &18.8\,(7.4)   &-19.6\,(7.5)  &     &     &                                                                                      \\
209  & 21:37:10.63 &  57:42:15.5$^{\rm r}$& 232  & 3232                      &                               &  16.03$^{\rm l}$&   15.4$^{\rm f}$&  14.25$^{\rm e}$&       &                 &11.946\,(23) &11.493\,(28) & 11.370\,(21)$^{\rm r}$&        F7$^{\rm e}$&                 &    2$^{\rm e}$&-1.7\,(3.8)   &4.4\,(3.8)    &1.6\,(7.4)    &11.4\,(7.4)   &-0.04&0.27 &                                                                                      \\
210  & 21:37:15.95 &  57:41:48.2$^{\rm r}$& 233  & 3233                      &                               &                 &                 &   14.4$^{\rm j}$&       &                 &10.805\,(22) &9.938 \,(26) & 9.733 \,(20)$^{\rm r}$&                    &                 &               &-4.8\,(4.7)   &-3.7\,(4.7)   &-2.6\,(7.5)   &-7.7\,(7.5)   &0.11 &0.13 &                                                                                      \\
211  & 21:37:34.33 &  57:40:40.8$^{\rm r}$& 234  & 3234                      &                               &  15.83$^{\rm l}$&  15.28$^{\rm f}$&  14.18$^{\rm e}$&       &                 &11.896\,(25) &11.424\,(31) & 11.325\,(25)$^{\rm r}$&        F9$^{\rm e}$&                 &  1.7$^{\rm e}$&-0.6\,(3.8)   &-6.4\,(3.8)   &-3\,(7.4)     &-21.7\,(7.5)  &0.29 &-0.36&                                                                                      \\
212  & 21:37:31.60 &  57:41:40.1$^{\rm r}$& 235  & 3235                      &                               &                 &                 &   15.1$^{\rm j}$&       &                 &12.932\,(27) &12.568\,(28) & 12.431\,(29)$^{\rm r}$&                    &                 &               &1.6\,(5.1)    &-60.6\,(5.1)  &-10.2\,(7.6)  &-22\,(7.7)    &     &     &                                                                                      \\
213  & 21:37:07.64 &  57:43:52.9$^{\rm r}$& 236  & 3236                      &                               &                 &                 &     15$^{\rm j}$&       &                 &12.638\,(25) &12.271\,(30) & 12.089\,(25)$^{\rm r}$&                    &                 &               &-4\,(3.8)     &-7.1\,(3.8)   &1.3\,(7.4)    &-23\,(7.5)    &     &     &                                                                                      \\
214  & 21:37:09.20 &  57:43:55.1$^{\rm j}$& 237  & 3237                      &                               &                 &                 &   14.2$^{\rm j}$&       &                 &             &             &                       &                    &                 &               &              &              &              &              &     &     &no star                                                                          \\
215  & 21:37:29.50 &  57:43:06.8$^{\rm r}$& 239  & 3239                      &                               &                 &                 &   14.8$^{\rm j}$&       &                 &13.002\,(26) &12.697\,(31) & 12.630\,(33)$^{\rm r}$&                    &                 &               &-1.5\,(3.8)   &-3\,(3.8)     &0.5\,(7.4)    &-12.2\,(7.4)  &     &     &                                                                                      \\
216  & 21:37:38.68 &  57:41:25.2$^{\rm r}$& 241  & 3241                      &                               &                 &                 &   14.4$^{\rm j}$&       &                 &12.490\,(23) &12.097\,(28) & 12.031\,(25)$^{\rm r}$&                    &                 &               &1.7\,(3.8)    &1.6\,(3.8)    &25.6\,(7.4)   &-15.6\,(7.6)  &0.11 &1.08 &                                                                                      \\
217  & 21:37:40.70 &  57:41:16.7$^{\rm r}$& 242  & 3242                      &                               &                 &                 &   14.6$^{\rm j}$&       &                 &12.457\,(23) &12.019\,(27) & 11.910\,(24)$^{\rm r}$&                    &                 &               &-3.3\,(3.8)   &-0.4\,(3.8)   &-10.2\,(7.3)  &-2.5\,(7.3)   &     &     &                                                                                      \\
218  & 21:37:52.53 &  57:40:45.9$^{\rm r}$& 243  & 3243                      &                               &                 &                 &   14.1$^{\rm j}$&       &                 &11.079\,(26) &10.294\,(29) & 10.071\,(18)$^{\rm r}$&                    &                 &               &-1.5\,(3.9)   &5.1\,(3.9)    &2\,(7.3)      &2.6\,(7.4)    &-0.26&0.47 &                                                                                      \\
219  & 21:37:55.80 &  57:40:15.5$^{\rm r}$& 244  & 3244                      &                               &                 &                 &   14.3$^{\rm j}$&       &                 &12.530\,(29) &12.153\,(33) & 11.995\,(24)$^{\rm r}$&                    &                 &               &-14.7\,(3.8)  &-7.1\,(3.8)   &-34.6\,(7.3)  &-5.3\,(7.4)   &0.95 &-0.93&                                                                                      \\
220  & 21:37:59.23 &  57:40:13.0$^{\rm r}$& 245  & 3245                      &                               &                 &                 &   14.2$^{\rm j}$&       &                 &12.629\,(43) &12.387\,(46) & 12.266\,(39)$^{\rm r}$&                    &                 &               &-8.7\,(5.1)   &-17\,(5.1)    &-91.5\,(7.6)  &-97.4\,(7.6)  &-0.03&-0.04&                                                                                      \\
221  & 21:38:05.11 &  57:39:49.1$^{\rm r}$& 246  & 3246                      &                               &                 &                 &   14.9$^{\rm j}$&       &                 &12.856\,(31) &12.378\,(31) & 12.307\,(26)$^{\rm r}$&                    &                 &               &22.9\,(5.4)   &-36.1\,(5.4)  &38.7\,(7.2)   &-61.8\,(7.2)  &     &     &                                                                                      \\
222  & 21:38:10.14 &  57:39:32.7$^{\rm r}$& 247  & 3247                      &                               &                 &                 &   14.9$^{\rm j}$&       &                 &12.072\,(26) &11.390\,(30) & 11.213\,(19)$^{\rm r}$&                    &                 &               &2.2\,(3.9)    &4.6\,(3.9)    &6\,(7.3)      &-1.6\,(7.3)   &     &     &                                                                                      \\
223  & 21:37:46.57 &  57:42:26.6$^{\rm r}$& 248  & 3248                      &                               &                 &                 &   14.7$^{\rm j}$&       &                 &12.509\,(26) &12.073\,(30) & 11.986\,(24)$^{\rm r}$&                    &                 &               &-14.2\,(3.8)  &-19.2\,(3.8)  &-21.9\,(7.5)  &-24.7\,(7.3)  &     &     &                                                                                      \\
224  & 21:37:33.95 &  57:43:42.2$^{\rm r}$& 249  & 3249                      &                               &                 &                 &   14.6$^{\rm j}$&       &                 &11.292\,(23) &10.505\,(26) & 10.309\,(21)$^{\rm r}$&                    &                 &               &0.6\,(3.8)    &-14.1\,(3.8)  &-0.5\,(7.4)   &-9.1\,(7.4)   &     &     &                                                                                      \\
225  & 21:37:27.41 &  57:44:17.6$^{\rm r}$& 251  & 3251                      &                               &                 &                 &   14.8$^{\rm j}$&       &                 &12.858\,(39) &12.519\,(44) & 12.447\,(43)$^{\rm r}$&                    &                 &               &1.5\,(3.8)    &8.5\,(3.8)    &20.3\,(7.6)   &58.1\,(7.6)   &     &     &                                                                                      \\
226  & 21:37:22.58 &  57:45:07.2$^{\rm r}$& 252  & 712                       &                               &  12.51$^{\rm l}$&  12.48$^{\rm f}$&  11.96$^{\rm e}$&       &                 &10.766\,(25) &10.653\,(26) & 10.616\,(20)$^{\rm r}$&        B7$^{\rm e}$&                 &    2$^{\rm e}$&-16.7\,(2.7)  &-4.8\,(2.7)   &-5.3\,(1.1)   &-3.3\,(2)     &-0.02&-0.28&                                                                                      \\
227  & 21:37:20.76 &  57:45:34.2$^{\rm r}$& 253  & 3253                      &                               &                 &                 &   12.7$^{\rm j}$&       &                 &10.885\,(23) &10.475\,(28) & 10.386\,(23)$^{\rm r}$&                    &                 &               &-1.7\,(3.9)   &9.9\,(3.9)    &5.8\,(8.3)    &13.9\,(8.3)   &-1.31&0.3  &                                                                                      \\
228  & 21:37:32.47 &  57:46:01.1$^{\rm r}$& 254  & 3254                      &                               &                 &                 &   13.1$^{\rm j}$&       &                 &10.299\,(23) &9.566 \,(27) & 9.407 \,(21)$^{\rm r}$&                    &                 &               &0.3\,(4.7)    &-0.5\,(4.7)   &6.5\,(8.3)    &-8.9\,(8.3)   &-0.28&-0.37&                                                                                      \\
229  & 21:37:42.91 &  57:45:04.3$^{\rm r}$& 255  & 3255                      &                               &                 &                 &   14.8$^{\rm j}$&       &                 &12.580\,(25) &12.097\,(27) & 12.004\,(23)$^{\rm r}$&                    &                 &               &-3\,(3.8)     &-9.5\,(3.8)   &-9.4\,(7.4)   &-0.5\,(7.4)   &     &     &                                                                                      \\
230  & 21:37:47.16 &  57:44:14.3$^{\rm r}$& 256  & 3256                      &                               &                 &                 &   11.2$^{\rm j}$&       &                 &4.573 \,(186)&3.492 \,(202)&3.048 \,(242)$^{\rm r}$&                    &                 &               &-0.1\,(2.8)   &9.2\,(2.8)    &2\,(1.3)      &5.3\,(1.5)    &-0.75&0.55 &                                                                                      \\
231  & 21:37:56.58 &  57:42:52.7$^{\rm r}$& 257  & 3257                      &                               &  14.79$^{\rm l}$&  14.99$^{\rm f}$&  14.28$^{\rm e}$& 13.85 &  13.33$^{\rm i}$&12.872\,(35) &12.685\,(41) & 12.561\,(33)$^{\rm r}$&        B2$^{\rm e}$&                 &  2.9$^{\rm e}$&-2.4\,(3.8)   &-2.4\,(3.8)   &-1\,(7.6)     &-20.7\,(7.3)  &0.03 &0.22 &                                                                                      \\
232  & 21:37:56.11 &  57:42:20.8$^{\rm r}$& 258  & 3258                      &                               &  14.69$^{\rm l}$&  14.27$^{\rm f}$&  13.66$^{\rm e}$& 13.35 &     13$^{\rm i}$&12.240\,(29) &11.975\,(31) & 11.844\,(21)$^{\rm r}$&        A2$^{\rm e}$&                 &  1.7$^{\rm e}$&-6.6\,(3.8)   &-1.3\,(3.8)   &-18.3\,(7.5)  &-11.3\,(7.5)  &0.33 &-0.25&                                                                                      \\
233  & 21:37:59.70 &  57:41:31.2$^{\rm r}$& 259  & 3259                      &                               &                 &                 &   13.6$^{\rm j}$&       &                 &11.833\,(26) &11.448\,(31) & 11.338\,(18)$^{\rm r}$&                    &                 &               &-12\,(10.8)   &-8.2\,(10.8)  &-19.3\,(7.4)  &-12.5\,(7.4)  &1.23 &-0.84&                                                                                      \\
234  & 21:38:04.71 &  57:41:00.0$^{\rm r}$& 260  & 3260                      &                               &                 &                 &   14.6$^{\rm j}$&       &                 &10.418\,(26) &9.442 \,(31) & 9.119 \,(19)$^{\rm r}$&                    &                 &               &6.1\,(4.8)    &-6.6\,(4.8)   &97.9\,(7.4)   &-8.4\,(7.4)   &0.88 &-0.36&                                                                                      \\
235  & 21:38:02.07 &  57:43:36.6$^{\rm r}$& 261  & 3261                      &                               &                 &                 &   14.7$^{\rm j}$&       &                 &12.927\,(29) &12.545\,(36) & 12.418\,(21)$^{\rm r}$&                    &                 &               &-0.9\,(3.9)   &-1.1\,(3.9)   &35.9\,(5.1)   &3.7\,(5.1)    &     &     &                                                                                      \\
236  & 21:38:13.68 &  57:11:24.0$^{\rm r}$& 263  & 3263                      &                               &                 &                 &   13.6$^{\rm j}$&       &                 &12.117\,(26) &11.882\,(32) & 11.730\,(19)$^{\rm r}$&                    &                 &               &-6\,(4)       &-3.6\,(4)     &-16.4\,(6.8)  &7.9\,(6.8)    &-0.03&-0.13&                                                                                      \\
237  & 21:38:25.97 &  57:12:02.3$^{\rm r}$& 264  & 3264                      &                               &                 &                 &   14.8$^{\rm j}$&       &                 &12.768\,(35) &12.380\,(38) & 12.196\,(35)$^{\rm r}$&                    &                 &               &-12.2\,(4)    &0.8\,(4)      &-11.8\,(6.8)  &32.7\,(6.8)   &     &     &                                                                                      \\
238  & 21:38:31.92 &  57:13:20.8$^{\rm r}$& 265  & 3265                      &                               &                 &                 &   14.6$^{\rm j}$&       &                 &10.755\,(26) &9.830 \,(29) & 9.564 \,(19)$^{\rm r}$&                    &                 &               &0.4\,(5.1)    &-4.6\,(5.1)   &              &              &-0.09&0.02 &                                                                                      \\
239  & 21:38:05.84 &  57:14:41.7$^{\rm r}$& 266  & 3266                      &                               &  14.52$^{\rm l}$&  14.02$^{\rm h}$&  13.07$^{\rm h}$& 13.36 &  13.02$^{\rm i}$&12.455\,(26) &12.267\,(29) & 12.172\,(21)$^{\rm r}$&        A1$^{\rm h}$&                 &  1.5$^{\rm h}$&-3.1\,(4)     &-2\,(4)       &0.8\,(6.8)    &-2.5\,(6.8)   &-0.07&-0.13&Dec [h] imprec.                                                               \\
240  & 21:38:17.12 &  57:14:37.6$^{\rm r}$& 267  & 3267                      &                               &                 &                 &   14.3$^{\rm j}$&       &                 &10.632\,(24) &9.797 \,(31) & 9.556 \,(19)$^{\rm r}$&                    &                 &               &-4.4\,(5.1)   &-2.9\,(5.1)   &-10.4\,(6.8)  &7.8\,(6.9)    &0.17 &-0.05&                                                                                      \\
241  & 21:38:22.66 &  57:14:30.1$^{\rm r}$& 268  & 461                       &                               &  11.35$^{\rm l}$&  10.28$^{\rm l}$&   9.08$^{\rm l}$&       &                 &6.930 \,(19) &6.401 \,(51) & 6.256 \,(23)$^{\rm r}$&     G8 K5$^{\rm q}$&                 &               &-0.1\,(1.2)   &-1.5\,(1.1)   &1.5\,(0.6)    &-1.2\,(0.7)   &0.05 &-0.02&                                                                                      \\
242  & 21:38:30.95 &  57:14:20.0$^{\rm r}$& 269  & 3269                      &                               &                 &                 &   14.8$^{\rm j}$&       &                 &13.312\,(24) &13.070\,(35) & 12.918\,(32)$^{\rm r}$&                    &                 &               &-2\,(4)       &-2.6\,(4)     &-0.3\,(6.9)   &-3.2\,(6.9)   &-0.59&0.09 &                                                                                      \\
243  & 21:38:45.75 &  57:11:58.0$^{\rm r}$& 270  & 3270                      &                               &                 &                 &   14.8$^{\rm j}$&       &                 &12.754\,(26) &12.405\,(32) & 12.271\,(25)$^{\rm r}$&                    &                 &               &-3.4\,(4)     &-1.7\,(4)     &3\,(6.8)      &5.7\,(6.8)    &-0.2 &-0.02&                                                                                      \\
244  & 21:38:52.84 &  57:13:30.2$^{\rm r}$& 272  & 3272                      &                               &                 &                 &   14.9$^{\rm j}$&       &                 &11.721\,(26) &10.959\,(28) & 10.750\,(23)$^{\rm r}$&                    &                 &               &-4.5\,(4)     &-6.2\,(4)     &-7.9\,(6.8)   &-0.8\,(6.9)   &     &     &                                                                                      \\
245  & 21:38:44.57 &  57:14:10.4$^{\rm r}$& 273  & 464                       &                               &                 &                 &   10.1$^{\rm j}$&       &                 &9.236 \,(23) &9.046 \,(29) & 8.970 \,(20)$^{\rm r}$&        F5$^{\rm q}$&                 &               &15.7\,(1.6)   &17.4\,(1.6)   &21.3\,(0.7)   &16.6\,(0.6)   &-2.63&2.31 &                                                                                      \\
246  & 21:38:39.11 &  57:14:46.2$^{\rm r}$& 274  & 3274                      &                               &                 &                 &   11.9$^{\rm j}$&       &                 &10.894\,(26) &10.508\,(30) & 10.429\,(21)$^{\rm r}$&                    &                 &               &13.9\,(2.7)   &14.1\,(2.7)   &19.8\,(0.9)   &20.1\,(1.1)   &-2.58&2.34 &                                                                                      \\
247  & 21:38:26.99 &  57:14:55.0$^{\rm r}$& 275  & 3275                      &                               &  16.27$^{\rm l}$&  15.57$^{\rm h}$&  14.68$^{\rm h}$& 14.07 &  13.51$^{\rm i}$&12.530\,(27) &12.178\,(29) & 12.033\,(23)$^{\rm r}$&        A0$^{\rm h}$&                 & 2.78$^{\rm h}$&-5.8\,(4)     &1.1\,(4)      &-8\,(6.8)     &7.3\,(6.9)    &-0.15&0.25 &Dec [h] imprec.                                                               \\
248  & 21:38:29.63 &  57:15:06.6$^{\rm r}$& 276  & 3276                      &                               &  15.30$^{\rm l}$&   14.8$^{\rm f}$&  13.82$^{\rm e}$& 13.2  &  12.61$^{\rm i}$&11.730\,(32) &11.310\,(35) & 11.178\,(29)$^{\rm r}$&        F4$^{\rm e}$&                 &  1.8$^{\rm e}$&-12.4\,(4)    &-8.3\,(4)     &-37.6\,(7)    &-29.1\,(7.1)  &-0.2 &-0.01&                                                                                      \\
249  & 21:38:29.63 &  57:15:06.6$^{\rm r}$& 276  & 3276                      &                               &  15.30$^{\rm l}$&   14.8$^{\rm f}$&  13.82$^{\rm e}$& 13.2  &  12.61$^{\rm i}$&11.730\,(32) &11.310\,(35) & 11.178\,(29)$^{\rm r}$&        F4$^{\rm e}$&                 &  1.8$^{\rm e}$&              &              &              &              &-0.2 &-0.01&                                                                                      \\
250  & 21:38:25.60 &  57:15:39.8$^{\rm r}$& 277  & 3277                      &                               &                 &                 &   12.8$^{\rm j}$&       &                 &11.403\,(26) &11.129\,(31) & 11.002\,(21)$^{\rm r}$&                    &                 &               &-11.4\,(4)    &-6.1\,(4)     &-15\,(7.1)    &-5.3\,(7.1)   &0.29 &-0.17&                                                                                      \\
251  & 21:38:40.06 &  57:15:47.8$^{\rm r}$& 278  & 3278                      &                               &                 &                 &   14.8$^{\rm j}$&       &                 &12.945\,(29) &12.582\,(29) & 12.475\,(29)$^{\rm r}$&                    &                 &               &4.3\,(4)      &-6.4\,(4)     &8.9\,(6.8)    &-1.4\,(6.8)   &-0.53&0.37 &                                                                                      \\
252  & 21:38:43.96 &  57:15:56.2$^{\rm r}$& 279  & 3279                      &                               &                 &                 &   14.9$^{\rm j}$&       &                 &13.307\,(29) &13.028\,(29) & 12.927\,(32)$^{\rm r}$&                    &                 &               &13\,(4)       &-3.8\,(4)     &71.9\,(6.8)   &4.2\,(6.8)    &     &     &                                                                                      \\
253  & 21:38:08.59 &  57:17:45.0$^{\rm r}$& 280  & 3280                      &                               &                 &                 &     15$^{\rm j}$&       &                 &13.039\,(27) &12.721\,(30) & 12.603\,(28)$^{\rm r}$&                    &                 &               &-7.6\,(4)     &3.6\,(4)      &-7.1\,(6.8)   &10.9\,(6.7)   &     &     &                                                                                      \\
254  & 21:38:12.97 &  57:17:21.0$^{\rm r}$& 281  & 3281                      &                               &                 &                 &     15$^{\rm j}$&       &                 &13.275\,(35) &12.843\,(40) &   12.481\,()$^{\rm r}$&                    &                 &               &-38.4\,(5.4)  &-34.6\,(5.4)  &              &              &0.11 &-0.92&2x[r]                                                                          \\
255  & 21:38:13.34 &  57:17:17.3$^{\rm r}$& 281  & 3281                      &                               &                 &                 &     15$^{\rm j}$&       &                 &16.066\,(114)&15.625\,(173)&   13.906\,()$^{\rm r}$&                    &                 &               &              &              &              &              &0.11 &-0.92&2x[r] (faint)                                                               \\
256  & 21:38:19.73 &  57:17:23.4$^{\rm r}$& 282  & 3282                      &                               &                 &                 &     15$^{\rm j}$&       &                 &13.115\,(29) &12.750\,(30) & 12.627\,(26)$^{\rm r}$&                    &                 &               &-9.7\,(4)     &-3\,(4)       &-10.8\,(6.8)  &2.6\,(6.9)    &     &     &                                                                                      \\
257  & 21:38:21.53 &  57:18:22.8$^{\rm r}$& 283  & 3283                      &                               &                 &                 &   15.1$^{\rm j}$&       &                 &13.350\,(26) &13.028\,(29) & 12.902\,(29)$^{\rm r}$&                    &                 &               &-8.7\,(4)     &2.7\,(4)      &-6.6\,(6.9)   &6.1\,(6.9)    &     &     &                                                                                      \\
258  & 21:38:30.40 &  57:18:13.2$^{\rm r}$& 284  & 3284                      &                               &                 &                 &     15$^{\rm j}$&       &                 &12.184\,(27) &11.514\,(31) & 11.335\,(23)$^{\rm r}$&                    &                 &               &-15.4\,(4)    &-9.5\,(4)     &-48.8\,(6.9)  &-19.8\,(6.8)  &-0.25&-0.51&                                                                                      \\
259  & 21:38:33.06 &  57:18:55.8$^{\rm r}$& 285  & 3285                      &                               &                 &                 &   14.8$^{\rm j}$&       &                 &11.558\,(26) &10.807\,(31) & 10.591\,(19)$^{\rm r}$&                    &                 &               &-2.6\,(4)     &-2.2\,(4)     &-0.1\,(6.9)   &3.5\,(6.8)    &-0.39&-0.06&                                                                                      \\
260  & 21:38:47.24 &  57:16:59.3$^{\rm r}$& 286  & 3286                      &                               &                 &                 &   12.6$^{\rm j}$&       &                 &10.138\,(23) &9.499 \,(31) & 9.294 \,(23)$^{\rm r}$&                    &                 &               &-9.7\,(5.1)   &-5\,(5.1)     &-8.3\,(7.3)   &-2.8\,(7.3)   &0.82 &-0.61&                                                                                      \\
261  & 21:38:50.70 &  57:15:29.9$^{\rm r}$& 287  & 3287                      &                               &                 &                 &   13.4$^{\rm j}$&       &                 &10.806\,(23) &10.179\,(28) & 10.030\,(23)$^{\rm r}$&                    &                 &               &-6.8\,(4)     &-5.3\,(4)     &-4.3\,(7.2)   &-8.3\,(7.1)   &0.2  &-0.27&                                                                                      \\
262  & 21:38:59.92 &  57:16:28.2$^{\rm r}$& 290  & 3290                      &                               &                 &                 &   14.1$^{\rm j}$&       &                 &12.479\,(26) &12.276\,(32) & 12.118\,(25)$^{\rm r}$&                    &                 &               &-3.7\,(4)     &-6.8\,(4)     &-6.2\,(6.7)   &-0.3\,(6.7)   &0.29 &-0.53&                                                                                      \\
263  & 21:39:06.39 &  57:14:31.4$^{\rm r}$& 291  & 3291                      &                               &                 &                 &   14.7$^{\rm j}$&       &                 &12.705\,(25) &12.312\,(29) & 12.178\,(26)$^{\rm r}$&                    &                 &               &-1.3\,(4)     &-4.8\,(4)     &-4.2\,(6.8)   &1.4\,(6.8)    &-0.26&0.37 &                                                                                      \\
264  & 21:39:13.08 &  57:13:50.5$^{\rm r}$& 292  & 3292                      &                               &                 &                 &   14.6$^{\rm j}$&       &                 &12.810\,(29) &12.457\,(31) & 12.348\,(28)$^{\rm r}$&                    &                 &               &-12.5\,(4)    &3.2\,(4)      &-21.5\,(6.8)  &14.1\,(6.8)   &-0.67&0.56 &                                                                                      \\
265  & 21:39:13.91 &  57:12:54.3$^{\rm r}$& 293  & 3293                      &                               &  11.92$^{\rm l}$&  11.97$^{\rm l}$&  11.69$^{\rm l}$&       &                 &11.065\,(22) &11.033\,(29) & 11.022\,(20)$^{\rm r}$&                    &                 &               &-6.4\,(1.7)   &-12\,(1.7)    &-3.5\,(0.8)   &-6.2\,(1.8)   &0.02 &-0.16&                                                                                      \\
266  & 21:39:10.77 &  57:12:00.6$^{\rm r}$& 294  & 3294                      &                               &                 &  13.73$^{\rm l}$&  11.72$^{\rm l}$&       &                 &7.715 \,(26) &6.829 \,(33) & 6.502 \,(21)$^{\rm r}$&                    &                 &               &-19.4\,(13)   &-25.9\,(13)   &-16\,(11.7)   &-11.7\,(16)   &-0.07&0.12 &                                                                                      \\
267  & 21:39:00.56 &  57:16:46.1$^{\rm r}$& 295  & 3295                      &                               &                 &                 &   12.9$^{\rm j}$&       &                 &11.606\,(25) &11.296\,(29) & 11.202\,(25)$^{\rm r}$&                    &                 &               &-15.5\,(4)    &-21.3\,(4)    &-5.6\,(7.2)   &-10.5\,(7.1)  &1.91 &-1.93&                                                                                      \\
268  & 21:38:58.00 &  57:17:00.8$^{\rm r}$& 296  & 3296                      &                               &                 &                 &   13.3$^{\rm j}$&       &                 &12.082\,(25) &11.856\,(29) & 11.772\,(22)$^{\rm r}$&                    &                 &               &-4.7\,(4)     &-1.7\,(4)     &1\,(6.9)      &8.5\,(6.9)    &-0.18&0.07 &                                                                                      \\
269  & 21:38:56.93 &  57:17:56.7$^{\rm r}$& 297  & 3297                      &                               &                 &                 &   12.2$^{\rm j}$&       &                 &11.090\,(23) &10.872\,(28) & 10.814\,(22)$^{\rm r}$&                    &                 &               &-0.4\,(12.5)  &-0.2\,(12.5)  &1\,(1)        &1.6\,(0.9)    &-0.34&0.54 &                                                                                      \\
270  & 21:38:39.85 &  57:19:44.9$^{\rm r}$& 298  & 3298                      &                               &                 &                 &   12.8$^{\rm j}$&       &                 &11.335\,(24) &10.964\,(31) & 10.849\,(19)$^{\rm r}$&                    &                 &               &-3.4\,(4)     &9.2\,(4)      &18.2\,(7.2)   &7.1\,(7.2)    &-0.32&0.79 &                                                                                      \\
271  & 21:38:17.22 &  57:19:50.5$^{\rm r}$& 299  & 3299                      &                               &  15.53$^{\rm l}$&  14.38$^{\rm l}$&  12.93$^{\rm l}$&       &                 &10.115\,(26) &9.378 \,(31) & 9.205 \,(21)$^{\rm r}$&                    &                 &               &-6.5\,(5.1)   &-2.8\,(5.1)   &-2.8\,(7.3)   &5.7\,(7.3)    &0.15 &0.04 &                                                                                      \\
272  & 21:38:04.83 &  57:20:23.1$^{\rm r}$& 300  & 3300                      &                               &                 &                 &   13.6$^{\rm j}$&       &                 &12.201\,()   &12.097\,()   & 12.133\,(35)$^{\rm r}$&                    &                 &               &-9.2\,(4)     &-15.6\,(4)    &-16.7\,(6.9)  &-32.1\,(6.9)  &-0.13&0.18 &                                                                                      \\
273  & 21:38:13.25 &  57:20:44.6$^{\rm r}$& 301  & 3301                      &                               &                 &                 &   12.2$^{\rm j}$&       &                 &10.089\,(26) &9.473 \,(30) & 9.339 \,(19)$^{\rm r}$&                    &                 &               &-10.6\,(5.1)  &14\,(5.1)     &-11\,(7.4)    &19.9\,(7.4)   &0.05 &1.9  &                                                                                      \\
274  & 21:39:09.39 &  57:15:34.9$^{\rm r}$& 302  & 3302                      &                               &                 &  16.45$^{\rm l}$&  14.84$^{\rm i}$& 13.97 &  13.16$^{\rm i}$&11.739\,(23) &11.101\,(28) & 10.900\,(22)$^{\rm r}$&                    &                 &               &-3.8\,(4)     &-3.6\,(4)     &-11.9\,(6.8)  &-3\,(6.8)     &-0.36&0.3  &                                                                                      \\
275  & 21:39:15.36 &  57:15:02.4$^{\rm r}$& 303  & 3303                      &                               &                 &                 &     15$^{\rm j}$&       &                 &15.746\,(85) &15.116\,(119)&15.035\,(144)$^{\rm r}$&                    &                 &               &              &              &-26.5\,(6.8)  &41.7\,(6.8)   &     &     &2x[r] (faint)                                                               \\
276  & 21:39:15.11 &  57:15:06.1$^{\rm r}$& 303  & 3303                      &                               &                 &                 &     15$^{\rm j}$&       &                 &13.331\,(32) &12.991\,(41) & 12.899\,(34)$^{\rm r}$&                    &                 &               &-9\,(4)       &7.3\,(4)      &-26.5\,(6.8)  &41.7\,(6.8)   &     &     &2x[r]                                                                          \\
277  & 21:39:12.95 &  57:14:14.2$^{\rm r}$& 304  & 3304                      &                               &                 &                 &     15$^{\rm j}$&       &                 &10.370\,(23) &9.235 \,(28) & 8.854 \,(20)$^{\rm r}$&                    &                 &               &4.1\,(6.6)    &67.5\,(6.6)   &11.6\,(7.1)   &44.5\,(7.2)   &     &     &                                                                                      \\
278  & 21:38:11.85 &  57:21:11.2$^{\rm r}$& 307  & 3307                      &                               &                 &                 &     15$^{\rm j}$&       &                 &13.109\,(24) &12.598\,(31) & 12.457\,(23)$^{\rm r}$&                    &                 &               &-21.4\,(4)    &2.6\,(4)      &-22\,(6.9)    &7.2\,(7.1)    &     &     &                                                                                      \\
279  & 21:38:13.90 &  57:21:16.0$^{\rm r}$& 308  & 3308                      &                               &                 &                 &   15.1$^{\rm j}$&       &                 &11.891\,(26) &11.123\,(31) & 10.818\,(18)$^{\rm r}$&                    &                 &               &-5\,(4)       &2.5\,(4)      &26.1\,(6.7)   &22.9\,(6.7)   &     &     &                                                                                      \\
280  & 21:38:19.46 &  57:21:11.8$^{\rm r}$& 309  & 3309                      &                               &                 &                 &     15$^{\rm j}$&       &                 &12.201\,(26) &11.261\,(31) & 10.947\,(23)$^{\rm r}$&                    &                 &               &-1.3\,(4)     &-4.1\,(4)     &-1.7\,(7.3)   &3.9\,(7.1)    &     &     &new coordinates                                                                       \\     
281  & 21:39:27.25 &  57:17:07.3$^{\rm r}$& 310  & 3310                      &                               &                 &                 &   14.7$^{\rm j}$&       &                 &12.082\,(25) &11.487\,(28) & 11.364\,(23)$^{\rm r}$&                    &                 &               &-2.9\,(4)     &4.6\,(4)      &-3.3\,(7.4)   &0.3\,(7.4)    &-0.17&0.6  &                                                                                      \\
282  & 21:39:34.56 &  57:15:19.2$^{\rm r}$& 311  & 3311                      &                               &                 &                 &   14.4$^{\rm j}$&       &                 &12.645\,(32) &12.327\,(40) & 12.214\,(32)$^{\rm r}$&                    &                 &               &1.6\,(17.5)   &-2.7\,(17.5)  &4.7\,(7)      &-6.4\,(7)     &0.39 &0.02 &                                                                                      \\
283  & 21:39:55.66 &  57:14:49.2$^{\rm r}$& 312  & 3312                      &                               &                 &  15.04$^{\rm h}$&  14.05$^{\rm h}$& 13.88 &   13.3$^{\rm i}$&12.445\,(24) &11.989\,(31) & 11.891\,(26)$^{\rm r}$&        F9$^{\rm h}$&                 & 1.28$^{\rm h}$&-5.3\,(4.1)   &-1.2\,(4.1)   &-2.8\,(6.8)   &0.6\,(6.8)    &0.22 &-0.28&Dec [h] imprec.                                                               \\
284  & 21:39:57.96 &  57:13:33.5$^{\rm j}$& 313  & 3313                      &                               &                 &                 &   14.4$^{\rm j}$&       &                 &             &             &                       &                    &                 &               &              &              &              &              &     &     &no star                                                                          \\
285  & 21:39:56.97 &  57:12:42.2$^{\rm r}$& 314  & 3314                      &                               &                 &                 &   12.6$^{\rm j}$&       &                 &10.180\,(24) &9.476 \,(30) & 9.295 \,(22)$^{\rm r}$&                    &                 &               &-0.6\,(5.1)   &-0.1\,(5.1)   &4.2\,(7.5)    &-1.4\,(7.5)   &-0.79&0.16 &                                                                                      \\
286  & 21:39:46.84 &  57:12:09.9$^{\rm r}$& 315  & 3315                      &                               &                 &                 &   14.6$^{\rm j}$&       &                 &12.916\,(28) &12.703\,(36) & 12.576\,(32)$^{\rm r}$&                    &                 &               &-6\,(4.1)     &4.4\,(4.1)    &-10.8\,(6.8)  &20.9\,(6.8)   &-0.17&-0.07&                                                                                      \\
287  & 21:40:08.48 &  57:12:04.9$^{\rm r}$& 316  & 3316                      &                               &                 &                 &   12.1$^{\rm j}$&       &                 &9.788 \,(23) &9.116 \,(30) & 8.975 \,(20)$^{\rm r}$&                    &                 &               &-18.5\,(12.7) &6.4\,(12.7)   &-15.4\,(8.2)  &5.9\,(8.1)    &-0.15&0.46 &                                                                                      \\
288  & 21:40:09.93 &  57:11:43.5$^{\rm r}$& 317  & 3317                      &                               &                 &                 &    9.6$^{\rm j}$&       &                 &6.391 \,(41) &5.640 \,(26) & 5.368 \,(18)$^{\rm r}$&                    &                 &               &3.9\,(1.6)    &3.2\,(1.6)    &5.2\,(1.1)    &0.4\,(1.1)    &-0.3 &0.36 &                                                                                      \\
289  & 21:40:04.45 &  57:14:41.7$^{\rm r}$& 318  & 483                       &                               &                 &                 &   11.2$^{\rm j}$&       &                 &10.415\,(23) &10.216\,(30) & 10.122\,(20)$^{\rm r}$&        F8$^{\rm q}$&                 &               &11.5\,(1.7)   &-7.3\,(1.7)   &13.7\,(0.8)   &-5.3\,(1)     &-1.57&-0.06&                                                                                      \\
290  & 21:39:38.59 &  57:17:05.1$^{\rm r}$& 319  & 478                       &                               &                 &                 &   11.1$^{\rm j}$&       &                 &10.457\,(24) &10.284\,(28) & 10.231\,(22)$^{\rm r}$&        A0$^{\rm q}$&                 &               &3\,(1.7)      &-8.7\,(1.7)   &4.1\,(0.7)    &-2.4\,(1.3)   &-0.61&0.33 &                                                                                      \\
291  & 21:39:46.27 &  57:16:59.4$^{\rm r}$& 320  & 3320                      &                               &                 &  14.42$^{\rm h}$&  13.51$^{\rm h}$& 12.95 &  13.46$^{\rm i}$&11.721\,(29) &11.443\,(38) & 11.324\,(28)$^{\rm r}$&        F6$^{\rm h}$&                 &  1.4$^{\rm h}$&-7.2\,(4.1)   &1.6\,(4.1)    &8\,(7.2)      &3.5\,(7.2)    &0.32 &0.2  &Dec [h] imprec.                                                               \\
292  & 21:39:51.42 &  57:16:56.6$^{\rm r}$& 321  & 3321                      &                               &                 &  14.99$^{\rm h}$&  14.27$^{\rm h}$& 13.86 &   13.4$^{\rm i}$&12.715\,(24) &12.476\,(36) & 12.350\,(29)$^{\rm r}$&        A0$^{\rm h}$&                 & 2.25$^{\rm h}$&-5.3\,(4.1)   &5.5\,(4.1)    &-15.8\,(6.8)  &12.4\,(6.8)   &-0.14&0.25 &                                                                                      \\
293  & 21:39:36.85 &  57:17:47.5$^{\rm r}$& 323  & 3323                      &                               &                 &                 &   14.8$^{\rm j}$&       &                 &12.563\,(24) &12.094\,(35) & 11.943\,(26)$^{\rm r}$&                    &                 &               &14.9\,(4.1)   &2.9\,(4.1)    &20.6\,(6.8)   &2.5\,(6.8)    &-2.07&0.72 &                                                                                      \\
294  & 21:39:50.47 &  57:18:35.6$^{\rm r}$& 325  & 3325                      &                               &                 &                 &   14.8$^{\rm j}$&       &                 &13.051\,(24) &12.694\,(27) & 12.631\,(32)$^{\rm r}$&                    &                 &               &-3.4\,(4.1)   &-0.2\,(4.1)   &-0.5\,(6.8)   &7\,(6.8)      &0.11 &-0.66&                                                                                      \\
295  & 21:39:55.16 &  57:17:42.5$^{\rm r}$& 326  & 3326                      &                               &                 &  16.75$^{\rm l}$&  14.38$^{\rm i}$& 12.94 &  11.53$^{\rm i}$&9.454 \,(24) &8.440 \,(32) & 7.993 \,(21)$^{\rm r}$&                    &                 &               &85\,(6.6)     &165.5\,(6.6)  &-2.1\,(7)     &18\,(7)       &0.1  &-0.09&                                                                                      \\
296  & 21:40:08.17 &  57:18:15.8$^{\rm r}$& 328  & 3328\footnote{also 5091}  &                               &                 &  17.43$^{\rm l}$&  14.67$^{\rm i}$& 13.71 &  12.81$^{\rm i}$&11.383\,(24) &10.659\,(32) & 10.417\,(23)$^{\rm r}$&        K3$^{\rm q}$&                 &               &-0.6\,(4.1)   &3.8\,(4.1)    &29.9\,(6.7)   &17.5\,(6.7)   &0.01 &-0.18&                                                                                      \\
297  & 21:40:19.69 &  57:18:09.7$^{\rm r}$& 329  & 3329                      &                               &                 &                 &     14$^{\rm j}$&       &                 &11.447\,(24) &10.986\,(33) & 10.844\,(19)$^{\rm r}$&                    &                 &               &-14.7\,(4.1)  &-18.5\,(4.1)  &-3.1\,(7.1)   &5\,(7.1)      &1.35 &-1.9 &                                                                                      \\
298  & 21:40:11.22 &  57:14:31.1$^{\rm r}$& 330  & 3330                      &                               &                 &                 &   13.3$^{\rm j}$&       &                 &10.606\,(24) &10.014\,(30) & 9.810 \,(22)$^{\rm r}$&                    &                 &               &-8.4\,(5.1)   &-2.1\,(5.1)   &-8.5\,(7.2)   &-7.6\,(7.2)   &0.1  &-0.1 &                                                                                      \\
299  & 21:41:00.55 &  57:11:39.9$^{\rm r}$& 335  & 3335                      &                               &  15.13$^{\rm l}$&  13.97$^{\rm l}$&  12.53$^{\rm l}$&       &                 &9.695 \,(24) &9.086 \,(32) & 8.896 \,(19)$^{\rm r}$&                    &                 &               &-4.8\,(5.1)   &-2.3\,(5.1)   &-5.3\,(7.8)   &4.8\,(7.8)    &-0.07&0.26 &                                                                                      \\
300  & 21:41:03.49 &  57:14:17.5$^{\rm r}$& 336  & 3336                      &                               &                 &                 &   13.9$^{\rm j}$&       &                 &11.438\,(36) &11.077\,()   &   10.571\,()$^{\rm r}$&                    &                 &               &5.1\,(4.1)    &11.7\,(4.1)   &-9.3\,(7.1)   &-6.9\,(7.2)   &-0.51&-0.02&                                                                                      \\
301  & 21:40:58.74 &  57:16:13.3$^{\rm r}$& 337  & 3337                      &                               &                 &  14.96$^{\rm l}$&  12.85$^{\rm l}$&       &                 &7.197 \,(26) &6.149 \,(51) & 5.735 \,(16)$^{\rm r}$&                    &                 &               &-8\,(5.1)     &-8.4\,(5.1)   &-11.7\,(7.7)  &-16.6\,(7.8)  &0.23 &-0.34&                                                                                      \\
302  & 21:39:43.17 &  57:20:29.7$^{\rm r}$& 338  & 3338                      &                               &                 &                 &   12.5$^{\rm j}$&       &                 &10.697\,(26) &10.323\,(32) & 10.183\,(23)$^{\rm r}$&                    &                 &               &-7.3\,(4.1)   &10.2\,(4.1)   &-1.8\,(4.1)   &1.5\,(6.5)    &0.05 &0.15 &                                                                                      \\
303  & 21:39:46.44 &  57:20:27.1$^{\rm r}$& 339  & 3339                      &                               &                 &                 &     11$^{\rm j}$&       &                 &6.417 \,(18) &5.430 \,(20) & 5.061 \,(23)$^{\rm r}$&                    &                 &               &0.9\,(5.1)    &4.9\,(5.1)    &-0.6\,(7.5)   &14.7\,(7.4)   &-0.66&0.32 &                                                                                      \\
304  & 21:39:44.10 &  57:21:03.8$^{\rm r}$& 340  & 3340                      &                               &                 &                 &   14.9$^{\rm j}$&       &                 &11.251\,(24) &10.309\,(28) & 10.068\,(22)$^{\rm r}$&                    &                 &               &1.2\,(4.1)    &-1.8\,(4.1)   &12.5\,(6.9)   &3.5\,(6.9)    &     &     &                                                                                      \\
305  & 21:39:47.56 &  57:21:03.3$^{\rm r}$& 341  & 3341                      &                               &                 &                 &   13.8$^{\rm j}$&       &                 &12.142\,(24) &11.707\,(28) & 11.574\,(23)$^{\rm r}$&                    &                 &               &-4\,(4.1)     &-13.8\,(4.1)  &4.3\,(7.1)    &-10.4\,(7.1)  &0.6  &-1.24&                                                                                      \\
306  & 21:39:50.44 &  57:21:29.0$^{\rm r}$& 342  & 3342                      &                               &                 &                 &   14.8$^{\rm j}$&       &                 &12.807\,()   &12.559\,()   &   12.447\,()$^{\rm r}$&                    &                 &               &-9.7\,(4.1)   &-8.8\,(4.1)   &-17.1\,(6.9)  &-33.4\,(6.9)  &0.05 &0.04 &                                                                                      \\
307  & 21:39:44.46 &  57:21:53.8$^{\rm r}$& 343  & 3343                      &                               &                 &                 &   12.7$^{\rm j}$&       &                 &11.708\,(23) &11.496\,(31) & 11.410\,(20)$^{\rm r}$&                    &                 &               &-9.5\,(4.1)   &-2.6\,(4.1)   &-9.1\,(0.5)   &-5.9\,(1.3)   &0.49 &-0.37&                                                                                      \\
308  & 21:39:47.45 &  57:22:05.9$^{\rm r}$& 344  & 3344                      &                               &                 &                 &   14.6$^{\rm j}$&       &                 &12.404\,(32) &12.000\,(40) & 11.838\,(26)$^{\rm r}$&                    &                 &               &2.3\,(7.2)    &-3.2\,(7.2)   &-30.7\,(6.4)  &-23.6\,(6.3)  &0.61 &-0.47&                                                                                      \\
309  & 21:39:52.71 &  57:22:00.1$^{\rm r}$& 345  & 3345                      &                               &                 &                 &   11.6$^{\rm j}$&       &                 &10.374\,(24) &10.134\,(27) & 9.986 \,(20)$^{\rm r}$&                    &                 &               &-6.1\,(5.1)   &-0.3\,(5.1)   &-11.4\,(1.9)  &-8.4\,(2.5)   &0.91 &-0.4 &                                                                                      \\
310  & 21:39:55.27 &  57:21:35.4$^{\rm r}$& 346  & 3346                      &                               &                 &                 &   14.8$^{\rm j}$&       &                 &13.034\,(26) &12.752\,(36) & 12.634\,(34)$^{\rm r}$&                    &                 &               &-4\,(4.1)     &1.1\,(4.1)    &2.1\,(7)      &9.5\,(6.8)    &0.15 &0.18 &                                                                                      \\
311  & 21:40:12.58 &  57:21:12.4$^{\rm r}$& 348  & 3348                      &                               &                 &                 &   13.2$^{\rm j}$&       &                 &12.152\,(23) &11.927\,(30) & 11.824\,(22)$^{\rm r}$&                    &                 &               &-14.3\,(4.1)  &-2.4\,(4.1)   &-1.4\,(7)     &6.5\,(7)      &-0.07&-0.14&                                                                                      \\
312  & 21:40:20.90 &  57:19:52.4$^{\rm r}$& 349  & 3349                      &                               &                 &                 &   15.1$^{\rm j}$&       &                 &13.099\,(27) &12.736\,(35) & 12.583\,(24)$^{\rm r}$&                    &                 &               &-4.4\,(4.1)   &-5.6\,(4.1)   &1.9\,(6.9)    &14.3\,(6.9)   &     &     &                                                                                      \\
313  & 21:40:10.60 &  57:18:57.8$^{\rm r}$& 350  & 3350                      &                               &                 &                 &     15$^{\rm j}$&       &                 &11.195\,(24) &10.280\,(31) & 9.963 \,(23)$^{\rm r}$&                    &                 &               &-10.6\,(5.1)  &-0.2\,(5.1)   &20.5\,(6.9)   &2.7\,(6.9)    &     &     &                                                                                      \\
314  & 21:40:23.68 &  57:19:41.2$^{\rm r}$& 352  & 3352                      &                               &                 &                 &   15.1$^{\rm j}$&       &                 &11.367\,(26) &10.449\,(32) & 10.129\,(21)$^{\rm r}$&                    &                 &               &-0.4\,(4.1)   &-10.6\,(4.1)  &7.8\,(6.9)    &-0.3\,(6.8)   &     &     &                                                                                      \\
315  & 21:40:27.18 &  57:19:05.4$^{\rm r}$& 353  & 3353                      &                               &                 &                 &   14.8$^{\rm j}$&       &                 &11.374\,()   &10.573\,()   & 10.468\,(45)$^{\rm r}$&                    &                 &               &              &              &              &              &-0.26&-0.16&3x[r]                                                                           \\
316  & 21:40:26.71 &  57:19:08.3$^{\rm r}$& 353  & 3353                      &                               &                 &                 &   14.8$^{\rm j}$&       &                 &12.872\,()   &12.514\,()   &   11.176\,()$^{\rm r}$&                    &                 &               &              &              &              &              &-0.26&-0.16&3x[r]                                                                           \\
317  & 21:40:27.08 &  57:19:08.9$^{\rm r}$& 353  & 3353                      &                               &                 &                 &   14.8$^{\rm j}$&       &                 &11.585\,()   &10.830\,()   &   12.152\,()$^{\rm r}$&                    &                 &               &24.3\,(4.1)   &28.5\,(4.1)   &              &              &-0.26&-0.16&3x[r]                                                                           \\
318  & 21:40:33.55 &  57:19:40.0$^{\rm r}$& 354  & 3354                      &                               &                 &                 &   13.2$^{\rm j}$&       &                 &11.719\,(27) &11.387\,(36) & 11.260\,(28)$^{\rm r}$&                    &                 &               &-5.8\,(15)    &-11.8\,(15)   &              &              &0.94 &-1.41&                                                                                      \\
319  & 21:40:32.62 &  57:19:42.3$^{\rm r}$& 355  & 3355                      &                               &                 &                 &     15$^{\rm j}$&       &                 &13.064\,(39) &12.641\,(40) & 12.524\,(33)$^{\rm r}$&                    &                 &               &              &              &              &              &0.66 &-0.48&                                                                                      \\
320  & 21:40:36.17 &  57:19:51.8$^{\rm r}$& 356  & 3356                      &                               &                 &                 &   14.9$^{\rm j}$&       &                 &11.383\,(24) &10.639\,(32) & 10.329\,(19)$^{\rm r}$&                    &                 &               &-6.6\,(4.1)   &-0.1\,(4.1)   &-4.5\,(6.8)   &5.8\,(7.1)    &0.51 &0.42 &                                                                                      \\
321  & 21:40:42.51 &  57:20:25.7$^{\rm r}$& 357  & 3357                      &                               &                 &                 &   14.8$^{\rm j}$&       &                 &12.797\,()   &12.541\,()   & 12.283\,(43)$^{\rm r}$&                    &                 &               &-2.8\,(4.1)   &-9.1\,(4.1)   &20.2\,(6.3)   &-9.3\,(6.3)   &0.5  &-0.06&                                                                                      \\
322  & 21:40:44.92 &  57:20:04.5$^{\rm r}$& 358  & 3358                      &                               &                 &                 &   15.1$^{\rm j}$&       &                 &11.676\,(24) &10.916\,(35) & 10.658\,(21)$^{\rm r}$&                    &                 &               &-8.2\,(4.1)   &-6.2\,(4.1)   &-10.3\,(6.9)  &7\,(6.8)      &     &     &                                                                                      \\
323  & 21:40:47.44 &  57:20:00.5$^{\rm r}$& 359  & 3359                      &                               &                 &                 &   13.8$^{\rm j}$&       &                 &10.055\,(24) &9.159 \,(33) & 8.863 \,(19)$^{\rm r}$&                    &                 &               &-4.8\,(5.1)   &-1.2\,(5.1)   &2.6\,(7.2)    &3.1\,(7.2)    &0.1  &-0.16&                                                                                      \\
324  & 21:40:53.99 &  57:18:47.9$^{\rm r}$& 360  & 3360                      &                               &                 &                 &     13$^{\rm j}$&       &                 &10.682\,(26) &10.079\,(33) & 9.870 \,(21)$^{\rm r}$&                    &                 &               &-0.2\,(5.1)   &5.1\,(5.1)    &2.7\,(7.2)    &5.4\,(7.2)    &-0.43&0.39 &                                                                                      \\
325  & 21:40:58.43 &  57:19:45.3$^{\rm r}$& 361  & 3361                      &                               &                 &                 &   13.8$^{\rm j}$&       &                 &12.141\,(27) &11.705\,(32) & 11.621\,(21)$^{\rm r}$&                    &                 &               &-13.3\,(12.7) &0.7\,(12.7)   &-34.9\,(6.7)  &26.4\,(6.7)   &.0.01&0.62 &                                                                                      \\
326  & 21:40:59.52 &  57:19:41.9$^{\rm r}$& 362  & 3362                      &                               &                 &                 &   13.8$^{\rm j}$&       &                 &12.582\,(27) &12.447\,(31) & 12.315\,(30)$^{\rm r}$&                    &                 &               &-12.5\,(10.9) &2.3\,(10.9)   &11\,(5.9)     &-1.2\,(5.8)   &.0.14&0.11 &                                                                                      \\
327  & 21:41:19.77 &  57:16:34.4$^{\rm r}$& 363  & 3363                      &                               &                 &                 &   12.6$^{\rm j}$&       &                 &11.600\,(24) &11.444\,(32) & 11.356\,(24)$^{\rm r}$&                    &                 &               &-10.8\,(4.1)  &-5.9\,(4.1)   &-10.4\,(0.7)  &-6.3\,(0.7)   &0.71 &-0.44&                                                                                      \\
328  & 21:41:12.64 &  57:20:47.1$^{\rm r}$& 364  & 3364                      &                               &                 &                 &     14$^{\rm j}$&       &                 &12.285\,(26) &11.975\,(33) & 11.862\,(23)$^{\rm r}$&                    &                 &               &3\,(4.1)      &3.3\,(4.1)    &10.3\,(6.9)   &12.1\,(6.9)   &-0.8 &0.23 &                                                                                      \\
329  & 21:41:29.31 &  57:16:28.6$^{\rm r}$& 365  & 3365                      &                               &                 &                 &   12.9$^{\rm j}$&       &                 &13.100\,(34) &12.724\,(37) & 12.588\,(30)$^{\rm r}$&                    &                 &               &-12\,(4.1)    &-6.3\,(4.1)   &-1.9\,(6.9)   &13.8\,(7)     &     &     &new coordinates\\%, small change                                                         \\     
330  & 21:41:31.74 &  57:18:13.9$^{\rm r}$& 366  & 3366                      &                               &                 &                 &   12.6$^{\rm j}$&       &                 &11.498\,(27) &11.323\,(32) & 11.191\,(24)$^{\rm r}$&                    &                 &               &-8.6\,(4.1)   &2.4\,(4.1)    &-17.4\,(7.4)  &34.5\,(7.4)   &0.17 &0.02 &                                                                                      \\
331  & 21:41:31.86 &  57:18:29.1$^{\rm r}$& 367  & 3367                      &                               &  13.86$^{\rm l}$&  13.04$^{\rm h}$&  12.07$^{\rm h}$&       &                 &11.347\,(26) &11.109\,(32) & 11.015\,(21)$^{\rm r}$&        A5$^{\rm h}$&                 & 1.84$^{\rm h}$&-5.3\,(4.1)   &3.2\,(4.1)    &-3.9\,(1.2)   &-0.7\,(2.9)   &0.19 &0.09 &Dec [h] imprec.                                                               \\
332  & 21:39:38.51 &  57:24:05.6$^{\rm r}$& 368  & 3368                      &                               &                 &                 &   14.3$^{\rm j}$&       &                 &9.474 \,(24) &8.387 \,(29) & 7.990 \,(21)$^{\rm r}$&                    &                 &               &-0.8\,(5.1)   &4\,(5.1)      &-10.4\,(7)    &26.4\,(7)     &0.16 &0.44 &                                                                                      \\
333  & 21:39:45.28 &  57:23:29.5$^{\rm r}$& 369  & 3369                      &                               &  12.72$^{\rm l}$&  12.34$^{\rm l}$&   11.8$^{\rm l}$&       &                 &10.648\,(23) &10.480\,(30) & 10.419\,(23)$^{\rm r}$&                    &                 &               &-5.9\,(2.7)   &-3.8\,(2.7)   &-6.1\,(0.7)   &-4.6\,(0.9)   &0.22 &0.03 &                                                                                      \\
334  & 21:39:34.23 &  57:22:31.4$^{\rm r}$& 370  & 3370                      &                               &                 &                 &   14.7$^{\rm j}$&       &                 &11.537\,(23) &10.839\,(32) & 10.610\,(22)$^{\rm r}$&                    &                 &               &-8.8\,(4.1)   &-5.6\,(4.1)   &-20.7\,(7.4)  &-13.2\,(7.4)  &0.1  &0.35 &                                                                                      \\
335  & 21:39:46.30 &  57:24:01.0$^{\rm r}$& 371  & 3371                      &                               &  14.50$^{\rm l}$&     14$^{\rm h}$&  13.32$^{\rm h}$&       &                 &11.933\,(24) &11.753\,(32) & 11.626\,(28)$^{\rm r}$&        A9$^{\rm h}$&                 & 1.28$^{\rm h}$&-1.6\,(4.1)   &3\,(4.1)      &5.4\,(7.2)    &24.1\,(7.1)   &-0.2 &0.25 &                                                                                      \\
336  & 21:39:50.46 &  57:24:44.9$^{\rm r}$& 372  & 3372                      &                               &                 &                 &   14.6$^{\rm j}$&       &                 &12.661\,(26) &12.274\,(30) & 12.172\,(26)$^{\rm r}$&                    &                 &               &-4.7\,(4.1)   &-0.4\,(4.1)   &-4.7\,(6.8)   &9.9\,(6.8)    &0.34 &0.15 &                                                                                      \\
337  & 21:39:54.82 &  57:24:23.7$^{\rm r}$& 373  & 3373                      &                               &                 &                 &   14.4$^{\rm j}$&       &                 &9.642 \,(24) &8.535 \,(31) & 8.149 \,(18)$^{\rm r}$&                    &                 &               &8.2\,(13.9)   &6.9\,(13.9)   &31.6\,(6.9)   &16.9\,(6.9)   &0.07 &-0.17&new coordinates                                                                       \\     
338  & 21:40:05.85 &  57:23:23.7$^{\rm r}$& 374  & 3374                      &                               &  14.93$^{\rm l}$&  13.65$^{\rm l}$&  12.17$^{\rm l}$&       &                 &9.363 \,(23) &8.736 \,(31) & 8.528 \,(20)$^{\rm r}$&                    &                 &               &-4.4\,(10.9)  &0.8\,(10.9)   &-2\,(6.8)     &-0.5\,(11.6)  &0.01 &0.06 &                                                                                      \\
339  & 21:40:02.42 &  57:23:05.7$^{\rm r}$& 375  & 3375                      &                               &                 &                 &   12.5$^{\rm j}$&       &                 &11.529\,(23) &11.330\,(31) & 11.226\,(22)$^{\rm r}$&                    &                 &               &-9.7\,(4.1)   &-13.1\,(4.1)  &-3.2\,(0.8)   &-12.4\,(12.7) &-0.13&-0.61&                                                                                      \\
340  & 21:40:16.09 &  57:22:32.9$^{\rm r}$& 376  & 3376                      &                               &                 &                 &     15$^{\rm j}$&       &                 &13.001\,()   &12.361\,()   &   12.694\,()$^{\rm r}$&                    &                 &               &-10.1\,(4.1)  &-15\,(4.1)    &              &              &     &     &near 340                                                                           \\
341  & 21:40:16.69 &  57:22:29.8$^{\rm r}$& 377  & 3377                      &                               &                 &                 &   15.1$^{\rm j}$&       &                 &15.138\,(61) &14.441\,(73) & 14.185\,(78)$^{\rm r}$&                    &                 &               &-46.1\,(7.7)  &15.4\,(7.7)   &-15.9\,(10.4) &-39.8\,(10.5) &     &     &near 339                                                                          \\
342  & 21:40:20.54 &  57:23:01.3$^{\rm r}$& 379  & 3379                      &                               &                 &                 &   13.1$^{\rm j}$&       &                 &8.671 \,(37) &7.580 \,(36) & 7.238 \,(33)$^{\rm r}$&                    &                 &               &-3.5\,(5.1)   &0.1\,(5.1)    &-4.7\,(7.3)   &5.1\,(7.3)    &-0.01&0.14 &                                                                                      \\
343  & 21:40:20.52 &  57:23:29.5$^{\rm r}$& 380  & 3380                      &                               &                 &                 &   15.1$^{\rm j}$&       &                 &13.071\,(27) &12.755\,(27) & 12.535\,(28)$^{\rm r}$&                    &                 &               &0.1\,(4.1)    &4.3\,(4.1)    &14.6\,(6.9)   &21.9\,(6.9)   &     &     &                                                                                      \\
344  & 21:40:31.18 &  57:23:33.0$^{\rm r}$& 381  & 3381                      &                               &                 &                 &   14.9$^{\rm j}$&       &                 &13.002\,(29) &12.551\,(31) & 12.461\,(26)$^{\rm r}$&                    &                 &               &-11.6\,(4.1)  &-3.6\,(4.1)   &-5.8\,(6.8)   &4.2\,(6.8)    &0.15 &-0.2 &                                                                                      \\
345  & 21:40:30.07 &  57:22:51.1$^{\rm r}$& 383  & 3383                      &                               &                 &                 &   14.2$^{\rm j}$&       &                 &12.381\,(27) &12.011\,(32) & 11.898\,(29)$^{\rm r}$&                    &                 &               &-8.3\,(4.1)   &10.7\,(4.1)   &-15.2\,(5.6)  &44.8\,(5.9)   &0.01 &0.32 &                                                                                      \\
346  & 21:40:32.67 &  57:23:10.5$^{\rm r}$& 384  & 3384                      &                               &                 &                 &   14.8$^{\rm j}$&       &                 &13.087\,(31) &12.761\,(35) & 12.667\,(28)$^{\rm r}$&                    &                 &               &2.2\,(4.1)    &-3.1\,(4.1)   &18.7\,(6.8)   &4.4\,(6.8)    &0.44 &0.27 &                                                                                      \\
347  & 21:40:39.21 &  57:22:01.7$^{\rm r}$& 385  & 3385                      &                               &                 &                 &   14.2$^{\rm j}$&       &                 &11.099\,(27) &10.309\,(30) & 10.127\,(21)$^{\rm r}$&                    &                 &               &-3.4\,(4.1)   &2.7\,(4.1)    &2.1\,(6.9)    &13.5\,(6.9)   &-0.39&0.47 &                                                                                      \\
348  & 21:40:41.97 &  57:23:04.5$^{\rm j}$& 386  & 3386                      &                               &                 &                 &   15.2$^{\rm j}$&       &                 &             &             &                       &                    &                 &               &              &              &              &              &     &     &no star                                                                          \\
349  & 21:40:50.88 &  57:20:32.8$^{\rm r}$& 389  & 3389                      &                               &                 &                 &   14.9$^{\rm j}$&       &                 &12.947\,(29) &12.515\,(38) & 12.414\,(32)$^{\rm r}$&                    &                 &               &-1.2\,(4.1)   &-1.3\,(4.1)   &-2.4\,(6.8)   &7.3\,(6.8)    &-0.27&0.55 &                                                                                      \\
350  & 21:40:57.52 &  57:22:09.3$^{\rm r}$& 390  & 3390                      &                               &                 &                 &   14.3$^{\rm j}$&       &                 &12.605\,(29) &12.269\,(33) & 12.173\,(24)$^{\rm r}$&                    &                 &               &-5.5\,(4.1)   &3.1\,(4.1)    &10\,(6.8)     &8.1\,(6.8)    &-0.13&0.34 &                                                                                      \\
351  & 21:41:01.54 &  57:22:59.2$^{\rm r}$& 391  & 3391                      &                               &                 &                 &   14.5$^{\rm j}$&       &                 &10.759\,(26) &9.870 \,(28) & 9.658 \,(24)$^{\rm r}$&                    &                 &               &-7.8\,(5.1)   &-2.7\,(5.1)   &47\,(6.8)     &-21.6\,(6.8)  &0.09 &0.43 &                                                                                      \\
352  & 21:39:27.81 &  57:26:12.2$^{\rm r}$& 392  & 3392                      &                               &                 &                 &     15$^{\rm j}$&       &                 &12.626\,(32) &12.164\,()   &   12.042\,()$^{\rm r}$&                    &                 &               &6.6\,(4)      &2.3\,(4)      &14.9\,(6.8)   &19.3\,(6.8)   &     &     &                                                                                      \\
353  & 21:39:35.48 &  57:25:33.7$^{\rm r}$& 393  & 3393                      &                               &                 &                 &   14.8$^{\rm j}$&       &                 &13.074\,(24) &12.798\,(32) & 12.652\,(28)$^{\rm r}$&                    &                 &               &-7\,(4.1)     &-4.2\,(4.1)   &-14.3\,(6.8)  &5.8\,(6.8)    &0.15 &-0.26&                                                                                      \\
354  & 21:39:39.55 &  57:25:33.2$^{\rm r}$& 394  & 3394                      &                               &                 &                 &   14.8$^{\rm j}$&       &                 &12.767\,(24) &12.281\,(32) & 12.171\,(26)$^{\rm r}$&                    &                 &               &-24.8\,(4.1)  &-10.3\,(4.1)  &-15.8\,(6.8)  &16.5\,(6.9)   &     &     &                                                                                      \\
355  & 21:39:41.88 &  57:26:00.6$^{\rm r}$& 395  & 3395                      &                               &                 &                 &   15.1$^{\rm j}$&       &                 &11.756\,(23) &10.930\,(31) & 10.718\,(22)$^{\rm r}$&                    &                 &               &2.3\,(4.1)    &2.9\,(4.1)    &3.2\,(6.8)    &9.6\,(7)      &     &     &                                                                                      \\
356  & 21:39:44.84 &  57:25:46.9$^{\rm j}$& 396  & 3396                      &                               &                 &                 &   14.2$^{\rm j}$&       &                 &             &             &                       &                    &                 &               &              &              &              &              &     &     &no star                                                                          \\
357  & 21:39:44.58 &  57:25:38.9$^{\rm r}$& 397  & 3397                      &                               &                 &                 &   14.2$^{\rm j}$&       &                 &14.345\,(38) &13.698\,(43) & 13.548\,(48)$^{\rm r}$&                    &                 &               &-7.7\,(4.1)   &10.4\,(4.1)   &              &              &     &     &no/faint star                                                               \\
358  & 21:39:46.21 &  57:26:10.3$^{\rm r}$& 398  & 3398\footnote{also 5089}  &                               &  14.22$^{\rm k}$&  14.31$^{\rm f}$&  13.68$^{\rm e}$&       &                 &12.291\,(21) &12.130\,(31) & 11.998\,(25)$^{\rm r}$&        A8$^{\rm e}$&                 &  1.3$^{\rm e}$&-5.7\,(4.1)   &-0.8\,(4.1)   &-0.6\,(6.9)   &8.7\,(6.9)    &0.14 &0.01 &                                                                                      \\
359  & 21:39:47.99 &  57:25:49.5$^{\rm r}$& 399  & 3399                      &                               &                 &                 &   14.8$^{\rm j}$&       &                 &11.460\,(24) &10.709\,(27) & 10.462\,(22)$^{\rm r}$&                    &                 &               &-0.2\,(4.1)   &0.2\,(4.1)    &-3\,(6.8)     &5.8\,(6.8)    &0.25 &-0.22&                                                                                      \\
360  & 21:38:12.01 &  57:21:39.4$^{\rm r}$& 400  & 3400                      &                               &                 &                 &   14.7$^{\rm j}$&       &                 &12.955\,(26) &12.560\,(29) & 12.445\,(21)$^{\rm r}$&                    &                 &               &-6.3\,(4)     &1.1\,(4)      &-7.5\,(6.8)   &8.5\,(6.8)    &0.12 &0.01 &                                                                                      \\
361  & 21:38:19.93 &  57:21:55.5$^{\rm r}$& 402  & 3402                      &                               &                 &                 &   12.3$^{\rm j}$&       &                 &11.154\,(29) &10.771\,(32) & 10.677\,(24)$^{\rm r}$&                    &                 &               &-10.5\,(4)    &29.5\,(4)     &-29.8\,(7.9)  &89.2\,(7.9)   &-1.52&1.18 &                                                                                      \\
362  & 21:38:00.39 &  57:22:47.9$^{\rm r}$& 403  & 3403                      &                               &                 &                 &   14.8$^{\rm j}$&       &                 &12.948\,(27) &12.563\,(31) & 12.446\,(27)$^{\rm r}$&                    &                 &               &-15\,(4)      &-5.8\,(4)     &-10.4\,(6.8)  &6.3\,(6.8)    &     &     &                                                                                      \\
363  & 21:38:07.68 &  57:23:24.4$^{\rm r}$& 404  & 3404                      &                               &                 &                 &   14.4$^{\rm j}$&       &                 &12.301\,(26) &11.777\,(31) & 11.673\,(18)$^{\rm r}$&                    &                 &               &0.7\,(4)      &-7.6\,(4)     &5.5\,(6.8)    &-0.6\,(6.8)   &-1.31&-1.08&                                                                                      \\
364  & 21:38:12.21 &  57:23:02.8$^{\rm r}$& 405  & 3405                      &                               &                 &                 &   14.1$^{\rm j}$&       &                 &12.494\,(26) &12.140\,(30) & 12.101\,(24)$^{\rm r}$&                    &                 &               &-2.3\,(4)     &-6.7\,(4)     &19.2\,(6.9)   &-55.2\,(6.9)  &0.11 &0.52 &                                                                                      \\
365  & 21:38:14.90 &  57:22:57.9$^{\rm r}$& 406  & 3406                      &                               &                 &                 &   14.5$^{\rm j}$&       &                 &11.284\,(26) &10.609\,(32) & 10.353\,(20)$^{\rm r}$&                    &                 &               &3.6\,(4)      &4.4\,(4)      &8\,(6.8)      &16.9\,(6.8)   &-0.37&-0.01&                                                                                      \\
366  & 21:38:16.48 &  57:22:30.5$^{\rm r}$& 407  & 3407                      &                               &                 &                 &   14.7$^{\rm j}$&       &                 &11.336\,(36) &10.553\,(35) & 10.343\,(28)$^{\rm r}$&                    &                 &               &-73.9\,(17.4) &-81.5\,(17.4) &              &              &0.2  &-0.17&                                                                                      \\
367  & 21:38:20.25 &  57:22:39.2$^{\rm r}$& 408  & 3408                      &                               &                 &                 &   14.2$^{\rm j}$&       &                 &12.653\,(26) &12.358\,(32) & 12.247\,(22)$^{\rm r}$&                    &                 &               &18.4\,(13.2)  &-9.4\,(13.2)  &64.5\,(6.1)   &16.6\,(6.1)   &0.04 &0.18 &                                                                                      \\
368  & 21:38:20.95 &  57:23:00.5$^{\rm r}$& 409  & 3409                      &                               &                 &                 &   14.8$^{\rm j}$&       &                 &11.690\,(26) &10.997\,(30) & 10.749\,(18)$^{\rm r}$&                    &                 &               &-1.9\,(4)     &-2.6\,(4)     &18\,(6.8)     &-19.5\,(6.8)  &     &     &                                                                                      \\
369  & 21:38:25.30 &  57:23:12.5$^{\rm r}$& 410  & 3410                      &                               &                 &  15.07$^{\rm f}$&   14.3$^{\rm e}$&       &                 &12.469\,(27) &12.086\,(30) & 11.975\,(22)$^{\rm r}$&        F5$^{\rm e}$&                 &  1.1$^{\rm e}$&-6.9\,(4)     &0.1\,(4)      &8.4\,(6.8)    &8.1\,(6.8)    &-0.28&-0.03&same star                                                                             \\
370  & 21:38:28.90 &  57:22:20.4$^{\rm r}$& 411  & 3411                      &                               &                 &                 &   13.8$^{\rm j}$&       &                 &10.830\,(27) &10.042\,(32) & 9.846 \,(20)$^{\rm r}$&                    &                 &               &0.5\,(5.1)    &-1.6\,(5.1)   &-20\,(7.2)    &-0.6\,(7.1)   &-0.1 &0.07 &                                                                                      \\
371  & 21:38:39.53 &  57:22:01.7$^{\rm r}$& 412  & 463                       &                               &   9.03$^{\rm l}$&    9.5$^{\rm l}$&    9.5$^{\rm g}$& 9.2   &   9.08$^{\rm g}$&8.203 \,(21) &8.203 \,(71) & 8.051 \,(23)$^{\rm r}$&        B1$^{\rm p}$&  III-V$^{\rm p}$&               &-4.3\,(1.3)   &-5\,(1.2)     &-5.8\,(0.6)   &-3.9\,(0.8)   &0.11 &0.01 &                                                                                      \\
372  & 21:38:40.09 &  57:23:27.5$^{\rm r}$& 413  & 3413                      &                               &                 &                 &   11.8$^{\rm j}$&       &                 &10.615\,(31) &10.278\,(31) & 10.206\,(24)$^{\rm r}$&                    &                 &               &-8.6\,(2.7)   &-12.5\,(2.7)  &-13.6\,(1.7)  &-8.9\,(2.3)   &0.83 &-0.65&                                                                                      \\
373  & 21:38:46.00 &  57:22:29.8$^{\rm r}$& 414  & 3414                      &                               &                 &                 &   13.3$^{\rm j}$&       &                 &11.829\,(25) &11.550\,(32) & 11.443\,(24)$^{\rm r}$&                    &                 &               &0.5\,(4)      &4\,(4)        &11.6\,(7.1)   &30.5\,(7.1)   &-0.35&-0.31&                                                                                      \\
374  & 21:38:51.39 &  57:20:10.5$^{\rm r}$& 415  & 3415                      &                               &                 &                 &   13.4$^{\rm j}$&       &                 &12.162\,(25) &11.962\,(29) & 11.839\,(23)$^{\rm r}$&                    &                 &               &0\,(4)        &1.4\,(4)      &9.3\,(6.9)    &8.9\,(6.9)    &-0.01&0.29 &                                                                                      \\
375  & 21:38:18.86 &  57:24:17.4$^{\rm r}$& 416  & 3416                      &                               &                 &                 &     15$^{\rm j}$&       &                 &13.570\,(39) &13.349\,(50) & 13.247\,(44)$^{\rm r}$&                    &                 &               &15.9\,(7.6)   &-10.7\,(7.6)  &-22.6\,(6.9)  &18.7\,(6.9)   &     &     &                                                                                      \\
376  & 21:38:43.60 &  57:24:03.2$^{\rm r}$& 417  & 3417                      &                               &                 &                 &   14.8$^{\rm j}$&       &                 &13.152\,(29) &12.826\,(32) & 12.795\,(33)$^{\rm r}$&                    &                 &               &-5.9\,(4)     &1.2\,(4)      &-21\,(6.7)    &-7.9\,(6.6)   &     &     &                                                                                      \\
377  & 21:38:47.86 &  57:23:47.0$^{\rm r}$& 418  & 3418                      &                               &                 &                 &   13.6$^{\rm j}$&       &                 &12.030\,(25) &11.645\,(29) & 11.540\,(23)$^{\rm r}$&                    &                 &               &43.6\,(18.1)  &-15.5\,(18.1) &12.7\,(6.8)   &-3.5\,(6.9)   &0.46 &-0.42&                                                                                      \\
378  & 21:39:04.35 &  57:21:14.3$^{\rm r}$& 419  & 3419                      &                               &                 &                 &   14.5$^{\rm j}$&       &                 &12.951\,(25) &12.784\,(35) & 12.692\,(29)$^{\rm r}$&                    &                 &               &-6.2\,(4)     &0.4\,(4)      &-3.7\,(6.8)   &9.4\,(6.7)    &-0.17&0.11 &                                                                                      \\
379  & 21:39:07.35 &  57:19:53.5$^{\rm r}$& 420  & 468                       &                               &  12.02$^{\rm l}$&  11.08$^{\rm h}$&  11.04$^{\rm h}$&       &                 &10.480\,(25) &10.392\,(29) & 10.297\,(21)$^{\rm r}$&        A1$^{\rm h}$&                 & 1.19$^{\rm h}$&-2.9\,(1.7)   &-2.7\,(1.7)   &-2.8\,(0.6)   &-2.7\,(0.9)   &-0.01&6.11 &Dec [h] imprec.                                                               \\
380  & 21:39:07.36 &  57:21:10.0$^{\rm r}$& 421  & 3421                      &                               &                 &                 &   13.3$^{\rm j}$&       &                 &11.866\,(25) &11.539\,(28) & 11.440\,(23)$^{\rm r}$&                    &                 &               &-1.7\,(4)     &1.2\,(4)      &8.7\,(6.9)    &6.3\,(6.9)    &-0.27&-0.07&                                                                                      \\
381  & 21:39:16.86 &  57:19:11.1$^{\rm r}$& 422  & 3422                      &                               &                 &                 &   11.6$^{\rm j}$&       &                 &9.053 \,(23) &8.392 \,(24) & 8.182 \,(26)$^{\rm r}$&                    &                 &               &1.9\,(11.3)   &-17.2\,(11.3) &-3.8\,(1.1)   &-11.3\,(0.5)  &0.46 &-0.55&                                                                                      \\
382  & 21:39:26.62 &  57:18:43.9$^{\rm r}$& 423  & 475                       &                               &  10.46$^{\rm l}$&  10.14$^{\rm l}$&   9.27$^{\rm l}$&       &                 &7.079 \,(24) &6.749 \,(36) & 6.591 \,(15)$^{\rm r}$&        A2$^{\rm p}$&     Ia$^{\rm p}$&               &-2.4\,(1.3)   &-3.9\,(1.2)   &-2.4\,(0.6)   &-3\,(0.7)     &0.09 &0.17 &                                                                                      \\
383  & 21:38:03.29 &  57:27:04.4$^{\rm r}$& 425  & 3425                      &                               &                 &                 &   15.1$^{\rm j}$&       &                 &11.979\,(26) &11.200\,(31) & 10.974\,(18)$^{\rm r}$&                    &                 &               &-4.8\,(4)     &0.3\,(4)      &-5.3\,(6.8)   &4.9\,(6.8)    &     &     &                                                                                      \\
384  & 21:38:08.45 &  57:26:47.7$^{\rm r}$& 426  & 457                       &                               &  12.11$^{\rm l}$&  11.93$^{\rm f}$&  11.59$^{\rm e}$&       &                 &10.510\,(26) &9.724 \,(30) & 8.767 \,(20)$^{\rm r}$&        B7$^{\rm e}$&                 &  1.4$^{\rm e}$&-3.1\,(1.5)   &-2.6\,(1.5)   &-3\,(0.5)     &-4.5\,(1.6)   &-0.07&-0.14&                                                                                      \\
385  & 21:38:16.94 &  57:26:23.7$^{\rm r}$& 427  & 3427                      &                               &                 &                 &   14.8$^{\rm j}$&       &                 &13.077\,(27) &12.638\,(35) & 12.564\,(27)$^{\rm r}$&                    &                 &               &-11.2\,(4)    &-1.9\,(4)     &-18.8\,(7.3)  &-2.6\,(7.3)   &     &     &                                                                                      \\
386  & 21:38:51.55 &  57:26:30.4$^{\rm r}$& 428  & 3428                      &                               &                 &                 &   14.9$^{\rm j}$&       &                 &13.472\,(32) &13.220\,(39) & 13.107\,(40)$^{\rm r}$&                    &                 &               &7.3\,(4)      &1.1\,(4)      &26.2\,(6.9)   &10.7\,(6.9)   &     &     &                                                                                      \\
387  & 21:39:09.69 &  57:24:55.7$^{\rm r}$& 430  & 3430                      &                               &  12.30$^{\rm l}$&   12.6$^{\rm h}$&  12.12$^{\rm h}$&       &                 &10.910\,(23) &10.784\,(28) & 10.710\,(21)$^{\rm r}$&        B2$^{\rm h}$&                 & 2.18$^{\rm h}$&-11.6\,(2.7)  &-1\,(2.7)     &-4.4\,(1.7)   &-5.8\,(0.5)   &0.13 &-0.04&[h] imprec.                                                            \\
388  & 21:39:14.65 &  57:23:14.9$^{\rm r}$& 431  & 3431                      &                               &  15.16$^{\rm l}$&  14.41$^{\rm l}$&  13.21$^{\rm l}$&       &                 &10.795\,(22) &10.265\,(29) & 10.127\,(23)$^{\rm r}$&                    &                 &               &1.4\,(4)      &2.4\,(4)      &1\,(7.3)      &5\,(7.3)      &-0.02&-0.1 &                                                                                      \\
389  & 21:39:21.05 &  57:24:01.3$^{\rm r}$& 432  & 3432\footnote{also 5084}  &                               &  15.37$^{\rm l}$&  14.75$^{\rm f}$&   13.9$^{\rm e}$&       &                 &11.984\,(23) &11.717\,(31) & 11.582\,(23)$^{\rm r}$&        F9$^{\rm e}$&                 &  0.9$^{\rm e}$&-6.8\,(4)     &-0.4\,(4)     &-2.9\,(6.8)   &5.7\,(6.8)    &0.11 &-0.28&                                                                                      \\
390  & 21:39:24.69 &  57:23:53.0$^{\rm r}$& 433  & 3433\footnote{also 5085}  &                               &  16.35$^{\rm l}$&  15.59$^{\rm f}$&  14.47$^{\rm e}$&       &                 &12.152\,(23) &11.681\,(31) & 11.558\,(25)$^{\rm r}$&        F7$^{\rm e}$&                 &  1.9$^{\rm e}$&10.6\,(4)     &7.6\,(4)      &15.7\,(6.8)   &10.3\,(6.8)   &-1.14&0.53 &                                                                                      \\
391  & 21:39:28.66 &  57:22:35.8$^{\rm r}$& 434  & 3434                      &                               &                 &                 &   14.3$^{\rm j}$&       &                 &12.981\,(39) &12.844\,(49) & 12.799\,(40)$^{\rm r}$&                    &                 &               &-37\,(4)      &37.1\,(4)     &              &              &-0.14&-0.25&                                                                                      \\
392  & 21:39:28.22 &  57:24:13.5$^{\rm r}$& 435  & 3435                      &                               &                 &                 &   13.5$^{\rm j}$&       &                 &11.890\,(23) &11.578\,(29) & 11.486\,(23)$^{\rm r}$&                    &                 &               &12.4\,(10.7)  &-8.3\,(10.7)  &4.8\,(7.3)    &-6.2\,(7.3)   &-0.31&0.19 &                                                                                      \\
393  & 21:39:36.43 &  57:23:20.2$^{\rm r}$& 436  & 3436                      &                               &  15.25$^{\rm l}$&  14.08$^{\rm h}$&  14.01$^{\rm h}$&       &                 &12.620\,(23) &12.433\,(31) & 12.270\,(26)$^{\rm r}$&        A8$^{\rm h}$&                 &  1.5$^{\rm h}$&-2.6\,(4.1)   &1.9\,(4.1)    &-4.1\,(6.9)   &3.7\,(6.9)    &0.25 &0.11 &                                                                                      \\
394  & 21:39:18.77 &  57:25:48.8$^{\rm r}$& 437  & 471                       &                               &  12.07$^{\rm l}$&  12.08$^{\rm f}$&  11.59$^{\rm e}$&       &                 &10.423\,(23) &10.195\,(28) & 10.104\,(21)$^{\rm r}$&        B7$^{\rm e}$&                 &  1.9$^{\rm e}$&-5.6\,(1.7)   &-3.6\,(1.7)   &-2.9\,(0.6)   &-6.1\,(1.7)   &0.07 &-0.19&                                                                                      \\
395  & 21:38:19.39 &  57:27:34.7$^{\rm r}$& 438  & 459                       &                               &                 &                 &   10.7$^{\rm j}$&       &                 &8.170 \,(43) &7.447 \,(51) & 7.268 \,(20)$^{\rm r}$&        G8$^{\rm q}$&                 &               &10.4\,(2.7)   &5\,(2.7)      &4.9\,(2.7)    &2.3\,(2.4)    &-1.15&1.15 &                                                                                      \\
396  & 21:38:13.75 &  57:27:21.9$^{\rm r}$& 439  & 3439                      &                               &                 &                 &   14.8$^{\rm j}$&       &                 &13.175\,(27) &12.914\,(33) & 12.764\,(30)$^{\rm r}$&                    &                 &               &-3.7\,(4)     &-5.5\,(4)     &-1.3\,(6.8)   &-10.8\,(6.8)  &     &     &                                                                                      \\
397  & 21:38:03.42 &  57:28:30.7$^{\rm r}$& 440  & 3440                      &                               &                 &                 &   14.5$^{\rm j}$&       &                 &12.443\,(27) &11.959\,(32) & 11.867\,(22)$^{\rm r}$&                    &                 &               &34.9\,(4)     &27.4\,(4)     &34.4\,(6.8)   &35.2\,(6.8)   &     &     &                                                                                      \\
398  & 21:38:01.91 &  57:29:00.2$^{\rm r}$& 441  & 3441                      &                               &                 &                 &   14.3$^{\rm j}$&       &                 &12.347\,(29) &11.941\,(31) & 11.871\,(25)$^{\rm r}$&                    &                 &               &-7.6\,(3.8)   &0.3\,(3.8)    &-11.4\,(6.9)  &9.9\,(6.8)    &-0.34&0.36 &                                                                                      \\
399  & 21:37:57.93 &  57:29:11.7$^{\rm r}$& 442  & 3442                      &                               &                 &                 &   14.8$^{\rm j}$&       &                 &12.737\,(27) &12.290\,(30) & 12.157\,(25)$^{\rm r}$&                    &                 &               &-3.2\,(3.8)   &-3\,(3.8)     &-2.9\,(6.8)   &12.7\,(6.8)   &     &     &                                                                                      \\
400  & 21:38:10.42 &  57:29:04.5$^{\rm r}$& 443  & 3443                      &                               &                 &                 &   14.8$^{\rm j}$&       &                 &11.799\,(24) &11.116\,(29) & 10.957\,(20)$^{\rm r}$&                    &                 &               &-5.6\,(4)     &-0.6\,(4)     &-1\,(6.8)     &5.7\,(6.8)    &0.38 &0.11 &                                                                                      \\
401  & 21:38:15.13 &  57:29:34.4$^{\rm r}$& 444  & 3444                      &                               &                 &                 &   14.8$^{\rm j}$&       &                 &13.412\,(26) &13.149\,(35) & 13.037\,(31)$^{\rm r}$&                    &                 &               &-9.3\,(3.9)   &2.2\,(3.9)    &-3.9\,(6.9)   &10.7\,(6.8)   &-0.15&-0.47&                                                                                      \\
402  & 21:38:20.73 &  57:29:13.8$^{\rm r}$& 445  & 3445                      &                               &                 &                 &   13.1$^{\rm j}$&       &                 &11.467\,(29) &11.005\,(33) & 10.909\,(22)$^{\rm r}$&                    &                 &               &24.3\,(10.8)  &13.7\,(10.8)  &59\,(7.4)     &-14.4\,(7.4)  &-2.14&3.08 &                                                                                      \\
403  & 21:38:27.23 &  57:28:55.0$^{\rm r}$& 446  & 3446                      &                               &                 &                 &   11.9$^{\rm j}$&       &                 &10.723\,(26) &10.450\,(29) & 10.354\,(20)$^{\rm r}$&                    &                 &               &-5.3\,(2.7)   &-24.3\,(2.7)  &-2\,(3.7)     &-22.8\,(3.9)  &-0.25&-2.06&                                                                                      \\
404  & 21:38:26.39 &  57:28:40.6$^{\rm r}$& 447  & 3447                      &                               &  12.94$^{\rm l}$&  12.44$^{\rm f}$&  11.75$^{\rm e}$&       &                 &10.310\,(27) &10.005\,(30) & 9.949 \,(20)$^{\rm r}$&        F0$^{\rm e}$&                 &  1.2$^{\rm e}$&-4.3\,(2)     &-8.1\,(2)     &-4.7\,(0.8)   &-5.1\,(1.2)   &-0.03&-0.11&                                                                                      \\
405  & 21:38:42.54 &  57:27:45.9$^{\rm r}$& 448  & 3448                      &                               &                 &                 &   14.6$^{\rm j}$&       &                 &12.789\,(30) &12.519\,(41) & 12.389\,(45)$^{\rm r}$&                    &                 &               &5.1\,(4)      &-5.1\,(4)     &38.7\,(6.7)   &-3.1\,(7)     &0.13 &0.18 &                                                                                      \\
406  & 21:38:49.95 &  57:27:32.8$^{\rm j}$& 449  & 3449                      &                               &                 &                 &   14.6$^{\rm j}$&       &                 &             &             &                       &                    &                 &               &-2.1\,(5.5)   &-5.7\,(5.5)   &-8.4\,(5.9)   &5.9\,(5.9)    &0.1  &0.07 &                                                                                      \\
407  & 21:38:51.20 &  57:27:34.8$^{\rm r}$& 450  & 3450                      &                               &                 &                 &   13.7$^{\rm j}$&       &                 &6.683 \,(23) &5.589 \,(29) & 5.076 \,(17)$^{\rm r}$&                    &                 &               &-4.3\,(4.9)   &-8.2\,(4.9)   &7.7\,(6.8)    &-2\,(6.9)     &-0.02&0.22 &                                                                                      \\
408  & 21:38:58.36 &  57:28:25.6$^{\rm r}$& 451  & 3451                      &                               &                 &                 &   11.6$^{\rm j}$&       &                 &10.627\,(25) &10.506\,(31) & 10.417\,(23)$^{\rm r}$&                    &                 &               &-3.3\,(2.7)   &-0.4\,(2.7)   &-4\,(0.8)     &-5.3\,(1.3)   &-0.06&-0.1 &                                                                                      \\
409  & 21:39:04.77 &  57:28:17.2$^{\rm r}$& 452  & 3452                      &                               &                 &                 &   14.7$^{\rm j}$&       &                 &11.700\,()   &11.341\,(61) & 11.164\,(43)$^{\rm r}$&                    &                 &               &20.1\,(3.9)   &-23.1\,(3.9)  &81.7\,(6.9)   &-17.7\,(6.9)  &     &     &                                                                                      \\
410  & 21:39:11.86 &  57:28:14.9$^{\rm r}$& 453  & 3453                      &                               &                 &                 &     15$^{\rm j}$&       &                 &13.050\,(31) &12.656\,(41) & 12.509\,(35)$^{\rm r}$&                    &                 &               &-8.8\,(3.9)   &-9.1\,(3.9)   &-27.3\,(7)    &39\,(6.9)     &     &     &                                                                                      \\
411  & 21:39:13.25 &  57:28:32.6$^{\rm j}$& 454  & 3454                      &                               &  13.22$^{\rm l}$&  12.85$^{\rm h}$&  12.34$^{\rm h}$&       &                 &             &             &                       &        A0$^{\rm h}$&                 & 1.59$^{\rm h}$&              &              &              &              &-0.05&-0.09&[b] No. 24 near\\%2 stars nearby                                                      \\
412  & 21:39:22.25 &  57:27:45.6$^{\rm r}$& 455  & 473                       &                               &  11.08$^{\rm l}$&  10.87$^{\rm h}$&  10.51$^{\rm h}$&       &                 &9.779 \,(23) &9.610 \,(29) & 9.592 \,(20)$^{\rm r}$&        F1$^{\rm h}$&                 & 0.09$^{\rm h}$&3.5\,(1.5)    &7.2\,(1.5)    &3.3\,(0.6)    &5.1\,(0.8)    &-0.61&0.98 &                                                                                      \\
413  & 21:38:07.00 &  57:30:29.4$^{\rm r}$& 456  & 3456                      &                               &                 &                 &   14.9$^{\rm j}$&       &                 &12.945\,(26) &12.498\,(31) & 12.396\,(22)$^{\rm r}$&                    &                 &               &-5.1\,(3.9)   &2\,(3.9)      &-0.8\,(7.2)   &1.3\,(7.3)    &     &     &                                                                                      \\
414  & 21:38:10.41 &  57:30:19.1$^{\rm r}$& 457  & 3457                      &                               &                 &                 &   14.3$^{\rm j}$&       &                 &12.359\,(27) &11.918\,(32) & 11.831\,(20)$^{\rm r}$&                    &                 &               &-11.5\,(3.9)  &-4\,(3.9)     &-11.5\,(7.3)  &-9.1\,(7.3)   &0.07 &-0.15&                                                                                      \\
415  & 21:38:17.99 &  57:30:27.5$^{\rm j}$& 458  & 3458                      &                               &                 &                 &   14.6$^{\rm j}$&       &                 &             &             &                       &                    &                 &               &-4.1\,(8.5)   &-14.8\,(8.5)  &              &              &     &     &no star                                                                          \\
416  & 21:38:19.48 &  57:30:10.6$^{\rm r}$& 459  & 3459                      &                               &                 &  16.79$^{\rm l}$&   14.5$^{\rm l}$&       &                 &9.736 \,(26) &8.618 \,(31) & 8.239 \,(29)$^{\rm r}$&                    &                 &               &-1.9\,(5.1)   &0.5\,(5.1)    &15.8\,(6.8)   &22.3\,(6.9)   &-0.23&0.13 &                                                                                      \\
417  & 21:38:57.61 &  57:29:20.5$^{\rm r}$& 460  & 466                       &                               &   5.17$^{\rm l}$&   5.89$^{\rm l}$&   9.12$^{\rm g}$& 8.66  &   8.42$^{\rm g}$&5.207 \,()   &5.254 \,(42) & 5.215 \,(17)$^{\rm r}$&        O6$^{\rm p}$&      V$^{\rm p}$&               &-1.6\,(0.3)   &-2.7\,(0.4)   &-6.2\,(2)     &-3.2\,(2.1)   &-0.37&-0.16&[m] colors inconsistent                                                             \\
418  & 21:38:56.71 &  57:29:39.1$^{\rm r}$& 462  & 3462                      &                               &   7.51$^{\rm l}$&    8.2$^{\rm l}$&   8.01$^{\rm l}$&       &                 &7.651 \,(24) &7.669 \,(33) & 7.704 \,(21)$^{\rm r}$&                    &                 &               &-1.4\,(1.9)   &-1.6\,(1.9)   &              &              &-0.01&-0.02&                                                                                      \\
419  & 21:39:11.99 &  57:29:57.2$^{\rm r}$& 463  & 470                       &                               &  12.76$^{\rm l}$&  12.42$^{\rm f}$&  11.92$^{\rm e}$&       &                 &10.750\,(25) &10.642\,(26) & 10.548\,(21)$^{\rm r}$&        A0$^{\rm e}$&                 &  1.6$^{\rm e}$&2.2\,(2)      &5.1\,(2)      &2.3\,(0.8)    &-3.9\,(1.8)   &-0.55&-0.18&                                                                                      \\
420  & 21:39:27.39 &  57:29:00.8$^{\rm r}$& 464  & 477                       &                               &   8.18$^{\rm l}$&   8.71$^{\rm l}$&    9.1$^{\rm g}$& 8.75  &   8.64$^{\rm g}$&7.506 \,(21) &7.452 \,(33) & 7.422 \,(20)$^{\rm r}$&        B0$^{\rm p}$&      V$^{\rm p}$&               &-3.4\,(1.2)   &-1\,(1.2)     &              &              &-0.04&0.03 &                                                                                      \\
421  & 21:39:27.06 &  57:28:40.0$^{\rm r}$& 465  & 3465                      &                               &                 &                 &   11.1$^{\rm j}$&       &                 &9.667 \,(23) &9.277 \,(26) & 8.702 \,(21)$^{\rm r}$&                    &                 &               &-11\,(6.6)    &-6.1\,(6.6)   &-3.1\,(0.6)   &-4.8\,(1)     &-0.01&-0.07&                                                                                      \\
422  & 21:38:55.68 &  57:30:33.8$^{\rm r}$& 466  & 3466                      &                               &                 &                 &   12.9$^{\rm j}$&       &                 &9.868 \,(23) &9.111 \,(28) & 8.924 \,(21)$^{\rm r}$&                    &                 &               &-13.2\,(4.9)  &-7\,(4.9)     &-20.5\,(6.4)  &11.1\,(6.5)   &0.7  &-0.69&                                                                                      \\
423  & 21:39:03.40 &  57:30:28.9$^{\rm r}$& 467  & 3467                      &                               &  15.69$^{\rm l}$&  15.46$^{\rm l}$&  14.26$^{\rm l}$&       &                 &11.504\,(23) &10.917\,(28) & 10.715\,(20)$^{\rm r}$&                    &                 &               &-0.3\,(3.8)   &-3.7\,(3.8)   &-0.7\,(6.7)   &11.2\,(6.7)   &0.18 &-0.1 &                                                                                      \\
424  & 21:38:08.16 &  57:31:26.8$^{\rm r}$& 468  & 456                       &                               &  10.88$^{\rm l}$&  10.91$^{\rm f}$&   10.6$^{\rm e}$&       &                 &9.868 \,(27) &9.738 \,(32) & 9.712 \,(18)$^{\rm r}$&        B7$^{\rm e}$&                 &  1.3$^{\rm e}$&1.3\,(1.4)    &-0.1\,(1.4)   &-2.4\,(0.6)   &-5.6\,(1.4)   &-0.14&-0.21&                                                                                      \\
425  & 21:38:17.32 &  57:31:22.0$^{\rm r}$& 469  & 3469\footnote{also 4633}  & 13-277                        &  16.99$^{\rm f}$&   16.4$^{\rm l}$&  13.92$^{\rm e}$& 12.96 &  12.15$^{\rm e}$&10.279\,(26) &9.329 \,(30) & 8.593 \,(20)$^{\rm r}$&        G1$^{\rm c}$&                 &  2.5$^{\rm e}$&-3\,(5.1)     &-3.2\,(5.1)   &-0.9\,(6.8)   &6.1\,(6.8)    &-0.19&-0.12&GM Cep, SB1:$^{\rm c}$                                                                                      \\
426  & 21:38:26.47 &  57:31:10.3$^{\rm r}$& 470  & 3470                      &                               &                 &                 &   14.8$^{\rm j}$&       &                 &12.626\,(29) &12.057\,(35) & 11.969\,(28)$^{\rm r}$&                    &                 &               &-31\,(5.4)    &1.3\,(5.4)    &-37.6\,(7.3)  &11.9\,(7)     &     &     &                                                                                      \\
427  & 21:38:11.10 &  57:32:50.9$^{\rm r}$& 471  & 458                       &                               &                 &                 &   11.5$^{\rm j}$&       &                 &10.548\,(27) &10.268\,(32) & 10.201\,(18)$^{\rm r}$&                    &                 &               &-4.1\,(1.7)   &-1.8\,(1.7)   &-0.9\,(8.2)   &-1\,(8.2)     &-0.75&0.46 &                                                                                      \\
428  & 21:38:19.22 &  57:31:56.8$^{\rm r}$& 472  & 3472                      &                               &  14.64$^{\rm l}$&  14.36$^{\rm f}$&  13.69$^{\rm e}$&       &                 &12.188\,(31) &11.960\,(38) & 11.839\,(28)$^{\rm r}$&        A8$^{\rm e}$&                 &  1.4$^{\rm e}$&3.1\,(3.8)    &-5.6\,(3.8)   &46.4\,(7.1)   &16.3\,(7.1)   &0.39 &-0.29&                                                                                      \\
429  & 21:38:22.48 &  57:32:12.4$^{\rm r}$& 473  & 3473                      &                               &                 &                 &   14.2$^{\rm j}$&       &                 &9.999 \,(26) &8.997 \,(32) & 8.692 \,(22)$^{\rm r}$&                    &                 &               &-0.7\,(4.7)   &-7\,(4.7)     &-10\,(7)      &3.6\,(6.9)    &-0.03&0.11 &                                                                                      \\
430  & 21:38:24.22 &  57:33:22.4$^{\rm r}$& 474  & 3474                      &                               &                 &                 &   12.5$^{\rm j}$&       &                 &11.595\,(26) &11.433\,(33) & 11.341\,(22)$^{\rm r}$&                    &                 &               &-9.4\,(2.7)   &-13.1\,(2.7)  &-6\,(1.4)     &1.3\,(1.3)    &-0.06&0.03 &                                                                                      \\
431  & 21:38:30.81 &  57:33:05.6$^{\rm r}$& 475  & 3475                      &                               &                 &                 &   15.2$^{\rm j}$&       &                 &13.750\,(35) &13.417\,(35) & 13.219\,(35)$^{\rm r}$&                    &                 &               &1.4\,(3.8)    &-8.1\,(3.8)   &-5.1\,(7.3)   &12\,(7.6)     &     &     &new coordinates                                                                       \\     
432  & 21:38:37.06 &  57:32:49.8$^{\rm r}$& 476  & 3476                      &                               &                 &                 &   15.2$^{\rm j}$&       &                 &12.352\,(29) &11.636\,(32) & 11.444\,(22)$^{\rm r}$&                    &                 &               &-6.4\,(3.9)   &4.3\,(3.9)    &-14.2\,(7.4)  &26.4\,(7.1)   &     &     &                                                                                      \\
433  & 21:38:40.65 &  57:32:35.9$^{\rm r}$& 477  & 3477                      &                               &                 &                 &   14.9$^{\rm j}$&       &                 &12.793\,(25) &12.366\,(31) & 12.201\,(25)$^{\rm r}$&                    &                 &               &-3.1\,(3.8)   &-6.2\,(3.8)   &-7.6\,(6.9)   &-0.7\,(7)     &     &     &                                                                                      \\
434  & 21:38:48.24 &  57:32:13.3$^{\rm r}$& 478  & 3478                      &                               &                 &                 &   14.7$^{\rm j}$&       &                 &11.274\,(25) &10.495\,(28) & 10.266\,(20)$^{\rm r}$&                    &                 &               &-12.8\,(18.1) &-19.9\,(18.1) &-45.6\,(6.6)  &-4.8\,(6.7)   &     &     &                                                                                      \\
435  & 21:38:59.80 &  57:31:58.2$^{\rm r}$& 479  & 3479                      &                               &                 &                 &   11.7$^{\rm j}$&       &                 &10.650\,(23) &10.424\,(26) & 10.358\,(18)$^{\rm r}$&                    &                 &               &-22.1\,(2.7)  &-17.1\,(2.7)  &-12\,(0.8)    &-11.9\,(2)    &1.03 &-0.74&                                                                                      \\
436  & 21:39:09.83 &  57:31:26.1$^{\rm r}$& 480  & 3480                      &                               &  14.51$^{\rm l}$&  14.01$^{\rm h}$&  13.05$^{\rm h}$&       &                 &12.016\,(23) &11.783\,(31) & 11.710\,(23)$^{\rm r}$&        F0$^{\rm h}$&                 & 1.15$^{\rm h}$&-3.9\,(3.9)   &-3.1\,(3.9)   &-0.3\,(7)     &2.7\,(7)      &-0.22&0.26 &Dec [h] imprec.                                                               \\
437  & 21:39:06.23 &  57:31:04.4$^{\rm r}$& 482  & 3482                      &                               &                 &                 &   14.9$^{\rm j}$&       &                 &11.952\,(27) &11.197\,(29) & 10.981\,(23)$^{\rm r}$&                    &                 &               &-2.1\,(3.8)   &-10.8\,(3.8)  &-6.4\,(6.9)   &-18.2\,(6.9)  &     &     &                                                                                      \\
438  & 21:38:21.87 &  57:34:17.5$^{\rm r}$& 483  & 3483                      &                               &                 &                 &     13$^{\rm j}$&       &                 &10.804\,(26) &10.174\,(29) & 10.061\,(22)$^{\rm r}$&                    &                 &               &-2.2\,(9.5)   &-22.8\,(9.5)  &-14.6\,(7.3)  &-30.5\,(7.3)  &0.6  &-2.45&                                                                                      \\
439  & 21:38:30.10 &  57:34:04.0$^{\rm r}$& 485  & 3485                      &                               &                 &                 &   15.2$^{\rm j}$&       &                 &12.014\,(26) &11.199\,(29) & 11.006\,(20)$^{\rm r}$&                    &                 &               &-1.6\,(3.8)   &-0.2\,(3.8)   &-8.6\,(7.5)   &-4.8\,(7.4)   &     &     &                                                                                      \\
440  & 21:38:34.08 &  57:35:00.5$^{\rm r}$& 486  & 3486                      &                               &                 &  15.46$^{\rm l}$&  13.83$^{\rm l}$&       &                 &10.651\,(26) &9.899 \,(29) & 9.688 \,(20)$^{\rm r}$&                    &                 &               &-4.4\,(4.8)   &-3.8\,(4.8)   &-7.3\,(7.4)   &-2.3\,(7.4)   &0.16 &0.14 &                                                                                      \\
441  & 21:38:42.28 &  57:33:46.7$^{\rm r}$& 488  & 3488                      &                               &                 &                 &     14$^{\rm j}$&       &                 &12.271\,(27) &11.956\,(40) & 11.832\,(26)$^{\rm r}$&                    &                 &               &-27.1\,(3.8)  &-11.8\,(3.8)  &-76\,(6.4)    &-15.2\,(6.4)  &1.39 &-1.07&                                                                                      \\
442  & 21:38:43.50 &  57:33:44.2$^{\rm r}$& 489  & 3489                      &                               &                 &                 &   14.8$^{\rm j}$&       &                 &11.730\,(23) &11.057\,(28) & 10.840\,(23)$^{\rm r}$&                    &                 &               &33.2\,(5)     &-17.8\,(5)    &39.4\,(4.9)   &-7.5\,(5.3)   &     &     &                                                                                      \\
443  & 21:38:49.81 &  57:33:05.1$^{\rm r}$& 490  & 3490                      &                               &                 &                 &   14.3$^{\rm j}$&       &                 &12.410\,(25) &12.137\,(33) & 11.986\,(24)$^{\rm r}$&                    &                 &               &-4.7\,(3.9)   &1.2\,(3.9)    &5.1\,(7)      &0.2\,(6.9)    &0.01 &0.1  &                                                                                      \\
444  & 21:38:48.50 &  57:34:14.6$^{\rm r}$& 492  & 3492                      &                               &                 &                 &     15$^{\rm j}$&       &                 &12.990\,(23) &12.583\,(29) & 12.510\,(24)$^{\rm r}$&                    &                 &               &-7.8\,(3.8)   &-6.1\,(3.8)   &2.3\,(7.3)    &-4.1\,(7.4)   &-6.41&2.78 &                                                                                      \\
445  & 21:38:57.22 &  57:33:13.1$^{\rm r}$& 493  & 3493                      &                               &                 &                 &   13.9$^{\rm j}$&       &                 &11.617\,(23) &11.075\,(28) & 10.918\,(18)$^{\rm r}$&                    &                 &               &-0.4\,(3.8)   &-0.3\,(3.8)   &0.7\,(7)      &8.6\,(7)      &-0.61&0.28 &                                                                                      \\
446  & 21:39:15.70 &  57:32:42.4$^{\rm r}$& 494  & 3494                      &                               &                 &                 &   15.1$^{\rm j}$&       &                 &11.818\,(23) &11.114\,(26) & 10.893\,(21)$^{\rm r}$&                    &                 &               &-9.8\,(3.8)   &-7.9\,(3.8)   &-5.4\,(6.9)   &7.4\,(6.9)    &     &     &                                                                                      \\
447  & 21:39:16.71 &  57:33:05.2$^{\rm r}$& 495  & 3495                      &                               &                 &                 &     15$^{\rm j}$&       &                 &13.065\,(29) &12.699\,(36) & 12.583\,(30)$^{\rm r}$&                    &                 &               &-12.4\,(3.9)  &0.4\,(3.9)    &-18.9\,(6.9)  &15.8\,(7)     &     &     &                                                                                      \\
448  & 21:39:22.32 &  57:31:48.9$^{\rm r}$& 497  & 3497                      &                               &  13.55$^{\rm l}$&  13.26$^{\rm f}$&  12.76$^{\rm e}$&       &                 &11.577\,(25) &11.372\,(33) & 11.269\,(24)$^{\rm r}$&        A1$^{\rm e}$&                 &  1.5$^{\rm e}$&-8.2\,(2.7)   &-10.6\,(2.7)  &-5.6\,(1.5)   &-1.4\,(2.8)   &0.2  &0.05 &                                                                                      \\
449  & 21:39:26.11 &  57:31:47.6$^{\rm r}$& 498  & 3498                      &                               &  15.59$^{\rm l}$&  14.78$^{\rm l}$&  13.58$^{\rm l}$&       &                 &11.084\,(23) &10.500\,(29) & 10.363\,(21)$^{\rm r}$&                    &                 &               &1.3\,(3.8)    &-9.2\,(3.8)   &22.7\,(7.2)   &-13.9\,(7.1)  &-0.01&-0.12&                                                                                      \\
450  & 21:39:30.39 &  57:30:00.2$^{\rm r}$& 499  & 3499                      &                               &                 &                 &   14.6$^{\rm j}$&       &                 &12.563\,(29) &12.153\,(37) & 12.061\,(29)$^{\rm r}$&                    &                 &               &-16.3\,(3.9)  &-0.9\,(3.9)   &-39.6\,(7.2)  &-16.8\,(7.1)  &     &     &                                                                                      \\
451  & 21:39:32.65 &  57:30:56.3$^{\rm r}$& 500  & 3500                      &                               &                 &                 &     15$^{\rm j}$&       &                 &11.490\,(23) &10.667\,(27) & 10.404\,(22)$^{\rm r}$&                    &                 &               &-7.1\,(3.9)   &-3.8\,(3.9)   &1.2\,(6.9)    &11.2\,(6.9)   &     &     &                                                                                      \\
452  & 21:38:32.85 &  57:36:01.2$^{\rm r}$& 502  & 3502                      &                               &                 &                 &   15.1$^{\rm j}$&       &                 &10.999\,(26) &9.923 \,(31) & 9.624 \,(22)$^{\rm r}$&                    &                 &               &-3.2\,(4.9)   &-1.6\,(4.9)   &-6\,(7.5)     &-10.4\,(8.2)  &     &     &                                                                                      \\
453  & 21:38:39.41 &  57:35:02.4$^{\rm r}$& 504  & 3504                      &                               &                 &                 &     15$^{\rm j}$&       &                 &13.030\,(32) &12.504\,(35) & 12.408\,(22)$^{\rm r}$&                    &                 &               &-2.9\,(3.9)   &1.4\,(3.9)    &13.4\,(7.4)   &-5.2\,(7.5)   &     &     &                                                                                      \\
454  & 21:39:13.25 &  57:34:28.6$^{\rm r}$& 507  & 3507                      &                               &                 &                 &     13$^{\rm j}$&       &                 &11.927\,(22) &11.764\,(29) & 11.665\,(23)$^{\rm r}$&                    &                 &               &-5.2\,(3.8)   &-4.9\,(3.8)   &-10.6\,(1.4)  &-5.3\,(2.1)   &0.16 &-0.12&                                                                                      \\
455  & 21:39:26.75 &  57:33:13.2$^{\rm r}$& 508  & 5086                      &                               &                 &  16.28$^{\rm f}$&  15.34$^{\rm e}$&       &                 &12.909\,(27) &12.539\,(32) & 12.457\,(30)$^{\rm r}$&        F9$^{\rm e}$&                 &  1.2$^{\rm e}$&-10.5\,(3.8)  &-5\,(3.8)     &-23.1\,(7)    &10.4\,(7)     &     &     &                                                                                      \\
456  & 21:39:24.96 &  57:33:29.5$^{\rm r}$& 509  & 3509                      &                               &                 &                 &   15.2$^{\rm j}$&       &                 &13.113\,(26) &12.747\,(31) & 12.588\,(29)$^{\rm r}$&                    &                 &               &-3.5\,(3.8)   &0.8\,(3.8)    &24.2\,(7)     &35.5\,(7)     &     &     &                                                                                      \\
457  & 21:39:24.37 &  57:34:05.3$^{\rm r}$& 511  & 3511                      &                               &                 &                 &   14.4$^{\rm j}$&       &                 &12.465\,(27) &12.224\,(36) & 12.066\,(30)$^{\rm r}$&                    &                 &               &-4.3\,(3.9)   &4\,(3.9)      &-27.5\,(7.4)  &-11.4\,(7.5)  &-0.9 &0.09 &                                                                                      \\
458  & 21:39:18.73 &  57:34:53.5$^{\rm r}$& 512  & 5083                      &                               &                 &  15.34$^{\rm f}$&  14.12$^{\rm e}$&       &                 &12.333\,(24) &11.900\,(28) & 11.818\,(26)$^{\rm r}$&        F9$^{\rm e}$&                 &  2.1$^{\rm e}$&-12.9\,(3.9)  &-10\,(3.9)    &-22.9\,(7.4)  &-18.5\,(7.4)  &0.8  &-0.25&                                                                                      \\
459  & 21:39:20.48 &  57:35:03.0$^{\rm r}$& 513  & 3513                      &                               &                 &                 &     14$^{\rm j}$&       &                 &12.228\,(22) &11.880\,(28) & 11.784\,(23)$^{\rm r}$&                    &                 &               &-8.3\,(4)     &-4.1\,(4)     &-7.2\,(6.8)   &4.6\,(6.8)    &0.75 &-0.61&                                                                                      \\
460  & 21:39:24.35 &  57:35:09.8$^{\rm r}$& 514  & 3514                      &                               &                 &  14.65$^{\rm l}$&  12.46$^{\rm l}$&       &                 &7.976 \,(24) &6.924 \,(44) & 6.598 \,(20)$^{\rm r}$&                    &                 &               &-8\,(6.4)     &-6\,(6.4)     &-42\,(6.5)    &-63.3\,(6.6)  &-0.06&0.11 &                                                                                      \\
461  & 21:39:25.23 &  57:35:17.2$^{\rm r}$& 515  & 3515                      &                               &                 &                 &   15.1$^{\rm j}$&       &                 &13.094\,(54) &12.846\,(94) & 12.640\,(79)$^{\rm r}$&                    &                 &               &0.6\,(5)      &6.6\,(5)      &50.6\,(5.9)   &47.3\,(6)     &     &     &                                                                                      \\
462  & 21:39:29.51 &  57:34:20.4$^{\rm r}$& 516  & 3516\footnote{also 5087}  &                               &  15.11$^{\rm l}$&  14.69$^{\rm f}$&     14$^{\rm e}$&       &                 &12.439\,(21) &12.205\,(27) & 12.117\,(25)$^{\rm r}$&        F0$^{\rm e}$&                 &  1.2$^{\rm e}$&-9.3\,(3.8)   &-6\,(3.8)     &-9.5\,(7.4)   &-10.8\,(7.4)  &0.4  &-0.22&                                                                                      \\
463  & 21:39:17.66 &  57:35:56.4$^{\rm r}$& 517  & 3517                      &                               &                 &                 &   14.8$^{\rm j}$&       &                 &11.362\,(22) &10.641\,(28) & 10.412\,(20)$^{\rm r}$&                    &                 &               &8.3\,(3.9)    &-6.3\,(3.9)   &35.2\,(7.4)   &-20.5\,(7.5)  &     &     &                                                                                      \\
464  & 21:38:43.99 &  57:36:03.7$^{\rm r}$& 518  & 3518                      &                               &                 &                 &   15.1$^{\rm j}$&       &                 &13.329\,(27) &12.922\,(28) & 12.742\,(26)$^{\rm r}$&                    &                 &               &-6.5\,(3.9)   &-0.3\,(3.9)   &-2.4\,(7.6)   &-13.1\,(7.7)  &     &     &                                                                                      \\
465  & 21:38:42.72 &  57:37:21.5$^{\rm r}$& 521  & 3521                      &                               &                 &                 &   12.2$^{\rm j}$&       &                 &9.839 \,(24) &9.195 \,(29) & 9.018 \,(22)$^{\rm r}$&                    &                 &               &-1.1\,(4.8)   &-3\,(4.8)     &2.8\,(7.6)    &-5.7\,(7.6)   &-0.16&-0.33&                                                                                      \\
466  & 21:38:59.50 &  57:36:50.7$^{\rm r}$& 523  & 3523                      &                               &  14.09$^{\rm l}$&  13.63$^{\rm f}$&  12.73$^{\rm e}$&       &                 &10.893\,(24) &10.520\,(28) & 10.445\,(22)$^{\rm r}$&        F7$^{\rm e}$&                 &  1.3$^{\rm e}$&-3.8\,(3.8)   &-1.8\,(3.8)   &-4.2\,(0.5)   &1.9\,(1.9)    &-0.23&0.06 &                                                                                      \\
467  & 21:39:11.42 &  57:36:33.1$^{\rm r}$& 525  & 3525                      &                               &                 &                 &   14.8$^{\rm j}$&       &                 &13.185\,(26) &12.854\,(31) & 12.811\,(26)$^{\rm r}$&                    &                 &               &-6.7\,(3.8)   &3.7\,(3.8)    &1.8\,(7.3)    &-10.7\,(7.5)  &     &     &                                                                                      \\
468  & 21:39:16.86 &  57:36:24.5$^{\rm j}$& 526  & 3526                      &                               &                 &                 &   14.8$^{\rm j}$&       &                 &             &             &                       &                    &                 &               &              &              &              &              &     &     &no star                                                                          \\
469  & 21:39:20.85 &  57:36:25.5$^{\rm r}$& 527  & 3527                      &                               &                 &                 &   13.9$^{\rm j}$&       &                 &12.128\,(22) &11.871\,(29) & 11.719\,(22)$^{\rm r}$&                    &                 &               &-3\,(3.8)     &-3.3\,(3.8)   &-3.7\,(7.3)   &-12.4\,(7.4)  &0.01 &0.03 &                                                                                      \\
470  & 21:38:12.33 &  57:38:38.8$^{\rm r}$& 528  & 3528\footnote{also 5075}  &                               &                 &                 &     15$^{\rm j}$&       &                 &11.826\,(27) &11.056\,(29) & 10.839\,(18)$^{\rm r}$&                    &                 &               &0.6\,(3.8)    &0.7\,(3.8)    &-2.2\,(7.4)   &-3\,(7.4)     &     &     &                                                                                      \\
471  & 21:38:19.70 &  57:38:46.7$^{\rm r}$& 529  & 3529                      &                               &                 &                 &     15$^{\rm j}$&       &                 &13.098\,(27) &12.687\,(32) & 12.586\,(24)$^{\rm r}$&                    &                 &               &-12.9\,(3.8)  &-11.6\,(3.8)  &-14.4\,(7.3)  &-8.8\,(7.4)   &     &     &                                                                                      \\
472  & 21:38:25.73 &  57:38:22.0$^{\rm r}$& 530  & 3530                      &                               &                 &                 &   15.1$^{\rm j}$&       &                 &13.025\,(44) &12.602\,(44) & 12.406\,(37)$^{\rm r}$&                    &                 &               &-114.6\,(5.4) &-102.2\,(5.4) &24.7\,(7.3)   &-63.9\,(7.6)  &     &     &                                                                                      \\
473  & 21:38:25.84 &  57:37:55.7$^{\rm j}$& 531  & 3531                      &                               &                 &                 &   15.1$^{\rm j}$&       &                 &             &             &                       &                    &                 &               &              &              &              &              &     &     &no star                                                                          \\
474  & 21:38:34.38 &  57:38:44.6$^{\rm r}$& 532  & 3532                      &                               &                 &                 &   13.4$^{\rm j}$&       &                 &10.681\,(26) &9.934 \,(30) & 9.766 \,(19)$^{\rm r}$&                    &                 &               &-2.3\,(4.7)   &-5.2\,(4.7)   &-7\,(7.4)     &-10.3\,(7.4)  &-0.03&-0.02&                                                                                      \\
475  & 21:38:46.35 &  57:38:49.3$^{\rm r}$& 534  & 3534                      &                               &                 &                 &     14$^{\rm j}$&       &                 &12.203\,(24) &11.823\,(28) & 11.708\,(20)$^{\rm r}$&                    &                 &               &-6.7\,(3.9)   &-1.9\,(3.9)   &-4.3\,(7.4)   &-27\,(7.5)    &0.14 &0.21 &                                                                                      \\
476  & 21:39:00.91 &  57:38:01.0$^{\rm r}$& 535  & 467                       &                               &  11.71$^{\rm l}$&   11.4$^{\rm f}$&   10.8$^{\rm e}$&       &                 &10.409\,(24) &10.249\,(24) & 10.247\,(22)$^{\rm r}$&        B9$^{\rm e}$&                 &  2.1$^{\rm e}$&-5.5\,(1.7)   &2.3\,(1.7)    &-3\,(0.6)     &-2.1\,(1.1)   &-0.25&0.18 &                                                                                      \\
477  & 21:39:16.68 &  57:37:21.3$^{\rm r}$& 537  & 3537                      &                               &  11.71$^{\rm l}$&  11.66$^{\rm l}$&  11.29$^{\rm l}$&       &                 &10.383\,(22) &9.477 \,(25) & 9.324 \,(22)$^{\rm r}$&     B8 B9$^{\rm q}$&                 &               &-4.7\,(4.7)   &-3.5\,(4.7)   &-7.9\,(7.3)   &-5.9\,(7.3)   &-0.02&0.06 &                                                                                      \\
478  & 21:39:18.86 &  57:37:23.5$^{\rm r}$& 538  & 3538                      &                               &                 &                 &   14.9$^{\rm j}$&       &                 &12.843\,(22) &12.262\,(28) & 12.204\,(23)$^{\rm r}$&                    &                 &               &-7.4\,(5.4)   &-23.7\,(5.4)  &33.5\,(7.4)   &23.4\,(7.4)   &     &     &                                                                                      \\
479  & 21:39:20.47 &  57:37:26.6$^{\rm r}$& 539  & 3539                      &                               &                 &                 &   13.6$^{\rm j}$&       &                 &12.101\,()   &12.289\,(47) & 12.188\,(44)$^{\rm r}$&                    &                 &               &-70.5\,(13.2) &-59.6\,(13.2) &              &              &-0.11&0.16 &                                                                                      \\
480  & 21:39:21.05 &  57:37:29.6$^{\rm r}$& 540  & 3540                      &                               &                 &                 &   14.9$^{\rm j}$&       &                 &11.134\,(28) &10.276\,(31) & 9.992 \,(22)$^{\rm r}$&                    &                 &               &              &              &              &              &     &     &                                                                                      \\
481  & 21:39:30.03 &  57:37:37.2$^{\rm r}$& 541  & 3541                      &                               &                 &                 &   15.1$^{\rm j}$&       &                 &13.188\,(26) &12.676\,(28) & 12.577\,(26)$^{\rm r}$&                    &                 &               &13.4\,(3.8)   &-7.1\,(3.8)   &13.5\,(7.5)   &-8.1\,(7.5)   &     &     &                                                                                      \\
482  & 21:39:32.47 &  57:38:03.0$^{\rm r}$& 542  & 3542                      &                               &                 &                 &   14.9$^{\rm j}$&       &                 &13.154\,(29) &12.789\,(33) & 12.729\,(19)$^{\rm r}$&                    &                 &               &-3.1\,(3.8)   &-3.2\,(3.8)   &0.6\,(7.4)    &-11\,(7.4)    &5.14 &0.9  &                                                                                      \\
483  & 21:39:12.20 &  57:38:49.3$^{\rm r}$& 544  & 3544                      &                               &                 &                 &     15$^{\rm j}$&       &                 &11.264\,(24) &10.407\,(28) & 10.148\,(23)$^{\rm r}$&                    &                 &               &5.7\,(3.8)    &0.4\,(3.8)    &11.6\,(7.4)   &15.8\,(7.4)   &     &     &                                                                                      \\
484  & 21:38:17.20 &  57:40:02.0$^{\rm r}$& 545  & 460                       &                               &  12.64$^{\rm l}$&  12.17$^{\rm f}$&  11.52$^{\rm e}$&       &                 &10.434\,(27) &10.215\,(32) & 10.146\,(21)$^{\rm r}$&        A7$^{\rm e}$&                 &  1.4$^{\rm e}$&-5.7\,(2)     &-9.8\,(2)     &-6.3\,(0.6)   &-6\,(1.1)     &0.29 &-0.31&                                                                                      \\
485  & 21:38:39.21 &  57:39:48.8$^{\rm r}$& 546  & 3546                      &                               &                 &                 &   12.5$^{\rm j}$&       &                 &10.053\,(27) &9.326 \,(31) & 9.165 \,(23)$^{\rm r}$&                    &                 &               &-4.2\,(6)     &-18.7\,(6)    &              &              &0.26 &-0.32&                                                                                      \\
486  & 21:38:38.53 &  57:40:38.9$^{\rm r}$& 547  & 3547                      &                               &                 &                 &   12.5$^{\rm j}$&       &                 &10.608\,(27) &10.076\,(31) & 9.920 \,(19)$^{\rm r}$&                    &                 &               &11.4\,(4.7)   &3.7\,(4.7)    &11.2\,(7.6)   &-10.5\,(7.6)  &-1.84&0.45 &                                                                                      \\
487  & 21:38:48.02 &  57:40:50.7$^{\rm r}$& 548  & 3548                      &                               &                 &                 &   13.6$^{\rm j}$&       &                 &10.796\,(21) &10.106\,(28) & 9.934 \,(19)$^{\rm r}$&                    &                 &               &-7.4\,(4.7)   &-0.7\,(4.7)   &-7.2\,(7.4)   &-10\,(7.4)    &0.17 &-0.04&                                                                                      \\
488  & 21:38:47.62 &  57:40:09.6$^{\rm r}$& 549  & 3549                      &                               &                 &                 &   14.1$^{\rm j}$&       &                 &15.033\,(53) &14.297\,(48) & 14.184\,(79)$^{\rm r}$&                    &                 &               &36.1\,(5.1)   &-37.7\,(5.1)  &              &              &     &     &no star                                                                          \\
489  & 21:38:51.32 &  57:39:51.1$^{\rm r}$& 550  & 465                       &                               &  13.37$^{\rm l}$&  12.74$^{\rm l}$&  11.72$^{\rm i}$& 11.1  &   10.5$^{\rm i}$&9.488 \,(22) &8.961 \,(28) & 8.842 \,(22)$^{\rm r}$&       F0:$^{\rm q}$&                 &               &-5.9\,(2.7)   &-5.6\,(2.7)   &-3.9\,(1.4)   &-5.4\,(1)     &0.12 &-0.29&                                                                                      \\
490  & 21:38:50.96 &  57:39:31.9$^{\rm r}$& 551  & 3551                      &                               &                 &                 &     15$^{\rm j}$&       &                 &12.364\,(24) &11.712\,(28) & 11.572\,(22)$^{\rm r}$&                    &                 &               &-13.1\,(3.8)  &-0.7\,(3.8)   &-1\,(7.4)     &-8\,(7.4)     &     &     &                                                                                      \\
491  & 21:39:04.09 &  57:39:54.1$^{\rm r}$& 552  & 3552                      &                               &                 &                 &   14.5$^{\rm j}$&       &                 &12.606\,(22) &12.251\,(26) & 12.157\,(26)$^{\rm r}$&                    &                 &               &9.1\,(3.8)    &3.9\,(3.8)    &5.9\,(7.3)    &-0.5\,(7.4)   &-0.66&0.74 &                                                                                      \\
492  & 21:39:02.65 &  57:39:19.5$^{\rm r}$& 553  & 3553                      &                               &                 &                 &   13.5$^{\rm j}$&       &                 &11.820\,(22) &11.456\,(28) & 11.346\,(20)$^{\rm r}$&                    &                 &               &-9.4\,(3.8)   &-1.4\,(3.8)   &-18.9\,(7.4)  &-10.5\,(7.5)  &0.14 &0.88 &                                                                                      \\
493  & 21:39:10.18 &  57:40:19.9$^{\rm r}$& 554  & 3554                      &                               &                 &                 &   11.4$^{\rm j}$&       &                 &8.598 \,(21) &7.781 \,(24) & 7.569 \,(20)$^{\rm r}$&                    &                 &               &-11.2\,(2.8)  &-2.6\,(2.8)   &-8.8\,(3.8)   &1\,(6.5)      &-0.44&0.5  &                                                                                      \\
494  & 21:39:21.91 &  57:39:11.9$^{\rm r}$& 555  & 474                       &                               &  10.92$^{\rm l}$&  10.13$^{\rm l}$&   9.09$^{\rm l}$&       &                 &7.221 \,(21) &6.759 \,(31) & 6.673 \,(24)$^{\rm r}$&     G8 K0$^{\rm q}$&                 &               &-3.2\,(1.3)   &-3\,(1.2)     &-2\,(0.6)     &-2.7\,(0.6)   &-0.21&-0.01&                                                                                      \\
495  & 21:39:25.60 &  57:40:26.3$^{\rm r}$& 556  & 476                       &                               &                 &                 &     11$^{\rm j}$&       &                 &10.626\,(22) &10.547\,(29) & 10.563\,(25)$^{\rm r}$&        A0$^{\rm q}$&                 &               &-3.2\,(1.7)   &0.3\,(1.7)    &-0.2\,(1.6)   &0.5\,(0.7)    &-0.42&0.48 &                                                                                      \\
496  & 21:39:31.93 &  57:40:39.2$^{\rm r}$& 557  & 3557                      &                               &                 &                 &   13.8$^{\rm j}$&       &                 &12.105\,(27) &11.781\,(31) & 11.661\,(23)$^{\rm r}$&                    &                 &               &8.7\,(3.8)    &3.4\,(3.8)    &9.3\,(7.5)    &-1.8\,(7.5)   &-0.88&0.9  &                                                                                      \\
497  & 21:38:29.18 &  57:41:22.7$^{\rm r}$& 558  & 462                       &                               &  11.67$^{\rm l}$&   11.2$^{\rm f}$&   10.6$^{\rm e}$&       &                 &9.700 \,(29) &9.547 \,(32) & 9.480 \,(18)$^{\rm r}$&        A4$^{\rm e}$&                 &  1.5$^{\rm e}$&-1.4\,(1.6)   &-6.4\,(1.6)   &-6.2\,(0.6)   &-5\,(0.6)     &0.23 &-0.32&[m] HIP\# wrong                                                                       \\
498  & 21:38:30.44 &  57:41:58.9$^{\rm r}$& 559  & 3559                      &                               &                 &                 &   14.6$^{\rm j}$&       &                 &12.674\,(29) &12.291\,(32) & 12.177\,(24)$^{\rm r}$&                    &                 &               &5.6\,(3.9)    &14.7\,(3.9)   &8.7\,(6.9)    &21.3\,(6.9)   &-1.17&1.41 &                                                                                      \\
499  & 21:38:34.95 &  57:42:10.4$^{\rm r}$& 560  & 3560                      &                               &                 &                 &   14.1$^{\rm j}$&       &                 &10.948\,(27) &10.235\,(31) & 10.023\,(19)$^{\rm r}$&                    &                 &               &-4.8\,(3.9)   &-5.7\,(3.9)   &-13.2\,(7.3)  &-11.7\,(7.3)  &0.35 &-0.26&new coordinates                                                                       \\     
500  & 21:38:46.71 &  57:41:59.0$^{\rm r}$& 561  & 3561                      &                               &                 &                 &   14.6$^{\rm j}$&       &                 &12.477\,(26) &12.171\,(29) & 11.987\,(22)$^{\rm r}$&                    &                 &               &9.6\,(3.9)    &6.6\,(3.9)    &9.6\,(6.6)    &8.1\,(6.6)    &0.32 &0.15 &                                                                                      \\
501  & 21:39:05.84 &  57:41:40.0$^{\rm r}$& 562  & 3562                      &                               &                 &                 &   15.1$^{\rm j}$&       &                 &12.981\,(59) &12.529\,(65) & 12.438\,(53)$^{\rm r}$&                    &                 &               &-84.1\,(12.6) &-99.7\,(12.6) &              &              &     &     &near 501                                                                          \\
502  & 21:39:06.23 &  57:41:43.1$^{\rm r}$& 563  & 3563                      &                               &                 &                 &   15.1$^{\rm j}$&       &                 &13.242\,(28) &12.803\,(41) & 12.746\,(40)$^{\rm r}$&                    &                 &               &27.6\,(3.8)   &21.8\,(3.8)   &              &              &     &     &near 500                                                                          \\
503  & 21:39:07.88 &  57:42:08.6$^{\rm r}$& 564  & 469                       &                               &  12.43$^{\rm l}$&   12.1$^{\rm f}$&  11.58$^{\rm e}$&       &                 &10.400\,(22) &10.241\,(28) & 10.154\,(20)$^{\rm r}$&        A9$^{\rm e}$&                 &  0.8$^{\rm e}$&-5.4\,(1.7)   &-8.4\,(1.7)   &-6.5\,(0.6)   &-6.7\,(0.8)   &0.35 &-0.3 &                                                                                      \\
504  & 21:39:16.28 &  57:42:13.4$^{\rm r}$& 565  & 3565                      &                               &  15.02$^{\rm l}$&  13.88$^{\rm l}$&  12.51$^{\rm l}$&       &                 &9.862 \,(21) &9.197 \,(25) & 9.054 \,(20)$^{\rm r}$&                    &                 &               &-5.1\,(4.7)   &-7\,(4.7)     &-4.1\,(7.6)   &-11.7\,(7.6)  &0.26 &-0.05&                                                                                      \\
505  & 21:39:18.88 &  57:42:29.1$^{\rm r}$& 566  & 472                       &                               &  11.65$^{\rm l}$&  11.33$^{\rm f}$&  10.85$^{\rm e}$&       &                 &9.837 \,(21) &9.687 \,(29) & 9.624 \,(22)$^{\rm r}$&        A1$^{\rm e}$&                 &  1.4$^{\rm e}$&-3.3\,(1.6)   &-9.5\,(1.6)   &-5.6\,(0.6)   &-6.9\,(0.7)   &0.27 &-0.28&                                                                                      \\
506  & 21:39:31.83 &  57:42:21.7$^{\rm r}$& 567  & 3567                      &                               &                 &                 &   13.6$^{\rm j}$&       &                 &11.486\,(29) &11.200\,()   & 10.887\,(23)$^{\rm r}$&                    &                 &               &7.7\,(11)     &-6.4\,(11)    &6.4\,(6.3)    &26.7\,(6.2)   &0.13 &-0.12&                                                                                      \\
507  & 21:38:28.78 &  57:42:45.7$^{\rm r}$& 568  & 3568                      &                               &                 &                 &   14.1$^{\rm j}$&       &                 &12.443\,(31) &12.159\,(32) & 12.088\,(21)$^{\rm r}$&                    &                 &               &-2.7\,(3.8)   &-5.3\,(3.8)   &-6.7\,(7.4)   &-4.2\,(7.5)   &0.32 &0.09 &                                                                                      \\
508  & 21:38:36.41 &  57:42:55.0$^{\rm r}$& 569  & 3569                      &                               &                 &                 &   12.8$^{\rm j}$&       &                 &10.096\,(27) &9.361 \,(30) & 9.151 \,(21)$^{\rm r}$&                    &                 &               &-2.9\,(4.7)   &-1.3\,(4.7)   &-7.5\,(7.4)   &-6.8\,(7.4)   &-0.11&0.12 &                                                                                      \\
509  & 21:39:25.73 &  57:43:38.5$^{\rm r}$& 570  & 3570                      &                               &                 &                 &   13.8$^{\rm j}$&       &                 &12.377\,(24) &12.056\,(26) & 12.021\,(22)$^{\rm r}$&                    &                 &               &-8.9\,(3.8)   &-6\,(3.8)     &1.3\,(7.4)    &-13.9\,(7.4)  &0.54 &0.1  &                                                                                      \\
510  & 21:38:28.55 &  57:43:34.3$^{\rm r}$& 571  & 3571                      &                               &                 &                 &   12.7$^{\rm j}$&       &                 &11.978\,(34) &11.845\,(35) & 11.771\,(28)$^{\rm r}$&                    &                 &               &1\,(2.7)      &-0.4\,(2.7)   &-5.8\,(3.3)   &-2.7\,(1.5)   &0.02 &0.1  &                                                                                      \\
511  & 21:38:29.47 &  57:43:41.9$^{\rm r}$& 572  & 3572                      &                               &                 &                 &   14.2$^{\rm j}$&       &                 &12.442\,(26) &12.005\,(33) & 11.906\,(23)$^{\rm r}$&                    &                 &               &-6.7\,(3.9)   &-0.6\,(3.9)   &13.5\,(6.5)   &6.5\,(6.4)    &0.02 &-0.34&                                                                                      \\
512  & 21:38:57.68 &  57:44:04.2$^{\rm r}$& 573  & 3573                      &                               &                 &                 &   14.7$^{\rm j}$&       &                 &10.902\,(22) &10.047\,(26) & 9.820 \,(23)$^{\rm r}$&                    &                 &               &-9.1\,(4.8)   &-4.4\,(4.8)   &-16.4\,(7.5)  &-13.6\,(7.5)  &0.92 &-0.33&                                                                                      \\
513  & 21:39:17.58 &  57:44:54.8$^{\rm r}$& 574  & 3574                      &                               &                 &                 &   14.2$^{\rm j}$&       &                 &12.315\,(24) &11.875\,(31) & 11.786\,(25)$^{\rm r}$&                    &                 &               &-2.4\,(3.8)   &-2.9\,(3.8)   &-1.6\,(7.7)   &-6.7\,(7.8)   &0.25 &-0.11&                                                                                      \\
514  & 21:39:26.19 &  57:45:12.6$^{\rm r}$& 575  & 3575                      &                               &  15.06$^{\rm l}$&  13.98$^{\rm l}$&   12.6$^{\rm l}$&       &                 &10.020\,(24) &9.410 \,(28) & 9.247 \,(23)$^{\rm r}$&                    &                 &               &-8.2\,(4.7)   &-7.8\,(4.7)   &-3.7\,(1.6)   &-8.6\,(4.9)   &-0.13&-0.13&                                                                                      \\
515  & 21:39:30.37 &  57:44:58.4$^{\rm r}$& 576  & 724                       &                               &  12.95$^{\rm l}$&  12.77$^{\rm l}$&   12.4$^{\rm l}$&       &                 &11.550\,(26) &11.344\,(28) & 11.333\,(21)$^{\rm r}$&        A0$^{\rm q}$&                 &               &-19.3\,(2.7)  &-11.4\,(2.7)  &-7.6\,(1.3)   &-5.6\,(1.8)   &0.43 &-0.09&                                                                                      \\
516  & 21:39:38.91 &  57:44:30.2$^{\rm r}$& 577  & 479                       &                               &                 &                 &   12.6$^{\rm j}$&       &                 &10.386\,(26) &9.729 \,(28) & 9.591 \,(21)$^{\rm r}$&        B8$^{\rm q}$&                 &               &-0.6\,(4.9)   &-3.4\,(4.9)   &40.4\,(8.6)   &2.2\,(8.6)    &0.14 &-0.32&                                                                                      \\
517  & 21:38:26.14 &  57:45:17.1$^{\rm r}$& 578  & 3578                      &                               &                 &                 &   13.7$^{\rm j}$&       &                 &12.106\,(26) &11.905\,(31) & 11.714\,(21)$^{\rm r}$&                    &                 &               &-6.1\,(3.8)   &-16\,(3.8)    &-20.5\,(7.4)  &-48.5\,(7.4)  &0.11 &-0.13&                                                                                      \\
518  & 21:38:18.09 &  57:45:38.5$^{\rm r}$& 579  & 3579                      &                               &                 &                 &   12.4$^{\rm j}$&       &                 &11.406\,(27) &11.072\,(30) & 11.011\,(19)$^{\rm r}$&                    &                 &               &12.8\,(2.7)   &-11.9\,(2.7)  &15.7\,(1.4)   &-4.6\,(1.6)   &-1.77&-0.74&                                                                                      \\
519  & 21:39:57.64 &  57:25:38.1$^{\rm r}$& 580  & 3580                      &                               &                 &                 &   14.3$^{\rm j}$&       &                 &13.105\,(24) &12.964\,(36) & 12.822\,(29)$^{\rm r}$&                    &                 &               &-10\,(4.1)    &-4\,(4.1)     &-14.8\,(5.9)  &-6\,(5.9)     &-0.02&-0.06&                                                                                      \\
520  & 21:39:58.56 &  57:25:43.1$^{\rm r}$& 581  & 3581                      &                               &                 &                 &   14.9$^{\rm j}$&       &                 &12.788\,(29) &12.340\,(33) & 12.194\,(29)$^{\rm r}$&                    &                 &               &8\,(4.1)      &-23.8\,(4.1)  &              &              &     &     &                                                                                      \\
521  & 21:40:04.33 &  57:25:22.0$^{\rm r}$& 582  & 3582                      &                               &                 &                 &   14.1$^{\rm j}$&       &                 &12.255\,(23) &11.725\,(31) & 11.640\,(22)$^{\rm r}$&                    &                 &               &-6.6\,(4.1)   &-1.9\,(4.1)   &-4.5\,(6.9)   &10.9\,(6.9)   &0.52 &-0.14&                                                                                      \\
522  & 21:40:11.13 &  57:25:51.6$^{\rm r}$& 583  & 484                       &                               &  10.77$^{\rm l}$&  11.04$^{\rm l}$&  10.81$^{\rm l}$&       &                 &10.292\,(24) &10.235\,(31) & 10.156\,(22)$^{\rm r}$&        B8$^{\rm q}$&                 &               &-4.7\,(1.4)   &-5\,(1.4)     &4.2\,(9.5)    &-1.8\,(9.4)   &-0.07&-0.02&                                                                                      \\
523  & 21:40:13.21 &  57:25:09.6$^{\rm r}$& 584  & 3584                      &                               &                 &                 &   14.2$^{\rm j}$&       &                 &12.808\,(26) &12.558\,(35) & 12.495\,(26)$^{\rm r}$&                    &                 &               &-7.7\,(7.6)   &-24.3\,(7.6)  &-17.6\,(6.9)  &3.9\,(6.9)    &-0.18&-0.12&                                                                                      \\
524  & 21:40:19.51 &  57:24:33.3$^{\rm r}$& 585  & 3585                      &                               &                 &                 &   13.8$^{\rm j}$&       &                 &12.489\,(38) &12.155\,()   & 12.023\,(37)$^{\rm r}$&                    &                 &               &2.6\,(4.1)    &-9.9\,(4.1)   &16.5\,(7)     &-10.3\,(7)    &0    &-0.05&                                                                                      \\
525  & 21:40:18.15 &  57:25:10.2$^{\rm r}$& 586  & 3586                      &                               &                 &                 &   14.8$^{\rm j}$&       &                 &12.895\,(34) &12.399\,(38) & 12.296\,(32)$^{\rm r}$&                    &                 &               &3.8\,(4.1)    &20\,(4.1)     &37\,(6.9)     &92.4\,(7.1)   &     &     &                                                                                      \\
526  & 21:40:22.35 &  57:25:24.9$^{\rm r}$& 587  & 3587                      &                               &                 &                 &   14.7$^{\rm j}$&       &                 &13.241\,(31) &13.005\,(38) & 12.883\,(33)$^{\rm r}$&                    &                 &               &-2.5\,(4.1)   &6.2\,(4.1)    &8.6\,(7)      &18.9\,(6.8)   &-0.26&0.01 &                                                                                      \\
527  & 21:40:29.06 &  57:25:06.1$^{\rm r}$& 588  & 3588                      &                               &                 &                 &   13.1$^{\rm j}$&       &                 &12.163\,(27) &11.945\,(31) & 11.892\,(24)$^{\rm r}$&                    &                 &               &-7.4\,(4.1)   &-1.8\,(4.1)   &-6.6\,(7.1)   &2.4\,(7.1)    &0.14 &0.1  &                                                                                      \\
528  & 21:40:39.33 &  57:24:37.4$^{\rm r}$& 589  & 3589                      &                               &                 &                 &     15$^{\rm j}$&       &                 &13.195\,(31) &12.763\,(36) & 12.701\,(32)$^{\rm r}$&                    &                 &               &-10.1\,(4.1)  &-1\,(4.1)     &-7.6\,(6.8)   &15.8\,(6.8)   &     &     &                                                                                      \\
529  & 21:40:55.34 &  57:24:17.9$^{\rm r}$& 590  & 3590                      &                               &                 &                 &   13.7$^{\rm j}$&       &                 &8.816 \,(35) &7.710 \,(33) & 7.348 \,(36)$^{\rm r}$&                    &                 &               &-6.5\,(5.1)   &-2.5\,(5.1)   &2.9\,(7.1)    &12.9\,(7.1)   &-0.03&0.23 &                                                                                      \\
530  & 21:41:05.69 &  57:23:58.3$^{\rm r}$& 591  & 3591                      &                               &                 &                 &   11.2$^{\rm j}$&       &                 &9.483 \,(26) &8.934 \,(28) & 8.865 \,(21)$^{\rm r}$&                    &                 &               &-8.8\,(10.4)  &-4.7\,(10.4)  &-6.6\,(0.7)   &-10.7\,(0.6)  &0.65 &-0.51&                                                                                      \\
531  & 21:41:23.85 &  57:24:10.9$^{\rm r}$& 592  & 3592                      &                               &                 &                 &   12.2$^{\rm j}$&       &                 &10.921\,(23) &10.627\,(28) & 10.543\,(23)$^{\rm r}$&                    &                 &               &-6.4\,(15)    &17.3\,(15)    &-99.4\,(6.8)  &-81.4\,(6.8)  &-0.13&-0.11&                                                                                      \\
532  & 21:41:22.62 &  57:24:30.1$^{\rm r}$& 593  & 3593                      &                               &                 &                 &   13.7$^{\rm j}$&       &                 &12.233\,(21) &11.958\,(27) & 11.861\,(25)$^{\rm r}$&                    &                 &               &-4.4\,(4.1)   &-2.8\,(4.1)   &-3.9\,(7.4)   &-12.7\,(7.4)  &-0.25&-0.25&                                                                                      \\
533  & 21:41:24.84 &  57:24:17.7$^{\rm r}$& 594  & 3594                      &                               &                 &                 &   13.8$^{\rm j}$&       &                 &11.655\,(28) &11.158\,(35) & 10.984\,(25)$^{\rm r}$&                    &                 &               &30.5\,(5.5)   &26.3\,(5.5)   &              &              &-1   &0.63 &                                                                                      \\
534  & 21:41:25.06 &  57:24:50.4$^{\rm r}$& 595  & 3595                      &                               &                 &                 &   13.8$^{\rm j}$&       &                 &12.273\,(26) &11.987\,(35) & 11.892\,(29)$^{\rm r}$&                    &                 &               &42\,(26)      &25.5\,(26)    &7.5\,(7.5)    &14.9\,(7.5)   &0.3  &-0.28&                                                                                      \\
535  & 21:39:54.00 &  57:27:01.9$^{\rm r}$& 596  & 3596                      &                               &                 &                 &   13.1$^{\rm j}$&       &                 &10.800\,(23) &10.192\,(28) & 10.008\,(22)$^{\rm r}$&                    &                 &               &21.8\,(3.9)   &-0.1\,(3.9)   &53.2\,(7.2)   &52.5\,(7.2)   &-2.27&0.48 &                                                                                      \\
536  & 21:39:37.34 &  57:27:45.0$^{\rm r}$& 597  & 3597                      &                               &                 &                 &   14.5$^{\rm j}$&       &                 &12.850\,(26) &12.649\,(30) & 12.516\,(25)$^{\rm r}$&                    &                 &               &-4.6\,(3.9)   &-4.1\,(3.9)   &-2.2\,(6.8)   &13.3\,(6.8)   &0.17 &0.18 &                                                                                      \\
537  & 21:39:48.06 &  57:28:43.7$^{\rm r}$& 598  & 3598                      &                               &  13.20$^{\rm l}$&  13.06$^{\rm h}$&   12.4$^{\rm h}$&       &                 &11.080\,(23) &10.827\,(28) & 10.793\,(23)$^{\rm r}$&        F5$^{\rm h}$&                 & 0.75$^{\rm h}$&-17.6\,(2.7)  &6.2\,(2.7)    &-12.2\,(1.7)  &-10.7\,(5.2)  &0.91 &-0.92&Dec [h] imprec.                                                               \\
538  & 21:39:58.13 &  57:28:33.6$^{\rm r}$& 599  & 3599                      &                               &                 &                 &   14.7$^{\rm j}$&       &                 &11.720\,(24) &10.968\,(31) & 10.422\,(22)$^{\rm r}$&                    &                 &               &-7.5\,(5.4)   &-6.9\,(5.4)   &-6.1\,(6.8)   &1.1\,(6.8)    &-0.14&-0.32&same star                                                                             \\
539  & 21:40:01.80 &  57:28:08.8$^{\rm r}$& 600  & 3600                      &                               &                 &                 &   13.1$^{\rm j}$&       &                 &11.652\,(24) &11.260\,(31) & 11.165\,(23)$^{\rm r}$&                    &                 &               &0.5\,(3.9)    &-1\,(3.9)     &-4.5\,(7.2)   &8.3\,(7.2)    &-1.2 &0.21 &                                                                                      \\
540  & 21:40:19.09 &  57:26:49.6$^{\rm r}$& 601  & 3601                      &                               &                 &  15.16$^{\rm f}$&  14.54$^{\rm e}$&       &                 &11.937\,(26) &11.568\,(31) & 11.381\,(23)$^{\rm r}$&        F3$^{\rm e}$&                 &  0.8$^{\rm e}$&3.9\,(4.1)    &-8.7\,(4.1)   &29.2\,(6.3)   &-19.3\,(6.3)  &0.07 &-0.03&same star                                                                             \\
541  & 21:40:17.98 &  57:28:09.5$^{\rm r}$& 602  & 3602                      &                               &                 &                 &   15.1$^{\rm j}$&       &                 &13.321\,(34) &12.895\,(38) & 12.747\,(29)$^{\rm r}$&                    &                 &               &-4.8\,(5.1)   &-0.4\,(5.1)   &-3\,(6.9)     &14.3\,(6.9)   &     &     &                                                                                      \\
542  & 21:40:24.79 &  57:27:45.3$^{\rm r}$& 604  & 3604                      &                               &                 &                 &  10.11$^{\rm j}$&       &                 &9.857 \,(27) &9.741 \,(31) & 9.715 \,(23)$^{\rm r}$&                    &                 &               &-1.3\,(1.4)   &4.9\,(1.4)    &-4.3\,(0.5)   &2.5\,(0.7)    &-0.03&0.75 &                                                                                      \\
543  & 21:40:29.29 &  57:27:58.4$^{\rm r}$& 605  & 3605                      &                               &                 &                 &   14.8$^{\rm j}$&       &                 &11.313\,(27) &10.481\,(30) & 10.254\,(21)$^{\rm r}$&                    &                 &               &-2.6\,(3.8)   &-7.6\,(3.8)   &18\,(6.9)     &20.3\,(6.9)   &-3.2 &2.11 &                                                                                      \\
544  & 21:40:30.50 &  57:27:42.7$^{\rm r}$& 606  & 3606                      &                               &                 &                 &   12.2$^{\rm j}$&       &                 &10.337\,(27) &9.728 \,(31) & 9.587 \,(21)$^{\rm r}$&                    &                 &               &3.9\,(5.1)    &-2.3\,(5.1)   &8.9\,(7.7)    &12.4\,(7.7)   &-0.61&0.34 &                                                                                      \\
545  & 21:40:37.36 &  57:27:43.8$^{\rm r}$& 607  & 3607                      &                               &                 &                 &   14.8$^{\rm j}$&       &                 &16.362\,(115)&15.673\,()   &   15.676\,()$^{\rm r}$&                    &                 &               &-2.5\,(4.6)   &-14.8\,(4.6)  &              &              &     &     &no star                                                                          \\
546  & 21:40:43.51 &  57:27:53.0$^{\rm r}$& 608  & 3608                      &                               &                 &                 &   13.7$^{\rm j}$&       &                 &12.522\,(34) &12.302\,(38) & 12.248\,(37)$^{\rm r}$&                    &                 &               &-18.5\,(3.9)  &-17.5\,(3.9)  &-66\,(7.1)    &-40.5\,(7.1)  &0.26 &-0.18&                                                                                      \\
547  & 21:40:43.25 &  57:27:21.9$^{\rm r}$& 609  & 3609                      &                               &                 &                 &   14.3$^{\rm j}$&       &                 &12.727\,(34) &12.448\,(37) & 12.359\,(41)$^{\rm r}$&                    &                 &               &-12.2\,(3.9)  &-9.4\,(3.9)   &              &              &-0.5 &0.16 &                                                                                      \\
548  & 21:40:53.23 &  57:27:11.9$^{\rm r}$& 610  & 3610                      &                               &                 &                 &     14$^{\rm j}$&       &                 &16.545\,(152)&15.455\,()   &15.346\,(213)$^{\rm r}$&                    &                 &               &-22.4\,(6.1)  &-13.3\,(6.1)  &              &              &     &     &no star                                                                          \\
549  & 21:40:51.31 &  57:26:19.0$^{\rm r}$& 611  & 3611                      &                               &                 &                 &   14.8$^{\rm j}$&       &                 &13.120\,(29) &12.719\,(36) & 12.666\,(35)$^{\rm r}$&                    &                 &               &-7.9\,(3.9)   &-7.5\,(3.9)   &-2\,(6.8)     &26.8\,(6.9)   &0.25 &-0.62&                                                                                      \\
550  & 21:41:01.76 &  57:27:26.5$^{\rm r}$& 612  & 3612                      &                               &                 &                 &   13.3$^{\rm j}$&       &                 &11.680\,(36) &11.368\,(37) & 11.222\,(33)$^{\rm r}$&                    &                 &               &8.6\,(3.9)    &33.6\,(3.9)   &              &              &-0.52&0.28 &                                                                                      \\
551  & 21:41:07.03 &  57:26:30.9$^{\rm r}$& 613  & 3613                      &                               &                 &                 &   11.3$^{\rm j}$&       &                 &9.549 \,(26) &9.035 \,(28) & 8.894 \,(23)$^{\rm r}$&                    &                 &               &14.5\,(10.4)  &-8.3\,(10.4)  &16.6\,(0.8)   &-7.8\,(1.2)   &-1.84&-0.37&                                                                                      \\
552  & 21:41:08.82 &  57:25:27.8$^{\rm r}$& 614  & 3614                      &                               &                 &                 &   14.4$^{\rm j}$&       &                 &12.603\,(24) &12.270\,(27) & 12.205\,(22)$^{\rm r}$&                    &                 &               &-7.9\,(4.1)   &6.4\,(4.1)    &-3.8\,(6.8)   &16.9\,(6.9)   &0.23 &0.12 &                                                                                      \\
553  & 21:41:24.07 &  57:25:30.5$^{\rm r}$& 615  & 3615                      &                               &  13.84$^{\rm l}$&   13.5$^{\rm l}$&  12.69$^{\rm l}$&       &                 &11.255\,(24) &10.947\,(30) & 10.866\,(22)$^{\rm r}$&                    &                 &               &-7.3\,(4.1)   &-1.5\,(4.1)   &-5.2\,(1.9)   &-0.1\,(0.7)   &0.12 &-0.02&                                                                                      \\
554  & 21:41:18.65 &  57:28:06.5$^{\rm r}$& 616  & 492                       &                               &                 &                 &   10.8$^{\rm j}$&       &                 &9.941 \,(28) &9.686 \,(32) & 9.611 \,(23)$^{\rm r}$&        F8$^{\rm q}$&                 &               &-3.2\,(3.1)   &13.2\,(3.3)   &-2.7\,(1.2)   &8.5\,(1.9)    &0.07 &1.36 &                                                                                      \\
555  & 21:41:19.41 &  57:28:00.2$^{\rm r}$& 617  & 3617                      &                               &                 &                 &   13.5$^{\rm j}$&       &                 &12.022\,(26) &11.749\,(32) & 11.641\,(23)$^{\rm r}$&                    &                 &               &              &              &              &              &-0.03&-0.42&                                                                                      \\
556  & 21:39:41.01 &  57:29:09.0$^{\rm r}$& 618  & 3618                      &                               &                 &                 &   14.4$^{\rm j}$&       &                 &13.409\,()   &14.945\,(99) &14.527\,(107)$^{\rm r}$&                    &                 &               &              &              &-28\,(6.8)    &13.6\,(6.8)   &0.61 &-0.56&2x[r]                                                                          \\
557  & 21:39:40.52 &  57:29:10.7$^{\rm r}$& 618  & 3618                      &                               &                 &                 &   14.4$^{\rm j}$&       &                 &12.543\,(29) &12.196\,(33) & 12.063\,(29)$^{\rm r}$&                    &                 &               &-18.8\,(3.9)  &-4\,(3.9)     &-28\,(6.8)    &13.6\,(6.8)   &0.61 &-0.56&2x[r]                                                                           \\
558  & 21:39:40.08 &  57:29:33.6$^{\rm r}$& 619  & 3619                      &                               &                 &                 &   14.4$^{\rm j}$&       &                 &12.082\,(23) &11.459\,(31) & 11.342\,(23)$^{\rm r}$&                    &                 &               &19.2\,(7)     &-370.7\,(7)   &12.1\,(6.7)   &11.6\,(6.7)   &-0.58&0    &                                                                                      \\
559  & 21:40:38.59 &  57:29:35.5$^{\rm r}$& 620  & 3620                      &                               &                 &                 &   14.5$^{\rm j}$&       &                 &11.094\,(26) &10.271\,(27) & 10.066\,(21)$^{\rm r}$&                    &                 &               &-8.7\,(3.8)   &-9.9\,(3.8)   &-19.5\,(6.9)  &1.6\,(6.9)    &0.25 &-0.02&                                                                                      \\
560  & 21:40:57.35 &  57:29:11.8$^{\rm r}$& 621  & 3621                      &                               &                 &                 &   13.4$^{\rm j}$&       &                 &12.361\,(32) &12.110\,(38) & 11.992\,(30)$^{\rm r}$&                    &                 &               &-21.4\,(3.9)  &10.4\,(3.9)   &-50.8\,(7.1)  &44.7\,(7.1)   &0.02 &0.2  &                                                                                      \\
561  & 21:40:59.59 &  57:28:53.7$^{\rm r}$& 622  & 3622                      &                               &                 &                 &   14.9$^{\rm j}$&       &                 &13.413\,(27) &13.072\,(33) & 13.087\,(23)$^{\rm r}$&                    &                 &               &-2.3\,(3.8)   &-11.3\,(3.8)  &8.6\,(6.8)    &8.5\,(6.8)    &-0.22&0.55 &                                                                                      \\
562  & 21:41:05.04 &  57:29:17.0$^{\rm r}$& 623  & 3623                      &                               &                 &                 &   14.6$^{\rm j}$&       &                 &12.632\,(27) &12.184\,(31) & 12.135\,(24)$^{\rm r}$&                    &                 &               &-4.6\,(3.9)   &-1.6\,(3.9)   &9.6\,(6.8)    &12.7\,(6.8)   &-0.27&0.01 &                                                                                      \\
563  & 21:41:08.01 &  57:28:52.9$^{\rm r}$& 624  & 3624                      &                               &                 &                 &   14.2$^{\rm j}$&       &                 &10.101\,(24) &9.121 \,(28) & 8.807 \,(22)$^{\rm r}$&                    &                 &               &-5.8\,(4.7)   &-8.8\,(4.7)   &-5.4\,(7)     &8.6\,(7)      &0.09 &0.22 &                                                                                      \\
564  & 21:41:25.00 &  57:29:27.9$^{\rm r}$& 625  & 5100                      &                               &                 &  15.56$^{\rm f}$&  14.57$^{\rm e}$&       &                 &12.714\,(24) &12.417\,(32) & 12.303\,(26)$^{\rm r}$&        F3$^{\rm e}$&                 &  1.9$^{\rm e}$&-24.9\,(5.1)  &-0.4\,(5.1)   &0\,(6.8)      &10.8\,(6.8)   &-0.15&-0.16&                                                                                      \\
565  & 21:39:40.04 &  57:30:34.9$^{\rm r}$& 626  & 3626                      &                               &                 &                 &   13.7$^{\rm j}$&       &                 &11.243\,(23) &10.564\,(31) & 10.399\,(22)$^{\rm r}$&                    &                 &               &-1.4\,(3.9)   &-3.7\,(3.9)   &-2.5\,(7.3)   &-6.5\,(7.3)   &0.26 &-0.16&                                                                                      \\
566  & 21:39:43.98 &  57:30:31.5$^{\rm r}$& 627  & 3627                      &                               &                 &                 &   13.9$^{\rm j}$&       &                 &12.805\,(23) &12.673\,(31) & 12.608\,(28)$^{\rm r}$&                    &                 &               &-7.8\,(4.1)   &-13.4\,(4.1)  &-11.9\,(7.2)  &-17.5\,(7.3)  &0.26 &0.03 &                                                                                      \\
567  & 21:39:44.79 &  57:30:39.9$^{\rm r}$& 628  & 3628                      &                               &                 &                 &   14.8$^{\rm j}$&       &                 &11.607\,(24) &10.868\,(32) & 10.532\,(25)$^{\rm r}$&                    &                 &               &7.4\,(3.9)    &9.7\,(3.9)    &-5.5\,(7.3)   &19.2\,(7.4)   &0.1  &-0.38&                                                                                      \\
568  & 21:39:50.69 &  57:30:56.6$^{\rm r}$& 629  & 3629                      &                               &                 &                 &     15$^{\rm j}$&       &                 &13.186\,(31) &12.827\,(33) & 12.732\,(33)$^{\rm r}$&                    &                 &               &-9.2\,(3.9)   &-10.4\,(3.9)  &-2.5\,(6.9)   &3.5\,(7.1)    &     &     &                                                                                      \\
569  & 21:40:02.37 &  57:31:02.6$^{\rm r}$& 630  & 480                       &                               &                 &                 &   14.9$^{\rm j}$&       &                 &12.739\,(23) &12.281\,(30) & 12.204\,(23)$^{\rm r}$&        A2$^{\rm q}$&                 &               &-5.2\,(3.9)   &-0.6\,(3.9)   &2.9\,(6.9)    &9.7\,(7)      &     &     &                                                                                      \\
570  & 21:40:12.14 &  57:31:14.4$^{\rm r}$& 631  & 3631                      &                               &                 &                 &     15$^{\rm j}$&       &                 &13.240\,(26) &12.797\,(27) & 12.680\,(34)$^{\rm r}$&                    &                 &               &5.9\,(3.8)    &1.8\,(3.8)    &31.2\,(6.9)   &5.5\,(6.9)    &     &     &                                                                                      \\
571  & 21:40:12.98 &  57:30:59.0$^{\rm r}$& 632  & 3632                      &                               &                 &                 &   14.1$^{\rm j}$&       &                 &12.307\,(24) &11.895\,(32) & 11.788\,(23)$^{\rm r}$&                    &                 &               &-1.9\,(3.8)   &-1.5\,(3.8)   &5.6\,(7.4)    &2.5\,(7.4)    &-0.65&0.59 &                                                                                      \\
572  & 21:40:18.37 &  57:30:39.4$^{\rm r}$& 633  & 486                       &                               &   9.90$^{\rm l}$&  10.34$^{\rm l}$&  10.11$^{\rm l}$&       &                 &9.691 \,(27) &9.570 \,(30) & 9.632 \,(21)$^{\rm r}$&        B2$^{\rm p}$&      V$^{\rm p}$&               &2.1\,(1.4)    &-7.3\,(1.4)   &-3\,(0.8)     &-6.1\,(0.6)   &-0.15&-0.04&                                                                                      \\
573  & 21:40:20.50 &  57:31:23.4$^{\rm r}$& 634  & 3634\footnote{also 5093}  &                               &  14.63$^{\rm l}$&  14.26$^{\rm f}$&  13.53$^{\rm e}$&       &                 &12.151\,(38) &11.874\,(44) & 11.779\,(39)$^{\rm r}$&        F1$^{\rm e}$&                 &  1.3$^{\rm e}$&16\,(18.5)    &19\,(18.5)    &7.7\,(7.1)    &77.7\,(7)     &-0.23&0.33 &                                                                                      \\
574  & 21:40:50.21 &  57:30:20.6$^{\rm r}$& 635  & 3635                      &                               &                 &                 &   14.5$^{\rm j}$&       &                 &12.571\,(27) &12.093\,(31) & 11.984\,(24)$^{\rm r}$&                    &                 &               &-3.4\,(3.8)   &-7.7\,(3.8)   &10.6\,(6.8)   &5.2\,(6.8)    &-0.12&0.17 &                                                                                      \\
575  & 21:40:59.39 &  57:30:08.2$^{\rm r}$& 636  & 3636                      &                               &                 &                 &   14.6$^{\rm j}$&       &                 &12.915\,(29) &12.556\,(31) & 12.481\,(26)$^{\rm r}$&                    &                 &               &-4\,(3.8)     &-8\,(3.8)     &0\,(6.8)      &8\,(6.8)      &0.33 &-0.13&                                                                                      \\
576  & 21:41:00.09 &  57:30:40.1$^{\rm r}$& 637  & 5097                      &                               &                 &  14.81$^{\rm f}$&  14.06$^{\rm e}$&       &                 &12.708\,()   &12.265\,(38) & 12.242\,(35)$^{\rm r}$&        F6$^{\rm e}$&                 &  0.9$^{\rm e}$&-3.8\,(3.8)   &-1.1\,(3.8)   &22.7\,(7.1)   &41.1\,(7.1)   &     &     &                                                                                      \\
577  & 21:41:14.49 &  57:30:35.2$^{\rm r}$& 638  & 491                       &                               &                 &                 &   11.1$^{\rm j}$&       &                 &10.070\,(23) &9.826 \,(30) & 9.733 \,(22)$^{\rm r}$&                    &                 &               &-9.8\,(2.7)   &1.8\,(2.7)    &-11.1\,(0.8)  &-5.1\,(1)     &1    &-0.13&                                                                                      \\
578  & 21:39:39.81 &  57:32:42.0$^{\rm r}$& 639  & 3639                      &                               &                 &                 &   14.8$^{\rm j}$&       &                 &13.019\,(24) &12.742\,(33) & 12.513\,(28)$^{\rm r}$&                    &                 &               &-3.1\,(3.9)   &-6\,(3.9)     &15.3\,(6.2)   &-20.6\,(6.3)  &0.79 &-0.15&                                                                                      \\
579  & 21:39:46.74 &  57:32:52.5$^{\rm r}$& 640  & 3640                      &                               &  11.80$^{\rm l}$&  11.44$^{\rm f}$&  10.97$^{\rm e}$&       &                 &9.879 \,(23) &9.738 \,(32) & 9.656 \,(22)$^{\rm r}$&        A7$^{\rm e}$&                 &  0.9$^{\rm e}$&-3.4\,(1.7)   &-6.4\,(1.7)   &-2\,(0.6)     &-5.6\,(0.9)   &-0.07&0.12 &                                                                                      \\
580  & 21:39:53.86 &  57:33:08.7$^{\rm r}$& 641  & 3641                      &                               &  14.67$^{\rm l}$&  14.86$^{\rm f}$&   14.2$^{\rm e}$&       &                 &12.796\,(24) &12.729\,(33) & 12.573\,(30)$^{\rm r}$&        B1$^{\rm e}$&                 &  2.9$^{\rm e}$&-11.9\,(5.1)  &166.4\,(5.1)  &2.4\,(7.3)    &6.4\,(7.4)    &0.06 &0.12 &                                                                                      \\
581  & 21:39:41.68 &  57:33:47.3$^{\rm r}$& 642  & 3642                      &                               &                 &                 &   14.5$^{\rm j}$&       &                 &12.671\,(24) &12.422\,(30) & 12.269\,(23)$^{\rm r}$&                    &                 &               &1.5\,(3.9)    &-0.8\,(3.9)   &16.2\,(6.3)   &3.7\,(6.3)    &-0.2 &-0.15&                                                                                      \\
582  & 21:39:59.34 &  57:33:29.4$^{\rm j}$& 643  & 3643                      &                               &                 &                 &   14.9$^{\rm j}$&       &                 &             &             &                       &                    &                 &               &              &              &              &              &     &     &no star                                                                          \\
583  & 21:40:04.53 &  57:34:09.6$^{\rm r}$& 644  & 3644                      &                               &                 &                 &   14.9$^{\rm j}$&       &                 &12.904\,(26) &12.520\,(32) & 12.386\,(26)$^{\rm r}$&                    &                 &               &-8.8\,(3.9)   &-7.9\,(3.9)   &1.4\,(6.9)    &-3.1\,(6.9)   &     &     &                                                                                      \\
584  & 21:40:09.26 &  57:33:23.5$^{\rm r}$& 646  & 3646                      &                               &                 &                 &   14.4$^{\rm j}$&       &                 &12.556\,(26) &12.383\,(35) & 12.185\,(28)$^{\rm r}$&                    &                 &               &-2.7\,(3.8)   &-10.4\,(3.8)  &7.6\,(6.8)    &-0.9\,(6.8)   &0.3  &0.12 &                                                                                      \\
585  & 21:40:29.76 &  57:33:38.5$^{\rm r}$& 647  & 3647                      &                               &  13.57$^{\rm l}$&  13.03$^{\rm h}$&  12.08$^{\rm h}$&       &                 &11.658\,(27) &11.464\,(28) & 11.391\,(23)$^{\rm r}$&        A0$^{\rm h}$&                 & 1.59$^{\rm h}$&-8\,(3.9)     &-7\,(3.9)     &-3.2\,(1.4)   &-1.8\,(1)     &-0.14&0.26 &Dec [h] imprec.                                                               \\
586  & 21:40:27.44 &  57:32:18.6$^{\rm r}$& 648  & 3648                      &                               &                 &                 &   14.6$^{\rm j}$&       &                 &12.857\,(29) &12.577\,(32) & 12.416\,(24)$^{\rm r}$&                    &                 &               &-6.5\,(3.8)   &-5.6\,(3.8)   &-13.3\,(7.4)  &-10.2\,(7.4)  &0.19 &-0.25&                                                                                      \\
587  & 21:40:37.57 &  57:33:06.5$^{\rm r}$& 649  & 3649                      &                               &                 &                 &   15.1$^{\rm j}$&       &                 &11.626\,(26) &10.844\,(30) & 10.610\,(19)$^{\rm r}$&                    &                 &               &-13\,(3.9)    &-8.8\,(3.9)   &-20.6\,(7.5)  &-9.1\,(7.5)   &     &     &                                                                                      \\
588  & 21:40:38.83 &  57:32:18.5$^{\rm r}$& 650  & 3650                      &                               &                 &                 &   14.8$^{\rm j}$&       &                 &13.282\,(27) &13.028\,(27) & 12.959\,(23)$^{\rm r}$&                    &                 &               &-7\,(3.8)     &-10.1\,(3.8)  &-15.7\,(7.5)  &-12.3\,(7.4)  &0.8  &0.25 &                                                                                      \\
589  & 21:40:47.87 &  57:32:47.3$^{\rm r}$& 651  & 3651                      &                               &                 &                 &   12.9$^{\rm j}$&       &                 &11.644\,(27) &11.292\,(28) & 11.196\,(24)$^{\rm r}$&                    &                 &               &18.4\,(3.9)   &8.5\,(3.9)    &              &              &-0.36&0.27 &                                                                                      \\
590  & 21:40:53.24 &  57:32:24.4$^{\rm r}$& 652  & 3652                      &                               &                 &                 &   14.1$^{\rm j}$&       &                 &12.390\,(26) &12.034\,(31) & 11.952\,(24)$^{\rm r}$&                    &                 &               &-0.7\,(3.9)   &-0.1\,(3.9)   &2.1\,(6.8)    &13.5\,(6.8)   &-0.75&0.1  &                                                                                      \\
591  & 21:41:00.25 &  57:33:06.3$^{\rm r}$& 653  & 3653                      &                               &                 &                 &   14.7$^{\rm j}$&       &                 &13.045\,(27) &12.650\,(32) & 12.618\,(24)$^{\rm r}$&                    &                 &               &-6.5\,(3.8)   &-3.9\,(3.8)   &-4.7\,(7.4)   &-13.2\,(7.6)  &-0.22&0.64 &                                                                                      \\
592  & 21:41:08.72 &  57:32:36.4$^{\rm r}$& 654  & 3654                      &                               &                 &                 &   13.7$^{\rm j}$&       &                 &11.964\,(51) &11.372\,(54) & 11.305\,(48)$^{\rm r}$&                    &                 &               &7.9\,(3.8)    &-22.3\,(3.8)  &47.7\,(7.2)   &-58.9\,(7.2)  &-0.82&0.63 &                                                                                      \\
593  & 21:41:23.07 &  57:32:45.5$^{\rm r}$& 655  & 3655                      &                               &                 &                 &   14.3$^{\rm j}$&       &                 &12.596\,(24) &12.184\,(31) & 12.109\,(22)$^{\rm r}$&                    &                 &               &-9.4\,(3.8)   &-6\,(3.8)     &-5.7\,(7.3)   &-10\,(7.4)    &0.4  &-0.06&                                                                                      \\
594  & 21:41:32.76 &  57:33:09.1$^{\rm r}$& 656  & 493                       &                               &                 &                 &   11.5$^{\rm j}$&       &                 &10.690\,(23) &10.570\,(30) & 10.435\,(22)$^{\rm r}$&        A3$^{\rm q}$&                 &               &-10.8\,(2)    &-6.1\,(2)     &-8.1\,(0.7)   &-5.9\,(0.9)   &0.48 &-0.36&                                                                                      \\
595  & 21:39:37.29 &  57:35:07.5$^{\rm r}$& 657  & 3657                      &                               &  13.99$^{\rm l}$&  13.67$^{\rm f}$&  13.07$^{\rm e}$&       &                 &11.843\,(27) &11.608\,(30) & 11.552\,(21)$^{\rm r}$&        F0$^{\rm e}$&                 &  0.9$^{\rm e}$&-5.2\,(3.9)   &-5.6\,(3.9)   &-4.5\,(0.5)   &-2.8\,(2.4)   &-0.05&-0.13&                                                                                      \\
596  & 21:39:41.02 &  57:34:55.6$^{\rm r}$& 658  & 3658                      &                               &  10.90$^{\rm l}$&  11.13$^{\rm l}$&  10.95$^{\rm l}$&       &                 &10.435\,(29) &10.319\,(30) & 10.353\,(21)$^{\rm r}$&                    &                 &               &-4.3\,(1.7)   &-8\,(1.6)     &-3.5\,(0.7)   &-5.5\,(1)     &-0.14&0.04 &                                                                                      \\
597  & 21:39:46.43 &  57:34:06.5$^{\rm r}$& 659  & 3659                      &                               &                 &                 &     15$^{\rm j}$&       &                 &11.549\,(29) &10.783\,(31) & 10.562\,(25)$^{\rm r}$&                    &                 &               &7\,(3.9)      &14.3\,(3.9)   &42.4\,(7.8)   &42.1\,(7.4)   &     &     &                                                                                      \\
598  & 21:39:40.87 &  57:35:09.0$^{\rm r}$& 660  & 3660                      &                               &  14.08$^{\rm l}$&  13.77$^{\rm f}$&  13.32$^{\rm e}$&       &                 &12.415\,(26) &12.269\,(31) & 12.173\,(28)$^{\rm r}$&        A2$^{\rm e}$&                 &  1.2$^{\rm e}$&-6.7\,(5.1)   &-3.4\,(5.1)   &-6.9\,(0.7)   &2.4\,(10.6)   &0.09 &-0.05&                                                                                      \\
599  & 21:39:47.74 &  57:36:13.1$^{\rm r}$& 662  & 481                       &                               &  10.78$^{\rm l}$&  10.87$^{\rm f}$&   10.6$^{\rm e}$&       &                 &9.931 \,(26) &9.730 \,(30) & 9.693 \,(21)$^{\rm r}$&        B9$^{\rm e}$&                 &    1$^{\rm e}$&-6.4\,(1.6)   &-1.1\,(1.6)   &-2.9\,(0.6)   &-5.1\,(0.9)   &-0.14&0.03 &                                                                                      \\
600  & 21:39:48.94 &  57:36:29.1$^{\rm r}$& 663  & 3663                      &                               &                 &                 &   13.9$^{\rm j}$&       &                 &11.568\,(24) &10.929\,(28) & 10.810\,(19)$^{\rm r}$&                    &                 &               &4.9\,(10.9)   &-3.9\,(10.9)  &16.4\,(7)     &21.7\,(7.1)   &0    &-0.34&                                                                                      \\
601  & 21:39:52.45 &  57:36:29.3$^{\rm r}$& 664  & 3664                      &                               &                 &                 &   14.2$^{\rm j}$&       &                 &12.752\,(27) &12.384\,(31) & 12.330\,(29)$^{\rm r}$&                    &                 &               &-5.7\,(3.9)   &0.6\,(3.9)    &-1.2\,(6.9)   &16.1\,(6.9)   &0.12 &0.17 &                                                                                      \\
602  & 21:40:00.99 &  57:37:01.9$^{\rm r}$& 665  & 3665                      &                               &                 &                 &   14.1$^{\rm j}$&       &                 &12.428\,(27) &12.097\,(31) & 11.995\,(23)$^{\rm r}$&                    &                 &               &-2.3\,(3.9)   &5.2\,(3.9)    &-5\,(7.4)     &3.1\,(7.4)    &-0.11&0.8  &                                                                                      \\
603  & 21:40:00.37 &  57:37:24.9$^{\rm r}$& 666  & 3666                      &                               &                 &                 &   14.7$^{\rm j}$&       &                 &11.862\,(29) &11.156\,(28) & 10.981\,(23)$^{\rm r}$&                    &                 &               &-11.2\,(5.1)  &-11.5\,(5.1)  &-10.3\,(7.3)  &-9.6\,(7.3)   &-0.29&-0.13&                                                                                      \\
604  & 21:39:58.98 &  57:37:32.5$^{\rm r}$& 667  & 3667                      &                               &                 &                 &   14.6$^{\rm j}$&       &                 &12.896\,(29) &12.481\,(35) & 12.384\,(26)$^{\rm r}$&                    &                 &               &5.5\,(3.8)    &3.1\,(3.8)    &32.7\,(7.3)   &-1.4\,(7.4)   &-0.11&0.66 &                                                                                      \\
605  & 21:40:17.60 &  57:35:39.9$^{\rm r}$& 668  & 3668                      &                               &                 &                 &   14.7$^{\rm j}$&       &                 &12.923\,(27) &12.611\,()   & 12.488\,(30)$^{\rm r}$&                    &                 &               &28.9\,(3.8)   &-12.2\,(3.8)  &              &              &0.31 &0    &                                                                                      \\
606  & 21:40:15.20 &  57:37:16.3$^{\rm r}$& 669  & 3669                      &                               &                 &  13.31$^{\rm f}$&  12.91$^{\rm e}$&       &                 &11.810\,(26) &11.618\,(28) & 11.553\,(21)$^{\rm r}$&        A2$^{\rm e}$&                 &  1.1$^{\rm e}$&-5.2\,(2.7)   &15.5\,(2.7)   &-7\,(0.5)     &-1.3\,(2.6)   &0.22 &0.02 &same star                                                                             \\
607  & 21:40:22.71 &  57:37:14.3$^{\rm r}$& 670  & 3670                      &                               &                 &                 &   12.4$^{\rm j}$&       &                 &10.289\,(27) &9.621 \,(31) & 9.442 \,(21)$^{\rm r}$&                    &                 &               &0.4\,(4.9)    &-6.3\,(4.9)   &21.3\,(7.5)   &-11\,(7.5)    &0.17 &-0.6 &                                                                                      \\
608  & 21:40:22.86 &  57:35:31.1$^{\rm r}$& 671  & 3671                      &                               &                 &                 &   12.6$^{\rm j}$&       &                 &10.222\,(26) &9.555 \,(28) & 9.399 \,(21)$^{\rm r}$&                    &                 &               &-3.9\,(4.9)   &2.1\,(4.9)    &32.9\,(6.1)   &89.2\,(6.1)   &-0.28&-0.61&                                                                                      \\
609  & 21:40:21.44 &  57:34:53.8$^{\rm r}$& 672  & 3672                      &                               &                 &                 &    7.6$^{\rm j}$&       &                 &6.571 \,(44) &6.388 \,(49) & 6.352 \,(20)$^{\rm r}$&                    &                 &               &              &              &              &              &4.34 &-1.12&                                                                                      \\
610  & 21:40:22.79 &  57:35:00.4$^{\rm r}$& 673  & 3673                      &                               &                 &                 &    8.1$^{\rm j}$&       &                 &7.588 \,(27) &7.340 \,(53) & 7.253 \,(27)$^{\rm r}$&                    &                 &               &-40.3\,(1.4)  &-17.6\,(1.3)  &-38.3\,(1.9)  &-17.6\,(2.5)  &4.22 &-1.1 &                                                                                      \\
611  & 21:40:23.08 &  57:34:53.7$^{\rm r}$& 674  & 3674                      &                               &                 &                 &   13.7$^{\rm j}$&       &                 &12.196\,(176)&11.816\,(104)& 11.768\,(89)$^{\rm r}$&                    &                 &               &              &              &              &              &0.53 &-1.95&                                                                                      \\
612  & 21:40:29.82 &  57:36:02.7$^{\rm r}$& 675  & 3675                      &                               &                 &                 &   14.3$^{\rm j}$&       &                 &12.623\,(27) &12.254\,(32) & 12.176\,(26)$^{\rm r}$&                    &                 &               &-20.3\,(3.8)  &-22.4\,(3.8)  &-17.3\,(7.1)  &-36.1\,(7.3)  &1.63 &-1.23&                                                                                      \\
613  & 21:40:46.72 &  57:35:28.5$^{\rm r}$& 676  & 3676                      &                               &                 &                 &   11.6$^{\rm j}$&       &                 &9.290 \,(27) &8.643 \,(30) & 8.492 \,(23)$^{\rm r}$&                    &                 &               &-5.7\,(4.9)   &-6.9\,(4.9)   &-6.3\,(1)     &-4.3\,(1)     &0.38 &0.22 &                                                                                      \\
614  & 21:40:47.93 &  57:36:13.0$^{\rm r}$& 677  & 3677                      &                               &                 &                 &   14.5$^{\rm j}$&       &                 &12.683\,(26) &12.340\,(32) & 12.210\,(23)$^{\rm r}$&                    &                 &               &-1.2\,(3.9)   &-2.1\,(3.9)   &-0.8\,(7.3)   &-9.2\,(7.3)   &-0.52&0.44 &                                                                                      \\
615  & 21:40:45.75 &  57:36:28.0$^{\rm r}$& 678  & 3678                      &                               &                 &                 &   14.7$^{\rm j}$&       &                 &12.969\,(26) &12.697\,(32) & 12.606\,(28)$^{\rm r}$&                    &                 &               &0\,(3.8)      &-2.5\,(3.8)   &-3.2\,(7.3)   &-8.9\,(7.3)   &-0.25&-0.19&                                                                                      \\
616  & 21:40:50.24 &  57:34:29.3$^{\rm r}$& 679  & 3679                      &                               &                 &                 &   14.8$^{\rm j}$&       &                 &12.761\,()   &12.717\,(58) &   12.314\,()$^{\rm r}$&                    &                 &               &1.1\,(3.8)    &-17.5\,(3.8)  &14.6\,(7.5)   &-71.1\,(7.5)  &0.1  &-0.05&                                                                                      \\
617  & 21:40:56.64 &  57:36:14.5$^{\rm r}$& 680  & 490                       &                               &                 &                 &   10.6$^{\rm j}$&       &                 &8.475 \,(20) &7.996 \,(76) & 7.759 \,(26)$^{\rm r}$&        G5$^{\rm q}$&                 &               &2.8\,(1.7)    &7.2\,(1.7)    &5.1\,(1)      &0\,(0.7)      &-0.67&0.32 &                                                                                      \\
618  & 21:41:14.24 &  57:35:09.1$^{\rm r}$& 681  & 3681                      &                               &                 &                 &   14.6$^{\rm j}$&       &                 &11.036\,(26) &10.259\,(31) & 10.022\,(25)$^{\rm r}$&                    &                 &               &-13\,(3.9)    &-4.1\,(3.9)   &-58.4\,(7.5)  &17.9\,(7.4)   &0.15 &0.18 &                                                                                      \\
619  & 21:41:14.13 &  57:35:00.3$^{\rm r}$& 682  & 3682                      &                               &                 &                 &   14.7$^{\rm j}$&       &                 &12.982\,(26) &12.711\,(36) & 12.584\,(25)$^{\rm r}$&                    &                 &               &-4.9\,(3.9)   &-18.2\,(3.9)  &-8.2\,(5.5)   &-27.7\,(5.5)  &     &     &                                                                                      \\
620  & 21:41:17.40 &  57:34:55.7$^{\rm r}$& 683  & 3683                      &                               &                 &                 &   14.5$^{\rm j}$&       &                 &12.475\,(24) &12.202\,(31) & 12.094\,(23)$^{\rm r}$&                    &                 &               &-5.6\,(5.1)   &-3.4\,(5.1)   &-19.5\,(7.5)  &2.6\,(7.4)    &0.06 &-0.13&                                                                                      \\
621  & 21:41:35.59 &  57:35:57.3$^{\rm r}$& 684  & 3684                      &                               &                 &                 &   13.5$^{\rm j}$&       &                 &11.750\,(23) &11.503\,(27) & 11.337\,(23)$^{\rm r}$&                    &                 &               &-4.1\,(14)    &0.3\,(14)     &6.4\,(7.3)    &-10.7\,(7.4)  &0.46 &-0.07&                                                                                      \\
622  & 21:40:30.36 &  57:37:23.4$^{\rm r}$& 685  & 3685                      &                               &                 &                 &   14.6$^{\rm j}$&       &                 &12.670\,(32) &12.276\,(36) & 12.222\,(37)$^{\rm r}$&                    &                 &               &7.5\,(3.8)    &3.1\,(3.8)    &28.6\,(8.2)   &22.4\,(8)     &-1.2 &1.04 &                                                                                      \\
623  & 21:40:44.11 &  57:37:53.1$^{\rm r}$& 686  & 3686                      &                               &                 &                 &   14.8$^{\rm j}$&       &                 &13.087\,(31) &12.701\,(37) & 12.628\,(33)$^{\rm r}$&                    &                 &               &-9\,(3.8)     &-3.8\,(3.8)   &              &              &4.51 &4.4  &                                                                                      \\
624  & 21:40:50.08 &  57:37:47.2$^{\rm r}$& 687  & 3687                      &                               &                 &                 &   14.8$^{\rm j}$&       &                 &12.553\,(29) &11.909\,(27) & 11.767\,(19)$^{\rm r}$&                    &                 &               &-15.4\,(3.8)  &2.4\,(3.8)    &-13.8\,(7.3)  &-4.7\,(7.3)   &     &     &                                                                                      \\
625  & 21:40:58.89 &  57:37:24.1$^{\rm r}$& 688  & 3688                      &                               &                 &                 &     13$^{\rm j}$&       &                 &12.152\,(26) &12.019\,(27) & 11.916\,(24)$^{\rm r}$&                    &                 &               &-6.4\,(3.8)   &-4\,(3.8)     &-3.6\,(2.3)   &-3.6\,(1.5)   &-0.06&-0.1 &                                                                                      \\
626  & 21:41:03.53 &  57:36:59.8$^{\rm r}$& 689  & 3689                      &                               &                 &                 &   14.9$^{\rm j}$&       &                 &13.149\,(29) &12.820\,(35) & 12.745\,(33)$^{\rm r}$&                    &                 &               &-9.8\,(3.8)   &7.6\,(3.8)    &3.6\,(7.4)    &-13\,(7.7)    &     &     &                                                                                      \\
627  & 21:41:25.11 &  57:36:57.1$^{\rm r}$& 690  & 3690                      &                               &                 &                 &   12.3$^{\rm j}$&       &                 &11.277\,(21) &11.038\,(31) & 10.943\,(20)$^{\rm r}$&                    &                 &               &-0.7\,(3.8)   &-12.4\,(3.8)  &-5.1\,(1.1)   &-8.3\,(2.6)   &0.14 &-0.59&                                                                                      \\
628  & 21:41:23.99 &  57:38:22.8$^{\rm r}$& 691  & 3691                      &                               &                 &                 &   12.2$^{\rm j}$&       &                 &9.977 \,(26) &9.310 \,(27) & 9.153 \,(23)$^{\rm r}$&                    &                 &               &1.6\,(4.7)    &-5.3\,(4.7)   &1\,(7.5)      &-11.3\,(7.5)  &-0.36&0.05 &                                                                                      \\
629  & 21:39:46.32 &  57:39:35.1$^{\rm r}$& 692  & 3692                      &                               &                 &                 &   11.8$^{\rm j}$&       &                 &5.813 \,(29) &4.724 \,()   &   4.215 \,()$^{\rm r}$&                    &                 &               &-6.6\,(6.1)   &-13.3\,(6.6)  &-12.4\,(7.5)  &-100.6\,(7.6) &0.36 &-0.18&                                                                                      \\
630  & 21:39:46.66 &  57:39:45.4$^{\rm r}$& 693  & 3693                      &                               &                 &                 &   14.5$^{\rm j}$&       &                 &9.122 \,(34) &7.935 \,(36) & 7.561 \,(17)$^{\rm r}$&                    &                 &               &-0.9\,(6.4)   &-7\,(6.4)     &              &              &0.16 &0.29 &                                                                                      \\
631  & 21:40:06.44 &  57:39:35.1$^{\rm r}$& 694  & 3694                      &                               &                 &                 &   14.9$^{\rm j}$&       &                 &11.277\,(26) &10.488\,(30) & 10.276\,(23)$^{\rm r}$&                    &                 &               &-8.9\,(3.8)   &-6.4\,(3.8)   &-5.3\,(7.4)   &-9.1\,(7.4)   &     &     &                                                                                      \\
632  & 21:40:11.13 &  57:37:47.7$^{\rm r}$& 695  & 3695                      &                               &                 &                 &   14.9$^{\rm j}$&       &                 &13.174\,(24) &12.790\,(27) & 12.712\,(32)$^{\rm r}$&                    &                 &               &1.1\,(3.8)    &-1.9\,(3.8)   &3.6\,(7.3)    &-3.7\,(7.4)   &     &     &                                                                                      \\
633  & 21:40:26.99 &  57:38:44.5$^{\rm r}$& 696  & 3696                      &                               &                 &                 &   14.1$^{\rm j}$&       &                 &12.616\,(24) &12.305\,(29) & 12.211\,(23)$^{\rm r}$&                    &                 &               &-0.9\,(3.8)   &-14.3\,(3.8)  &0.4\,(7.4)    &-23.7\,(7.4)  &-0.4 &-0.61&                                                                                      \\
634  & 21:40:27.25 &  57:39:14.7$^{\rm r}$& 697  & 3697                      &                               &                 &                 &   11.4$^{\rm j}$&       &                 &8.631 \,(26) &7.861 \,(76) & 7.590 \,(21)$^{\rm r}$&                    &                 &               &-4.5\,(4.7)   &-3.2\,(4.7)   &-1.5\,(1.3)   &1.6\,(1.6)    &-0.02&0.17 &                                                                                      \\
635  & 21:40:28.41 &  57:39:45.6$^{\rm r}$& 698  & 3698                      &                               &                 &                 &   12.9$^{\rm j}$&       &                 &11.648\,(24) &11.293\,(31) & 11.155\,(23)$^{\rm r}$&                    &                 &               &-9.1\,(3.9)   &-6.8\,(3.9)   &-9.3\,(7.4)   &-10\,(7.4)    &0.09 &-0.75&                                                                                      \\
636  & 21:40:33.53 &  57:40:15.5$^{\rm r}$& 699  & 3699                      &                               &                 &                 &   14.3$^{\rm j}$&       &                 &12.385\,(24) &11.888\,(28) & 11.789\,(22)$^{\rm r}$&                    &                 &               &-3.7\,(3.9)   &-8.1\,(3.9)   &-10.1\,(7.4)  &-6.1\,(7.4)   &0.6  &-0.33&                                                                                      \\
637  & 21:40:44.12 &  57:39:22.0$^{\rm r}$& 700  & 489                       &                               &                 &                 &   10.8$^{\rm j}$&       &                 &10.284\,(23) &10.172\,(28) & 10.125\,(22)$^{\rm r}$&        A2$^{\rm q}$&                 &               &-0.9\,(1.7)   &-11.8\,(1.7)  &-1.8\,(1)     &-9.7\,(1.2)   &-0.01&-0.48&                                                                                      \\
638  & 21:40:48.21 &  57:39:20.1$^{\rm r}$& 701  & 3701                      &                               &                 &                 &   14.8$^{\rm j}$&       &                 &12.899\,(28) &12.566\,(35) & 12.472\,(32)$^{\rm r}$&                    &                 &               &3.4\,(3.8)    &-12.4\,(3.8)  &22.4\,(6.8)   &-11.6\,(6.8)  &-0.42&0.35 &                                                                                      \\
639  & 21:40:54.17 &  57:39:42.2$^{\rm r}$& 702  & 3702                      &                               &                 &                 &   12.3$^{\rm j}$&       &                 &11.719\,(23) &11.588\,(29) & 11.518\,(23)$^{\rm r}$&                    &                 &               &-8.2\,(2.7)   &6.8\,(2.7)    &-7.5\,(0.9)   &-3.8\,(1.7)   &0.35 &-0.1 &                                                                                      \\
640  & 21:41:05.36 &  57:39:36.3$^{\rm r}$& 703  & 3703                      &                               &                 &                 &   13.5$^{\rm j}$&       &                 &12.528\,(28) &12.351\,(35) & 12.265\,(26)$^{\rm r}$&                    &                 &               &-5.3\,(3.8)   &1.2\,(3.8)    &-4.9\,(7.2)   &10.4\,(7.3)   &-0.2 &0.05 &                                                                                      \\
641  & 21:39:50.44 &  57:41:12.6$^{\rm r}$& 704  & 3704                      &                               &                 &                 &   14.6$^{\rm j}$&       &                 &12.892\,(26) &12.533\,(27) & 12.369\,(28)$^{\rm r}$&                    &                 &               &-2.2\,(3.8)   &-2.8\,(3.8)   &-1.2\,(7.3)   &-4.6\,(7.5)   &-0.04&0.44 &                                                                                      \\
642  & 21:39:57.51 &  57:42:22.1$^{\rm r}$& 705  & 3705                      &                               &                 &                 &   12.2$^{\rm j}$&       &                 &11.108\,(29) &10.810\,(33) & 10.756\,(26)$^{\rm r}$&                    &                 &               &0.2\,(3.8)    &6.1\,(3.8)    &2.6\,(0.7)    &0.8\,(11.7)   &-0.49&0.33 &                                                                                      \\
643  & 21:39:55.60 &  57:43:05.7$^{\rm r}$& 706  & 3706                      &                               &                 &                 &   13.6$^{\rm j}$&       &                 &12.041\,(24) &11.556\,(27) & 11.497\,(23)$^{\rm r}$&                    &                 &               &-9.3\,(3.9)   &-12.5\,(3.9)  &-3\,(7.3)     &-12.1\,(7.3)  &0.46 &-0.71&                                                                                      \\
644  & 21:40:01.19 &  57:42:53.1$^{\rm r}$& 707  & 3707                      &                               &                 &                 &   12.7$^{\rm j}$&       &                 &9.969 \,(27) &9.203 \,(30) & 8.993 \,(23)$^{\rm r}$&                    &                 &               &-8.4\,(4.8)   &-4.8\,(4.8)   &-8.8\,(7.3)   &8.9\,(7.4)    &-0.07&-0.15&                                                                                      \\
645  & 21:40:02.61 &  57:42:33.5$^{\rm r}$& 708  & 3708                      &                               &                 &                 &   13.1$^{\rm j}$&       &                 &11.886\,(27) &11.611\,(27) & 11.517\,(23)$^{\rm r}$&                    &                 &               &-7.8\,(3.8)   &-1.6\,(3.8)   &-3.3\,(7.3)   &5.6\,(7.4)    &0.38 &-0.06&                                                                                      \\
646  & 21:40:04.21 &  57:42:28.5$^{\rm r}$& 709  & 3709                      &                               &                 &                 &   14.3$^{\rm j}$&       &                 &12.058\,(31) &11.629\,(39) & 11.475\,(37)$^{\rm r}$&                    &                 &               &-7.3\,(3.9)   &5.1\,(3.9)    &-3.3\,(7.3)   &0.6\,(7.4)    &0.31 &0.38 &                                                                                      \\
647  & 21:40:05.09 &  57:42:43.3$^{\rm r}$& 710  & 3710                      &                               &                 &                 &   14.4$^{\rm j}$&       &                 &11.853\,(29) &11.308\,(33) & 11.190\,(28)$^{\rm r}$&                    &                 &               &-16\,(6)      &-24.6\,(6)    &              &              &0.27 &-0.15&                                                                                      \\
648  & 21:40:05.88 &  57:42:48.9$^{\rm r}$& 711  & 3711                      &                               &                 &                 &   12.5$^{\rm j}$&       &                 &11.442\,(27) &11.032\,(32) & 10.939\,(23)$^{\rm r}$&                    &                 &               &51.8\,(19.1)  &57.2\,(19.1)  &11.4\,(1.8)   &5.2\,(1.6)    &-1.48&0.82 &                                                                                      \\
649  & 21:40:15.08 &  57:42:29.9$^{\rm r}$& 712  & 3712                      &                               &                 &  15.35$^{\rm l}$&  13.02$^{\rm l}$&       &                 &8.514 \,(18) &7.513 \,(36) & 7.174 \,(20)$^{\rm r}$&                    &                 &               &-0.2\,(4.9)   &-1\,(4.9)     &16.3\,(7.3)   &27.4\,(7.4)   &0.12 &0.11 &                                                                                      \\
650  & 21:40:12.28 &  57:41:46.8$^{\rm r}$& 713  & 485                       &                               &  11.15$^{\rm l}$&  11.49$^{\rm l}$&  11.15$^{\rm l}$&       &                 &10.200\,(26) &9.989 \,(28) & 9.836 \,(21)$^{\rm r}$&        B5$^{\rm q}$&                 &               &1.8\,(1.7)    &1.4\,(1.7)    &-3.4\,(0.5)   &-2\,(1.2)     &-0.09&0.15 &                                                                                      \\
651  & 21:40:14.31 &  57:41:51.0$^{\rm r}$& 714  & 3714                      &                               &                 &                 &   12.6$^{\rm j}$&       &                 &11.923\,(26) &11.764\,(28) & 11.721\,(23)$^{\rm r}$&                    &                 &               &3.5\,(11)     &-5.7\,(11)    &-4.1\,(0.6)   &0.1\,(0.8)    &-0.03&0.18 &                                                                                      \\
652  & 21:40:15.09 &  57:40:51.4$^{\rm r}$& 715  & 3715                      &                               &                 &  13.85$^{\rm f}$&   13.2$^{\rm e}$&       &                 &12.145\,(35) &11.922\,(46) & 11.881\,(51)$^{\rm r}$&        B9$^{\rm e}$&                 &  2.2$^{\rm e}$&23\,(13.5)    &-12.4\,(13.5) &-4.5\,(6.6)   &-0.2\,(0.7)   &0.03 &0.11 &near 1807                                                                          \\
653  & 21:40:45.26 &  57:40:43.4$^{\rm r}$& 716  & 3716                      &                               &                 &                 &   13.8$^{\rm j}$&       &                 &12.453\,(21) &12.174\,(31) & 12.091\,(22)$^{\rm r}$&                    &                 &               &-5.6\,(3.8)   &-9.5\,(3.8)   &-7.9\,(7.4)   &-15.3\,(7.4)  &0.35 &-0.4 &                                                                                      \\
654  & 21:40:55.84 &  57:41:01.6$^{\rm r}$& 717  & 3717                      &                               &                 &                 &   13.1$^{\rm j}$&       &                 &11.820\,(23) &11.440\,(25) & 11.396\,(22)$^{\rm r}$&                    &                 &               &3.4\,(3.8)    &14.9\,(3.8)   &4.2\,(7.4)    &14.8\,(7.4)   &-0.66&2.1  &                                                                                      \\
655  & 21:41:00.63 &  57:40:50.4$^{\rm r}$& 718  & 3718                      &                               &                 &                 &   12.2$^{\rm j}$&       &                 &11.177\,(24) &10.810\,(28) & 10.737\,(23)$^{\rm r}$&                    &                 &               &2.6\,(2.7)    &-12.5\,(2.7)  &-0.7\,(1.9)   &-12.6\,(1.8)  &-0.15&-0.96&                                                                                      \\
656  & 21:41:09.59 &  57:41:50.6$^{\rm r}$& 719  & 3719                      &                               &                 &                 &   14.3$^{\rm j}$&       &                 &12.741\,(35) &12.537\,(38) & 12.424\,(33)$^{\rm r}$&                    &                 &               &-10.9\,(3.9)  &-17.5\,(3.9)  &-25.9\,(7.3)  &-88.3\,(7.4)  &0.29 &0.05 &                                                                                      \\
657  & 21:40:40.74 &  57:42:08.4$^{\rm r}$& 720  & 3720                      &                               &                 &                 &   14.2$^{\rm j}$&       &                 &12.091\,(23) &11.468\,(28) & 11.411\,(22)$^{\rm r}$&                    &                 &               &35\,(19.7)    &20.5\,(19.7)  &16.1\,(7.3)   &2.9\,(7.4)    &-2.3 &1.47 &                                                                                      \\
658  & 21:40:48.56 &  57:42:51.5$^{\rm r}$& 721  & 3721                      &                               &                 &                 &   14.4$^{\rm j}$&       &                 &12.777\,(24) &12.511\,(28) & 12.398\,(26)$^{\rm r}$&                    &                 &               &1.7\,(3.9)    &-4.9\,(3.9)   &17\,(7.2)     &5.3\,(7.3)    &-0.13&0.09 &                                                                                      \\
659  & 21:41:00.93 &  57:43:04.5$^{\rm r}$& 722  & 3722                      &                               &                 &                 &   14.4$^{\rm j}$&       &                 &12.304\,(23) &11.953\,(29) & 11.849\,(23)$^{\rm r}$&                    &                 &               &16\,(13.9)    &-26.7\,(13.9) &21.4\,(7.4)   &-57.4\,(7.4)  &-0.52&0.15 &                                                                                      \\
660  & 21:41:06.85 &  57:42:57.1$^{\rm r}$& 723  & 3723                      &                               &                 &                 &   13.2$^{\rm j}$&       &                 &11.931\,(23) &11.626\,(29) & 11.594\,(23)$^{\rm r}$&                    &                 &               &-4.3\,(3.8)   &1.2\,(3.8)    &-4.3\,(7.4)   &1.5\,(7.4)    &-0.31&0.82 &                                                                                      \\
661  & 21:39:46.81 &  57:44:40.3$^{\rm r}$& 724  & 3724                      &                               &                 &                 &   14.4$^{\rm j}$&       &                 &12.384\,(27) &11.982\,(31) & 11.897\,(23)$^{\rm r}$&                    &                 &               &-2.7\,(3.9)   &3.4\,(3.9)    &0.9\,(7.4)    &-18.4\,(7.5)  &-0.56&0.93 &                                                                                      \\
662  & 21:40:02.14 &  57:44:17.9$^{\rm r}$& 725  & 730                       &                               &                 &                 &   11.4$^{\rm j}$&       &                 &10.854\,(26) &10.710\,(30) & 10.634\,(21)$^{\rm r}$&        A0$^{\rm q}$&                 &               &-3.3\,(2)     &-1.4\,(2)     &-1.9\,(0.8)   &-3.3\,(2)     &-0.06&-0.01&                                                                                      \\
663  & 21:40:19.87 &  57:44:07.5$^{\rm r}$& 726  & 5092                      &                               &                 &  15.97$^{\rm f}$&  14.77$^{\rm e}$&       &                 &12.746\,(24) &12.340\,(29) & 12.184\,(23)$^{\rm r}$&        F9$^{\rm e}$&                 &    2$^{\rm e}$&-3.9\,(3.8)   &-5.7\,(3.8)   &-6.5\,(7.4)   &-10.7\,(7.5)  &     &     &                                                                                      \\
664  & 21:40:34.89 &  57:44:08.9$^{\rm j}$& 727  & 3727                      &                               &                 &                 &   14.4$^{\rm j}$&       &                 &             &             &                       &                    &                 &               &              &              &              &              &     &     &no star                                                                          \\
665  & 21:40:41.83 &  57:43:58.5$^{\rm r}$& 728  & 740                       &                               &                 &                 &   11.7$^{\rm j}$&       &                 &11.263\,(23) &11.164\,(28) & 11.121\,(22)$^{\rm r}$&        A0$^{\rm q}$&                 &               &-0.8\,(2.7)   &-3\,(2.7)     &-2.6\,(1.1)   &-7.7\,(1.5)   &0.01 &-0.52&                                                                                      \\
666  & 21:40:48.50 &  57:43:49.3$^{\rm r}$& 729  & 3729                      &                               &                 &                 &   14.2$^{\rm j}$&       &                 &12.328\,(24) &11.920\,(29) & 11.802\,(25)$^{\rm r}$&                    &                 &               &-0.4\,(3.8)   &14.8\,(3.8)   &-13.8\,(7.4)  &29.8\,(7.4)   &-0.9 &1.68 &                                                                                      \\
667  & 21:41:04.45 &  57:43:44.3$^{\rm r}$& 730  & 3730                      &                               &                 &                 &   14.7$^{\rm j}$&       &                 &12.816\,(21) &12.477\,(28) & 12.372\,(26)$^{\rm r}$&                    &                 &               &-1.1\,(3.9)   &-3.9\,(3.9)   &-0.5\,(7.3)   &-13.7\,(7.3)  &-0.21&-0.22&                                                                                      \\
668  & 21:41:06.40 &  57:43:57.5$^{\rm r}$& 731  & 3731                      &                               &                 &                 &   13.9$^{\rm j}$&       &                 &12.254\,(28) &11.867\,(31) & 11.773\,(23)$^{\rm r}$&                    &                 &               &-6.9\,(3.8)   &-7.9\,(3.8)   &-17\,(7.5)    &-8.3\,(7.5)   &-0.1 &-0.37&new coordinates                                                                       \\    
669  & 21:39:49.26 &  57:46:17.5$^{\rm r}$& 732  & 727                       &                               &                 &                 &   11.3$^{\rm j}$&       &                 &10.594\,(27) &10.456\,(33) & 10.376\,(24)$^{\rm r}$&        A0$^{\rm q}$&                 &               &2\,(1.7)      &-4.6\,(1.7)   &-2.7\,(0.8)   &-3.9\,(1)     &-0.16&0.16 &                                                                                      \\
670  & 21:39:58.85 &  57:45:33.8$^{\rm r}$& 733  & 3733                      &                               &                 &                 &   14.3$^{\rm j}$&       &                 &12.421\,(24) &12.105\,(33) & 11.973\,(21)$^{\rm r}$&                    &                 &               &-4.4\,(3.9)   &2.1\,(3.9)    &-8.9\,(7.7)   &-7.8\,(7.8)   &0.42 &-0.09&                                                                                      \\
671  & 21:40:13.75 &  57:44:58.6$^{\rm r}$& 734  & 3734                      &                               &                 &                 &   13.7$^{\rm j}$&       &                 &12.078\,(27) &11.713\,(31) & 11.631\,(23)$^{\rm r}$&                    &                 &               &5.5\,(3.8)    &6.1\,(3.8)    &4\,(7.4)      &-2.7\,(7.4)   &-0.85&0.95 &                                                                                      \\
672  & 21:40:27.18 &  57:45:30.1$^{\rm r}$& 735  & 3735                      &                               &                 &                 &   14.6$^{\rm j}$&       &                 &12.512\,(32) &12.170\,(35) & 12.092\,(29)$^{\rm r}$&                    &                 &               &5.5\,(7.2)    &-22.7\,(7.2)  &              &              &0.53 &-0.37&                                                                                      \\
673  & 21:40:31.07 &  57:45:29.5$^{\rm r}$& 736  & 737                       &                               &                 &                 &   11.5$^{\rm j}$&       &                 &11.067\,(24) &10.938\,(32) & 10.859\,(25)$^{\rm r}$&        A0$^{\rm q}$&                 &               &-0.6\,(2)     &-7.8\,(2)     &-3.3\,(0.7)   &-5.1\,(1)     &0.13 &-0.07&                                                                                      \\
674  & 21:40:45.56 &  57:45:31.5$^{\rm r}$& 737  & 741                       &                               &                 &                 &   11.8$^{\rm j}$&       &                 &10.760\,(28) &10.424\,(29) & 10.379\,(25)$^{\rm r}$&                    &                 &               &-281.8\,(5.1) &302.2\,(5.1)  &7\,(2.8)      &1\,(2.7)      &-1.13&0.31 &                                                                                      \\
675  & 21:40:46.45 &  57:45:24.0$^{\rm r}$& 738  & 3738                      &                               &                 &                 &   14.4$^{\rm j}$&       &                 &10.972\,(23) &10.243\,(28) & 10.020\,(22)$^{\rm r}$&                    &                 &               &15.2\,(11.6)  &-18.4\,(11.6) &40.3\,(5.3)   &-53.7\,(5.6)  &0.09 &-0.34&                                                                                      \\
676  & 21:40:55.81 &  57:45:13.9$^{\rm r}$& 739  & 3739                      &                               &                 &                 &   14.3$^{\rm j}$&       &                 &12.635\,(26) &12.210\,(29) & 12.216\,(28)$^{\rm r}$&                    &                 &               &-25.8\,(3.8)  &-13.1\,(3.8)  &-37.6\,(7.4)  &-14.3\,(7.4)  &1.37 &-0.48&                                                                                      \\
677  & 21:41:15.39 &  57:45:30.0$^{\rm r}$& 740  & 3740                      &                               &                 &                 &   12.8$^{\rm j}$&       &                 &10.992\,(24) &10.515\,(27) & 10.390\,(21)$^{\rm r}$&                    &                 &               &-4.6\,(3.8)   &-3.8\,(3.8)   &-2.8\,(7.5)   &-8.9\,(7.5)   &-0.07&-0.12&                                                                                      \\
678  & 21:41:22.88 &  57:45:16.2$^{\rm r}$& 741  & 744                       &                               &  10.46$^{\rm l}$&  10.69$^{\rm l}$&  10.52$^{\rm l}$&       &                 &10.089\,(26) &10.071\,(30) & 10.031\,(19)$^{\rm r}$&        A0$^{\rm q}$&                 &               &-5.9\,(2)     &-5.2\,(2)     &-3.4\,(0.5)   &-4.5\,(0.6)   &0.11 &-0.2 &                                                                                      \\
679  & 21:41:24.94 &  57:44:59.9$^{\rm r}$& 742  & 3742                      &                               &                 &                 &   14.5$^{\rm j}$&       &                 &12.771\,(32) &12.483\,(32) & 12.391\,(32)$^{\rm r}$&                    &                 &               &-6.3\,(3.8)   &-12.3\,(3.8)  &14.5\,(7.1)   &-30.5\,(7.1)  &0.48 &-0.11&                                                                                      \\
680  & 21:41:28.13 &  57:44:44.6$^{\rm r}$& 743  & 3743                      &                               &                 &                 &     14$^{\rm j}$&       &                 &10.773\,(24) &10.044\,(31) & 9.832 \,(19)$^{\rm r}$&                    &                 &               &31.1\,(6.6)   &-377.2\,(6.6) &-3.6\,(7.2)   &-7.5\,(7.3)   &0.16 &0.2  &                                                                                      \\
681  & 21:41:28.14 &  57:43:59.3$^{\rm r}$& 744  & 3744                      &                               &                 &                 &   14.1$^{\rm j}$&       &                 &9.994 \,(26) &9.090 \,(28) & 8.851 \,(19)$^{\rm r}$&                    &                 &               &-3.4\,(4.9)   &-3.2\,(4.9)   &-9.9\,(7.3)   &-12.1\,(7.3)  &0.25 &-0.04&                                                                                      \\
682  & 21:41:23.48 &  57:43:17.1$^{\rm r}$& 745  & 3745                      &                               &                 &                 &   14.5$^{\rm j}$&       &                 &12.010\,(26) &11.366\,(27) & 11.264\,(21)$^{\rm r}$&                    &                 &               &15.9\,(3.8)   &-4.4\,(3.8)   &13.9\,(7.3)   &-10.3\,(7.4)  &-2.35&0.44 &                                                                                      \\
683  & 21:41:12.69 &  57:46:30.2$^{\rm r}$& 746  & 743                       &                               &   9.51$^{\rm l}$&   9.97$^{\rm l}$&   9.85$^{\rm l}$&       &                 &9.446 \,(26) &9.386 \,(30) & 9.328 \,(21)$^{\rm r}$&        B8$^{\rm q}$&                 &               &-4.5\,(2)     &-7.4\,(2)     &-4.7\,(0.6)   &-5.6\,(0.5)   &-0.19&-0.06&                                                                                      \\
684  & 21:39:57.59 &  57:36:16.4$^{\rm r}$& 747  & 482                       &                               &   9.36$^{\rm l}$&   9.83$^{\rm l}$&   9.71$^{\rm l}$&       &                 &9.463 \,(26) &9.461 \,(28) & 9.452 \,(23)$^{\rm r}$&        B8$^{\rm q}$&                 &               &-1.6\,(1.3)   &-7.4\,(1.2)   &-3.1\,(0.5)   &-5.2\,(0.7)   &-0.23&-0.14&                                                                                      \\
685  & 21:40:31.59 &  57:16:40.9$^{\rm r}$& 800  & 488                       &                               &  12.05$^{\rm l}$&  11.16$^{\rm l}$&  10.03$^{\rm l}$&       &                 &7.916 \,(21) &7.413 \,(42) & 7.262 \,(26)$^{\rm r}$&        G5$^{\rm q}$&                 &               &-3.2\,(1.7)   &-10.2\,(1.7)  &-2.4\,(0.9)   &-8.7\,(1.2)   &0.02 &-0.23&                                                                                      \\
686  & 21:41:27.49 &  57:13:07.6$^{\rm r}$& 801  & 3801                      &                               &                 &                 &   12.6$^{\rm j}$&       &                 &10.164\,(27) &9.524 \,(32) & 9.320 \,(21)$^{\rm r}$&                    &                 &               &-6.3\,(5.1)   &-5\,(5.1)     &-13.3\,(7.5)  &-7.3\,(7.5)   &-0.05&-0.1 &                                                                                      \\
687  & 21:41:33.26 &  57:13:09.2$^{\rm r}$& 802  & 3802                      &                               &                 &                 &     13$^{\rm j}$&       &                 &11.518\,(31) &11.246\,(40) & 11.076\,(33)$^{\rm r}$&                    &                 &               &-7\,(4.1)     &10.5\,(4.1)   &2.1\,(7.1)    &28.1\,(7.2)   &-0.8 &0.51 &                                                                                      \\
688  & 21:41:40.23 &  57:18:54.0$^{\rm r}$& 803  & 3803                      &                               &                 &                 &   12.9$^{\rm j}$&       &                 &11.703\,(27) &11.538\,(33) & 11.418\,(21)$^{\rm r}$&                    &                 &               &-4\,(4.1)     &0.7\,(4.1)    &5.6\,(7.1)    &8.3\,(7.1)    &0.01 &0.01 &                                                                                      \\
689  & 21:41:38.59 &  57:19:46.9$^{\rm r}$& 804  & 3804                      &                               &                 &                 &   13.6$^{\rm j}$&       &                 &12.137\,(29) &11.911\,(36) & 11.783\,(26)$^{\rm r}$&                    &                 &               &6\,(4.1)      &11.1\,(4.1)   &29.1\,(7.1)   &62.1\,(7.1)   &-0.79&0.47 &                                                                                      \\
690  & 21:41:36.21 &  57:22:25.8$^{\rm r}$& 805  & 3805                      &                               &  10.99$^{\rm l}$&   11.4$^{\rm f}$&     11$^{\rm e}$&       &                 &10.470\,()   &10.346\,()   & 10.475\,(52)$^{\rm r}$&        B6$^{\rm e}$&                 &  1.7$^{\rm e}$&-5.1\,(2)     &-5.6\,(2)     &-2.3\,(0.6)   &-5.7\,(0.9)   &0.01 &0.01 &                                                                                      \\
691  & 21:41:32.85 &  57:22:46.4$^{\rm r}$& 806  & 3806                      &                               &                 &                 &   13.5$^{\rm j}$&       &                 &10.740\,(21) &10.103\,(30) & 9.894 \,(20)$^{\rm r}$&                    &                 &               &-6.2\,(5.1)   &-8.3\,(5.1)   &12.1\,(7.2)   &-14.7\,(7.2)  &0.55 &-0.51&                                                                                      \\
692  & 21:41:35.26 &  57:23:43.2$^{\rm r}$& 807  & 3807                      &                               &                 &                 &   12.3$^{\rm j}$&       &                 &11.221\,(24) &10.951\,(30) & 10.883\,(26)$^{\rm r}$&                    &                 &               &-13.2\,(5.5)  &-30.7\,(5.5)  &-0.1\,(0.7)   &-0.4\,(0.9)   &-0.4 &0.13 &                                                                                      \\
693  & 21:41:35.64 &  57:23:49.5$^{\rm r}$& 808  & 3808                      &                               &                 &                 &   14.4$^{\rm j}$&       &                 &12.636\,(35) &12.214\,(32) & 12.148\,(32)$^{\rm r}$&                    &                 &               &              &              &              &              &-1.26&-0.13&                                                                                      \\
694  & 21:41:40.79 &  57:25:46.8$^{\rm r}$& 809  & 3809                      &                               &                 &                 &   13.8$^{\rm j}$&       &                 &12.318\,(28) &11.995\,(36) & 11.906\,(28)$^{\rm r}$&                    &                 &               &-6.4\,(4.1)   &-6.6\,(4.1)   &9.6\,(6.9)    &11\,(6.9)     &0.3  &0.11 &                                                                                      \\
695  & 21:41:35.95 &  57:31:53.2$^{\rm r}$& 810  & 3810                      &                               &                 &                 &   12.7$^{\rm j}$&       &                 &9.870 \,(23) &9.093 \,(31) & 8.882 \,(20)$^{\rm r}$&                    &                 &               &20.9\,(8.4)   &-9.3\,(8.4)   &              &              &-0.23&-0.04&                                                                                      \\
696  & 21:41:37.05 &  57:31:48.9$^{\rm r}$& 811  & 495                       &                               &                 &                 &   11.1$^{\rm j}$&       &                 &10.733\,(23) &10.560\,(27) & 10.491\,(20)$^{\rm r}$&        A5$^{\rm q}$&                 &               &3.5\,(2)      &6.8\,(2)      &4.7\,(1.1)    &3.6\,(2)      &-0.81&0.63 &                                                                                      \\
697  & 21:41:38.41 &  57:32:35.5$^{\rm r}$& 812  & 496                       &                               &                 &                 &   12.1$^{\rm j}$&       &                 &11.269\,(21) &11.149\,(30) & 11.050\,(19)$^{\rm r}$&        A0$^{\rm q}$&                 &               &-1.6\,(11)    &-6.8\,(11)    &-0.3\,(1.3)   &2.2\,(1.2)    &-0.39&0.41 &                                                                                      \\
698  & 21:41:39.95 &  57:33:08.9$^{\rm r}$& 813  & 3813                      &                               &                 &                 &   13.8$^{\rm j}$&       &                 &12.569\,(21) &12.345\,(27) & 12.248\,(19)$^{\rm r}$&                    &                 &               &-11.2\,(3.8)  &-5\,(3.8)     &-5.9\,(7.3)   &-13.9\,(7.4)  &0.52 &-0.3 &                                                                                      \\
699  & 21:41:40.16 &  57:37:32.3$^{\rm r}$& 814  & 3814                      &                               &                 &                 &   13.4$^{\rm j}$&       &                 &11.989\,(23) &11.625\,(31) & 11.526\,(22)$^{\rm r}$&                    &                 &               &-5\,(3.8)     &-3.8\,(3.8)   &2.7\,(7.3)    &-13.1\,(7.4)  &0.2  &-0.32&                                                                                      \\
700  & 21:41:34.86 &  57:39:25.7$^{\rm r}$& 815  & 3815                      &                               &                 &                 &   14.6$^{\rm j}$&       &                 &12.854\,(34) &12.476\,(33) & 12.428\,(35)$^{\rm r}$&                    &                 &               &-3\,(3.9)     &-3.4\,(3.9)   &-3.7\,(7.3)   &-14.1\,(7.3)  &0.62 &-0.32&                                                                                      \\
701  & 21:41:50.80 &  57:18:23.5$^{\rm r}$& 816  & 3816                      &                               &                 &                 &   14.3$^{\rm j}$&       &                 &12.443\,(24) &12.167\,(33) & 12.089\,(24)$^{\rm r}$&                    &                 &               &-5.3\,(4.1)   &0.1\,(4.1)    &2.9\,(6.9)    &7.7\,(7)      &-0.02&0.19 &                                                                                      \\
702  & 21:41:44.37 &  57:20:22.8$^{\rm r}$& 817  & 3817                      &                               &                 &                 &   14.2$^{\rm j}$&       &                 &11.158\,(27) &10.487\,(32) & 10.278\,(21)$^{\rm r}$&                    &                 &               &17.7\,(4.1)   &-9.5\,(4.1)   &73.2\,(6.9)   &-20\,(7)      &-0.21&0.11 &                                                                                      \\
703  & 21:41:53.16 &  57:19:20.0$^{\rm r}$& 818  & 3818                      &                               &                 &                 &   12.4$^{\rm j}$&       &                 &9.783 \,(27) &9.039 \,(32) & 8.821 \,(19)$^{\rm r}$&                    &                 &               &6.1\,(6.5)    &-0.8\,(6.5)   &-14.6\,(7.5)  &14.7\,(7.5)   &-0.39&0.76 &                                                                                      \\
704  & 21:41:59.28 &  57:19:19.7$^{\rm r}$& 819  & 3819                      &                               &                 &                 &   13.6$^{\rm j}$&       &                 &11.967\,(26) &11.666\,(32) & 11.489\,(23)$^{\rm r}$&                    &                 &               &2.4\,(4.1)    &-2\,(4.1)     &3.5\,(7.1)    &7.4\,(7.1)    &-0.61&-0.49&                                                                                      \\
705  & 21:41:59.37 &  57:20:50.6$^{\rm r}$& 820  & 3820                      &                               &                 &                 &   13.5$^{\rm j}$&       &                 &16.267\,(129)&14.704\,()   &   14.812\,()$^{\rm r}$&                    &                 &               &-12\,(5.5)    &-25.6\,(5.5)  &              &              &     &     &no/faint star                                                               \\
706  & 21:42:19.95 &  57:15:17.1$^{\rm r}$& 821  & 3821                      &                               &                 &                 &   13.9$^{\rm j}$&       &                 &12.199\,(24) &11.958\,(29) & 11.879\,(22)$^{\rm r}$&                    &                 &               &-0.3\,(4.1)   &-2.7\,(4.1)   &4.3\,(6.9)    &10.6\,(6.9)   &-0.26&0.11 &                                                                                      \\
707  & 21:42:23.81 &  57:20:46.6$^{\rm r}$& 822  & 500                       &                               &  11.90$^{\rm l}$&  11.09$^{\rm h}$&  11.05$^{\rm h}$&       &                 &10.582\,(39) &10.436\,(43) & 10.318\,(37)$^{\rm r}$&        B9$^{\rm h}$&                 &  1.4$^{\rm h}$&-4.5\,(2)     &-2\,(2)       &-3.1\,(0.6)   &-4.3\,(0.8)   &0.13 &-0.12&                                                                                      \\
708  & 21:42:25.76 &  57:21:29.0$^{\rm r}$& 823  & 3823                      &                               &                 &                 &   13.4$^{\rm j}$&       &                 &11.045\,(26) &10.478\,(31) & 10.279\,(24)$^{\rm r}$&                    &                 &               &-10.9\,(4.1)  &-1.8\,(4.1)   &-6\,(7)       &14.7\,(7)     &0.85 &-0.49&                                                                                      \\
709  & 21:42:20.65 &  57:22:45.5$^{\rm r}$& 824  & 3824                      &                               &                 &                 &   14.4$^{\rm j}$&       &                 &10.795\,(24) &10.062\,(31) & 9.825 \,(21)$^{\rm r}$&                    &                 &               &-3\,(5.1)     &-3.4\,(5.1)   &3.5\,(7)      &12.9\,(7)     &0.03 &0.25 &                                                                                      \\
710  & 21:42:15.03 &  57:23:14.3$^{\rm r}$& 825  & 3825                      &                               &                 &                 &   12.8$^{\rm j}$&       &                 &11.384\,(26) &11.097\,(32) & 10.980\,(21)$^{\rm r}$&                    &                 &               &-19\,(4.1)    &-11\,(4.1)    &-47.5\,(7.2)  &-33.3\,(7.3)  &0.78 &-0.48&                                                                                      \\
711  & 21:42:06.85 &  57:25:24.5$^{\rm r}$& 826  & 3826                      &                               &                 &                 &   12.3$^{\rm j}$&       &                 &11.206\,(24) &10.812\,(30) & 10.774\,(23)$^{\rm r}$&                    &                 &               &24.8\,(5.1)   &-5.9\,(5.1)   &43.4\,(7.4)   &5.3\,(7.3)    &-3.81&0.17 &                                                                                      \\
712  & 21:42:16.34 &  57:26:44.4$^{\rm r}$& 827  & 3827                      &                               &                 &                 &   14.1$^{\rm j}$&       &                 &11.826\,(26) &11.439\,(31) & 11.249\,(21)$^{\rm r}$&                    &                 &               &12.4\,(19.8)  &68.6\,(19.8)  &6.9\,(6.9)    &11.6\,(7.2)   &-0.28&-0.12&                                                                                      \\
713  & 21:42:00.96 &  57:26:26.5$^{\rm r}$& 828  & 3828                      &                               &                 &                 &   14.6$^{\rm j}$&       &                 &12.968\,(26) &12.616\,(30) & 12.529\,(24)$^{\rm r}$&                    &                 &               &15.7\,(7.1)   &43.7\,(7.1)   &-4.9\,(6.9)   &10.2\,(6.8)   &0.37 &-0.22&                                                                                      \\
714  & 21:41:54.15 &  57:26:32.4$^{\rm r}$& 829  & 3829                      &                               &                 &                 &   14.4$^{\rm j}$&       &                 &12.310\,(26) &11.885\,(32) & 11.784\,(25)$^{\rm r}$&                    &                 &               &-12.7\,(3.9)  &-1.4\,(3.9)   &3.6\,(6.8)    &17.7\,(7)     &0.1  &0.36 &                                                                                      \\
715  & 21:41:59.14 &  57:27:50.9$^{\rm r}$& 830  & 3830                      &                               &                 &                 &   12.2$^{\rm j}$&       &                 &11.307\,(26) &11.152\,(33) & 11.065\,(24)$^{\rm r}$&                    &                 &               &-7.4\,(3.8)   &-5.3\,(3.8)   &-6.1\,(1.6)   &-4.1\,(1.1)   &0.03 &-0.13&                                                                                      \\
716  & 21:42:17.78 &  57:27:26.9$^{\rm r}$& 831  & 3831                      &                               &                 &                 &     13$^{\rm j}$&       &                 &10.123\,(24) &9.442 \,(31) & 9.190 \,(21)$^{\rm r}$&                    &                 &               &-8\,(4.7)     &-3.7\,(4.7)   &-2.6\,(7.3)   &19.8\,(7.3)   &0.4  &0.02 &                                                                                      \\
717  & 21:41:59.16 &  57:28:05.0$^{\rm r}$& 832  & 3832                      &                               &                 &                 &   14.6$^{\rm j}$&       &                 &12.557\,(26) &12.054\,(31) & 11.944\,(21)$^{\rm r}$&                    &                 &               &-4.1\,(3.8)   &0.3\,(3.8)    &19.3\,(7.4)   &28.6\,(6.9)   &0.02 &-0.36&                                                                                      \\
718  & 21:42:00.65 &  57:28:12.1$^{\rm r}$& 833  & 3833                      &                               &                 &                 &   14.5$^{\rm j}$&       &                 &12.347\,()   &12.013\,()   &   11.809\,()$^{\rm r}$&                    &                 &               &-10.3\,(5.1)  &23.3\,(5.1)   &87.6\,(7.3)   &23.9\,(7.3)   &-2.65&2.27 &                                                                                      \\
719  & 21:42:00.30 &  57:28:25.2$^{\rm r}$& 834  & 3834                      &                               &                 &                 &   14.6$^{\rm j}$&       &                 &12.672\,(29) &12.181\,(33) &   12.099\,()$^{\rm r}$&                    &                 &               &-4.2\,(3.9)   &17.7\,(3.9)   &1.3\,(6.8)    &53.4\,(6.8)   &-0.61&1.75 &                                                                                      \\
720  & 21:42:09.26 &  57:29:40.1$^{\rm r}$& 835  & 3835                      &                               &                 &                 &   14.6$^{\rm j}$&       &                 &12.867\,(26) &12.501\,(37) & 12.409\,(24)$^{\rm r}$&                    &                 &               &-7.8\,(3.9)   &-12.4\,(3.9)  &3.6\,(6.9)    &8.6\,(6.8)    &-0.22&-0.65&                                                                                      \\
721  & 21:41:49.98 &  57:29:28.6$^{\rm r}$& 836  & 3836                      &                               &                 &                 &   13.1$^{\rm j}$&       &                 &11.893\,(23) &11.678\,(30) & 11.574\,(25)$^{\rm r}$&                    &                 &               &-9.7\,(4.1)   &-6.5\,(4.1)   &-0.2\,(7.1)   &4.7\,(7)      &0.21 &-0.13&                                                                                      \\
722  & 21:41:55.56 &  57:30:08.8$^{\rm r}$& 837  & 3837                      &                               &                 &                 &   13.5$^{\rm j}$&       &                 &8.549 \,(19) &7.396 \,(33) & 7.032 \,(23)$^{\rm r}$&                    &                 &               &-3.7\,(4.8)   &1.3\,(4.8)    &0.7\,(7.1)    &13.9\,(7.1)   &-0.07&0.9  &                                                                                      \\
723  & 21:42:15.94 &  57:31:14.5$^{\rm r}$& 838  & 3838                      &                               &                 &                 &   13.1$^{\rm j}$&       &                 &11.724\,(26) &11.416\,(30) & 11.331\,(24)$^{\rm r}$&                    &                 &               &-5.6\,(3.8)   &-5.2\,(3.8)   &-1.1\,(7.3)   &-12.4\,(7.4)  &0.01 &-0.42&                                                                                      \\
724  & 21:41:55.88 &  57:32:52.4$^{\rm r}$& 839  & 3839                      &                               &                 &                 &   13.6$^{\rm j}$&       &                 &12.035\,(26) &11.790\,(33) & 11.669\,(26)$^{\rm r}$&                    &                 &               &-10.6\,(3.9)  &-7\,(3.9)     &-3.8\,(7.3)   &-17.4\,(7.3)  &0.22 &-0.34&                                                                                      \\
725  & 21:41:42.05 &  57:34:04.7$^{\rm r}$& 840  & 3840                      &                               &                 &                 &   13.6$^{\rm j}$&       &                 &11.870\,(21) &11.510\,(28) & 11.395\,(20)$^{\rm r}$&                    &                 &               &-18.3\,(3.8)  &-12.7\,(3.8)  &-10.2\,(7.4)  &-19\,(7.4)    &0.8  &-0.99&                                                                                      \\
726  & 21:41:59.88 &  57:33:43.8$^{\rm r}$& 841  & 3841                      &                               &                 &                 &   10.7$^{\rm j}$&       &                 &9.850 \,(32) &9.492 \,()   &   9.426 \,()$^{\rm r}$&                    &                 &               &14.3\,(2)     &16.6\,(2)     &13.9\,(1.5)   &9.6\,(2.1)    &-1.75&1.33 &                                                                                      \\
727  & 21:42:01.87 &  57:33:53.9$^{\rm r}$& 842  & 3842                      &                               &  10.48$^{\rm l}$&  10.74$^{\rm l}$&  10.44$^{\rm l}$&       &                 &9.534 \,(27) &9.225 \,(30) & 9.164 \,(23)$^{\rm r}$&                    &                 &               &-3.5\,(2)     &-0.5\,(2)     &-2.8\,(1)     &-4.1\,(0.8)   &-0.06&-0.09&                                                                                      \\
728  & 21:41:50.02 &  57:34:48.6$^{\rm r}$& 843  & 3843                      &                               &                 &                 &     12$^{\rm j}$&       &                 &10.736\,(21) &10.392\,(27) & 10.276\,(19)$^{\rm r}$&                    &                 &               &-19.6\,(3.8)  &-2.8\,(3.8)   &-1.2\,(0.7)   &-9\,(1.3)     &-0.16&-0.67&                                                                                      \\
729  & 21:41:52.56 &  57:37:20.0$^{\rm r}$& 844  & 3844                      &                               &                 &                 &   14.2$^{\rm j}$&       &                 &12.340\,(29) &12.000\,(28) & 11.890\,(29)$^{\rm r}$&                    &                 &               &7\,(3.9)      &-10.2\,(3.9)  &44.6\,(7)     &-14.5\,(7.1)  &0.31 &-0.02&                                                                                      \\
730  & 21:41:51.50 &  57:37:30.3$^{\rm r}$& 845  & 3845                      &                               &                 &                 &   11.8$^{\rm j}$&       &                 &10.686\,(28) &10.408\,(31) & 10.346\,(28)$^{\rm r}$&                    &                 &               &54\,(8.3)     &-28.9\,(8.3)  &-9\,(2)       &-7.7\,(3.8)   &0.28 &-0.61&                                                                                      \\
731  & 21:41:51.17 &  57:37:37.0$^{\rm r}$& 846  & 497                       &                               &                 &                 &   11.3$^{\rm j}$&       &                 &10.116\,(45) &9.834 \,(42) & 9.764 \,(30)$^{\rm r}$&                    &                 &               &-4.6\,(6.7)   &-13.5\,(6.7)  &-7.1\,(0.6)   &-7.4\,(1.1)   &0.19 &-0.59&                                                                                      \\
732  & 21:41:58.50 &  57:37:52.9$^{\rm r}$& 847  & 498                       &                               &                 &                 &   10.4$^{\rm j}$&       &                 &8.075 \,(29) &7.444 \,(42) & 7.282 \,(18)$^{\rm r}$&        G5$^{\rm q}$&                 &               &-12.9\,(3.3)  &93\,(4.1)     &-9.2\,(1)     &-0.5\,(1.2)   &0.95 &-0.23&near 732                                                                          \\
733  & 21:41:58.50 &  57:37:52.9$^{\rm r}$& 848  & 3848                      &                               &                 &                 &   10.3$^{\rm j}$&       &                 &8.075 \,(29) &7.444 \,(42) & 7.282 \,(18)$^{\rm r}$&                    &                 &               &-12.9\,(3.3)  &93\,(4.1)     &-9.2\,(1)     &-0.5\,(1.2)   &     &     &near 731                                                                           \\
734  & 21:42:02.39 &  57:37:25.8$^{\rm r}$& 849  & 3849                      &                               &                 &                 &   14.7$^{\rm j}$&       &                 &12.668\,(62) &12.141\,(57) & 12.041\,(57)$^{\rm r}$&                    &                 &               &17.4\,(3.9)   &33.6\,(3.9)   &              &              &-2.65&1.49 &                                                                                      \\
735  & 21:42:04.01 &  57:39:16.6$^{\rm r}$& 850  & 499                       &                               &  12.43$^{\rm l}$&  12.02$^{\rm h}$&  11.09$^{\rm h}$&       &                 &11.112\,(26) &11.052\,(32) & 10.986\,(21)$^{\rm r}$&        A0$^{\rm h}$&                 & 1.34$^{\rm h}$&4.8\,(2)      &-0.3\,(2)     &0.7\,(0.9)    &-2.1\,(0.9)   &-0.2 &0.02 &                                                                                      \\
736  & 21:42:18.64 &  57:37:06.4$^{\rm r}$& 851  & 3851                      &                               &                 &                 &   14.4$^{\rm j}$&       &                 &13.010\,(31) &12.807\,(40) & 12.620\,(37)$^{\rm r}$&                    &                 &               &-3.3\,(3.8)   &-1.8\,(3.8)   &5.2\,(7.3)    &-19\,(7.3)    &-0.17&0.15 &                                                                                      \\
737  & 21:42:21.68 &  57:37:50.6$^{\rm r}$& 852  & 5107                      &                               &                 &                 &   14.3$^{\rm j}$&       &                 &12.584\,(26) &12.337\,(32) & 12.181\,(24)$^{\rm r}$&                    &                 &               &-4.6\,(3.8)   &0.1\,(3.8)    &6.1\,(7.3)    &-15.6\,(7.3)  &-0.13&0.44 &                                                                                      \\
738  & 21:42:24.56 &  57:37:53.1$^{\rm r}$& 853  & 3853                      &                               &                 &                 &   13.9$^{\rm j}$&       &                 &10.690\,(24) &9.976 \,(31) & 9.732 \,(19)$^{\rm r}$&                    &                 &               &0.3\,(4.7)    &-6.3\,(4.7)   &0.9\,(7.2)    &-9.8\,(7.2)   &-0.01&-0.3 &                                                                                      \\
739  & 21:42:26.13 &  57:36:28.7$^{\rm r}$& 854  & 3854                      &                               &                 &                 &   13.4$^{\rm j}$&       &                 &12.175\,(26) &11.934\,(32) & 11.838\,(24)$^{\rm r}$&                    &                 &               &-35.5\,(7)    &-4.7\,(7)     &-1.1\,(7.1)   &-80.8\,(7.1)  &0.16 &0.13 &                                                                                      \\
740  & 21:41:42.96 &  57:40:38.0$^{\rm r}$& 855  & 3855                      &                               &                 &                 &   12.6$^{\rm j}$&       &                 &10.012\,(26) &9.253 \,(30) & 9.054 \,(19)$^{\rm r}$&                    &                 &               &-4.6\,(4.9)   &-6\,(4.9)     &-3.7\,(7.4)   &-10\,(7.4)    &0.3  &-0.15&                                                                                      \\
741  & 21:42:16.62 &  57:39:53.8$^{\rm r}$& 856  & 3856                      &                               &                 &                 &   13.9$^{\rm j}$&       &                 &10.757\,(26) &9.978 \,(30) & 9.772 \,(21)$^{\rm r}$&                    &                 &               &-7.1\,(4.7)   &-7\,(4.7)     &-4.8\,(7.3)   &-12.4\,(7.3)  &0.32 &-0.2 &                                                                                      \\
742  & 21:42:18.08 &  57:41:09.7$^{\rm r}$& 857  & 3857                      &                               &                 &                 &   14.2$^{\rm j}$&       &                 &12.359\,(26) &12.105\,(32) & 11.971\,(21)$^{\rm r}$&                    &                 &               &-6.9\,(12.8)  &6.8\,(12.8)   &-4.7\,(7.3)   &-6.1\,(7.3)   &0.2  &0.09 &                                                                                      \\
743  & 21:41:44.48 &  57:42:32.4$^{\rm r}$& 858  & 3858                      &                               &                 &                 &   13.7$^{\rm j}$&       &                 &11.279\,(26) &10.653\,(28) & 10.514\,(19)$^{\rm r}$&                    &                 &               &-10.8\,(3.8)  &-11.4\,(3.8)  &-20.2\,(7.4)  &-24.7\,(7.4)  &0.86 &-0.71&                                                                                      \\
744  & 21:41:40.29 &  57:43:25.0$^{\rm r}$& 859  & 3859                      &                               &                 &                 &   13.5$^{\rm j}$&       &                 &11.380\,(24) &10.816\,(31) & 10.702\,(21)$^{\rm r}$&                    &                 &               &13.9\,(3.8)   &-2.5\,(3.8)   &              &              &-0.35&-0.46&                                                                                      \\
745  & 21:42:02.51 &  57:44:42.6$^{\rm r}$& 861  & 3861                      &                               &                 &                 &   14.4$^{\rm j}$&       &                 &13.020\,(26) &12.798\,(32) & 12.688\,(24)$^{\rm r}$&                    &                 &               &-7.4\,(3.8)   &-5.7\,(3.8)   &0.8\,(7.3)    &-15.7\,(7.3)  &0.21 &-0.07&                                                                                      \\
746  & 21:41:50.94 &  57:45:36.8$^{\rm r}$& 862  & 3862                      &                               &                 &                 &   14.5$^{\rm j}$&       &                 &12.861\,(27) &12.667\,(28) & 12.521\,(19)$^{\rm r}$&                    &                 &               &-15.5\,(3.8)  &-3.9\,(3.8)   &-31.4\,(7.2)  &3.1\,(7.1)    &0.49 &-0.01&                                                                                      \\
747  & 21:42:14.09 &  57:43:09.9$^{\rm r}$& 863  & 3863                      &                               &                 &                 &   12.1$^{\rm j}$&       &                 &10.813\,(26) &10.436\,(30) & 10.335\,(21)$^{\rm r}$&                    &                 &               &-1\,(11.6)    &7.3\,(11.6)   &8.3\,(1.1)    &5.7\,(1.4)    &-0.93&0.76 &                                                                                      \\
748  & 21:42:24.18 &  57:44:09.9$^{\rm r}$& 864  & 750                       &                               &   6.28$^{\rm l}$&   7.09$^{\rm l}$&   6.86$^{\rm l}$&       &                 &6.072 \,(24) &5.914 \,(33) & 5.567 \,(17)$^{\rm r}$&        B0$^{\rm p}$&      V$^{\rm p}$&               &-2.3\,(0.5)   &-4.6\,(0.5)   &-1\,(2.9)     &3.2\,(2.4)    &-0.49&-0.39&                                                                                      \\
749  & 21:42:32.89 &  57:13:05.8$^{\rm r}$& 865  & 3865                      &                               &                 &                 &   12.1$^{\rm j}$&       &                 &9.525 \,(26) &8.813 \,(33) & 8.561 \,(22)$^{\rm r}$&                    &                 &               &-2.2\,(5.1)   &0.8\,(5.1)    &3.8\,(7.4)    &14.6\,(7.5)   &-0-38&0.33 &                                                                                      \\
750  & 21:42:43.58 &  57:12:00.1$^{\rm r}$& 866  & 3866                      &                               &                 &                 &   10.7$^{\rm j}$&       &                 &7.095 \,(20) &6.247 \,(23) & 5.946 \,(16)$^{\rm r}$&                    &                 &               &-6.3\,(13.3)  &7.1\,(13.3)   &-4.1\,(1.4)   &-8.2\,(0.6)   &0.6  &-0.21&                                                                                      \\
751  & 21:43:31.88 &  57:13:22.0$^{\rm r}$& 867  & 3867                      &                               &                 &                 &   13.3$^{\rm j}$&       &                 &12.163\,(29) &11.992\,(32) & 11.868\,(21)$^{\rm r}$&                    &                 &               &-4.9\,(4.1)   &-3.8\,(4.1)   &-8.3\,(1.9)   &-2.3\,(0.9)   &     &     &                                                                                      \\
752  & 21:43:17.56 &  57:15:47.0$^{\rm r}$& 868  & 3868                      &                               &                 &                 &   13.2$^{\rm j}$&       &                 &10.232\,(32) &9.559 \,(36) & 9.330 \,(28)$^{\rm r}$&                    &                 &               &2.6\,(5.1)    &-5.4\,(5.1)   &6.1\,(7.3)    &51.2\,(7.4)   &-0.37&-0.19&                                                                                      \\
753  & 21:43:17.39 &  57:18:36.0$^{\rm r}$& 869  & 3869                      &                               &                 &                 &   12.7$^{\rm j}$&       &                 &11.438\,(25) &11.150\,(30) & 11.074\,(23)$^{\rm r}$&                    &                 &               &0.5\,(4.1)    &4.4\,(4.1)    &7.3\,(7.1)    &14.5\,(7)     &-0.34&0.81 &                                                                                      \\
754  & 21:42:52.60 &  57:20:12.2$^{\rm r}$& 870  & 3870                      &                               &                 &                 &   14.3$^{\rm j}$&       &                 &10.755\,(27) &9.948 \,(29) & 9.698 \,(21)$^{\rm r}$&                    &                 &               &-5.3\,(5.1)   &-2.3\,(5.1)   &4.1\,(6.9)    &15.4\,(6.9)   &-0.3 &0.36 &[m] imprec.                                                          \\
755  & 21:42:45.73 &  57:20:13.4$^{\rm r}$& 871  & 3871                      &                               &                 &                 &   13.8$^{\rm j}$&       &                 &12.427\,(26) &12.202\,(31) & 12.080\,(23)$^{\rm r}$&                    &                 &               &-5.8\,(4.1)   &0.1\,(4.1)    &6.1\,(6.9)    &12.2\,(6.9)   &0.28 &0.01 &                                                                                      \\
756  & 21:42:30.58 &  57:20:13.3$^{\rm r}$& 872  & 3872                      &                               &                 &                 &   12.6$^{\rm j}$&       &                 &11.167\,(27) &10.821\,(31) & 10.702\,(23)$^{\rm r}$&                    &                 &               &-10.9\,(4.1)  &-0.8\,(4.1)   &-5.5\,(1.5)   &-0.8\,(2.1)   &0.15 &-0.04&                                                                                      \\
757  & 21:42:38.71 &  57:21:12.1$^{\rm r}$& 873  & 3873\footnote{also 5112}  &                               &  16.39$^{\rm k}$&  14.94$^{\rm k}$&   13.9$^{\rm j}$&       &                 &12.272\,(26) &11.994\,(32) & 11.886\,(24)$^{\rm r}$&        F4$^{\rm q}$&                 &               &-6.5\,(4.1)   &5.2\,(4.1)    &6\,(6.9)      &25.4\,(6.9)   &0.29 &0.42 &                                                                                      \\
758  & 21:42:58.64 &  57:21:38.5$^{\rm r}$& 874  & 3874                      &                               &                 &                 &   13.5$^{\rm j}$&       &                 &11.284\,(46) &10.828\,(52) & 10.682\,(50)$^{\rm r}$&                    &                 &               &9.3\,(4.1)    &1.2\,(4.1)    &47.1\,(7.1)   &23.5\,(7.1)   &-0.51&0.55 &                                                                                      \\
759  & 21:43:01.72 &  57:23:27.6$^{\rm r}$& 875  & 3875                      &                               &                 &                 &   14.6$^{\rm j}$&       &                 &10.817\,(27) &10.062\,(29) & 9.815 \,(21)$^{\rm r}$&                    &                 &               &-5.5\,(5.1)   &-4.4\,(5.1)   &-3.2\,(6.9)   &10.8\,(6.9)   &-0.39&-0.23&                                                                                      \\
760  & 21:43:01.75 &  57:23:54.4$^{\rm j}$& 876  & 3876                      &                               &                 &                 &   14.6$^{\rm j}$&       &                 &             &             &                       &                    &                 &               &-5.6\,(5.7)   &-2.5\,(5.7)   &92.5\,(7.1)   &47.9\,(7.1)   &-0.11&-0.32&                                                                                      \\
761  & 21:43:04.80 &  57:24:01.1$^{\rm r}$& 877  & 3877                      &                               &                 &                 &   13.3$^{\rm j}$&       &                 &11.763\,(27) &11.363\,(31) & 11.288\,(21)$^{\rm r}$&                    &                 &               &-16\,(10)     &-10.5\,(10)   &-4.1\,(6.9)   &9.7\,(6.9)    &0.81 &-0.82&                                                                                      \\
762  & 21:43:06.74 &  57:24:15.6$^{\rm r}$& 878  & 3878                      &                               &                 &                 &   12.4$^{\rm j}$&       &                 &11.188\,(25) &10.895\,(31) & 10.807\,(19)$^{\rm r}$&                    &                 &               &-10.4\,(4.1)  &3.3\,(4.1)    &-11.4\,(0.9)  &3.4\,(1.5)    &0.65 &0.23 &                                                                                      \\
763  & 21:43:09.50 &  57:24:54.2$^{\rm r}$& 879  & 3879                      &                               &                 &                 &   14.1$^{\rm j}$&       &                 &12.176\,(27) &11.742\,(31) & 11.643\,(19)$^{\rm r}$&                    &                 &               &24\,(3.9)     &-2.6\,(3.9)   &45\,(6.9)     &15.8\,(6.9)   &-3.06&0.46 &                                                                                      \\
764  & 21:42:49.20 &  57:26:09.0$^{\rm r}$& 880  & 3880                      &                               &                 &                 &   12.9$^{\rm j}$&       &                 &11.836\,(27) &11.702\,(31) & 11.578\,(23)$^{\rm r}$&                    &                 &               &-1.8\,(3.9)   &-0.5\,(3.9)   &-2.9\,(7.1)   &14.1\,(7)     &-0.37&0.31 &                                                                                      \\
765  & 21:43:06.62 &  57:28:22.3$^{\rm r}$& 881  & 506                       &                               &                 &                 &   10.4$^{\rm j}$&       &                 &9.657 \,(27) &9.486 \,(31) & 9.437 \,(19)$^{\rm r}$&        F0$^{\rm q}$&                 &               &11.1\,(1.6)   &9.5\,(1.6)    &12.4\,(1.5)   &6.1\,(0.8)    &-1.48&1.1  &                                                                                      \\
766  & 21:42:59.84 &  57:28:11.3$^{\rm r}$& 882  & 3882                      &                               &                 &                 &   14.2$^{\rm j}$&       &                 &12.441\,(29) &12.082\,(31) & 11.965\,(24)$^{\rm r}$&                    &                 &               &-6.8\,(3.8)   &-7\,(3.8)     &-7.3\,(7.4)   &-19.4\,(7.4)  &0.3  &-0.84&                                                                                      \\
767  & 21:42:50.18 &  57:28:32.1$^{\rm r}$& 883  & 502                       &                               &                 &                 &   10.7$^{\rm j}$&       &                 &8.609 \,(24) &8.121 \,(27) & 7.871 \,(26)$^{\rm r}$&                    &                 &               &-9.8\,(2.7)   &-2.3\,(2.7)   &0.4\,(1.1)    &-7\,(0.8)     &0.01 &-0.37&                                                                                      \\
768  & 21:42:45.29 &  57:29:23.6$^{\rm r}$& 884  & 3884                      &                               &                 &                 &   12.9$^{\rm j}$&       &                 &11.238\,(26) &10.827\,(32) & 10.747\,(19)$^{\rm r}$&                    &                 &               &-4.1\,(3.9)   &-10\,(3.9)    &-4.6\,(7.3)   &-23.5\,(7.3)  &-0.09&-0.51&                                                                                      \\
769  & 21:42:29.79 &  57:29:26.2$^{\rm r}$& 885  & 3885                      &                               &                 &                 &   12.4$^{\rm j}$&       &                 &11.465\,(27) &11.338\,(32) & 11.246\,(24)$^{\rm r}$&                    &                 &               &-1.6\,(3.9)   &-0.3\,(3.9)   &3.6\,(7.4)    &1\,(7.4)      &-0.32&0.56 &                                                                                      \\
770  & 21:42:36.90 &  57:30:04.3$^{\rm j}$& 886  & 3886                      &                               &                 &                 &   14.2$^{\rm j}$&       &                 &             &             &                       &                    &                 &               &              &              &              &              &     &     &no star                                                                          \\
771  & 21:42:35.52 &  57:30:57.5$^{\rm r}$& 888  & 3888                      &                               &                 &                 &   14.6$^{\rm j}$&       &                 &12.821\,(27) &12.533\,(32) & 12.419\,(28)$^{\rm r}$&                    &                 &               &-57\,(5.1)    &-8.9\,(5.1)   &-11.8\,(7.3)  &-25.1\,(7.3)  &     &     &                                                                                      \\
772  & 21:42:42.10 &  57:30:14.3$^{\rm r}$& 889  & 3889                      &                               &                 &                 &   13.3$^{\rm j}$&       &                 &10.365\,(27) &9.665 \,(37) & 9.424 \,(26)$^{\rm r}$&                    &                 &               &-19\,(11.6)   &22.7\,(11.6)  &-7.6\,(6.7)   &64\,(6.8)     &-0.4 &0.63 &                                                                                      \\
773  & 21:42:42.61 &  57:30:02.8$^{\rm r}$& 890  & 3890                      &                               &                 &                 &   14.5$^{\rm j}$&       &                 &9.144 \,(32) &7.981 \,(33) & 7.601 \,(27)$^{\rm r}$&                    &                 &               &6.9\,(4.9)    &-11.6\,(4.9)  &17.7\,(6.4)   &-20.9\,(6.5)  &0.3  &0.56 &                                                                                      \\
774  & 21:42:57.36 &  57:30:45.5$^{\rm r}$& 891  & 3891                      &                               &                 &                 &   12.7$^{\rm j}$&       &                 &9.917 \,(27) &9.184 \,(30) & 8.920 \,(21)$^{\rm r}$&                    &                 &               &-4.8\,(4.7)   &-5.7\,(4.7)   &-3.2\,(7.4)   &-7.5\,(7.4)   &0.25 &-0.14&                                                                                      \\
775  & 21:42:47.07 &  57:31:52.2$^{\rm r}$& 892  & 3892                      &                               &                 &                 &   13.7$^{\rm j}$&       &                 &11.970\,(27) &11.562\,(30) & 11.458\,(23)$^{\rm r}$&                    &                 &               &-19.7\,(11)   &-1.5\,(11)    &-3\,(7.3)     &-15.7\,(7.4)  &-0.07&-0.2 &                                                                                      \\
776  & 21:42:21.53 &  57:33:02.2$^{\rm r}$& 893  & 3893                      &                               &                 &                 &   14.7$^{\rm j}$&       &                 &12.509\,(24) &12.160\,(30) & 11.920\,(21)$^{\rm r}$&                    &                 &               &-10.3\,(3.8)  &4.7\,(3.8)    &-2.4\,(7.5)   &-8.1\,(7.3)   &0.15 &0.38 &                                                                                      \\
777  & 21:42:33.57 &  57:32:58.0$^{\rm r}$& 894  & 3894                      &                               &                 &                 &   14.3$^{\rm j}$&       &                 &12.530\,(26) &12.116\,(31) & 12.006\,(23)$^{\rm r}$&                    &                 &               &-18.8\,(3.9)  &-13.1\,(3.9)  &-22.5\,(7.3)  &-28.2\,(7.3)  &1.35 &-1.4 &                                                                                      \\
778  & 21:42:39.98 &  57:33:18.0$^{\rm r}$& 895  & 3895                      &                               &                 &                 &   13.3$^{\rm j}$&       &                 &10.251\,(26) &9.467 \,(31) & 9.215 \,(21)$^{\rm r}$&                    &                 &               &-15.2\,(10)   &10.8\,(10)    &-2.9\,(7.3)   &-12.3\,(7.4)  &0.28 &0.11 &                                                                                      \\
779  & 21:42:47.35 &  57:34:15.8$^{\rm r}$& 896  & 3896                      &                               &                 &                 &   14.6$^{\rm j}$&       &                 &12.580\,(30) &12.176\,(32) & 12.085\,(24)$^{\rm r}$&                    &                 &               &2.1\,(3.8)    &10.6\,(3.8)   &-6.9\,(7.3)   &14.1\,(7.3)   &-1.26&0.76 &2x[r]                                                                          \\
780  & 21:42:47.82 &  57:34:11.9$^{\rm r}$& 896  & 3896                      &                               &                 &                 &   14.6$^{\rm j}$&       &                 &15.398\,(72) &14.741\,(79) & 14.435\,(84)$^{\rm r}$&                    &                 &               &6\,(5.1)      &18.3\,(5.1)   &-6.9\,(7.3)   &14.1\,(7.3)   &-1.26&0.76 &2x[r] (faint)                                                               \\
781  & 21:42:55.08 &  57:33:45.1$^{\rm r}$& 897  & 3897                      &                               &                 &                 &   12.8$^{\rm j}$&       &                 &9.130 \,(37) &8.336 \,(63) & 7.975 \,(23)$^{\rm r}$&                    &                 &               &-16.5\,(4.7)  &9\,(4.7)      &7.5\,(7.3)    &-3.3\,(7.3)   &0.09 &0.02 &                                                                                      \\
782  & 21:43:07.63 &  57:34:49.1$^{\rm r}$& 898  & 505                       &                               &  12.75$^{\rm l}$&  11.98$^{\rm l}$&  10.82$^{\rm l}$&       &                 &8.711 \,(46) &8.193 \,(59) & 8.026 \,(31)$^{\rm r}$&       G2:$^{\rm q}$&                 &               &-8.2\,(2)     &-2.5\,(2)     &-5\,(1.6)     &-3.5\,(1)     &0.43 &-0.22&                                                                                      \\
783  & 21:42:54.78 &  57:36:11.0$^{\rm r}$& 899  & 3899                      &                               &                 &                 &   13.3$^{\rm j}$&       &                 &11.712\,(27) &11.407\,(32) & 11.271\,(24)$^{\rm r}$&                    &                 &               &0.6\,(3.8)    &-1.7\,(3.8)   &              &              &-0.1 &0    &                                                                                      \\
784  & 21:42:57.79 &  57:36:36.2$^{\rm r}$& 900  & 3900                      &                               &                 &                 &   14.1$^{\rm j}$&       &                 &12.404\,(30) &12.045\,(35) & 11.965\,(30)$^{\rm r}$&                    &                 &               &15.6\,(3.8)   &3.6\,(3.8)    &46.3\,(7.2)   &14.2\,(7.3)   &-1.73&0.64 &                                                                                      \\
785  & 21:42:53.00 &  57:37:05.2$^{\rm r}$& 901  & 503                       &                               &                 &                 &     10$^{\rm j}$&       &                 &6.880 \,(23) &6.021 \,(38) & 5.804 \,(23)$^{\rm r}$&                    &                 &               &14.9\,(2.7)   &29.4\,(2.7)   &10.7\,(1)     &16.9\,(1.3)   &-1.08&1.77 &                                                                                      \\
786  & 21:43:07.13 &  57:37:45.7$^{\rm r}$& 902  & 3902                      &                               &                 &                 &     14$^{\rm j}$&       &                 &10.746\,(29) &9.939 \,(32) & 9.712 \,(19)$^{\rm r}$&                    &                 &               &-3.9\,(4.7)   &-4.5\,(4.7)   &-2.5\,(7.3)   &-6.5\,(7.3)   &0.27 &-0.32&                                                                                      \\
787  & 21:42:40.71 &  57:37:39.7$^{\rm r}$& 903  & 501                       &                               &                 &                 &    9.9$^{\rm j}$&       &                 &7.626 \,(26) &7.031 \,(26) & 6.891 \,(21)$^{\rm r}$&       gG8$^{\rm q}$&                 &               &-13\,(2)      &3\,(2)        &-10.3\,(0.7)  &-5.4\,(1.4)   &1    &-0.34&                                                                                      \\
788  & 21:43:02.88 &  57:40:21.5$^{\rm r}$& 906  & 3906                      &                               &                 &                 &   13.4$^{\rm j}$&       &                 &11.955\,(33) &11.700\,(40) & 11.569\,(26)$^{\rm r}$&                    &                 &               &-3.7\,(3.8)   &-11.7\,(3.8)  &1.8\,(7.4)    &-31.8\,(7.4)  &0.27 &-0.84&                                                                                      \\
789  & 21:42:59.67 &  57:42:42.6$^{\rm r}$& 907  & 3907                      &                               &                 &                 &   13.3$^{\rm j}$&       &                 &11.906\,(27) &11.653\,(30) & 11.488\,(18)$^{\rm r}$&                    &                 &               &-0.6\,(3.8)   &-2.1\,(3.8)   &11.4\,(7.3)   &1.5\,(7.3)    &0.39 &-0.27&                                                                                      \\
790  & 21:42:53.98 &  57:43:47.5$^{\rm r}$& 908  & 752                       &                               &                 &                 &   11.6$^{\rm j}$&       &                 &10.977\,(29) &10.823\,(32) & 10.732\,(19)$^{\rm r}$&                    &                 &               &-7.3\,(2)     &-4.7\,(2)     &-8\,(0.6)     &-5.2\,(1)     &0.74 &-0.44&                                                                                      \\
791  & 21:42:46.96 &  57:44:38.2$^{\rm r}$& 909  & 3909                      &                               &                 &                 &   12.3$^{\rm j}$&       &                 &11.340\,(25) &11.156\,(31) & 11.071\,(23)$^{\rm r}$&                    &                 &               &-4.8\,(3.9)   &-3.8\,(3.9)   &-3.3\,(3.5)   &-1\,(1.5)     &0.19 &-0.07&                                                                                      \\
792  & 21:42:43.14 &  57:45:22.8$^{\rm r}$& 910  & 3910                      &                               &  13.47$^{\rm l}$&  12.74$^{\rm l}$&  11.67$^{\rm l}$&       &                 &9.377 \,(27) &8.803 \,(32) & 8.609 \,(21)$^{\rm r}$&                    &                 &               &-3.5\,(2.7)   &10.6\,(2.7)   &3.7\,(0.9)    &-0.4\,(2.2)   &-0.35&0.16 &                                                                                      \\
793  & 21:42:35.60 &  57:46:18.0$^{\rm r}$& 911  & 3911                      &                               &                 &                 &   11.9$^{\rm j}$&       &                 &11.518\,(26) &11.376\,(31) & 11.365\,(24)$^{\rm r}$&                    &                 &               &-11.3\,(3.8)  &-7.1\,(3.8)   &-6.2\,(0.6)   &-6.1\,(0.8)   &0.68 &-0.22&                                                                                      \\
794  & 21:42:31.75 &  57:46:39.2$^{\rm r}$& 912  & 3912                      &                               &                 &                 &     12$^{\rm j}$&       &                 &11.143\,(34) &10.842\,(40) & 10.767\,(30)$^{\rm r}$&                    &                 &               &-47.7\,(13.4) &40.9\,(13.4)  &-1.6\,(9.2)   &9.7\,(15.9)   &-0.29&0.76 &                                                                                      \\
795  & 21:43:35.45 &  57:14:20.3$^{\rm r}$& 914  & 3914                      &                               &                 &                 &   12.8$^{\rm j}$&       &                 &11.211\,(29) &10.872\,(32) & 10.770\,(24)$^{\rm r}$&                    &                 &               &-4\,(4.1)     &-10.1\,(4.1)  &6.5\,(7.3)    &1.1\,(7.3)    &0.47 &-0.58&                                                                                      \\
796  & 21:43:37.03 &  57:14:26.0$^{\rm r}$& 915  & 3915                      &                               &                 &                 &   12.7$^{\rm j}$&       &                 &11.268\,(27) &10.975\,(31) & 10.858\,(21)$^{\rm r}$&                    &                 &               &-1.7\,(4.1)   &-1\,(4.1)     &3.9\,(7.1)    &8.9\,(7.1)    &0.25 &0.11 &                                                                                      \\
797  & 21:43:49.21 &  57:15:22.1$^{\rm r}$& 916  & 3916                      &                               &                 &                 &     14$^{\rm j}$&       &                 &12.189\,(23) &11.930\,(29) & 11.800\,(21)$^{\rm r}$&                    &                 &               &-5.4\,(4.1)   &-6.4\,(4.1)   &1.2\,(6.9)    &5.4\,(6.9)    &0.85 &-0.52&                                                                                      \\
798  & 21:44:06.64 &  57:15:52.5$^{\rm r}$& 917  & 3917                      &                               &                 &                 &   12.1$^{\rm j}$&       &                 &9.411 \,(21) &8.687 \,(49) & 8.450 \,(19)$^{\rm r}$&                    &                 &               &-4.3\,(5.1)   &-8.9\,(5.1)   &-1.8\,(7.5)   &6\,(7.5)      &0.16 &-0.48&                                                                                      \\
799  & 21:44:14.98 &  57:15:20.7$^{\rm r}$& 918  & 3918                      &                               &                 &                 &     12$^{\rm j}$&       &                 &10.757\,(21) &10.422\,(29) & 10.296\,(19)$^{\rm r}$&                    &                 &               &-26\,(3.1)    &2.7\,(3.1)    &-16.8\,(0.6)  &-4.4\,(0.8)   &1.61 &-0.46&                                                                                      \\
800  & 21:44:17.33 &  57:15:14.0$^{\rm r}$& 919  & 3919                      &                               &                 &                 &   13.3$^{\rm j}$&       &                 &11.583\,(23) &11.291\,(31) & 11.166\,(23)$^{\rm r}$&                    &                 &               &15.1\,(4.1)   &-23\,(4.1)    &58.1\,(7.4)   &-83\,(7.4)    &-0.81&0.25 &                                                                                      \\
801  & 21:44:17.86 &  57:14:13.0$^{\rm r}$& 920  & 3920                      &                               &                 &                 &   13.7$^{\rm j}$&       &                 &11.947\,(24) &11.711\,(35) & 11.601\,(28)$^{\rm r}$&                    &                 &               &17.9\,(19)    &31.5\,(19)    &              &              &0.28 &-0.07&                                                                                      \\
802  & 21:44:38.04 &  57:16:28.2$^{\rm r}$& 921  & 512                       &                               &                 &                 &   10.6$^{\rm j}$&       &                 &9.955 \,(32) &9.754 \,(37) & 9.618 \,(26)$^{\rm r}$&        A5$^{\rm q}$&                 &               &9.1\,(1.8)    &3.3\,(1.8)    &5.2\,(0.8)    &1.8\,(1.1)    &-0.17&0.95 &                                                                                      \\
803  & 21:44:18.76 &  57:17:39.7$^{\rm r}$& 922  & 3922                      &                               &                 &                 &   12.5$^{\rm j}$&       &                 &10.191\,(23) &9.805 \,(31) & 9.571 \,(19)$^{\rm r}$&                    &                 &               &-4.6\,(5.1)   &-2.2\,(5.1)   &-9.3\,(7.3)   &20.7\,(7.3)   &0.4  &0.12 &                                                                                      \\
804  & 21:43:38.07 &  57:19:19.6$^{\rm r}$& 923  & 3923                      &                               &                 &                 &     14$^{\rm j}$&       &                 &12.315\,(23) &12.064\,(33) & 11.908\,(21)$^{\rm r}$&                    &                 &               &-5.9\,(4.1)   &-2.3\,(4.1)   &8\,(6.9)      &8.5\,(6.9)    &0.22 &-0.12&                                                                                      \\
805  & 21:44:13.93 &  57:20:04.9$^{\rm r}$& 924  & 3924                      &                               &                 &                 &   12.7$^{\rm j}$&       &                 &11.592\,(24) &11.367\,(31) & 11.231\,(23)$^{\rm r}$&                    &                 &               &-8.3\,(13.2)  &5.1\,(13.2)   &3.7\,(7.1)    &26.6\,(7.1)   &0.64 &-0.11&                                                                                      \\
806  & 21:44:11.71 &  57:20:16.7$^{\rm r}$& 925  & 3925                      &                               &                 &                 &   14.3$^{\rm j}$&       &                 &12.416\,(23) &12.122\,(31) & 11.977\,(24)$^{\rm r}$&                    &                 &               &-6.9\,(4.1)   &-2.4\,(4.1)   &3.5\,(6.8)    &16.4\,(6.8)   &-0.14&-0.23&                                                                                      \\
807  & 21:44:34.95 &  57:20:09.8$^{\rm r}$& 926  & 3926                      &                               &                 &                 &   13.1$^{\rm j}$&       &                 &11.040\,()   &11.093\,(68) & 10.941\,(64)$^{\rm r}$&                    &                 &               &14.1\,(18.3)  &-15.4\,(18.3) &15.4\,(7.1)   &-16.4\,(7.1)  &0.32 &0.07 &                                                                                      \\
808  & 21:44:34.10 &  57:20:32.9$^{\rm r}$& 927  & 3927                      &                               &                 &                 &   14.2$^{\rm j}$&       &                 &12.476\,(38) &12.135\,(37) & 12.064\,(30)$^{\rm r}$&                    &                 &               &-4.3\,(4.1)   &7.3\,(4.1)    &15.1\,(7)     &51.2\,(7)     &-0.15&0.35 &                                                                                      \\
809  & 21:43:33.30 &  57:22:15.7$^{\rm r}$& 928  & 3928                      &                               &                 &                 &   13.8$^{\rm j}$&       &                 &11.914\,(27) &11.495\,(31) & 11.414\,(21)$^{\rm r}$&                    &                 &               &-0.2\,(4.1)   &5\,(4.1)      &11.2\,(7)     &15.2\,(7)     &-0.63&0.43 &                                                                                      \\
810  & 21:43:48.82 &  57:24:29.9$^{\rm r}$& 929  & 3929                      &                               &                 &                 &   14.2$^{\rm j}$&       &                 &12.507\,(24) &12.261\,(33) & 12.115\,(24)$^{\rm r}$&                    &                 &               &-4\,(4.1)     &-4.2\,(4.1)   &-2.5\,(7)     &24.7\,(7)     &0.32 &-0.02&                                                                                      \\
811  & 21:44:11.72 &  57:23:34.1$^{\rm r}$& 930  & 3930                      &                               &                 &                 &   12.9$^{\rm j}$&       &                 &11.424\,(24) &11.071\,(31) & 11.003\,(23)$^{\rm r}$&                    &                 &               &2.6\,(4.1)    &-12\,(4.1)    &29.2\,(7.1)   &-21.3\,(7.1)  &0.44 &-0.17&                                                                                      \\
812  & 21:44:20.60 &  57:22:25.1$^{\rm r}$& 931  & 3931                      &                               &                 &                 &   14.1$^{\rm j}$&       &                 &12.494\,(26) &12.243\,(35) & 12.096\,(26)$^{\rm r}$&                    &                 &               &-3\,(4.1)     &1.4\,(4.1)    &5.1\,(6.9)    &21.4\,(6.8)   &-0.16&0.02 &                                                                                      \\
813  & 21:44:33.25 &  57:22:44.5$^{\rm r}$& 932  & 3932                      &                               &                 &                 &   13.9$^{\rm j}$&       &                 &10.766\,(27) &10.083\,(34) & 9.873 \,(20)$^{\rm r}$&                    &                 &               &-6\,(5.1)     &-4.9\,(5.1)   &5.8\,(7.1)    &6.6\,(7)      &0.04 &-0.22&                                                                                      \\
814  & 21:44:37.97 &  57:22:41.5$^{\rm j}$& 933  & 3933                      &                               &                 &                 &     14$^{\rm j}$&       &                 &             &             &                       &                    &                 &               &              &              &              &              &     &     &no star                                                                          \\
815  & 21:44:09.86 &  57:25:18.0$^{\rm r}$& 935  & 3935                      &                               &                 &                 &   13.7$^{\rm j}$&       &                 &12.078\,(23) &11.886\,(31) & 11.668\,(21)$^{\rm r}$&                    &                 &               &-3.4\,(4.1)   &-0.1\,(4.1)   &9.9\,(6.9)    &15.4\,(6.9)   &-0.42&0.04 &                                                                                      \\
816  & 21:44:03.07 &  57:26:19.2$^{\rm r}$& 936  & 509                       &                               &  11.09$^{\rm l}$&  10.93$^{\rm l}$&  10.47$^{\rm l}$&       &                 &11.025\,()   &13.068\,(88) & 12.772\,(58)$^{\rm r}$&        A0$^{\rm q}$&                 &               &              &              &-1.6\,(1.5)   &-7.6\,(1)     &0.09 &-0.02&2x[r]                                                                          \\
817  & 21:44:02.57 &  57:26:21.8$^{\rm r}$& 936  & 509                       &                               &  11.09$^{\rm l}$&  10.93$^{\rm l}$&  10.47$^{\rm l}$&       &                 &9.493 \,(26) &9.434 \,(33) & 9.341 \,(28)$^{\rm r}$&        A0$^{\rm q}$&                 &               &4.4\,(1.3)    &-10.3\,(1.3)  &-1.6\,(1.5)   &-7.6\,(1)     &0.09 &-0.02&2x[r]                                                                           \\
818  & 21:44:13.04 &  57:27:22.6$^{\rm r}$& 937  & 3937                      &                               &                 &                 &   12.8$^{\rm j}$&       &                 &11.758\,(24) &11.555\,(30) & 11.460\,(23)$^{\rm r}$&                    &                 &               &-5.7\,(4.1)   &-6.1\,(4.1)   &-1.5\,(7.4)   &-15.3\,(7.4)  &-0.22&-0.36&                                                                                      \\
819  & 21:44:35.70 &  57:27:12.5$^{\rm r}$& 938  & 511                       &                               &                 &                 &    9.1$^{\rm j}$&       &                 &6.500 \,(26) &5.973 \,(51) & 5.882 \,(21)$^{\rm r}$&       gG8$^{\rm q}$&                 &               &23.2\,(1.2)   &14.7\,(1.2)   &25\,(0.7)     &13.2\,(1.1)   &-2.13&2.01 &                                                                                      \\
820  & 21:43:19.72 &  57:29:52.3$^{\rm r}$& 939  & 3939                      &                               &                 &                 &   13.3$^{\rm j}$&       &                 &10.358\,(25) &9.589 \,(29) & 9.373 \,(19)$^{\rm r}$&                    &                 &               &-5\,(4.7)     &-8.6\,(4.7)   &-1.2\,(7.3)   &-27.9\,(7.3)  &0.55 &-0.38&                                                                                      \\
821  & 21:43:27.23 &  57:28:54.8$^{\rm r}$& 940  & 3940                      &                               &                 &                 &   14.4$^{\rm j}$&       &                 &12.511\,(25) &12.228\,(32) & 12.095\,(24)$^{\rm r}$&                    &                 &               &-2.1\,(3.8)   &-8.2\,(3.8)   &2.7\,(7.3)    &-16\,(7.5)    &0.05 &-0.37&                                                                                      \\
822  & 21:43:20.07 &  57:31:35.5$^{\rm r}$& 941  & 3941                      &                               &                 &                 &     13$^{\rm j}$&       &                 &11.745\,(25) &11.509\,(31) & 11.378\,(21)$^{\rm r}$&                    &                 &               &-3.8\,(3.8)   &-3.7\,(3.8)   &0.6\,(7.3)    &-5.8\,(7.4)   &0.19 &0.06 &                                                                                      \\
823  & 21:43:15.49 &  57:32:10.9$^{\rm r}$& 942  & 3942                      &                               &                 &                 &   14.2$^{\rm j}$&       &                 &8.477 \,(30) &7.235 \,(69) & 6.790 \,(20)$^{\rm r}$&                    &                 &               &-202.6\,(7.2) &-64.8\,(7.2)  &-10.9\,(7.2)  &14.1\,(7.2)   &0.2  &-0.2 &                                                                                      \\
824  & 21:43:59.05 &  57:29:34.2$^{\rm r}$& 943  & 5123                      &                               &                 &                 &   14.6$^{\rm j}$&       &                 &12.317\,(32) &11.896\,(43) & 11.752\,(30)$^{\rm r}$&                    &                 &               &-33.2\,(7.1)  &-13.4\,(7.1)  &-69.4\,(7.3)  &-70.5\,(7.4)  &-0.2 &0.4  &                                                                                      \\
825  & 21:43:37.45 &  57:32:22.6$^{\rm r}$& 944  & 3944                      &                               &                 &                 &   14.2$^{\rm j}$&       &                 &12.683\,(32) &12.408\,(36) & 12.271\,(32)$^{\rm r}$&                    &                 &               &-11.1\,(3.9)  &-3.5\,(3.9)   &3.4\,(7.3)    &-17.9\,(7.3)  &0.2  &-0.35&                                                                                      \\
826  & 21:43:39.80 &  57:33:11.7$^{\rm r}$& 945  & 507                       &                               &                 &                 &   11.9$^{\rm j}$&       &                 &11.472\,(24) &11.415\,(32) & 11.303\,(19)$^{\rm r}$&        A0$^{\rm q}$&                 &               &8.8\,(3.8)    &-14.2\,(3.8)  &-4.6\,(6.2)   &-7.3\,(12.9)  &0.25 &-0.36&                                                                                      \\
827  & 21:43:17.82 &  57:34:30.8$^{\rm r}$& 946  & 3946                      &                               &                 &                 &   12.1$^{\rm j}$&       &                 &9.558 \,(25) &8.848 \,(33) & 8.655 \,(19)$^{\rm r}$&                    &                 &               &-14.4\,(10)   &13.2\,(10)    &-6.5\,(7.5)   &-3.4\,(7.5)   &0.43 &0.49 &                                                                                      \\
828  & 21:43:25.74 &  57:34:43.3$^{\rm r}$& 947  & 3947                      &                               &                 &                 &   14.5$^{\rm j}$&       &                 &12.512\,(29) &12.148\,(36) & 12.037\,(28)$^{\rm r}$&                    &                 &               &-53.2\,(5.1)  &44.8\,(5.1)   &              &              &0.12 &0.49 &                                                                                      \\
829  & 21:43:44.29 &  57:34:07.0$^{\rm r}$& 948  & 3948                      &                               &                 &                 &   13.5$^{\rm j}$&       &                 &11.941\,(24) &11.569\,(33) & 11.404\,(23)$^{\rm r}$&                    &                 &               &-30.6\,(5.2)  &6.7\,(5.2)    &-7\,(7.4)     &-13.9\,(7.5)  &0.09 &-0.6 &                                                                                      \\
830  & 21:43:53.79 &  57:35:59.1$^{\rm r}$& 949  & 3949                      &                               &                 &                 &   12.9$^{\rm j}$&       &                 &10.693\,(21) &10.250\,(30) & 10.060\,(19)$^{\rm r}$&                    &                 &               &-4.5\,(4)     &0.7\,(4)      &1.1\,(7.4)    &-6.8\,(7.4)   &-0.31&0.13 &                                                                                      \\
831  & 21:43:56.70 &  57:35:55.1$^{\rm r}$& 950  & 508                       &                               &                 &                 &   11.3$^{\rm j}$&       &                 &10.340\,(21) &10.091\,(30) & 9.970 \,(21)$^{\rm r}$&                    &                 &               &1.8\,(2)      &-8.9\,(2)     &-9.6\,(1.4)   &-7.4\,(1.4)   &0.95 &-0.75&                                                                                      \\
832  & 21:44:17.90 &  57:34:44.2$^{\rm r}$& 952  & 3952                      &                               &                 &                 &     14$^{\rm j}$&       &                 &12.158\,(23) &11.809\,(31) & 11.682\,(18)$^{\rm r}$&                    &                 &               &-6.7\,(3.8)   &13.2\,(3.8)   &13.7\,(7.3)   &-0.4\,(7.4)   &-1.3 &1.05 &                                                                                      \\
833  & 21:44:31.31 &  57:33:30.0$^{\rm r}$& 953  & 3953                      &                               &                 &                 &   14.4$^{\rm j}$&       &                 &11.846\,(25) &11.441\,(30) & 11.272\,(22)$^{\rm r}$&                    &                 &               &-11.5\,(3.9)  &-4.6\,(3.9)   &3\,(7.2)      &-16.6\,(7.3)  &0.34 &-0.02&                                                                                      \\
834  & 21:44:32.71 &  57:34:19.5$^{\rm r}$& 954  & 3954                      &                               &                 &                 &   14.3$^{\rm j}$&       &                 &12.133\,(27) &11.725\,(32) & 11.612\,(18)$^{\rm r}$&                    &                 &               &-3.7\,(4.1)   &-15.7\,(4.1)  &-0.9\,(7.3)   &-25.6\,(7.4)  &0.46 &-1.43&                                                                                      \\
835  & 21:44:36.08 &  57:35:01.4$^{\rm r}$& 955  & 3955                      &                               &                 &                 &   12.4$^{\rm j}$&       &                 &11.236\,(25) &10.891\,(32) & 10.857\,(22)$^{\rm r}$&                    &                 &               &-4.7\,(4.1)   &-15.2\,(4.1)  &1\,(7.4)      &-25.7\,(7.4)  &0.44 &-1.5 &                                                                                      \\
836  & 21:44:14.18 &  57:36:35.9$^{\rm r}$& 956  & 3956                      &                               &                 &                 &   13.5$^{\rm j}$&       &                 &11.541\,(23) &11.142\,(31) & 10.966\,(21)$^{\rm r}$&                    &                 &               &10.9\,(11.5)  &3.4\,(11.5)   &14.8\,(7.4)   &-7.1\,(7.4)   &-1.17&0.44 &                                                                                      \\
837  & 21:44:07.36 &  57:37:01.2$^{\rm r}$& 957  & 3957                      &                               &                 &                 &   14.5$^{\rm j}$&       &                 &12.317\,(23) &11.995\,(32) & 11.766\,(21)$^{\rm r}$&                    &                 &               &1.5\,(4)      &2.2\,(4)      &8.1\,(7.4)    &-1.2\,(7.3)   &-0.82&0.25 &                                                                                      \\
838  & 21:43:57.70 &  57:37:18.2$^{\rm r}$& 958  & 3958                      &                               &                 &                 &   14.4$^{\rm j}$&       &                 &11.750\,(21) &11.168\,(32) & 10.976\,(19)$^{\rm r}$&                    &                 &               &-11.7\,(4)    &0.1\,(4)      &-4.1\,(7.3)   &-7.8\,(7.4)   &0.34 &0.2  &                                                                                      \\
839  & 21:43:22.33 &  57:39:53.8$^{\rm r}$& 959  & 3959                      &                               &                 &                 &   12.5$^{\rm j}$&       &                 &11.848\,(27) &11.671\,(33) & 11.617\,(24)$^{\rm r}$&                    &                 &               &-20.2\,(11.6) &-7.2\,(11.6)  &-1\,(4)       &-2.9\,(2.3)   &0.21 &-0.4 &                                                                                      \\
840  & 21:43:58.34 &  57:39:13.6$^{\rm r}$& 960  & 3960                      &                               &                 &                 &   13.5$^{\rm j}$&       &                 &11.488\,(23) &11.054\,(31) & 10.886\,(23)$^{\rm r}$&                    &                 &               &-13.8\,(3.8)  &-6.2\,(3.8)   &-9.5\,(7.3)   &-20.4\,(7.3)  &0.67 &-0.91&                                                                                      \\
841  & 21:43:20.36 &  57:42:21.9$^{\rm r}$& 961  & 3961                      &                               &                 &                 &     13$^{\rm j}$&       &                 &10.127\,(27) &9.386 \,(31) & 9.156 \,(19)$^{\rm r}$&                    &                 &               &-1.3\,(4.7)   &-2.5\,(4.7)   &-0.5\,(7.5)   &-4.4\,(7.5)   &-0.04&-0.04&                                                                                      \\
842  & 21:43:24.69 &  57:42:41.2$^{\rm r}$& 962  & 3962                      &                               &                 &                 &   13.8$^{\rm j}$&       &                 &11.833\,(30) &11.465\,(32) & 11.276\,(21)$^{\rm r}$&                    &                 &               &24.6\,(19.8)  &-44.3\,(19.8) &              &              &0    &-0.13&                                                                                      \\
843  & 21:43:52.66 &  57:42:38.0$^{\rm j}$& 963  & 3963                      &                               &                 &                 &   13.2$^{\rm j}$&       &                 &             &             &                       &                    &                 &               &              &              &              &              &     &     &no star                                                                          \\
844  & 21:44:09.03 &  57:40:48.7$^{\rm r}$& 964  & 510                       &                               &                 &                 &   11.7$^{\rm j}$&       &                 &14.433\,(93) &13.523\,(66) & 13.076\,(50)$^{\rm r}$&        A5$^{\rm q}$&                 &               &              &              &3.2\,(2.7)    &2.3\,(1.4)    &-0.34&0.09 &2x[r] (faint)                                                               \\
845  & 21:44:08.56 &  57:40:52.2$^{\rm r}$& 964  & 510                       &                               &                 &                 &   11.7$^{\rm j}$&       &                 &10.709\,(28) &10.459\,(36) & 10.374\,(33)$^{\rm r}$&        A5$^{\rm q}$&                 &               &-8.3\,(2.7)   &-6\,(2.7)     &3.2\,(2.7)    &2.3\,(1.4)    &-0.34&0.09 &2x[r]                                                                           \\
846  & 21:44:15.67 &  57:41:09.0$^{\rm r}$& 965  & 3965                      &                               &                 &                 &   12.1$^{\rm j}$&       &                 &11.176\,(23) &10.889\,(31) & 10.742\,(21)$^{\rm r}$&                    &                 &               &41.5\,(14.3)  &20\,(14.3)    &34.5\,(1.7)   &29.9\,(1.5)   &-3.86&3.15 &                                                                                      \\
847  & 21:44:25.15 &  57:41:37.7$^{\rm r}$& 966  & 3966                      &                               &                 &                 &   10.9$^{\rm j}$&       &                 &8.682 \,(19) &8.052 \,(31) & 7.900 \,(21)$^{\rm r}$&                    &                 &               &-12.6\,(2.8)  &12.9\,(2.8)   &-4.8\,(1.1)   &-3.9\,(2.3)   &0.83 &-0.55&                                                                                      \\
848  & 21:44:21.52 &  57:44:14.3$^{\rm r}$& 967  & 763                       &                               &   9.33$^{\rm l}$&   9.32$^{\rm l}$&      9$^{\rm l}$&       &                 &8.247 \,(24) &8.146 \,(23) & 8.158 \,(29)$^{\rm r}$&        B8$^{\rm p}$&    III$^{\rm p}$&               &1.5\,(0.7)    &2\,(0.8)      &2\,(0.6)      &1.6\,(0.6)    &-0.01&-0.07&                                                                                      \\
849  & 21:43:11.02 &  57:35:42.8$^{\rm r}$& 968  & 3968                      &                               &                 &                 &     13$^{\rm j}$&       &                 &11.988\,(29) &11.811\,(33) & 11.718\,(21)$^{\rm r}$&                    &                 &               &-6.2\,(5.1)   &-3.3\,(5.1)   &-6.9\,(1.9)   &-8.1\,(1.6)   &0.55 &-0.74&                                                                                      \\
850  & 21:42:26.19 &  57:47:03.3$^{\rm r}$& 1000 & 4000                      &                               &                 &                 &   12.2$^{\rm j}$&       &                 &10.155\,(24) &9.518 \,(30) & 9.368 \,(23)$^{\rm r}$&                    &                 &               &-9.7\,(4.7)   &-0.1\,(4.7)   &7.5\,(6.8)    &9.9\,(6.8)    &0.32 &-0.1 &                                                                                      \\
851  & 21:40:07.50 &  57:46:49.0$^{\rm r}$& 1001 & 733                       &                               &                 &                 &   11.6$^{\rm j}$&       &                 &10.844\,(26) &10.562\,(30) & 10.521\,(18)$^{\rm r}$&        A3$^{\rm q}$&                 &               &-6.3\,(3.9)   &-5.2\,(3.9)   &-4.3\,(1)     &-6.4\,(0.9)   &-0.09&-0.19&                                                                                      \\
852  & 21:40:51.74 &  57:46:34.6$^{\rm r}$& 1002 & 4002                      &                               &                 &                 &   14.1$^{\rm j}$&       &                 &10.333\,(24) &9.394 \,(26) & 9.183 \,(23)$^{\rm r}$&                    &                 &               &-15.8\,(4.7)  &3.2\,(4.7)    &-107.2\,(7.6) &48.1\,(7.6)   &0.03 &0.35 &                                                                                      \\
853  & 21:39:47.30 &  57:46:48.5$^{\rm r}$& 1003 & 4003                      &                               &                 &                 &   13.2$^{\rm j}$&       &                 &10.410\,(24) &9.664 \,(28) & 9.488 \,(19)$^{\rm r}$&                    &                 &               &-3.3\,(4.7)   &-4.6\,(4.7)   &-4.5\,(7.7)   &-7.7\,(7.8)   &0.2  &-0.36&                                                                                      \\
854  & 21:41:50.36 &  57:47:04.4$^{\rm r}$& 1004 & 4004                      &                               &                 &                 &   12.5$^{\rm j}$&       &                 &8.676 \,(19) &7.681 \,(57) & 7.393 \,(34)$^{\rm r}$&                    &                 &               &-433\,(7.3)   &-287.3\,(6.9) &-7.9\,(7.1)   &-4.7\,(7.2)   &0.14 &-0.18&                                                                                      \\
855  & 21:38:48.16 &  57:46:51.2$^{\rm r}$& 1005 & 4005                      &                               &                 &                 &   12.9$^{\rm j}$&       &                 &10.205\,(22) &9.438 \,(28) & 9.227 \,(22)$^{\rm r}$&                    &                 &               &-1.5\,(4.7)   &-2.1\,(4.7)   &3.4\,(7.4)    &9.1\,(7.4)    &0.06 &0.16 &                                                                                      \\
856  & 21:38:42.32 &  57:46:50.9$^{\rm r}$& 1006 & 4006                      &                               &                 &                 &   11.9$^{\rm j}$&       &                 &10.911\,(24) &10.654\,(28) & 10.586\,(20)$^{\rm r}$&                    &                 &               &-2.7\,(2.7)   &-5\,(2.7)     &-7.1\,(1.2)   &-5.1\,(1.5)   &0.25 &-0.36&                                                                                      \\
857  & 21:38:30.29 &  57:46:26.6$^{\rm r}$& 1007 & 716                       &                               &                 &   9.96$^{\rm f}$&    9.5$^{\rm e}$&       &                 &8.310 \,(26) &8.132 \,(51) & 8.100 \,(23)$^{\rm r}$&        A7$^{\rm e}$&                 &  0.8$^{\rm e}$&-3.4\,(1.6)   &-4.2\,(1.6)   &-5.8\,(0.7)   &-7.1\,(1)     &-0.06&-0.4 &                                                                                      \\
858  & 21:37:58.30 &  57:46:43.3$^{\rm r}$& 1008 & 4008                      &                               &                 &                 &   12.4$^{\rm j}$&       &                 &9.895 \,(27) &9.188 \,(33) & 8.961 \,(21)$^{\rm r}$&                    &                 &               &-18.8\,(4.7)  &8.3\,(4.7)    &-15.4\,(8.5)  &12.3\,(8.6)   &0.96 &1.07 &                                                                                      \\
859  & 21:38:16.01 &  57:47:06.0$^{\rm r}$& 1009 & 4009                      &                               &                 &                 &   13.5$^{\rm j}$&       &                 &11.896\,(26) &11.544\,(31) & 11.434\,(24)$^{\rm r}$&                    &                 &               &-9.7\,(10.8)  &-29.4\,(10.8) &6.4\,(7.3)    &-49.4\,(7.4)  &-0.05&-3.08&                                                                                      \\
860  & 21:37:59.67 &  57:47:09.2$^{\rm r}$& 1010 & 4010                      &                               &                 &                 &   12.1$^{\rm j}$&       &                 &9.670 \,(26) &9.037 \,(29) & 8.900 \,(18)$^{\rm r}$&                    &                 &               &-0.6\,(4.7)   &-3.8\,(4.7)   &10.9\,(7.7)   &-2.9\,(7.8)   &-0.22&-0.92&                                                                                      \\
861  & 21:38:38.50 &  57:48:18.9$^{\rm r}$& 1011 & 4011                      &                               &                 &                 &   11.7$^{\rm j}$&       &                 &9.259 \,(27) &8.505 \,(47) & 8.296 \,(20)$^{\rm r}$&                    &                 &               &-0.5\,(11.3)  &-2.5\,(11.3)  &2.1\,(6.9)    &-34.7\,(7.1)  &-0.07&0.17 &                                                                                      \\
862  & 21:38:08.75 &  57:49:24.4$^{\rm r}$& 1012 & 4012                      &                               &                 &                 &   12.9$^{\rm j}$&       &                 &11.433\,(26) &11.163\,(32) & 11.027\,(21)$^{\rm r}$&                    &                 &               &-2.9\,(3.8)   &1.2\,(3.8)    &-5.6\,(7.5)   &3.2\,(7.5)    &0.03 &-0.13&                                                                                      \\
863  & 21:38:32.23 &  57:49:59.6$^{\rm r}$& 1014 & 4014                      &                               &                 &                 &   13.3$^{\rm j}$&       &                 &11.371\,(26) &11.039\,(31) & 10.836\,(19)$^{\rm r}$&                    &                 &               &10.2\,(4)     &0.1\,(4)      &14.9\,(7.4)   &5.7\,(7.5)    &-1.82&0.34 &                                                                                      \\
864  & 21:38:52.32 &  57:50:26.1$^{\rm r}$& 1015 & 4015                      &                               &                 &                 &   13.7$^{\rm j}$&       &                 &10.614\,(24) &9.855 \,(28) & 9.684 \,(22)$^{\rm r}$&                    &                 &               &-1\,(4.7)     &-1.7\,(4.7)   &-7.1\,(7.5)   &2.6\,(7.5)    &-0.21&0.21 &                                                                                      \\
865  & 21:38:53.28 &  57:51:19.6$^{\rm r}$& 1016 & 719                       &                               &                 &                 &   11.4$^{\rm j}$&       &                 &10.479\,(27) &10.215\,(33) & 10.099\,(22)$^{\rm r}$&        G5$^{\rm q}$&                 &               &-4.8\,(2)     &-13.1\,(1.9)  &-6.7\,(0.7)   &-12.9\,(0.6)  &0.28 &-1.13&                                                                                      \\
866  & 21:38:45.58 &  57:51:49.5$^{\rm r}$& 1017 & 4017                      &                               &                 &                 &   13.6$^{\rm j}$&       &                 &11.940\,(26) &11.657\,(33) & 11.526\,(25)$^{\rm r}$&                    &                 &               &3.1\,(3.8)    &-2.4\,(3.8)   &9.3\,(7.4)    &-12.8\,(7.4)  &-0.35&-0.09&                                                                                      \\
867  & 21:37:56.44 &  57:53:13.8$^{\rm r}$& 1018 & 4018                      &                               &                 &                 &   12.8$^{\rm j}$&       &                 &11.578\,(27) &11.299\,(31) & 11.192\,(19)$^{\rm r}$&                    &                 &               &-7.4\,(3.8)   &-3.6\,(3.8)   &-5\,(7.4)     &0.4\,(7.4)    &0.18 &-0.27&                                                                                      \\
868  & 21:38:07.83 &  57:55:21.8$^{\rm r}$& 1020 & 4020                      &                               &                 &                 &   13.3$^{\rm j}$&       &                 &12.240\,(26) &12.077\,(31) & 11.972\,(19)$^{\rm r}$&                    &                 &               &-0.7\,(3.8)   &-8.2\,(3.8)   &-9.4\,(0.7)   &-8.7\,(0.5)   &0.26 &-0.36&                                                                                      \\
869  & 21:38:18.76 &  57:55:16.1$^{\rm r}$& 1021 & 4021                      &                               &                 &                 &   13.7$^{\rm j}$&       &                 &11.951\,(26) &11.651\,(31) & 11.500\,(19)$^{\rm r}$&                    &                 &               &-8.3\,(3.8)   &-12.7\,(3.8)  &-8.8\,(7.3)   &-8.6\,(7.3)   &0.42 &-0.25&                                                                                      \\
870  & 21:38:45.24 &  57:54:41.7$^{\rm r}$& 1022 & 717                       &                               &                 &                 &   10.4$^{\rm j}$&       &                 &9.299 \,(22) &8.991 \,(27) & 8.952 \,(19)$^{\rm r}$&        F8$^{\rm q}$&                 &               &6.9\,(1.6)    &21.9\,(1.6)   &5.5\,(0.5)    &18.9\,(0.9)   &-1.29&2.11 &                                                                                      \\
871  & 21:38:53.88 &  57:52:34.9$^{\rm r}$& 1023 & 4023                      &                               &                 &                 &   13.8$^{\rm j}$&       &                 &11.868\,(24) &11.518\,(28) & 11.392\,(22)$^{\rm r}$&                    &                 &               &2.6\,(3.8)    &6.1\,(3.8)    &2.6\,(7.4)    &-1\,(7.5)     &-0.37&0.4  &                                                                                      \\
872  & 21:38:59.18 &  57:53:00.9$^{\rm r}$& 1024 & 4024                      &                               &                 &                 &   12.1$^{\rm j}$&       &                 &9.649 \,(22) &8.931 \,(30) & 8.767 \,(22)$^{\rm r}$&                    &                 &               &-6.5\,(4.7)   &6.3\,(4.7)    &12.9\,(7)     &-50.7\,(7.2)  &-0.1 &-0.25&                                                                                      \\
873  & 21:37:53.82 &  57:57:50.3$^{\rm r}$& 1025 & 4025                      &                               &                 &                 &     12$^{\rm j}$&       &                 &10.170\,(34) &9.623 \,(35) & 9.494 \,(24)$^{\rm r}$&                    &                 &               &-2.4\,(4.7)   &0.3\,(4.7)    &-9.1\,(7.3)   &24.1\,(7.3)   &-0.33&-0.8 &                                                                                      \\
874  & 21:38:19.41 &  57:58:01.1$^{\rm r}$& 1026 & 4026                      &                               &                 &                 &   11.8$^{\rm j}$&       &                 &9.495 \,(26) &8.838 \,(31) & 8.681 \,(18)$^{\rm r}$&                    &                 &               &-6.2\,(4.7)   &16\,(4.7)     &-1.1\,(8.6)   &19.6\,(8.7)   &0.01 &2.35 &                                                                                      \\
875  & 21:38:25.44 &  57:56:24.5$^{\rm r}$& 1027 & 4027                      &                               &                 &                 &   13.8$^{\rm j}$&       &                 &10.422\,(26) &9.662 \,(31) & 9.402 \,(21)$^{\rm r}$&                    &                 &               &1.7\,(4.7)    &-2.7\,(4.7)   &-3.8\,(7.8)   &21.5\,(7.8)   &0.02 &0.4  &                                                                                      \\
876  & 21:38:45.82 &  57:56:47.5$^{\rm r}$& 1028 & 4028                      &                               &                 &                 &   13.4$^{\rm j}$&       &                 &13.549\,(27) &13.100\,(36) & 12.923\,(34)$^{\rm r}$&                    &                 &               &45.1\,(5)     &-172.8\,(5)   &-2.1\,(7.6)   &-0.5\,(7.6)   &     &     &                                                                                      \\
877  & 21:38:00.09 &  57:59:36.6$^{\rm r}$& 1029 & 4029                      &                               &                 &                 &   13.1$^{\rm j}$&       &                 &11.251\,(26) &10.662\,(32) & 10.519\,(19)$^{\rm r}$&                    &                 &               &81.6\,(3.8)   &65.2\,(3.8)   &75.1\,(7.2)   &58.2\,(7.2)   &     &     &                                                                                      \\
878  & 21:38:20.11 &  57:59:31.8$^{\rm r}$& 1030 & 715                       &                               &                 &                 &   11.1$^{\rm j}$&       &                 &10.097\,(32) &9.864 \,(42) & 9.763 \,(28)$^{\rm r}$&        A0$^{\rm q}$&                 &               &-8.7\,(2)     &-13.1\,(2)    &-5.8\,(5.1)   &10.8\,(5.4)   &0.3  &-0.22&                                                                                      \\
879  & 21:38:37.74 &  57:58:19.7$^{\rm r}$& 1031 & 4031                      &                               &                 &                 &   13.7$^{\rm j}$&       &                 &10.947\,(29) &10.389\,(37) & 10.217\,(26)$^{\rm r}$&                    &                 &               &-20.2\,(3.8)  &-13\,(3.8)    &-79.5\,(8)    &-44.1\,(8)    &0.22 &-0.39&                                                                                      \\
880  & 21:38:44.49 &  57:58:32.2$^{\rm r}$& 1032 & 4032                      &                               &                 &                 &   12.8$^{\rm j}$&       &                 &10.317\,(22) &9.844 \,(30) & 9.660 \,(20)$^{\rm r}$&                    &                 &               &-4.3\,(4.7)   &0.7\,(4.7)    &-14.1\,(8.1)  &-0.7\,(8.1)   &-0.11&-0.13&                                                                                      \\
881  & 21:37:58.68 &  58:01:38.4$^{\rm r}$& 1033 & 4033                      &                               &                 &                 &   13.2$^{\rm j}$&       &                 &10.459\,(26) &9.717 \,(31) & 9.529 \,(21)$^{\rm r}$&                    &                 &               &-3\,(4.7)     &7.8\,(4.7)    &15.1\,(8.2)   &14.2\,(8.2)   &-0.66&1.43 &                                                                                      \\
882  & 21:38:54.76 &  57:58:36.1$^{\rm r}$& 1034 & 718                       &                               &                 &                 &   11.3$^{\rm j}$&       &                 &9.616 \,(24) &9.023 \,(28) & 8.915 \,(22)$^{\rm r}$&      dK3:$^{\rm q}$&                 &               &129\,(2.7)    &147.8\,(2.7)  &128\,         &150\,         &     &     &[j] imprec.                                                          \\
883  & 21:38:14.76 &  58:05:27.6$^{\rm r}$& 1035 & 4035                      &                               &                 &                 &   12.8$^{\rm j}$&       &                 &10.683\,(26) &10.230\,(30) & 10.101\,(19)$^{\rm r}$&                    &                 &               &-0.4\,(3.8)   &-6.8\,(3.8)   &1.8\,(7.2)    &-10.3\,(7.1)  &0.39 &-0.09&                                                                                      \\
884  & 21:38:45.13 &  58:04:30.2$^{\rm r}$& 1036 & 4036                      &                               &                 &                 &   12.4$^{\rm j}$&       &                 &11.271\,(22) &11.058\,(30) & 10.960\,(23)$^{\rm r}$&                    &                 &               &-0.3\,(3.8)   &-0.1\,(3.8)   &-8.6\,(1.5)   &-1.2\,(2.7)   &-0.02&0.02 &                                                                                      \\
885  & 21:39:14.07 &  57:48:04.8$^{\rm r}$& 1037 & 721                       &                               &  10.13$^{\rm l}$&  10.46$^{\rm l}$&  10.13$^{\rm l}$&       &                 &9.022 \,(24) &8.827 \,(28) & 8.624 \,(22)$^{\rm r}$&        B3$^{\rm p}$&      V$^{\rm p}$&               &-4.8\,(1.2)   &-3.7\,(1.3)   &-3.4\,(0.6)   &-2.5\,(0.9)   &-0.17&0.06 &                                                                                      \\
886  & 21:39:34.04 &  57:47:17.1$^{\rm r}$& 1038 & 4038                      &                               &                 &                 &   11.7$^{\rm j}$&       &                 &9.355 \,(26) &8.591 \,(28) & 8.419 \,(21)$^{\rm r}$&                    &                 &               &-7.4\,(4.7)   &-0.1\,(4.7)   &-9.5\,(2.7)   &-2.1\,(1)     &0.07 &0.07 &                                                                                      \\
887  & 21:39:20.66 &  57:49:37.2$^{\rm r}$& 1039 & 4039                      &                               &                 &                 &   11.8$^{\rm j}$&       &                 &10.868\,(24) &10.483\,(29) & 10.443\,(23)$^{\rm r}$&                    &                 &               &10.2\,(2.7)   &0.3\,(2.7)    &6.5\,(1.2)    &-1\,(0.8)     &-1.04&0    &                                                                                      \\
888  & 21:39:33.59 &  57:49:17.0$^{\rm r}$& 1040 & 4040                      &                               &                 &                 &   13.7$^{\rm j}$&       &                 &11.679\,()   &11.473\,()   & 11.474\,(36)$^{\rm r}$&                    &                 &               &10.5\,(3.8)   &5.7\,(3.8)    &41.2\,(7.4)   &33.7\,(7.4)   &0.44 &-0.15&                                                                                      \\
889  & 21:39:46.37 &  57:49:11.0$^{\rm r}$& 1041 & 4041                      &                               &                 &                 &   13.3$^{\rm j}$&       &                 &12.049\,(26) &11.795\,(28) & 11.713\,(23)$^{\rm r}$&                    &                 &               &-10.3\,(3.9)  &2.7\,(3.9)    &-5.7\,(7.5)   &1.3\,(7.5)    &-0.05&0.01 &                                                                                      \\
890  & 21:39:57.46 &  57:49:45.8$^{\rm r}$& 1042 & 4042                      &                               &                 &                 &   12.2$^{\rm j}$&       &                 &10.704\,(26) &10.309\,(30) & 10.224\,(23)$^{\rm r}$&                    &                 &               &-10.2\,(12.6) &-13.8\,(12.6) &-18.9\,(6.2)  &-9.7\,(7.2)   &1.48 &-0.7 &                                                                                      \\
891  & 21:39:58.73 &  57:49:52.3$^{\rm r}$& 1043 & 4043                      &                               &                 &                 &     12$^{\rm j}$&       &                 &11.406\,(34) &11.143\,()   &   11.114\,()$^{\rm r}$&                    &                 &               &-4.5\,(2.7)   &-1.6\,(2.7)   &-7.7\,(0.7)   &-3\,(1.7)     &0.1  &0.07 &                                                                                      \\
892  & 21:40:08.81 &  57:48:11.8$^{\rm r}$& 1044 & 734                       &                               &                 &                 &   11.8$^{\rm j}$&       &                 &11.306\,(26) &11.178\,(27) & 11.132\,(21)$^{\rm r}$&        A0$^{\rm q}$&                 &               &-6.9\,(2.7)   &1.5\,(2.7)    &-5.2\,(0.7)   &-3.5\,(1)     &-0.06&-0.02&                                                                                      \\
893  & 21:39:19.13 &  57:52:43.1$^{\rm r}$& 1045 & 4045                      &                               &                 &                 &   13.7$^{\rm j}$&       &                 &11.958\,(35) &11.667\,(44) & 11.519\,(34)$^{\rm r}$&                    &                 &               &-7.3\,(5.1)   &2.3\,(5)      &              &              &-0.26&-0.02&                                                                                      \\
894  & 21:39:21.36 &  57:52:40.1$^{\rm r}$& 1046 & 722                       &                               &                 &                 &    8.5$^{\rm j}$&       &                 &6.171 \,(19) &5.662 \,(33) & 5.582 \,(16)$^{\rm r}$&      g:G8$^{\rm q}$&                 &               &-18.5\,(1.2)  &-4.3\,(1.2)   &-15.3\,(0.6)  &-6.7\,(0.7)   &0.96 &-0.37&                                                                                      \\
895  & 21:39:28.27 &  57:51:18.0$^{\rm r}$& 1047 & 4047                      &                               &                 &                 &   12.4$^{\rm j}$&       &                 &11.488\,(22) &11.327\,(32) & 11.239\,(23)$^{\rm r}$&                    &                 &               &-6.8\,(2.7)   &-10.6\,(2.7)  &-6.7\,(1.3)   &-5.2\,(5.1)   &0.09 &-0.15&                                                                                      \\
896  & 21:39:36.80 &  57:52:43.0$^{\rm j}$& 1048 & 4048                      &                               &                 &                 &   12.5$^{\rm j}$&       &                 &             &             &                       &                    &                 &               &              &              &              &              &0.94 &1.07 &new coordinates, no star                                             \\
897  & 21:39:33.90 &  57:53:04.9$^{\rm r}$& 1049 & 4049                      &                               &                 &                 &   13.8$^{\rm j}$&       &                 &9.599 \,(23) &8.595 \,(21) & 8.364 \,(23)$^{\rm r}$&                    &                 &               &-3.1\,(4.8)   &-2.5\,(4.8)   &0.4\,(7.8)    &9.5\,(7.9)    &-0.05&0.11 &                                                                                      \\
898  & 21:39:31.41 &  57:53:30.1$^{\rm r}$& 1050 & 726                       &                               &                 &                 &   11.1$^{\rm j}$&       &                 &10.248\,(24) &9.948 \,(30) & 9.855 \,(22)$^{\rm r}$&        G2$^{\rm q}$&                 &               &-29.8\,(2)    &-27.4\,(2)    &-31.5\,(0.7)  &-24.7\,(0.9)  &2.75 &-1.88&                                                                                      \\
899  & 21:39:04.04 &  57:55:04.9$^{\rm r}$& 1051 & 4051                      &                               &                 &                 &   12.2$^{\rm j}$&       &                 &9.857 \,(22) &9.252 \,(30) & 9.066 \,(20)$^{\rm r}$&                    &                 &               &-0.7\,(4.7)   &0.3\,(4.7)    &-3\,(8.6)     &2.7\,(8.6)    &-0.39&0.83 &                                                                                      \\
900  & 21:39:46.08 &  57:55:40.8$^{\rm r}$& 1052 & 4052                      &                               &                 &                 &   12.9$^{\rm j}$&       &                 &9.883 \,(23) &9.161 \,(27) & 8.944 \,(20)$^{\rm r}$&                    &                 &               &-8.3\,(12.8)  &-7.7\,(12.8)  &-4.4\,(7.3)   &-5\,(7.3)     &-0.19&0.06 &                                                                                      \\
901  & 21:39:40.53 &  57:57:14.8$^{\rm r}$& 1053 & 4053                      &                               &                 &                 &   13.6$^{\rm j}$&       &                 &11.067\,(21) &10.434\,(28) & 10.078\,(20)$^{\rm r}$&                    &                 &               &-2.8\,(3.8)   &-2.4\,(3.8)   &-5.3\,(7.3)   &-2.1\,(7.3)   &-0.19&-0.26&                                                                                      \\
902  & 21:39:33.70 &  57:59:38.8$^{\rm r}$& 1054 & 4054                      &                               &                 &                 &   11.3$^{\rm j}$&       &                 &8.548 \,(29) &7.782 \,(17) & 7.623 \,(17)$^{\rm r}$&                    &                 &               &-0.7\,(2.8)   &-13.1\,(2.8)  &-5.3\,(2.5)   &-7.7\,(1.1)   &0.04 &-0.39&                                                                                      \\
903  & 21:39:06.72 &  58:01:00.2$^{\rm r}$& 1055 & 4055                      &                               &                 &                 &   13.7$^{\rm j}$&       &                 &10.055\,(22) &9.173 \,(28) & 8.896 \,(20)$^{\rm r}$&                    &                 &               &-1.4\,(4.7)   &-5.7\,(4.7)   &-6.6\,(7.5)   &-4.4\,(7.5)   &-0.46&0.32 &                                                                                      \\
904  & 21:39:19.32 &  58:01:02.1$^{\rm r}$& 1056 & 4056                      &                               &                 &                 &   13.8$^{\rm j}$&       &                 &10.589\,(22) &9.871 \,(28) & 9.658 \,(23)$^{\rm r}$&                    &                 &               &1.9\,(6)      &6.2\,(6)      &              &              &-0.02&0.26 &                                                                                      \\
905  & 21:39:20.20 &  58:01:48.3$^{\rm r}$& 1057 & 723                       &                               &                 &                 &   10.9$^{\rm j}$&       &                 &9.795 \,(22) &9.567 \,(30) & 9.459 \,(20)$^{\rm r}$&        A3$^{\rm q}$&                 &               &-14.1\,(2)    &-9.6\,(2)     &-12.4\,(1.4)  &-9.3\,(1.5)   &0.54 &-0.2 &                                                                                      \\
906  & 21:39:40.59 &  57:59:55.3$^{\rm r}$& 1058 & 4058                      &                               &                 &                 &   13.9$^{\rm j}$&       &                 &12.121\,(24) &11.815\,(30) & 11.695\,(22)$^{\rm r}$&                    &                 &               &-1.7\,(3.8)   &-4\,(3.8)     &-2.5\,(7.2)   &-14.3\,(7.3)  &0.07 &-0.26&                                                                                      \\
907  & 21:39:37.40 &  58:00:50.7$^{\rm r}$& 1059 & 4059                      &                               &                 &                 &   13.3$^{\rm j}$&       &                 &9.403 \,(23) &8.428 \,(65) & 8.137 \,(24)$^{\rm r}$&                    &                 &               &-2.7\,(4.7)   &-4\,(4.7)     &-5.9\,(8.4)   &-9.4\,(8.4)   &-0.28&0.13 &                                                                                      \\
908  & 21:40:02.95 &  57:56:16.2$^{\rm r}$& 1060 & 732                       &                               &                 &                 &   11.4$^{\rm j}$&       &                 &10.043\,(23) &9.836 \,(28) & 9.663 \,(20)$^{\rm r}$&        A0$^{\rm q}$&                 &               &-7.2\,(2.7)   &-15.9\,(2.7)  &-5.9\,(0.9)   &-6.5\,(2.5)   &0.07 &-0.07&                                                                                      \\
909  & 21:40:02.19 &  57:57:18.3$^{\rm r}$& 1061 & 731                       &                               &                 &   11.7$^{\rm f}$&   11.3$^{\rm e}$&       &                 &9.363 \,(23) &9.217 \,(27) & 9.070 \,(22)$^{\rm r}$&        A1$^{\rm e}$&                 &  1.2$^{\rm e}$&-6.4\,(1.7)   &-9.9\,(1.7)   &-8\,(0.8)     &-6.8\,(0.6)   &0.11 &-0.19&                                                                                      \\
910  & 21:39:52.58 &  58:00:40.0$^{\rm r}$& 1062 & 4062                      &                               &                 &                 &   12.2$^{\rm j}$&       &                 &8.177 \,(20) &7.288 \,(42) & 6.911 \,(18)$^{\rm r}$&                    &                 &               &-3\,(4.7)     &-2.1\,(4.7)   &-9.7\,(7.7)   &-2.7\,(7.8)   &-0.09&0.12 &                                                                                      \\
911  & 21:40:09.07 &  58:00:50.2$^{\rm r}$& 1063 & 735                       &                               &                 &                 &    9.7$^{\rm j}$&       &                 &8.872 \,(23) &8.796 \,(30) & 8.812 \,(22)$^{\rm r}$&        B8$^{\rm q}$&                 &               &6\,(1.6)      &-1.8\,(1.6)   &-4.2\,(0.7)   &-4.5\,(0.8)   &-0.4 &0.07 &                                                                                      \\
912  & 21:39:45.40 &  58:03:33.4$^{\rm r}$& 1064 & 4064                      &                               &                 &                 &     11$^{\rm j}$&       &                 &8.311 \,(26) &7.591 \,(29) & 7.425 \,(20)$^{\rm r}$&                    &                 &               &0.9\,(2.8)    &-12.3\,(2.8)  &1\,(2.3)      &-9.4\,(1.5)   &-0.6 &-0.28&                                                                                      \\
913  & 21:39:30.13 &  58:05:11.4$^{\rm r}$& 1065 & 4065                      &                               &                 &                 &   13.6$^{\rm j}$&       &                 &11.594\,(24) &11.248\,(31) & 11.114\,(26)$^{\rm r}$&                    &                 &               &-4.7\,(3.8)   &-1.6\,(3.8)   &-2.4\,(7.2)   &-8.8\,(7.3)   &0.37 &-0.47&                                                                                      \\
914  & 21:39:32.25 &  58:05:52.6$^{\rm r}$& 1066 & 4066                      &                               &                 &                 &   11.3$^{\rm j}$&       &                 &10.645\,(26) &10.410\,(32) & 10.324\,(22)$^{\rm r}$&                    &                 &               &2.5\,(2.7)    &0.7\,(2.7)    &1.1\,(0.9)    &0.6\,(0.8)    &-0.53&0.37 &                                                                                      \\
915  & 21:39:49.96 &  58:06:21.5$^{\rm r}$& 1068 & 4068                      &                               &                 &                 &   11.2$^{\rm j}$&       &                 &10.477\,(24) &10.267\,(31) & 10.207\,(22)$^{\rm r}$&                    &                 &               &-7.8\,(2)     &-5.1\,(2)     &-6.9\,(0.6)   &-6.3\,(0.6)   &0.05 &-0.06&                                                                                      \\
916  & 21:40:21.07 &  58:02:10.9$^{\rm r}$& 1069 & 736                       &                               &                 &                 &   11.3$^{\rm j}$&       &                 &10.390\,(23) &10.197\,(28) & 10.135\,(20)$^{\rm r}$&        A3$^{\rm q}$&                 &               &-13.8\,(2.7)  &-1.1\,(2.7)   &-7.8\,(0.7)   &-6.1\,(0.6)   &0.22 &-0.19&                                                                                      \\
917  & 21:40:23.40 &  57:51:20.1$^{\rm r}$& 1070 & 4070                      &                               &                 &                 &   12.3$^{\rm j}$&       &                 &11.091\,(24) &10.824\,(28) & 10.729\,(23)$^{\rm r}$&                    &                 &               &-8.7\,(3.8)   &-0.3\,(3.8)   &-7\,(1.3)     &-4.3\,(1.8)   &0.13 &0.13 &                                                                                      \\
918  & 21:40:57.73 &  57:49:26.4$^{\rm r}$& 1071 & 4071                      &                               &                 &                 &   13.8$^{\rm j}$&       &                 &11.899\,(24) &11.536\,(29) & 11.396\,(26)$^{\rm r}$&                    &                 &               &-13.3\,(3.8)  &-11.4\,(3.8)  &-3.5\,(7.4)   &-23.7\,(7.4)  &1.13 &-0.77&                                                                                      \\
919  & 21:41:04.36 &  57:51:17.3$^{\rm r}$& 1072 & 4072                      &                               &                 &                 &     12$^{\rm j}$&       &                 &7.510 \,(20) &6.354 \,(49) & 6.003 \,(21)$^{\rm r}$&                    &                 &               &-2.8\,(4.9)   &-4.2\,(4.9)   &-0.8\,(6.6)   &-1.1\,(6.7)   &-0.36&0.07 &                                                                                      \\
920  & 21:41:30.03 &  57:48:13.7$^{\rm r}$& 1073 & 4073                      &                               &                 &                 &   13.4$^{\rm j}$&       &                 &12.026\,(26) &11.624\,(30) & 11.556\,(23)$^{\rm r}$&                    &                 &               &-5.9\,(13.5)  &5.5\,(13.5)   &-7\,(7.4)     &6.1\,(7.4)    &-0.06&1.51 &                                                                                      \\
921  & 21:41:30.13 &  57:49:31.9$^{\rm r}$& 1074 & 4074                      &                               &                 &                 &   12.2$^{\rm j}$&       &                 &11.293\,(26) &11.035\,(28) & 10.965\,(19)$^{\rm r}$&                    &                 &               &-20.3\,(3.8)  &-5.9\,(3.8)   &-14.9\,(1.4)  &-4.6\,(1.3)   &1.21 &-0.5 &                                                                                      \\
922  & 21:40:58.71 &  57:52:22.8$^{\rm j}$& 1075 & 4075                      &                               &                 &                 &   12.3$^{\rm j}$&       &                 &             &             &                       &                    &                 &               &              &              &              &              &     &     &no star                                                                          \\
923  & 21:41:38.48 &  57:50:40.8$^{\rm r}$& 1076 & 4076                      &                               &                 &                 &   13.3$^{\rm j}$&       &                 &10.588\,(27) &9.864 \,(28) & 9.693 \,(21)$^{\rm r}$&                    &                 &               &-2.4\,(4.7)   &-1.8\,(4.7)   &5.8\,(7.5)    &1\,(7.5)      &0.06 &0.37 &                                                                                      \\
924  & 21:40:51.27 &  57:56:30.3$^{\rm r}$& 1077 & 4077                      &                               &                 &                 &   12.8$^{\rm j}$&       &                 &11.409\,(23) &10.973\,(32) & 10.899\,(23)$^{\rm r}$&                    &                 &               &-0.2\,(3.8)   &-6.5\,(3.8)   &15.5\,(7.8)   &-13\,(7.8)    &-0.28&-0.62&                                                                                      \\
925  & 21:41:19.81 &  57:54:54.5$^{\rm r}$& 1078 & 4078                      &                               &                 &                 &   13.2$^{\rm j}$&       &                 &11.917\,(27) &11.630\,(32) & 11.537\,(21)$^{\rm r}$&                    &                 &               &-5.7\,(3.8)   &4.2\,(3.8)    &-4\,(7.3)     &-2.1\,(7.4)   &-0.25&0.62 &                                                                                      \\
926  & 21:41:39.84 &  57:54:22.6$^{\rm r}$& 1079 & 4079                      &                               &                 &                 &   12.1$^{\rm j}$&       &                 &13.697\,(32) &13.228\,(40) & 13.075\,(37)$^{\rm r}$&                    &                 &               &-3.8\,(3.8)   &-2.6\,(3.8)   &-1.3\,(7.3)   &-5.1\,(7.5)   &     &     &new coordinates                                                          \\
927  & 21:41:36.16 &  57:54:31.9$^{\rm r}$& 1080 & 4080                      &                               &                 &                 &   12.9$^{\rm j}$&       &                 &12.867\,(27) &12.538\,(33) & 12.351\,(28)$^{\rm r}$&                    &                 &               &-8.1\,(3.8)   &-10.5\,(3.8)  &-4.2\,(7.2)   &-10.1\,(7.3)  &     &     &new coordinates                                                          \\
928  & 21:41:38.70 &  57:55:26.3$^{\rm r}$& 1081 & 4081                      &                               &                 &                 &     12$^{\rm j}$&       &                 &11.065\,(27) &10.832\,(31) & 10.714\,(21)$^{\rm r}$&                    &                 &               &-4.2\,(3.8)   &3.9\,(3.8)    &1.3\,(0.5)    &7.2\,(2.6)    &-0.5 &0.81 &                                                                                      \\
929  & 21:40:57.21 &  58:01:04.9$^{\rm r}$& 1082 & 742                       &                               &                 &                 &   10.8$^{\rm j}$&       &                 &10.089\,(24) &9.848 \,(32) & 9.863 \,(23)$^{\rm r}$&        F2$^{\rm q}$&                 &               &9.9\,(1.7)    &-1.9\,(1.7)   &3\,(1.7)      &-3.1\,(0.7)   &-0.67&0.2  &                                                                                      \\
930  & 21:41:15.31 &  58:00:42.3$^{\rm r}$& 1083 & 4083                      &                               &                 &                 &   13.3$^{\rm j}$&       &                 &11.025\,(25) &10.446\,(29) & 10.238\,(19)$^{\rm r}$&                    &                 &               &-6\,(3.8)     &-25.6\,(3.8)  &-9.8\,(8.3)   &-100.6\,(8.3) &0.03 &0.02 &                                                                                      \\
931  & 21:41:02.14 &  58:01:52.4$^{\rm r}$& 1084 & 4084                      &                               &                 &                 &   12.7$^{\rm j}$&       &                 &11.558\,(24) &11.333\,(30) & 11.182\,(23)$^{\rm r}$&                    &                 &               &-5.2\,(3.8)   &1.1\,(3.8)    &-7.6\,(0.8)   &-2.3\,(1.8)   &-0.07&-0.02&                                                                                      \\
932  & 21:41:01.33 &  58:04:49.5$^{\rm r}$& 1085 & 4085                      &                               &                 &                 &   12.8$^{\rm j}$&       &                 &11.195\,(21) &10.846\,(31) & 10.723\,(22)$^{\rm r}$&                    &                 &               &-3.3\,(11.2)  &-6.5\,(11.2)  &-25\,(7.2)    &-1\,(7.2)     &-0.09&0.09 &                                                                                      \\
933  & 21:41:21.70 &  58:03:41.2$^{\rm r}$& 1086 & 4086                      &                               &                 &                 &   13.4$^{\rm j}$&       &                 &11.955\,(29) &11.700\,(32) & 11.564\,(23)$^{\rm r}$&                    &                 &               &-14.3\,(3.8)  &-5.6\,(3.8)   &-15.9\,(7.3)  &-11.5\,(7.3)  &0.55 &-0.37&                                                                                      \\
934  & 21:41:24.75 &  58:03:26.9$^{\rm r}$& 1087 & 4087                      &                               &                 &                 &   13.8$^{\rm j}$&       &                 &11.178\,(27) &10.536\,(32) & 10.332\,(21)$^{\rm r}$&                    &                 &               &-4.8\,(3.8)   &-20.9\,(3.8)  &-16.4\,(7.5)  &-67.2\,(7.5)  &-0.29&-0.07&                                                                                      \\
935  & 21:41:49.90 &  58:02:21.0$^{\rm r}$& 1088 & 4088                      &                               &                 &                 &   14.1$^{\rm j}$&       &                 &12.504\,(27) &12.300\,(30) & 12.156\,(23)$^{\rm r}$&                    &                 &               &2.1\,(3.8)    &-24\,(3.8)    &              &              &-0.18&0.1  &                                                                                      \\
936  & 21:41:28.83 &  58:04:57.9$^{\rm r}$& 1090 & 4090                      &                               &                 &                 &   12.1$^{\rm j}$&       &                 &9.739 \,(27) &9.056 \,(32) & 8.863 \,(21)$^{\rm r}$&                    &                 &               &-2\,(4.7)     &-6.6\,(4.7)   &-0.7\,(7.6)   &-24.7\,(7.6)  &-0.36&0.32 &                                                                                      \\
937  & 21:41:53.28 &  57:51:35.9$^{\rm r}$& 1091 & 747                       &                               &                 &                 &    8.9$^{\rm j}$&       &                 &6.439 \,(32) &5.853 \,(42) & 5.719 \,(18)$^{\rm r}$&       gK0$^{\rm q}$&                 &               &14.9\,(1.3)   &15.4\,(1.3)   &15.4\,(0.8)   &16.1\,(0.8)   &-2.17&1.81 &                                                                                      \\
938  & 21:42:02.38 &  57:51:18.0$^{\rm r}$& 1092 & 4092                      &                               &                 &                 &   13.6$^{\rm j}$&       &                 &10.752\,(24) &10.067\,(32) & 9.904 \,(21)$^{\rm r}$&                    &                 &               &-16.4\,(4.7)  &2.7\,(4.7)    &-21\,(7.4)    &-3.6\,(7.4)   &0.77 &0.39 &                                                                                      \\
939  & 21:42:07.91 &  57:50:22.2$^{\rm r}$& 1093 & 4093                      &                               &                 &                 &   11.5$^{\rm j}$&       &                 &10.219\,(24) &9.821 \,(30) & 9.740 \,(21)$^{\rm r}$&                    &                 &               &-4.4\,(2.7)   &-5.5\,(2.7)   &-4.6\,(1.8)   &-4.2\,(1.5)   &-0.02&-0.22&                                                                                      \\
940  & 21:42:12.52 &  57:47:34.2$^{\rm r}$& 1094 & 4094                      &                               &                 &                 &   11.3$^{\rm j}$&       &                 &10.429\,(24) &10.085\,(32) & 10.007\,(21)$^{\rm r}$&                    &                 &               &-9.1\,(2.7)   &-1.2\,(2.7)   &-5.2\,(0.7)   &3.9\,(1.7)    &0.3  &0.66 &                                                                                      \\
941  & 21:42:27.62 &  57:48:45.0$^{\rm r}$& 1095 & 4095                      &                               &                 &                 &   14.2$^{\rm j}$&       &                 &12.506\,(24) &12.159\,(31) & 12.071\,(23)$^{\rm r}$&                    &                 &               &-5.2\,(3.8)   &-7.6\,(3.8)   &-8.4\,(7.9)   &-10.4\,(8)    &0.42 &-0.6 &                                                                                      \\
942  & 21:42:27.81 &  57:49:37.7$^{\rm r}$& 1096 & 4096                      &                               &                 &                 &   13.6$^{\rm j}$&       &                 &12.201\,(26) &11.972\,(31) & 11.870\,(24)$^{\rm r}$&                    &                 &               &-21.2\,(3.9)  &-6.3\,(3.9)   &-13\,(7.8)    &-41.2\,(7.8)  &0.63 &-0.25&                                                                                      \\
943  & 21:42:21.45 &  57:49:52.5$^{\rm r}$& 1097 & 749                       &                               &                 &                 &   10.4$^{\rm j}$&       &                 &9.600 \,(24) &9.430 \,(31) & 9.347 \,(21)$^{\rm r}$&        A3$^{\rm q}$&                 &               &-10\,(2)      &-12.9\,(2)    &-6.3\,(0.9)   &-15.3\,(0.7)  &0.5  &-1   &                                                                                      \\
944  & 21:42:29.11 &  57:50:50.2$^{\rm r}$& 1098 & 751                       &                               &   9.97$^{\rm l}$&  10.15$^{\rm l}$&     10$^{\rm l}$&       &                 &9.580 \,(24) &9.574 \,(30) & 9.551 \,(21)$^{\rm r}$&        B8$^{\rm q}$&                 &               &-5.7\,(2)     &-5.8\,(2)     &-5.7\,(0.6)   &-4.7\,(0.7)   &0.15 &0    &                                                                                      \\
945  & 21:41:59.48 &  57:53:55.2$^{\rm r}$& 1099 & 4099                      &                               &                 &                 &   13.5$^{\rm j}$&       &                 &12.234\,()   &12.146\,()   &   12.019\,()$^{\rm r}$&                    &                 &               &-8.7\,(3.8)   &-3\,(3.8)     &-23.5\,(7.8)  &-12.8\,(7.8)  &-0.01&-0.07&                                                                                      \\
946  & 21:42:20.17 &  57:53:06.7$^{\rm r}$& 1100 & 4100                      &                               &                 &                 &   13.7$^{\rm j}$&       &                 &11.832\,(29) &11.506\,(31) & 11.355\,(24)$^{\rm r}$&                    &                 &               &-6.2\,(3.8)   &-5.2\,(3.8)   &-12.2\,(7.4)  &-10.5\,(7.4)  &0.2  &-0.26&                                                                                      \\
947  & 21:42:27.91 &  57:53:12.6$^{\rm r}$& 1101 & 4101                      &                               &                 &                 &   13.7$^{\rm j}$&       &                 &12.094\,(27) &11.773\,(31) & 11.693\,(23)$^{\rm r}$&                    &                 &               &-25.4\,(18.5) &-9.6\,(18.5)  &-11.3\,(7)    &-22\,(7)      &0.05 &0.11 &                                                                                      \\
948  & 21:42:32.01 &  57:53:56.8$^{\rm r}$& 1102 & 4102                      &                               &                 &                 &   12.3$^{\rm j}$&       &                 &11.345\,(26) &11.052\,(31) & 10.996\,(21)$^{\rm r}$&                    &                 &               &-14.8\,(3.8)  &-5.2\,(3.8)   &-9.1\,(7.2)   &-7.1\,(4.6)   &0.51 &-0.19&                                                                                      \\
949  & 21:41:44.06 &  57:55:56.3$^{\rm r}$& 1103 & 4103                      &                               &                 &                 &   13.6$^{\rm j}$&       &                 &10.530\,(27) &9.796 \,(30) & 9.564 \,(21)$^{\rm r}$&                    &                 &               &4.1\,(4.7)    &-0.6\,(4.7)   &-1.5\,(7.2)   &-15.1\,(7.2)  &-0.32&0.02 &                                                                                      \\
950  & 21:42:09.72 &  57:57:17.1$^{\rm r}$& 1104 & 4104                      &                               &                 &                 &   13.1$^{\rm j}$&       &                 &10.680\,(26) &10.021\,(30) & 9.846 \,(21)$^{\rm r}$&                    &                 &               &-13.4\,(4.7)  &-7\,(4.7)     &-11.1\,(7.7)  &-11.7\,(7.7)  &0.49 &-0.33&                                                                                      \\
951  & 21:42:32.16 &  57:56:32.9$^{\rm r}$& 1105 & 4105                      &                               &                 &                 &   12.4$^{\rm j}$&       &                 &9.084 \,(21) &8.140 \,(38) & 7.874 \,(18)$^{\rm r}$&                    &                 &               &-9.1\,(4.7)   &0.5\,(4.7)    &2.3\,(7.1)    &-8.2\,(7.2)   &0.09 &0.11 &                                                                                      \\
952  & 21:42:37.73 &  57:56:52.8$^{\rm r}$& 1106 & 4106                      &                               &                 &                 &   12.5$^{\rm j}$&       &                 &11.863\,(26) &11.685\,(31) & 11.645\,(19)$^{\rm r}$&                    &                 &               &-3.5\,(3.8)   &1.6\,(3.8)    &-8.9\,(3.2)   &-3.7\,(2.1)   &0.19 &-0.13&                                                                                      \\
953  & 21:42:52.48 &  57:56:00.1$^{\rm r}$& 1107 & 4107                      &                               &                 &                 &   12.8$^{\rm j}$&       &                 &11.479\,(29) &11.177\,(32) & 11.034\,(23)$^{\rm r}$&                    &                 &               &202\,(7.4)    &-43.3\,(7.4)  &-7.3\,(3.8)   &-9.9\,(5)     &0.47 &.0.44&                                                                                      \\
954  & 21:41:48.34 &  57:58:41.4$^{\rm r}$& 1108 & 4108                      &                               &                 &                 &   13.6$^{\rm j}$&       &                 &11.023\,(29) &10.338\,(31) & 10.149\,(21)$^{\rm r}$&                    &                 &               &12.5\,(3.8)   &5.7\,(3.8)    &14.1\,(7.3)   &11.6\,(7.3)   &-1.79&0.52 &                                                                                      \\
955  & 21:41:53.26 &  58:00:15.1$^{\rm r}$& 1109 & 4109                      &                               &                 &                 &   13.8$^{\rm j}$&       &                 &12.167\,(30) &11.814\,(30) & 11.704\,(23)$^{\rm r}$&                    &                 &               &-12.4\,(3.8)  &-34.6\,(3.8)  &              &              &0.88 &-0.79&                                                                                      \\
956  & 21:42:17.73 &  57:59:06.5$^{\rm r}$& 1110 & 748                       &                               &                 &                 &   10.1$^{\rm j}$&       &                 &9.141 \,(32) &8.869 \,(31) & 8.833 \,(19)$^{\rm r}$&        F8$^{\rm q}$&                 &               &-33.5\,(1.3)  &-37\,(1.4)    &-33.6\,(0.8)  &-36.4\,(0.6)  &3.22 &-2.86&                                                                                      \\
957  & 21:42:15.09 &  58:00:03.7$^{\rm r}$& 1111 & 4111                      &                               &                 &                 &   13.5$^{\rm j}$&       &                 &12.080\,(24) &11.719\,(29) & 11.658\,(23)$^{\rm r}$&                    &                 &               &-44\,(3.8)    &2.9\,(3.8)    &-38.8\,(7.3)  &-11.1\,(7.3)  &3.37 &0.33 &                                                                                      \\
958  & 21:42:37.82 &  57:59:08.0$^{\rm r}$& 1112 & 4112                      &                               &                 &                 &   12.3$^{\rm j}$&       &                 &11.160\,(26) &10.796\,(29) & 10.684\,(21)$^{\rm r}$&                    &                 &               &-10\,(3.8)    &-15.9\,(3.8)  &-3.3\,(1.9)   &-20.3\,(2.5)  &0.15 &-1.3 &                                                                                      \\
959  & 21:42:32.99 &  57:59:24.5$^{\rm r}$& 1113 & 4113                      &                               &                 &                 &   12.9$^{\rm j}$&       &                 &10.068\,(26) &9.279 \,(31) & 9.070 \,(19)$^{\rm r}$&                    &                 &               &-15\,(4.7)    &2.1\,(4.7)    &5.4\,(7.3)    &-5.8\,(7.3)   &0.4  &0.37 &                                                                                      \\
960  & 21:42:07.06 &  58:00:53.7$^{\rm r}$& 1114 & 4114                      &                               &                 &                 &   12.9$^{\rm j}$&       &                 &11.560\,(27) &11.330\,(30) & 11.168\,(23)$^{\rm r}$&                    &                 &               &-6.6\,(3.8)   &-7.2\,(3.8)   &-11.7\,(7.2)  &-27\,(7.2)    &-0.34&0.09 &                                                                                      \\
961  & 21:41:47.35 &  58:07:20.2$^{\rm r}$& 1115 & 4115                      &                               &                 &                 &   12.1$^{\rm j}$&       &                 &9.668 \,(29) &8.924 \,(31) & 8.722 \,(19)$^{\rm r}$&                    &                 &               &-17.7\,(4.7)  &-4\,(4.7)     &-17.9\,(7.3)  &-10.9\,(7.3)  &0.6  &0.22 &                                                                                      \\
962  & 21:42:00.21 &  58:06:18.3$^{\rm r}$& 1116 & 4116                      &                               &                 &                 &   13.5$^{\rm j}$&       &                 &11.796\,(27) &11.471\,(31) & 11.361\,(21)$^{\rm r}$&                    &                 &               &-12.2\,(3.8)  &-15.2\,(3.8)  &-17.9\,(7.6)  &-19.5\,(7.6)  &0.86 &-0.69&                                                                                      \\
963  & 21:42:16.60 &  58:06:58.7$^{\rm r}$& 1117 & 4117                      &                               &                 &                 &   13.3$^{\rm j}$&       &                 &11.703\,(27) &11.446\,(31) & 11.304\,(21)$^{\rm r}$&                    &                 &               &-4.8\,(3.8)   &5.4\,(3.8)    &3.9\,(7.1)    &-3.5\,(7.2)   &-0.49&0.58 &                                                                                      \\
964  & 21:42:55.39 &  57:51:28.9$^{\rm r}$& 1118 & 4118                      &                               &                 &                 &   13.1$^{\rm j}$&       &                 &10.848\,(25) &10.463\,(30) & 10.429\,(23)$^{\rm r}$&                    &                 &               &-0.4\,(3.8)   &-0.2\,(3.8)   &0.1\,(0.7)    &-2.5\,(0.9)   &-0.78&-0.11&                                                                                      \\
965  & 21:42:48.36 &  57:47:34.0$^{\rm r}$& 1119 & 4119                      &                               &                 &                 &   12.7$^{\rm j}$&       &                 &10.262\,(27) &9.559 \,(29) & 9.376 \,(23)$^{\rm r}$&                    &                 &               &-8.3\,(4.7)   &4.7\,(4.7)    &14.4\,(7)     &-14.2\,(7)    &0.22 &-0.11&                                                                                      \\
966  & 21:42:53.68 &  57:48:46.6$^{\rm r}$& 1120 & 4120                      &                               &                 &                 &   13.7$^{\rm j}$&       &                 &10.494\,(25) &9.823 \,(30) & 9.621 \,(21)$^{\rm r}$&                    &                 &               &-20.2\,(11.6) &7\,(11.6)     &5.9\,(7.4)    &-5.6\,(7.5)   &-0.61&0.21 &                                                                                      \\
967  & 21:43:10.00 &  57:59:55.9$^{\rm r}$& 1121 & 4121                      &                               &                 &                 &   12.7$^{\rm j}$&       &                 &11.236\,(25) &10.925\,(33) & 10.812\,(21)$^{\rm r}$&                    &                 &               &2.5\,(3.8)    &7.6\,(3.8)    &3.3\,(2.5)    &1.9\,(0.6)    &     &     &                                                                                      \\
968  & 21:42:46.26 &  58:05:36.4$^{\rm r}$& 1122 & 4122                      &                               &                 &                 &   13.3$^{\rm j}$&       &                 &10.273\,(25) &9.651 \,(36) & 9.398 \,(21)$^{\rm r}$&                    &                 &               &-4.5\,(4.7)   &-13\,(4.7)    &-2.6\,(7.5)   &-14.1\,(7.6)  &     &     &                                                                                      \\
969  & 21:37:44.06 &  57:46:15.0$^{\rm j}$& 1150 & 4150                      &                               &                 &                 &   13.4$^{\rm j}$&       &                 &             &             &                       &                    &                 &               &              &              &              &              &     &     &no star                                                                          \\
970  & 21:36:37.03 &  57:46:15.6$^{\rm r}$& 1151 & 4151                      &                               &                 &                 &   13.9$^{\rm j}$&       &                 &11.919\,(24) &11.638\,(31) & 11.461\,(25)$^{\rm r}$&                    &                 &               &-5.8\,(4)     &4.3\,(4)      &-7.3\,(7.4)   &4.4\,(7.5)    &0.35 &-0.58&                                                                                      \\
971  & 21:36:19.20 &  57:45:54.0$^{\rm r}$& 1152 & 4152                      &                               &                 &                 &   11.2$^{\rm j}$&       &                 &10.210\,(23) &9.940 \,(29) & 9.856 \,(22)$^{\rm r}$&                    &                 &               &4.6\,(2.7)    &30.5\,(2.7)   &3.9\,(0.7)    &22.1\,(1.2)   &-0.95&2.02 &                                                                                      \\
972  & 21:36:07.71 &  57:39:44.4$^{\rm r}$& 1153 & 4153                      &                               &                 &                 &   12.3$^{\rm j}$&       &                 &10.375\,(27) &9.825 \,(36) & 9.613 \,(23)$^{\rm r}$&                    &                 &               &44.4\,(4.7)   &12.9\,(4.7)   &31.1\,(7.5)   &-5.5\,(7.5)   &-4.84&0.38 &                                                                                      \\
973  & 21:37:51.24 &  57:59:28.1$^{\rm r}$& 1154 & 713                       &                               &                 &                 &   10.8$^{\rm j}$&       &                 &9.567 \,(26) &9.179 \,(31) & 9.144 \,(19)$^{\rm r}$&       dK0$^{\rm q}$&                 &               &14\,(1.7)     &-25.4\,(1.7)  &12.4\,(0.6)   &-21.1\,(1.4)  &-1.47&-2.15&                                                                                      \\
974  & 21:34:51.92 &  57:30:52.2$^{\rm r}$& 1155 & 4155                      &                               &                 &                 &   13.6$^{\rm j}$&       &                 &11.997\,(24) &11.728\,(29) & 11.608\,(23)$^{\rm r}$&                    &                 &               &-4.4\,(3.8)   &-2\,(3.8)     &-16.1\,(7.5)  &6.4\,(7.5)    &-0.52&0.44 &                                                                                      \\
975  & 21:35:09.46 &  57:31:04.0$^{\rm r}$& 1156 & 4156                      &                               &                 &                 &   12.7$^{\rm j}$&       &                 &11.532\,(21) &11.249\,(29) & 11.155\,(20)$^{\rm r}$&                    &                 &               &-14.9\,(3.8)  &-12.3\,(3.8)  &-19.8\,(7.2)  &-24.4\,(7.1)  &-0.04&-0.6 &                                                                                      \\
976  & 21:35:11.35 &  57:31:34.7$^{\rm r}$& 1157 & 4157                      &                               &                 &                 &   13.6$^{\rm j}$&       &                 &12.172\,(22) &11.817\,(25) & 11.774\,(23)$^{\rm r}$&                    &                 &               &-15.5\,(3.8)  &-9.2\,(3.8)   &-18\,(7.4)    &-12.9\,(7.5)  &1.1  &-1.01&                                                                                      \\
977  & 21:35:23.46 &  57:31:28.1$^{\rm r}$& 1158 & 425                       &                               &                 &                 &   10.1$^{\rm j}$&       &                 &9.796 \,(23) &9.584 \,(28) & 9.504 \,(21)$^{\rm r}$&        A0$^{\rm q}$&                 &               &2.8\,(1.6)    &3.1\,(1.6)    &-2.8\,(1.2)   &-3.4\,(0.9)   &-0.27&-0.32&                                                                                      \\
978  & 21:35:24.28 &  57:32:15.0$^{\rm r}$& 1159 & 4159                      &                               &                 &                 &   11.6$^{\rm j}$&       &                 &10.477\,(23) &10.256\,(28) & 10.162\,(23)$^{\rm r}$&                    &                 &               &-5.7\,(3.8)   &-4\,(3.8)     &-6.1\,(1.5)   &-7.1\,(1.2)   &0.3  &-0.54&                                                                                      \\
979  & 21:35:38.01 &  57:33:02.6$^{\rm r}$& 1160 & 4160                      &                               &                 &                 &     14$^{\rm j}$&       &                 &10.225\,(23) &9.319 \,(27) & 9.077 \,(21)$^{\rm r}$&                    &                 &               &-3.6\,(4.9)   &-3.9\,(4.9)   &-4\,(7.3)     &15\,(7.3)     &0.05 &0    &                                                                                      \\
980  & 21:35:41.74 &  57:32:52.5$^{\rm r}$& 1161 & 4161                      &                               &                 &                 &   13.3$^{\rm j}$&       &                 &11.492\,(25) &11.182\,(32) & 11.067\,(25)$^{\rm r}$&                    &                 &               &-18.6\,(4)    &-32.2\,(4)    &              &              &0.14 &-0.32&new coordinates\footnote[26]{in finding chart [j] wrong? (1611 $\rightarrow$ 1161)}                        \\
981  & 21:35:59.08 &  57:33:03.7$^{\rm r}$& 1162 & 4162                      &                               &                 &                 &   13.7$^{\rm j}$&       &                 &11.835\,(23) &11.459\,(28) & 11.331\,(24)$^{\rm r}$&                    &                 &               &-9.6\,(4)     &16\,(4)       &10.2\,(6.7)   &21.5\,(6.7)   &0.26 &1.61 &                                                                                      \\
982  & 21:36:02.37 &  57:33:37.1$^{\rm r}$& 1163 & 4163                      &                               &                 &                 &   14.3$^{\rm j}$&       &                 &11.765\,()   &11.343\,()   &   11.193\,()$^{\rm r}$&                    &                 &               &-6\,(4)       &7.4\,(4)      &-10.3\,(6.8)  &48\,(6.9)     &0.05 &0.13 &                                                                                      \\
983  & 21:35:15.59 &  57:33:56.2$^{\rm r}$& 1164 & 4164                      &                               &                 &                 &   14.2$^{\rm j}$&       &                 &12.018\,()   &11.459\,()   & 11.385\,(32)$^{\rm r}$&                    &                 &               &36.4\,(10)    &31.9\,(10)    &46.7\,(8.6)   &20.6\,(9.9)   &-2.91&1.11 &                                                                                      \\
984  & 21:35:22.68 &  57:34:04.4$^{\rm r}$& 1165 & 4165                      &                               &                 &                 &   14.4$^{\rm j}$&       &                 &10.579\,(22) &9.655 \,(28) & 9.394 \,(22)$^{\rm r}$&                    &                 &               &-1.1\,(4.7)   &-3\,(4.7)     &-8.1\,(7.4)   &-6.2\,(7.4)   &0.14 &0.09 &new coordinates                                                                       \\     
985  & 21:35:06.97 &  57:35:03.8$^{\rm r}$& 1166 & 424                       &                               &                 &                 &   11.7$^{\rm j}$&       &                 &10.188\,(21) &9.872 \,(29) & 9.579 \,(21)$^{\rm r}$&                    &                 &               &-5.6\,(10)    &22.3\,(10)    &-3.8\,(2)     &0.2\,(3.7)    &-0.03&-0.05&                                                                                      \\
986  & 21:35:47.29 &  57:34:51.9$^{\rm r}$& 1167 & 4167                      &                               &                 &                 &   13.5$^{\rm j}$&       &                 &10.561\,(23) &9.850 \,(27) & 9.663 \,(21)$^{\rm r}$&                    &                 &               &-8.2\,(4.9)   &-6.6\,(4.9)   &-6\,(6.8)     &-0.7\,(6.8)   &0.44 &-0.53&                                                                                      \\
987  & 21:35:41.92 &  57:35:15.9$^{\rm r}$& 1168 & 4168                      &                               &                 &                 &     12$^{\rm j}$&       &                 &10.785\,(25) &10.588\,(27) & 10.497\,(21)$^{\rm r}$&                    &                 &               &-5.9\,(4)     &2.6\,(4)      &-2.8\,(0.6)   &-0.7\,(1.1)   &0.06 &-0.12&                                                                                      \\
988  & 21:35:37.79 &  57:35:49.5$^{\rm r}$& 1169 & 4169                      &                               &                 &                 &   12.8$^{\rm j}$&       &                 &9.449 \,(23) &8.603 \,(27) & 8.328 \,(24)$^{\rm r}$&                    &                 &               &-3.5\,(4.9)   &-3.5\,(4.9)   &-2.9\,(7.3)   &5.9\,(7.3)    &0.23 &-0.12&                                                                                      \\
989  & 21:35:39.39 &  57:36:29.3$^{\rm r}$& 1170 & 4170                      &                               &                 &                 &   12.7$^{\rm j}$&       &                 &9.690 \,(23) &8.958 \,(28) & 8.738 \,(21)$^{\rm r}$&                    &                 &               &-4.9\,(4.7)   &-0.5\,(4.7)   &-0.5\,(7.3)   &5.5\,(7.3)    &0.62 &;0.07&                                                                                      \\
990  & 21:35:53.49 &  57:35:16.7$^{\rm r}$& 1171 & 4171                      &                               &                 &                 &   12.6$^{\rm j}$&       &                 &11.240\,(25) &10.942\,(27) & 10.863\,(24)$^{\rm r}$&                    &                 &               &-7.9\,(3.8)   &-3.4\,(3.8)   &3.6\,(7)      &-4.7\,(7)     &0.68 &-0.45&                                                                                      \\
991  & 21:36:02.14 &  57:34:57.8$^{\rm r}$& 1172 & 4172                      &                               &                 &                 &   11.9$^{\rm j}$&       &                 &9.054 \,(23) &8.431 \,(27) & 8.202 \,(23)$^{\rm r}$&                    &                 &               &-4.5\,(4.7)   &-3.6\,(4.7)   &-5.1\,(7.9)   &12.7\,(7.9)   &0.02 &-0.33&                                                                                      \\
992  & 21:35:19.16 &  57:36:38.3$^{\rm r}$& 1173 & 4173                      &                               &                 &                 &     13$^{\rm j}$&       &                 &11.665\,(24) &10.973\,(28) & 10.224\,(20)$^{\rm r}$&                    &                 &               &-6.7\,(3.8)   &2.9\,(3.8)    &-4.4\,(6.9)   &5.4\,(6.9)    &0.27 &-0.38&                                                                                      \\
993  & 21:35:54.81 &  57:37:52.8$^{\rm r}$& 1174 & 4174                      &                               &                 &                 &   12.4$^{\rm j}$&       &                 &11.550\,(25) &11.426\,(33) & 11.368\,(25)$^{\rm r}$&                    &                 &               &0.1\,(4)      &7.9\,(4)      &-4\,(0.7)     &-1.2\,(0.7)   &0.14 &-0.19&                                                                                      \\
994  & 21:35:48.69 &  57:39:31.9$^{\rm r}$& 1175 & 4175                      &                               &                 &                 &   14.2$^{\rm j}$&       &                 &12.298\,(24) &11.953\,(29) & 11.840\,(24)$^{\rm r}$&                    &                 &               &1.2\,(4)      &-10.5\,(4)    &7.5\,(7.4)    &-9.3\,(7.4)   &-0.3 &-1.32&                                                                                      \\
995  & 21:35:54.57 &  57:40:24.2$^{\rm r}$& 1176 & 4176                      &                               &                 &                 &   12.6$^{\rm j}$&       &                 &11.032\,(24) &10.647\,(29) & 10.558\,(21)$^{\rm r}$&                    &                 &               &-0.5\,(3.8)   &5.9\,(3.8)    &-2.9\,(1.3)   &-0.9\,(1.1)   &-0.39&0.07 &                                                                                      \\
996  & 21:35:18.53 &  57:40:18.2$^{\rm r}$& 1177 & 4177                      &                               &                 &                 &   13.4$^{\rm j}$&       &                 &11.657\,(26) &11.270\,(29) & 11.175\,(20)$^{\rm r}$&                    &                 &               &6.4\,(3.8)    &10.3\,(3.8)   &4\,(7.3)      &3.7\,(7.3)    &-0.85&0.52 &                                                                                      \\
997  & 21:35:12.10 &  57:41:00.2$^{\rm r}$& 1178 & 4178                      &                               &                 &                 &   12.8$^{\rm j}$&       &                 &11.622\,(28) &11.245\,(31) & 11.152\,(23)$^{\rm r}$&                    &                 &               &35.6\,(3.8)   &30.9\,(3.8)   &48.5\,(4.7)   &30.8\,(1.7)   &-5.23&2.35 &                                                                                      \\
998  & 21:36:02.13 &  57:41:36.2$^{\rm r}$& 1179 & 4179                      &                               &                 &                 &   14.7$^{\rm j}$&       &                 &12.361\,(22) &11.908\,(29) & 11.803\,(24)$^{\rm r}$&                    &                 &               &12.8\,(4)     &5.7\,(4)      &27.6\,(7.4)   &6.6\,(7.4)    &-1.56&1.33 &                                                                                      \\
999  & 21:35:53.94 &  57:42:47.5$^{\rm r}$& 1180 & 5058                      &                               &                 &  15.77$^{\rm f}$&  14.61$^{\rm e}$&       &                 &12.664\,(24) &12.300\,(29) & 12.180\,(21)$^{\rm r}$&        F6$^{\rm e}$&                 &  2.2$^{\rm e}$&11.8\,(4)     &1.9\,(4)      &8\,(7.3)      &-3.5\,(7.5)   &     &     &                                                                                      \\
1000 & 21:35:40.62 &  57:42:49.8$^{\rm r}$& 1181 & 4181                      &                               &                 &                 &   14.5$^{\rm j}$&       &                 &12.518\,(24) &12.164\,(31) & 12.051\,(28)$^{\rm r}$&                    &                 &               &9.7\,(5.1)    &2.9\,(5.1)    &5.2\,(5)      &-8.5\,(5)     &-0.07&-2.8 &                                                                                      \\
1001 & 21:35:32.35 &  57:43:31.4$^{\rm r}$& 1182 & 4182                      &                               &                 &                 &   14.5$^{\rm j}$&       &                 &12.691\,(22) &12.249\,(32) & 12.152\,(24)$^{\rm r}$&                    &                 &               &-14.3\,(5.6)  &-32\,(5.6)    &-24.2\,(7.4)  &-29.4\,(7.4)  &1.71 &-2.62&                                                                                      \\
1002 & 21:35:22.33 &  57:43:52.3$^{\rm r}$& 1183 & 703                       &                               &                 &                 &   12.5$^{\rm j}$&       &                 &11.195\,(22) &10.936\,(28) & 10.831\,(21)$^{\rm r}$&        B8$^{\rm q}$&                 &               &-8.8\,(10)    &7.8\,(10)     &2\,(7.2)      &2.7\,(7.3)    &-0.04&0.17 &                                                                                      \\
1003 & 21:35:13.68 &  57:44:36.6$^{\rm r}$& 1184 & 4184                      &                               &                 &                 &   14.3$^{\rm j}$&       &                 &12.663\,(26) &12.268\,(31) & 12.165\,(24)$^{\rm r}$&                    &                 &               &-15.9\,(3.8)  &-23\,(3.8)    &-5.9\,(7.5)   &-25.5\,(7.6)  &     &     &                                                                                      \\
1004 & 21:35:05.91 &  57:44:24.0$^{\rm r}$& 1185 & 4185                      &                               &                 &                 &   10.3$^{\rm j}$&       &                 &9.140 \,(24) &9.062 \,(29) & 8.926 \,(21)$^{\rm r}$&                    &                 &               &-0.9\,(2)     &5.3\,(2)      &-5.5\,(0.6)   &-3.8\,(0.7)   &0.23 &-0.79&                                                                                      \\
1005 & 21:35:11.37 &  57:46:08.1$^{\rm r}$& 1186 & 4186                      &                               &                 &                 &   12.1$^{\rm j}$&       &                 &11.192\,(43) &11.045\,(33) & 10.948\,(24)$^{\rm r}$&                    &                 &               &-5.1\,(3.8)   &11\,(3.8)     &0.1\,(1.4)    &-1.4\,(9.4)   &-0.31&-0.65&2x[r]                                                                           \\
1006 & 21:35:11.76 &  57:46:06.1$^{\rm r}$& 1186 & 4186                      &                               &                 &                 &   12.1$^{\rm j}$&       &                 &12.306\,(32) &11.108\,(76) & 10.667\,(44)$^{\rm r}$&                    &                 &               &              &              &0.1\,(1.4)    &-1.4\,(9.4)   &-0.31&-0.65&2x[r]                                                                          \\
1007 & 21:36:02.73 &  57:45:33.7$^{\rm r}$& 1188 & 705                       &                               &                 &                 &   10.7$^{\rm j}$&       &                 &7.858 \,(20) &7.159 \,(31) & 6.979 \,(18)$^{\rm r}$&        G8$^{\rm q}$&                 &               &-1.1\,(12)    &-20.4\,(12)   &6.5\,(1.5)    &-11.6\,(1.3)  &-0.95&-1.8 &                                                                                      \\
1008 & 21:36:25.06 &  57:46:42.0$^{\rm r}$& 1189 & 4189                      &                               &                 &                 &   14.3$^{\rm j}$&       &                 &11.724\,(26) &11.208\,(31) & 11.085\,(23)$^{\rm r}$&                    &                 &               &7.4\,(3.8)    &19.4\,(3.8)   &30.9\,(7.3)   &39.8\,(7.4)   &     &     &                                                                                      \\
1009 & 21:35:47.86 &  57:47:43.0$^{\rm r}$& 1190 & 4190                      &                               &                 &                 &   14.5$^{\rm j}$&       &                 &12.306\,(24) &11.884\,(29) & 11.750\,(21)$^{\rm r}$&                    &                 &               &1.6\,(4)      &-6.6\,(4)     &-0.1\,(7.5)   &-10.5\,(7.4)  &     &     &                                                                                      \\
1010 & 21:35:52.37 &  57:52:03.8$^{\rm r}$& 1191 & 704                       &                               &                 &                 &   11.1$^{\rm j}$&       &                 &9.930 \,(22) &9.573 \,(28) & 9.499 \,(21)$^{\rm r}$&        G5$^{\rm q}$&                 &               &0.3\,(13.3)   &52.4\,(13.3)  &-10.3\,(1.4)  &41.1\,(1.4)   &0.53 &4.04 &                                                                                      \\
1011 & 21:34:58.38 &  57:54:36.3$^{\rm r}$& 1192 & 4192                      &                               &                 &                 &   12.2$^{\rm j}$&       &                 &11.108\,(23) &10.962\,(28) & 10.867\,(27)$^{\rm r}$&                    &                 &               &-0.4\,(3.8)   &-6.8\,(3.8)   &1.9\,(1.5)    &1.9\,(4.2)    &-0.63&-0.4 &                                                                                      \\
1012 & 21:34:56.10 &  57:54:52.5$^{\rm r}$& 1193 & 4193                      &                               &                 &                 &   14.1$^{\rm j}$&       &                 &12.045\,(29) &11.692\,(28) & 11.572\,(27)$^{\rm r}$&                    &                 &               &10.7\,(3.8)   &-5.3\,(3.8)   &1.3\,(7.7)    &-26\,(7.7)    &     &     &                                                                                      \\
1013 & 21:34:58.50 &  57:54:51.0$^{\rm r}$& 1194 & 4194                      &                               &                 &                 &   12.2$^{\rm j}$&       &                 &11.186\,(23) &11.040\,(26) & 10.927\,(25)$^{\rm r}$&                    &                 &               &-2.4\,(3.8)   &1.7\,(3.8)    &-2.3\,(0.9)   &-6.3\,(2.3)   &0.09 &-0.99&                                                                                      \\
1014 & 21:34:49.32 &  57:55:47.6$^{\rm r}$& 1195 & 4195                      &                               &                 &                 &   14.4$^{\rm j}$&       &                 &12.342\,(25) &12.047\,(32) & 11.914\,(27)$^{\rm r}$&                    &                 &               &10.7\,(3.8)   &1.6\,(3.8)    &4.5\,(7.3)    &-3.2\,(7.4)   &     &     &                                                                                      \\
1015 & 21:35:17.14 &  57:56:22.9$^{\rm r}$& 1196 & 4196                      &                               &                 &                 &   14.5$^{\rm j}$&       &                 &12.484\,(26) &11.953\,(31) & 11.808\,(28)$^{\rm r}$&                    &                 &               &43.4\,(3.8)   &53\,(3.8)     &33.3\,(8)     &51\,(8)       &     &     &                                                                                      \\
1016 & 21:35:27.65 &  57:57:16.5$^{\rm r}$& 1197 & 4197                      &                               &                 &                 &   13.1$^{\rm j}$&       &                 &11.713\,(23) &11.508\,(27) & 11.439\,(23)$^{\rm r}$&                    &                 &               &1.7\,(13.3)   &11.9\,(13.3)  &13.3\,(7.2)   &0.9\,(7.3)    &0.16 &-0.9 &                                                                                      \\
1017 & 21:35:48.42 &  57:55:16.7$^{\rm r}$& 1198 & 4198                      &                               &                 &                 &   13.8$^{\rm j}$&       &                 &10.593\,(23) &9.844 \,(26) & 9.625 \,(21)$^{\rm r}$&                    &                 &               &5.5\,(4.7)    &6\,(4.7)      &7.8\,(7.2)    &2.9\,(7.3)    &-0.95&0.5  &                                                                                      \\
1018 & 21:35:02.52 &  57:59:30.5$^{\rm r}$& 1199 & 4199                      &                               &                 &                 &   13.4$^{\rm j}$&       &                 &11.758\,(23) &11.333\,(28) & 11.200\,(26)$^{\rm r}$&                    &                 &               &-7.7\,(3.8)   &-8.8\,(3.8)   &-21.6\,(7.2)  &-4.2\,(7.2)   &1.17 &-1.01&                                                                                      \\
1019 & 21:35:07.13 &  58:03:56.6$^{\rm r}$& 1200 & 4200                      &                               &                 &                 &   13.7$^{\rm j}$&       &                 &11.891\,(23) &11.556\,(28) & 11.458\,(25)$^{\rm r}$&                    &                 &               &-1.9\,(3.8)   &-5.3\,(3.8)   &1\,(7.3)      &-9.2\,(7.3)   &-0.2 &-0.02&                                                                                      \\
1020 & 21:35:13.98 &  58:03:59.4$^{\rm r}$& 1201 & 4201                      &                               &                 &                 &   12.7$^{\rm j}$&       &                 &11.545\,(25) &11.269\,(30) & 11.213\,(26)$^{\rm r}$&                    &                 &               &-11.5\,(3.8)  &0.3\,(3.8)    &-8.6\,(11.2)  &-1.1\,(2.8)   &0.03 &-0.14&                                                                                      \\
1021 & 21:36:30.00 &  57:49:46.3$^{\rm r}$& 1202 & 4202                      &                               &                 &                 &   14.4$^{\rm j}$&       &                 &10.883\,(24) &10.101\,(29) & 9.866 \,(22)$^{\rm r}$&                    &                 &               &1.6\,(4.9)    &-4.1\,(4.9)   &-3.2\,(7.4)   &-2.2\,(7.4)   &     &     &                                                                                      \\
1022 & 21:36:09.98 &  57:52:35.8$^{\rm r}$& 1203 & 4203                      &                               &                 &                 &   14.4$^{\rm j}$&       &                 &10.641\,(22) &9.731 \,(28) & 9.462 \,(21)$^{\rm r}$&                    &                 &               &-3.1\,(4.7)   &0.5\,(4.7)    &6.6\,(7.3)    &-5.4\,(7.3)   &     &     &                                                                                      \\
1023 & 21:36:07.65 &  57:53:11.1$^{\rm r}$& 1204 & 4204                      &                               &                 &                 &   13.1$^{\rm j}$&       &                 &11.532\,(22) &11.187\,(28) & 11.085\,(21)$^{\rm r}$&                    &                 &               &-0.4\,(3.8)   &-1.2\,(3.8)   &3.5\,(7.3)    &-2.3\,(7.3)   &-0.34&-0.52&                                                                                      \\
1024 & 21:36:18.56 &  57:52:45.8$^{\rm r}$& 1205 & 4205                      &                               &                 &                 &   14.6$^{\rm j}$&       &                 &12.760\,(35) &12.425\,(41) & 12.289\,(35)$^{\rm r}$&                    &                 &               &18.9\,(5.2)   &-27.2\,(5.2)  &-34.5\,(7.5)  &57.4\,(7.5)   &     &     &                                                                                      \\
1025 & 21:36:20.53 &  57:54:22.1$^{\rm r}$& 1206 & 4206                      &                               &                 &                 &   13.7$^{\rm j}$&       &                 &12.308\,(26) &12.110\,(29) & 11.980\,(25)$^{\rm r}$&                    &                 &               &-6.7\,(3.8)   &-5.3\,(3.8)   &0.6\,(7.5)    &-13.4\,(7.5)  &0.09 &-0.71&new coordinates                                                                       \\     
1026 & 21:36:02.30 &  57:55:01.3$^{\rm r}$& 1207 & 4207                      &                               &                 &                 &   14.6$^{\rm j}$&       &                 &12.720\,(25) &12.255\,(30) & 12.127\,(24)$^{\rm r}$&                    &                 &               &11.6\,(3.8)   &-1.5\,(3.8)   &3.2\,(7.2)    &-3.8\,(7.2)   &     &     &                                                                                      \\
1027 & 21:36:36.41 &  57:53:11.2$^{\rm r}$& 1208 & 4208                      &                               &                 &                 &   13.8$^{\rm j}$&       &                 &12.258\,(26) &11.998\,(29) & 11.875\,(28)$^{\rm r}$&                    &                 &               &-1.5\,(3.8)   &0.3\,(3.8)    &-6.9\,(7.4)   &-0.5\,(7.4)   &0.35 &-0.21&                                                                                      \\
1028 & 21:36:32.75 &  57:55:15.2$^{\rm r}$& 1209 & 5066                      &                               &                 &                 &   14.5$^{\rm j}$&       &                 &10.612\,(23) &9.662 \,(29) & 9.421 \,(23)$^{\rm r}$&       F1:$^{\rm q}$&                 &               &-3.5\,(4.7)   &-3.1\,(4.7)   &1.4\,(7.3)    &-5.1\,(7.3)   &     &     &                                                                                      \\
1029 & 21:36:32.97 &  57:56:10.2$^{\rm r}$& 1210 & 4210                      &                               &                 &                 &   14.3$^{\rm j}$&       &                 &12.487\,(24) &12.138\,(32) & 12.080\,(26)$^{\rm r}$&                    &                 &               &-4.3\,(3.8)   &2\,(3.8)      &-5.1\,(7.3)   &3.9\,(7.3)    &0.09 &-0.04&                                                                                      \\
1030 & 21:36:17.22 &  57:57:29.0$^{\rm r}$& 1211 & 4211                      &                               &                 &                 &   12.3$^{\rm j}$&       &                 &11.010\,(26) &10.681\,()   &   10.591\,()$^{\rm r}$&                    &                 &               &1.5\,(3.8)    &-3.6\,(3.8)   &-7.6\,(7.1)   &-8.6\,(13.3)  &0.61 &-1.64&\\%faint star nearby                                                                     \\
1031 & 21:35:59.92 &  57:57:51.6$^{\rm r}$& 1212 & 4212                      &                               &                 &                 &   13.4$^{\rm j}$&       &                 &10.470\,(23) &9.764 \,(27) & 9.563 \,(20)$^{\rm r}$&                    &                 &               &-2.3\,(4.7)   &-3.9\,(4.7)   &-17.1\,(7.3)  &-0.5\,(7.4)   &0.02 &-0.21&                                                                                      \\
1032 & 21:36:15.61 &  58:00:14.4$^{\rm r}$& 1213 & 4213                      &                               &                 &                 &   12.1$^{\rm j}$&       &                 &10.860\,(24) &10.702\,(29) & 10.589\,(22)$^{\rm r}$&                    &                 &               &-9.7\,(3.8)   &-8.7\,(3.8)   &-9\,(1.7)     &-4.8\,(7.4)   &0.25 &-0.58&                                                                                      \\
1033 & 21:35:56.57 &  58:01:21.5$^{\rm r}$& 1214 & 4214                      &                               &                 &                 &   12.4$^{\rm j}$&       &                 &11.707\,(23) &11.594\,(28) & 11.539\,(23)$^{\rm r}$&                    &                 &               &-5.5\,(3.8)   &-2.7\,(3.8)   &-2.8\,(0.9)   &1.4\,(2.3)    &-0.06&-0.45&                                                                                      \\
1034 & 21:36:09.13 &  58:01:36.5$^{\rm r}$& 1215 & 4215                      &                               &                 &                 &   12.9$^{\rm j}$&       &                 &11.368\,(25) &11.068\,(31) & 10.963\,(21)$^{\rm r}$&                    &                 &               &-4.9\,(3.8)   &-5.9\,(3.8)   &-3.4\,(7.3)   &-10.7\,(7.4)  &0.03 &-0.47&                                                                                      \\
1035 & 21:36:17.96 &  58:01:49.8$^{\rm r}$& 1216 & 708                       &                               &                 &                 &   11.9$^{\rm j}$&       &                 &10.643\,(23) &10.346\,(29) & 10.264\,(22)$^{\rm r}$&       F8:$^{\rm q}$&                 &               &5.8\,(13.3)   &26.4\,(13.3)  &2.6\,(1)      &7.6\,(1.4)    &-0.61&0.7  &                                                                                      \\
1036 & 21:36:13.19 &  58:03:42.2$^{\rm r}$& 1217 & 4217                      &                               &                 &                 &   12.4$^{\rm j}$&       &                 &10.832\,(24) &10.582\,(29) & 10.447\,(22)$^{\rm r}$&                    &                 &               &-4.7\,(3.8)   &0.2\,(3.8)    &-4.4\,(0.6)   &-2.2\,(2)     &-0.18&-0.1 &                                                                                      \\
1037 & 21:37:02.86 &  57:47:00.3$^{\rm r}$& 1219 & 4219                      &                               &                 &                 &   12.6$^{\rm j}$&       &                 &11.509\,(23) &11.195\,(30) & 11.155\,(21)$^{\rm r}$&                    &                 &               &4.8\,(3.8)    &27.8\,(3.8)   &13.1\,(0.5)   &24.3\,(0.7)   &-1.82&1.75 &                                                                                      \\
1038 & 21:37:30.53 &  57:48:13.8$^{\rm r}$& 1220 & 4220                      &                               &                 &                 &   14.3$^{\rm j}$&       &                 &11.179\,(23) &10.526\,(28) & 10.352\,(21)$^{\rm r}$&                    &                 &               &25.3\,(3.8)   &-0.7\,(3.8)   &24.3\,(7.4)   &5.5\,(7.5)    &     &     &                                                                                      \\
1039 & 21:37:32.26 &  57:48:48.5$^{\rm r}$& 1221 & 4221                      &                               &                 &                 &   14.4$^{\rm j}$&       &                 &11.893\,(23) &11.465\,(30) & 11.302\,(21)$^{\rm r}$&                    &                 &               &-8.4\,(3.9)   &-1.2\,(3.9)   &-5.2\,(7.4)   &-5.6\,(7.4)   &     &     &                                                                                      \\
1040 & 21:36:55.90 &  57:49:36.4$^{\rm r}$& 1222 & 4222                      &                               &                 &                 &   13.8$^{\rm j}$&       &                 &9.983 \,(24) &9.065 \,(28) & 8.797 \,(22)$^{\rm r}$&                    &                 &               &-9.1\,(4.9)   &-5.7\,(4.9)   &-9\,(7.3)     &-7.2\,(7.3)   &0.21 &-0.77&                                                                                      \\
1041 & 21:37:19.99 &  57:50:30.6$^{\rm r}$& 1223 & 4223                      &                               &                 &                 &   14.5$^{\rm j}$&       &                 &11.092\,(23) &10.296\,(28) & 10.062\,(23)$^{\rm r}$&                    &                 &               &1\,(3.8)      &1.2\,(3.8)    &9.2\,(7.4)    &-0.5\,(7.5)   &     &     &                                                                                      \\
1042 & 21:37:45.79 &  57:51:29.2$^{\rm r}$& 1224 & 4224                      &                               &                 &                 &   13.4$^{\rm j}$&       &                 &12.143\,(26) &11.934\,(28) & 11.849\,(26)$^{\rm r}$&                    &                 &               &-2.3\,(3.8)   &0.3\,(3.8)    &7.2\,(7.4)    &-8.3\,(7.4)   &-0.31&-0.56&                                                                                      \\
1043 & 21:37:50.35 &  57:52:18.8$^{\rm r}$& 1225 & 4225                      &                               &                 &                 &     14$^{\rm j}$&       &                 &12.677\,(29) &12.415\,(30) & 12.407\,(26)$^{\rm r}$&                    &                 &               &-1.3\,(3.8)   &-3.6\,(3.8)   &-5.6\,(7.8)   &2.1\,(7.8)    &0.01 &-0.28&                                                                                      \\
1044 & 21:37:09.54 &  57:53:09.5$^{\rm r}$& 1226 & 711                       &                               &                 &                 &   11.5$^{\rm j}$&       &                 &10.678\,(25) &10.527\,(32) &   10.430\,()$^{\rm r}$&        A0$^{\rm q}$&                 &               &-17.8\,(2.7)  &0.8\,(2.7)    &-5.9\,(0.9)   &-1.3\,(0.8)   &0.12 &-0.22&                                                                                      \\
1045 & 21:36:37.69 &  57:55:36.5$^{\rm r}$& 1227 & 4227                      &                               &                 &                 &   13.8$^{\rm j}$&       &                 &9.750 \,(23) &8.650 \,(42) & 8.387 \,(22)$^{\rm r}$&                    &                 &               &-0.3\,(4.7)   &-1.9\,(4.7)   &43.2\,(7.3)   &-3.5\,(7.4)   &0.25 &0.1  &                                                                                      \\
1046 & 21:36:51.21 &  57:56:45.4$^{\rm r}$& 1228 & 4228                      &                               &                 &                 &   13.4$^{\rm j}$&       &                 &12.197\,(23) &11.968\,(28) & 11.909\,(23)$^{\rm r}$&                    &                 &               &-0.9\,(3.8)   &-6\,(3.8)     &-1.6\,(7.3)   &-6.2\,(7.3)   &-0.16&-0.6 &                                                                                      \\
1047 & 21:37:15.16 &  57:54:47.4$^{\rm r}$& 1229 & 4229                      &                               &                 &                 &   14.5$^{\rm j}$&       &                 &13.075\,(25) &12.867\,(38) & 12.801\,(38)$^{\rm r}$&                    &                 &               &-3.2\,(3.8)   &-0.7\,(3.8)   &2.8\,(7.3)    &-3.7\,(7.3)   &     &     &                                                                                      \\
1048 & 21:37:18.61 &  57:54:42.9$^{\rm r}$& 1230 & 4230                      &                               &                 &                 &   14.4$^{\rm j}$&       &                 &11.210\,(21) &10.494\,(30) & 10.218\,(26)$^{\rm r}$&                    &                 &               &3.4\,(3.8)    &-12.9\,(3.8)  &12.9\,(7.7)   &-24.1\,(7.8)  &     &     &                                                                                      \\
1049 & 21:37:29.86 &  57:55:49.2$^{\rm r}$& 1231 & 4231                      &                               &                 &                 &   14.4$^{\rm j}$&       &                 &11.716\,(21) &11.071\,(30) & 10.889\,(26)$^{\rm r}$&                    &                 &               &-15.9\,(3.8)  &-8.4\,(3.8)   &-18\,(7.9)    &-11.1\,(7.3)  &     &     &                                                                                      \\
1050 & 21:37:47.40 &  57:55:21.8$^{\rm r}$& 1232 & 4232                      &                               &                 &                 &   13.3$^{\rm j}$&       &                 &11.513\,(22) &11.040\,(30) & 10.945\,(27)$^{\rm r}$&                    &                 &               &17.2\,(3.8)   &29.5\,(3.8)   &18.6\,(6.9)   &37.5\,(6.9)   &-2.01&2.56 &                                                                                      \\
1051 & 21:37:09.69 &  57:57:19.2$^{\rm r}$& 1233 & 4233                      &                               &                 &                 &   14.5$^{\rm j}$&       &                 &12.600\,(24) &12.216\,(33) & 12.096\,(31)$^{\rm r}$&                    &                 &               &-2.7\,(3.8)   &1.5\,(3.8)    &-6.6\,(7.3)   &2.3\,(7.2)    &     &     &                                                                                      \\
1052 & 21:37:15.08 &  57:57:34.5$^{\rm r}$& 1234 & 4234                      &                               &                 &                 &   14.7$^{\rm j}$&       &                 &12.714\,(25) &12.284\,(36) & 12.208\,(32)$^{\rm r}$&                    &                 &               &-1.7\,(3.8)   &1.5\,(3.8)    &-6.8\,(8.1)   &0.6\,(8.8)    &     &     &                                                                                      \\
1053 & 21:37:16.54 &  57:58:21.3$^{\rm r}$& 1235 & 4235                      &                               &                 &                 &   14.7$^{\rm j}$&       &                 &12.940\,(24) &12.612\,(35) & 12.528\,(34)$^{\rm r}$&                    &                 &               &1.9\,(3.8)    &-0.4\,(3.8)   &0.7\,(7.2)    &3.4\,(7.5)    &-0.55&0.31 &                                                                                      \\
1054 & 21:37:24.92 &  57:58:48.3$^{\rm r}$& 1236 & 4236                      &                               &                 &                 &   14.3$^{\rm j}$&       &                 &9.851 \,(21) &8.809 \,(29) & 8.445 \,(24)$^{\rm r}$&                    &                 &               &-5.8\,(4.7)   &-5.3\,(4.7)   &-29.4\,(7.1)  &-21.8\,(7.2)  &0.44 &-0.46&                                                                                      \\
1055 & 21:37:30.92 &  57:57:39.2$^{\rm r}$& 1237 & 4237                      &                               &                 &                 &   12.2$^{\rm j}$&       &                 &9.991 \,(21) &9.381 \,(28) & 9.211 \,(24)$^{\rm r}$&                    &                 &               &14.2\,(4.7)   &0.8\,(4.7)    &61.9\,(7.1)   &15.8\,(7.2)   &-1.48&-0.5 &                                                                                      \\
1056 & 21:37:41.39 &  57:57:28.7$^{\rm r}$& 1238 & 4238                      &                               &                 &                 &   14.6$^{\rm j}$&       &                 &12.809\,(24) &12.563\,(33) & 12.476\,(31)$^{\rm r}$&                    &                 &               &-4.2\,(3.8)   &1\,(3.8)      &-10.1\,(7.2)  &3.3\,(7.2)    &     &     &                                                                                      \\
1057 & 21:37:47.32 &  57:57:39.1$^{\rm r}$& 1239 & 4239                      &                               &                 &                 &   14.3$^{\rm j}$&       &                 &7.796 \,(21) &6.668 \,(20) & 6.188 \,(23)$^{\rm r}$&                    &                 &               &2.1\,(4.7)    &-0.2\,(4.7)   &7.1\,(7.2)    &-0.9\,(7.3)   &     &     &                                                                                      \\
1058 & 21:37:27.27 &  57:59:05.7$^{\rm r}$& 1240 & 4240                      &                               &                 &                 &   14.7$^{\rm j}$&       &                 &12.843\,(25) &12.486\,(35) & 12.350\,(33)$^{\rm r}$&                    &                 &               &-4.1\,(3.8)   &-7\,(3.8)     &-25.2\,(8.2)  &-24.9\,(8.1)  &     &     &                                                                                      \\
1059 & 21:37:32.59 &  57:59:47.1$^{\rm r}$& 1241 & 4241                      &                               &                 &                 &   14.6$^{\rm j}$&       &                 &12.336\,(49) &11.895\,(57) & 11.771\,(47)$^{\rm r}$&                    &                 &               &0.4\,(7)      &-0.2\,(7)     &              &              &     &     &                                                                                      \\
1060 & 21:37:44.25 &  57:59:40.1$^{\rm r}$& 1242 & 4242                      &                               &                 &                 &   14.7$^{\rm j}$&       &                 &11.548\,(21) &10.847\,(30) & 10.657\,(26)$^{\rm r}$&                    &                 &               &0.9\,(3.8)    &1.3\,(3.8)    &10.8\,(7.8)   &12.9\,(7.9)   &     &     &                                                                                      \\
1061 & 21:36:38.81 &  57:58:47.9$^{\rm r}$& 1243 & 4243                      &                               &                 &                 &   13.2$^{\rm j}$&       &                 &11.860\,(23) &11.472\,(28) & 11.370\,(23)$^{\rm r}$&                    &                 &               &1.7\,(3.8)    &-6\,(3.8)     &2.3\,(7.2)    &-4.4\,(7.3)   &-0.72&-1.17&                                                                                      \\
1062 & 21:36:32.67 &  58:02:15.8$^{\rm r}$& 1244 & 709                       &                               &                 &                 &   10.4$^{\rm j}$&       &                 &9.707 \,(23) &9.628 \,(28) & 9.603 \,(23)$^{\rm r}$&        A0$^{\rm q}$&                 &               &-4.6\,(2)     &7.9\,(2)      &-0.2\,(0.7)   &1.4\,(1.1)    &-0.68&0.14 &                                                                                      \\
1063 & 21:36:48.87 &  58:02:36.5$^{\rm r}$& 1245 & 4245                      &                               &                 &                 &   13.7$^{\rm j}$&       &                 &10.575\,(23) &9.953 \,(26) & 9.748 \,(23)$^{\rm r}$&                    &                 &               &-2\,(4.7)     &-9.2\,(4.7)   &-18\,(7.2)    &-21.4\,(7.3)  &-0.04&0.14 &                                                                                      \\
1064 & 21:36:56.99 &  58:01:57.6$^{\rm r}$& 1246 & 4246                      &                               &                 &                 &     13$^{\rm j}$&       &                 &11.781\,(24) &11.369\,(28) & 11.328\,(23)$^{\rm r}$&                    &                 &               &6.5\,(3.8)    &3.9\,(3.8)    &3.1\,(7.3)    &1.6\,(7.4)    &-1.03&0.53 &                                                                                      \\
1065 & 21:37:09.59 &  58:02:50.2$^{\rm r}$& 1247 & 4247                      &                               &                 &                 &     13$^{\rm j}$&       &                 &10.468\,(19) &9.796 \,(28) & 9.650 \,(24)$^{\rm r}$&                    &                 &               &-7.8\,(4.7)   &-7\,(4.7)     &-40.2\,(8.3)  &12.1\,(8.3)   &-0.13&-0-30&                                                                                      \\
1066 & 21:37:26.83 &  58:01:09.9$^{\rm r}$& 1248 & 4248                      &                               &                 &                 &     14$^{\rm j}$&       &                 &12.311\,(21) &12.088\,(30) & 11.936\,(27)$^{\rm r}$&                    &                 &               &-1.6\,(3.8)   &-0.7\,(3.8)   &-2.9\,(7.9)   &-8.9\,(7.9)   &-0.12&0.11 &                                                                                      \\
1067 & 21:37:12.68 &  58:04:20.8$^{\rm r}$& 1249 & 4249                      &                               &                 &                 &   14.3$^{\rm j}$&       &                 &12.292\,(29) &11.855\,(48) & 11.749\,(30)$^{\rm r}$&                    &                 &               &-3\,(3.8)     &5.1\,(3.8)    &-17.9\,(7.3)  &40.4\,(7.4)   &-0.8 &0.19 &                                                                                      \\
1068 & 21:37:22.91 &  58:04:38.9$^{\rm r}$& 1250 & 4250                      &                               &                 &                 &   12.8$^{\rm j}$&       &                 &8.559 \,(26) &7.696 \,(38) & 7.411 \,(18)$^{\rm r}$&                    &                 &               &-1.5\,(4.7)   &4.7\,(4.7)    &11\,(8)       &74.7\,(8.1)   &-0.69&-0.06&                                                                                      \\
1069 & 21:34:18.55 &  57:11:11.4$^{\rm r}$& 1251 & 4251                      &                               &                 &                 &   13.8$^{\rm j}$&       &                 &11.700\,(23) &11.415\,(32) & 11.308\,(27)$^{\rm r}$&                    &                 &               &22.7\,(4.1)   &-14.2\,(4.1)  &              &              &     &     &                                                                                      \\
1070 & 21:34:17.01 &  57:12:46.8$^{\rm r}$& 1253 & 4253                      &                               &                 &                 &   13.5$^{\rm j}$&       &                 &11.012\,(22) &10.398\,(30) & 10.242\,(24)$^{\rm r}$&                    &                 &               &52.3\,(4.1)   &-1.1\,(4.1)   &58.2\,(7)     &10.8\,(7)     &     &     &                                                                                      \\
1071 & 21:33:37.57 &  57:14:50.9$^{\rm r}$& 1254 & 4254                      &                               &                 &                 &   13.5$^{\rm j}$&       &                 &11.364\,(22) &11.029\,(26) & 10.923\,(22)$^{\rm r}$&                    &                 &               &-3.4\,(4.1)   &7.7\,(4.1)    &-3.5\,(7)     &12\,(7)       &-0.3 &0.42 &                                                                                      \\
1072 & 21:34:12.74 &  57:14:45.6$^{\rm r}$& 1255 & 4255                      &                               &                 &                 &   12.4$^{\rm j}$&       &                 &11.423\,(22) &11.282\,(29) & 11.194\,(27)$^{\rm r}$&                    &                 &               &-1.6\,(2.7)   &0\,(2.7)      &-5.4\,(0.6)   &-3.8\,(0.5)   &-0.12&0    &                                                                                      \\
1073 & 21:34:18.80 &  57:14:31.1$^{\rm r}$& 1256 & 4256                      &                               &                 &                 &   14.5$^{\rm j}$&       &                 &12.815\,(31) &12.258\,(49) & 12.210\,(36)$^{\rm r}$&                    &                 &               &1.4\,(5.6)    &-1.6\,(5.6)   &              &              &-0.34&0.5  &near 1664                                                                          \\
1074 & 21:34:27.83 &  57:14:12.9$^{\rm r}$& 1257 & 4257                      &                               &                 &                 &  14.11$^{\rm j}$&       &                 &12.291\,(24) &11.757\,(31) & 11.636\,(29)$^{\rm r}$&                    &                 &               &-31.6\,(4.1)  &-7.7\,(4.1)   &-33.8\,(6.8)  &-4.7\,(6.8)   &3.16 &0.2  &                                                                                      \\
1075 & 21:34:10.57 &  57:15:20.7$^{\rm r}$& 1258 & 4258                      &                               &                 &                 &   13.6$^{\rm j}$&       &                 &11.634\,(34) &11.296\,(44) & 11.179\,(35)$^{\rm r}$&                    &                 &               &-33.7\,(4.1)  &-33.8\,(4.1)  &              &              &-1.49&0.52 &                                                                                      \\
1076 & 21:33:37.17 &  57:18:08.2$^{\rm r}$& 1259 & 4259                      &                               &                 &                 &   13.9$^{\rm j}$&       &                 &10.358\,(22) &9.556 \,(28) & 9.341 \,(22)$^{\rm r}$&                    &                 &               &-5.8\,(5.2)   &-2.5\,(5.2)   &-37.5\,(5.7)  &21.8\,(5.9)   &0.29 &-0.51&                                                                                      \\
1077 & 21:33:48.76 &  57:17:55.0$^{\rm r}$& 1260 & 4260                      &                               &                 &                 &   12.5$^{\rm j}$&       &                 &11.070\,(22) &10.862\,(29) & 10.747\,(26)$^{\rm r}$&                    &                 &               &5.7\,(2.7)    &0.6\,(2.7)    &2.5\,(6.3)    &1.2\,(3.8)    &-0.54&0.42 &                                                                                      \\
1078 & 21:33:54.71 &  57:17:37.0$^{\rm r}$& 1261 & 414                       &                               &                 &                 &    9.5$^{\rm j}$&       &                 &8.977 \,(22) &8.920 \,(29) & 8.921 \,(24)$^{\rm r}$&        A0$^{\rm q}$&                 &               &1.4\,(1.2)    &0.1\,(1.2)    &-0.2\,(0.7)   &-1.5\,(0.6)   &-1.16&0.15 &                                                                                      \\
1079 & 21:34:04.45 &  57:17:56.4$^{\rm r}$& 1262 & 416                       &                               &                 &                 &   11.8$^{\rm j}$&       &                 &10.910\,(23) &10.642\,(31) & 10.550\,(26)$^{\rm r}$&        F0$^{\rm q}$&                 &               &12.9\,(2.7)   &2.5\,(2.7)    &2.7\,(0.7)    &1.4\,(0.6)    &-1.16&0.11 &                                                                                      \\
1080 & 21:34:09.93 &  57:17:02.4$^{\rm r}$& 1263 & 4263                      &                               &                 &                 &   14.3$^{\rm j}$&       &                 &10.221\,(22) &9.293 \,(29) & 9.049 \,(24)$^{\rm r}$&                    &                 &               &-6.5\,(5.1)   &-1.1\,(5.1)   &34.7\,(6.9)   &3\,(6.9)      &-0.25&0.53 &                                                                                      \\
1081 & 21:34:26.22 &  57:16:30.9$^{\rm r}$& 1264 & 4264                      &                               &                 &                 &     13$^{\rm j}$&       &                 &11.567\,(28) &11.454\,(43) & 11.292\,(36)$^{\rm r}$&                    &                 &               &9.5\,(2.7)    &3\,(2.7)      &-4.6\,(1.7)   &-2.3\,(1.3)   &-0.07&-0.1 &                                                                                      \\
1082 & 21:34:39.65 &  57:16:06.3$^{\rm r}$& 1265 & 4265                      &                               &                 &                 &     12$^{\rm j}$&       &                 &10.512\,(24) &10.174\,(26) & 10.049\,(22)$^{\rm r}$&                    &                 &               &-14.1\,(2.7)  &-11.5\,(2.7)  &-8.4\,(0.8)   &-7.7\,(2)     &0.18 &-0.48&                                                                                      \\
1083 & 21:33:52.89 &  57:20:22.5$^{\rm r}$& 1266 & 4266                      &                               &                 &                 &   12.5$^{\rm j}$&       &                 &11.019\,(22) &10.813\,(29) & 10.713\,(26)$^{\rm r}$&                    &                 &               &11.7\,(2.7)   &1.1\,(2.7)    &-7.3\,(2.2)   &-3.2\,(4.9)   &-0.23&0.05 &                                                                                      \\
1084 & 21:33:52.42 &  57:20:31.5$^{\rm r}$& 1267 & 4267                      &                               &                 &                 &   14.1$^{\rm j}$&       &                 &12.148\,(24) &11.760\,(31) & 11.624\,(27)$^{\rm r}$&                    &                 &               &              &              &              &              &0.35 &-0.18&                                                                                      \\
1085 & 21:34:25.34 &  57:18:44.4$^{\rm r}$& 1268 & 4268                      &                               &                 &                 &   13.5$^{\rm j}$&       &                 &11.869\,(23) &11.678\,(33) & 11.573\,(28)$^{\rm r}$&                    &                 &               &-12.3\,(4.1)  &-3.1\,(4.1)   &-31.4\,(6.7)  &-22.6\,(6.7)  &-0.23&-0.03&                                                                                      \\
1086 & 21:34:00.52 &  57:21:04.0$^{\rm r}$& 1269 & 4269                      &                               &                 &                 &   13.9$^{\rm j}$&       &                 &11.897\,(23) &11.364\,(29) & 11.264\,(26)$^{\rm r}$&                    &                 &               &9.9\,(4.1)    &9.7\,(4.1)    &14.1\,(6.8)   &8.8\,(6.8)    &     &     &                                                                                      \\
1087 & 21:34:33.78 &  57:19:54.4$^{\rm r}$& 1270 & 420                       &                               &                 &                 &   11.2$^{\rm j}$&       &                 &10.705\,(24) &10.607\,(26) & 10.577\,(23)$^{\rm r}$&        A0$^{\rm q}$&                 &               &13.4\,(2)     &-1.5\,(2)     &5\,(1.3)      &0.9\,(0.9)    &-1.11&0.64 &RA [m] wrong                                                                        \\
1088 & 21:33:39.65 &  57:22:50.3$^{\rm r}$& 1271 & 4271                      &                               &                 &                 &   12.2$^{\rm j}$&       &                 &10.070\,(24) &9.450 \,(28) & 9.296 \,(20)$^{\rm r}$&                    &                 &               &-1.1\,(5.1)   &-3.8\,(5.1)   &-9.6\,(7.9)   &-84.2\,(7.9)  &-0.34&-0.11&                                                                                      \\
1089 & 21:34:13.58 &  57:22:42.9$^{\rm r}$& 1272 & 417                       &                               &                 &                 &   10.3$^{\rm j}$&       &                 &9.519 \,(22) &9.436 \,(28) & 9.334 \,(20)$^{\rm r}$&        A0$^{\rm q}$&                 &               &3.1\,(1.3)    &1.8\,(1.3)    &4.4\,(0.7)    &-0.9\,(0.6)   &-1.2 &0.31 &                                                                                      \\
1090 & 21:34:23.09 &  57:23:06.5$^{\rm r}$& 1273 & 4273                      &                               &                 &                 &   14.3$^{\rm j}$&       &                 &12.391\,(24) &12.008\,(27) & 11.929\,(23)$^{\rm r}$&                    &                 &               &2.8\,(4.1)    &-2.5\,(4.1)   &0.6\,(6.8)    &-2.4\,(6.7)   &     &     &                                                                                      \\
1091 & 21:34:26.03 &  57:23:03.9$^{\rm r}$& 1274 & 4274                      &                               &                 &                 &   13.1$^{\rm j}$&       &                 &10.411\,(24) &9.763 \,(28) & 9.555 \,(21)$^{\rm r}$&                    &                 &               &0.6\,(5.7)    &5.4\,(5.7)    &38.5\,(6.5)   &28\,(6.5)     &-0.39&0.14 &                                                                                      \\
1092 & 21:34:03.93 &  57:24:46.7$^{\rm r}$& 1275 & 4275                      &                               &                 &                 &   13.7$^{\rm j}$&       &                 &11.219\,(46) &10.641\,(46) &   10.333\,()$^{\rm r}$&                    &                 &               &19\,(4.1)     &-12.2\,(4.1)  &73.3\,(7.3)   &-42.8\,(7.3)  &-0.31&-0.45&                                                                                      \\
1093 & 21:34:34.84 &  57:23:53.9$^{\rm r}$& 1276 & 4276                      &                               &                 &                 &   14.1$^{\rm j}$&       &                 &12.125\,(26) &11.767\,(29) & 11.627\,(23)$^{\rm r}$&                    &                 &               &-8\,(4.1)     &2.2\,(4.1)    &-12.9\,(6.7)  &2\,(6.7)      &0.15 &-0.01&                                                                                      \\
1094 & 21:34:28.30 &  57:25:34.1$^{\rm r}$& 1277 & 419                       &                               &                 &                 &     11$^{\rm j}$&       &                 &10.385\,(24) &10.271\,(27) & 10.249\,(23)$^{\rm r}$&        B8$^{\rm q}$&                 &               &-4.1\,(1.5)   &-4.3\,(1.5)   &4.7\,(9.7)    &-2.6\,(9.6)   &-0.32&-0.2 &                                                                                      \\
1095 & 21:34:20.43 &  57:26:12.3$^{\rm r}$& 1278 & 4278                      &                               &                 &                 &   14.6$^{\rm j}$&       &                 &12.750\,(24) &12.382\,(28) & 12.282\,(24)$^{\rm r}$&                    &                 &               &0.3\,(4.1)    &1.1\,(4.1)    &-7.2\,(7.6)   &9.1\,(7.6)    &     &     &                                                                                      \\
1096 & 21:34:04.18 &  57:28:23.6$^{\rm r}$& 1279 & 4279                      &                               &                 &                 &   14.6$^{\rm j}$&       &                 &12.604\,(22) &12.084\,(28) & 11.992\,(24)$^{\rm r}$&                    &                 &               &-3.9\,(4.1)   &-2.6\,(4.1)   &-0.9\,(7.2)   &-0.5\,(7.3)   &     &     &                                                                                      \\
1097 & 21:34:09.08 &  57:29:19.4$^{\rm r}$& 1280 & 4280                      &                               &                 &                 &     13$^{\rm j}$&       &                 &11.787\,(24) &11.457\,(27) & 11.401\,(23)$^{\rm r}$&                    &                 &               &-7.9\,(4)     &7.9\,(4)      &-32.8\,(6.5)  &30\,(6.5)     &0.23 &0.09 &                                                                                      \\
1098 & 21:33:42.81 &  57:31:27.3$^{\rm r}$& 1281 & 4281                      &                               &                 &                 &   13.5$^{\rm j}$&       &                 &11.954\,(24) &11.701\,(31) & 11.608\,(24)$^{\rm r}$&                    &                 &               &-3.7\,(4)     &-0.1\,(4)     &-8.1\,(6.7)   &5.6\,(6.7)    &-0.48&-0.21&                                                                                      \\
1099 & 21:34:08.99 &  57:34:01.5$^{\rm r}$& 1282 & 4282                      &                               &                 &                 &   12.6$^{\rm j}$&       &                 &11.160\,(27) &10.848\,(31) & 10.776\,(24)$^{\rm r}$&                    &                 &               &7.9\,(4)      &19.3\,(4)     &8.9\,(0.7)    &13\,(18.3)    &-1.31&0.67 &                                                                                      \\
1100 & 21:33:32.24 &  57:35:32.5$^{\rm r}$& 1283 & 4283                      &                               &                 &                 &   12.5$^{\rm j}$&       &                 &11.264\,(23) &10.854\,(27) & 10.804\,(23)$^{\rm r}$&                    &                 &               &15.2\,(3.8)   &11.9\,(3.8)   &19.8\,(6.9)   &48.6\,(6.9)   &-1.42&1.16 &                                                                                      \\
1101 & 21:34:00.71 &  57:35:49.0$^{\rm r}$& 1284 & 415                       &                               &                 &                 &   10.9$^{\rm j}$&       &                 &9.830 \,(24) &9.543 \,(26) & 9.495 \,(21)$^{\rm r}$&        F8$^{\rm q}$&                 &               &0.3\,(2)      &-0.4\,(2)     &-2.1\,(1)     &2.1\,(0.8)    &-0.38&0.33 &                                                                                      \\
1102 & 21:33:50.53 &  57:36:38.0$^{\rm r}$& 1285 & 4285                      &                               &                 &                 &   12.7$^{\rm j}$&       &                 &11.234\,(22) &10.919\,(28) & 10.800\,(21)$^{\rm r}$&                    &                 &               &-5.8\,(4)     &12.7\,(4)     &-5.3\,(12.2)  &-3.8\,(12.2)  &-0.48&0.81 &                                                                                      \\
1103 & 21:34:02.45 &  57:36:38.3$^{\rm r}$& 1286 & 4286                      &                               &                 &                 &   13.6$^{\rm j}$&       &                 &12.035\,(24) &11.736\,(30) & 11.642\,(24)$^{\rm r}$&                    &                 &               &-3.3\,(4)     &-0.7\,(4)     &-11.5\,(6.7)  &11.8\,(6.7)   &0.02 &-0.13&                                                                                      \\
1104 & 21:34:04.30 &  57:36:55.2$^{\rm r}$& 1287 & 4287                      &                               &                 &                 &   13.6$^{\rm j}$&       &                 &11.934\,(22) &11.710\,(27) & 11.573\,(23)$^{\rm r}$&                    &                 &               &-3.9\,(4)     &4.8\,(4)      &-4.5\,(6.7)   &7.1\,(6.7)    &-0.19&0.07 &                                                                                      \\
1105 & 21:33:29.49 &  57:40:31.9$^{\rm r}$& 1288 & 413                       &                               &                 &                 &    9.8$^{\rm j}$&       &                 &9.035 \,(25) &8.959 \,(28) & 8.956 \,(21)$^{\rm r}$&        B8$^{\rm q}$&                 &               &9.3\,(1.6)    &-3.7\,(1.6)   &-3.9\,(0.7)   &-4.1\,(1.2)   &-0.34&-0.39&                                                                                      \\
1106 & 21:33:36.61 &  57:41:03.1$^{\rm r}$& 1289 & 4289                      &                               &                 &                 &   13.9$^{\rm j}$&       &                 &11.907\,(27) &11.481\,(33) & 11.382\,(25)$^{\rm r}$&                    &                 &               &1\,(3.8)      &5.5\,(3.8)    &-60.6\,(7)    &33.8\,(7.1)   &-1.02&0.18 &                                                                                      \\
1107 & 21:33:55.43 &  57:40:45.0$^{\rm r}$& 1290 & 4290                      &                               &                 &  13.01$^{\rm l}$&  11.27$^{\rm l}$&       &                 &7.567 \,(20) &6.811 \,(23) & 6.540 \,(20)$^{\rm r}$&                    &                 &               &-6.2\,(10.7)  &15.7\,(10.7)  &-7.2\,(0.6)   &-1.5\,(1.3)   &0.31 &-0.2 &                                                                                      \\
1108 & 21:34:09.08 &  57:41:40.7$^{\rm r}$& 1291 & 4291                      &                               &                 &                 &   12.1$^{\rm j}$&       &                 &10.715\,(24) &10.283\,(26) & 10.178\,(23)$^{\rm r}$&                    &                 &               &35.2\,(5.5)   &17.1\,(5.5)   &34.8\,(0.5)   &20.6\,(4.1)   &-4.02&2.09 &                                                                                      \\
1109 & 21:34:06.89 &  57:42:29.7$^{\rm r}$& 1292 & 4292                      &                               &                 &                 &   13.8$^{\rm j}$&       &                 &11.749\,(22) &11.364\,(26) & 11.254\,(20)$^{\rm r}$&                    &                 &               &-2.9\,(4)     &13.3\,(4)     &-5.5\,(7.4)   &3.5\,(7.4)    &     &     &                                                                                      \\
1110 & 21:34:21.19 &  57:41:25.7$^{\rm r}$& 1294 & 4294                      &                               &                 &                 &   14.2$^{\rm j}$&       &                 &10.411\,(21) &9.590 \,(29) & 9.347 \,(22)$^{\rm r}$&                    &                 &               &-9.9\,(11.4)  &1.4\,(11.4)   &-38.3\,(7.2)  &17.5\,(7.3)   &1.1  &-0.79&                                                                                      \\
1111 & 21:34:17.72 &  57:43:00.9$^{\rm r}$& 1295 & 418                       &                               &                 &                 &     10$^{\rm j}$&       &                 &8.031 \,(26) &7.491 \,(23) & 7.413 \,(24)$^{\rm r}$&        K0$^{\rm q}$&                 &               &-45.6\,(1.7)  &6.6\,(1.7)    &-44.6\,(0.7)  &1.5\,(1.3)    &     &     &                                                                                      \\
1112 & 21:34:54.72 &  57:15:40.8$^{\rm r}$& 1298 & 4298                      &                               &                 &                 &   11.9$^{\rm j}$&       &                 &9.530 \,(22) &8.852 \,(26) & 8.724 \,(22)$^{\rm r}$&                    &                 &               &10.4\,(2.7)   &-7.5\,(2.7)   &5.9\,(9.8)    &-2.4\,(8.2)   &-1.28&0.95 &                                                                                      \\
1113 & 21:35:03.17 &  57:15:26.8$^{\rm r}$& 1299 & 4299                      &                               &                 &                 &   13.4$^{\rm j}$&       &                 &12.023\,(55) &11.811\,(31) & 11.697\,(26)$^{\rm r}$&                    &                 &               &-20.6\,(4.1)  &-14.1\,(4.1)  &-69.8\,(6.9)  &-55.6\,(6.9)  &-0.03&0.14 &                                                                                      \\
1114 & 21:35:13.78 &  57:14:03.3$^{\rm r}$& 1300 & 4300                      &                               &                 &                 &   13.4$^{\rm j}$&       &                 &10.597\,(24) &9.845 \,(26) & 9.725 \,(22)$^{\rm r}$&                    &                 &               &-0.2\,(5.1)   &5\,(5.1)      &4.6\,(7.1)    &4.7\,(7.1)    &0.21 &0.35 &                                                                                      \\
1115 & 21:35:14.74 &  57:12:50.3$^{\rm j}$& 1301 & 4301                      &                               &                 &                 &   14.5$^{\rm j}$&       &                 &             &             &                       &                    &                 &               &-13.3\,(7.4)  &-5.8\,(7.4)   &              &              &     &     &no star                                                                          \\
1116 & 21:35:24.04 &  57:10:52.9$^{\rm r}$& 1302 & 4302                      &                               &                 &                 &   12.8$^{\rm j}$&       &                 &10.102\,(24) &9.318 \,(29) & 9.154 \,(21)$^{\rm r}$&                    &                 &               &9.9\,(5.1)    &11.9\,(5.1)   &-1.5\,(7.2)   &-10.5\,(7.2)  &     &     &                                                                                      \\
1117 & 21:35:30.17 &  57:10:25.9$^{\rm r}$& 1303 & 4303                      &                               &                 &                 &   12.9$^{\rm j}$&       &                 &11.348\,(26) &11.035\,(28) & 10.944\,(23)$^{\rm r}$&                    &                 &               &3.6\,(4.1)    &9.1\,(4.1)    &27.8\,(7.4)   &45.9\,(7.4)   &0.16 &-0.5 &                                                                                      \\
1118 & 21:35:34.00 &  57:12:27.6$^{\rm r}$& 1304 & 4304                      &                               &                 &                 &   12.1$^{\rm j}$&       &                 &11.161\,(22) &10.932\,(26) & 10.842\,(21)$^{\rm r}$&                    &                 &               &3.2\,(4.1)    &-29.4\,(4.1)  &-5.6\,(1)     &-11.6\,(1.4)  &0.11 &-0.9 &                                                                                      \\
1119 & 21:35:45.75 &  57:13:20.2$^{\rm r}$& 1305 & 4305                      &                               &                 &                 &   12.2$^{\rm j}$&       &                 &11.443\,(24) &11.287\,(28) & 11.236\,(21)$^{\rm r}$&                    &                 &               &-5.4\,(4.1)   &4.5\,(4.1)    &-4.5\,(1.1)   &-4.6\,(3.1)   &-0.33&-0.33&                                                                                      \\
1120 & 21:35:51.20 &  57:14:49.0$^{\rm r}$& 1306 & 4306                      &                               &                 &                 &   14.4$^{\rm j}$&       &                 &11.265\,(22) &10.541\,(26) & 10.332\,(21)$^{\rm r}$&                    &                 &               &-8.6\,(4.1)   &1.1\,(4.1)    &-10.8\,(6.8)  &6\,(6.8)      &0.42 &-0.38&                                                                                      \\
1121 & 21:35:36.66 &  57:16:12.0$^{\rm r}$& 1307 & 4307                      &                               &                 &                 &   13.1$^{\rm j}$&       &                 &10.223\,(22) &9.490 \,(28) & 9.292 \,(21)$^{\rm r}$&                    &                 &               &-4.7\,(5.1)   &3.3\,(5.1)    &-3.5\,(7)     &1.8\,(7)      &0.02 &0.36 &                                                                                      \\
1122 & 21:34:52.09 &  57:17:55.5$^{\rm r}$& 1308 & 4308                      &                               &                 &                 &   14.2$^{\rm j}$&       &                 &12.062\,(22) &11.656\,(29) & 11.573\,(23)$^{\rm r}$&                    &                 &               &-18.7\,(4.1)  &1.6\,(4.1)    &-22.4\,(6.8)  &-1.5\,(6.8)   &1.16 &0.42 &                                                                                      \\
1123 & 21:34:51.93 &  57:19:33.0$^{\rm r}$& 1309 & 4309                      &                               &                 &                 &   12.7$^{\rm j}$&       &                 &11.248\,(24) &10.932\,(28) & 10.865\,(25)$^{\rm r}$&                    &                 &               &5\,(2.7)      &-6.4\,(2.7)   &2.4\,(5.3)    &-8.7\,(3.8)   &0.27 &-0.79&                                                                                      \\
1124 & 21:35:24.90 &  57:19:00.9$^{\rm r}$& 1310 & 4310                      &                               &                 &                 &   14.2$^{\rm j}$&       &                 &12.661\,(24) &12.426\,(31) & 12.286\,(24)$^{\rm r}$&                    &                 &               &0.3\,(4.1)    &7.5\,(4.1)    &4.3\,(6.7)    &8.6\,(6.7)    &-0.52&0.27 &                                                                                      \\
1125 & 21:35:27.87 &  57:19:06.8$^{\rm r}$& 1311 & 4311                      &                               &                 &                 &   14.3$^{\rm j}$&       &                 &10.229\,(24) &9.284 \,(28) & 8.968 \,(20)$^{\rm r}$&                    &                 &               &-4.9\,(5.1)   &-1.9\,(5.1)   &-8.4\,(6.9)   &-7\,(6.9)     &0.22 &-0.12&                                                                                      \\
1126 & 21:35:56.55 &  57:20:52.9$^{\rm r}$& 1312 & 4312                      &                               &  10.37$^{\rm l}$&  10.67$^{\rm f}$&  10.34$^{\rm e}$&       &                 &9.718 \,(31) &9.636 \,(37) &   9.594 \,()$^{\rm r}$&        B4$^{\rm e}$&                 &  1.5$^{\rm e}$&-1.8\,(2)     &3\,(2)        &-3\,(0.8)     &-4.6\,(1.7)   &-0.25&-0.19&                                                                                      \\
1127 & 21:36:05.71 &  57:20:05.6$^{\rm r}$& 1313 & 4313                      &                               &                 &  15.05$^{\rm l}$&   12.8$^{\rm l}$&       &                 &7.314 \,(18) &6.140 \,(27) & 5.774 \,(18)$^{\rm r}$&                    &                 &               &-5\,(5.1)     &1.1\,(5.1)    &-0.7\,(5.9)   &-2.8\,(6)     &0.31 &-0.33&                                                                                      \\
1128 & 21:36:07.68 &  57:20:34.8$^{\rm r}$& 1314 & 4314                      &                               &                 &                 &   12.7$^{\rm j}$&       &                 &10.465\,(22) &9.849 \,(29) & 9.690 \,(20)$^{\rm r}$&                    &                 &               &4.3\,(5.1)    &11\,(5.1)     &-5.2\,(6.8)   &0.7\,(6.8)    &-1.33&1.07 &                                                                                      \\
1129 & 21:34:55.98 &  57:22:54.2$^{\rm r}$& 1315 & 423                       &                               &                 &                 &   10.3$^{\rm j}$&       &                 &8.273 \,(30) &7.763 \,(53) & 7.599 \,(20)$^{\rm r}$&        G5$^{\rm q}$&                 &               &-11.9\,(1.5)  &-9.4\,(1.5)   &-9.2\,(0.5)   &-9.3\,(1)     &0.61 &-0.66&                                                                                      \\
1130 & 21:34:55.23 &  57:23:11.5$^{\rm r}$& 1316 & 4316                      &                               &                 &                 &   13.5$^{\rm j}$&       &                 &12.145\,(24) &11.857\,(32) & 11.768\,(25)$^{\rm r}$&                    &                 &               &-16.7\,(11.4) &3.2\,(11.4)   &-42.7\,(6.6)  &7.5\,(6.8)    &0.7  &-0.45&                                                                                      \\
1131 & 21:35:07.23 &  57:22:27.7$^{\rm r}$& 1317 & 4317                      &                               &                 &                 &     13$^{\rm j}$&       &                 &11.638\,(22) &11.271\,(31) & 11.208\,(22)$^{\rm r}$&                    &                 &               &-10.6\,(4.1)  &-3.7\,(4.1)   &-9.2\,(7.3)   &-4.9\,(7.3)   &0.39 &-0.56&                                                                                      \\
1132 & 21:35:20.37 &  57:21:46.5$^{\rm r}$& 1318 & 4318                      &                               &  12.20$^{\rm l}$&  12.04$^{\rm l}$&  11.66$^{\rm l}$&       &                 &10.727\,(24) &10.639\,(29) & 10.571\,(23)$^{\rm r}$&                    &                 &               &-9.1\,(11.4)  &-7.7\,(11.4)  &-5.6\,(1.1)   &-3.7\,(1.6)   &-0.23&0.13 &                                                                                      \\
1133 & 21:35:24.16 &  57:21:10.1$^{\rm r}$& 1319 & 4319                      &                               &                 &                 &   14.4$^{\rm j}$&       &                 &12.636\,(22) &12.326\,(28) & 12.207\,(23)$^{\rm r}$&                    &                 &               &-8.1\,(4.1)   &1.7\,(4.1)    &-9.5\,(6.7)   &0.4\,(6.8)    &-0.07&.0.01&                                                                                      \\
1134 & 21:35:26.72 &  57:22:00.0$^{\rm r}$& 1320 & 4320                      &                               &                 &                 &   14.4$^{\rm j}$&       &                 &12.585\,(23) &12.164\,(27) & 12.099\,(24)$^{\rm r}$&                    &                 &               &-18.6\,(4.1)  &-5.3\,(4.1)   &-21.8\,(6.7)  &-7.8\,(6.7)   &1.45 &-1.2 &new coordinates                                                                       \\     
1135 & 21:35:13.42 &  57:22:56.5$^{\rm r}$& 1321 & 4321                      &                               &                 &                 &   13.3$^{\rm j}$&       &                 &11.831\,(22) &11.404\,(31) & 11.287\,(23)$^{\rm r}$&                    &                 &               &-27.6\,(4.1)  &-16.9\,(4.1)  &-37\,(7.3)    &-1.6\,(7.3)   &2.25 &-1.54&                                                                                      \\
1136 & 21:35:46.14 &  57:22:19.5$^{\rm r}$& 1322 & 428                       &                               &                 &                 &   10.1$^{\rm j}$&       &                 &9.222 \,(22) &8.968 \,(26) & 8.970 \,(20)$^{\rm r}$&        F5$^{\rm q}$&                 &               &-17.3\,(1.6)  &4.4\,(1.6)    &-15.9\,(0.8)  &-5.3\,(1.5)   &1.09 &-0.13&                                                                                      \\
1137 & 21:35:51.50 &  57:23:39.5$^{\rm r}$& 1323 & 429                       &                               &                 &                 &   10.6$^{\rm j}$&       &                 &9.473 \,(23) &9.307 \,(27) & 9.215 \,(20)$^{\rm r}$&        B8$^{\rm q}$&                 &               &-10.7\,(1.7)  &-5.6\,(1.7)   &-4.3\,(1.4)   &-5.2\,(0.6)   &-0.21&-0.2 &                                                                                      \\
1138 & 21:34:39.47 &  57:26:27.1$^{\rm r}$& 1324 & 4324                      &                               &                 &                 &   13.2$^{\rm j}$&       &                 &11.452\,(22) &11.092\,(29) & 11.017\,(22)$^{\rm r}$&                    &                 &               &0.3\,(4.1)    &5.1\,(4.1)    &11.3\,(7.3)   &10.3\,(7.3)   &-0.66&0.36 &                                                                                      \\
1139 & 21:34:49.51 &  57:26:02.7$^{\rm r}$& 1325 & 4325                      &                               &                 &                 &   14.5$^{\rm j}$&       &                 &12.400\,(24) &12.025\,(26) & 11.941\,(25)$^{\rm r}$&                    &                 &               &1.7\,(4.1)    &-0.4\,(4.1)   &-1.5\,(7.6)   &0.4\,(7.6)    &-0.7 &-0.2 &                                                                                      \\
1140 & 21:35:01.00 &  57:26:37.0$^{\rm r}$& 1326 & 4326                      &                               &                 &                 &   13.2$^{\rm j}$&       &                 &11.688\,(26) &11.384\,(31) & 11.246\,(25)$^{\rm r}$&                    &                 &               &2.6\,(4.1)    &8.9\,(4.1)    &24.4\,(7.2)   &32.4\,(7.2)   &-0.5 &0.35 &                                                                                      \\
1141 & 21:35:27.81 &  57:27:15.6$^{\rm r}$& 1327 & 4327                      &                               &                 &                 &   14.2$^{\rm j}$&       &                 &12.530\,(25) &12.193\,(28) & 12.111\,(26)$^{\rm r}$&                    &                 &               &-6\,(4.1)     &-0.4\,(4.1)   &-11.3\,(7.6)  &12.9\,(7.6)   &0.15 &-0.04&                                                                                      \\
1142 & 21:35:43.84 &  57:28:03.5$^{\rm r}$& 1328 & 427                       &                               &   8.23$^{\rm l}$&   8.78$^{\rm l}$&   8.41$^{\rm l}$&       &                 &7.544 \,(19) &7.539 \,(46) & 7.489 \,(18)$^{\rm r}$&      B0.5$^{\rm p}$&      V$^{\rm p}$&               &-0.4\,(1.2)   &-3.1\,(1.2)   &-4.3\,(0.7)   &-5\,(1.1)     &-0.31&-0.4 &                                                                                      \\
1143 & 21:35:54.98 &  57:28:54.4$^{\rm r}$& 1329 & 4329                      &                               &                 &                 &   12.7$^{\rm j}$&       &                 &9.765 \,(23) &8.984 \,(27) & 8.789 \,(21)$^{\rm r}$&                    &                 &               &-7.2\,(5.1)   &-4.1\,(5.1)   &-36.9\,(6.6)  &-67.2\,(6.5)  &0.53 &-0.58&new coordinates                                                                       \\     
1144 & 21:35:59.05 &  57:28:53.1$^{\rm r}$& 1330 & 4330                      &                               &                 &                 &   14.3$^{\rm j}$&       &                 &10.640\,(23) &9.826 \,(27) & 9.560 \,(21)$^{\rm r}$&                    &                 &               &-1.7\,(4.9)   &1.6\,(4.9)    &8.1\,(7)      &18.5\,(7)     &-0.31&0.29 &                                                                                      \\
1145 & 21:36:01.48 &  57:28:14.4$^{\rm r}$& 1331 & 431                       &                               &   9.85$^{\rm l}$&   9.84$^{\rm l}$&   9.42$^{\rm l}$&       &                 &8.810 \,(27) &8.648 \,(27) & 8.636 \,(23)$^{\rm r}$&        F3$^{\rm p}$&    III$^{\rm p}$&               &19.9\,(1.7)   &-4.6\,(1.7)   &21.6\,(3)     &-3.8\,(1.1)   &-2.28&-0.6 &                                                                                      \\
1146 & 21:36:02.76 &  57:28:14.0$^{\rm r}$& 1332 & 432                       &                               &                 &                 &    9.5$^{\rm j}$&       &                 &8.364 \,(27) &8.177 \,(42) & 8.170 \,(31)$^{\rm r}$&        F3$^{\rm q}$&                 &               &21\,(1.3)     &-4.2\,(1.3)   &20.2\,(1.8)   &-3.7\,(1.8)   &-2.23&-0.59&                                                                                      \\
1147 & 21:34:17.29 &  57:30:27.4$^{\rm r}$& 1333 & 4333                      &                               &                 &                 &   14.6$^{\rm j}$&       &                 &12.749\,(26) &12.414\,(33) & 12.308\,(28)$^{\rm r}$&                    &                 &               &-2.7\,(5.5)   &0.9\,(5.5)    &              &              &     &     &                                                                                      \\
1148 & 21:34:40.73 &  57:31:08.9$^{\rm r}$& 1334 & 4334                      &                               &                 &                 &   13.9$^{\rm j}$&       &                 &11.949\,(22) &11.561\,(31) & 11.446\,(22)$^{\rm r}$&                    &                 &               &-16\,(4)      &-2.1\,(4)     &-51.5\,(7.3)  &-59.5\,(7.3)  &0.48 &0.13 &                                                                                      \\
1149 & 21:34:42.42 &  57:31:17.5$^{\rm r}$& 1335 & 4335                      &                               &                 &                 &  13.11$^{\rm j}$&       &                 &11.997\,(24) &11.678\,(28) & 11.602\,(25)$^{\rm r}$&                    &                 &               &-14.8\,(3.8)  &-14.9\,(3.8)  &-9.7\,(6.7)   &-7.5\,(6.7)   &0.56 &-0.9 &                                                                                      \\
1150 & 21:34:42.07 &  57:33:05.5$^{\rm r}$& 1336 & 422                       &                               &                 &                 &   11.2$^{\rm j}$&       &                 &10.543\,(24) &10.473\,(26) & 10.455\,(22)$^{\rm r}$&        A0$^{\rm q}$&                 &               &8\,(2)        &1.1\,(2)      &-4.5\,(0.7)   &-3.8\,(1.7)   &-0.14&-0.3 &                                                                                      \\
1151 & 21:34:29.80 &  57:34:14.1$^{\rm r}$& 1337 & 4337                      &                               &                 &                 &   13.7$^{\rm j}$&       &                 &10.458\,(24) &9.600 \,(27) & 9.340 \,(20)$^{\rm r}$&                    &                 &               &1.1\,(4.7)    &1.1\,(4.7)    &13.9\,(6.8)   &16.7\,(6.8)   &0.25 &0.26 &                                                                                      \\
1152 & 21:34:33.03 &  57:40:41.1$^{\rm r}$& 1338 & 4338                      &                               &                 &                 &   14.1$^{\rm j}$&       &                 &12.417\,(24) &11.953\,(28) & 11.942\,(23)$^{\rm r}$&                    &                 &               &-0.1\,(3.8)   &4\,(3.8)      &-6.4\,(7.3)   &-3.1\,(7.5)   &0.01 &0.31 &                                                                                      \\
1153 & 21:36:10.14 &  57:12:53.3$^{\rm r}$& 1340 & 4340                      &                               &                 &                 &   10.9$^{\rm j}$&       &                 &10.095\,(22) &9.924 \,(29) & 9.879 \,(21)$^{\rm r}$&                    &                 &               &27.6\,(2)     &20.8\,(2)     &26.2\,(0.6)   &9.5\,(1.3)    &-3.45&1.52 &                                                                                      \\
1154 & 21:36:10.47 &  56:52:30.0$^{\rm r}$& 1341 & 4341                      &                               &                 &                 &   11.6$^{\rm j}$&       &                 &10.459\,(22) &10.051\,(28) & 9.955 \,(21)$^{\rm r}$&                    &                 &               &15.3\,(5.1)   &22.2\,(5.1)   &15.1\,(1.3)   &21.2\,(0.6)   &     &     &                                                                                      \\
1155 & 21:36:24.19 &  56:52:50.2$^{\rm r}$& 1342 & 4342                      &                               &                 &                 &     12$^{\rm j}$&       &                 &10.499\,(23) &10.172\,(27) & 10.047\,(20)$^{\rm r}$&                    &                 &               &3.3\,(4.1)    &-5.4\,(4.1)   &7.7\,(2.6)    &-12.3\,(2)    &-1.45&-0.98&                                                                                      \\
1156 & 21:35:48.68 &  56:54:22.1$^{\rm r}$& 1344 & 4344                      &                               &                 &                 &   14.2$^{\rm j}$&       &                 &12.109\,(22) &11.794\,(31) & 11.656\,(24)$^{\rm r}$&                    &                 &               &0.2\,(4.1)    &4.8\,(4.1)    &5.1\,(6.8)    &8\,(6.8)      &-0.86&0.37 &                                                                                      \\
1157 & 21:36:06.85 &  56:53:40.4$^{\rm r}$& 1345 & 4345                      &                               &                 &                 &   14.5$^{\rm j}$&       &                 &12.512\,(24) &12.212\,(31) & 12.137\,(25)$^{\rm r}$&                    &                 &               &0.3\,(4.1)    &-0.2\,(4.1)   &-6.1\,(5.9)   &-2.7\,(5.9)   &-0.69&0.87 &                                                                                      \\
1158 & 21:35:34.88 &  56:57:16.9$^{\rm r}$& 1346 & 4346                      &                               &                 &                 &   12.7$^{\rm j}$&       &                 &11.366\,(22) &11.023\,(28) & 10.934\,(21)$^{\rm r}$&                    &                 &               &-10.8\,(4.1)  &-26.7\,(4.1)  &-7.3\,(7.1)   &-22.3\,(7)    &0.62 &-2.01&                                                                                      \\
1159 & 21:36:25.44 &  56:57:10.6$^{\rm r}$& 1347 & 4347                      &                               &                 &                 &   14.5$^{\rm j}$&       &                 &10.479\,(23) &9.585 \,(27) & 9.288 \,(20)$^{\rm r}$&                    &                 &               &-1.2\,(5.1)   &-7.8\,(5.1)   &-6\,(6.8)     &-0.1\,(6.8)   &     &     &                                                                                      \\
1160 & 21:35:50.57 &  56:59:48.9$^{\rm r}$& 1348 & 4348                      &                               &                 &                 &   14.2$^{\rm j}$&       &                 &12.576\,(24) &12.429\,(28) & 12.278\,(23)$^{\rm r}$&                    &                 &               &-4.1\,(4.1)   &-0.8\,(4.1)   &-5.4\,(6.8)   &5.6\,(6.7)    &-0.39&0.13 &                                                                                      \\
1161 & 21:35:52.37 &  56:59:50.0$^{\rm r}$& 1349 & 4349                      &                               &                 &                 &   13.3$^{\rm j}$&       &                 &10.186\,(22) &9.440 \,(28) & 9.212 \,(21)$^{\rm r}$&                    &                 &               &2.6\,(5.1)    &2.3\,(5.1)    &-2.6\,(6.9)   &8.4\,(7)      &-0.93&0.79 &                                                                                      \\
1162 & 21:36:09.54 &  56:59:37.9$^{\rm r}$& 1350 & 4350                      &                               &                 &                 &   14.6$^{\rm j}$&       &                 &11.219\,(24) &10.516\,(29) & 10.263\,(21)$^{\rm r}$&                    &                 &               &-1.4\,(4.1)   &3.9\,(4.1)    &2.5\,(6.8)    &41.2\,(6.8)   &-0.14&0.06 &                                                                                      \\
1163 & 21:36:24.73 &  57:00:25.7$^{\rm r}$& 1351 & 4351                      &                               &                 &                 &   12.3$^{\rm j}$&       &                 &9.680 \,(23) &9.060 \,(27) & 8.862 \,(21)$^{\rm r}$&                    &                 &               &-16.9\,(5.1)  &-1.3\,(5.1)   &-12.7\,(7.7)  &5.8\,(7.7)    &0.44 &-0.29&                                                                                      \\
1164 & 21:35:54.49 &  57:02:23.4$^{\rm r}$& 1352 & 4352                      &                               &                 &                 &     13$^{\rm j}$&       &                 &10.153\,(24) &9.497 \,(29) & 9.265 \,(21)$^{\rm r}$&                    &                 &               &-6.6\,(5.1)   &-5.7\,(5.1)   &-34.7\,(7.2)  &-20.2\,(7.2)  &0.03 &-0.37&                                                                                      \\
1165 & 21:36:28.20 &  57:01:58.3$^{\rm r}$& 1353 & 4353                      &                               &                 &                 &   14.2$^{\rm j}$&       &                 &12.145\,(25) &11.819\,(33) & 11.743\,(26)$^{\rm r}$&                    &                 &               &-6.1\,(4.1)   &-2\,(4.1)     &-15.6\,(6.8)  &1.3\,(6.8)    &-0.15&0.26 &                                                                                      \\
1166 & 21:35:42.77 &  57:03:59.5$^{\rm r}$& 1354 & 4354                      &                               &                 &                 &   10.7$^{\rm j}$&       &                 &9.781 \,(24) &9.684 \,(28) & 9.597 \,(20)$^{\rm r}$&                    &                 &               &-0.8\,(2)     &-8.7\,(2)     &-2.6\,(1)     &-6.1\,(1)     &-0.38&-0.11&                                                                                      \\
1167 & 21:35:57.15 &  57:03:14.8$^{\rm r}$& 1355 & 4355                      &                               &                 &                 &   14.5$^{\rm j}$&       &                 &12.529\,(22) &12.176\,(32) & 12.080\,(25)$^{\rm r}$&                    &                 &               &-3.3\,(4.1)   &-6.4\,(4.1)   &-3.3\,(6.8)   &-4.2\,(6.8)   &-0.35&0.31 &                                                                                      \\
1168 & 21:36:11.32 &  57:03:31.3$^{\rm r}$& 1356 & 4356                      &                               &                 &                 &   14.3$^{\rm j}$&       &                 &11.021\,(22) &10.236\,(29) & 10.023\,(20)$^{\rm r}$&                    &                 &               &2.9\,(4.1)    &-4.9\,(4.1)   &1.3\,(6.8)    &9.4\,(6.8)    &-1.1 &-0.88&                                                                                      \\
1169 & 21:36:16.72 &  57:03:09.9$^{\rm r}$& 1357 & 4357                      &                               &                 &                 &   13.1$^{\rm j}$&       &                 &11.797\,()   &11.596\,(34) & 11.486\,(29)$^{\rm r}$&                    &                 &               &-11\,(4.1)    &11.8\,(4.1)   &-5.4\,(2)     &-1.4\,(3.4)   &-0.01&0.06 &2x[r]                                                                           \\
1170 & 21:36:16.85 &  57:03:06.2$^{\rm r}$& 1357 & 4357                      &                               &                 &                 &   13.1$^{\rm j}$&       &                 &13.864\,(55) &12.006\,()   &   11.956\,()$^{\rm r}$&                    &                 &               &              &              &-5.4\,(2)     &-1.4\,(3.4)   &-0.01&0.06 &2x[r]                                                                           \\
1171 & 21:36:19.66 &  57:03:31.4$^{\rm r}$& 1358 & 4358                      &                               &                 &                 &   14.5$^{\rm j}$&       &                 &12.454\,(24) &12.162\,(28) & 12.013\,(25)$^{\rm r}$&                    &                 &               &-6.5\,(4.1)   &0.7\,(4.1)    &-8.1\,(6.7)   &6.3\,(6.7)    &-0.28&0.33 &                                                                                      \\
1172 & 21:36:22.45 &  57:03:20.6$^{\rm r}$& 1359 & 4359                      &                               &                 &                 &   12.8$^{\rm j}$&       &                 &11.368\,(22) &11.161\,(27) & 11.065\,(23)$^{\rm r}$&                    &                 &               &-5.9\,(4.1)   &1.3\,(4.1)    &-10.6\,(6.8)  &5.3\,(6.8)    &-0.18&0.01 &                                                                                      \\
1173 & 21:35:33.06 &  57:06:31.4$^{\rm r}$& 1360 & 158                       &                               &                 &                 &     10$^{\rm j}$&       &                 &8.682 \,(23) &8.299 \,(44) &   8.293 \,()$^{\rm r}$&       dK0$^{\rm q}$&                 &               &-84\,(2)      &61.2\,(2)     &-83\,(0.7)    &64\,(0.7)     &     &     &[j] imprec.                                                          \\
1174 & 21:35:50.78 &  57:06:08.2$^{\rm r}$& 1361 & 4361                      &                               &                 &                 &   13.1$^{\rm j}$&       &                 &11.414\,(24) &11.042\,(31) & 10.818\,(21)$^{\rm r}$&                    &                 &               &1.2\,(4.1)    &-11.2\,(4.1)  &3\,(7.1)      &-15\,(7.1)    &-0.1 &-0.06&                                                                                      \\
1175 & 21:36:10.38 &  57:05:45.5$^{\rm r}$& 1362 & 4362                      &                               &                 &                 &   14.8$^{\rm j}$&       &                 &12.853\,(24) &12.453\,(32) & 12.345\,(25)$^{\rm r}$&                    &                 &               &-6.3\,(4.1)   &4.4\,(4.1)    &-9.6\,(6.8)   &14.9\,(6.8)   &     &     &                                                                                      \\
1176 & 21:36:20.40 &  57:05:26.4$^{\rm r}$& 1363 & 4363                      &                               &                 &                 &   13.9$^{\rm j}$&       &                 &10.807\,(22) &10.072\,(27) & 9.878 \,(21)$^{\rm r}$&                    &                 &               &-5.9\,(5.1)   &0.2\,(5.1)    &-27.3\,(6.9)  &-0.8\,(7)     &0.03 &-0.05&                                                                                      \\
1177 & 21:35:47.63 &  57:07:05.2$^{\rm j}$& 1364 & 4364                      &                               &                 &                 &   14.3$^{\rm j}$&       &                 &             &             &                       &                    &                 &               &              &              &              &              &     &     &no star                                                                          \\
1178 & 21:35:52.56 &  57:07:46.0$^{\rm r}$& 1365 & 4365                      &                               &                 &                 &   12.2$^{\rm j}$&       &                 &9.584 \,(27) &8.594 \,(78) & 8.380 \,(69)$^{\rm r}$&                    &                 &               &-12.1\,(5.2)  &-8.5\,(5.2)   &-46.7\,(7.8)  &-62.8\,(7.8)  &-0.03&0.34 &                                                                                      \\
1179 & 21:36:15.01 &  57:07:57.1$^{\rm r}$& 1366 & 4366                      &                               &                 &                 &   14.6$^{\rm j}$&       &                 &13.124\,(26) &12.893\,(30) & 12.841\,(31)$^{\rm r}$&                    &                 &               &-5\,(4.1)     &-0.3\,(4.1)   &-6.9\,(6.8)   &-8.2\,(6.8)   &     &     &                                                                                      \\
1180 & 21:36:18.59 &  57:07:28.1$^{\rm r}$& 1367 & 4367                      &                               &                 &                 &   14.5$^{\rm j}$&       &                 &11.461\,(22) &10.810\,(28) & 10.611\,(20)$^{\rm r}$&                    &                 &               &6.6\,(4.1)    &28.2\,(4.1)   &0.4\,(6.8)    &36\,(6.8)     &-2.02&3.25 &                                                                                      \\
1181 & 21:36:21.54 &  57:07:56.0$^{\rm r}$& 1368 & 4368                      &                               &                 &                 &   14.8$^{\rm j}$&       &                 &12.863\,(39) &12.542\,(44) & 12.424\,(36)$^{\rm r}$&                    &                 &               &-19.9\,(4.1)  &5.8\,(4.1)    &              &              &     &     &                                                                                      \\
1182 & 21:35:36.71 &  57:08:21.4$^{\rm r}$& 1370 & 4370                      &                               &                 &                 &  13.11$^{\rm j}$&       &                 &11.633\,(26) &11.385\,(29) & 11.222\,(23)$^{\rm r}$&                    &                 &               &-10.1\,(4.1)  &3.5\,(4.1)    &-11.8\,(7)    &26.4\,(7)     &0.16 &0.06 &                                                                                      \\
1183 & 21:36:07.08 &  57:09:32.7$^{\rm r}$& 1371 & 4371                      &                               &                 &                 &   14.4$^{\rm j}$&       &                 &10.914\,(24) &10.194\,(29) & 9.922 \,(21)$^{\rm r}$&                    &                 &               &-3.4\,(5.1)   &-4.5\,(5.1)   &-2.3\,(6.9)   &-19.3\,(6.9)  &0.2  &0.36 &                                                                                      \\
1184 & 21:36:10.22 &  57:09:33.9$^{\rm r}$& 1372 & 4372                      &                               &                 &                 &   14.2$^{\rm j}$&       &                 &12.366\,(26) &12.086\,(31) & 11.970\,(25)$^{\rm r}$&                    &                 &               &-8.2\,(4.1)   &0.5\,(4.1)    &-19\,(6.7)    &24.3\,(6.7)   &0.27 &0.06 &                                                                                      \\
1185 & 21:36:47.51 &  56:53:37.5$^{\rm r}$& 1373 & 4373                      &                               &                 &                 &   12.8$^{\rm j}$&       &                 &10.415\,(22) &9.806 \,(27) & 9.629 \,(20)$^{\rm r}$&                    &                 &               &-2.4\,(5.1)   &-1.7\,(5.1)   &2.1\,(7)      &4.6\,(7)      &-0.13&-0.09&                                                                                      \\
1186 & 21:37:05.38 &  56:52:53.5$^{\rm r}$& 1375 & 4375                      &                               &                 &                 &   14.6$^{\rm j}$&       &                 &11.605\,(22) &10.850\,(26) & 10.689\,(21)$^{\rm r}$&                    &                 &               &-0.2\,(4.1)   &-5.4\,(4.1)   &-3.2\,(6.8)   &-2.2\,(6.8)   &     &     &                                                                                      \\
1187 & 21:37:02.23 &  56:53:33.6$^{\rm r}$& 1376 & 4376                      &                               &                 &                 &   14.6$^{\rm j}$&       &                 &13.083\,(22) &12.809\,(27) & 12.832\,(33)$^{\rm r}$&                    &                 &               &0.2\,(4.1)    &1.3\,(4.1)    &-2.7\,(6.8)   &2.1\,(6.8)    &0.12 &0.27 &                                                                                      \\
1188 & 21:37:16.80 &  56:52:29.1$^{\rm r}$& 1378 & 4378                      &                               &                 &                 &   12.2$^{\rm j}$&       &                 &8.438 \,(20) &7.492 \,(57) & 7.141 \,(23)$^{\rm r}$&                    &                 &               &-1.4\,(5.1)   &-1.9\,(5.1)   &-48.2\,(7.2)  &17.3\,(7.3)   &     &     &                                                                                      \\
1189 & 21:37:21.60 &  56:53:14.8$^{\rm r}$& 1379 & 4379                      &                               &                 &                 &   14.3$^{\rm j}$&       &                 &12.945\,(22) &12.665\,(33) & 12.626\,(26)$^{\rm r}$&                    &                 &               &-1.1\,(4.1)   &-2.1\,(4.1)   &-5.4\,(6.8)   &7\,(6.7)      &-0.22&-0.33&                                                                                      \\
1190 & 21:37:25.23 &  56:53:02.5$^{\rm r}$& 1380 & 4380                      &                               &                 &                 &   13.2$^{\rm j}$&       &                 &11.870\,()   &11.641\,(30) & 11.567\,(24)$^{\rm r}$&                    &                 &               &3.7\,(4.1)    &-3\,(4.1)     &2.3\,(6.9)    &-21.9\,(7)    &-1.11&0.75 &                                                                                      \\
1191 & 21:37:31.97 &  56:53:41.1$^{\rm r}$& 1381 & 4381                      &                               &                 &                 &     13$^{\rm j}$&       &                 &11.492\,(22) &10.960\,(28) & 10.896\,(20)$^{\rm r}$&                    &                 &               &20.9\,(4.1)   &25.1\,(4.1)   &19.8\,(7)     &27.5\,(7)     &-3.03&2.89 &                                                                                      \\
1192 & 21:37:00.24 &  56:55:40.2$^{\rm r}$& 1382 & 4382                      &                               &                 &                 &   12.1$^{\rm j}$&       &                 &9.740 \,(22) &9.161 \,(27) & 8.972 \,(20)$^{\rm r}$&                    &                 &               &-7.9\,(5.1)   &-2.1\,(5.1)   &-2.3\,(7.4)   &-0.7\,(7.4)   &0.1  &-0.15&                                                                                      \\
1193 & 21:36:58.02 &  56:56:14.3$^{\rm r}$& 1383 & 4383                      &                               &                 &                 &   14.4$^{\rm j}$&       &                 &12.576\,(24) &12.296\,(30) & 12.195\,(24)$^{\rm r}$&                    &                 &               &-2.5\,(4.1)   &-6.6\,(4.1)   &-4.7\,(6.8)   &-0.9\,(6.7)   &-0.36&-0.05&                                                                                      \\
1194 & 21:37:26.83 &  56:56:46.3$^{\rm r}$& 1384 & 4384                      &                               &                 &                 &   12.9$^{\rm j}$&       &                 &9.979 \,(22) &9.198 \,(26) & 8.998 \,(20)$^{\rm r}$&                    &                 &               &1.1\,(5.1)    &-3.3\,(5.1)   &-4.1\,(7)     &-0.9\,(7)     &-0.06&0.16 &                                                                                      \\
1195 & 21:37:39.71 &  56:55:08.6$^{\rm r}$& 1385 & 4385                      &                               &                 &                 &     13$^{\rm j}$&       &                 &15.080\,(68) &14.301\,(66) & 14.243\,(82)$^{\rm r}$&                    &                 &               &              &              &-35.5\,(7.1)  &22.4\,(7)     &0.21 &-0.6 &2x[r]  (faint)                                                               \\
1196 & 21:37:39.02 &  56:55:09.3$^{\rm r}$& 1385 & 4385                      &                               &                 &                 &     13$^{\rm j}$&       &                 &11.517\,(26) &11.152\,(32) & 11.050\,(24)$^{\rm r}$&                    &                 &               &-5.7\,(4.1)   &2.7\,(4.1)    &-35.5\,(7.1)  &22.4\,(7)     &0.21 &-0.6 &2x[r]                                                                           \\
1197 & 21:37:41.03 &  56:53:57.0$^{\rm r}$& 1386 & 4386                      &                               &                 &                 &   11.5$^{\rm j}$&       &                 &8.632 \,(23) &7.875 \,(49) & 7.627 \,(21)$^{\rm r}$&                    &                 &               &-0.4\,(5.1)   &-4.9\,(5.1)   &70.4\,(7.5)   &-4.5\,(7.5)   &0.19 &-0.37&                                                                                      \\
1198 & 21:37:50.59 &  56:53:16.4$^{\rm r}$& 1387 & 4387                      &                               &                 &                 &   14.2$^{\rm j}$&       &                 &10.990\,(23) &10.405\,(31) & 10.217\,(18)$^{\rm r}$&                    &                 &               &-4\,(4)       &2.5\,(4)      &-0.1\,(6.9)   &0.4\,(6.9)    &-1.58&0.54 &                                                                                      \\
1199 & 21:37:54.87 &  56:54:22.7$^{\rm r}$& 1388 & 4388                      &                               &                 &                 &   10.6$^{\rm j}$&       &                 &9.157 \,(29) &8.925 \,(30) & 8.934 \,(20)$^{\rm r}$&                    &                 &               &18.3\,(1.3)   &-5.2\,(1.2)   &17.9\,(0.5)   &-2\,(0.5)     &-2.83&-0.31&                                                                                      \\
1200 & 21:36:57.78 &  56:57:36.1$^{\rm r}$& 1389 & 4389                      &                               &                 &                 &   12.1$^{\rm j}$&       &                 &7.075 \,(20) &5.962 \,(36) & 5.552 \,(20)$^{\rm r}$&                    &                 &               &-0.1\,(5.1)   &-2.7\,(5.1)   &-9.2\,(7.8)   &-0.1\,(7.8)   &0    &-0.11&                                                                                      \\
1201 & 21:37:35.89 &  56:58:46.4$^{\rm r}$& 1391 & 4391                      &                               &                 &                 &   13.6$^{\rm j}$&       &                 &12.080\,(22) &11.774\,(27) & 11.742\,(25)$^{\rm r}$&                    &                 &               &5.6\,(4.1)    &1.6\,(4.1)    &0.3\,(6.9)    &17.3\,(6.9)   &-0.78&0.31 &                                                                                      \\
1202 & 21:37:43.54 &  56:59:11.1$^{\rm r}$& 1392 & 4392                      &                               &                 &                 &     10$^{\rm j}$&       &                 &8.591 \,(27) &8.230 \,(29) & 8.240 \,(23)$^{\rm r}$&                    &                 &               &3.2\,(1.3)    &-50\,(1.2)    &8.9\,(7.2)    &-34.5\,(7.1)  &-0.32&-5.1 &                                                                                      \\
1203 & 21:37:50.71 &  56:59:47.8$^{\rm r}$& 1393 & 4393                      &                               &                 &                 &   14.3$^{\rm j}$&       &                 &12.169\,(24) &11.831\,(32) & 11.729\,(20)$^{\rm r}$&                    &                 &               &-8.6\,(4)     &0.9\,(4)      &0.3\,(6.8)    &14.8\,(6.8)   &0.49 &-0.15&                                                                                      \\
1204 & 21:36:37.59 &  57:00:57.3$^{\rm r}$& 1394 & 4394                      &                               &                 &                 &   13.8$^{\rm j}$&       &                 &7.933 \,(21) &6.827 \,(27) & 6.365 \,(24)$^{\rm r}$&                    &                 &               &-2.9\,(5.1)   &-3\,(5.1)     &-8.9\,(6.7)   &2.8\,(6.8)    &-0.1 &0.07 &                                                                                      \\
1205 & 21:37:00.77 &  57:00:45.0$^{\rm r}$& 1395 & 4395                      &                               &                 &                 &     13$^{\rm j}$&       &                 &11.417\,(24) &10.964\,(31) & 10.827\,(21)$^{\rm r}$&                    &                 &               &40\,(18.4)    &-38.7\,(18.4) &              &              &-2.4 &2.14 &                                                                                      \\
1206 & 21:37:34.16 &  57:01:57.7$^{\rm r}$& 1396 & 4396                      &                               &                 &                 &   11.5$^{\rm j}$&       &                 &8.588 \,(23) &7.741 \,(31) & 7.515 \,(16)$^{\rm r}$&                    &                 &               &-3.1\,(13.2)  &-18.2\,(13.2) &16\,(9.6)     &-0.2\,(8.9)   &0.31 &0.05 &                                                                                      \\
1207 & 21:37:33.90 &  57:02:09.6$^{\rm r}$& 1397 & 4397                      &                               &                 &                 &   12.6$^{\rm j}$&       &                 &11.383\,(27) &11.059\,(31) & 10.984\,(25)$^{\rm r}$&                    &                 &               &-8.1\,(4.1)   &10.1\,(4.1)   &              &              &0.16 &0.49 &                                                                                      \\
1208 & 21:37:53.02 &  57:01:33.7$^{\rm r}$& 1398 & 4398                      &                               &                 &                 &   13.3$^{\rm j}$&       &                 &9.083 \,(21) &8.096 \,(80) & 7.755 \,(29)$^{\rm r}$&                    &                 &               &-3.9\,(5.1)   &-2\,(5.1)     &-9.4\,(7)     &3.3\,(7)      &0.22 &-0.11&                                                                                      \\
1209 & 21:37:46.42 &  57:01:53.2$^{\rm r}$& 1399 & 4399                      &                               &                 &                 &   14.6$^{\rm j}$&       &                 &13.281\,(26) &13.008\,(34) & 12.918\,(34)$^{\rm r}$&                    &                 &               &-4.8\,(4.1)   &-7.3\,(4.1)   &-6.7\,(6.8)   &1.5\,(6.8)    &     &     &                                                                                      \\
1210 & 21:37:53.14 &  57:01:59.7$^{\rm r}$& 1400 & 4400                      &                               &                 &                 &   12.7$^{\rm j}$&       &                 &10.246\,(21) &9.664 \,(30) & 9.527 \,(18)$^{\rm r}$&                    &                 &               &-2.8\,(5.1)   &-2.2\,(5.1)   &-15.7\,(7.2)  &-23.5\,(7.2)  &0.05 &0.4  &                                                                                      \\
1211 & 21:36:34.88 &  57:02:12.5$^{\rm r}$& 1401 & 4401                      &                               &                 &                 &   13.2$^{\rm j}$&       &                 &12.068\,(27) &11.920\,(33) & 11.887\,(28)$^{\rm r}$&                    &                 &               &-72.7\,(7.6)  &26.5\,(7.6)   &-21\,(6.9)    &43.1\,(6.9)   &-0.25&0.15 &                                                                                      \\
1212 & 21:36:40.64 &  57:01:38.1$^{\rm r}$& 1402 & 4402                      &                               &                 &                 &   11.8$^{\rm j}$&       &                 &7.207 \,(21) &6.159 \,(20) & 5.789 \,(17)$^{\rm r}$&                    &                 &               &-3.1\,(5.1)   &-3\,(5.1)     &-7.6\,(7.8)   &0.7\,(7.8)    &-0.14&-0.46&                                                                                      \\
1213 & 21:36:54.60 &  57:02:53.7$^{\rm r}$& 1403 & 4403                      &                               &                 &                 &   12.2$^{\rm j}$&       &                 &8.555 \,(19) &7.579 \,(18) & 7.310 \,(16)$^{\rm r}$&                    &                 &               &-1.6\,(5.1)   &-4.2\,(5.1)   &-7.4\,(7.6)   &-3.4\,(7.7)   &-0.28&-0.19&                                                                                      \\
1214 & 21:36:53.74 &  57:03:04.4$^{\rm r}$& 1404 & 4404                      &                               &                 &                 &   13.2$^{\rm j}$&       &                 &12.012\,(26) &11.815\,(36) & 11.721\,(29)$^{\rm r}$&                    &                 &               &-29.2\,(7.6)  &399.5\,(8.1)  &-14.8\,(6.3)  &16\,(6.4)     &-0.38&-0.01&                                                                                      \\
1215 & 21:37:01.98 &  57:02:52.6$^{\rm r}$& 1405 & 4405                      &                               &                 &                 &   13.7$^{\rm j}$&       &                 &11.680\,(22) &11.260\,(27) & 11.146\,(21)$^{\rm r}$&                    &                 &               &-5.7\,(4.1)   &-4.5\,(4.1)   &-1.6\,(6.8)   &-3\,(6.8)     &-0.09&0.07 &                                                                                      \\
1216 & 21:37:11.39 &  57:03:08.0$^{\rm r}$& 1406 & 4406                      &                               &                 &                 &   13.3$^{\rm j}$&       &                 &9.124 \,(24) &8.064 \,(51) & 7.721 \,(26)$^{\rm r}$&                    &                 &               &-4\,(5.1)     &-4.9\,(5.1)   &-10.7\,(7)    &4.4\,(7)      &0.02 &-0.16&                                                                                      \\
1217 & 21:37:12.37 &  57:03:28.4$^{\rm r}$& 1407 & 4407                      &                               &                 &                 &   12.9$^{\rm j}$&       &                 &11.670\,(31) &11.362\,(46) & 11.275\,(38)$^{\rm r}$&                    &                 &               &-7.4\,(4.1)   &3.2\,(4.1)    &-37.2\,(7)    &23.7\,(7)     &-0.07&0.05 &                                                                                      \\
1218 & 21:37:17.61 &  57:04:36.2$^{\rm r}$& 1408 & 4408                      &                               &                 &                 &   10.9$^{\rm j}$&       &                 &8.317 \,(29) &7.645 \,(24) & 7.476 \,(23)$^{\rm r}$&                    &                 &               &5.5\,(2.8)    &8.3\,(2.8)    &6.5\,(1.6)    &7\,(1.1)      &-1   &1.2  &                                                                                      \\
1219 & 21:37:25.17 &  57:04:23.7$^{\rm r}$& 1409 & 4409                      &                               &                 &                 &   14.6$^{\rm j}$&       &                 &12.481\,(26) &12.036\,(32) & 11.924\,(28)$^{\rm r}$&                    &                 &               &-0.6\,(4.1)   &-2.5\,(4.1)   &3.4\,(6.9)    &-13.4\,(6.9)  &-0.35&0.22 &                                                                                      \\
1220 & 21:37:30.80 &  57:04:11.2$^{\rm r}$& 1410 & 4410                      &                               &                 &                 &   14.3$^{\rm j}$&       &                 &12.729\,(26) &12.378\,(31) & 12.253\,(29)$^{\rm r}$&                    &                 &               &6.1\,(4.1)    &0.7\,(4.1)    &10.6\,(6.8)   &22.1\,(6.8)   &-1.04&0.56 &                                                                                      \\
1221 & 21:37:42.03 &  57:03:20.6$^{\rm r}$& 1411 & 4411                      &                               &                 &                 &   12.7$^{\rm j}$&       &                 &11.555\,(24) &11.298\,(30) & 11.240\,(23)$^{\rm r}$&                    &                 &               &8.6\,(4.1)    &5.6\,(4.1)    &3.5\,(1.2)    &-0.8\,(5.2)   &-1   &0.48 &                                                                                      \\
1222 & 21:37:43.06 &  57:04:17.4$^{\rm r}$& 1412 & 4412                      &                               &                 &                 &   14.7$^{\rm j}$&       &                 &10.656\,(24) &9.597 \,(26) & 9.366 \,(21)$^{\rm r}$&                    &                 &               &0\,(5.1)      &-1.2\,(5.1)   &-4.2\,(6.8)   &-1.9\,(6.8)   &     &     &                                                                                      \\
1223 & 21:37:45.16 &  57:04:30.2$^{\rm r}$& 1413 & 4413                      &                               &                 &                 &     14$^{\rm j}$&       &                 &12.547\,(24) &12.219\,(28) & 12.113\,(23)$^{\rm r}$&                    &                 &               &-7.2\,(4.1)   &-6.2\,(4.1)   &-7.1\,(6.8)   &2.5\,(6.8)    &0.29 &-0.54&                                                                                      \\
1224 & 21:36:54.55 &  57:05:18.7$^{\rm r}$& 1414 & 4414                      &                               &                 &                 &   13.5$^{\rm j}$&       &                 &12.062\,(22) &11.761\,(30) & 11.662\,(21)$^{\rm r}$&                    &                 &               &-3.4\,(4.1)   &1.2\,(4.1)    &-7.1\,(6.8)   &6.5\,(6.8)    &-0.01&0.13 &                                                                                      \\
1225 & 21:36:32.81 &  57:06:33.5$^{\rm r}$& 1415 & 4415                      &                               &                 &                 &   13.3$^{\rm j}$&       &                 &11.980\,(24) &11.805\,(30) & 11.707\,(23)$^{\rm r}$&                    &                 &               &-1.2\,(4.1)   &4.5\,(4.1)    &-1.7\,(6.8)   &5.7\,(6.8)    &-0.63&0.4  &                                                                                      \\
1226 & 21:36:45.23 &  57:06:53.6$^{\rm r}$& 1416 & 4416                      &                               &                 &                 &   14.1$^{\rm j}$&       &                 &12.356\,(24) &12.061\,(28) & 11.937\,(21)$^{\rm r}$&                    &                 &               &4.6\,(4.1)    &3.4\,(4.1)    &39.1\,(6.2)   &26.7\,(6.2)   &-0.23&-0.02&                                                                                      \\
1227 & 21:36:59.83 &  57:06:31.6$^{\rm j}$& 1417 & 4417                      &                               &                 &                 &   12.6$^{\rm j}$&       &                 &             &             &                       &                    &                 &               &              &              &              &              &     &     &no star                                                                          \\
1228 & 21:37:35.89 &  57:06:09.6$^{\rm r}$& 1418 & 4418                      &                               &                 &                 &   13.8$^{\rm j}$&       &                 &11.545\,(24) &10.900\,(26) & 10.798\,(21)$^{\rm r}$&                    &                 &               &-13.4\,(4.1)  &-1.4\,(4.1)   &-14.5\,(6.8)  &5.7\,(6.8)    &0.68 &0.07 &                                                                                      \\
1229 & 21:37:36.37 &  57:06:23.6$^{\rm r}$& 1419 & 4419                      &                               &                 &                 &   14.7$^{\rm j}$&       &                 &12.662\,(27) &12.238\,(32) & 12.158\,(28)$^{\rm r}$&                    &                 &               &-4.1\,(4.1)   &9.3\,(4.1)    &-2.2\,(6.8)   &19.3\,(6.9)   &     &     &                                                                                      \\
1230 & 21:37:46.76 &  57:07:05.6$^{\rm r}$& 1420 & 166                       &                               &                 &                 &   10.3$^{\rm j}$&       &                 &7.712 \,(34) &7.080 \,(31) & 6.934 \,(33)$^{\rm r}$&        G8$^{\rm q}$&                 &               &17.8\,(1.7)   &3.8\,(1.7)    &11.9\,(0.9)   &2.2\,(0.6)    &-1.47&0.86 &                                                                                      \\
1231 & 21:36:33.90 &  57:08:20.2$^{\rm r}$& 1421 & 4421                      &                               &                 &                 &   11.9$^{\rm j}$&       &                 &8.882 \,(26) &8.110 \,(26) & 7.839 \,(21)$^{\rm r}$&                    &                 &               &-11.7\,(13.8) &8.9\,(13.8)   &-11.3\,(7.9)  &-14.5\,(5.3)  &-0.49&0.13 &                                                                                      \\
1232 & 21:37:03.24 &  57:08:09.9$^{\rm r}$& 1422 & 4422                      &                               &                 &                 &   13.4$^{\rm j}$&       &                 &11.516\,(24) &11.082\,(28) & 10.984\,(21)$^{\rm r}$&                    &                 &               &-3.2\,(4.1)   &1.5\,(4.1)    &-1.3\,(6.8)   &1.7\,(6.8)    &-0.16&-0.11&                                                                                      \\
1233 & 21:37:28.93 &  57:08:52.8$^{\rm r}$& 1423 & 4423                      &                               &                 &                 &   14.3$^{\rm j}$&       &                 &10.895\,(24) &10.073\,(27) & 9.820 \,(21)$^{\rm r}$&                    &                 &               &-2.7\,(5.1)   &-3.7\,(5.1)   &-13.8\,(6.8)  &13.8\,(6.8)   &     &     &                                                                                      \\
1234 & 21:37:57.85 &  57:07:26.2$^{\rm r}$& 1425 & 4425                      &                               &                 &                 &   13.3$^{\rm j}$&       &                 &11.214\,(26) &10.665\,(32) & 10.558\,(19)$^{\rm r}$&                    &                 &               &-15.3\,(4)    &-1.6\,(4)     &-11.9\,(7)    &-25.3\,(7.1)  &0.83 &0.94 &                                                                                      \\
1235 & 21:38:01.16 &  57:08:58.4$^{\rm r}$& 1426 & 4426                      &                               &                 &                 &   14.2$^{\rm j}$&       &                 &12.297\,(24) &12.020\,(31) & 11.856\,(18)$^{\rm r}$&                    &                 &               &-9.2\,(4)     &-2.7\,(4)     &-22.7\,(6.7)  &-0.7\,(6.8)   &-0.13&0.12 &                                                                                      \\
1236 & 21:36:46.05 &  57:10:17.4$^{\rm r}$& 1427 & 4427                      &                               &                 &                 &   12.4$^{\rm j}$&       &                 &9.072 \,(26) &8.191 \,(38) & 7.910 \,(21)$^{\rm r}$&                    &                 &               &-2.8\,(5.1)   &-1.4\,(5.1)   &2.5\,(7.2)    &2\,(7.3)      &0.12 &0.13 &                                                                                      \\
1237 & 21:37:13.37 &  57:10:22.3$^{\rm r}$& 1428 & 4428                      &                               &                 &                 &   11.5$^{\rm j}$&       &                 &8.566 \,(21) &7.749 \,(29) & 7.558 \,(20)$^{\rm r}$&                    &                 &               &-5.8\,(2.8)   &-2.2\,(2.8)   &-3.7\,(1.6)   &2.2\,(1.8)    &-0.17&0.71 &                                                                                      \\
1238 & 21:37:40.47 &  57:09:48.3$^{\rm r}$& 1429 & 4429                      &                               &                 &                 &   13.9$^{\rm j}$&       &                 &12.501\,(24) &12.182\,(28) & 12.112\,(24)$^{\rm r}$&                    &                 &               &-8.4\,(4.1)   &1.5\,(4.1)    &-10.9\,(6.8)  &-1.4\,(6.9)   &0.27 &0.23 &                                                                                      \\
1239 & 21:38:45.92 &  56:53:05.1$^{\rm r}$& 1431 & 4431                      &                               &                 &                 &     13$^{\rm j}$&       &                 &11.657\,(29) &11.392\,(32) & 11.281\,(23)$^{\rm r}$&                    &                 &               &-7.9\,(4)     &-25.2\,(4)    &-7.9\,(7.3)   &-96\,(7.3)    &0.06 &0.19 &                                                                                      \\
1240 & 21:39:07.49 &  56:53:31.3$^{\rm r}$& 1434 & 4434                      &                               &                 &                 &   14.2$^{\rm j}$&       &                 &11.491\,(21) &10.859\,(27) & 10.680\,(20)$^{\rm r}$&                    &                 &               &-3.6\,(4)     &0.1\,(4)      &-3.4\,(6.8)   &5.8\,(6.8)    &-0.11&0.45 &                                                                                      \\
1241 & 21:38:20.66 &  56:56:09.0$^{\rm r}$& 1436 & 4436                      &                               &                 &                 &   13.9$^{\rm j}$&       &                 &11.607\,(23) &11.123\,(33) & 10.996\,(24)$^{\rm r}$&                    &                 &               &10.5\,(4)     &-8.9\,(4)     &30.6\,(7.1)   &-42.3\,(7.1)  &-0.94&1.13 &                                                                                      \\
1242 & 21:38:28.51 &  56:57:14.5$^{\rm r}$& 1437 & 4437                      &                               &                 &                 &   13.1$^{\rm j}$&       &                 &12.007\,(21) &11.542\,(29) & 11.541\,(18)$^{\rm r}$&                    &                 &               &7.4\,(4)      &-2.1\,(4)     &8.3\,(7.1)    &-19.4\,(7.1)  &-1.11&-0.13&                                                                                      \\
1243 & 21:38:49.68 &  56:55:07.1$^{\rm r}$& 1438 & 4438                      &                               &                 &                 &   13.3$^{\rm j}$&       &                 &12.285\,(24) &12.049\,(32) & 11.973\,(26)$^{\rm r}$&                    &                 &               &              &              &              &              &-0.05&-0.21&                                                                                      \\
1244 & 21:38:26.21 &  56:58:18.8$^{\rm r}$& 1439 & 4439                      &                               &                 &                 &   12.8$^{\rm j}$&       &                 &10.227\,(54) &9.799 \,(38) & 9.612 \,(18)$^{\rm r}$&                    &                 &               &              &              &              &              &0.39 &0.29 &                                                                                      \\
1245 & 21:38:25.76 &  56:58:36.8$^{\rm r}$& 1441 & 4441                      &                               &                 &                 &   12.6$^{\rm j}$&       &                 &11.504\,(36) &11.159\,(40) & 11.114\,(30)$^{\rm r}$&                    &                 &               &              &              &              &              &-1.58&1.35 &                                                                                      \\
1246 & 21:38:34.22 &  56:58:49.1$^{\rm r}$& 1442 & 4442                      &                               &                 &                 &   14.1$^{\rm j}$&       &                 &10.645\,(21) &9.823 \,(30) & 9.621 \,(20)$^{\rm r}$&                    &                 &               &0.6\,(6.6)    &-2.5\,(6.6)   &10.6\,(6.8)   &1.6\,(6.9)    &0.06 &-0.18&                                                                                      \\
1247 & 21:38:50.83 &  56:57:30.3$^{\rm r}$& 1443 & 4443                      &                               &                 &                 &   13.1$^{\rm j}$&       &                 &11.878\,(24) &11.432\,(27) & 11.364\,(23)$^{\rm r}$&                    &                 &               &3.9\,(4)      &2.7\,(4)      &2.9\,(7)      &9.5\,(7)      &-1.27&0.64 &                                                                                      \\
1248 & 21:39:04.74 &  56:56:59.5$^{\rm r}$& 1444 & 174                       &                               &   8.92$^{\rm l}$&   9.38$^{\rm l}$&   8.34$^{\rm g}$& 7.88  &   7.76$^{\rm g}$&8.560 \,(24) &8.528 \,(49) & 8.494 \,(20)$^{\rm r}$&        B2$^{\rm p}$&   IV-V$^{\rm p}$&               &-2.5\,(0.7)   &-6.1\,(0.8)   &-1.2\,(0.6)   &-5.1\,(0.5)   &0.03 &-0.25&                                                                                      \\
1249 & 21:39:06.88 &  56:56:27.9$^{\rm r}$& 1445 & 4445                      &                               &                 &                 &   11.9$^{\rm j}$&       &                 &9.222 \,(26) &8.467 \,(31) & 8.269 \,(23)$^{\rm r}$&                    &                 &               &-2.1\,(5.1)   &4\,(5.1)      &              &              &0.12 &0.14 &                                                                                      \\
1250 & 21:39:12.41 &  56:55:51.9$^{\rm r}$& 1446 & 4446                      &                               &                 &                 &     12$^{\rm j}$&       &                 &10.766\,(23) &10.472\,(31) & 10.408\,(23)$^{\rm r}$&                    &                 &               &-0.8\,(2.7)   &5.6\,(2.7)    &-2.8\,(0.7)   &4.9\,(1.3)    &-0.15&0.91 &                                                                                      \\
1251 & 21:39:17.93 &  56:54:41.4$^{\rm r}$& 1447 & 4447                      &                               &                 &                 &   14.5$^{\rm j}$&       &                 &13.089\,(23) &12.930\,(36) & 12.785\,(30)$^{\rm r}$&                    &                 &               &-8.5\,(4)     &2.2\,(4)      &-4.4\,(6.8)   &11.1\,(6.8)   &-0.4 &0.1  &                                                                                      \\
1252 & 21:39:20.35 &  56:54:51.5$^{\rm r}$& 1448 & 4448                      &                               &                 &                 &   14.1$^{\rm j}$&       &                 &12.083\,(23) &11.710\,(33) & 11.541\,(20)$^{\rm r}$&                    &                 &               &-5\,(4)       &-0.8\,(4)     &-5.8\,(6.8)   &-2.7\,(6.8)   &0.01 &-0.12&                                                                                      \\
1253 & 21:37:58.73 &  57:00:09.3$^{\rm r}$& 1449 & 4449                      &                               &                 &                 &   14.5$^{\rm j}$&       &                 &10.160\,(21) &9.175 \,(31) & 8.869 \,(20)$^{\rm r}$&                    &                 &               &-2.6\,(5.1)   &-1.1\,(5.1)   &-4.2\,(6.8)   &7.6\,(6.8)    &-0.51&0.44 &                                                                                      \\
1254 & 21:38:09.12 &  57:00:50.5$^{\rm r}$& 1450 & 4450                      &                               &                 &                 &   13.7$^{\rm j}$&       &                 &9.803 \,(21) &8.934 \,(30) & 8.663 \,(20)$^{\rm r}$&                    &                 &               &-4.2\,(5.1)   &-3.7\,(5.1)   &-5.1\,(7)     &1\,(7)        &0.33 &0.18 &                                                                                      \\
1255 & 21:38:14.12 &  57:01:33.1$^{\rm r}$& 1451 & 4451                      &                               &                 &                 &   14.3$^{\rm j}$&       &                 &12.870\,(23) &12.653\,(29) & 12.584\,(20)$^{\rm r}$&                    &                 &               &-5\,(4)       &-4.4\,(4)     &-9.6\,(6.8)   &-10.8\,(6.8)  &-0.21&0.06 &                                                                                      \\
1256 & 21:38:21.36 &  57:01:52.4$^{\rm r}$& 1452 & 4452                      &                               &                 &                 &   13.9$^{\rm j}$&       &                 &12.000\,(21) &11.767\,(31) & 11.577\,(24)$^{\rm r}$&                    &                 &               &-3.5\,(4)     &-5.3\,(4)     &              &              &0.23 &0.03 &                                                                                      \\
1257 & 21:38:24.37 &  57:02:08.8$^{\rm r}$& 1453 & 4453                      &                               &                 &                 &   13.6$^{\rm j}$&       &                 &11.595\,(21) &11.251\,(29) & 11.114\,(18)$^{\rm r}$&                    &                 &               &-7.7\,(4)     &-10\,(4)      &-12.2\,(6.9)  &0.8\,(7)      &0.13 &0.04 &                                                                                      \\
1258 & 21:38:22.34 &  57:02:29.3$^{\rm r}$& 1454 & 4454                      &                               &                 &                 &   13.5$^{\rm j}$&       &                 &11.848\,(29) &11.544\,(40) & 11.418\,(31)$^{\rm r}$&                    &                 &               &-29.2\,(4)    &-15.6\,(4)    &              &              &0.04 &-0.67&2x[r]                                                                           \\
1259 & 21:38:23.00 &  57:02:28.2$^{\rm r}$& 1454 & 4454                      &                               &                 &                 &   13.5$^{\rm j}$&       &                 &14.092\,(44) &13.650\,(47) & 13.480\,(54)$^{\rm r}$&                    &                 &               &261.1\,(9.4)  &-76.7\,(9.4)  &              &              &0.04 &-0.67&2x[r] (faint)                                                               \\
1260 & 21:38:39.84 &  57:02:57.7$^{\rm r}$& 1455 & 4455                      &                               &                 &                 &   14.1$^{\rm j}$&       &                 &10.861\,(23) &10.132\,(31) & 9.908 \,(22)$^{\rm r}$&                    &                 &               &-4.6\,(6.5)   &-1.2\,(7.2)   &              &              &0.1  &0.21 &2x[r]                                                                           \\
1261 & 21:38:40.34 &  57:02:53.7$^{\rm r}$& 1455 & 4455                      &                               &                 &                 &   14.1$^{\rm j}$&       &                 &14.670\,(82) &14.185\,(92) &14.032\,(114)$^{\rm r}$&                    &                 &               &-16.6\,(5.3)  &14.6\,(5.3)   &              &              &0.1  &0.21 &2x[r]  (faint)                                                               \\
1262 & 21:38:14.86 &  57:03:04.8$^{\rm r}$& 1456 & 4456                      &                               &                 &                 &   14.2$^{\rm j}$&       &                 &11.315\,(21) &10.694\,(29) & 10.453\,(18)$^{\rm r}$&                    &                 &               &-3.1\,(4)     &-6.2\,(4)     &0\,(6.9)      &4.3\,(6.9)    &-0.26&0.03 &                                                                                      \\
1263 & 21:38:00.65 &  57:05:27.2$^{\rm r}$& 1457 & 4457                      &                               &                 &                 &   14.5$^{\rm j}$&       &                 &12.509\,(24) &12.338\,(37) & 12.114\,(20)$^{\rm r}$&                    &                 &               &1.1\,(4)      &-0.6\,(4)     &-3.7\,(6.8)   &-4.6\,(6.8)   &-0.74&0.46 &                                                                                      \\
1264 & 21:38:04.10 &  57:05:16.9$^{\rm r}$& 1458 & 4458                      &                               &                 &                 &   13.1$^{\rm j}$&       &                 &10.283\,(23) &9.585 \,(31) & 9.348 \,(18)$^{\rm r}$&                    &                 &               &-4.3\,(5.1)   &-0.1\,(5.1)   &18.9\,(7.1)   &29.1\,(7.2)   &0.23 &0.39 &                                                                                      \\
1265 & 21:38:18.16 &  57:06:48.2$^{\rm r}$& 1459 & 169                       &                               &                 &  10.77$^{\rm h}$&  10.35$^{\rm h}$&       &                 &9.289 \,(26) &9.052 \,(30) & 8.993 \,(19)$^{\rm r}$&        F0$^{\rm h}$&                 & 0.34$^{\rm h}$&8.2\,(1.6)    &-13.8\,(1.6)  &6.6\,(0.5)    &-10.2\,(0.5)  &-0.92&-0.6 &                                                                                      \\
1266 & 21:38:26.39 &  57:06:19.6$^{\rm r}$& 1461 & 4461                      &                               &                 &                 &   14.7$^{\rm j}$&       &                 &10.641\,(26) &9.619 \,(30) & 9.294 \,(19)$^{\rm r}$&                    &                 &               &-5.8\,(5.1)   &-3.2\,(5.1)   &-24.4\,(6.9)  &15.9\,(6.9)   &     &     &                                                                                      \\
1267 & 21:38:32.64 &  57:06:05.9$^{\rm r}$& 1462 & 4462                      &                               &                 &                 &   12.1$^{\rm j}$&       &                 &11.594\,(24) &11.310\,(30) & 11.231\,(19)$^{\rm r}$&                    &                 &               &-6.4\,(4)     &-1.8\,(4)     &-7.3\,(1.9)   &-3.3\,(2.1)   &0.27 &-0.12&                                                                                      \\
1268 & 21:38:21.91 &  57:07:22.8$^{\rm r}$& 1463 & 4463                      &                               &                 &                 &     13$^{\rm j}$&       &                 &11.682\,(24) &11.229\,(30) & 11.162\,(19)$^{\rm r}$&                    &                 &               &-13.8\,(4)    &-37.6\,(4)    &-15.1\,(7)    &-29.7\,(7)    &1.18 &-3.22&                                                                                      \\
1269 & 21:38:24.29 &  57:08:26.3$^{\rm r}$& 1464 & 4464                      &                               &                 &                 &   13.3$^{\rm j}$&       &                 &11.963\,(24) &11.737\,(30) & 11.602\,(23)$^{\rm r}$&                    &                 &               &-1.4\,(4)     &-3.6\,(4)     &9.8\,(7)      &1.5\,(7)      &0    &-0.04&                                                                                      \\
1270 & 21:38:16.36 &  57:10:11.6$^{\rm r}$& 1465 & 168                       &                               &                 &   11.7$^{\rm f}$&   11.5$^{\rm e}$&       &                 &9.988 \,(24) &9.919 \,(29) & 9.888 \,(18)$^{\rm r}$&        B7$^{\rm e}$&                 &    1$^{\rm e}$&-2.4\,(2)     &-7.7\,(1.9)   &-2.1\,(0.5)   &-7.7\,(0.8)   &-0.04&-0.34&                                                                                      \\
1271 & 21:38:28.57 &  57:10:44.4$^{\rm r}$& 1466 & 4466                      &                               &                 &                 &   12.2$^{\rm j}$&       &                 &10.695\,(29) &10.274\,(31) & 10.184\,(18)$^{\rm r}$&                    &                 &               &-2.7\,(2.7)   &-6.9\,(2.7)   &-8.9\,(3.2)   &-1\,(1.9)     &0.59 &0.23 &                                                                                      \\
1272 & 21:38:33.02 &  57:09:56.6$^{\rm r}$& 1467 & 172                       &                               &                 &                 &    9.5$^{\rm j}$&       &                 &5.640 \,(34) &4.835 \,(47) &   4.494 \,()$^{\rm r}$&        K5$^{\rm q}$&                 &               &-2.8\,(1.3)   &-13\,(1.4)    &0.2\,(1.1)    &-12.7\,(0.7)  &0.21 &-0.93&                                                                                      \\
1273 & 21:38:40.98 &  57:09:21.4$^{\rm r}$& 1468 & 4468                      &                               &                 &                 &   14.5$^{\rm j}$&       &                 &12.837\,(27) &12.533\,(33) & 12.435\,(30)$^{\rm r}$&                    &                 &               &-15.5\,(4)    &0.9\,(4)      &-15.4\,(6.8)  &-11.1\,(6.8)  &-0.1 &-0.05&                                                                                      \\
1274 & 21:38:46.48 &  57:10:07.7$^{\rm r}$& 1469 & 4469                      &                               &                 &                 &   13.5$^{\rm j}$&       &                 &12.219\,(23) &11.935\,(29) & 11.868\,(23)$^{\rm r}$&                    &                 &               &-8.8\,(4)     &-3.7\,(4)     &-11\,(6.9)    &1.1\,(6.9)    &0.23 &-0.09&                                                                                      \\
1275 & 21:38:43.16 &  57:09:20.9$^{\rm r}$& 1470 & 4470                      &                               &                 &                 &   13.6$^{\rm j}$&       &                 &9.177 \,(23) &8.062 \,(49) & 7.715 \,(21)$^{\rm r}$&                    &                 &               &14.2\,(13)    &3\,(13)       &              &              &0.05 &0.29 &                                                                                      \\
1276 & 21:38:47.70 &  57:07:15.7$^{\rm r}$& 1471 & 4471                      &                               &                 &                 &   13.4$^{\rm j}$&       &                 &11.601\,(26) &11.269\,(31) & 11.141\,(23)$^{\rm r}$&                    &                 &               &-25\,(4)      &7.2\,(4)      &-80.6\,(7.1)  &28\,(7.1)     &-0.04&-0.22&                                                                                      \\
1277 & 21:38:52.45 &  57:07:54.7$^{\rm r}$& 1472 & 4472                      &                               &                 &                 &   13.5$^{\rm j}$&       &                 &11.981\,(22) &11.666\,(29) & 11.539\,(25)$^{\rm r}$&                    &                 &               &-2.4\,(4)     &2.5\,(4)      &0.5\,(6.9)    &9.5\,(6.9)    &-0.29&0.25 &                                                                                      \\
1278 & 21:38:56.66 &  57:09:07.4$^{\rm r}$& 1473 & 4473                      &                               &                 &                 &   14.2$^{\rm j}$&       &                 &12.405\,(34) &12.048\,(45) & 11.922\,(35)$^{\rm r}$&                    &                 &               &6.2\,(4)      &5\,(4)        &-20\,(6.9)    &16.1\,(6.9)   &-2.42&1.15 &                                                                                      \\
1279 & 21:39:43.29 &  56:54:19.7$^{\rm r}$& 1474 & 4474                      &                               &                 &                 &   13.1$^{\rm j}$&       &                 &10.507\,(26) &9.762 \,(28) & 9.574 \,(20)$^{\rm r}$&                    &                 &               &-0.1\,(5.2)   &0.2\,(5.2)    &-2.8\,(7)     &2.6\,(7.1)    &-0.13&0.2  &                                                                                      \\
1280 & 21:39:47.99 &  56:54:49.5$^{\rm r}$& 1475 & 4475                      &                               &                 &                 &   13.7$^{\rm j}$&       &                 &10.854\,(24) &10.156\,(27) & 10.005\,(19)$^{\rm r}$&                    &                 &               &-2.9\,(4.1)   &-4\,(4.1)     &-8.5\,(6.9)   &-1.8\,(6.9)   &0.56 &-0.18&new coordinates                                                                       \\     
1281 & 21:39:32.64 &  56:55:12.9$^{\rm r}$& 1476 & 4476                      &                               &                 &                 &   13.2$^{\rm j}$&       &                 &10.433\,(26) &9.685 \,(28) & 9.525 \,(22)$^{\rm r}$&                    &                 &               &0.2\,(5.2)    &-4.1\,(5.2)   &-28.5\,(7)    &-2.9\,(7)     &0.12 &-0.15&                                                                                      \\
1282 & 21:40:00.84 &  56:54:48.6$^{\rm r}$& 1477 & 4477                      &                               &                 &                 &   12.1$^{\rm j}$&       &                 &11.269\,(26) &11.094\,(28) & 11.015\,(22)$^{\rm r}$&                    &                 &               &-8.3\,(2.7)   &-4.3\,(2.7)   &-6.9\,(0.9)   &-5.4\,(0.9)   &0.46 &-0.1 &                                                                                      \\
1283 & 21:40:10.11 &  56:54:36.7$^{\rm r}$& 1478 & 4478                      &                               &                 &                 &   11.9$^{\rm j}$&       &                 &11.160\,(24) &11.004\,(30) & 10.903\,(20)$^{\rm r}$&                    &                 &               &-5.2\,(4.1)   &1\,(4.1)      &-4\,(1.1)     &-3.7\,(1.3)   &0.31 &0.07 &                                                                                      \\
1284 & 21:40:11.63 &  56:55:12.4$^{\rm r}$& 1479 & 4479                      &                               &                 &                 &   12.7$^{\rm j}$&       &                 &11.406\,(24) &11.132\,(27) & 10.999\,(18)$^{\rm r}$&                    &                 &               &-3.4\,(4.1)   &1.6\,(4.1)    &-2.1\,(2.3)   &-0.3\,(1.9)   &-0.31&0.66 &                                                                                      \\
1285 & 21:40:06.84 &  56:55:34.8$^{\rm r}$& 1480 & 4480                      &                               &                 &                 &   13.1$^{\rm j}$&       &                 &11.579\,(26) &11.224\,(31) & 11.102\,(25)$^{\rm r}$&                    &                 &               &15.6\,(4.1)   &0.1\,(4.1)    &21.7\,(7)     &4\,(7)        &-1.23&-0.16&                                                                                      \\
1286 & 21:40:09.77 &  56:56:11.1$^{\rm r}$& 1481 & 4481                      &                               &                 &                 &   13.6$^{\rm j}$&       &                 &11.898\,(24) &11.526\,(28) & 11.436\,(23)$^{\rm r}$&                    &                 &               &-6.2\,(4.1)   &-1.2\,(4.1)   &-6.9\,(6.9)   &3\,(7)        &0.22 &0.52 &                                                                                      \\
1287 & 21:40:20.52 &  56:55:07.5$^{\rm r}$& 1482 & 4482                      &                               &                 &                 &   11.5$^{\rm j}$&       &                 &11.053\,(26) &10.987\,(32) & 10.875\,(18)$^{\rm r}$&                    &                 &               &-3.7\,(1.7)   &-9.6\,(1.7)   &-6.4\,(1)     &-3.8\,(1.4)   &0.68 &0.34 &                                                                                      \\
1288 & 21:39:22.82 &  56:56:48.4$^{\rm r}$& 1483 & 4483                      &                               &                 &                 &   12.7$^{\rm j}$&       &                 &14.850\,(46) &9.856 \,()   &   9.238 \,()$^{\rm r}$&                    &                 &               &-2.6\,(5.1)   &-3.2\,(5.1)   &-13.1\,(7.2)  &5.3\,(7.3)    &0.47 &0.1  &2x[r] (faint)                                                               \\
1289 & 21:39:22.15 &  56:56:48.7$^{\rm r}$& 1483 & 4483                      &                               &                 &                 &   12.7$^{\rm j}$&       &                 &8.299 \,(19) &7.265 \,(33) & 6.880 \,(27)$^{\rm r}$&                    &                 &               &-2.6\,(5.1)   &-3.2\,(5.1)   &-13.1\,(7.2)  &5.3\,(7.3)    &0.47 &0.1  &2x[r]                                                                          \\
1290 & 21:39:34.95 &  56:57:17.1$^{\rm r}$& 1484 & 4484                      &                               &                 &                 &   12.7$^{\rm j}$&       &                 &10.511\,(26) &9.880 \,(28) & 9.698 \,(20)$^{\rm r}$&                    &                 &               &8.7\,(5.2)    &-4.9\,(5.2)   &5.6\,(7.1)    &1.1\,(7.1)    &-0.71&0    &                                                                                      \\
1291 & 21:39:34.03 &  56:57:59.5$^{\rm r}$& 1485 & 4485                      &                               &                 &                 &   13.5$^{\rm j}$&       &                 &12.316\,(29) &12.084\,(32) & 11.923\,(26)$^{\rm r}$&                    &                 &               &4.7\,(4.1)    &-6.7\,(4.1)   &26.7\,(7.3)   &-39.3\,(7.3)  &0.02 &0.09 &                                                                                      \\
1292 & 21:39:08.71 &  56:59:27.9$^{\rm r}$& 1486 & 4486                      &                               &                 &                 &   13.3$^{\rm j}$&       &                 &12.192\,(24) &11.992\,(31) & 11.892\,(23)$^{\rm r}$&                    &                 &               &11.7\,(4)     &-3.7\,(4)     &23.4\,(6.8)   &0.1\,(6.8)    &-0.1 &0.13 &                                                                                      \\
1293 & 21:39:20.41 &  56:58:49.5$^{\rm r}$& 1487 & 4487                      &                               &                 &                 &   12.6$^{\rm j}$&       &                 &11.272\,(35) &10.956\,(43) & 10.893\,(34)$^{\rm r}$&                    &                 &               &5.5\,(4)      &17.2\,(4)     &0.7\,(1.8)    &-2.9\,(2.4)   &-0.37&0.05 &                                                                                      \\
1294 & 21:39:30.68 &  56:59:10.5$^{\rm r}$& 1488 & 4488                      &                               &                 &                 &   14.3$^{\rm j}$&       &                 &11.829\,(24) &11.414\,(28) & 10.926\,(20)$^{\rm r}$&                    &                 &               &-6.2\,(4.1)   &0.5\,(4.1)    &-7.3\,(6.8)   &3.6\,(6.8)    &0.07 &0.09 &                                                                                      \\
1295 & 21:39:34.17 &  56:59:27.0$^{\rm r}$& 1489 & 4489                      &                               &                 &                 &   12.6$^{\rm j}$&       &                 &11.132\,(24) &10.612\,(30) & 10.493\,(19)$^{\rm r}$&                    &                 &               &32.7\,(4.1)   &-7.1\,(4.1)   &35.6\,(7.1)   &-1.4\,(7.1)   &-3.78&-0.47&                                                                                      \\
1296 & 21:39:10.02 &  57:00:04.8$^{\rm r}$& 1490 & 4490                      &                               &                 &                 &   14.2$^{\rm j}$&       &                 &15.288\,(117)&14.711\,(109)&14.383\,(130)$^{\rm r}$&                    &                 &               &              &              &-57.7\,(6.8)  &-0.9\,(6.9)   &0.35 &-0.65&2x[r] (faint)                                                               \\
1297 & 21:39:09.38 &  57:00:05.3$^{\rm r}$& 1490 & 4490                      &                               &                 &                 &   14.2$^{\rm j}$&       &                 &11.177\,(23) &10.480\,(30) & 10.222\,(23)$^{\rm r}$&                    &                 &               &-19.3\,(4)    &-8\,(4)       &-57.7\,(6.8)  &-0.9\,(6.9)   &0.35 &-0.65&2x[r]                                                                           \\
1298 & 21:39:26.15 &  57:00:09.3$^{\rm r}$& 1491 & 4491                      &                               &                 &                 &   12.6$^{\rm j}$&       &                 &11.602\,(21) &11.333\,(28) & 11.278\,(19)$^{\rm r}$&                    &                 &               &5.9\,(2.7)    &-1.6\,(2.7)   &-0.5\,(0.8)   &-4.8\,(0.4)   &0.03 &-0.17&                                                                                      \\
1299 & 21:39:21.84 &  57:00:31.7$^{\rm r}$& 1492 & 4492                      &                               &                 &                 &   12.3$^{\rm j}$&       &                 &10.770\,(26) &10.547\,(28) & 10.461\,(25)$^{\rm r}$&                    &                 &               &2\,(2.7)      &-4.1\,(2.7)   &-1\,(0.7)     &-7.6\,(0.8)   &-1.05&-0.28&                                                                                      \\
1300 & 21:39:22.14 &  57:00:47.6$^{\rm r}$& 1493 & 4493                      &                               &                 &                 &     14$^{\rm j}$&       &                 &10.529\,(23) &9.675 \,(28) & 9.445 \,(23)$^{\rm r}$&                    &                 &               &-1.6\,(5.1)   &-3.5\,(5.1)   &3.3\,(7)      &5.7\,(7)      &-0.01&0.19 &                                                                                      \\
1301 & 21:38:59.97 &  57:02:52.0$^{\rm r}$& 1494 & 4494                      &                               &                 &                 &   14.5$^{\rm j}$&       &                 &12.355\,(23) &12.047\,(28) & 11.875\,(19)$^{\rm r}$&                    &                 &               &-2.3\,(4)     &4\,(4)        &-4.2\,(6.8)   &12.6\,(6.8)   &-0.36&0.6  &                                                                                      \\
1302 & 21:39:09.41 &  57:02:13.4$^{\rm r}$& 1495 & 4495                      &                               &                 &                 &   12.9$^{\rm j}$&       &                 &11.271\,(23) &10.872\,(26) & 10.786\,(20)$^{\rm r}$&                    &                 &               &-7.1\,(4)     &-3.9\,(4)     &-11.1\,(7.2)  &-0.9\,(7.2)   &0.38 &-0.33&                                                                                      \\
1303 & 21:39:33.01 &  57:01:51.3$^{\rm r}$& 1496 & 4496                      &                               &                 &                 &   10.9$^{\rm j}$&       &                 &8.808 \,(41) &8.262 \,(29) & 8.114 \,(21)$^{\rm r}$&                    &                 &               &5.4\,(2.8)    &-10.2\,(2.8)  &1.5\,(1.2)    &-8.7\,(0.9)   &-0.03&-0.49&                                                                                      \\
1304 & 21:39:38.37 &  57:02:10.2$^{\rm r}$& 1497 & 4497                      &                               &                 &                 &   13.5$^{\rm j}$&       &                 &10.794\,(26) &10.095\,(30) & 9.844 \,(20)$^{\rm r}$&                    &                 &               &-2.3\,(5.2)   &-2.1\,(5.2)   &-13.2\,(6.9)  &-55.2\,(7)    &0.09 &0.67 &                                                                                      \\
1305 & 21:39:50.34 &  57:00:58.6$^{\rm r}$& 1498 & 4498                      &                               &                 &                 &   14.3$^{\rm j}$&       &                 &11.073\,(24) &10.366\,(30) & 10.096\,(20)$^{\rm r}$&                    &                 &               &-3.6\,(4.1)   &2.2\,(4.1)    &-9.1\,(6.9)   &15.3\,(6.9)   &-0.29&-0.21&                                                                                      \\
1306 & 21:39:54.41 &  57:00:43.5$^{\rm r}$& 1499 & 4499                      &                               &                 &                 &   13.2$^{\rm j}$&       &                 &10.338\,(24) &9.580 \,(27) & 9.401 \,(20)$^{\rm r}$&                    &                 &               &-1.6\,(5.2)   &-6.1\,(5.2)   &-8\,(7.2)     &-0.9\,(7.2)   &-0.1 &-0.12&                                                                                      \\
1307 & 21:40:03.02 &  56:59:36.8$^{\rm r}$& 1500 & 4500                      &                               &                 &                 &   13.6$^{\rm j}$&       &                 &12.000\,(38) &11.648\,(38) &   11.533\,()$^{\rm r}$&                    &                 &               &8.5\,(4.1)    &1.9\,(4.1)    &26.2\,(7)     &4.4\,(7)      &-0.75&0.36 &                                                                                      \\
1308 & 21:40:12.80 &  56:59:39.2$^{\rm r}$& 1501 & 4501                      &                               &                 &                 &   14.3$^{\rm j}$&       &                 &11.220\,(26) &10.423\,(28) & 10.259\,(20)$^{\rm r}$&                    &                 &               &-6.9\,(4.1)   &2\,(4.1)      &-6.3\,(6.9)   &7.7\,(6.9)    &-0.14&0.12 &                                                                                      \\
1309 & 21:39:53.81 &  57:02:03.1$^{\rm r}$& 1502 & 4502                      &                               &                 &                 &     14$^{\rm j}$&       &                 &12.434\,(27) &12.041\,(27) & 11.962\,(22)$^{\rm r}$&                    &                 &               &-2\,(4.1)     &-8.8\,(4.1)   &-0.3\,(6.9)   &-10.1\,(6.9)  &0.64 &-0.85&                                                                                      \\
1310 & 21:39:57.22 &  57:02:25.4$^{\rm r}$& 1503 & 4503                      &                               &                 &                 &   13.1$^{\rm j}$&       &                 &11.814\,(32) &11.556\,(43) & 11.492\,(22)$^{\rm r}$&                    &                 &               &-3.3\,(5.9)   &0.9\,(5.9)    &              &              &-1.36&0.67 &                                                                                      \\
1311 & 21:39:57.99 &  57:02:21.3$^{\rm r}$& 1504 & 4504                      &                               &                 &                 &   13.2$^{\rm j}$&       &                 &11.858\,(23) &11.674\,(27) & 11.528\,(20)$^{\rm r}$&                    &                 &               &              &              &              &              &0.02 &0.14 &                                                                                      \\
1312 & 21:40:12.02 &  57:01:05.4$^{\rm r}$& 1505 & 4505                      &                               &                 &                 &     14$^{\rm j}$&       &                 &9.912 \,(26) &8.906 \,(28) & 8.605 \,(20)$^{\rm r}$&                    &                 &               &-2.1\,(5.1)   &-3.7\,(5.1)   &45.2\,(7.2)   &-28.8\,(7.2)  &-0.36&0.13 &                                                                                      \\
1313 & 21:40:10.16 &  57:01:39.8$^{\rm r}$& 1506 & 4506                      &                               &                 &                 &   11.9$^{\rm j}$&       &                 &10.722\,(26) &10.455\,(30) & 10.307\,(20)$^{\rm r}$&                    &                 &               &-5.8\,(2.7)   &17\,(2.7)     &-2.1\,(1)     &3\,(2.5)      &-0.21&0.79 &                                                                                      \\
1314 & 21:39:04.10 &  57:04:04.2$^{\rm r}$& 1507 & 4507                      &                               &                 &                 &   12.2$^{\rm j}$&       &                 &11.239\,(23) &11.074\,(29) & 11.000\,(20)$^{\rm r}$&                    &                 &               &0.2\,(4)      &5.5\,(4)      &1.7\,(0.8)    &-0.7\,(4.9)   &-0.75&0.33 &                                                                                      \\
1315 & 21:39:28.01 &  57:03:23.9$^{\rm r}$& 1508 & 4508                      &                               &                 &                 &   14.4$^{\rm j}$&       &                 &9.700 \,(22) &8.656 \,(26) & 8.294 \,(26)$^{\rm r}$&                    &                 &               &              &              &              &              &.0.12&0.07 &                                                                                      \\
1316 & 21:39:28.12 &  57:03:32.9$^{\rm r}$& 1509 & 4509                      &                               &                 &                 &   13.4$^{\rm j}$&       &                 &11.403\,(26) &10.917\,(32) & 10.873\,(30)$^{\rm r}$&                    &                 &               &-13.2\,(4)    &-20.5\,(4)    &              &              &1.07 &-2.72&                                                                                      \\
1317 & 21:39:37.94 &  57:04:03.2$^{\rm r}$& 1510 & 4510                      &                               &  13.91$^{\rm l}$&  13.42$^{\rm l}$&  12.56$^{\rm l}$&       &                 &10.387\,(21) &10.070\,(30) & 9.921 \,(22)$^{\rm r}$&                    &                 &               &-4.1\,(5.1)   &-1.4\,(5.1)   &-3.9\,(1.5)   &-0.7\,(2.1)   &-0.01&0.26 &                                                                                      \\
1318 & 21:40:04.98 &  57:03:20.8$^{\rm r}$& 1511 & 4511                      &                               &                 &                 &   14.3$^{\rm j}$&       &                 &12.709\,(23) &12.394\,(33) & 12.311\,(30)$^{\rm r}$&                    &                 &               &-3.9\,(5.9)   &-2.3\,(5.9)   &-8.5\,(6.8)   &3.9\,(6.8)    &0.25 &0.27 &                                                                                      \\
1319 & 21:38:57.31 &  57:05:19.2$^{\rm r}$& 1512 & 4512                      &                               &                 &                 &   14.4$^{\rm j}$&       &                 &12.736\,(26) &12.453\,(29) & 12.462\,(23)$^{\rm r}$&                    &                 &               &-1.5\,(4)     &-2.6\,(4)     &-5.5\,(6.9)   &4.3\,(6.8)    &-0.15&-0.05&                                                                                      \\
1320 & 21:39:02.46 &  57:05:02.6$^{\rm r}$& 1513 & 4513                      &                               &                 &                 &   14.2$^{\rm j}$&       &                 &12.560\,(25) &12.169\,(32) & 12.097\,(26)$^{\rm r}$&                    &                 &               &-11.9\,(4)    &-4.5\,(4)     &-24.7\,(6.8)  &-3\,(6.8)     &0.96 &-0.55&                                                                                      \\
1321 & 21:39:10.64 &  57:04:49.9$^{\rm r}$& 1514 & 4514                      &                               &                 &                 &   14.2$^{\rm j}$&       &                 &11.002\,(25) &10.284\,(26) & 10.078\,(22)$^{\rm r}$&                    &                 &               &-24.3\,(18)   &15.5\,(18)    &-31\,(7.2)    &54.4\,(7.2)   &-0.31&0.27 &                                                                                      \\
1322 & 21:39:14.19 &  57:04:41.6$^{\rm r}$& 1515 & 4515                      &                               &                 &                 &   14.5$^{\rm j}$&       &                 &12.720\,(27) &12.437\,(32) & 12.294\,(28)$^{\rm r}$&                    &                 &               &              &              &              &              &-0.44&0.31 &near 1322                                                                           \\
1323 & 21:39:14.15 &  57:04:46.1$^{\rm r}$& 1516 & 4516                      &                               &                 &                 &   14.2$^{\rm j}$&       &                 &12.348\,(52) &11.990\,(61) &   11.871\,()$^{\rm r}$&                    &                 &               &-6.2\,(4)     &54.5\,(4)     &              &              &0.25 &-0.26&near 1321                                                                          \\
1324 & 21:39:18.52 &  57:05:40.1$^{\rm r}$& 1517 & 4517                      &                               &                 &                 &   13.4$^{\rm j}$&       &                 &12.056\,(22) &11.717\,(26) & 11.610\,(22)$^{\rm r}$&                    &                 &               &-6.4\,(4)     &-3.9\,(4)     &-15.5\,(6.9)  &-3.2\,(6.9)   &0.01 &-0.21&                                                                                      \\
1325 & 21:38:53.70 &  57:06:28.8$^{\rm r}$& 1518 & 4518                      &                               &                 &                 &   14.4$^{\rm j}$&       &                 &12.530\,(57) &12.211\,(41) &   12.083\,()$^{\rm r}$&                    &                 &               &-16.4\,(4)    &20.8\,(4)     &-69.5\,(7.2)  &69.7\,(7.2)   &-0.36&0.05 &2x[r]                                                                           \\
1326 & 21:38:54.21 &  57:06:27.5$^{\rm r}$& 1518 & 4518                      &                               &                 &                 &   14.4$^{\rm j}$&       &                 &12.949\,(31) &12.158\,(50) & 11.946\,(29)$^{\rm r}$&                    &                 &               &              &              &              &              &-0.36&0.05 &2x[r]                                                                           \\
1327 & 21:38:59.34 &  57:06:31.1$^{\rm r}$& 1519 & 4519                      &                               &                 &                 &   14.6$^{\rm j}$&       &                 &12.776\,()   &12.606\,(61) &   12.411\,()$^{\rm r}$&                    &                 &               &5.2\,(4)      &7.1\,(4)      &21.9\,(6.8)   &37.1\,(6.8)   &0.11 &0.06 &                                                                                      \\
1328 & 21:39:16.71 &  57:07:09.4$^{\rm r}$& 1520 & 4520                      &                               &                 &                 &   13.1$^{\rm j}$&       &                 &10.457\,(22) &9.746 \,(29) & 9.562 \,(22)$^{\rm r}$&                    &                 &               &-0.6\,(5.1)   &-4.3\,(5.1)   &67.6\,(7.3)   &-12\,(7.3)    &0.04 &-0.31&                                                                                      \\
1329 & 21:39:11.98 &  57:07:28.1$^{\rm r}$& 1521 & 4521                      &                               &                 &                 &   13.9$^{\rm j}$&       &                 &12.278\,(25) &11.886\,(28) & 11.771\,(26)$^{\rm r}$&                    &                 &               &-15.7\,(4)    &-4.1\,(4)     &-13.5\,(6.8)  &-0.6\,(6.8)   &0.93 &-0.4 &                                                                                      \\
1330 & 21:40:00.10 &  57:05:26.9$^{\rm r}$& 1522 & 4522                      &                               &                 &                 &   14.5$^{\rm j}$&       &                 &11.197\,(21) &10.439\,(27) & 10.274\,(20)$^{\rm r}$&                    &                 &               &-6.3\,(4.1)   &-7.7\,(4.1)   &-2.8\,(6.9)   &1.1\,(7)      &0.43 &-0.16&                                                                                      \\
1331 & 21:40:05.08 &  57:06:27.1$^{\rm r}$& 1523 & 4523                      &                               &                 &                 &   14.4$^{\rm j}$&       &                 &10.275\,(23) &9.252 \,(27) & 8.983 \,(20)$^{\rm r}$&                    &                 &               &-4.2\,(5.2)   &-1.8\,(5.2)   &9.9\,(7.1)    &33.3\,(7.2)   &-0.13&0.3  &                                                                                      \\
1332 & 21:39:56.32 &  57:07:16.1$^{\rm r}$& 1524 & 4524                      &                               &                 &                 &   13.3$^{\rm j}$&       &                 &9.773 \,(21) &8.998 \,(69) & 8.578 \,(20)$^{\rm r}$&                    &                 &               &-11.3\,(5.2)  &6.9\,(5.2)    &-59\,(7.4)    &51.9\,(7.5)   &-0.19&0.44 &2x[r]                                                                           \\
1333 & 21:39:56.77 &  57:07:13.0$^{\rm r}$& 1524 & 4524                      &                               &                 &                 &   13.3$^{\rm j}$&       &                 &13.796\,(113)&10.597\,()   &   10.092\,()$^{\rm r}$&                    &                 &               &              &              &              &              &-0.19&0.44 &2x[r] (faint)                                                               \\
1334 & 21:39:49.55 &  57:07:39.3$^{\rm r}$& 1525 & 4525                      &                               &                 &                 &   14.5$^{\rm j}$&       &                 &12.625\,(21) &12.353\,(28) & 12.207\,(26)$^{\rm r}$&                    &                 &               &10.5\,(4.1)   &2.3\,(4.1)    &12.4\,(6.9)   &6.9\,(6.8)    &-1.04&0.39 &                                                                                      \\
1335 & 21:39:21.44 &  57:08:30.5$^{\rm r}$& 1526 & 4526                      &                               &                 &                 &   14.6$^{\rm j}$&       &                 &12.812\,(25) &12.514\,(32) & 12.400\,(28)$^{\rm r}$&                    &                 &               &-6.9\,(4)     &0.7\,(4)      &-1.2\,(6.8)   &-5.9\,(6.8)   &0.65 &-0.02&                                                                                      \\
1336 & 21:39:06.86 &  57:09:24.9$^{\rm j}$& 1527 & 4527                      &                               &                 &                 &   14.6$^{\rm j}$&       &                 &             &             &                       &                    &                 &               &              &              &              &              &     &     &no star                                                                          \\
1337 & 21:39:02.80 &  57:10:08.7$^{\rm r}$& 1528 & 4528                      &                               &                 &                 &   14.8$^{\rm j}$&       &                 &10.345\,(23) &9.272 \,(29) & 8.975 \,(23)$^{\rm r}$&                    &                 &               &-14\,(5.1)    &6.8\,(5.1)    &-56.3\,(6.9)  &59.7\,(6.9)   &     &     &                                                                                      \\
1338 & 21:39:18.25 &  57:09:45.4$^{\rm r}$& 1529 & 176                       &                               &                 &                 &   10.5$^{\rm j}$&       &                 &9.536 \,(22) &9.346 \,(26) & 9.332 \,(23)$^{\rm r}$&        F0$^{\rm q}$&                 &               &-0.8\,(1.6)   &-14.4\,(1.6)  &4.1\,(0.9)    &-12.7\,(0.6)  &-0.53&-0.95&                                                                                      \\
1339 & 21:39:33.90 &  57:08:45.5$^{\rm r}$& 1530 & 4530                      &                               &                 &                 &   14.3$^{\rm j}$&       &                 &13.124\,(23) &12.992\,(30) & 12.888\,(25)$^{\rm r}$&                    &                 &               &-1.6\,(4.1)   &-0.3\,(4.1)   &-0.4\,(6.8)   &6.1\,(6.7)    &0.16 &0.17 &                                                                                      \\
1340 & 21:39:27.23 &  57:09:37.6$^{\rm r}$& 1531 & 4531                      &                               &                 &                 &   13.3$^{\rm j}$&       &                 &11.851\,(22) &11.528\,(29) & 11.463\,(23)$^{\rm r}$&                    &                 &               &8.7\,(4)      &3.6\,(4)      &14.1\,(6.9)   &7.6\,(6.9)    &-1.2 &0.77 &                                                                                      \\
1341 & 21:39:23.17 &  57:10:13.0$^{\rm r}$& 1532 & 4532                      &                               &                 &                 &   12.5$^{\rm j}$&       &                 &11.243\,(23) &10.818\,(29) & 10.744\,(23)$^{\rm r}$&                    &                 &               &-18.3\,(4)    &11\,(4)       &-18.2\,(0.5)  &2.8\,(4.1)    &1.3  &0.76 &                                                                                      \\
1342 & 21:39:50.14 &  57:08:24.1$^{\rm r}$& 1533 & 4533                      &                               &                 &                 &   13.2$^{\rm j}$&       &                 &11.915\,(23) &11.611\,(28) & 11.554\,(23)$^{\rm r}$&                    &                 &               &2.5\,(4.1)    &-1\,(4.1)     &-1.8\,(7)     &3.8\,(7)      &-0.37&-0.65&                                                                                      \\
1343 & 21:39:47.93 &  57:09:52.3$^{\rm r}$& 1534 & 4534                      &                               &                 &                 &   13.5$^{\rm j}$&       &                 &12.102\,(23) &11.822\,(31) & 11.725\,(23)$^{\rm r}$&                    &                 &               &-1.7\,(4.1)   &4.7\,(4.1)    &-4.7\,(7)     &14.2\,(7)     &0.12 &0.02 &                                                                                      \\
1344 & 21:39:46.62 &  57:10:33.7$^{\rm r}$& 1535 & 179                       &                               &                 &                 &   11.6$^{\rm j}$&       &                 &10.880\,(23) &10.778\,(30) & 10.690\,(22)$^{\rm r}$&        A0$^{\rm q}$&                 &               &-6\,(1.7)     &-1.5\,(1.7)   &-5.3\,(1.1)   &-3.6\,(1.1)   &0.36 &-0.05&                                                                                      \\
1345 & 21:40:01.51 &  57:09:28.9$^{\rm j}$& 1536 & 4536                      &                               &                 &                 &   14.1$^{\rm j}$&       &                 &             &             &                       &                    &                 &               &              &              &              &              &     &     &no star                                                                          \\
1346 & 21:39:55.97 &  57:10:50.4$^{\rm r}$& 1537 & 181                       &                               &                 &                 &   11.3$^{\rm j}$&       &                 &10.365\,(21) &10.118\,(27) & 10.065\,(20)$^{\rm r}$&        G2$^{\rm q}$&                 &               &16.1\,(1.7)   &25.5\,(1.7)   &12.6\,(0.9)   &26.1\,(0.6)   &-1.64&3.19 &                                                                                      \\
1347 & 21:39:49.65 &  57:11:11.5$^{\rm r}$& 1538 & 4538                      &                               &                 &                 &   13.5$^{\rm j}$&       &                 &12.097\,(21) &11.837\,(30) & 11.703\,(22)$^{\rm r}$&                    &                 &               &10.7\,(4.1)   &5.9\,(4.1)    &11.2\,(6.9)   &18.1\,(6.9)   &-1.46&0.92 &                                                                                      \\
1348 & 21:40:19.25 &  57:10:59.5$^{\rm r}$& 1539 & 182                       &                               &                 &                 &     12$^{\rm j}$&       &                 &9.754 \,(26) &9.183 \,(31) & 8.982 \,(19)$^{\rm r}$&       gK2$^{\rm q}$&                 &               &-11.1\,(5.1)  &-8.3\,(5.1)   &-3.2\,(7.8)   &-1.5\,(7.8)   &0.32 &-0.29&                                                                                      \\
1349 & 21:40:33.27 &  57:05:40.4$^{\rm r}$& 1540 & 4540                      &                               &                 &                 &     12$^{\rm j}$&       &                 &9.662 \,(26) &9.040 \,(31) & 8.836 \,(24)$^{\rm r}$&                    &                 &               &-42.6\,(8.2)  &38\,(8.2)     &5.6\,(1)      &17.3\,(2.6)   &-1.09&2.18 &                                                                                      \\
1350 & 21:40:36.55 &  57:01:34.9$^{\rm r}$& 1541 & 4541                      &                               &                 &                 &   13.8$^{\rm j}$&       &                 &11.074\,(27) &10.429\,(35) & 10.190\,(23)$^{\rm r}$&                    &                 &               &-28.5\,(12.7) &-16.6\,(12.7) &-28.4\,(7.2)  &-56.4\,(7.2)  &-0.1 &-0.23&                                                                                      \\
1351 & 21:40:28.94 &  57:12:42.5$^{\rm r}$& 1542 & 4542                      &                               &                 &                 &   13.5$^{\rm j}$&       &                 &12.000\,(24) &11.769\,(30) & 11.656\,(24)$^{\rm r}$&                    &                 &               &-6.1\,(4.1)   &-4.2\,(4.1)   &-24.7\,(6.6)  &1.3\,(6.6)    &-0.06&0.1  &                                                                                      \\
1352 & 21:40:29.59 &  57:13:09.1$^{\rm r}$& 1543 & 4543                      &                               &                 &                 &   13.8$^{\rm j}$&       &                 &12.299\,(29) &12.029\,(33) & 11.911\,(24)$^{\rm r}$&                    &                 &               &-23.8\,(5.4)  &-6.4\,(5.4)   &              &              &0.62 &-0.21&                                                                                      \\
1353 & 21:40:30.34 &  57:13:26.3$^{\rm r}$& 1544 & 185                       &                               &                 &   11.5$^{\rm f}$&     11$^{\rm e}$&       &                 &9.821 \,(24) &9.735 \,(31) & 9.625 \,(23)$^{\rm r}$&        A1$^{\rm e}$&                 &  1.5$^{\rm e}$&3.6\,(1.7)    &-2.7\,(1.6)   &6.2\,(0.5)    &1.4\,(0.6)    &-0.98&0.78 &                                                                                      \\
1354 & 21:40:29.54 &  57:17:21.6$^{\rm r}$& 1545 & 4545                      &                               &                 &                 &   12.6$^{\rm j}$&       &                 &11.393\,(26) &10.990\,(30) & 10.907\,(21)$^{\rm r}$&                    &                 &               &24.2\,(5.4)   &24.4\,(5.4)   &22.1\,(7.4)   &31.4\,(7.4)   &-3.53&2.26 &                                                                                      \\
1355 & 21:40:46.92 &  57:00:54.0$^{\rm r}$& 1546 & 4546                      &                               &                 &                 &   13.4$^{\rm j}$&       &                 &10.896\,(24) &10.299\,(31) & 10.107\,(21)$^{\rm r}$&                    &                 &               &-6.9\,(4.1)   &-1.7\,(4.1)   &-4.9\,(7)     &5.7\,(7)      &0.07 &-0.07&                                                                                      \\
1356 & 21:40:54.57 &  57:00:45.1$^{\rm r}$& 1547 & 4547                      &                               &                 &                 &   12.3$^{\rm j}$&       &                 &11.024\,(26) &10.718\,(31) & 10.642\,(23)$^{\rm r}$&                    &                 &               &-16.8\,(5.8)  &-20.5\,(5.8)  &-2.1\,(1.2)   &-10.2\,(0.6)  &-0.09&-1.11&                                                                                      \\
1357 & 21:40:48.14 &  57:04:30.9$^{\rm r}$& 1549 & 4549                      &                               &                 &                 &   13.8$^{\rm j}$&       &                 &12.611\,(26) &12.453\,(31) & 12.329\,(28)$^{\rm r}$&                    &                 &               &-5.6\,(4.1)   &0.2\,(4.1)    &-3.7\,(6.9)   &7.7\,(6.9)    &-0.19&0.26 &                                                                                      \\
1358 & 21:40:48.43 &  57:04:56.0$^{\rm r}$& 1550 & 4550                      &                               &                 &                 &   12.4$^{\rm j}$&       &                 &9.716 \,(24) &9.077 \,(30) & 8.819 \,(19)$^{\rm r}$&                    &                 &               &-7.8\,(5.1)   &-2.9\,(5.1)   &-0.6\,(7.4)   &2.1\,(7.4)    &0.01 &-0.03&                                                                                      \\
1359 & 21:41:00.06 &  57:04:00.4$^{\rm r}$& 1551 & 4551                      &                               &                 &                 &     11$^{\rm j}$&       &                 &8.788 \,(18) &8.368 \,(29) & 8.071 \,(21)$^{\rm r}$&                    &                 &               &-7.2\,(2)     &-4.8\,(2)     &-5.1\,(0.5)   &-3.5\,(0.8)   &0.29 &0.04 &                                                                                      \\
1360 & 21:41:00.48 &  57:03:51.2$^{\rm r}$& 1552 & 4552                      &                               &                 &                 &   14.3$^{\rm j}$&       &                 &12.645\,(34) &12.300\,(36) & 12.192\,(35)$^{\rm r}$&                    &                 &               &              &              &              &              &0.32 &0.19 &                                                                                      \\
1361 & 21:41:05.67 &  57:02:20.4$^{\rm r}$& 1553 & 4553                      &                               &                 &                 &   14.4$^{\rm j}$&       &                 &12.745\,(39) &12.434\,(40) & 12.332\,(37)$^{\rm r}$&                    &                 &               &-21.7\,(4.1)  &1.9\,(4.1)    &-41.1\,(6.8)  &23.9\,(6.8)   &0.1  &0.09 &                                                                                      \\
1362 & 21:41:18.15 &  57:02:07.1$^{\rm r}$& 1554 & 4554                      &                               &                 &                 &   14.2$^{\rm j}$&       &                 &12.789\,(24) &12.487\,(27) & 12.441\,(26)$^{\rm r}$&                    &                 &               &-3.9\,(4.1)   &-2\,(4.1)     &-3.3\,(6.8)   &6.1\,(6.8)    &-0.06&-0.04&                                                                                      \\
1363 & 21:41:21.11 &  57:03:25.1$^{\rm r}$& 1556 & 4556                      &                               &                 &                 &   10.2$^{\rm j}$&       &                 &5.477 \,(21) &4.741 \,()   & 4.277 \,(33)$^{\rm r}$&                    &                 &               &-11.5\,(2)    &-3.9\,(2)     &-7.6\,(1.5)   &-6.1\,(0.8)   &0.81 &-31  &                                                                                      \\
1364 & 21:41:29.90 &  57:02:51.5$^{\rm r}$& 1557 & 4557                      &                               &                 &                 &   11.7$^{\rm j}$&       &                 &10.867\,(24) &10.673\,(28) & 10.676\,(23)$^{\rm r}$&                    &                 &               &-15\,(2)      &-1\,(2)       &-9.4\,(0.6)   &-6.7\,(1.6)   &0.7  &-0.29&                                                                                      \\
1365 & 21:41:02.67 &  57:05:01.7$^{\rm r}$& 1558 & 4558                      &                               &                 &                 &   13.4$^{\rm j}$&       &                 &12.075\,(24) &11.866\,(32) & 11.749\,(21)$^{\rm r}$&                    &                 &               &-6.1\,(4.1)   &-4.3\,(4.1)   &-3.1\,(6.9)   &6.4\,(6.9)    &0.47 &-0.02&                                                                                      \\
1366 & 21:40:49.42 &  57:06:00.8$^{\rm r}$& 1559 & 4559                      &                               &                 &                 &   13.8$^{\rm j}$&       &                 &12.689\,(27) &12.521\,(36) & 12.424\,(30)$^{\rm r}$&                    &                 &               &-10.9\,(8.6)  &-0.4\,(8.6)   &-21.5\,(7)    &18.7\,(7)     &-0.01&0·35 &                                                                                      \\
1367 & 21:41:22.98 &  57:06:32.1$^{\rm r}$& 1560 & 4560                      &                               &                 &                 &   14.6$^{\rm j}$&       &                 &12.524\,(29) &12.019\,(36) & 11.846\,(28)$^{\rm r}$&                    &                 &               &-5.7\,(4.1)   &-0.6\,(4.1)   &-7.2\,(6.8)   &15.8\,(6.8)   &-0.05&0.32 &                                                                                      \\
1368 & 21:40:56.13 &  57:08:17.0$^{\rm r}$& 1561 & 4561                      &                               &                 &                 &   14.4$^{\rm j}$&       &                 &12.775\,(27) &12.516\,(33) & 12.361\,(26)$^{\rm r}$&                    &                 &               &-3.2\,(4.1)   &-10.6\,(4.1)  &19.8\,(6.7)   &-35\,(6.8)    &-0.64&0.01 &                                                                                      \\
1369 & 21:40:48.45 &  57:09:05.3$^{\rm r}$& 1562 & 4562                      &                               &                 &                 &   14.6$^{\rm j}$&       &                 &12.913\,(24) &12.642\,(32) & 12.516\,(23)$^{\rm r}$&                    &                 &               &-6.5\,(4.1)   &-3.9\,(4.1)   &-4.2\,(6.8)   &3.5\,(6.8)    &0.15 &-0.56&                                                                                      \\
1370 & 21:41:01.56 &  57:08:18.7$^{\rm r}$& 1563 & 188                       &                               &                 &                 &    8.9$^{\rm j}$&       &                 &6.923 \,(23) &6.390 \,(38) & 6.333 \,(17)$^{\rm r}$&        G8$^{\rm q}$&                 &               &17.5\,(1.3)   &33.8\,(1.3)   &19.1\,(0.6)   &33.6\,(0.6)   &-2.3 &3.97 &                                                                                      \\
1371 & 21:41:17.70 &  57:07:24.8$^{\rm r}$& 1564 & 4564                      &                               &                 &                 &   12.6$^{\rm j}$&       &                 &11.961\,(26) &11.826\,(32) & 11.728\,(23)$^{\rm r}$&                    &                 &               &-4.4\,(4.1)   &1.1\,(4.1)    &-0.1\,(7.3)   &6.6\,(7.2)    &-0.14&-0.14&                                                                                      \\
1372 & 21:41:01.31 &  57:09:23.2$^{\rm r}$& 1565 & 4565                      &                               &                 &                 &   13.6$^{\rm j}$&       &                 &10.444\,(24) &9.669 \,(29) & 9.416 \,(21)$^{\rm r}$&                    &                 &               &-6.3\,(5.1)   &-2.9\,(5.1)   &-6.8\,(7.1)   &6.9\,(7.1)    &0.16 &0.03 &                                                                                      \\
1373 & 21:41:21.93 &  57:08:10.1$^{\rm r}$& 1566 & 4566                      &                               &                 &                 &   12.8$^{\rm j}$&       &                 &10.569\,(26) &9.989 \,(32) & 9.786 \,(23)$^{\rm r}$&                    &                 &               &-17.4\,(5.1)  &-9.6\,(5.1)   &-47.8\,(7.4)  &9.6\,(7.4)    &1.18 &-0.58&                                                                                      \\
1374 & 21:41:44.08 &  57:05:39.0$^{\rm r}$& 1567 & 4567                      &                               &                 &                 &   11.3$^{\rm j}$&       &                 &10.696\,(26) &10.568\,(31) & 10.433\,(18)$^{\rm r}$&                    &                 &               &-3.8\,(2)     &-6.3\,(2)     &-3.7\,(0.8)   &-5.6\,(0.7)   &0.25 &-0.06&                                                                                      \\
1375 & 21:41:47.33 &  57:04:40.6$^{\rm r}$& 1568 & 4568                      &                               &                 &                 &   13.9$^{\rm j}$&       &                 &10.834\,(26) &10.070\,(27) & 9.840 \,(21)$^{\rm r}$&                    &                 &               &5.2\,(5.1)    &2.2\,(5.1)    &10.6\,(7)     &15.3\,(7)     &-0.63&0.28 &                                                                                      \\
1376 & 21:41:48.33 &  57:02:47.2$^{\rm r}$& 1569 & 4569                      &                               &                 &                 &   14.1$^{\rm j}$&       &                 &12.367\,(29) &12.000\,(30) & 11.900\,(23)$^{\rm r}$&                    &                 &               &-4.2\,(4.1)   &-12.9\,(4.1)  &-3.5\,(6.9)   &-19.3\,(6.9)  &-0.49&-0.14&                                                                                      \\
1377 & 21:41:47.71 &  57:07:26.3$^{\rm r}$& 1571 & 4571                      &                               &                 &                 &   14.3$^{\rm j}$&       &                 &12.754\,(27) &12.596\,(35) & 12.528\,(26)$^{\rm r}$&                    &                 &               &-5.6\,(4.1)   &0.4\,(4.1)    &-0.3\,(6.8)   &11.1\,(6.9)   &-0.15&0.07 &                                                                                      \\
1378 & 21:41:44.18 &  57:08:09.4$^{\rm r}$& 1572 & 4572                      &                               &                 &                 &   13.3$^{\rm j}$&       &                 &12.255\,(24) &12.142\,(35) & 12.001\,(24)$^{\rm r}$&                    &                 &               &2.9\,(4.1)    &-3.4\,(4.1)   &39.9\,(7.1)   &-3\,(7.1)     &0.26 &0.05 &                                                                                      \\
1379 & 21:41:42.04 &  57:08:41.6$^{\rm r}$& 1573 & 4573                      &                               &                 &                 &   12.8$^{\rm j}$&       &                 &11.327\,(36) &10.936\,(44) & 10.800\,(36)$^{\rm r}$&                    &                 &               &-23.1\,(4.1)  &-0.7\,(4.1)   &-28.9\,(7.6)  &59.9\,(7.6)   &2.12 &-1.62&                                                                                      \\
1380 & 21:41:39.26 &  57:09:44.2$^{\rm r}$& 1574 & 4574                      &                               &                 &                 &  14.11$^{\rm j}$&       &                 &12.864\,(27) &12.628\,(36) & 12.541\,(24)$^{\rm r}$&                    &                 &               &-4.8\,(4.1)   &1.7\,(4.1)    &3.4\,(6.8)    &4.2\,(6.9)    &-0.18&-0.04&                                                                                      \\
1381 & 21:40:46.56 &  57:10:48.5$^{\rm r}$& 1575 & 4575                      &                               &                 &                 &   14.2$^{\rm j}$&       &                 &12.512\,(31) &12.234\,(38) & 12.122\,(30)$^{\rm r}$&                    &                 &               &17.2\,(4.1)   &0.3\,(4.1)    &47.9\,(6.9)   &33.8\,(6.9)   &-0.3 &0.13 &                                                                                      \\
1382 & 21:41:41.06 &  57:10:44.0$^{\rm r}$& 1576 & 4576                      &                               &                 &                 &   14.1$^{\rm j}$&       &                 &12.649\,(24) &12.515\,(33) & 12.333\,(24)$^{\rm r}$&                    &                 &               &-2.4\,(4.1)   &2.5\,(4.1)    &2.9\,(6.9)    &9.3\,(6.9)    &-0.15&0.27 &                                                                                      \\
1383 & 21:41:33.06 &  57:11:45.3$^{\rm r}$& 1577 & 4577                      &                               &                 &                 &   13.8$^{\rm j}$&       &                 &10.629\,(26) &9.923 \,(31) & 9.619 \,(21)$^{\rm r}$&                    &                 &               &3.9\,(5.1)    &2.7\,(5.1)    &3.9\,(7)      &16\,(7.1)     &-0.67&0.56 &                                                                                      \\
1384 & 21:41:39.17 &  57:11:33.3$^{\rm r}$& 1578 & 4578                      &                               &                 &                 &   12.7$^{\rm j}$&       &                 &11.456\,(26) &11.219\,(31) & 11.098\,(23)$^{\rm r}$&                    &                 &               &0.7\,(4.1)    &-1.8\,(4.1)   &8.6\,(7.2)    &5.5\,(7.2)    &-0.53&0.35 &                                                                                      \\
1385 & 21:42:10.19 &  57:03:20.3$^{\rm r}$& 1579 & 4579                      &                               &                 &                 &     11$^{\rm j}$&       &                 &9.979 \,(24) &9.666 \,(31) & 9.609 \,(22)$^{\rm r}$&                    &                 &               &              &              &              &              &1.86 &-3.12&                                                                                      \\
1386 & 21:42:03.86 &  57:07:59.8$^{\rm r}$& 1580 & 4580                      &                               &                 &                 &   12.2$^{\rm j}$&       &                 &10.674\,(26) &10.340\,(32) & 10.174\,(22)$^{\rm r}$&                    &                 &               &6.6\,(6.3)    &-14.3\,(6.3)  &-19.6\,(7.7)  &-1.5\,(7.7)   &0.72 &-0.15&                                                                                      \\
1387 & 21:42:12.63 &  57:08:03.2$^{\rm r}$& 1581 & 4581                      &                               &                 &                 &   14.2$^{\rm j}$&       &                 &12.238\,(29) &11.852\,(32) & 11.694\,(23)$^{\rm r}$&                    &                 &               &-2\,(4.1)     &-4.5\,(4.1)   &4.7\,(7)      &-17.5\,(7)    &-0.21&-0.06&                                                                                      \\
1388 & 21:42:21.73 &  57:05:07.9$^{\rm r}$& 1582 & 201                       &                               &                 &                 &   10.7$^{\rm j}$&       &                 &10.187\,(26) &10.112\,(32) & 10.054\,(22)$^{\rm r}$&        B8$^{\rm q}$&                 &               &-1.3\,(1.3)   &-6.9\,(1.5)   &-2.7\,(0.9)   &-5\,(0.9)     &0.6  &0    &                                                                                      \\
1389 & 21:42:45.35 &  57:01:12.7$^{\rm r}$& 1583 & 4583                      &                               &                 &                 &   13.3$^{\rm j}$&       &                 &11.526\,(27) &10.944\,(30) & 10.889\,(21)$^{\rm r}$&                    &                 &               &-69\,(4.1)    &-9.5\,(4.1)   &-61\,(7.1)    &-0.1\,(7.1)   &     &     &                                                                                      \\
1390 & 21:42:46.76 &  57:01:47.4$^{\rm r}$& 1584 & 204                       &                               &   9.06$^{\rm l}$&   9.62$^{\rm l}$&   9.41$^{\rm l}$&       &                 &10.206\,(26) &9.983 \,(32) & 9.893 \,(19)$^{\rm r}$&        B2$^{\rm p}$&      V$^{\rm p}$&               &11.3\,(2)     &3.5\,(2)      &8.4\,(1.1)    &4.1\,(1.5)    &-0.78&1.11 &[m] HIP\# wrong                                                                       \\
1391 & 21:42:46.69 &  57:01:59.3$^{\rm r}$& 1585 & 4585                      &                               &                 &                 &   11.8$^{\rm j}$&       &                 &10.779\,(27) &10.524\,(32) & 10.414\,(19)$^{\rm r}$&                    &                 &               &10.9\,(12.8)  &28.9\,(12.8)  &5.4\,(1.6)    &4.7\,(1.4)    &-1.01&0.96 &                                                                                      \\
1392 & 21:42:35.91 &  57:03:34.9$^{\rm r}$& 1586 & 4586                      &                               &                 &                 &   14.2$^{\rm j}$&       &                 &10.936\,(26) &10.171\,(31) & 9.962 \,(19)$^{\rm r}$&                    &                 &               &-8.4\,(5.1)   &-7.1\,(5.1)   &-4.3\,(7)     &4.5\,(7)      &0.65 &-0.17&                                                                                      \\
1393 & 21:42:21.88 &  57:05:27.5$^{\rm r}$& 1587 & 4587                      &                               &                 &                 &     14$^{\rm j}$&       &                 &12.157\,(24) &11.871\,(33) & 11.667\,(25)$^{\rm r}$&                    &                 &               &-2.6\,(4.1)   &-1.3\,(4.1)   &7.7\,(6.7)    &1.4\,(6.7)    &-0.35&0.17 &                                                                                      \\
1394 & 21:42:36.29 &  57:05:13.7$^{\rm r}$& 1588 & 4588                      &                               &                 &                 &   14.4$^{\rm j}$&       &                 &11.341\,(26) &10.641\,(31) & 10.418\,(22)$^{\rm r}$&                    &                 &               &-4.9\,(4.1)   &-8.2\,(4.1)   &1.1\,(6.9)    &5.1\,(6.9)    &0.29 &-0.53&                                                                                      \\
1395 & 21:42:20.12 &  57:05:59.0$^{\rm r}$& 1589 & 199                       &                               &                 &                 &   11.7$^{\rm j}$&       &                 &10.626\,(26) &10.452\,(31) & 10.324\,(22)$^{\rm r}$&        A5$^{\rm q}$&                 &               &-1.4\,(2.7)   &1\,(2.7)      &-1.1\,(0.7)   &0.1\,(1.2)    &-0.09&0.43 &                                                                                      \\
1396 & 21:42:23.80 &  57:08:56.8$^{\rm r}$& 1590 & 4590                      &                               &                 &                 &   14.3$^{\rm j}$&       &                 &12.583\,(26) &12.376\,(37) & 12.228\,(26)$^{\rm r}$&                    &                 &               &-6.8\,(4.1)   &-3.9\,(4.1)   &-1.2\,(6.8)   &6.5\,(6.8)    &0.45 &-0.58&                                                                                      \\
1397 & 21:42:19.89 &  57:09:27.6$^{\rm r}$& 1591 & 4591                      &                               &                 &                 &   12.7$^{\rm j}$&       &                 &11.638\,(26) &11.443\,(31) & 11.328\,(23)$^{\rm r}$&                    &                 &               &-4.2\,(4.1)   &2.3\,(4.1)    &-2.4\,(0.6)   &-0.1\,(3.5)   &-0.19&0.07 &                                                                                      \\
1398 & 21:42:12.89 &  57:10:18.1$^{\rm r}$& 1592 & 4592                      &                               &                 &                 &   14.2$^{\rm j}$&       &                 &12.342\,(26) &11.978\,(30) & 11.857\,(22)$^{\rm r}$&                    &                 &               &6.2\,(4.1)    &-0.7\,(4.1)   &10.5\,(6.9)   &9.4\,(6.9)    &-1.18&-0.03&                                                                                      \\
1399 & 21:42:00.00 &  57:11:01.3$^{\rm r}$& 1593 & 4593                      &                               &                 &                 &   14.7$^{\rm j}$&       &                 &12.996\,(24) &12.749\,(29) & 12.624\,(26)$^{\rm r}$&                    &                 &               &-7.1\,(4.1)   &-6.5\,(4.1)   &-2.8\,(7)     &5.2\,(7)      &     &     &                                                                                      \\
1400 & 21:42:08.07 &  57:11:21.3$^{\rm r}$& 1594 & 4594                      &                               &                 &                 &   14.9$^{\rm j}$&       &                 &13.067\,(27) &12.731\,(33) & 12.573\,(26)$^{\rm r}$&                    &                 &               &-10\,(4.1)    &-4.4\,(4.1)   &-2.2\,(7)     &-2\,(7)       &     &     &                                                                                      \\
1401 & 21:42:20.11 &  57:10:14.4$^{\rm r}$& 1595 & 4595                      &                               &                 &                 &   14.3$^{\rm j}$&       &                 &12.400\,(31) &12.028\,(35) & 11.866\,(26)$^{\rm r}$&                    &                 &               &-6.4\,(4.1)   &-20.1\,(4.1)  &2.7\,(7)      &-51\,(7)      &0.94 &-0.77&                                                                                      \\
1402 & 21:42:20.12 &  57:10:32.2$^{\rm r}$& 1596 & 4596                      &                               &                 &                 &   14.4$^{\rm j}$&       &                 &9.117 \,(26) &7.923 \,(36) & 7.529 \,(20)$^{\rm r}$&                    &                 &               &-2.5\,(5.1)   &-4.6\,(5.1)   &9.7\,(7.1)    &15.5\,(7.1)   &-0.16&0.15 &                                                                                      \\
1403 & 21:42:51.81 &  57:09:17.9$^{\rm r}$& 1597 & 4597                      &                               &                 &                 &   13.5$^{\rm j}$&       &                 &11.829\,(27) &11.476\,(32) & 11.404\,(21)$^{\rm r}$&                    &                 &               &-1\,(4.1)     &-5\,(4.1)     &7.8\,(7)      &2.7\,(7)      &0    &-0.25&                                                                                      \\
1404 & 21:42:48.26 &  57:09:47.2$^{\rm r}$& 1598 & 4598                      &                               &                 &                 &   13.8$^{\rm j}$&       &                 &10.652\,(27) &9.874 \,(29) & 9.692 \,(21)$^{\rm r}$&                    &                 &               &-6.7\,(5.1)   &-2\,(5.1)     &3.1\,(7)      &13.6\,(7)     &0.31 &-0.34&                                                                                      \\
1405 & 21:42:48.38 &  57:10:11.4$^{\rm r}$& 1599 & 4599                      &                               &                 &                 &   14.2$^{\rm j}$&       &                 &11.057\,(29) &10.399\,(31) & 10.208\,(19)$^{\rm r}$&                    &                 &               &-6.7\,(4.1)   &-5.5\,(4.1)   &-8.2\,(6.9)   &1.6\,(6.9)    &0.01 &-0.11&                                                                                      \\
1406 & 21:42:47.86 &  57:10:38.6$^{\rm r}$& 1600 & 4600                      &                               &                 &                 &   13.1$^{\rm j}$&       &                 &10.579\,(27) &9.895 \,(31) & 9.758 \,(21)$^{\rm r}$&                    &                 &               &3.7\,(5.6)    &-13.2\,(5.1)  &-52.5\,(6.9)  &18\,(7)       &     &     &                                                                                      \\
1407 & 21:37:08.44 &  57:22:48.4$^{\rm r}$&      &                           & 11-1067                       &                 &                 &  18.07$^{\rm e}$& 16.92 &   15.7$^{\rm e}$&14.402\,(35) &13.614\,(37) & 13.475\,(48)$^{\rm r}$&      M0.5$^{\rm c}$&                 &  1.2$^{\rm e}$&-7.1\,(4.1)   &5\,(4.1)      &              &              &     &     &                                                                                      \\
1408 & 21:37:30.61 &  57:23:17.4$^{\rm r}$&      & 4615                      & 11-1180                       &                 &                 &  18.86$^{\rm f}$& 17.7  &  16.52$^{\rm f}$&14.900\,(51) &14.078\,(55) & 13.796\,(67)$^{\rm r}$&       G-K$^{\rm f}$&                 &               &-8.8\,(4.1)   &-16.7\,(4.1)  &              &              &     &     &                                                                                      \\
1409 & 21:37:44.87 &  57:24:13.5$^{\rm r}$&      &                           & 11-1384                       &                 &                 &  17.13$^{\rm e}$& 16.03 &   14.9$^{\rm e}$&13.351\,(30) &12.593\,(34) & 12.387\,(29)$^{\rm r}$&      K6.5$^{\rm c}$&                 &  1.7$^{\rm e}$&-4.2\,(4.1)   &-14.6\,(4.1)  &4.8\,(8.7)    &-20.7\,(7.4)  &     &     &                                                                                      \\
1410 & 21:37:01.40 &  57:24:45.9$^{\rm r}$&      &                           & 11-1499                       &                 &                 &  17.23$^{\rm e}$& 16.19 &  14.85$^{\rm e}$&13.305\,(40) &12.459\,(55) & 12.206\,(55)$^{\rm r}$&      M1.5$^{\rm c}$&                 &  0.6$^{\rm e}$&-14.2\,(4.1)  &-16.9\,(4.1)  &              &              &     &     &                                                                                      \\
1411 & 21:37:11.84 &  57:24:48.7$^{\rm r}$&      &                           & 11-1513                       &                 &                 &  17.17$^{\rm e}$& 15.98 &  14.76$^{\rm e}$&13.162\,(29) &12.348\,(30) & 12.145\,(28)$^{\rm r}$&      K7.5$^{\rm c}$&                 &  1.8$^{\rm e}$&4.1\,(4.1)    &-2.2\,(4.1)   &-3\,(9.7)     &-12.4\,(8.5)  &     &     &                                                                                      \\
1412 & 21:37:00.89 &  57:25:22.4$^{\rm r}$&      &                           & 11-1659                       &                 &                 &  17.24$^{\rm e}$& 16.16 &  15.12$^{\rm e}$&13.611\,(42) &12.844\,(44) & 12.656\,(39)$^{\rm r}$&        K5$^{\rm c}$&                 &  1.9$^{\rm e}$&6.4\,(4.1)    &-1.4\,(4.1)   &              &              &     &     &                                                                                      \\
1413 & 21:37:41.14 &  57:25:40.6$^{\rm r}$&      & 4618                      & 11-1721                       &                 &                 &   17.5$^{\rm f}$& 16.35 &  15.22$^{\rm f}$&13.446\,(29) &12.572\,(31) & 12.310\,(31)$^{\rm r}$&        K5$^{\rm f}$&                 & 1.98$^{\rm f}$&-6.3\,(4.1)   &-2.4\,(4.1)   &              &              &     &     &                                                                                      \\
1414 & 21:37:34.20 &  57:26:15.4$^{\rm r}$&      & 4616                      & 11-1864                       &                 &                 &  17.55$^{\rm e}$& 16.47 &  15.44$^{\rm e}$&14.065\,()   &13.686\,(70) &   13.198\,()$^{\rm r}$&       G-K$^{\rm c}$&                 &  1.7$^{\rm e}$&-6.5\,(4.1)   &7\,(4.1)      &              &              &     &     &SB1$^{\rm c}$, \textit{JHK} in [f] different                                                                         \\
1415 & 21:37:02.55 &  57:26:14.5$^{\rm r}$&      &                           & 11-1871                       &                 &                 &  18.15$^{\rm e}$& 17.08 &  15.68$^{\rm e}$&14.074\,(46) &13.331\,(44) & 13.129\,(45)$^{\rm r}$&        M2$^{\rm c}$&                 &  0.8$^{\rm e}$&52.7\,(5.5)   &-14.9\,(5.5)  &              &              &     &     &                                                                                      \\
1416 & 21:37:15.92 &  57:26:59.2$^{\rm r}$&      & 4610                      & 11-2031                       &  18.08$^{\rm f}$&                 &  15.48$^{\rm e}$& 14.58 &  13.69$^{\rm e}$&12.499\,(25) &11.534\,(28) & 10.856\,(20)$^{\rm r}$&        K2$^{\rm c}$&                 &  1.7$^{\rm e}$&-0.8\,(4.1)   &-4.3\,(4.1)   &-7.3\,(7.1)   &4.9\,(7)      &     &     &\textit{JHK} in [f] slightly different                                                                        \\
1417 & 21:37:07.03 &  57:27:00.8$^{\rm r}$&      & 4606                      & 11-2037                       &  18.12$^{\rm f}$&                 &  16.03$^{\rm e}$& 15.08 &  14.12$^{\rm e}$&12.649\,(26) &11.752\,(27) & 11.258\,(20)$^{\rm r}$&      K4.5$^{\rm c}$&                 &  1.6$^{\rm e}$&-3.5\,(4.1)   &-3.3\,(4.1)   &-0.5\,(8.6)   &-0.1\,(8.3)   &     &     &                                                                                      \\
1418 & 21:37:12.16 &  57:27:26.2$^{\rm r}$&      & 4609                      & 11-2131                       &  19.82$^{\rm f}$&                 &  17.72$^{\rm e}$& 16.43 &  15.25$^{\rm e}$&13.393\,(27) &12.214\,(31) & 11.523\,(25)$^{\rm r}$&      K6.5$^{\rm c}$&                 &  2.3$^{\rm e}$&-3.2\,(4.1)   &-0.7\,(4.1)   &2.6\,(12.5)   &-11.8\,(12.7) &     &     &                                                                                      \\
1419 & 21:36:57.67 &  57:27:33.1$^{\rm r}$&      & 4602                      & 11-2146                       &  18.16$^{\rm f}$&                 &  16.92$^{\rm e}$& 15.69 &  14.41$^{\rm e}$&12.442\,()   &11.327\,(32) & 10.640\,(25)$^{\rm r}$&        K6$^{\rm c}$&                 &  2.6$^{\rm e}$&-1\,(4.1)     &-5.7\,(4.1)   &8.2\,(7.5)    &21.2\,(7.4)   &     &     &                                                                                      \\
1420 & 21:37:45.21 &  57:28:17.4$^{\rm r}$&      & 4620                      & 11-2318                       &                 &                 &     18$^{\rm e}$& 16.86 &  15.66$^{\rm e}$&14.176\,(35) &13.248\,(36) & 13.005\,(37)$^{\rm r}$&      M0.0$^{\rm e}$&                 &  1.1$^{\rm e}$&1.6\,(4.1)    &-8.8\,(4.1)   &              &              &     &     &                                                                                      \\
1421 & 21:37:01.92 &  57:28:22.3$^{\rm r}$&      & 4604                      & 11-2322                       &  18.69$^{\rm f}$&                 &  16.69$^{\rm e}$& 15.59 &  14.39$^{\rm e}$&12.854\,(29) &11.969\,(33) & 11.560\,(25)$^{\rm r}$&        M1$^{\rm c}$&                 &  0.9$^{\rm e}$&-4.3\,(3.9)   &6\,(3.9)      &              &              &     &     &                                                                                      \\
1422 & 21:37:14.50 &  57:28:40.9$^{\rm r}$&      &                           & 11-2397                       &                 &                 &  18.54$^{\rm e}$& 17.48 &  16.34$^{\rm e}$&14.511\,(39) &13.559\,(41) & 12.954\,(34)$^{\rm r}$&      K7.0$^{\rm e}$&                 &  1.2$^{\rm e}$&-6\,(3.9)     &-12.2\,(3.9)  &              &              &     &     &                                                                                      \\
1423 & 21:37:14.98 &  57:29:12.3$^{\rm r}$&      &                           & 11-2487                       &                 &                 &  17.61$^{\rm e}$& 16.48 &  15.27$^{\rm e}$&13.726\,(32) &12.946\,(33) & 12.734\,(35)$^{\rm r}$&        K7$^{\rm c}$&                 &  1.9$^{\rm e}$&-7.3\,(3.9)   &1.3\,(3.9)    &              &              &     &     &                                                                                      \\
1424 & 21:37:28.05 &  57:29:15.6$^{\rm r}$&      &                           & 11-2503                       &                 &                 &  18.07$^{\rm e}$& 16.69 &  15.22$^{\rm e}$&13.440\,(27) &12.556\,(33) & 12.329\,(29)$^{\rm r}$&      M0.0$^{\rm e}$&                 &  2.3$^{\rm e}$&-12.3\,(3.8)  &-9.1\,(3.8)   &              &              &     &     &                                                                                      \\
1425 & 21:37:45.14 &  57:19:42.4$^{\rm r}$&      & 4619                      & 11-383                        &                 &                 &  17.61$^{\rm e}$& 16.49 &  15.35$^{\rm e}$&13.903\,(26) &12.971\,(33) & 12.582\,(26)$^{\rm r}$&        K5$^{\rm c}$&                 &  0.8$^{\rm e}$&-7.2\,(4.1)   &-2.5\,(4.1)   &              &              &     &     &                                                                                      \\
1426 & 21:37:28.29 &  57:20:32.6$^{\rm r}$&      & 4612                      & 11-581                        &  19.96$^{\rm f}$&                 &  16.73$^{\rm e}$& 15.63 &   14.6$^{\rm e}$&13.095\,(44) &12.332\,(45) & 12.104\,(39)$^{\rm r}$&         G$^{\rm e}$&                 &  1.5$^{\rm e}$&-23.1\,(4.1)  &-35.5\,(4.1)  &              &              &     &     &                                                                                      \\
1427 & 21:38:17.50 &  57:22:30.8$^{\rm r}$&      &                           & 12-1009                       &                 &                 &  16.17$^{\rm e}$& 15.34 &  14.39$^{\rm e}$&13.119\,(45) &12.361\,(43) & 12.130\,(37)$^{\rm r}$&      K5.5$^{\rm c}$&                 &  0.9$^{\rm e}$&29.8\,(5.4)   &-13.9\,(5.4)  &              &              &     &     &                                                                                      \\
1428 & 21:38:50.29 &  57:22:28.3$^{\rm r}$&      &                           & 12-1010                       &                 &                 &  18.19$^{\rm e}$& 17.19 &  15.95$^{\rm e}$&14.372\,(40) &13.507\,()   &   13.076\,()$^{\rm r}$&        M2$^{\rm c}$&                 &  0.3$^{\rm e}$&-1.3\,(4)     &-20.3\,(4)    &              &              &     &     &                                                                                      \\
1429 & 21:38:15.09 &  57:21:55.5$^{\rm r}$&      &                           & 12-1017                       &                 &                 &  16.83$^{\rm e}$& 15.82 &  14.84$^{\rm e}$&13.283\,()   &12.654\,()   & 12.554\,(28)$^{\rm r}$&      K5.5$^{\rm c}$&                 &  1.4$^{\rm e}$&-19.3\,(5.4)  &-0.9\,(5.4)   &              &              &     &     &                                                                                      \\
1430 & 21:39:03.19 &  57:22:31.8$^{\rm r}$&      & 4641                      & 12-1027                       &                 &                 &  18.58$^{\rm e}$& 17.51 &  15.95$^{\rm e}$&14.249\,(41) &13.561\,(48) & 13.308\,(40)$^{\rm r}$&        M0$^{\rm c}$&                 &  1.3$^{\rm e}$&-6.1\,(5.4)   &-9.9\,(5.4)   &              &              &     &     &                                                                                      \\
1431 & 21:38:05.94 &  57:22:43.9$^{\rm r}$&      & 4627                      & 12-1081                       &                 &                 &  18.35$^{\rm e}$& 17.15 &  15.95$^{\rm e}$&14.354\,(35) &13.533\,(33) & 13.271\,(31)$^{\rm r}$&      M0.5$^{\rm c}$&                 &  1.3$^{\rm e}$&-4.5\,(4)     &-5.8\,(4)     &              &              &     &     &                                                                                      \\
1432 & 21:37:57.62 &  57:22:47.7$^{\rm r}$&      & 4623                      & 12-1091                       &  18.61$^{\rm f}$&                 &  16.59$^{\rm e}$& 15.65 &  14.73$^{\rm e}$&13.182\,(24) &12.215\,(30) & 11.673\,(20)$^{\rm r}$&      G2.5$^{\rm c}$&                 &  2.8$^{\rm e}$&-3.6\,(4)     &-3.8\,(4)     &0.1\,(9.2)    &-11.3\,(9.2)  &     &     &\textit{JHK} in [f] slightly different                                                                         \\
1433 & 21:37:57.57 &  57:24:19.7$^{\rm r}$&      & 4622                      & 12-1422                       &  20.23$^{\rm f}$&                 &  18.77$^{\rm e}$& 17.48 &  16.09$^{\rm e}$&14.458\,(45) &13.606\,(43) & 13.337\,(43)$^{\rm r}$&        M0$^{\rm c}$&                 &  1.9$^{\rm e}$&-0.5\,(5.4)   &-5.2\,(5.4)   &              &              &     &     &                                                                                      \\
1434 & 21:38:47.07 &  57:24:20.7$^{\rm r}$&      &                           & 12-1423                       &                 &                 &  17.04$^{\rm e}$& 15.94 &  14.83$^{\rm e}$&13.250\,(26) &12.459\,(31) & 12.288\,(23)$^{\rm r}$&        K7$^{\rm c}$&                 &  1.5$^{\rm e}$&-4.3\,(4)     &-2.6\,(4)     &-12.6\,(11.1) &7\,(11.7)     &     &     &                                                                                      \\
1435 & 21:38:08.48 &  57:25:11.9$^{\rm r}$&      & 4629                      & 12-1613                       &                 &                 &  18.42$^{\rm e}$& 17.15 &  15.78$^{\rm e}$&14.062\,(32) &13.241\,(40) & 12.951\,(35)$^{\rm r}$&        M1$^{\rm c}$&                 &  1.3$^{\rm e}$&-5.8\,(4)     &-4\,(4)       &              &              &     &     &                                                                                      \\
1436 & 21:39:04.68 &  57:25:12.8$^{\rm r}$&      & 4642                      & 12-1617                       &  19.72$^{\rm f}$&                 &  17.86$^{\rm e}$& 16.53 &  15.18$^{\rm e}$&13.542\,(26) &12.627\,(28) & 12.129\,(24)$^{\rm r}$&        M1$^{\rm c}$&                 &  1.6$^{\rm e}$&-5\,(4)       &-3.7\,(4)     &              &              &     &     &                                                                                      \\
1437 & 21:39:04.71 &  57:25:21.5$^{\rm r}$&      &                           & 12-1650                       &                 &                 &                 &       &                 &14.406\,(37) &13.888\,(49) & 13.722\,(66)$^{\rm r}$&                    &                 &               &-7.4\,(5.4)   &-44.5\,(5.4)  &              &              &     &     &                                                                                      \\
1438 & 21:38:33.82 &  57:26:05.3$^{\rm r}$&      & 4637                      & 12-1825                       &                 &                 &  19.23$^{\rm e}$& 17.86 &  16.52$^{\rm e}$&14.976\,(63) &14.141\,(47) & 13.877\,(66)$^{\rm r}$&      M0.0$^{\rm e}$&                 &  2.1$^{\rm e}$&-16.8\,(5.4)  &-0.5\,(5.4)   &              &              &     &     &                                                                                      \\
1439 & 21:38:44.47 &  57:18:09.1$^{\rm r}$&      & 4639                      & 12-44\footnote{also 54-1547}  &                 &                 &  16.65$^{\rm e}$& 15.69 &  14.81$^{\rm e}$&13.421\,(27) &12.647\,(35) & 12.180\,(26)$^{\rm r}$&      K5.5$^{\rm c}$&                 &  1.2$^{\rm e}$&-4.9\,(4)     &-3.9\,(4)     &              &              &     &     &2 diffent SHB-2004                                               \\
1440 & 21:38:26.92 &  57:26:38.5$^{\rm r}$&      & 4634                      & 12-1955                       &                 &                 &  17.53$^{\rm e}$& 16.46 &  15.42$^{\rm e}$&14.096\,(43) &13.320\,(44) & 13.122\,(42)$^{\rm r}$&      K6.5$^{\rm c}$&                 &  1.4$^{\rm e}$&-6.9\,(4)     &-4.4\,(4)     &              &              &     &     &                                                                                      \\
1441 & 21:37:54.88 &  57:26:42.5$^{\rm r}$&      &                           & 12-1968                       &                 &                 &   16.5$^{\rm e}$& 15.51 &  14.55$^{\rm e}$&13.185\,(38) &12.344\,(32) & 12.026\,(27)$^{\rm r}$&        K6$^{\rm c}$&                 &  1.2$^{\rm e}$&11.7\,(4)     &-6.1\,(4)     &35.4\,(7.6)   &7.2\,(7.6)    &     &     &SB2:$^{\rm c}$                                                                                      \\
1442 & 21:37:50.23 &  57:25:48.8$^{\rm r}$&      &                           & 12-1984                       &                 &                 &   16.6$^{\rm e}$& 15.69 &   14.8$^{\rm e}$&13.288\,(29) &12.451\,(31) & 12.251\,(22)$^{\rm r}$&        K6$^{\rm c}$&                 &  0.8$^{\rm e}$&-1.6\,(4)     &-7.5\,(4)     &-7.9\,(8)     &-3.1\,(8.6)   &     &     &                                                                                      \\
1443 & 21:37:58.41 &  57:18:04.7$^{\rm r}$&      & 4625                      & 12-94                         &                 &                 &  17.39$^{\rm e}$& 16.28 &  15.31$^{\rm e}$&13.780\,(26) &12.899\,(33) & 12.673\,(26)$^{\rm r}$&      K4.0$^{\rm e}$&                 &  1.8$^{\rm e}$&4.4\,(4)      &2.9\,(4)      &              &              &     &     &                                                                                      \\
1444 & 21:38:52.53 &  57:27:18.5$^{\rm r}$&      &                           & 12-2098                       &                 &                 &  18.43$^{\rm e}$& 17.18 &  15.73$^{\rm e}$&14.137\,(53) &13.365\,(56) & 13.156\,(50)$^{\rm r}$&      M2.5$^{\rm c}$&                 &    1$^{\rm e}$&-2\,(5.3)     &-0.9\,(5.3)   &              &              &     &     &                                                                                      \\
1445 & 21:38:27.43 &  57:27:20.8$^{\rm r}$&      & 4636                      & 12-2113                       &  18.79$^{\rm f}$&                 &  17.14$^{\rm e}$& 15.92 &  14.67$^{\rm e}$&12.856\,(31) &11.986\,(35) & 11.506\,(24)$^{\rm r}$&        K6$^{\rm c}$&                 &  1.2$^{\rm e}$&4.4\,(4)      &-4.7\,(4)     &16.9\,(7.1)   &5.2\,(7.2)    &     &     &SB1$^{\rm c}$                                                                                      \\
1446 & 21:38:45.44 &  57:28:23.1$^{\rm r}$&      &                           & 12-2363                       &                 &                 &  17.68$^{\rm e}$& 16.36 &  15.05$^{\rm e}$&13.523\,(32) &12.733\,(39) & 12.500\,(33)$^{\rm r}$&      M0.5$^{\rm c}$&                 &  1.9$^{\rm e}$&-0.2\,(3.9)   &-8.9\,(3.9)   &              &              &     &     &SB1:$^{\rm c}$                                                                                      \\
1447 & 21:38:00.59 &  57:28:25.4$^{\rm r}$&      &                           & 12-2373                       &                 &                 &  17.66$^{\rm e}$& 16.48 &  15.24$^{\rm e}$&13.552\,(32) &12.728\,(50) & 12.477\,(33)$^{\rm r}$&        M1$^{\rm c}$&                 &  1.2$^{\rm e}$&-7.7\,(4)     &-0.7\,(4)     &              &              &     &     &                                                                                      \\
1448 & 21:37:51.07 &  57:27:50.2$^{\rm r}$&      &                           & 12-2519                       &                 &                 &  17.27$^{\rm e}$& 16.23 &  15.21$^{\rm e}$&13.647\,(31) &12.816\,(36) & 12.460\,(22)$^{\rm r}$&      K5.5$^{\rm c}$&                 &  1.6$^{\rm e}$&-1.3\,(3.9)   &-18.8\,(3.9)  &              &              &     &     &                                                                                      \\
1449 & 21:37:58.28 &  57:20:35.5$^{\rm r}$&      & 4624                      & 12-583                        &                 &                 &  17.48$^{\rm e}$& 16.26 &  14.93$^{\rm e}$&13.427\,(40) &12.574\,(38) & 12.306\,(29)$^{\rm r}$&        M0$^{\rm c}$&                 &  1.6$^{\rm e}$&-15.1\,(4)    &-5.1\,(4)     &              &              &     &     &                                                                                      \\
1450 & 21:38:46.23 &  57:20:38.0$^{\rm r}$&      & 4640                      & 12-595                        &                 &                 &  18.51$^{\rm e}$& 17.43 &  16.35$^{\rm e}$&15.375\,(66) &14.634\,(82) & 14.327\,(81)$^{\rm r}$&        K7$^{\rm c}$&                 &  1.2$^{\rm e}$&-21.9\,(4.1)  &-18.5\,(4.1)  &              &              &     &     &                                                                                      \\
1451 & 21:39:12.88 &  57:21:08.8$^{\rm r}$&      &                           & 12-705                        &                 &                 &  17.59$^{\rm e}$& 16.37 &  15.13$^{\rm e}$&13.489\,(26) &12.685\,(32) & 12.425\,(25)$^{\rm r}$&        M1$^{\rm c}$&                 &  1.3$^{\rm e}$&-3.7\,(4)     &-4.8\,(4)     &              &              &     &     &                                                                                      \\
1452 & 21:39:14.24 &  57:22:13.0$^{\rm r}$&      &                           & 12-942                        &                 &                 &  17.15$^{\rm e}$& 15.98 &  14.79$^{\rm e}$&13.238\,(29) &12.431\,(37) & 12.196\,(28)$^{\rm r}$&      K7.5$^{\rm c}$&                 &  1.7$^{\rm e}$&-14.5\,(4)    &-7.9\,(4)     &-30.9\,(8.3)  &0.3\,(8.1)    &     &     &SB1$^{\rm c}$                                                                                      \\
1453 & 21:39:10.89 &  57:35:18.1$^{\rm r}$&      &                           & 13-1048                       &                 &                 &   18.9$^{\rm e}$& 17.58 &  16.42$^{\rm e}$&13.695\,(24) &12.714\,(28) & 12.165\,(26)$^{\rm r}$&        M0$^{\rm c}$&                 &  1.7$^{\rm e}$&3\,(3.8)      &-4\,(3.8)     &              &              &     &     &                                                                                      \\
1454 & 21:38:55.42 &  57:35:29.9$^{\rm r}$&      &                           & 13-1087                       &                 &                 &  16.04$^{\rm e}$& 15.11 &  14.23$^{\rm e}$&12.878\,(24) &12.110\,(28) & 11.943\,(23)$^{\rm r}$&        K4$^{\rm c}$&                 &  1.7$^{\rm e}$&-2.2\,(3.9)   &-10\,(3.9)    &-2\,(7.8)     &-39.7\,(7.8)  &     &     &                                                                                      \\
1455 & 21:37:58.53 &  57:35:47.9$^{\rm r}$&      &                           & 13-1143                       &                 &                 &                 &       &                 &13.950\,(34) &13.205\,(40) & 13.022\,(35)$^{\rm r}$&                    &                 &               &3.2\,(3.8)    &-3.1\,(3.8)   &              &              &     &     &                                                                                      \\
1456 & 21:38:07.72 &  57:35:53.3$^{\rm r}$&      & 4628                      & 13-1161                       &                 &                 &  18.05$^{\rm e}$& 16.8  &  15.64$^{\rm e}$&14.111\,(39) &13.249\,(38) & 12.973\,(28)$^{\rm r}$&        M0$^{\rm c}$&                 &  1.5$^{\rm e}$&-10.5\,(4)    &-9.6\,(4)     &              &              &     &     &                                                                                      \\
1457 & 21:37:59.26 &  57:36:16.2$^{\rm r}$&      & 4626                      & 13-1238                       &  19.15$^{\rm f}$&                 &  18.36$^{\rm e}$& 16.83 &  15.37$^{\rm e}$&13.321\,(27) &12.400\,(32) & 11.889\,(18)$^{\rm r}$&        M1$^{\rm c}$&                 &  2.6$^{\rm e}$&-2\,(3.9)     &-5.2\,(3.9)   &              &              &     &     &                                                                                      \\
1458 & 21:39:12.13 &  57:36:16.5$^{\rm r}$&      & 4644                      & 13-1250                       &  18.68$^{\rm f}$&                 &  16.05$^{\rm e}$& 15.09 &  14.24$^{\rm e}$&12.920\,(24) &12.119\,(29) & 11.799\,(23)$^{\rm r}$&      K4.5$^{\rm c}$&                 &  1.4$^{\rm e}$&-6.3\,(3.8)   &-6.4\,(3.8)   &-8.2\,(8.5)   &-16.6\,(8.4)  &     &     &                                                                                      \\
1459 & 21:38:08.56 &  57:37:07.6$^{\rm r}$&      & 4630                      & 13-1426                       &                 &                 &  19.67$^{\rm e}$& 18.06 &  16.43$^{\rm e}$&14.306\,(29) &13.425\,(33) & 12.901\,(30)$^{\rm r}$&        M0$^{\rm c}$&                 &  3.2$^{\rm e}$&-2.2\,(3.9)   &-6.2\,(3.9)   &              &              &     &     &                                                                                      \\
1460 & 21:38:28.04 &  57:30:46.5$^{\rm r}$&      &                           & 13-157                        &                 &                 &  16.23$^{\rm e}$& 15.25 &  14.32$^{\rm e}$&12.912\,(31) &11.979\,(36) & 11.249\,(52)$^{\rm r}$&      K5.5$^{\rm c}$&                 &  1.2$^{\rm e}$&              &              &              &              &     &     &                                                                                      \\
1461 & 21:38:40.38 &  57:38:37.4$^{\rm r}$&      &                           & 13-1709                       &                 &                 &   16.7$^{\rm e}$& 15.7  &  14.82$^{\rm e}$&13.563\,(27) &12.715\,(32) & 12.507\,(23)$^{\rm r}$&      K5.5$^{\rm c}$&                 &  1.2$^{\rm e}$&0.9\,(3.8)    &-6.1\,(3.8)   &-8.2\,(8.8)   &-7\,(8.9)     &     &     &                                                                                      \\
1462 & 21:38:17.03 &  57:39:26.6$^{\rm r}$&      &                           & 13-1877                       &                 &                 &  16.83$^{\rm e}$& 15.59 &  14.45$^{\rm e}$&12.935\,()   &11.954\,()   &   11.263\,()$^{\rm r}$&        K7$^{\rm c}$&                 &    2$^{\rm e}$&3.2\,(3.8)    &-5.4\,(3.8)   &              &              &     &     &                                                                                      \\
1463 & 21:38:40.02 &  57:39:30.3$^{\rm r}$&      & 4638                      & 13-1891                       &                 &                 &  18.16$^{\rm e}$& 17.06 &  15.89$^{\rm e}$&14.509\,(27) &13.600\,(42) & 13.215\,(30)$^{\rm r}$&        M0$^{\rm c}$&                 &    1$^{\rm e}$&10.6\,(3.9)   &-12.3\,(3.9)  &              &              &     &     &                                                                                      \\
1464 & 21:38:17.50 &  57:41:02.0$^{\rm r}$&      &                           & 13-2236                       &                 &                 &  17.54$^{\rm e}$& 16.4  &  15.35$^{\rm e}$&13.873\,(34) &13.004\,(37) & 12.767\,(35)$^{\rm r}$&      K6.5$^{\rm c}$&                 &  1.6$^{\rm e}$&-8.2\,(3.9)   &-4.9\,(3.9)   &              &              &     &     &                                                                                      \\
1465 & 21:38:28.35 &  57:31:07.2$^{\rm r}$&      &                           & 13-232                        &                 &                 &  17.36$^{\rm e}$& 16.2  &  15.11$^{\rm e}$&13.647\,(43) &12.804\,(42) & 12.561\,(40)$^{\rm r}$&        M0$^{\rm c}$&                 &  1.1$^{\rm e}$&-5.4\,(3.9)   &-16.7\,(3.9)  &              &              &     &     &                                                                                      \\
1466 & 21:38:27.42 &  57:31:08.2$^{\rm r}$&      & 4635                      & 13-236                        &  17.56$^{\rm f}$&                 &  15.64$^{\rm e}$& 14.7  &  13.85$^{\rm e}$&12.355\,(36) &11.363\,(40) & 10.774\,(24)$^{\rm r}$&        K2$^{\rm c}$&                 &  1.8$^{\rm e}$&24.9\,(3.9)   &-14.3\,(3.9)  &29.5\,(7.3)   &1.1\,(7.3)    &     &     &                                                                                      \\
1467 & 21:39:10.25 &  57:31:06.6$^{\rm r}$&      & 4643                      & 13-238                        &                 &                 &   19.5$^{\rm f}$& 17.5  &   15.9$^{\rm f}$&14.084\,(31) &13.222\,(35) & 12.980\,(28)$^{\rm r}$&        K1$^{\rm q}$&                 &               &-1.3\,(5.4)   &-1.6\,(5.4)   &              &              &     &     &                                                                                      \\
1468 & 21:37:58.13 &  57:31:20.0$^{\rm r}$&      &                           & 13-269                        &                 &                 &  16.95$^{\rm e}$& 15.8  &  14.65$^{\rm e}$&12.829\,(27) &12.001\,(32) & 11.724\,(20)$^{\rm r}$&      K6.5$^{\rm c}$&                 &  1.9$^{\rm e}$&-5.6\,(3.9)   &-2.8\,(3.9)   &-10.7\,(9.1)  &21.4\,(8.5)   &     &     &                                                                                      \\
1469 & 21:38:13.85 &  57:31:41.5$^{\rm r}$&      & 4632                      & 13-350                        &                 &                 &  18.08$^{\rm e}$& 16.97 &  15.81$^{\rm e}$&14.475\,(66) &13.621\,(60) & 13.389\,(43)$^{\rm r}$&        M1$^{\rm c}$&                 &  0.7$^{\rm e}$&17.6\,(3.8)   &-32.2\,(3.8)  &              &              &     &     &                                                                                      \\
1470 & 21:38:32.55 &  57:30:16.1$^{\rm r}$&      &                           & 13-52                         &                 &                 &  17.27$^{\rm e}$& 16.19 &  15.15$^{\rm e}$&13.738\,(47) &12.842\,(40) & 12.607\,(31)$^{\rm r}$&        K7$^{\rm c}$&                 &  1.3$^{\rm e}$&-2.2\,(4)     &-6.9\,(4)     &              &              &     &     &                                                                                      \\
1471 & 21:38:34.81 &  57:32:50.0$^{\rm r}$&      &                           & 13-566                        &                 &                 &  18.05$^{\rm e}$& 16.81 &  15.47$^{\rm e}$&13.899\,(29) &13.049\,(35) & 12.789\,(22)$^{\rm r}$&      K5.5$^{\rm c}$&                 &  2.4$^{\rm e}$&-2.8\,(3.8)   &-6.4\,(3.8)   &              &              &     &     &                                                                                      \\
1472 & 21:38:09.28 &  57:33:26.2$^{\rm r}$&      & 4631                      & 13-669                        &  18.29$^{\rm f}$&                 &  15.84$^{\rm e}$& 14.86 &  13.93$^{\rm e}$&12.391\,(29) &11.566\,(30) & 11.195\,(20)$^{\rm r}$&        K1$^{\rm c}$&                 &  2.2$^{\rm e}$&0\,(4)        &4.3\,(4)      &3.3\,(8.5)    &14.8\,(7.4)   &     &     &                                                                                      \\
1473 & 21:38:25.97 &  57:34:09.4$^{\rm r}$&      &                           & 13-819                        &                 &                 &  16.44$^{\rm e}$& 15.42 &  14.45$^{\rm e}$&13.032\,(27) &12.234\,(29) & 11.993\,(20)$^{\rm r}$&      K5.5$^{\rm c}$&                 &  1.4$^{\rm e}$&1.8\,(3.8)    &-5.2\,(3.8)   &-10.3\,(8.4)  &-13.1\,(8.4)  &     &     &                                                                                      \\
1474 & 21:38:11.21 &  57:34:18.2$^{\rm r}$&      &                           & 13-838                        &                 &                 &                 &       &                 &14.113\,(34) &13.407\,(33) & 13.199\,(33)$^{\rm r}$&                    &                 &               &-4.2\,(3.9)   &-3.8\,(3.9)   &              &              &     &     &                                                                                      \\
1475 & 21:37:50.19 &  57:33:40.4$^{\rm r}$&      & 4621                      & 13-924                        &                 &                 &  16.93$^{\rm e}$& 15.93 &  14.92$^{\rm e}$&13.225\,(27) &12.360\,(36) & 12.108\,(24)$^{\rm r}$&        K5$^{\rm c}$&                 &  1.6$^{\rm e}$&-3.9\,(3.8)   &-4.1\,(3.8)   &-7.9\,(12.6)  &14.3\,(9.1)   &     &     &                                                                                      \\
1476 & 21:37:28.94 &  57:36:04.3$^{\rm r}$&      & 4613                      & 14-1017                       &                 &                 &  18.66$^{\rm e}$& 17.3  &  15.92$^{\rm e}$&13.994\,(31) &13.027\,(30) & 12.630\,(26)$^{\rm r}$&        M0$^{\rm c}$&                 &  2.1$^{\rm e}$&0.7\,(3.8)    &-9\,(3.8)     &              &              &     &     &                                                                                      \\
1477 & 21:37:19.76 &  57:31:04.4$^{\rm r}$&      &                           & 14-103                        &                 &                 &  17.83$^{\rm e}$& 16.93 &  15.82$^{\rm e}$&14.283\,(27) &13.387\,(31) & 13.193\,(40)$^{\rm r}$&        K7$^{\rm c}$&                 &    1$^{\rm e}$&-4.2\,(3.9)   &2.3\,(3.9)    &              &              &     &     &                                                                                      \\
1478 & 21:37:10.32 &  57:30:18.9$^{\rm r}$&      &                           & 14-11                         &                 &                 &  17.53$^{\rm e}$& 16.15 &  14.78$^{\rm e}$&13.057\,(29) &12.216\,(34) & 11.956\,(26)$^{\rm r}$&      M1.5$^{\rm c}$&                 &  1.8$^{\rm e}$&1.6\,(3.8)    &-10.7\,(3.8)  &              &              &     &     &                                                                                      \\
1479 & 21:36:55.79 &  57:36:53.3$^{\rm r}$&      &                           & 14-1229                       &                 &                 &  17.81$^{\rm e}$& 16.74 &  15.74$^{\rm e}$&14.361\,(29) &13.501\,(41) & 13.351\,(34)$^{\rm r}$&        K6$^{\rm c}$&                 &  1.3$^{\rm e}$&-2.8\,(4)     &-0.4\,(4)     &              &              &     &     &                                                                                      \\
1480 & 21:37:10.54 &  57:31:12.5$^{\rm r}$&      & 4607                      & 14-125                        &  20.03$^{\rm f}$&                 &  16.74$^{\rm e}$& 15.73 &   14.7$^{\rm e}$&13.090\,(26) &12.170\,(30) & 11.732\,(21)$^{\rm r}$&        K5$^{\rm c}$&                 &  1.7$^{\rm e}$&1\,(3.8)      &-14.3\,(3.8)  &              &              &     &     &                                                                                      \\
1481 & 21:36:49.42 &  57:31:22.1$^{\rm r}$&      & 4601                      & 14-141                        &  19.35$^{\rm f}$&                 &  15.81$^{\rm e}$& 14.68 &  13.52$^{\rm e}$&11.918\,(22) &10.914\,(29) & 10.355\,(21)$^{\rm r}$&        K6$^{\rm c}$&                 &  1.7$^{\rm e}$&-1.4\,(3.8)   &-18.5\,(3.8)  &              &              &     &     &\textit{JHK} in [f] different                                                                         \\
1482 & 21:37:27.33 &  57:31:29.5$^{\rm r}$&      & 4611                      & 14-160                        &  20.23$^{\rm f}$&                 &  16.91$^{\rm e}$& 15.84 &  14.83$^{\rm e}$&13.218\,(37) &12.349\,(41) & 11.954\,(33)$^{\rm r}$&        K5$^{\rm c}$&                 &  1.8$^{\rm e}$&0.8\,(3.8)    &-9.8\,(3.8)   &-14.2\,(12)   &0.6\,(12.2)   &     &     &                                                                                      \\
1483 & 21:37:11.24 &  57:39:16.9$^{\rm r}$&      & 4608                      & 14-1827                       &  18.05$^{\rm f}$&                 &  15.67$^{\rm f}$& 14.92 &  14.19$^{\rm f}$&13.171\,(22) &12.515\,(28) & 12.399\,(25)$^{\rm r}$&         G$^{\rm f}$&                 &               &7.5\,(3.8)    &8.5\,(3.8)    &6.6\,(7.6)    &1.7\,(7.6)    &     &     &                                                                                      \\
1484 & 21:37:38.49 &  57:31:40.8$^{\rm r}$&      & 4617                      & 14-183                        &  20.84$^{\rm f}$&                 &  16.61$^{\rm e}$& 15.37 &  14.21$^{\rm e}$&13.303\,(29) &12.229\,(31) &   11.668\,()$^{\rm r}$& K7.0(K5)R$^{\rm c}$&                 &    2$^{\rm e}$&              &              &              &              &     &     &SB1:$^{\rm c}$                                                                                      \\
1485 & 21:37:23.68 &  57:31:53.9$^{\rm r}$&      &                           & 14-197                        &                 &                 &  17.07$^{\rm e}$& 16.01 &  14.99$^{\rm e}$&13.446\,(25) &12.635\,(31) & 12.473\,(26)$^{\rm r}$&      K5.5$^{\rm c}$&                 &  1.7$^{\rm e}$&3.8\,(5.1)    &-3.1\,(5.1)   &-10.8\,(12.8) &-2.7\,(13.2)  &     &     &SB1:$^{\rm c}$                                                                                      \\
1486 & 21:37:41.85 &  57:40:40.1$^{\rm r}$&      &                           & 14-2148                       &                 &                 &  18.25$^{\rm e}$& 16.93 &  15.62$^{\rm e}$&13.985\,(30) &13.087\,(34) & 12.832\,(30)$^{\rm r}$&      M1.5$^{\rm c}$&                 &  1.5$^{\rm e}$&-1\,(3.8)     &-10.3\,(3.8)  &              &              &     &     &                                                                                      \\
1487 & 21:37:06.07 &  57:32:01.6$^{\rm r}$&      & 4605                      & 14-222                        &  18.52$^{\rm f}$&                 &  15.81$^{\rm e}$& 14.74 &  13.65$^{\rm e}$&12.105\,(22) &11.273\,(26) & 11.077\,(21)$^{\rm r}$&        K7$^{\rm c}$&                 &  1.2$^{\rm e}$&5.4\,(3.9)    &-2.6\,(3.9)   &              &              &     &     &                                                                                      \\
1488 & 21:37:06.50 &  57:32:31.7$^{\rm r}$&      &                           & 14-287                        &                 &                 &  17.85$^{\rm e}$& 16.51 &  15.17$^{\rm e}$&13.321\,(23) &12.327\,(28) & 11.909\,(23)$^{\rm r}$&        M0$^{\rm c}$&                 &  2.2$^{\rm e}$&3.2\,(3.8)    &-14.2\,(3.8)  &              &              &     &     &                                                                                      \\
1489 & 21:36:26.77 &  57:32:37.5$^{\rm r}$&      &                           & 14-306                        &                 &                 &   18.4$^{\rm e}$& 17.29 &  16.12$^{\rm e}$&14.258\,(40) &13.294\,(42) & 12.936\,(34)$^{\rm r}$&      K6.5$^{\rm c}$&                 &  1.9$^{\rm e}$&19.2\,(4)     &22.2\,(4)     &              &              &     &     &                                                                                      \\
1490 & 21:37:29.16 &  57:32:53.5$^{\rm r}$&      & 4614                      & 14-335                        &                 &                 &  17.14$^{\rm e}$& 16.09 &  14.98$^{\rm e}$&13.236\,(35) &12.142\,(31) & 11.553\,(26)$^{\rm r}$&      K6.5$^{\rm c}$&                 &  1.5$^{\rm e}$&8.1\,(3.8)    &3.9\,(3.8)    &42\,(13.1)    &103.6\,(14.2) &     &     &                                                                                      \\
1491 & 21:37:39.88 &  57:36:03.0$^{\rm r}$&      &                           & 14-995                        &                 &                 &                 &       &                 &14.542\,(43) &13.692\,(47) & 13.454\,(47)$^{\rm r}$&                    &                 &               &18.9\,(3.8)   &12.1\,(3.8)   &              &              &     &     &                                                                                      \\
1492 & 21:39:15.84 &  57:24:35.0$^{\rm r}$&      &                           & 21-1189                       &                 &                 &                 &       &                 &14.606\,(41) &14.059\,(63) & 13.847\,(64)$^{\rm r}$&                    &                 &               &-11\,(4)      &-7.5\,(4)     &              &              &     &     &                                                                                      \\
1493 & 21:39:45.71 &  57:26:24.3$^{\rm r}$&      &                           & 21-1536                       &                 &                 &  18.85$^{\rm e}$& 17.61 &  16.16$^{\rm e}$&14.561\,(40) &13.756\,(44) & 13.224\,(42)$^{\rm r}$&      M0.0$^{\rm e}$&                 &  1.8$^{\rm e}$&-4.4\,(4.1)   &3.4\,(4.1)    &              &              &     &     &faint star                                                                        \\
1494 & 21:39:47.94 &  57:26:42.8$^{\rm r}$&      & 4648                      & 21-1586                       &                 &                 &  18.23$^{\rm e}$& 17.14 &  15.84$^{\rm e}$&14.479\,(41) &13.615\,(38) & 13.375\,(38)$^{\rm r}$&        K7$^{\rm c}$&                 &  1.5$^{\rm e}$&-8.2\,(4.1)   &-3.2\,(4.1)   &              &              &     &     &                                                                                      \\
1495 & 21:39:15.54 &  57:26:44.1$^{\rm r}$&      &                           & 21-1590                       &                 &                 &   18.1$^{\rm e}$& 16.79 &  15.62$^{\rm e}$&14.053\,(30) &13.252\,(39) & 12.941\,(34)$^{\rm r}$&        K7$^{\rm c}$&                 &  2.2$^{\rm e}$&-1.9\,(3.9)   &-19.7\,(3.9)  &              &              &     &     &                                                                                      \\
1496 & 21:40:01.28 &  57:27:18.5$^{\rm r}$&      & 4649                      & 21-1692                       &                 &                 &  18.58$^{\rm e}$& 17.22 &  15.94$^{\rm e}$&14.305\,(31) &13.460\,(40) & 13.243\,(40)$^{\rm r}$&        M1$^{\rm c}$&                 &  1.7$^{\rm e}$&-6.2\,(4.1)   &-3.9\,(4.1)   &              &              &     &     &                                                                                      \\
1497 & 21:40:09.25 &  57:27:39.3$^{\rm r}$&      &                           & 21-1762                       &                 &                 &     17$^{\rm e}$& 15.96 &  14.94$^{\rm e}$&13.391\,(24) &12.727\,(33) & 12.413\,(28)$^{\rm r}$&        K5$^{\rm c}$&                 &  1.8$^{\rm e}$&-7.5\,(3.9)   &-9.1\,(3.9)   &-6\,(8.2)     &4.5\,(8.7)    &     &     &                                                                                      \\
1498 & 21:39:58.62 &  57:28:40.5$^{\rm r}$&      &                           & 21-1974                       &                 &                 &  16.77$^{\rm e}$& 15.79 &  14.86$^{\rm e}$&13.446\,(32) &12.707\,(44) & 12.497\,(32)$^{\rm r}$&      G7.5$^{\rm e}$&                 &  2.5$^{\rm e}$&-0.7\,(3.8)   &1.3\,(3.8)    &0.2\,(9.6)    &30.9\,(8.8)   &     &     &                                                                                      \\
1499 & 21:40:13.91 &  57:28:48.2$^{\rm r}$&      &                           & 21-2006                       &                 &                 &  16.99$^{\rm e}$& 16    &  15.02$^{\rm e}$&13.704\,(29) &12.920\,(38) & 12.488\,(30)$^{\rm r}$&      K5.0$^{\rm e}$&                 &  1.5$^{\rm e}$&-0.6\,(4.1)   &-4.3\,(4.1)   &30.1\,(14.1)  &12.2\,(14.2)  &     &     &                                                                                      \\
1500 & 21:39:47.55 &  57:25:21.1$^{\rm r}$&      &                           & 21-2251                       &                 &                 &  18.05$^{\rm e}$& 16.73 &  15.31$^{\rm e}$&13.800\,(28) &12.995\,(32) & 12.709\,(30)$^{\rm r}$&        M2$^{\rm c}$&                 &  1.5$^{\rm e}$&-0.5\,(4.1)   &-12.3\,(4.1)  &              &              &     &     &SB1:$^{\rm c}$                                                                                      \\
1501 & 21:39:41.69 &  57:19:27.4$^{\rm r}$&      &                           & 21-230                        &                 &                 &  18.17$^{\rm e}$& 16.94 &  15.69$^{\rm e}$&14.210\,(38) &13.345\,(36) & 13.174\,(38)$^{\rm r}$&      M0.5$^{\rm c}$&                 &  1.5$^{\rm e}$&2\,(4.1)      &0.3\,(4.1)    &              &              &     &     &                                                                                      \\
1502 & 21:39:35.62 &  57:18:22.1$^{\rm r}$&      & 4646                      & 21-33                         &                 &                 &  18.68$^{\rm e}$& 17.4  &  16.14$^{\rm e}$&14.501\,(39) &13.581\,(40) & 13.220\,(44)$^{\rm r}$&        M0$^{\rm c}$&                 &  1.7$^{\rm e}$&-9.5\,(4.1)   &-10.4\,(4.1)  &              &              &     &     &                                                                                      \\
1503 & 21:40:02.60 &  57:22:09.0$^{\rm r}$&      &                           & 21-763                        &                 &                 &  17.65$^{\rm e}$& 16.5  &  15.37$^{\rm e}$&13.849\,()   &13.164\,()   & 12.968\,(35)$^{\rm r}$&        M0$^{\rm c}$&                 &  1.2$^{\rm e}$&-6.7\,(4.1)   &-8.4\,(4.1)   &              &              &     &     &                                                                                      \\
1504 & 21:39:39.26 &  57:22:32.1$^{\rm r}$&      &                           & 21-833                        &                 &                 &  18.09$^{\rm e}$& 16.86 &  15.51$^{\rm e}$&13.836\,(38) &13.098\,(44) & 12.797\,(32)$^{\rm r}$&      M0.0$^{\rm e}$&                 &  1.6$^{\rm e}$&-18.7\,(4.1)  &21\,(4.1)     &              &              &     &     &                                                                                      \\
1505 & 21:39:10.13 &  57:22:32.3$^{\rm r}$&      &                           & 21-840                        &                 &                 &  18.98$^{\rm e}$& 17.54 &  16.09$^{\rm e}$&14.435\,(51) &13.562\,(58) & 13.233\,(51)$^{\rm r}$&      M1.0$^{\rm e}$&                 &  2.2$^{\rm e}$&-11\,(5.4)    &-12.2\,(5.4)  &              &              &     &     &                                                                                      \\
1506 & 21:38:55.04 &  57:20:42.3$^{\rm r}$&      &                           & 21-851                        &                 &                 &                 &       &                 &14.736\,(49) &14.257\,(66) & 13.836\,(60)$^{\rm r}$&                    &                 &               &-4.6\,(4)     &3\,(4)        &              &              &     &     &                                                                                      \\
1507 & 21:40:03.21 &  57:22:50.5$^{\rm r}$&      &                           & 21-895                        &                 &                 &  16.39$^{\rm e}$& 15.43 &  14.48$^{\rm e}$&13.216\,(26) &12.508\,(33) & 12.306\,(28)$^{\rm r}$&        K5$^{\rm c}$&                 &  1.4$^{\rm e}$&-12.9\,(4.1)  &-7.4\,(4.1)   &-20.1\,(8)    &-1.1\,(7.4)   &     &     &                                                                                      \\
1508 & 21:39:34.81 &  57:23:27.8$^{\rm r}$&      &                           & 21-998                        &                 &                 &  17.48$^{\rm e}$& 16.4  &   15.3$^{\rm e}$&13.850\,(29) &12.972\,(33) & 12.476\,(28)$^{\rm r}$&      K5.5$^{\rm c}$&                 &  1.9$^{\rm e}$&-2.3\,(4.1)   &-7.9\,(4.1)   &              &              &     &     &SB1:$^{\rm c}$                                                                                      \\
1509 & 21:40:22.87 &  57:27:33.0$^{\rm r}$&      &                           & 22-1418                       &                 &                 &  17.59$^{\rm e}$& 17.68 &  15.25$^{\rm e}$&14.134\,(32) &13.174\,(30) & 12.591\,(28)$^{\rm r}$&      M1.5$^{\rm c}$&                 &  0.7$^{\rm e}$&-3.8\,(3.9)   &-10.7\,(3.9)  &              &              &     &     &                                                                                      \\
1510 & 21:41:02.13 &  57:28:22.0$^{\rm r}$&      &                           & 22-1526                       &                 &                 &  17.76$^{\rm e}$& 16.86 &  15.56$^{\rm e}$&14.383\,(62) &13.476\,(59) & 13.382\,(51)$^{\rm r}$&        M1$^{\rm c}$&                 &  0.6$^{\rm e}$&-16.3\,(3.9)  &-13.8\,(3.9)  &              &              &     &     &star elongated                                                                        \\
1511 & 21:41:18.38 &  57:28:43.3$^{\rm r}$&      &                           & 22-1569                       &                 &                 &  17.58$^{\rm e}$& 16.7  &  15.09$^{\rm e}$&13.607\,(32) &12.826\,(33) & 12.578\,(30)$^{\rm r}$&        M1$^{\rm c}$&                 &  1.4$^{\rm e}$&-0.4\,(3.9)   &-9.8\,(3.9)   &9\,(11.3)     &-5.9\,(11.3)  &     &     &                                                                                      \\
1512 & 21:40:21.30 &  57:26:57.9$^{\rm r}$&      &                           & 22-2651                       &                 &                 &  18.43$^{\rm e}$& 17.1  &  16.01$^{\rm e}$&14.500\,(38) &13.405\,(32) & 12.624\,(28)$^{\rm r}$&      M1.5$^{\rm c}$&                 &  0.9$^{\rm e}$&-0.8\,(4.1)   &-5.2\,(4.1)   &              &              &     &     &                                                                                      \\
1513 & 21:41:09.71 &  57:20:50.8$^{\rm r}$&      &                           & 22-404                        &                 &                 &  16.02$^{\rm e}$& 15.37 &  14.51$^{\rm e}$&13.479\,(27) &12.892\,(35) & 12.681\,(35)$^{\rm r}$&      G7.0$^{\rm e}$&                 &  1.4$^{\rm e}$&-4.8\,(4.1)   &-2.7\,(4.1)   &-5.8\,(7.2)   &7.5\,(7.2)    &     &     &                                                                                      \\
1514 & 21:40:58.70 &  57:21:09.6$^{\rm r}$&      &                           & 22-445                        &                 &                 &   17.9$^{\rm e}$& 16.83 &   15.5$^{\rm e}$&13.906\,(34) &13.104\,(42) & 12.818\,(23)$^{\rm r}$&      M0.5$^{\rm c}$&                 &  1.3$^{\rm e}$&-9\,(4.1)     &-11\,(4.1)    &              &              &     &     &                                                                                      \\
1515 & 21:40:48.84 &  57:24:18.6$^{\rm r}$&      &                           & 22-939                        &                 &                 &  17.58$^{\rm e}$& 16.54 &  15.49$^{\rm e}$&14.266\,(35) &13.507\,(38) & 13.210\,(39)$^{\rm r}$&      K6.0$^{\rm e}$&                 &  1.3$^{\rm e}$&-7.2\,(4.1)   &-2.9\,(4.1)   &              &              &     &     &                                                                                      \\
1516 & 21:41:15.24 &  57:24:25.6$^{\rm r}$&      &                           & 22-960                        &                 &                 &  18.23$^{\rm e}$& 17.1  &  15.67$^{\rm e}$&14.222\,(50) &13.364\,(49) & 13.107\,(46)$^{\rm r}$&      M2.5$^{\rm c}$&                 &  0.6$^{\rm e}$&-1\,(5.4)     &-50.3\,(5.4)  &              &              &     &     &                                                                                      \\
1517 & 21:40:55.43 &  57:39:55.6$^{\rm r}$&      &                           & 23-1161                       &                 &                 &  17.81$^{\rm e}$& 16.65 &  15.74$^{\rm e}$&14.407\,(32) &13.602\,(39) & 13.425\,(39)$^{\rm r}$&      K6.5$^{\rm e}$&                 &  1.4$^{\rm e}$&-2.1\,(3.8)   &-9.2\,(3.8)   &              &              &     &     &                                                                                      \\
1518 & 21:41:05.51 &  57:41:03.3$^{\rm r}$&      &                           & 23-1282                       &                 &                 &                 &       &                 &13.664\,(31) &12.918\,(36) & 12.729\,(30)$^{\rm r}$&                    &                 &               &-3.5\,(3.8)   &-6.2\,(3.8)   &3.8\,(15.2)   &-9.8\,(12.5)  &     &     &                                                                                      \\
1519 & 21:40:44.50 &  57:31:31.4$^{\rm r}$&      &                           & 23-162                        &                 &                 &  17.97$^{\rm e}$& 16.7  &  15.68$^{\rm e}$&14.044\,(31) &13.144\,(37) & 12.588\,(29)$^{\rm r}$&        K7$^{\rm c}$&                 &  1.7$^{\rm e}$&-1.5\,(3.9)   &-5.4\,(3.9)   &              &              &     &     &                                                                                      \\
1520 & 21:41:32.36 &  57:32:24.6$^{\rm r}$&      &                           & 23-259                        &                 &                 &                 &       &                 &13.304\,(26) &12.664\,(36) & 12.453\,(25)$^{\rm r}$&                    &                 &               &-21.8\,(3.9)  &11.5\,(3.9)   &-27.1\,(8.7)  &13.7\,(10.2)  &     &     &                                                                                      \\
1521 & 21:40:31.35 &  57:33:41.8$^{\rm r}$&      & 4652                      & 23-405                        &                 &                 &  16.46$^{\rm e}$& 15.47 &  14.65$^{\rm e}$&13.389\,(32) &12.504\,(33) & 12.125\,(28)$^{\rm r}$&        K5$^{\rm c}$&                 &  1.1$^{\rm e}$&-4.7\,(3.8)   &-2.9\,(3.8)   &-12.4\,(9.6)  &-13.1\,(9.6)  &     &     &                                                                                      \\
1522 & 21:40:35.75 &  57:34:55.1$^{\rm r}$&      &                           & 23-570                        &                 &                 &  16.85$^{\rm e}$& 15.72 &  14.88$^{\rm e}$&13.477\,(31) &12.556\,(30) & 12.194\,(28)$^{\rm r}$&        K6$^{\rm c}$&                 &  1.3$^{\rm e}$&-2\,(3.9)     &-2.2\,(3.9)   &              &              &     &     &                                                                                      \\
1523 & 21:40:53.92 &  57:36:19.9$^{\rm r}$&      &                           & 23-753                        &                 &                 &  18.08$^{\rm e}$& 16.98 &  16.01$^{\rm e}$&14.664\,(27) &13.855\,(43) & 13.663\,(54)$^{\rm r}$&   M0.5-M0$^{\rm c}$&                 &  0.5$^{\rm e}$&-4.9\,(3.8)   &-14.4\,(3.8)  &              &              &     &     &                                                                                      \\
1524 & 21:41:28.65 &  57:36:43.3$^{\rm r}$&      & 4653                      & 23-798                        &  19.72$^{\rm f}$&                 &  18.76$^{\rm e}$& 17.43 &  16.36$^{\rm e}$&14.410\,(23) &13.533\,(35) & 12.820\,(30)$^{\rm r}$&        K6$^{\rm c}$&                 &  2.2$^{\rm e}$&-4.3\,(3.8)   &-4.9\,(3.8)   &              &              &     &     &faint star                                                                        \\
1525 & 21:41:14.98 &  57:38:14.9$^{\rm r}$&      &                           & 23-969                        &                 &                 &  16.39$^{\rm e}$& 15.29 &  14.52$^{\rm e}$&13.144\,(26) &12.325\,(32) & 11.919\,(21)$^{\rm r}$&      K5.5$^{\rm c}$&                 &  1.2$^{\rm e}$&-1.7\,(3.8)   &-7.8\,(3.8)   &-21\,(8.2)    &-16.2\,(8.2)  &     &     &                                                                                      \\
1526 & 21:39:50.88 &  57:36:16.8$^{\rm r}$&      &                           & 24-1047                       &                 &                 &                 &       &                 &13.237\,(26) &12.395\,(31) & 12.252\,(28)$^{\rm r}$&                    &                 &               &3\,(3.8)      &-6.4\,(3.8)   &15.6\,(8.3)   &21.8\,(8.3)   &     &     &                                                                                      \\
1527 & 21:39:03.90 &  57:31:03.8$^{\rm r}$&      &                           & 24-108                        &                 &                 &  17.27$^{\rm e}$& 16.22 &  15.12$^{\rm e}$&13.540\,(29) &12.739\,(35) & 12.471\,(28)$^{\rm r}$&      K5.5$^{\rm c}$&                 &  1.9$^{\rm e}$&1.2\,(3.8)    &-7.9\,(3.8)   &              &              &     &     &                                                                                      \\
1528 & 21:39:36.13 &  57:31:28.9$^{\rm r}$&      & 4647                      & 24-170                        &                 &                 &   17.4$^{\rm e}$& 16.27 &   15.1$^{\rm e}$&13.444\,(32) &12.648\,(35) & 12.398\,(30)$^{\rm r}$&      K7.5$^{\rm c}$&                 &  1.5$^{\rm e}$&4.5\,(3.8)    &-4.1\,(3.8)   &              &              &     &     &                                                                                      \\
1529 & 21:40:11.35 &  57:39:51.8$^{\rm r}$&      & 4651                      & 24-1736                       &  19.72$^{\rm f}$&                 &  19.07$^{\rm e}$& 18.12 &  16.35$^{\rm e}$&14.309\,(24) &13.443\,(36) & 12.968\,(28)$^{\rm r}$&        M1$^{\rm c}$&                 &    1$^{\rm e}$&-4.7\,(3.8)   &-4.9\,(3.8)   &              &              &     &     &                                                                                      \\
1530 & 21:40:11.83 &  57:40:12.2$^{\rm r}$&      &                           & 24-1796                       &                 &                 &   17.4$^{\rm e}$& 16.37 &  15.32$^{\rm e}$&13.876\,(35) &13.076\,(39) & 12.734\,(37)$^{\rm r}$&        K7$^{\rm c}$&                 &  1.2$^{\rm e}$&-8.9\,(3.8)   &-6.1\,(3.8)   &              &              &     &     &SB2:$^{\rm c}$                                                                                      \\
1531 & 21:40:10.23 &  57:32:51.2$^{\rm r}$&      &                           & 24-382                        &                 &                 &  17.91$^{\rm e}$& 16.81 &  15.67$^{\rm e}$&14.132\,(31) &13.317\,(36) & 13.091\,(35)$^{\rm r}$&      K7.5$^{\rm c}$&                 &  1.4$^{\rm e}$&-3.2\,(3.8)   &-5.5\,(3.8)   &              &              &     &     &                                                                                      \\
1532 & 21:39:38.05 &  57:30:44.0$^{\rm r}$&      &                           & 24-48                         &                 &                 &  18.11$^{\rm e}$& 16.81 &  15.63$^{\rm e}$&14.129\,(32) &13.278\,(27) & 13.073\,(29)$^{\rm r}$&      M0.5$^{\rm c}$&                 &  1.5$^{\rm e}$&-17\,(3.8)    &-6.1\,(3.8)   &              &              &     &     &                                                                                      \\
1533 & 21:39:34.07 &  57:33:31.6$^{\rm r}$&      &                           & 24-515                        &                 &                 &   17.9$^{\rm e}$& 16.7  &  15.58$^{\rm e}$&14.057\,(28) &13.125\,(33) & 12.704\,(28)$^{\rm r}$&      M0.5$^{\rm c}$&                 &  1.1$^{\rm e}$&1.5\,(3.9)    &-13.1\,(3.9)  &              &              &     &     &                                                                                      \\
1534 & 21:39:29.57 &  57:33:41.7$^{\rm r}$&      & 4645                      & 24-542                        &  18.54$^{\rm f}$&                 &  15.88$^{\rm e}$& 14.97 &   14.1$^{\rm e}$&12.823\,(24) &12.053\,(31) & 11.867\,(23)$^{\rm r}$&        K4$^{\rm c}$&                 &    1$^{\rm e}$&-1.5\,(3.8)   &-7.7\,(3.8)   &-6\,(8.3)     &-5.4\,(8.4)   &     &     &                                                                                      \\
1535 & 21:39:00.55 &  57:34:28.1$^{\rm r}$&      &                           & 24-692                        &                 &                 &  18.43$^{\rm e}$& 17.13 &  15.88$^{\rm e}$&14.294\,(35) &13.402\,(33) & 13.138\,(39)$^{\rm r}$&        M1$^{\rm c}$&                 &  1.5$^{\rm e}$&-6.7\,(5)     &-12.5\,(5)    &              &              &     &     &SB1$^{\rm c}$                                                                                      \\
1536 & 21:39:03.47 &  57:30:52.8$^{\rm r}$&      &                           & 24-77                         &                 &                 &  17.53$^{\rm e}$& 16.6  &  15.44$^{\rm e}$&13.771\,(29) &12.987\,(35) & 12.687\,(28)$^{\rm r}$&      K6.5$^{\rm c}$&                 &  1.4$^{\rm e}$&1\,(5)        &-8.1\,(5)     &              &              &     &     &                                                                                      \\
1537 & 21:39:49.37 &  57:30:54.7$^{\rm r}$&      &                           & 24-78                         &                 &                 &  17.91$^{\rm e}$& 16.59 &  15.27$^{\rm e}$&13.561\,(23) &12.767\,(30) & 12.465\,(25)$^{\rm r}$&        M2$^{\rm c}$&                 &  1.3$^{\rm e}$&-2.8\,(3.9)   &-8.4\,(3.9)   &              &              &     &     &                                                                                      \\
1538 & 21:40:02.73 &  57:35:05.0$^{\rm r}$&      & 4650                      & 24-817                        &  19.03$^{\rm f}$&                 &  17.82$^{\rm e}$& 16.62 &  15.49$^{\rm e}$&13.968\,(26) &13.122\,(30) & 12.916\,(28)$^{\rm r}$&      K6.5$^{\rm c}$&                 &    2$^{\rm e}$&-2.7\,(3.8)   &-10\,(3.8)    &              &              &     &     &                                                                                      \\
1539 & 21:39:47.47 &  57:35:06.0$^{\rm r}$&      &                           & 24-820                        &                 &                 &  18.35$^{\rm e}$& 17.28 &  16.32$^{\rm e}$&15.104\,(36) &14.363\,(49) & 14.069\,(60)$^{\rm r}$&      K6.5$^{\rm c}$&                 &  1.2$^{\rm e}$&0.9\,(3.9)    &-9\,(3.9)     &              &              &     &     &                                                                                      \\
1540 & 21:43:49.33 &  57:19:20.9$^{\rm r}$&      &                           & 43-795                        &                 &                 &   17.1$^{\rm e}$& 16.04 &  15.02$^{\rm e}$&13.376\,(26) &12.582\,(31) & 12.331\,(19)$^{\rm r}$&      K5.5$^{\rm e}$&                 &  1.7$^{\rm e}$&-4.4\,(4.1)   &-6.2\,(4.1)   &9.8\,(12.2)   &12\,(11.8)    &     &     &                                                                                      \\
1541 & 21:39:46.44 &  57:05:07.3$^{\rm r}$&      &                           & 52-1649                       &                 &                 &  17.22$^{\rm e}$& 16.25 &  15.38$^{\rm e}$&14.213\,(41) &13.557\,(50) & 13.310\,(42)$^{\rm r}$&      K5.0$^{\rm e}$&                 &  1.1$^{\rm e}$&17.3\,(5.4)   &49.3\,(5.5)   &              &              &     &     &                                                                                      \\
1542 & 21:40:55.93 &  57:17:59.2$^{\rm r}$&      &                           & 53-1561                       &                 &                 &  17.86$^{\rm e}$& 16.59 &  15.45$^{\rm e}$&13.814\,(27) &12.830\,(35) & 12.245\,(24)$^{\rm r}$&        K6$^{\rm c}$&                 &  2.1$^{\rm e}$&-8.3\,(4.1)   &-6.4\,(4.1)   &              &              &     &     &                                                                                      \\
1543 & 21:39:50.29 &  57:19:17.7$^{\rm r}$&      &                           & 53-1762                       &                 &                 &  17.94$^{\rm e}$& 16.73 &  15.62$^{\rm e}$&14.134\,(24) &13.293\,(32) & 12.984\,(30)$^{\rm r}$&        M0$^{\rm c}$&                 &  1.3$^{\rm e}$&-4.5\,(4.1)   &-0.7\,(4.1)   &              &              &     &     &SB1:$^{\rm c}$                                                                                      \\
1544 & 21:39:38.03 &  57:19:33.2$^{\rm r}$&      &                           & 53-1803                       &                 &                 &  18.25$^{\rm e}$& 17.12 &  16.16$^{\rm e}$&14.643\,(43) &13.744\,(44) & 13.448\,(48)$^{\rm r}$&      K6.5$^{\rm c}$&                 &  1.4$^{\rm e}$&-2.6\,(4.1)   &4.6\,(4.1)    &              &              &     &     &SB1$^{\rm c}$                                                                                      \\
1545 & 21:40:35.92 &  57:19:39.9$^{\rm r}$&      &                           & 53-1843                       &                 &                 &  17.65$^{\rm e}$& 16.55 &  15.51$^{\rm e}$&13.945\,(27) &13.172\,(41) & 12.938\,(26)$^{\rm r}$&      M0.5$^{\rm c}$&                 &  0.7$^{\rm e}$&-1.9\,(4.1)   &-6.2\,(4.1)   &              &              &     &     &                                                                                      \\
1546 & 21:39:17.48 &  57:17:47.4$^{\rm r}$&      &                           & 54-1488                       &                 &                 &  16.48$^{\rm e}$& 15.65 &  14.83$^{\rm e}$&13.774\,(47) &13.013\,(46) & 12.799\,(44)$^{\rm r}$&      K7.0$^{\rm e}$&                 &  0.1$^{\rm e}$&-29.8\,(4)    &25.2\,(4)     &-5.6\,(11.6)  &-64.9\,(11.5) &     &     &                                                                                      \\
1547 & 21:38:43.32 &  57:18:36.0$^{\rm r}$&      &                           & 54-1613                       &                 &                 &  16.58$^{\rm e}$& 15.6  &  14.73$^{\rm e}$&13.446\,(27) &12.688\,(29) & 12.457\,(25)$^{\rm r}$&        K5$^{\rm c}$&                 &  1.2$^{\rm e}$&-3.2\,(4)     &-2.6\,(4)     &-12.5\,(7.4)  &5.1\,(7.4)    &     &     &                                                                                      \\
1548 & 21:38:16.13 &  57:19:35.8$^{\rm r}$&      &                           & 54-1781                       &                 &                 &  18.18$^{\rm e}$& 17    &  15.65$^{\rm e}$&13.991\,(32) &13.092\,(32) & 12.792\,(30)$^{\rm r}$&        M1$^{\rm c}$&                 &  1.2$^{\rm e}$&-1.9\,(5.4)   &-4.6\,(5.4)   &              &              &     &     &                                                                                      \\
1549 & 21:36:26.15 &  57:01:29.3$^{\rm r}$&      &                           & 61-413                        &                 &                 &                 &       &                 &14.123\,(27) &13.494\,(33) & 13.344\,(34)$^{\rm r}$&                    &                 &               &-3.3\,(4.1)   &-6.2\,(4.1)   &              &              &     &     &                                                                                      \\
1550 & 21:35:50.70 &  57:03:57.1$^{\rm r}$&      &                           & 61-608                        &                 &                 &                 &       &                 &14.250\,(41) &13.575\,(49) & 13.366\,(45)$^{\rm r}$&                    &                 &               &-6.7\,(4.1)   &-2.4\,(4.1)   &-51.8\,(13.6) &17.2\,(13.1)  &     &     &                                                                                      \\
1551 & 21:36:00.91 &  57:07:12.9$^{\rm r}$&      &                           & 61-893                        &                 &                 &                 &       &                 &14.188\,(34) &13.291\,()   & 13.092\,(37)$^{\rm r}$&                    &                 &               &0\,(4.1)      &-4.7\,(4.1)   &              &              &     &     &                                                                                      \\
1552 & 21:35:18.05 &  57:09:44.1$^{\rm r}$&      &                           & 64-156                        &                 &                 &                 &       &                 &13.559\,(40) &13.266\,(58) & 13.161\,(51)$^{\rm r}$&                    &                 &               &-1.6\,(4.1)   &2.9\,(4.1)    &-8.8\,(7.1)   &13.3\,(7.1)   &     &     &                                                                                      \\
1553 & 21:35:50.83 &  57:12:07.1$^{\rm r}$&      &                           & 64-376                        &                 &                 &                 &       &                 &14.222\,(36) &13.539\,(39) & 13.341\,(38)$^{\rm r}$&                    &                 &               &17\,(4.1)     &4.4\,(4.1)    &18.9\,(11.6)  &8.5\,(11.6)   &     &     &                                                                                      \\
1554 & 21:34:09.74 &  57:29:55.0$^{\rm r}$&      &                           & 71-1309                       &                 &                 &                 &       &                 &14.697\,(38) &13.939\,(46) & 13.737\,(51)$^{\rm r}$&                    &                 &               &9.3\,(4)      &-1.7\,(4)     &              &              &     &     &                                                                                      \\
1555 & 21:35:16.28 &  57:28:22.2$^{\rm r}$&      &                           & 72-1427                       &                 &                 &  18.08$^{\rm e}$& 16.82 &  15.58$^{\rm e}$&14.005\,(31) &13.067\,(29) & 12.673\,(28)$^{\rm r}$&        M1$^{\rm c}$&                 &  1.4$^{\rm e}$&-6.3\,(5.5)   &-12.7\,(5.5)  &              &              &     &     &                                                                                      \\
1556 & 21:35:14.82 &  57:21:23.3$^{\rm r}$&      &                           & 72-489                        &                 &                 &  18.11$^{\rm e}$& 17.05 &  16.06$^{\rm e}$&14.607\,(38) &13.837\,(43) & 13.543\,(44)$^{\rm r}$&     K5.0?$^{\rm e}$&                 &  1.7$^{\rm e}$&-14.1\,(4.1)  &-16.9\,(4.1)  &              &              &     &     &                                                                                      \\
1557 & 21:35:49.75 &  57:24:04.2$^{\rm r}$&      &                           & 72-875                        &                 &                 &  18.37$^{\rm e}$& 17.38 &  16.18$^{\rm e}$&14.092\,(31) &13.006\,(27) & 12.500\,(26)$^{\rm r}$&      M0.5$^{\rm c}$&                 &  0.8$^{\rm e}$&-17.3\,(5.5)  &-4.2\,(5.5)   &              &              &     &     &SB1:$^{\rm c}$                                                                                      \\
1558 & 21:35:23.86 &  57:38:14.6$^{\rm r}$&      &                           & 73-1059                       &                 &                 &                 &       &                 &14.402\,(35) &13.693\,(52) & 13.452\,(39)$^{\rm r}$&                    &                 &               &-33.7\,(5.5)  &-483.2\,(5.5) &              &              &     &     &                                                                                      \\
1559 & 21:35:52.23 &  57:32:14.5$^{\rm r}$&      &                           & 73-194                        &                 &                 &  18.06$^{\rm e}$& 16.95 &  15.94$^{\rm e}$&14.734\,(42) &13.957\,(52) & 13.764\,(50)$^{\rm r}$&      K6.5$^{\rm c}$&                 &  1.5$^{\rm e}$&-4.4\,(4)     &-8.5\,(4)     &              &              &     &     &                                                                                      \\
1560 & 21:35:24.51 &  57:33:01.1$^{\rm r}$&      &                           & 73-311                        &                 &                 &  18.38$^{\rm e}$& 17.13 &  15.89$^{\rm e}$&14.014\,(34) &13.115\,(37) & 12.674\,(30)$^{\rm r}$&      M1.5$^{\rm c}$&                 &  1.1$^{\rm e}$&-0.7\,(3.8)   &-5.6\,(3.8)   &              &              &     &     &                                                                                      \\
1561 & 21:35:18.61 &  57:34:09.2$^{\rm r}$&      &                           & 73-472                        &                 &                 &  16.85$^{\rm e}$& 15.75 &   14.8$^{\rm e}$&13.219\,(24) &12.299\,(31) & 11.829\,(22)$^{\rm r}$&        K5$^{\rm c}$&                 &  1.7$^{\rm e}$&-3.9\,(4)     &0\,(4)        &-16.1\,(8.3)  &1.8\,(8.4)    &     &     &                                                                                      \\
1562 & 21:36:07.24 &  57:34:32.4$^{\rm r}$&      &                           & 73-537                        &                 &                 &   17.3$^{\rm e}$& 16.27 &  15.25$^{\rm e}$&14.047\,(52) &13.159\,(39) &   12.674\,()$^{\rm r}$&      G1.5$^{\rm c}$&                 &  3.3$^{\rm e}$&              &              &              &              &     &     &near 1766                                                                           \\
1563 & 21:35:20.77 &  57:35:28.9$^{\rm r}$&      &                           & 73-674                        &                 &                 &                 &       &                 &12.762\,(27) &11.988\,(29) & 11.762\,(21)$^{\rm r}$&                    &                 &               &-5.7\,(4.1)   &9.4\,(4.1)    &-16.3\,(6.9)  &23\,(6.7)     &     &     &                                                                                      \\
1564 & 21:35:30.21 &  57:31:16.5$^{\rm r}$&      &                           & 73-71                         &                 &                 &  16.99$^{\rm e}$& 15.81 &  14.69$^{\rm e}$&12.979\,(34) &12.105\,(37) & 11.715\,(29)$^{\rm r}$&        K6$^{\rm c}$&                 &  2.1$^{\rm e}$&-4.6\,(4)     &-6.1\,(4)     &              &              &     &     &                                                                                      \\
1565 & 21:35:08.35 &  57:36:02.9$^{\rm r}$&      &                           & 73-758                        &                 &                 &  17.06$^{\rm e}$& 15.84 &  14.76$^{\rm e}$&13.330\,(28) &12.533\,(32) & 12.163\,(28)$^{\rm r}$&      K6.5$^{\rm c}$&                 &  1.9$^{\rm e}$&-0.5\,(3.8)   &-5.5\,(3.8)   &1.1\,(11.5)   &-21\,(9.5)    &     &     &                                                                                      \\
1566 & 21:34:47.30 &  57:31:14.9$^{\rm r}$&      &                           & 74-48                         &                 &                 &                 &       &                 &13.102\,(26) &12.210\,(32) & 11.957\,(25)$^{\rm r}$&                    &                 &               &-4.5\,(4)     &-0.2\,(4)     &              &              &     &     &                                                                                      \\
1567 & 21:35:17.46 &  57:48:22.3$^{\rm r}$&      &                           & 81-541                        &                 &                 &  17.52$^{\rm e}$& 16.32 &  15.18$^{\rm e}$&13.399\,(31) &12.469\,(33) & 12.037\,(26)$^{\rm r}$&      K5.5$^{\rm c}$&                 &  2.3$^{\rm e}$&2.2\,(3.8)    &-1.5\,(3.8)   &              &              &     &     &                                                                                      \\
1568 & 21:38:03.51 &  57:41:35.0$^{\rm r}$&      &                           & 82-272                        &                 &                 &  17.26$^{\rm e}$& 16.17 &     15$^{\rm e}$&12.502\,(26) &11.483\,(29) & 10.847\,(19)$^{\rm r}$&        G9$^{\rm c}$&                 &  3.6$^{\rm e}$&-4.4\,(4)     &-1.7\,(4)     &-10\,(8.5)    &-14.4\,(8.5)  &     &     &SB2$^{\rm c}$                                                                                      \\
1569 & 21:37:36.96 &  57:55:14.9$^{\rm r}$&      &                           & 83-343                        &                 &                 &  16.67$^{\rm e}$& 15.55 &  14.43$^{\rm e}$&13.006\,()   &12.277\,(45) &   11.940\,()$^{\rm r}$&      M0.5$^{\rm c}$&                 &  0.9$^{\rm e}$&-9.8\,(5.1)   &8.5\,(5.1)    &-27.7\,(14.2) &47.2\,(14.2)  &     &     &                                                                                      \\
1570 & 21:36:12.81 &  57:53:00.4$^{\rm r}$&      &                           & 84-23                         &                 &                 &                 &       &                 &13.904\,(32) &13.422\,(48) & 13.308\,(40)$^{\rm r}$&                    &                 &               &-0.7\,(3.8)   &-3.5\,(3.8)   &5.7\,(10.1)   &-14.6\,(9.9)  &     &     &                                                                                      \\
1571 & 21:38:34.71 &  57:41:27.4$^{\rm r}$&      &                           & 91-155                        &                 &                 &  18.31$^{\rm e}$& 16.95 &  15.52$^{\rm e}$&13.741\,(35) &12.875\,(48) & 12.489\,(36)$^{\rm r}$&      M2.5$^{\rm c}$&                 &  1.2$^{\rm e}$&-6.6\,(3.9)   &-2.5\,(3.9)   &              &              &     &     &                                                                                      \\
1572 & 21:38:58.07 &  57:43:34.4$^{\rm r}$&      &                           & 91-506                        &                 &                 &  16.84$^{\rm e}$& 15.76 &  14.75$^{\rm e}$&13.386\,(31) &12.521\,(36) & 12.040\,(22)$^{\rm r}$&      K6.5$^{\rm c}$&                 &  1.4$^{\rm e}$&-1.8\,(5.4)   &-1.9\,(5.4)   &46.6\,(9.5)   &-6.6\,(11.2)  &     &     &                                                                                      \\
1573 & 21:39:14.65 &  57:45:17.7$^{\rm r}$&      &                           & 91-815                        &                 &                 &   18.3$^{\rm e}$& 17.03 &  15.65$^{\rm e}$&14.142\,(52) &13.379\,(61) & 13.091\,(52)$^{\rm r}$&        M2$^{\rm c}$&                 &  1.3$^{\rm e}$&-20.4\,(3.9)  &-10.1\,(3.9)  &              &              &     &     &                                                                                      \\
1574 & 21:40:22.74 &  57:46:24.1$^{\rm r}$&      &                           & 92-1103                       &                 &                 &  16.52$^{\rm e}$& 15.39 &   14.3$^{\rm e}$&12.816\,(32) &11.935\,(33) & 11.553\,(26)$^{\rm r}$&      K5.5$^{\rm c}$&                 &    2$^{\rm e}$&17.2\,(3.9)   &-3.6\,(3.9)   &              &              &     &     &                                                                                      \\
1575 & 21:39:49.75 &  57:46:46.8$^{\rm r}$&      &                           & 92-1162                       &                 &                 &  18.16$^{\rm e}$& 16.84 &  15.46$^{\rm e}$&13.791\,(27) &12.897\,(30) & 12.563\,(28)$^{\rm r}$&        M2$^{\rm c}$&                 &  1.4$^{\rm e}$&-4.3\,(3.9)   &-7.2\,(3.9)   &              &              &     &     &                                                                                      \\
1576 & 21:39:40.10 &  57:46:56.2$^{\rm r}$&      &                           & 92-1198                       &                 &                 &                 &       &                 &13.101\,(32) &12.287\,(28) & 12.060\,(26)$^{\rm r}$&                    &                 &               &-5.1\,(3.8)   &-3.8\,(3.8)   &-2.6\,(8.1)   &-1.3\,(8.5)   &     &     &                                                                                      \\
1577 & 21:39:44.09 &  57:42:16.0$^{\rm r}$&      &                           & 92-393                        &                 &                 &  18.27$^{\rm e}$& 16.82 &  15.35$^{\rm e}$&13.619\,(26) &12.765\,(28) & 12.505\,(23)$^{\rm r}$&        M2$^{\rm c}$&                 &    2$^{\rm e}$&2.7\,(3.8)    &-0.7\,(3.8)   &              &              &     &     &                                                                                      \\
1578 & 21:40:25.93 &  57:43:27.2$^{\rm r}$&      &                           & 92-582                        &                 &                 &                 &       &                 &14.903\,(45) &14.221\,(58) & 14.039\,(76)$^{\rm r}$&                    &                 &               &-7.7\,(4.1)   &-5.1\,(4.1)   &              &              &     &     &                                                                                      \\
1579 & 21:40:41.51 &  57:45:22.0$^{\rm r}$&      &                           & 92-926                        &                 &                 &                 &       &                 &13.200\,(26) &12.542\,(28) & 12.353\,(23)$^{\rm r}$&                    &                 &               &1.8\,(3.8)    &-4.9\,(3.8)   &11\,(8.7)     &9.2\,(8.3)    &     &     &                                                                                      \\
1580 & 21:40:40.61 &  57:54:06.4$^{\rm r}$&      &                           & 93-168                        &                 &                 &   16.6$^{\rm e}$& 15.56 &   14.6$^{\rm e}$&13.423\,(24) &12.520\,(30) & 12.092\,(26)$^{\rm r}$&      K6.5$^{\rm c}$&                 &  1.2$^{\rm e}$&-4.3\,(3.8)   &-9.3\,(3.8)   &              &              &     &     &                                                                                      \\
1581 & 21:39:52.37 &  57:56:18.7$^{\rm r}$&      &                           & 93-361                        &                 &                 &  16.74$^{\rm e}$& 15.6  &   14.6$^{\rm e}$&12.847\,()   &11.793\,()   & 10.760\,(25)$^{\rm r}$&        G1$^{\rm c}$&                 &  3.6$^{\rm e}$&-4\,(3.8)     &16\,(3.8)     &              &              &     &     &                                                                                      \\
1582 & 21:40:35.86 &  57:58:13.0$^{\rm r}$&      &                           & 93-540                        &                 &                 &  18.47$^{\rm e}$& 17.07 &  15.81$^{\rm e}$&14.176\,(31) &13.177\,(37) & 12.646\,(32)$^{\rm r}$&        M0$^{\rm c}$&                 &  2.2$^{\rm e}$&-0.7\,(3.8)   &-7.5\,(3.8)   &              &              &     &     &                                                                                      \\
1583 & 21:40:09.99 &  58:00:03.7$^{\rm r}$&      &                           & 93-720                        &                 &                 &                 &       &                 &12.709\,(24) &11.580\,(30) & 10.940\,(22)$^{\rm r}$&                    &                 &               &-5.8\,(3.8)   &-7.6\,(3.8)   &-8.9\,(7.4)   &-13.5\,(7.6)  &     &     &                                                                                      \\
1584 & 21:38:26.69 &  58:02:37.8$^{\rm r}$&      &                           & 94-1050                       &                 &                 &                 &       &                 &14.328\,(39) &13.568\,(42) & 13.235\,(40)$^{\rm r}$&                    &                 &               &-3.8\,(3.8)   &-4.9\,(3.8)   &              &              &     &     &                                                                                      \\
1585 & 21:38:18.62 &  58:03:28.3$^{\rm r}$&      &                           & 94-1119                       &                 &                 &                 &       &                 &13.126\,(26) &12.240\,(30) & 11.927\,(23)$^{\rm r}$&                    &                 &               &-11.4\,(3.8)  &-7.1\,(3.8)   &              &              &     &     &                                                                                      \\
1586 & 21:40:37.22 &  57:29:12.7$^{\rm r}$&      &                           &                               &                 &                 &                 &       &                 &14.640\,(53) &13.815\,(53) & 13.512\,(61)$^{\rm r}$&      K7.5$^{\rm e}$&                 &               &-4.6\,(3.8)   &-10.9\,(3.8)  &              &              &     &     &faint star                                                                        \\
1587 & 21:40:21.92 &  57:30:05.4$^{\rm r}$&      &                           &                               &                 &                 &                 &       &                 &13.608\,(39) &12.754\,(38) & 12.538\,(32)$^{\rm r}$&        K6$^{\rm c}$&                 &               &0.1\,(5.1)    &6.2\,(5.1)    &11.2\,(7.8)   &64.9\,(7.7)   &     &     &                                                                                      \\
1588 & 21:40:04.52 &  57:28:36.4$^{\rm r}$&      &                           &                               &                 &                 &                 &       &                 &13.011\,(32) &12.210\,(36) & 11.810\,(29)$^{\rm r}$&      K5.0$^{\rm e}$&                 &               &-11\,(4.1)    &-6.6\,(4.1)   &-7.4\,(7.6)   &1.4\,(7)      &     &     &                                                                                      \\
1589 & 21:39:03.21 &  57:30:42.1$^{\rm r}$&      &                           &                               &                 &                 &                 &       &                 &14.338\,(46) &13.500\,(46) & 13.146\,(42)$^{\rm r}$&      K7.0$^{\rm e}$&                 &               &25.4\,(5)     &-8.4\,(5)     &              &              &     &     &                                                                                      \\
1590 & 21:38:56.69 &  57:30:48.4$^{\rm r}$&      &                           &                               &                 &                 &                 &       &                 &14.919\,(53) &14.163\,(55) & 13.902\,(65)$^{\rm r}$&      K5.0$^{\rm e}$&                 &               &-9.1\,(5.4)   &15.5\,(5.4)   &              &              &     &     &                                                                                      \\
1591 & 21:38:43.51 &  57:27:27.1$^{\rm r}$&      &                           &                               &                 &                 &                 &       &                 &14.246\,(34) &13.447\,(37) & 13.065\,(34)$^{\rm r}$&        M2$^{\rm c}$&                 &               &1.1\,(5.4)    &-3.9\,(5.4)   &              &              &     &     &very faint opt. cp.                                                        \\
1592 & 21:38:32.17 &  57:26:35.9$^{\rm r}$&      &                           &                               &                 &                 &                 &       &                 &14.811\,(45) &13.860\,(44) & 13.175\,(34)$^{\rm r}$&        M0$^{\rm c}$&                 &               &              &              &              &              &     &     &faint star, ok                                                                        \\
1593 & 21:37:42.76 &  57:33:25.1$^{\rm r}$&      &                           &                               &                 &                 &                 &       &                 &12.561\,(23) &11.308\,(26) & 10.388\,(21)$^{\rm r}$&        F9$^{\rm c}$&                 &               &18.4\,(4.1)   &-29.2\,(4.1)  &13\,(18.6)    &-2.9\,(19.3)  &     &     &                                                                                      \\
1594 & 21:37:24.48 &  57:31:36.0$^{\rm r}$&      &                           &                               &                 &                 &                 &       &                 &14.543\,(32) &13.790\,(46) & 13.278\,(38)$^{\rm r}$&      M3.5$^{\rm e}$&                 &               &              &              &              &              &     &     &no opt. cp.                                                                \\
1595 & 21:37:09.37 &  57:29:48.4$^{\rm r}$&      &                           &                               &                 &                 &                 &       &                 &13.385\,(27) &12.325\,(47) & 11.837\,(34)$^{\rm r}$&      M0.5$^{\rm e}$&                 &               &0.4\,(4.1)    &3.4\,(4.1)    &              &              &     &     &                                                                                      \\
1596 & 21:36:59.47 &  57:31:34.9$^{\rm r}$&      &                           &                               &                 &                 &                 &       &                 &14.520\,(34) &13.403\,(39) & 12.765\,(28)$^{\rm r}$&        M0$^{\rm c}$&                 &               &-16.3\,(4)    &-9\,(4)       &              &              &     &     &faint star, ok                                                                        \\
1597 & 21:36:47.63 &  57:29:54.1$^{\rm r}$&      &                           &                               &                 &                 &                 &       &                 &13.568\,()   &12.342\,(42) & 11.655\,(33)$^{\rm r}$&        K6$^{\rm c}$&                 &               &-12.6\,(3.8)  &-0.7\,(3.8)   &              &              &     &     &                                                                                      \\
1598 & 21:36:45.97 &  57:29:33.9$^{\rm r}$&      &                           &                               &                 &                 &                 &       &                 &14.211\,(39) &12.416\,(36) & 11.189\,(25)$^{\rm r}$&                    &                 &               &-10.4\,(5.5)  &0.1\,(5.5)    &              &              &     &     &faint star, ok                                                                        \\
1599 & 21:36:25.08 &  57:27:50.3$^{\rm r}$&      &                           &                               &                 &                 &                 &       &                 &14.952\,(44) &14.031\,(48) & 13.518\,(40)$^{\rm r}$&        M0$^{\rm c}$&                 &               &48.2\,(8.6)   &73\,(8.6)     &              &              &     &     &SB1$^{\rm c}$, very faint opt. cp.                                                                \\
1600 & 21:30:45.93 &  57:12:00.1$^{\rm r}$&      & 136                       &                               &   8.41$^{\rm l}$&   9.02$^{\rm l}$&    8.6$^{\rm g}$& 8.27  &   8.13$^{\rm g}$&8.500 \,(24) &8.581 \,(27) & 8.579 \,(21)$^{\rm r}$&      B1.5$^{\rm p}$&      V$^{\rm p}$&               &-3.3\,(0.7)   &-4.7\,(0.6)   &-2.4\,(0.5)   &-4.1\,(0.8)   &     &     &outFoV                                                                  \\
1601 & 21:38:26.29 &  56:58:25.3$^{\rm r}$&      & 171                       &                               &   6.76$^{\rm l}$&   7.54$^{\rm l}$&   7.42$^{\rm l}$&       &                 &7.187 \,(24) &7.193 \,(42) & 7.234 \,(31)$^{\rm r}$&   O9.5-B0$^{\rm p}$&      V$^{\rm p}$&               &-5.7\,(1.3)   &-5.9\,(1.3)   &-4.6\,(1.7)   &-0.9\,(3.1)   &     &     &                                                                                      \\
1602 & 21:43:24.46 &  57:01:23.3$^{\rm r}$&      & 207                       &                               &   8.56$^{\rm l}$&   9.18$^{\rm l}$&   8.92$^{\rm l}$&       &                 &8.354 \,(24) &8.344 \,(44) & 8.350 \,(44)$^{\rm r}$&        B1$^{\rm p}$&      V$^{\rm p}$&               &-2.6\,(0.7)   &-5.9\,(0.7)   &-2.4\,(0.8)   &-5.6\,(0.8)   &     &     &                                                                                      \\
1603 & 21:46:22.58 &  56:55:02.0$^{\rm r}$&      & 225                       &                               &   9.54$^{\rm l}$&   9.89$^{\rm l}$&   9.21$^{\rm l}$&       &                 &7.620 \,(21) &7.460 \,(44) & 7.363 \,(20)$^{\rm r}$&      B0.5$^{\rm p}$&      V$^{\rm p}$&               &-5.2\,(0.8)   &-2.4\,(0.7)   &-6.4\,(0.6)   &-2.7\,(0.6)   &     &     &outFoV                                                                  \\
1604 & 21:31:38.40 &  57:30:09.1$^{\rm r}$&      & 401                       &                               &   7.54$^{\rm l}$&      8$^{\rm l}$&   7.42$^{\rm l}$&       &                 &5.914 \,(24) &5.741 \,(31) & 5.589 \,(20)$^{\rm r}$&        B0$^{\rm p}$&     Ib$^{\rm p}$&               &-4.6\,(0.5)   &-2.9\,(0.5)   &-2.2\,(1.9)   &-2.8\,(3.8)   &     &     &outFoV                                                                  \\
1605 & 21:34:40.91 &  57:28:56.7$^{\rm r}$&      & 421                       &                               &   9.49$^{\rm l}$&   9.43$^{\rm l}$&   9.32$^{\rm l}$&       &                 &8.993 \,(39) &8.970 \,(26) & 8.962 \,(20)$^{\rm r}$&        A1$^{\rm p}$&      V$^{\rm p}$&               &0\,(1.2)      &6.5\,(1.2)    &-2.3\,(0.7)   &4.8\,(0.8)    &     &     &                                                                                      \\
1606 & 21:29:53.46 &  57:48:57.2$^{\rm r}$&      & 677                       &                               &   9.17$^{\rm l}$&    9.6$^{\rm l}$&   8.42$^{\rm g}$& 7.97  &   7.74$^{\rm g}$&8.707 \,(18) &8.747 \,(28) & 8.710 \,(22)$^{\rm r}$&        B3$^{\rm p}$&     IV$^{\rm p}$&               &-4.2\,(0.8)   &-3.5\,(0.8)   &-4.5\,(0.7)   &-4.5\,(1.2)   &     &     &outFoV                                                                  \\
1607 & 21:30:33.41 &  58:01:51.3$^{\rm r}$&      & 682                       &                               &   9.84$^{\rm l}$&  10.27$^{\rm l}$&   9.97$^{\rm l}$&       &                 &9.366 \,(22) &9.369 \,(29) & 9.330 \,(18)$^{\rm r}$&        B5$^{\rm p}$&    III$^{\rm p}$&               &-12.4\,(2)    &-0.3\,(2)     &-4.5\,(1.2)   &-3.6\,(0.6)   &     &     &outFoV                                                                  \\
1608 & 21:31:25.94 &  57:53:56.5$^{\rm r}$&      & 686                       &                               &   9.33$^{\rm l}$&   9.75$^{\rm l}$&   9.56$^{\rm g}$& 9.08  &   8.52$^{\rm g}$&9.044 \,(25) &9.058 \,(32) &   9.057 \,()$^{\rm r}$&        B3$^{\rm p}$&      V$^{\rm p}$&               &-4.9\,(1.3)   &-2.3\,(1.2)   &              &              &     &     &outFoV                                                                  \\
1609 & 21:36:59.64 &  58:08:24.6$^{\rm r}$&      & 710                       &                               &   8.72$^{\rm l}$&   9.03$^{\rm l}$&   8.61$^{\rm l}$&       &                 &7.542 \,(21) &7.478 \,(18) & 7.284 \,(18)$^{\rm r}$&      B2-3$^{\rm p}$&   IV-V$^{\rm p}$&               &-6.6\,(0.7)   &-7.5\,(0.7)   &-6.4\,(0.6)   &-8\,(0.6)     &     &     &                                                                                      \\
1610 & 21:46:16.60 &  58:03:45.0$^{\rm r}$&      & 780                       &                               &  11.12$^{\rm l}$&    9.3$^{\rm l}$&   7.39$^{\rm l}$&       &                 &3.210 \,(240)&2.246 \,(198)&1.788 \,(226)$^{\rm r}$&        M3$^{\rm p}$&     Ia$^{\rm p}$&               &-0.6\,(1.2)   &-7.4\,(1.2)   &1.6\,(1)      &-5.1\,(0.7)   &     &     &outFoV                                                                  \\
1611 & 21:39:44.43 &  58:14:43.8$^{\rm r}$&      & 1025                      &                               &   8.82$^{\rm l}$&   9.33$^{\rm l}$&   9.03$^{\rm l}$&       &                 &8.468 \,(34) &8.419 \,(38) & 8.380 \,(20)$^{\rm r}$&        B2$^{\rm p}$&     IV$^{\rm p}$&               &-2.1\,(1.2)   &-4.3\,(1.2)   &              &              &     &     &                                                                                      \\
1612 & 21:41:58.56 &  58:30:00.2$^{\rm r}$&      & 1036                      &                               &   8.39$^{\rm l}$&   8.86$^{\rm l}$&   7.93$^{\rm g}$& 7.24  &   6.75$^{\rm g}$&8.003 \,(39) &7.963 \,(46) & 7.941 \,(31)$^{\rm r}$&        B2$^{\rm p}$&     IV$^{\rm p}$&               &-4.7\,(0.7)   &-4.7\,(0.7)   &-3.5\,(0.6)   &-4.6\,(0.9)   &     &     &outFoV                                                                  \\
1613 & 22:02:15.32 &  59:39:36.3$^{\rm r}$&      & 1044                      &                               &   9.66$^{\rm l}$&  10.04$^{\rm l}$&   9.74$^{\rm l}$&       &                 &6.782 \,(20) &6.268 \,(23) & 6.121 \,(23)$^{\rm r}$&      B2.5$^{\rm p}$&      V$^{\rm p}$&               &              &              &              &              &     &     &[m] HIP\# wrong, outFoV                                                 \\
1614 & 22:03:42.91 &  59:10:00.1$^{\rm r}$&      & 1055                      &                               &   8.22$^{\rm l}$&   8.91$^{\rm l}$&   8.75$^{\rm l}$&       &                 &7.254 \,(30) &7.076 \,(26) & 7.056 \,(21)$^{\rm r}$&        B1$^{\rm p}$&      V$^{\rm p}$&               &60.9\,(1.2)   &10.1\,(1.2)   &60.5\,(0.6)   &8.4\,(0.8)    &     &     &[m] HIP\# wrong, outFoV                                                 \\
1615 & 21:28:57.77 &  58:44:23.3$^{\rm r}$&      & 1240                      &                               &   8.60$^{\rm l}$&   8.73$^{\rm l}$&   9.29$^{\rm g}$& 8.94  &    8.8$^{\rm g}$&6.472 \,(20) &6.356 \,(21) & 6.319 \,(16)$^{\rm r}$&   O9.5-B0$^{\rm p}$&      V$^{\rm p}$&               &-1.6\,(1.2)   &-5.1\,(1.2)   &              &              &     &     &[m] colors inconsistent, outFoV                                       \\
1616 & 21:37:38.77 &  58:45:30.2$^{\rm r}$&      & 1289                      &                               &   7.80$^{\rm l}$&   7.78$^{\rm l}$&   7.67$^{\rm l}$&       &                 &7.241 \,(24) &7.212 \,(46) & 7.191 \,(17)$^{\rm r}$&      B9.5$^{\rm p}$&      V$^{\rm p}$&               &-2.5\,(0.5)   &-7.3\,(0.5)   &-2.5\,(0.6)   &-7.6\,(0.6)   &     &     &outFoV                                                                  \\
1617 & 21:38:14.06 &  59:10:03.1$^{\rm r}$&      & 1293                      &                               &  10.08$^{\rm l}$&  10.22$^{\rm l}$&   8.82$^{\rm g}$& 8.53  &   8.48$^{\rm g}$&8.196 \,(29) &8.045 \,(49) & 8.036 \,(31)$^{\rm r}$&        B2$^{\rm p}$&     IV$^{\rm p}$&               &-4.6\,(0.8)   &-4.5\,(0.8)   &-4.4\,(0.6)   &-5.1\,(1.6)   &     &     &outFoV                                                                  \\
1618 & 21:43:22.61 &  58:50:42.3$^{\rm r}$&      & 1318                      &                               &   9.86$^{\rm l}$&  10.22$^{\rm l}$&   9.89$^{\rm l}$&       &                 &9.044 \,(23) &8.975 \,(40) & 8.979 \,(22)$^{\rm r}$&        B2$^{\rm p}$&      V$^{\rm p}$&               &-4.4\,(1.3)   &-3.5\,(1.2)   &-5\,(0.6)     &-2.7\,(0.7)   &     &     &outFoV                                                                  \\
1619 & 21:43:30.45 &  58:46:48.1$^{\rm r}$&      & 1319                      &                               &   8.79$^{\rm l}$&   6.42$^{\rm l}$&   4.07$^{\rm l}$& 2.07  &   0.34$^{\rm o}$&-0.326\,(204)&-1.264\,(180)&-1.620\,(160)$^{\rm r}$&     G5/M2$^{\rm p}$&  Ib/Ia$^{\rm p}$&               &3.5\,(1.3)    &-9.4\,(1.3)   &              &              &     &     &outFoV                                                                  \\
1620 & 21:44:34.01 &  59:03:25.7$^{\rm r}$&      & 1331                      &                               &   9.22$^{\rm l}$&   9.76$^{\rm l}$&   9.51$^{\rm l}$&       &                 &8.850 \,(20) &8.819 \,(33) & 8.691 \,(18)$^{\rm r}$&        B2$^{\rm p}$& III-IV$^{\rm p}$&               &0.6\,(0.8)    &0.7\,(0.8)    &-1.9\,(1)     &1.1\,(1.2)    &     &     &outFoV                                                                  \\
1621 & 21:48:40.74 &  58:59:01.2$^{\rm r}$&      & 1354                      &                               &                 &                 &                 &       &                 &10.299\,(28) &10.255\,(36) & 10.184\,(26)$^{\rm r}$&        B9$^{\rm p}$&      V$^{\rm p}$&               &-6.7\,(2)     &-6.6\,(2.1)   &-3.8\,(2)     &-6.7\,(0.9)   &     &     &outFoV                                                                  \\
1622 & 21:27:32.59 &  59:17:40.7$^{\rm r}$&      & 1492                      &                               &                 &                 &                 &       &                 &8.010 \,(21) &7.918 \,(27) &   7.749 \,()$^{\rm r}$&        B2$^{\rm p}$&      V$^{\rm p}$&               &-2.8\,(0.7)   &-3.4\,(0.7)   &-2.2\,(0.6)   &-3.9\,(0.6)   &     &     &outFoV                                                                  \\
1623 & 21:32:20.70 &  59:34:21.0$^{\rm r}$&      & 1513                      &                               &                 &                 &                 &       &                 &7.128 \,(24) &7.034 \,(27) & 7.043 \,(17)$^{\rm r}$&      B1.5$^{\rm p}$&      V$^{\rm p}$&               &-3.2\,(1.3)   &-0.7\,(1.4)   &              &              &     &     &outFoV                                                                  \\
1624 & 21:34:22.58 &  59:28:43.9$^{\rm r}$&      & 1522                      &                               &                 &                 &                 &       &                 &7.733 \,(27) &7.510 \,(34) & 7.222 \,(33)$^{\rm r}$&        B2$^{\rm p}$&    III$^{\rm p}$&               &-2\,(1.2)     &-2\,(1.3)     &-3.7\,(0.7)   &-3.3\,(1.7)   &     &     &outFoV                                                                  \\
1625 & 21:47:39.80 &  59:42:01.4$^{\rm r}$&      & 1588                      &                               &   6.98$^{\rm l}$&   7.62$^{\rm l}$&   7.29$^{\rm l}$&       &                 &6.761 \,(41) &6.756 \,(24) & 6.760 \,(15)$^{\rm r}$&   O9.5-B0$^{\rm p}$&      V$^{\rm p}$&               &-2.3\,(0.6)   &-2.6\,(0.5)   &-2.9\,(0.6)   &-1.6\,(0.6)   &     &     &outFoV                                                                  \\
1626 & 21:25:58.39 &  60:09:42.8$^{\rm r}$&      & 1732                      &                               &                 &                 &                 &       &                 &9.321 \,(39) &9.273 \,(42) & 9.192 \,(32)$^{\rm r}$&        B5$^{\rm p}$&      V$^{\rm p}$&               &-25.6\,(1.6)  &-9.4\,(1.2)   &-13\,(3.2)    &-6\,(1.4)     &     &     &outFoV                                                                  \\
1627 & 21:25:26.63 &  58:09:06.4$^{\rm r}$&      & 5001                      &                               &  16.21$^{\rm k}$&  15.63$^{\rm k}$&  14.61$^{\rm n}$&       &                 &12.305\,(27) &11.906\,(29) & 11.747\,(42)$^{\rm r}$&                    &                 &               &-24.2\,(3.9)  &-12.1\,(3.9)  &-62.8\,(7.2)  &-17.6\,(7.3)  &     &     &outFoV                                                                  \\
1628 & 21:25:51.06 &  58:11:03.4$^{\rm r}$&      & 5002                      &                               &                 &  16.33$^{\rm k}$&  14.92$^{\rm n}$&       &                 &11.374\,(23) &10.608\,(23) & 10.364\,(19)$^{\rm r}$&                    &                 &               &0\,(3.9)      &3.3\,(3.9)    &-9.7\,(7.3)   &-1.2\,(7.2)   &     &     &outFoV                                                                  \\
1629 & 21:26:24.84 &  57:52:51.4$^{\rm r}$&      & 5003                      &                               &                 &  17.34$^{\rm k}$&  15.62$^{\rm n}$&       &                 &11.397\,(22) &10.472\,(19) & 10.174\,(17)$^{\rm r}$&                    &                 &               &-2.9\,(3.9)   &2.5\,(3.9)    &-9.1\,(7.3)   &23.7\,(7.4)   &     &     &outFoV                                                                  \\
1630 & 21:26:37.78 &  57:57:31.8$^{\rm r}$&      & 5004                      &                               &                 &   16.5$^{\rm k}$&   15.9$^{\rm n}$&       &                 &13.441\,(39) &12.954\,(42) & 12.839\,(47)$^{\rm r}$&                    &                 &               &-7.3\,(3.9)   &-4.7\,(3.9)   &              &              &     &     &outFoV                                                                  \\
1631 & 21:26:39.92 &  57:46:17.8$^{\rm r}$&      & 5005                      &                               &                 &   17.5$^{\rm k}$&   16.5$^{\rm n}$&       &                 &11.209\,(21) &10.179\,(17) & 9.873 \,(15)$^{\rm r}$&                    &                 &               &3.7\,(4.7)    &-1.5\,(4.7)   &-22.3\,(7.2)  &4.9\,(7)      &     &     &outFoV                                                                  \\
1632 & 21:27:58.14 &  57:08:57.8$^{\rm r}$&      & 5006                      &                               &                 &  17.32$^{\rm k}$&  15.52$^{\rm n}$&       &                 &10.693\,(20) &9.655 \,(18) & 9.312 \,(19)$^{\rm r}$&                    &                 &               &0.5\,(5.1)    &-5.3\,(5.1)   &-13\,(6.9)    &0.4\,(6.9)    &     &     &outFoV                                                                  \\
1633 & 21:28:04.33 &  57:02:32.7$^{\rm r}$&      & 5007                      &                               &  16.17$^{\rm k}$&  16.12$^{\rm k}$&  14.83$^{\rm n}$&       &                 &12.455\,(23) &11.945\,(21) & 11.799\,(25)$^{\rm r}$&                    &                 &               &5.3\,(4.2)    &4\,(4.2)      &-16.9\,(6.8)  &-1.4\,(6.8)   &     &     &outFoV                                                                  \\
1634 & 21:28:33.73 &  56:28:39.9$^{\rm r}$&      & 5008                      &                               &                 &  15.81$^{\rm k}$&   14.1$^{\rm n}$&       &                 &10.880\,(23) &10.128\,(19) & 9.881 \,(15)$^{\rm r}$&        K7$^{\rm q}$&                 &               &5.5\,(5.2)    &-2\,(5.2)     &8.1\,(7.2)    &-1.8\,(7.1)   &     &     &outFoV                                                                  \\
1635 & 21:29:03.33 &  57:53:27.2$^{\rm m}$&      & 5009                      &                               &                 &  17.36$^{\rm k}$&  15.07$^{\rm n}$&       &                 &             &             &                       &                    &                 &               &              &              &              &              &     &     &outFoV                                                                  \\
1636 & 21:29:08.17 &  57:30:24.4$^{\rm r}$&      & 5010                      &                               &  16.43$^{\rm k}$&  15.94$^{\rm k}$&  14.58$^{\rm n}$&       &                 &11.715\,(21) &11.215\,(19) & 11.005\,(19)$^{\rm r}$&                    &                 &               &6.8\,(4)      &0.5\,(4)      &12.2\,(6.3)   &17\,(6.3)     &     &     &outFoV                                                                  \\
1637 & 21:29:06.58 &  58:33:35.2$^{\rm r}$&      & 5011                      &                               &                 &   17.4$^{\rm k}$&  15.97$^{\rm n}$&       &                 &12.348\,(20) &11.459\,(18) & 11.255\,(17)$^{\rm r}$&                    &                 &               &1.3\,(3.9)    &1.4\,(3.9)    &-5.3\,(7.7)   &-0.2\,(7.8)   &     &     &outFoV                                                                  \\
1638 & 21:29:36.54 &  58:30:07.2$^{\rm r}$&      & 5012                      &                               &                 &  16.21$^{\rm k}$&  14.63$^{\rm n}$&       &                 &11.648\,(22) &10.998\,(28) & 10.802\,(21)$^{\rm r}$&                    &                 &               &-2.9\,(3.8)   &-3.2\,(3.8)   &-10.5\,(7.3)  &-6.5\,(7.2)   &     &     &outFoV                                                                  \\
1639 & 21:29:47.51 &  57:26:52.0$^{\rm r}$&      & 5013                      &                               &  16.80$^{\rm k}$&   15.6$^{\rm k}$&  13.92$^{\rm n}$&       &                 &10.992\,(23) &10.316\,(28) & 10.141\,(22)$^{\rm r}$&                    &                 &               &3.2\,(4.1)    &12.2\,(4.1)   &1.7\,(7.2)    &5.3\,(7.2)    &     &     &outFoV                                                                  \\
1640 & 21:29:57.92 &  56:26:57.9$^{\rm r}$&      & 5014                      &                               &                 &                 &   15.4$^{\rm n}$&       &                 &10.317\,(22) &9.271 \,(26) & 8.909 \,(22)$^{\rm r}$&                    &                 &               &2.4\,(5.1)    &-5.8\,(5.1)   &-6.5\,(7.1)   &-6.2\,(7.5)   &     &     &outFoV                                                                  \\
1641 & 21:30:15.27 &  56:58:48.0$^{\rm r}$&      & 5015                      &                               &                 &  17.29$^{\rm k}$&  15.23$^{\rm n}$&       &                 &10.968\,(22) &10.093\,(28) & 9.750 \,(22)$^{\rm r}$&        M3$^{\rm q}$&                 &               &1.5\,(5.1)    &-1.1\,(5.1)   &1.9\,(6.8)    &-23\,(6.8)    &     &     &outFoV                                                                  \\
1642 & 21:04:17.62 &  57:03:06.0$^{\rm m}$&      & 5016                      &                               &  15.95$^{\rm k}$&  15.54$^{\rm k}$&  14.47$^{\rm n}$&       &                 &             &             &                       &                    &                 &               &0.7\,(8)      &0\,(8)        &              &              &     &     &[m] wrong, outFoV                                       \\
1643 & 21:30:27.20 &  56:58:40.3$^{\rm r}$&      & 5017                      &                               &  16.61$^{\rm k}$&   16.1$^{\rm k}$&  14.74$^{\rm n}$&       &                 &12.815\,(28) &12.433\,(38) & 12.257\,(34)$^{\rm r}$&        F4$^{\rm q}$&                 &               &-61.9\,(19.1) &-40.6\,(19.1) &              &              &     &     &outFoV                                                                  \\
1644 & 21:30:29.24 &  58:31:07.7$^{\rm r}$&      & 5018                      &                               &                 &     17$^{\rm k}$&  15.16$^{\rm n}$&       &                 &10.680\,(23) &9.617 \,(28) & 9.274 \,(21)$^{\rm r}$&                    &                 &               &1\,(4.7)      &-4.1\,(4.7)   &-3.2\,(7.3)   &-6.5\,(7.4)   &     &     &outFoV                                                                  \\
1645 & 21:30:46.23 &  59:04:09.8$^{\rm r}$&      & 5019                      &                               &                 &  16.12$^{\rm k}$&  15.47$^{\rm n}$&       &                 &11.906\,(22) &11.207\,(28) & 10.965\,(23)$^{\rm r}$&                    &                 &               &-1.5\,(3.8)   &-5.8\,(3.8)   &-7.6\,(7.6)   &-2.6\,(7.6)   &     &     &outFoV                                                                  \\
1646 & 21:30:50.93 &  57:21:43.2$^{\rm r}$&      & 5020                      &                               &  16.11$^{\rm k}$&   15.5$^{\rm k}$&  14.82$^{\rm n}$&       &                 &12.747\,(25) &12.471\,(28) & 12.326\,(24)$^{\rm r}$&                    &                 &               &1.1\,(4.1)    &7.1\,(4.1)    &1.9\,(6.8)    &6.1\,(6.8)    &     &     &outFoV                                                                  \\
1647 & 21:31:02.33 &  59:06:05.0$^{\rm r}$&      & 5021                      &                               &                 &  16.78$^{\rm k}$&  15.78$^{\rm n}$&       &                 &13.228\,(26) &12.736\,(33) & 12.618\,(29)$^{\rm r}$&                    &                 &               &-6.6\,(3.8)   &-10.1\,(3.8)  &-13.5\,(8.1)  &-7.9\,(10)    &     &     &outFoV                                                                  \\
1648 & 21:31:26.37 &  59:08:10.4$^{\rm r}$&      & 5022                      &                               &                 &   16.2$^{\rm k}$&  15.27$^{\rm n}$&       &                 &11.812\,(24) &11.046\,(32) & 10.773\,(23)$^{\rm r}$&                    &                 &               &-0.3\,(3.8)   &-31.9\,(3.8)  &-18.1\,(7.7)  &-34.8\,(7.2)  &     &     &outFoV                                                                  \\
1649 & 21:31:34.39 &  57:12:52.1$^{\rm m}$&      & 5023                      &                               &  15.20$^{\rm k}$&   14.3$^{\rm k}$&  13.14$^{\rm n}$&       &                 &             &             &                       &                    &                 &               &              &              &              &              &     &     &outFoV                                                                  \\
1650 & 21:31:35.83 &  56:57:47.7$^{\rm r}$&      & 5024                      &                               &  16.63$^{\rm k}$&  16.13$^{\rm k}$&   14.7$^{\rm n}$&       &                 &12.112\,(21) &11.676\,(26) & 11.495\,(22)$^{\rm r}$&                    &                 &               &3.2\,(4.1)    &0.2\,(4.1)    &2.9\,(6.8)    &10.7\,(6.8)   &     &     &outFoV                                                                  \\
1651 & 21:31:43.32 &  56:37:55.9$^{\rm r}$&      & 5025                      &                               &                 &   16.5$^{\rm k}$&  14.45$^{\rm n}$&       &                 &10.406\,(21) &9.495 \,(31) & 9.156 \,(20)$^{\rm r}$&        M0$^{\rm q}$&                 &               &5.5\,(5.1)    &-5.5\,(5.1)   &53.7\,(6.9)   &-37\,(6.9)    &     &     &outFoV                                                                  \\
1652 & 21:31:37.63 &  59:08:29.3$^{\rm r}$&      & 5026                      &                               &                 &                 &     17$^{\rm n}$&       &                 &13.143\,(24) &12.515\,(32) & 12.249\,(25)$^{\rm r}$&                    &                 &               &-2.6\,(3.8)   &-3.3\,(3.8)   &-3.7\,(8.2)   &-13.5\,(8.3)  &     &     &outFoV                                                                  \\
1653 & 21:31:46.08 &  56:37:53.5$^{\rm r}$&      & 5027                      &                               &                 &  16.68$^{\rm k}$&  14.94$^{\rm n}$&       &                 &11.269\,(23) &10.557\,(30) & 10.284\,(20)$^{\rm r}$&                    &                 &               &-0.9\,(4.1)   &-0.2\,(4.1)   &-5.4\,(6.8)   &2.3\,(6.8)    &     &     &outFoV                                                                  \\
1654 & 21:32:02.65 &  56:36:18.9$^{\rm r}$&      & 5028                      &                               &                 &  16.56$^{\rm k}$&  14.59$^{\rm n}$&       &                 &10.928\,(23) &10.132\,(31) & 9.830 \,(22)$^{\rm r}$&                    &                 &               &1.5\,(5.1)    &-9.3\,(5.1)   &9.5\,(7)      &-43.5\,(6.9)  &     &     &outFoV                                                                  \\
1655 & 21:32:28.79 &  57:43:31.8$^{\rm r}$&      & 5029                      &                               &  16.22$^{\rm k}$&  16.79$^{\rm k}$&  15.74$^{\rm n}$&       &                 &13.165\,(32) &12.709\,(42) & 12.485\,(33)$^{\rm r}$&                    &                 &               &-11.7\,(4)    &-9.1\,(4)     &-56.9\,(7.1)  &-23\,(9.1)    &     &     &outFoV                                                                  \\
1656 & 21:33:07.73 &  56:16:19.1$^{\rm m}$&      & 5030                      &                               &                 &  17.18$^{\rm k}$&  15.71$^{\rm n}$&       &                 &             &             &                       &                    &                 &               &              &              &              &              &     &     &outFoV                                                                  \\
1657 & 21:33:11.98 &  56:19:56.2$^{\rm r}$&      & 5031                      &                               &  16.03$^{\rm k}$&  15.02$^{\rm k}$&  14.52$^{\rm n}$&       &                 &11.724\,(25) &11.436\,(30) & 11.300\,(24)$^{\rm r}$&                    &                 &               &1.9\,(4.1)    &2.7\,(4.1)    &4\,(6.9)      &20.1\,(6.9)   &     &     &outFoV                                                                  \\
1658 & 21:33:19.74 &  56:25:34.8$^{\rm r}$&      & 5032                      &                               &                 &                 &  14.91$^{\rm n}$&       &                 &16.328\,(124)&15.579\,(158)&15.456\,(196)$^{\rm r}$&                    &                 &               &              &              &              &              &     &     &outFoV                                                                  \\
1659 & 21:33:29.70 &  57:56:28.2$^{\rm r}$&      & 5033                      &                               &  16.05$^{\rm k}$&  15.75$^{\rm k}$&  14.52$^{\rm n}$&       &                 &12.368\,(24) &11.986\,(28) & 11.862\,(23)$^{\rm r}$&                    &                 &               &-1\,(3.8)     &0.4\,(3.8)    &-3.1\,(7.3)   &-0.1\,(7.3)   &     &     &                                                                                      \\
1660 & 21:33:33.90 &  56:22:55.9$^{\rm r}$&      & 5034                      &                               &                 &                 &     17$^{\rm n}$&       &                 &10.003\,(23) &8.768 \,(28) & 8.314 \,(21)$^{\rm r}$&                    &                 &               &1.4\,(5.1)    &-6.4\,(5.1)   &              &              &     &     &outFoV                                                                  \\
1661 & 21:33:42.90 &  56:48:27.8$^{\rm r}$&      & 5035                      &                               &                 &                 &  15.19$^{\rm n}$&       &                 &11.352\,(23) &10.529\,(28) & 10.242\,(21)$^{\rm r}$&                    &                 &               &-3.9\,(4.1)   &-1.4\,(4.1)   &-12.6\,(7)    &2.4\,(6.9)    &     &     &                                                                                      \\
1662 & 21:33:47.96 &  56:51:11.9$^{\rm r}$&      & 5036                      &                               &                 &  16.34$^{\rm k}$&  14.89$^{\rm n}$&       &                 &12.907\,(26) &12.374\,(32) & 12.292\,(25)$^{\rm r}$&                    &                 &               &1.3\,(4.1)    &-5.4\,(4.1)   &5.3\,(6.8)    &-1.5\,(6.9)   &     &     &                                                                                      \\
1663 & 21:34:03.33 &  58:05:17.8$^{\rm r}$&      & 5037                      &                               &  15.99$^{\rm k}$&  15.59$^{\rm k}$&  14.87$^{\rm n}$&       &                 &12.809\,(26) &12.493\,(35) & 12.304\,(28)$^{\rm r}$&                    &                 &               &-7.3\,(3.8)   &-5.5\,(3.8)   &-19.7\,(8.2)  &-14.9\,(8.3)  &     &     &                                                                                      \\
1664 & 21:34:13.33 &  57:00:48.6$^{\rm r}$&      & 5038                      &                               &  16.05$^{\rm k}$&  15.87$^{\rm k}$&  14.83$^{\rm n}$&       &                 &12.638\,(25) &12.238\,(34) & 12.099\,(26)$^{\rm r}$&                    &                 &               &2.9\,(4.1)    &2.6\,(4.1)    &4.2\,(6.8)    &14.6\,(6.8)   &     &     &                                                                                      \\
1665 & 21:34:18.21 &  57:14:32.6$^{\rm r}$&      & 5039                      &                               &  14.66$^{\rm k}$&  14.25$^{\rm k}$&  14.14$^{\rm n}$&       &                 &12.670\,(52) &12.469\,(35) & 12.412\,(31)$^{\rm r}$&                    &                 &               &-149.5\,(13.3)&48.8\,(13.3)  &              &              &     &     &near 1072                                                                          \\
1666 & 21:34:25.33 &  56:48:58.7$^{\rm r}$&      & 5040                      &                               &                 &   16.8$^{\rm k}$&  15.12$^{\rm n}$&       &                 &11.368\,(23) &10.520\,(27) & 10.320\,(24)$^{\rm r}$&                    &                 &               &-1.5\,(4.1)   &-1.9\,(4.1)   &-7.6\,(6.9)   &4.1\,(6.8)    &     &     &                                                                                      \\
1667 & 21:34:36.16 &  56:32:54.5$^{\rm r}$&      & 5041                      &                               &                 &  14.76$^{\rm k}$&  13.67$^{\rm n}$&       &                 &12.205\,(28) &11.819\,(31) & 11.713\,(25)$^{\rm r}$&        F8$^{\rm q}$&                 &               &25.4\,(4.1)   &27\,(4.1)     &              &              &     &     &outFoV                                                                  \\
1668 & 21:34:34.81 &  57:14:16.4$^{\rm r}$&      & 5042                      &                               &                 &                 &     17$^{\rm n}$&       &                 &12.502\,(24) &11.788\,(31) & 11.578\,(26)$^{\rm r}$&                    &                 &               &-6.2\,(4.1)   &-1.4\,(4.1)   &-17.1\,(7)    &5.2\,(7.1)    &     &     &                                                                                      \\
1669 & 21:34:39.84 &  57:35:54.5$^{\rm r}$&      & 5043                      &                               &                 &                 &   16.5$^{\rm n}$&       &                 &11.706\,(22) &10.751\,(26) & 10.447\,(21)$^{\rm r}$&                    &                 &               &2.3\,(3.8)    &-1.6\,(3.8)   &-11.1\,(7.1)  &5.5\,(7.1)    &     &     &                                                                                      \\
1670 & 21:34:47.44 &  56:34:30.1$^{\rm r}$&      & 5044                      &                               &  16.75$^{\rm k}$&  15.32$^{\rm k}$&  14.26$^{\rm n}$&       &                 &11.747\,(24) &11.373\,(28) & 11.226\,(22)$^{\rm r}$&                    &                 &               &-5.8\,(4.1)   &-4.8\,(4.1)   &-3.7\,(6.8)   &2\,(6.8)      &     &     &Simbad wrong, outFoV                                                    \\
1671 & 21:35:03.30 &  56:38:11.6$^{\rm r}$&      & 5045                      &                               &  16.31$^{\rm k}$&   15.9$^{\rm k}$&  14.04$^{\rm n}$&       &                 &12.131\,(21) &11.679\,(29) & 11.528\,(23)$^{\rm r}$&                    &                 &               &-5\,(4.1)     &1.3\,(4.1)    &-13.9\,(6.8)  &6.4\,(6.8)    &     &     &outFoV                                                                  \\
1672 & 21:35:04.46 &  56:12:11.7$^{\rm m}$&      & 5046                      &                               &                 &  16.04$^{\rm k}$&  15.47$^{\rm n}$&       &                 &             &             &                       &                    &                 &               &              &              &              &              &     &     &outFoV                                                                  \\
1673 & 21:35:19.02 &  56:34:38.6$^{\rm r}$&      & 5047                      &                               &  16.18$^{\rm k}$&  15.42$^{\rm k}$&  14.21$^{\rm n}$&       &                 &11.867\,(24) &11.422\,(33) & 11.282\,(26)$^{\rm r}$&                    &                 &               &17.9\,(4.1)   &0.9\,(4.1)    &58.8\,(6.8)   &16.1\,(6.8)   &     &     &outFoV                                                                  \\
1674 & 21:35:16.75 &  57:45:10.1$^{\rm r}$&      & 5048                      &                               &                 &                 &   15.1$^{\rm n}$&       &                 &12.537\,(24) &11.691\,(29) & 11.413\,(23)$^{\rm r}$&                    &                 &               &-5.9\,(4)     &-6.6\,(4)     &-0.8\,(8.8)   &2.2\,(8.6)    &     &     &                                                                                      \\
1675 & 21:35:21.63 &  56:18:20.4$^{\rm r}$&      & 5049                      &                               &  16.92$^{\rm k}$&  16.41$^{\rm k}$&  15.31$^{\rm n}$&       &                 &12.885\,(48) &12.472\,(68) & 12.285\,(30)$^{\rm r}$&                    &                 &               &-16\,(4.1)    &16.4\,(4.1)   &-59.1\,(7.1)  &71.5\,(7)     &     &     &outFoV                                                                  \\
1676 & 21:35:21.56 &  57:45:14.4$^{\rm r}$&      & 5050                      &                               &                 &                 &   15.1$^{\rm n}$&       &                 &12.000\,(24) &11.142\,(28) & 10.827\,(21)$^{\rm r}$&                    &                 &               &-7.3\,(3.8)   &-3\,(3.8)     &-5.6\,(7.7)   &-12.8\,(8.7)  &     &     &                                                                                      \\
1677 & 21:35:32.31 &  56:23:21.4$^{\rm r}$&      & 5051                      &                               &                 &  17.04$^{\rm k}$&  15.12$^{\rm n}$&       &                 &11.241\,(24) &10.442\,(30) & 10.187\,(20)$^{\rm r}$&                    &                 &               &6.5\,(4.1)    &-1.7\,(4.1)   &-2.5\,(6.9)   &3.4\,(6.9)    &     &     &outFoV                                                                  \\
1678 & 21:35:33.34 &  56:50:51.5$^{\rm r}$&      & 5052                      &                               &  16.27$^{\rm k}$&  15.73$^{\rm k}$&  14.09$^{\rm n}$&       &                 &12.098\,(32) &11.669\,(41) & 11.519\,(24)$^{\rm r}$&                    &                 &               &27.3\,(4.1)   &8.7\,(4.1)    &              &              &     &     &                                                                                      \\
1679 & 21:35:39.15 &  57:02:04.3$^{\rm r}$&      & 5053                      &                               &  16.06$^{\rm k}$&  15.37$^{\rm k}$&  14.08$^{\rm n}$&       &                 &12.560\,(27) &12.216\,(32) & 12.064\,(25)$^{\rm r}$&                    &                 &               &200.5\,(7.6)  &-268.8\,(7.6) &-0.2\,(6.8)   &-10.6\,(6.8)  &     &     &                                                                                      \\
1680 & 21:35:42.03 &  56:07:07.4$^{\rm r}$&      & 5054                      &                               &                 &   16.5$^{\rm k}$&  15.97$^{\rm n}$&       &                 &11.251\,(24) &10.377\,(29) & 10.040\,(21)$^{\rm r}$&                    &                 &               &1.6\,(4.1)    &-5.5\,(4.1)   &-6\,(7.1)     &8.5\,(7.2)    &     &     &outFoV                                                                  \\
1681 & 21:35:42.67 &  56:49:19.1$^{\rm r}$&      & 5055                      &                               &  16.03$^{\rm k}$&  15.83$^{\rm k}$&  14.48$^{\rm n}$&       &                 &12.403\,()   &12.004\,(43) &   11.809\,()$^{\rm r}$&                    &                 &               &-2.2\,(4.1)   &2.5\,(4.1)    &              &              &     &     &                                                                                      \\
1682 & 21:35:48.64 &  57:20:28.3$^{\rm r}$&      & 5056                      &                               &  16.13$^{\rm k}$&  16.32$^{\rm f}$&  15.16$^{\rm e}$&       &                 &13.341\,(24) &12.953\,(32) & 12.831\,(29)$^{\rm r}$&        F9$^{\rm e}$&                 &  1.9$^{\rm e}$&-5.1\,(4.1)   &2.7\,(4.1)    &-3.7\,(6.9)   &2\,(7)        &     &     &                                                                                      \\
1683 & 21:35:51.43 &  57:15:19.7$^{\rm r}$&      & 5057                      &                               &                 &                 &   16.5$^{\rm n}$&       &                 &13.800\,(31) &13.395\,(39) & 13.268\,(42)$^{\rm r}$&                    &                 &               &-3\,(4.1)     &1.8\,(4.1)    &-2.4\,(6.8)   &25.7\,(7.1)   &     &     &                                                                                      \\
1684 & 21:35:56.56 &  57:05:04.6$^{\rm r}$&      & 5059                      &                               &  16.31$^{\rm k}$&   15.4$^{\rm k}$&  14.25$^{\rm n}$&       &                 &12.511\,(24) &12.047\,(29) & 11.914\,(23)$^{\rm r}$&                    &                 &               &11.4\,(4.1)   &2\,(4.1)      &4.6\,(6.8)    &35.7\,(6.8)   &     &     &                                                                                      \\
1685 & 21:35:58.75 &  57:03:14.5$^{\rm m}$&      & 5060                      &                               &  15.86$^{\rm k}$&  15.41$^{\rm k}$&  14.48$^{\rm n}$&       &                 &             &             &                       &                    &                 &               &              &              &              &              &     &     &no star                                                                               \\
1686 & 21:35:58.74 &  59:11:07.5$^{\rm r}$&      & 5061                      &                               &  17.30$^{\rm k}$&  16.31$^{\rm k}$&  15.47$^{\rm n}$&       &                 &12.985\,(24) &12.585\,(29) & 12.450\,(28)$^{\rm r}$&                    &                 &               &-3.4\,(3.8)   &-5.3\,(3.8)   &-14.9\,(7.8)  &-3.5\,(7.6)   &     &     &outFoV                                                                  \\
1687 & 21:36:13.29 &  56:15:47.2$^{\rm r}$&      & 5062                      &                               &                 &  16.16$^{\rm k}$&  15.42$^{\rm n}$&       &                 &12.074\,()   &11.465\,()   & 11.262\,(27)$^{\rm r}$&                    &                 &               &0.3\,(4.1)    &0.5\,(4.1)    &2.5\,(6.2)    &18.8\,(6.8)   &     &     &outFoV                                                                  \\
1688 & 21:36:13.90 &  59:04:37.2$^{\rm r}$&      & 5063                      &                               &                 &  15.97$^{\rm k}$&  14.84$^{\rm n}$&       &                 &12.845\,(45) &12.469\,(51) & 12.339\,(45)$^{\rm r}$&                    &                 &               &-24.5\,(3.8)  &1.2\,(3.8)    &              &              &     &     &outFoV                                                                  \\
1689 & 21:36:17.27 &  59:06:52.4$^{\rm r}$&      & 5064                      &                               &                 &                 &  15.58$^{\rm n}$&       &                 &11.524\,(24) &10.562\,(27) & 10.284\,(21)$^{\rm r}$&                    &                 &               &1\,(3.8)      &-3.3\,(3.8)   &-6.8\,(7.6)   &0.5\,(8.4)    &     &     &outFoV                                                                  \\
1690 & 21:36:19.58 &  58:48:48.4$^{\rm r}$&      & 5065                      &                               &                 &   17.1$^{\rm k}$&  15.89$^{\rm n}$&       &                 &11.924\,(23) &11.052\,(27) & 10.800\,(20)$^{\rm r}$&                    &                 &               &-0.1\,(3.8)   &-21.9\,(3.8)  &-0.3\,(7.5)   &-16.2\,(7.6)  &     &     &outFoV                                                                  \\
1691 & 21:36:33.23 &  59:01:02.1$^{\rm r}$&      & 5067                      &                               &                 &   15.9$^{\rm k}$&  15.16$^{\rm n}$&       &                 &12.744\,(27) &12.439\,(32) & 12.317\,(26)$^{\rm r}$&                    &                 &               &-1.2\,(3.8)   &-3\,(3.8)     &-4\,(7.3)     &-3.3\,(7.4)   &     &     &outFoV                                                                  \\
1692 & 21:36:51.82 &  57:59:05.7$^{\rm r}$&      & 5069                      &                               &                 &                 &   16.3$^{\rm n}$&       &                 &12.111\,(26) &11.260\,(29) & 11.016\,(22)$^{\rm r}$&                    &                 &               &6.2\,(3.8)    &1.4\,(3.8)    &1.9\,(7.8)    &14.8\,(7.7)   &     &     &                                                                                      \\
1693 & 21:37:00.91 &  58:47:30.3$^{\rm r}$&      & 5070                      &                               &  16.38$^{\rm k}$&  15.21$^{\rm k}$&   14.1$^{\rm n}$&       &                 &12.105\,(39) &11.721\,(46) & 11.578\,(38)$^{\rm r}$&                    &                 &               &-33.3\,(3.8)  &-1.5\,(3.8)   &              &              &     &     &outFoV                                                                  \\
1694 & 21:37:07.08 &  58:07:18.3$^{\rm r}$&      & 5071                      &                               &                 &   16.7$^{\rm k}$&  15.05$^{\rm n}$&       &                 &11.330\,(19) &10.535\,(28) & 10.268\,(26)$^{\rm r}$&                    &                 &               &-3.4\,(3.8)   &-9.6\,(3.8)   &-11.1\,(10.4) &8.4\,(9.7)    &     &     &                                                                                      \\
1695 & 21:37:54.88 &  56:38:38.3$^{\rm r}$&      & 5072                      &                               &  14.53$^{\rm k}$&  14.78$^{\rm k}$&  14.08$^{\rm n}$&       &                 &12.521\,(27) &12.277\,(29) & 12.175\,(18)$^{\rm r}$&                    &                 &               &-2.8\,(4)     &-0.1\,(4)     &-7.7\,(6.8)   &4.1\,(6.8)    &     &     &outFoV                                                                  \\
1696 & 21:37:59.28 &  58:41:24.1$^{\rm r}$&      & 5074                      &                               &                 &                 &   16.4$^{\rm n}$&       &                 &11.923\,(26) &11.037\,(30) & 10.697\,(18)$^{\rm r}$&                    &                 &               &-8\,(3.8)     &-9.5\,(3.8)   &-5\,(7.5)     &-5.8\,(7.5)   &     &     &outFoV                                                                  \\
1697 & 21:38:11.68 &  59:10:48.7$^{\rm r}$&      & 5076                      &                               &                 &                 &   16.5$^{\rm n}$&       &                 &11.897\,(27) &10.853\,(27) & 10.572\,(20)$^{\rm r}$&                    &                 &               &-2.9\,(3.8)   &-6.8\,(3.8)   &-4.9\,(8.1)   &-2.6\,(9.9)   &     &     &outFoV, [m] wrong                                                     \\
1698 & 21:38:26.03 &  56:10:27.2$^{\rm r}$&      & 5077                      &                               &                 &                 &   16.3$^{\rm n}$&       &                 &13.161\,(41) &12.752\,(51) & 12.565\,(39)$^{\rm r}$&                    &                 &               &9.2\,(4)      &12\,(4)       &              &              &     &     &outFoV                                                                  \\
1699 & 21:38:22.13 &  59:10:40.2$^{\rm r}$&      & 5078                      &                               &  16.96$^{\rm k}$&  16.04$^{\rm k}$&  14.58$^{\rm n}$&       &                 &11.951\,(26) &11.442\,(28) & 11.306\,(24)$^{\rm r}$&                    &                 &               &-5.3\,(3.8)   &-9.1\,(3.8)   &-7.7\,(7.1)   &-7.9\,(7.2)   &     &     &outFoV                                                                  \\
1700 & 21:38:40.21 &  58:11:10.4$^{\rm r}$&      & 5079                      &                               &                 &                 &  15.61$^{\rm n}$&       &                 &11.241\,(26) &10.247\,(32) & 9.932 \,(19)$^{\rm r}$&                    &                 &               &-2.1\,(4.7)   &-5.9\,(4.7)   &-9.4\,(8.1)   &-14.9\,(7.4)  &     &     &                                                                                      \\
1701 & 21:38:42.32 &  57:30:27.8$^{\rm r}$&      & 5080                      &                               &                 &                 &   15.5$^{\rm n}$&       &                 &9.748 \,(23) &8.087 \,(40) & 6.960 \,(18)$^{\rm r}$&                    &                 &               &-6.2\,(4.9)   &0\,(4.9)      &-7.7\,(6.4)   &2.5\,(6.5)    &     &     &                                                                                      \\
1702 & 21:39:07.73 &  55:57:58.8$^{\rm r}$&      & 5081                      &                               &                 &  15.54$^{\rm k}$&  14.49$^{\rm n}$&       &                 &13.222\,()   &12.950\,(41) & 12.804\,(45)$^{\rm r}$&                    &                 &               &6.3\,(4)      &-35.5\,(4)    &              &              &     &     &outFoV                                                                  \\
1703 & 21:39:13.28 &  56:05:24.4$^{\rm r}$&      & 5082                      &                               &  15.06$^{\rm k}$&   15.3$^{\rm k}$&  14.49$^{\rm n}$&       &                 &12.999\,(26) &12.895\,(33) & 12.711\,(36)$^{\rm r}$&                    &                 &               &-0.4\,(4)     &2.3\,(4)      &-3.7\,(6.8)   &0.7\,(6.8)    &     &     &outFoV                                                                  \\
1704 & 21:39:40.12 &  58:15:01.7$^{\rm r}$&      & 5088                      &                               &                 &  16.85$^{\rm k}$&  15.61$^{\rm n}$&       &                 &12.305\,()   &11.456\,()   &   11.180\,()$^{\rm r}$&                    &                 &               &2.8\,(3.8)    &-6.2\,(3.8)   &-12.6\,(7.8)  &10.7\,(8.3)   &     &     &                                                                                      \\
1705 & 21:40:00.16 &  58:55:11.5$^{\rm r}$&      & 5090                      &                               &                 &  16.85$^{\rm k}$&  15.05$^{\rm n}$&       &                 &11.232\,(24) &10.392\,(29) & 10.097\,(21)$^{\rm r}$&                    &                 &               &-3.8\,(3.8)   &-10.1\,(3.8)  &-8.8\,(7.3)   &-13.2\,(7.3)  &     &     &outFoV                                                                  \\
1706 & 21:40:27.41 &  57:31:45.3$^{\rm r}$&      & 5094                      &                               &                 &   16.6$^{\rm k}$&  14.98$^{\rm n}$&       &                 &10.937\,(24) &9.924 \,(28) & 9.619 \,(21)$^{\rm r}$&        M1$^{\rm q}$&                 &               &-0.6\,(4.9)   &-9.2\,(4.9)   &-7.7\,(7.3)   &-13.2\,(7.4)  &     &     &                                                                                      \\
1707 & 21:40:47.53 &  56:40:45.2$^{\rm r}$&      & 5095                      &                               &                 &  16.74$^{\rm k}$&  15.16$^{\rm n}$&       &                 &11.742\,(26) &10.916\,(28) & 10.684\,(19)$^{\rm r}$&                    &                 &               &-3\,(4.1)     &1.5\,(4.1)    &-7.8\,(6.9)   &9\,(6.9)      &     &     &outFoV                                                                  \\
1708 & 21:40:48.51 &  57:40:01.2$^{\rm r}$&      & 5096                      &                               &  16.22$^{\rm k}$&  16.46$^{\rm k}$&  15.04$^{\rm n}$&       &                 &13.020\,(24) &12.579\,(35) & 12.499\,(30)$^{\rm r}$&                    &                 &               &33\,(3.9)     &-71.6\,(3.9)  &14.3\,(7.4)   &-47.8\,(7.8)  &     &     &                                                                                      \\
1709 & 21:41:02.55 &  57:59:32.3$^{\rm r}$&      & 5098                      &                               &                 &  16.55$^{\rm k}$&  15.34$^{\rm n}$&       &                 &11.883\,(21) &11.068\,(31) & 10.846\,(22)$^{\rm r}$&       G2:$^{\rm q}$&                 &               &-1.3\,(3.8)   &-10.7\,(3.8)  &19.3\,(8.2)   &-85.3\,(8.8)  &     &     &[m] wrong                                                                           \\
1710 & 21:41:04.08 &  58:59:34.8$^{\rm r}$&      & 5099                      &                               &                 &  15.73$^{\rm k}$&   14.5$^{\rm n}$&       &                 &11.741\,(21) &11.076\,(32) & 10.837\,(20)$^{\rm r}$&                    &                 &               &6.4\,(3.8)    &-3.2\,(3.8)   &38.5\,(7.3)   &-7.6\,(7.4)   &     &     &outFoV                                                                  \\
1711 & 21:41:21.52 &  59:20:55.2$^{\rm r}$&      & 5101                      &                               &                 &  15.13$^{\rm k}$&  14.49$^{\rm n}$&       &                 &12.797\,(25) &12.269\,(29) & 12.201\,(23)$^{\rm r}$&                    &                 &               &7.3\,(3.8)    &-13.2\,(3.8)  &9.7\,(8)      &-11\,(8)      &     &     &outFoV                                                                  \\
1712 & 21:41:24.48 &  58:12:56.4$^{\rm r}$&      & 5102                      &                               &                 &   16.6$^{\rm k}$&  15.54$^{\rm n}$&       &                 &12.945\,(24) &12.490\,(31) & 12.299\,(24)$^{\rm r}$&                    &                 &               &-3.7\,(3.8)   &-5\,(3.8)     &-8\,(7.6)     &-11.4\,(7.4)  &     &     &                                                                                      \\
1713 & 21:41:39.56 &  58:13:21.9$^{\rm r}$&      & 5103                      &                               &                 &  16.86$^{\rm k}$&  14.67$^{\rm n}$&       &                 &10.232\,(26) &9.244 \,(32) & 8.865 \,(21)$^{\rm r}$&        G9$^{\rm q}$&                 &               &6.3\,(4.8)    &-5.1\,(4.8)   &9.5\,(6.8)    &-22.2\,(6.8)  &     &     &                                                                                      \\
1714 & 21:41:39.80 &  58:11:50.3$^{\rm r}$&      & 5104                      &                               &  14.71$^{\rm k}$&  14.99$^{\rm k}$&  13.81$^{\rm n}$&       &                 &11.726\,(24) &11.300\,(32) & 11.163\,(23)$^{\rm r}$&        G0$^{\rm q}$&                 &               &-12.2\,(3.8)  &9.3\,(3.8)    &-3.8\,(7.2)   &2.4\,(7.2)    &     &     &                                                                                      \\
1715 & 21:41:55.21 &  58:09:27.4$^{\rm r}$&      & 5105                      &                               &                 &     17$^{\rm k}$&     16$^{\rm n}$&       &                 &13.587\,(32) &13.135\,(35) & 12.958\,(32)$^{\rm r}$&                    &                 &               &-1.3\,(3.8)   &-12.7\,(3.8)  &12.2\,(7.9)   &-27.2\,(7.9)  &     &     &                                                                                      \\
1716 & 21:42:04.06 &  57:03:59.1$^{\rm r}$&      & 5106                      &                               &                 &  16.86$^{\rm k}$&  14.77$^{\rm n}$&       &                 &11.180\,(26) &10.408\,(32) & 10.100\,(22)$^{\rm r}$&                    &                 &               &-0.7\,(4.1)   &-7.8\,(4.1)   &-2.9\,(6.9)   &5.6\,(6.9)    &     &     &                                                                                      \\
1717 & 21:42:11.63 &  57:37:08.5$^{\rm r}$&      & 5108                      &                               &                 &                 &     15$^{\rm n}$&       &                 &11.127\,(24) &10.121\,(30) & 9.757 \,(21)$^{\rm r}$&                    &                 &               &-4.2\,(4.7)   &-3.4\,(4.7)   &-6.9\,(8.3)   &-13.4\,(7.7)  &     &     &                                                                                      \\
1718 & 21:42:15.33 &  57:03:51.4$^{\rm r}$&      & 5109                      &                               &  16.13$^{\rm k}$&     16$^{\rm k}$&  14.95$^{\rm n}$&       &                 &13.213\,(29) &12.932\,(32) & 12.828\,(34)$^{\rm r}$&                    &                 &               &-9.9\,(4.1)   &-6.1\,(4.1)   &-19.3\,(7)    &5.1\,(7)      &     &     &                                                                                      \\
1719 & 21:42:16.06 &  58:01:12.9$^{\rm r}$&      & 5110                      &                               &                 &  17.13$^{\rm k}$&  15.36$^{\rm n}$&       &                 &11.613\,(26) &10.778\,(32) & 10.504\,(21)$^{\rm r}$&        K0$^{\rm q}$&                 &               &6.1\,(3.8)    &2.9\,(3.8)    &48.3\,(8.3)   &15.3\,(8.9)   &     &     &                                                                                      \\
1720 & 21:42:35.03 &  56:28:10.5$^{\rm r}$&      & 5111                      &                               &                 &   17.1$^{\rm k}$&   15.2$^{\rm n}$&       &                 &11.364\,(25) &10.588\,(31) & 10.300\,(20)$^{\rm r}$&                    &                 &               &-7.5\,(4.1)   &-10.6\,(4.1)  &-11.2\,(7)    &-4.5\,(6.9)   &     &     &outFoV                                                                  \\
1721 & 21:42:44.27 &  58:59:50.6$^{\rm r}$&      & 5113                      &                               &                 &  16.05$^{\rm k}$&  15.36$^{\rm n}$&       &                 &13.356\,(39) &12.847\,(47) & 12.636\,(37)$^{\rm r}$&                    &                 &               &-5.2\,(3.8)   &4.1\,(3.8)    &-35.3\,(7.6)  &15.1\,(7.6)   &     &     &outFoV                                                                  \\
1722 & 21:43:01.87 &  56:27:42.1$^{\rm r}$&      & 5114                      &                               &  15.95$^{\rm k}$&  15.17$^{\rm k}$&  14.34$^{\rm n}$&       &                 &12.540\,(29) &12.122\,(30) & 12.044\,(23)$^{\rm r}$&        F4$^{\rm q}$&                 &               &-2.3\,(4.1)   &1\,(4.1)      &-11\,(6.9)    &9.5\,(7.1)    &     &     &outFoV                                                                  \\
1723 & 21:43:04.43 &  56:28:52.0$^{\rm r}$&      & 5115                      &                               &  17.28$^{\rm k}$&  15.82$^{\rm k}$&  14.16$^{\rm n}$&       &                 &11.065\,(27) &10.351\,(30) & 10.146\,(21)$^{\rm r}$&        K3$^{\rm q}$&                 &               &-7.3\,(4.1)   &-11\,(4.1)    &-5.5\,(6.9)   &-9.7\,(6.9)   &     &     &outFoV                                                                  \\
1724 & 21:43:05.90 &  57:39:00.1$^{\rm r}$&      & 5116                      &                               &  16.51$^{\rm k}$&  16.47$^{\rm k}$&  15.55$^{\rm n}$&       &                 &13.342\,(27) &12.835\,(35) & 12.701\,(30)$^{\rm r}$&                    &                 &               &-18.5\,(3.8)  &-15.1\,(3.8)  &-17.9\,(7.5)  &-18.8\,(7.5)  &     &     &                                                                                      \\
1725 & 21:43:20.72 &  56:24:20.0$^{\rm r}$&      & 5117                      &                               &                 &  16.69$^{\rm k}$&  15.36$^{\rm n}$&       &                 &10.984\,(29) &10.033\,(31) & 9.710 \,(21)$^{\rm r}$&                    &                 &               &-6.5\,(5.2)   &-9.5\,(5.2)   &-73\,(7.4)    &-6.2\,(7)     &     &     &outFoV                                                                  \\
1726 & 21:43:21.48 &  57:30:15.7$^{\rm r}$&      & 5118                      &                               &                 &  16.61$^{\rm k}$&  14.79$^{\rm n}$&       &                 &9.059 \,(26) &7.611 \,(36) & 7.010 \,(23)$^{\rm r}$&                    &                 &               &-8.3\,(6.3)   &1.1\,(6.3)    &              &              &     &     &                                                                                      \\
1727 & 21:43:32.55 &  57:24:01.6$^{\rm r}$&      & 5119                      &                               &  16.04$^{\rm k}$&  16.09$^{\rm k}$&  14.95$^{\rm n}$&       &                 &12.416\,(25) &11.797\,(31) & 11.689\,(21)$^{\rm r}$&                    &                 &               &-22.2\,(4.1)  &-20.9\,(4.1)  &-11.8\,(7)    &-3\,(6.9)     &     &     &                                                                                      \\
1728 & 21:43:45.56 &  57:07:56.7$^{\rm r}$&      & 5120                      &                               &  15.72$^{\rm k}$&  15.51$^{\rm k}$&  14.54$^{\rm n}$&       &                 &12.686\,(23) &12.307\,(30) & 12.142\,(21)$^{\rm r}$&                    &                 &               &-9.1\,(4.1)   &-5.3\,(4.1)   &0.7\,(6.9)    &9.2\,(6.9)    &     &     &                                                                                      \\
1729 & 21:43:46.87 &  59:08:13.4$^{\rm r}$&      & 5121                      &                               &  16.71$^{\rm k}$&  16.58$^{\rm k}$&  15.24$^{\rm n}$&       &                 &11.831\,(23) &11.086\,(32) & 10.754\,(21)$^{\rm r}$&                    &                 &               &-7.2\,(3.8)   &-5.4\,(3.8)   &5.9\,(7.6)    &-1.9\,(7.2)   &     &     &outFoV                                                                  \\
1730 & 21:43:51.15 &  57:30:25.9$^{\rm r}$&      & 5122                      &                               &  16.17$^{\rm k}$&  15.94$^{\rm k}$&  15.05$^{\rm n}$&       &                 &13.020\,(24) &12.705\,(32) & 12.549\,(24)$^{\rm r}$&                    &                 &               &-4.1\,(4)     &-1.8\,(4)     &-2.4\,(7.3)   &-20.1\,(7.3)  &     &     &                                                                                      \\
1731 & 21:44:07.97 &  59:11:03.5$^{\rm r}$&      & 5124                      &                               &                 &                 &   16.5$^{\rm n}$&       &                 &12.363\,(36) &11.554\,(46) & 11.219\,(32)$^{\rm r}$&                    &                 &               &10.2\,(3.8)   &4.4\,(3.8)    &40.6\,(8.3)   &19.8\,(8.2)   &     &     &outFoV                                                                  \\
1732 & 21:44:15.38 &  58:25:08.4$^{\rm r}$&      & 5125                      &                               &  16.93$^{\rm k}$&  16.23$^{\rm k}$&   14.9$^{\rm n}$&       &                 &13.349\,(26) &13.006\,(35) & 12.820\,(23)$^{\rm r}$&                    &                 &               &0.9\,(3.8)    &-7.3\,(3.8)   &-0.8\,(7.4)   &-17\,(8.2)    &     &     &outFoV                                                                  \\
1733 & 21:44:25.90 &  57:08:28.0$^{\rm r}$&      & 5126                      &                               &                 &   17.4$^{\rm k}$&   15.8$^{\rm n}$&       &                 &9.721 \,(23) &8.444 \,(24) & 8.012 \,(16)$^{\rm r}$&                    &                 &               &-4.9\,(5.1)   &-4.8\,(5.1)   &33.4\,(7)     &13.5\,(7.1)   &     &     &                                                                                      \\
1734 & 21:44:39.59 &  57:20:00.6$^{\rm r}$&      & 5127                      &                               &  15.91$^{\rm k}$&  15.57$^{\rm k}$&  15.08$^{\rm n}$&       &                 &13.020\,(29) &12.602\,(31) & 12.462\,(28)$^{\rm r}$&                    &                 &               &2.1\,(4.1)    &-3.4\,(4.1)   &22.3\,(7.5)   &11.6\,(7)     &     &     &                                                                                      \\
1735 & 21:44:35.58 &  59:05:25.5$^{\rm r}$&      & 5128                      &                               &                 &   16.2$^{\rm k}$&  15.36$^{\rm n}$&       &                 &12.232\,(27) &11.470\,(32) & 11.235\,(23)$^{\rm r}$&                    &                 &               &-13.5\,(3.8)  &-8.4\,(3.8)   &-11.6\,(7.5)  &-18\,(7.5)    &     &     &outFoV                                                                  \\
1736 & 21:44:52.38 &  58:30:43.5$^{\rm r}$&      & 5129                      &                               &                 &  16.86$^{\rm k}$&   15.3$^{\rm n}$&       &                 &11.643\,(34) &10.815\,(36) & 10.562\,(29)$^{\rm r}$&                    &                 &               &2.2\,(3.8)    &-9.8\,(3.8)   &-2.6\,(7.2)   &-13.3\,(7.3)  &     &     &outFoV                                                                  \\
1737 & 21:45:27.46 &  56:32:25.1$^{\rm r}$&      & 5130                      &                               &  18.24$^{\rm k}$&  17.35$^{\rm k}$&  15.66$^{\rm n}$&       &                 &13.283\,(26) &12.884\,(32) & 12.763\,(28)$^{\rm r}$&                    &                 &               &-5.3\,(4.1)   &-4.2\,(4.1)   &-19.7\,(6.8)  &7.9\,(7)      &     &     &outFoV                                                                  \\
1738 & 21:45:27.22 &  58:45:00.4$^{\rm r}$&      & 5131                      &                               &                 &  17.06$^{\rm k}$&  15.42$^{\rm n}$&       &                 &11.752\,(27) &11.071\,(31) & 10.886\,(22)$^{\rm r}$&                    &                 &               &-4.4\,(3.8)   &-1.4\,(3.8)   &-2.2\,(7.3)   &-5.3\,(7.3)   &     &     &outFoV                                                                  \\
1739 & 21:45:29.65 &  58:45:36.4$^{\rm r}$&      & 5132                      &                               &                 &  16.67$^{\rm k}$&  15.46$^{\rm n}$&       &                 &12.595\,(29) &11.999\,(33) & 11.775\,(24)$^{\rm r}$&                    &                 &               &-2.2\,(3.8)   &0\,(3.8)      &-2.1\,(7.6)   &0.3\,(7.5)    &     &     &outFoV                                                                  \\
1740 & 21:45:37.16 &  56:39:43.2$^{\rm r}$&      & 5133                      &                               &  16.34$^{\rm k}$&  15.51$^{\rm k}$&  14.73$^{\rm n}$&       &                 &12.777\,(26) &12.407\,(29) & 12.273\,(19)$^{\rm r}$&                    &                 &               &-5.9\,(4.1)   &-8\,(4.1)     &-24.6\,(6.5)  &-9.6\,(6.6)   &     &     &outFoV                                                                  \\
1741 & 21:45:39.18 &  56:39:28.5$^{\rm r}$&      & 5134                      &                               &  16.81$^{\rm k}$&  16.02$^{\rm k}$&  14.72$^{\rm n}$&       &                 &12.768\,(24) &12.213\,(32) & 12.133\,(21)$^{\rm r}$&                    &                 &               &-2.1\,(4.1)   &-39.5\,(4.1)  &-15.7\,(7.1)  &-27.2\,(6.6)  &     &     &outFoV                                                                  \\
1742 & 21:45:34.65 &  58:46:20.2$^{\rm r}$&      & 5135                      &                               &  16.24$^{\rm k}$&  15.62$^{\rm k}$&  15.23$^{\rm n}$&       &                 &13.481\,(25) &13.173\,(30) & 13.107\,(27)$^{\rm r}$&                    &                 &               &-2.5\,(3.8)   &-4.4\,(3.8)   &-2.6\,(7.5)   &-10\,(7.3)    &     &     &outFoV                                                                  \\
1743 & 21:45:37.56 &  58:47:02.4$^{\rm r}$&      & 5136                      &                               &                 &  15.45$^{\rm k}$&   13.9$^{\rm n}$&       &                 &11.831\,(38) &11.058\,(38) & 10.833\,(27)$^{\rm r}$&                    &                 &               &5\,(5.2)      &-9.7\,(5.2)   &              &              &     &     &outFoV                                                                  \\
1744 & 21:46:09.84 &  56:22:14.8$^{\rm r}$&      & 5137                      &                               &  16.87$^{\rm k}$&   15.4$^{\rm k}$&  14.41$^{\rm n}$&       &                 &12.498\,(32) &11.999\,(32) & 11.925\,(26)$^{\rm r}$&                    &                 &               &17.2\,(4.1)   &1\,(4.1)      &              &              &     &     &outFoV                                                                  \\
1745 & 21:46:16.67 &  56:21:10.7$^{\rm r}$&      & 5138                      &                               &  16.48$^{\rm k}$&   15.5$^{\rm k}$&  14.29$^{\rm n}$&       &                 &12.510\,(42) &12.202\,(47) & 11.934\,(30)$^{\rm r}$&                    &                 &               &              &              &              &              &     &     &outFoV                                                                  \\
1746 & 21:46:40.60 &  58:02:25.6$^{\rm r}$&      & 5139                      &                               &  15.62$^{\rm k}$&  15.38$^{\rm k}$&  14.15$^{\rm n}$&       &                 &11.853\,(29) &11.284\,(31) & 11.186\,(18)$^{\rm r}$&                    &                 &               &-7.9\,(3.8)   &-31.4\,(3.8)  &4.3\,(9.7)    &-8.3\,(10)    &     &     &outFoV                                                                  \\
1747 & 21:46:41.49 &  58:15:15.6$^{\rm r}$&      & 5140                      &                               &  16.55$^{\rm k}$&  15.94$^{\rm k}$&  14.75$^{\rm n}$&       &                 &13.184\,(50) &12.728\,(49) & 12.613\,(47)$^{\rm r}$&                    &                 &               &-24.6\,(5.1)  &13.2\,(5.1)   &-101.7\,(8.9) &57.9\,(12.1)  &     &     &outFoV                                                                  \\
1748 & 21:46:42.88 &  58:46:35.9$^{\rm r}$&      & 5141                      &                               &  16.75$^{\rm k}$&  15.94$^{\rm k}$&  14.87$^{\rm n}$&       &                 &12.864\,(28) &12.310\,(32) & 12.226\,(23)$^{\rm r}$&                    &                 &               &11.5\,(3.8)   &-21.7\,(3.8)  &41.4\,(7)     &15.5\,(7)     &     &     &outFoV                                                                  \\
1749 & 21:46:47.32 &  58:25:11.3$^{\rm r}$&      & 5142                      &                               &  16.32$^{\rm k}$&  16.06$^{\rm k}$&   14.9$^{\rm n}$&       &                 &12.923\,(26) &12.573\,(31) & 12.377\,(19)$^{\rm r}$&                    &                 &               &-7.4\,(3.8)   &-14.3\,(3.8)  &-0.3\,(7.3)   &-19.6\,(7.3)  &     &     &outFoV                                                                  \\
1750 & 21:46:56.82 &  57:10:27.3$^{\rm r}$&      & 5143                      &                               &                 &   16.4$^{\rm k}$&   15.2$^{\rm n}$&       &                 &13.145\,(27) &12.662\,(31) & 12.568\,(26)$^{\rm r}$&                    &                 &               &-1.5\,(4.1)   &-9.7\,(4.1)   &-15.8\,(6.8)  &8\,(6.9)      &     &     &outFoV                                                                  \\
1751 & 21:47:00.76 &  57:15:20.2$^{\rm m}$&      & 5144                      &                               &                 &  16.41$^{\rm k}$&  14.59$^{\rm n}$&       &                 &             &             &                       &                    &                 &               &              &              &              &              &     &     &outFoV                                                                  \\
1752 & 21:47:15.86 &  58:00:40.3$^{\rm r}$&      & 5145                      &                               &                 &                 &     15$^{\rm n}$&       &                 &12.084\,(23) &11.381\,(31) & 11.180\,(22)$^{\rm r}$&                    &                 &               &-3.3\,(3.8)   &-2.6\,(3.8)   &3.1\,(7.7)    &-4.2\,(7.3)   &     &     &outFoV                                                                  \\
1753 & 21:48:44.08 &  57:02:10.6$^{\rm r}$&      & 5146                      &                               &                 &  16.19$^{\rm k}$&  14.74$^{\rm n}$&       &                 &12.280\,(24) &11.765\,(28) & 11.607\,(23)$^{\rm r}$&                    &                 &               &-1.2\,(4.1)   &-5.5\,(4.1)   &-9.8\,(6.8)   &3.4\,(6.5)    &     &     &outFoV                                                                  \\
1754 & 21:48:45.09 &  57:01:07.9$^{\rm r}$&      & 5147                      &                               &  17.12$^{\rm k}$&  16.46$^{\rm k}$&  14.98$^{\rm n}$&       &                 &13.000\,(24) &12.551\,(30) & 12.445\,(24)$^{\rm r}$&                    &                 &               &-2.8\,(4.1)   &-0.6\,(4.1)   &-10.8\,(6.5)  &7.3\,(6.5)    &     &     &outFoV                                                                  \\
1755 & 21:49:03.24 &  56:59:26.6$^{\rm r}$&      & 5148                      &                               &  16.03$^{\rm k}$&  15.98$^{\rm k}$&  14.67$^{\rm n}$&       &                 &12.626\,(27) &12.189\,(30) & 11.995\,(25)$^{\rm r}$&                    &                 &               &-12.1\,(4.1)  &-12.7\,(4.1)  &-26.3\,(6.5)  &-16\,(6.5)    &     &     &outFoV                                                                  \\
1756 & 21:49:01.55 &  58:16:21.9$^{\rm r}$&      & 5149                      &                               &  16.18$^{\rm k}$&  16.18$^{\rm k}$&  14.28$^{\rm n}$&       &                 &12.615\,(26) &12.192\,(28) & 12.109\,(23)$^{\rm r}$&                    &                 &               &8.5\,(3.8)    &-0.2\,(3.8)   &2.6\,(7.2)    &-11.6\,(7.2)  &     &     &outFoV                                                                  \\
1757 & 21:49:11.36 &  58:12:29.9$^{\rm r}$&      & 5150                      &                               &  16.32$^{\rm k}$&  16.09$^{\rm k}$&   14.1$^{\rm n}$&       &                 &11.755\,(26) &10.773\,(31) & 10.425\,(20)$^{\rm r}$&                    &                 &               &6.2\,(3.8)    &-3.1\,(3.8)   &11\,(7.7)     &-20.2\,(10)   &     &     &outFoV                                                                  \\
1758 & 21:49:11.85 &  58:15:50.2$^{\rm r}$&      & 5151                      &                               &                 &  16.51$^{\rm k}$&   15.2$^{\rm n}$&       &                 &11.806\,(26) &11.060\,(28) & 10.777\,(22)$^{\rm r}$&                    &                 &               &9.3\,(3.8)    &-4.8\,(3.8)   &13.8\,(7.3)   &-12.9\,(7.6)  &     &     &outFoV                                                                  \\
1759 & 21:49:21.84 &  58:01:34.4$^{\rm r}$&      & 5152                      &                               &  14.40$^{\rm k}$&  14.39$^{\rm k}$&  13.57$^{\rm n}$&       &                 &11.857\,(23) &11.561\,(28) & 11.457\,(23)$^{\rm r}$&                    &                 &               &-7.9\,(3.8)   &-5.4\,(3.8)   &5.1\,(7.3)    &-17.3\,(7.3)  &     &     &outFoV                                                                  \\
1760 & 21:49:44.70 &  57:01:08.6$^{\rm r}$&      & 5153                      &                               &                 &  16.75$^{\rm k}$&   15.7$^{\rm n}$&       &                 &12.941\,(26) &12.529\,(30) & 12.352\,(24)$^{\rm r}$&                    &                 &               &-4.9\,(4.1)   &-3.1\,(4.1)   &-21.7\,(7.6)  &1.4\,(7.8)    &     &     &outFoV                                                                  \\
1761 & 21:50:30.35 &  56:33:54.3$^{\rm r}$&      & 5154                      &                               &                 &  16.43$^{\rm k}$&  15.19$^{\rm n}$&       &                 &12.573\,(26) &12.039\,(31) & 11.882\,(26)$^{\rm r}$&                    &                 &               &-19.4\,(4.1)  &-16.3\,(4.1)  &-20.7\,(6.5)  &-12.7\,(6.5)  &     &     &outFoV                                                                  \\
1762 & 21:50:46.44 &  56:27:17.7$^{\rm r}$&      & 5155                      &                               &                 &  16.32$^{\rm k}$&   14.7$^{\rm n}$&       &                 &11.761\,(26) &11.025\,(29) & 10.844\,(19)$^{\rm r}$&                    &                 &               &-1.2\,(4.1)   &5.5\,(4.1)    &-4.5\,(6.4)   &9.1\,(6.4)    &     &     &outFoV                                                                  \\
1763 & 21:35:57.93 &  57:29:09.9$^{\rm r}$&      &                           &                               &                 &                 &                 &       &                 &15.910\,(90) &15.182\,(107)&   14.354\,()$^{\rm r}$&                    &                 &               &1.1\,(3.8)    &-3\,(4)       &              &              &     &     &                                                                                      \\
1764 & 21:35:59.06 &  57:30:23.3$^{\rm r}$&      &                           &                               &                 &                 &                 &       &                 &14.499\,(47) &13.864\,(57) & 13.402\,(46)$^{\rm r}$&                    &                 &               &13.3\,(4.1)   &16.8\,(4.1)   &              &              &     &     &                                                                                      \\
1765 & 21:36:03.89 &  57:27:12.2$^{\rm r}$&      &                           &                               &                 &                 &                 &       &                 &14.700\,(42) &13.077\,(37) & 12.297\,(24)$^{\rm r}$&                    &                 &               &              &              &              &              &     &     &no opt. cp.                                                                \\
1766 & 21:36:06.06 &  57:26:34.2$^{\rm r}$&      &                           &                               &                 &                 &                 &       &                 &17.635\,()   &15.540\,()   & 14.450\,(82)$^{\rm r}$&                    &                 &               &              &              &              &              &     &     &no opt. cp.                                                                \\
1767 & 21:36:07.45 &  57:34:29.7$^{\rm r}$&      &                           &                               &                 &                 &                 &       &                 &14.806\,(43) &13.832\,()   &   12.914\,()$^{\rm r}$&                    &                 &               &7.1\,(4)      &-21.2\,(4)    &              &              &     &     &near 1561                                                                             \\
1768 & 21:36:07.98 &  57:26:37.1$^{\rm r}$&      &                           &                               &                 &                 &                 &       &                 &18.366\,()   &17.484\,()   & 14.507\,(88)$^{\rm r}$&                    &                 &               &              &              &              &              &     &     &no opt. cp.                                                                \\
1769 & 21:36:14.20 &  57:27:57.8$^{\rm r}$&      &                           &                               &                 &                 &                 &       &                 &15.265\,(51) &14.520\,(56) & 14.395\,(74)$^{\rm r}$&                    &                 &               &-31.5\,(4.2)  &11.2\,(4.2)   &              &              &     &     &                                                                                      \\
1770 & 21:36:16.65 &  57:28:40.5$^{\rm r}$&      &                           &                               &                 &                 &                 &       &                 &17.017\,(198)&14.958\,(86) & 14.026\,(56)$^{\rm r}$&                    &                 &               &              &              &              &              &     &     &no opt. cp.                                                                \\
1771 & 21:36:17.00 &  57:26:39.9$^{\rm r}$&      &                           &                               &                 &                 &                 &       &                 &14.604\,(31) &13.783\,(40) & 13.405\,(37)$^{\rm r}$&                    &                 &               &              &              &              &              &     &     &very faint opt. cp.                                                        \\
1772 & 21:36:23.69 &  57:32:45.2$^{\rm r}$&      &                           &                               &                 &                 &                 &       &                 &14.685\,(45) &13.650\,(45) & 13.105\,(35)$^{\rm r}$&                    &                 &               &-9\,(5.2)     &-11.3\,(5.2)  &              &              &     &     &                                                                                      \\
1773 & 21:36:33.00 &  57:28:49.4$^{\rm r}$&      &                           &                               &                 &                 &                 &       &                 &15.816\,(76) &14.827\,(70) & 14.220\,(64)$^{\rm r}$&                    &                 &               &              &              &              &              &     &     &no opt. cp.                                                                \\
1774 & 21:36:35.32 &  57:29:31.2$^{\rm r}$&      &                           &                               &                 &                 &                 &       &                 &14.059\,(44) &12.825\,()   &   12.223\,()$^{\rm r}$&                    &                 &               &20.5\,(4)     &-13\,(4)      &              &              &     &     &                                                                                      \\
1775 & 21:36:36.91 &  57:31:32.7$^{\rm r}$&      &                           &                               &                 &                 &                 &       &                 &13.754\,(31) &12.631\,(37) & 12.021\,(26)$^{\rm r}$&                    &                 &               &-30\,(5.5)    &-2.3\,(5.5)   &              &              &     &     &                                                                                      \\
1776 & 21:36:38.42 &  57:29:17.5$^{\rm r}$&      &                           &                               &                 &                 &                 &       &                 &14.265\,(27) &13.013\,(31) & 12.303\,(25)$^{\rm r}$&                    &                 &               &              &              &              &              &     &     &no opt. cp.                                                                \\
1777 & 21:36:39.15 &  57:29:53.3$^{\rm r}$&      &                           &                               &                 &                 &                 &       &                 &11.924\,(21) &10.422\,(28) & 9.392 \,(21)$^{\rm r}$&                    &                 &               &-2.9\,(5)     &-23.4\,(5)    &              &              &     &     &                                                                                      \\
1778 & 21:36:41.46 &  57:30:27.8$^{\rm r}$&      &                           &                               &                 &                 &                 &       &                 &15.921\,(81) &14.293\,(58) & 13.542\,(40)$^{\rm r}$&                    &                 &               &              &              &              &              &     &     &no opt. cp.                                                                \\
1779 & 21:36:41.65 &  57:32:17.5$^{\rm r}$&      &                           &                               &                 &                 &                 &       &                 &16.147\,(102)&14.901\,(85) & 13.796\,(46)$^{\rm r}$&                    &                 &               &              &              &              &              &     &     &no opt. cp.                                                                \\
1780 & 21:36:43.98 &  57:29:28.7$^{\rm r}$&      &                           &                               &                 &                 &                 &       &                 &14.232\,(27) &13.012\,(31) & 12.364\,(24)$^{\rm r}$&                    &                 &               &-7.7\,(5.6)   &0.4\,(5.6)    &              &              &     &     &no opt. cp.                                                                \\
1781 & 21:36:44.01 &  57:28:46.8$^{\rm r}$&      &                           &                               &                 &                 &                 &       &                 &12.949\,(21) &12.559\,(26) & 12.501\,(24)$^{\rm r}$&                    &                 &               &-11.5\,(4)    &1.6\,(4)      &              &              &     &     &                                                                                      \\
1782 & 21:36:46.60 &  57:29:38.5$^{\rm r}$&      &                           &                               &                 &                 &                 &       &                 &16.458\,()   &15.194\,(187)& 12.673\,(38)$^{\rm r}$&                    &                 &               &              &              &              &              &     &     &no opt. cp.                                                                \\
1783 & 21:36:47.14 &  57:28:53.0$^{\rm d}$&      &                           &                               &                 &                 &                 &       &                 &             &             &                       &                    &                 &               &              &              &              &              &     &     &no opt. cp.                                                                \\
1784 & 21:36:47.89 &  57:31:30.7$^{\rm r}$&      &                           &                               &                 &                 &                 &       &                 &16.504\,(140)&14.697\,(77) & 13.822\,(65)$^{\rm r}$&                    &                 &               &              &              &              &              &     &     &no opt. cp.                                                                \\
1785 & 21:36:52.81 &  57:29:43.8$^{\rm r}$&      &                           &                               &                 &                 &                 &       &                 &15.666\,(68) &13.956\,(43) & 13.286\,(34)$^{\rm r}$&                    &                 &               &              &              &              &              &     &     &no opt. cp.                                                                \\
1786 & 21:36:54.50 &  57:30:05.2$^{\rm r}$&      &                           &                               &                 &                 &                 &       &                 &13.948\,(27) &12.045\,(31) & 10.926\,(23)$^{\rm r}$&                    &                 &               &              &              &              &              &     &     &no opt. cp.                                                                \\
1787 & 21:36:54.75 &  57:31:45.1$^{\rm r}$&      &                           &                               &                 &                 &                 &       &                 &16.017\,(95) &14.796\,(81) & 13.909\,(57)$^{\rm r}$&                    &                 &               &              &              &              &              &     &     &no opt. cp.                                                                \\
1788 & 21:36:54.90 &  57:30:00.4$^{\rm r}$&      &                           &                               &                 &                 &                 &       &                 &16.341\,()   &15.700\,(151)& 13.837\,(60)$^{\rm r}$&                    &                 &               &              &              &              &              &     &     &no opt. cp.                                                                \\
1789 & 21:36:55.21 &  57:30:30.1$^{\rm r}$&      &                           &                               &                 &                 &                 &       &                 &14.714\,(78) &12.665\,(63) & 11.382\,(34)$^{\rm r}$&                    &                 &               &              &              &              &              &     &     &no opt. cp.                                                                \\
1790 & 21:36:55.43 &  57:31:39.1$^{\rm r}$&      &                           &                               &                 &                 &                 &       &                 &14.199\,(32) &12.456\,(29) & 11.723\,(21)$^{\rm r}$&                    &                 &               &              &              &              &              &     &     &no opt. cp.                                                                \\
1791 & 21:36:56.99 &  57:29:22.7$^{\rm r}$&      &                           &                               &                 &                 &                 &       &                 &16.991\,(187)&14.564\,(57) & 13.195\,(30)$^{\rm r}$&                    &                 &               &              &              &              &              &     &     &no opt. cp.                                                                \\
1792 & 21:36:57.84 &  57:30:56.1$^{\rm r}$&      &                           &                               &                 &                 &                 &       &                 &17.425\,()   &15.365\,(135)& 14.000\,(72)$^{\rm r}$&                    &                 &               &              &              &              &              &     &     &no opt. cp.                                                                \\
1793 & 21:36:57.93 &  57:29:10.7$^{\rm r}$&      &                           &                               &                 &                 &                 &       &                 &13.621\,(28) &12.025\,(32) & 11.215\,(23)$^{\rm r}$&                    &                 &               &-10.6\,(5.6)  &1.7\,(5.6)    &              &              &     &     &no opt. cp.                                                                \\
1794 & 21:36:58.91 &  57:30:29.3$^{\rm r}$&      &                           &                               &                 &                 &                 &       &                 &17.551\,()   &16.181\,(218)& 14.615\,(93)$^{\rm r}$&                    &                 &               &              &              &              &              &     &     &no opt. cp.                                                                \\
1795 & 21:37:02.00 &  57:31:55.3$^{\rm r}$&      &                           &                               &                 &                 &                 &       &                 &15.871\,(87) &13.808\,(44) & 12.881\,(30)$^{\rm r}$&                    &                 &               &              &              &              &              &     &     &no opt. cp.                                                                \\
1796 & 21:37:02.32 &  57:31:15.3$^{\rm r}$&      &                           &                               &                 &                 &                 &       &                 &16.946\,()   &15.588\,()   & 13.254\,(45)$^{\rm r}$&                    &                 &               &              &              &              &              &     &     &no opt. cp.                                                                \\
1797 & 21:37:05.20 &  57:30:02.2$^{\rm r}$&      &                           &                               &                 &                 &                 &       &                 &16.308\,()   &15.599\,()   &15.124\,(155)$^{\rm r}$&                    &                 &               &              &              &              &              &     &     &no opt. cp.                                                                \\
1798 & 21:37:07.71 &  57:32:11.0$^{\rm r}$&      &                           &                               &                 &                 &                 &       &                 &14.606\,(35) &13.785\,(39) & 13.522\,(47)$^{\rm r}$&                    &                 &               &              &              &              &              &     &     &very faint opt. cp.                                                        \\
1799 & 21:37:08.02 &  57:34:09.5$^{\rm r}$&      &                           &                               &                 &                 &                 &       &                 &9.617 \,(22) &8.373 \,(28) & 7.860 \,(20)$^{\rm r}$&                    &                 &               &              &              &-32.7\,(6.4)  &126.3\,(6.8)  &     &     &                                                                                      \\
1800 & 21:37:09.44 &  57:30:36.7$^{\rm r}$&      &                           &                               &                 &                 &                 &       &                 &13.884\,(30) &13.036\,(34) & 12.648\,(33)$^{\rm r}$&                    &                 &               &              &              &              &              &     &     &                                                                                      \\
1801 & 21:37:10.14 &  57:31:26.6$^{\rm r}$&      &                           &                               &                 &                 &                 &       &                 &14.972\,(46) &14.241\,(55) & 13.806\,(59)$^{\rm r}$&                    &                 &               &              &              &              &              &     &     &no opt. cp.                                                                \\
1802 & 21:37:24.10 &  57:24:11.5$^{\rm r}$&      &                           &                               &                 &                 &                 &       &                 &14.198\,(66) &13.293\,(77) & 12.821\,(47)$^{\rm r}$&                    &                 &               &-21.7\,(4.1)  &-31.8\,(4.1)  &              &              &     &     &                                                                                      \\
1803 & 21:37:48.93 &  57:23:21.0$^{\rm r}$&      &                           &                               &                 &                 &                 &       &                 &14.657\,(55) &13.682\,(63) & 13.252\,(49)$^{\rm r}$&                    &                 &               &1.7\,(5.4)    &-0.7\,(5.4)   &              &              &     &     &                                                                                      \\
1804 & 21:38:09.25 &  57:20:19.9$^{\rm r}$&      &                           &                               &                 &                 &                 &       &                 &13.918\,(31) &12.958\,(33) & 12.477\,(23)$^{\rm r}$&                    &                 &               &-4.8\,(4)     &-5.3\,(4)     &              &              &     &     &                                                                                      \\
1805 & 21:38:09.79 &  57:29:42.8$^{\rm r}$&      &                           &                               &                 &                 &                 &       &                 &14.161\,(40) &13.356\,(44) & 12.974\,(42)$^{\rm r}$&                    &                 &               &              &              &              &              &     &     &                                                                                      \\
1806 & 21:39:25.41 &  57:33:20.3$^{\rm r}$&      &                           &                               &                 &                 &                 &       &                 &12.590\,(25) &11.686\,(29) & 11.312\,(18)$^{\rm r}$&                    &                 &               &-0.9\,(3.8)   &-8.9\,(3.8)   &2.1\,(10.4)   &6\,(10.4)     &     &     &                                                                                      \\
1807 & 21:39:31.05 &  57:47:14.0$^{\rm r}$&      &                           &                               &                 &                 &                 &       &                 &12.410\,(27) &11.491\,(28) & 11.023\,(23)$^{\rm r}$&                    &                 &               &              &              &9.1\,(8.7)    &3.4\,(8.4)    &     &     &                                                                                      \\
1808 & 21:40:14.38 &  57:40:50.8$^{\rm r}$&      &                           &                               &                 &                 &                 &       &                 &13.530\,(40) &12.481\,(32) & 12.053\,(26)$^{\rm r}$&                    &                 &               &              &              &              &              &     &     &near 651                                                                            \\
1809 & 21:36:07.46 &  57:26:43.6$^{\rm d}$&      &                           &                               &                 &                 &                 &       &                 &             &             &                       &                    &                 &               &              &              &              &              &     &     &no opt. cp.                                                                \\
1810 & 21:36:18.20 &  57:28:31.0$^{\rm d}$&      &                           &                               &                 &                 &                 &       &                 &             &             &                       &                    &                 &               &              &              &              &              &     &     &no opt. cp.                                                                \\
1811 & 21:36:19.20 &  57:28:38.0$^{\rm d}$&      &                           &                               &                 &                 &                 &       &                 &             &             &                       &                    &                 &               &              &              &              &              &     &     &no opt. cp.                                                                \\
1812 & 21:36:47.16 &  57:28:44.2$^{\rm d}$&      &                           &                               &                 &                 &                 &       &                 &             &             &                       &                    &                 &               &              &              &              &              &     &     &no opt. cp.                                                                \\
1813 & 21:36:59.45 &  57:31:30.6$^{\rm d}$&      &                           &                               &                 &                 &                 &       &                 &             &             &                       &                    &                 &               &              &              &              &              &     &     &no opt. cp.                                                                \\
1814 & 21:37:01.05 &  57:30:39.7$^{\rm d}$&      &                           &                               &                 &                 &                 &       &                 &             &             &                       &                    &                 &               &              &              &              &              &     &     &no opt. cp.                                                                \\
1815 & 21:37:07.18 &  57:31:27.8$^{\rm d}$&      &                           &                               &                 &                 &                 &       &                 &             &             &                       &                    &                 &               &              &              &              &              &     &     &no opt. cp.                                                                \\
1816 & 21:39:25.71 &  57:29:45.6$^{\rm r}$&      &                           &                               &                 &                 &                 &       &                 &13.942\,(32) &12.996\,(32) & 12.700\,(28)$^{\rm r}$&                    &                 &               &-11\,(3.9)    &-6.1\,(3.9)   &              &              &     &     &                                                                                      \\
1817 & 21:39:26.15 &  57:00:09.3$^{\rm r}$&      &                           &                               &                 &  13.34$^{\rm f}$&  12.77$^{\rm e}$&       &                 &11.602\,(21) &11.333\,(28) & 11.278\,(19)$^{\rm r}$&        F1$^{\rm e}$&                 &  0.8$^{\rm e}$&5.9\,(2.7)    &-1.6\,(2.7)   &-0.5\,(0.8)   &-4.8\,(0.4)   &     &     &                                                                                      \\
1818 & 21:38:31.05 &  57:28:00.5$^{\rm r}$&      &                           &                               &                 &                 &                 &       &                 &13.658\,(27) &13.010\,(37) & 12.825\,(25)$^{\rm r}$&                    &                 &               &-1.6\,(5.4)   &-3.5\,(5.4)   &              &              &     &     &                                                                                      \\
1819 & 21:38:32.85 &  57:29:18.4$^{\rm r}$&      &                           &                               &                 &                 &                 &       &                 &14.524\,(53) &13.749\,(40) & 13.447\,(42)$^{\rm r}$&                    &                 &               &-0.6\,(4)     &-4.8\,(4)     &              &              &     &     &                                                                                      \\
1820 & 21:38:42.83 &  57:28:54.8$^{\rm r}$&      &                           &                               &                 &                 &                 &       &                 &12.174\,(23) &11.442\,(28) & 11.198\,(21)$^{\rm r}$&                    &                 &               &6.3\,(3.8)    &-11.5\,(3.8)  &17.6\,(6.9)   &0\,(6.9)      &     &     &                                                                                      \\
1821 & 21:38:43.70 &  57:31:03.3$^{\rm r}$&      &                           &                               &                 &                 &                 &       &                 &14.483\,(45) &13.617\,(41) & 13.374\,(42)$^{\rm r}$&                    &                 &               &-6.2\,(5.4)   &-6.5\,(5.4)   &              &              &     &     &                                                                                      \\
1822 & 21:38:49.68 &  57:31:55.6$^{\rm r}$&      &                           &                               &                 &                 &                 &       &                 &13.474\,()   &13.032\,(48) & 12.790\,(40)$^{\rm r}$&                    &                 &               &              &              &              &              &     &     &                                                                                      \\
1823 & 21:38:50.41 &  57:30:05.1$^{\rm r}$&      &                           &                               &                 &                 &                 &       &                 &13.472\,(25) &12.683\,(35) & 12.499\,(25)$^{\rm r}$&                    &                 &               &-4.9\,(3.9)   &-6.9\,(3.9)   &5.2\,(8.8)    &-15.8\,(8.6)  &     &     &                                                                                      \\
1824 & 21:38:50.99 &  57:28:42.7$^{\rm r}$&      &                           &                               &                 &                 &                 &       &                 &13.331\,(34) &12.592\,(40) & 12.311\,(30)$^{\rm r}$&                    &                 &               &-19.2\,(3.9)  &-10.4\,(3.9)  &-97.5\,(11.2) &32.6\,(10.3)  &     &     &                                                                                      \\
1825 & 21:38:54.65 &  57:29:25.0$^{\rm r}$&      &                           &                               &                 &                 &                 &       &                 &13.932\,(91) &12.775\,()   &   12.659\,()$^{\rm r}$&                    &                 &               &              &              &              &              &     &     &                                                                                      \\
1826 & 21:38:56.18 &  57:28:58.6$^{\rm r}$&      &                           &                               &                 &                 &                 &       &                 &13.714\,(50) &13.042\,(41) & 12.813\,(33)$^{\rm r}$&                    &                 &               &              &              &              &              &     &     &                                                                                      \\
1827 & 21:38:57.62 &  57:30:06.1$^{\rm b}$&      &                           &                               &                 &                 &                 &       &                 &             &             &                       &                    &                 &               &              &              &              &              &     &     &                                                                                      \\
1828 & 21:38:58.24 &  57:28:15.1$^{\rm r}$&      &                           &                               &                 &                 &                 &       &                 &12.984\,(27) &12.211\,(32) & 12.015\,(24)$^{\rm r}$&                    &                 &               &5.9\,(18)     &-26\,(18)     &6.4\,(8.2)    &34.7\,(8.2)   &     &     &                                                                                      \\
1829 & 21:38:58.88 &  57:29:14.6$^{\rm r}$&      &                           &                               &                 &                 &                 &       &                 &7.585 \,(20) &7.632 \,(36) & 7.595 \,(21)$^{\rm r}$&                    &                 &               &-5.1\,(1.6)   &-2.4\,(1.5)   &              &              &     &     &                                                                                      \\
1830 & 21:38:59.63 &  57:30:08.1$^{\rm r}$&      &                           &                               &                 &                 &                 &       &                 &13.485\,(23) &12.600\,(29) & 12.108\,(25)$^{\rm r}$&                    &                 &               &              &              &3.6\,(9.9)    &6.7\,(10)     &     &     &                                                                                      \\
1831 & 21:39:03.76 &  57:29:41.7$^{\rm r}$&      &                           &                               &                 &                 &                 &       &                 &12.722\,(31) &12.064\,(40) & 11.880\,(29)$^{\rm r}$&                    &                 &               &-1.9\,(3.9)   &-6.1\,(3.9)   &-76.7\,(7.2)  &12.8\,(7.2)   &     &     &                                                                                      \\
1832 & 21:39:06.24 &  57:28:10.7$^{\rm r}$&      &                           &                               &                 &                 &                 &       &                 &15.024\,(58) &13.895\,(61) & 13.252\,(35)$^{\rm r}$&                    &                 &               &              &              &              &              &     &     &                                                                                      \\
1833 & 21:39:09.19 &  57:30:50.3$^{\rm r}$&      &                           &                               &                 &                 &                 &       &                 &12.569\,(26) &11.831\,(29) & 11.641\,(25)$^{\rm r}$&                    &                 &               &-10.7\,(4)    &5.2\,(4)      &-13.2\,(6.6)  &22.8\,(6.7)   &     &     &                                                                                      \\
1834 & 21:39:13.42 &  57:28:38.8$^{\rm r}$&      &                           &                               &                 &                 &                 &       &                 &14.180\,(53) &13.349\,(55) & 13.032\,(47)$^{\rm r}$&                    &                 &               &              &              &              &              &     &     &                                                                                      \\
1835 & 21:39:16.38 &  57:31:18.8$^{\rm r}$&      &                           &                               &                 &                 &                 &       &                 &13.555\,(30) &12.944\,(31) & 12.787\,(25)$^{\rm r}$&                    &                 &               &-1.3\,(3.8)   &-4.9\,(3.8)   &2.9\,(7.5)    &9.3\,(7.5)    &     &     &                                                                                      \\
1836 & 21:36:15.20 &  57:25:28.0$^{\rm r}$&      &                           &                               &                 &                 &                 &       &                 &15.422\,(76) &14.797\,(87) &   14.292\,()$^{\rm r}$&                    &                 &               &-182.6\,(5.6) &-241\,(5.6)   &              &              &     &     &                                                                                      \\
1837 & 21:36:42.47 &  57:25:23.2$^{\rm r}$&      &                           &                               &                 &                 &                 &       &                 &14.806\,(43) &14.089\,(47) & 13.758\,(49)$^{\rm r}$&                    &                 &               &              &              &              &              &     &     &                                                                                      \\
1838 & 21:36:40.33 &  57:25:45.5$^{\rm r}$&      &                           &                               &                 &                 &                 &       &                 &16.271\,(94) &15.203\,(99) &14.867\,(113)$^{\rm r}$&                    &                 &               &              &              &              &              &     &     &                                                                                      \\
1839 & 21:36:45.86 &  57:26:22.8$^{\rm r}$&      &                           &                               &                 &                 &                 &       &                 &15.192\,(48) &14.582\,(67) & 14.378\,(70)$^{\rm r}$&                    &                 &               &-0.9\,(4.2)   &-13.3\,(4.2)  &              &              &     &     &                                                                                      \\
1840 & 21:36:38.02 &  57:26:58.0$^{\rm r}$&      &                           &                               &                 &                 &                 &       &                 &15.604\,(72) &14.893\,(73) & 14.523\,(80)$^{\rm r}$&                    &                 &               &              &              &              &              &     &     &                                                                                      \\
1841 & 21:36:54.72 &  57:27:26.7$^{\rm a}$&      &                           &                               &                 &                 &                 &       &                 &             &             &                       &                    &                 &               &              &              &              &              &     &     &                                                                                      \\
1842 & 21:36:33.20 &  57:27:51.8$^{\rm r}$&      &                           &                               &                 &                 &                 &       &                 &15.078\,(62) &14.372\,(56) & 14.155\,(63)$^{\rm r}$&                    &                 &               &3.1\,(4.2)    &0\,(4.2)      &              &              &     &     &                                                                                      \\
1843 & 21:35:51.09 &  57:28:12.5$^{\rm r}$&      &                           &                               &                 &                 &                 &       &                 &15.465\,(62) &14.983\,(92) & 14.523\,(84)$^{\rm r}$&                    &                 &               &-6.8\,(5.6)   &-0.3\,(5.6)   &              &              &     &     &                                                                                      \\
1844 & 21:35:58.05 &  57:28:50.3$^{\rm r}$&      &                           &                               &                 &                 &                 &       &                 &15.609\,(127)&14.810\,(164)&14.783\,(177)$^{\rm r}$&                    &                 &               &-4.3\,(5.7)   &-6.4\,(5.7)   &              &              &     &     &                                                                                      \\
1845 & 21:36:18.97 &  57:29:05.1$^{\rm a}$&      &                           &                               &                 &                 &                 &       &                 &             &             &                       &                    &                 &               &              &              &              &              &     &     &                                                                                      \\
1846 & 21:35:58.50 &  57:29:15.0$^{\rm r}$&      &                           &                               &                 &                 &                 &       &                 &16.582\,(167)&15.373\,(125)&14.632\,(100)$^{\rm r}$&                    &                 &               &              &              &              &              &     &     &                                                                                      \\
1847 & 21:37:17.37 &  57:29:20.7$^{\rm r}$&      &                           &                               &                 &                 &                 &       &                 &14.091\,()   &13.557\,(71) &   13.130\,()$^{\rm r}$&                    &                 &               &              &              &              &              &     &     &                                                                                      \\
1848 & 21:37:17.42 &  57:29:27.3$^{\rm r}$&      &                           &                               &                 &                 &                 &       &                 &14.122\,(43) &13.135\,(43) & 12.583\,(37)$^{\rm r}$&                    &                 &               &-1.2\,(4.1)   &-10.8\,(4.1)  &              &              &     &     &                                                                                      \\
1849 & 21:36:36.95 &  57:29:28.6$^{\rm r}$&      &                           &                               &                 &                 &                 &       &                 &15.670\,(55) &14.979\,(78) & 14.664\,(98)$^{\rm r}$&                    &                 &               &              &              &              &              &     &     &                                                                                      \\
1850 & 21:36:37.64 &  57:29:31.7$^{\rm r}$&      &                           &                               &                 &                 &                 &       &                 &15.555\,(58) &15.022\,(78) & 14.685\,(99)$^{\rm r}$&                    &                 &               &-32.9\,(5.3)  &-3.9\,(5.3)   &              &              &     &     &                                                                                      \\
1851 & 21:35:53.11 &  57:29:37.0$^{\rm r}$&      &                           &                               &                 &                 &                 &       &                 &15.300\,(69) &14.774\,(83) & 14.513\,(85)$^{\rm r}$&                    &                 &               &-7.9\,(4.1)   &-11.9\,(4.1)  &              &              &     &     &                                                                                      \\
1852 & 21:35:55.41 &  57:29:42.7$^{\rm r}$&      &                           &                               &                 &                 &                 &       &                 &15.587\,(74) &15.194\,(111)&14.874\,(113)$^{\rm r}$&                    &                 &               &-5.5\,(4.1)   &-4\,(4.1)     &              &              &     &     &                                                                                      \\
1853 & 21:36:17.03 &  57:29:48.1$^{\rm r}$&      &                           &                               &                 &                 &                 &       &                 &15.858\,(87) &14.704\,(65) & 14.288\,(66)$^{\rm r}$&                    &                 &               &              &              &              &              &     &     &                                                                                      \\
1854 & 21:36:49.03 &  57:29:49.0$^{\rm r}$&      &                           &                               &                 &                 &                 &       &                 &17.026\,()   &15.595\,(143)& 14.523\,(83)$^{\rm r}$&                    &                 &               &              &              &              &              &     &     &                                                                                      \\
1855 & 21:36:10.98 &  57:29:50.7$^{\rm r}$&      &                           &                               &                 &                 &                 &       &                 &15.547\,(69) &14.614\,(65) & 14.468\,(80)$^{\rm r}$&                    &                 &               &-4\,(5.6)     &11.6\,(5.6)   &              &              &     &     &                                                                                      \\
1856 & 21:36:56.27 &  57:29:52.4$^{\rm r}$&      &                           &                               &                 &                 &                 &       &                 &18.527\,()   &16.098\,()   &15.200\,(141)$^{\rm r}$&                    &                 &               &              &              &              &              &     &     &                                                                                      \\
1857 & 21:36:47.16 &  57:29:52.6$^{\rm r}$&      &                           &                               &                 &                 &                 &       &                 &14.184\,()   &14.073\,(58) & 12.630\,(28)$^{\rm r}$&                    &                 &               &              &              &              &              &     &     &                                                                                      \\
1858 & 21:37:10.56 &  57:29:52.7$^{\rm r}$&      &                           &                               &                 &                 &                 &       &                 &15.135\,(49) &14.288\,(56) & 14.027\,(79)$^{\rm r}$&                    &                 &               &-1.8\,(3.9)   &-3.8\,(3.9)   &              &              &     &     &                                                                                      \\
1859 & 21:35:55.61 &  57:30:03.4$^{\rm r}$&      &                           &                               &                 &                 &                 &       &                 &15.436\,(73) &14.973\,(83) & 14.547\,(85)$^{\rm r}$&                    &                 &               &-8.6\,(3.9)   &-2.3\,(3.9)   &              &              &     &     &                                                                                      \\
1860 & 21:36:13.37 &  57:30:16.2$^{\rm r}$&      &                           &                               &                 &                 &                 &       &                 &15.593\,(75) &14.999\,(82) &14.778\,(109)$^{\rm r}$&                    &                 &               &-9.6\,(4)     &-9.5\,(4)     &              &              &     &     &                                                                                      \\
1861 & 21:36:17.95 &  57:30:16.4$^{\rm r}$&      &                           &                               &                 &                 &                 &       &                 &15.873\,(93) &15.358\,(129)&14.995\,(122)$^{\rm r}$&                    &                 &               &17.7\,(4.1)   &26.6\,(4.2)   &              &              &     &     &                                                                                      \\
1862 & 21:36:53.16 &  57:30:19.3$^{\rm r}$&      &                           &                               &                 &                 &                 &       &                 &17.860\,()   &16.019\,()   &15.125\,(145)$^{\rm r}$&                    &                 &               &              &              &              &              &     &     &                                                                                      \\
1863 & 21:36:40.48 &  57:30:25.8$^{\rm r}$&      &                           &                               &                 &                 &                 &       &                 &16.008\,(104)&14.335\,(53) &   13.753\,()$^{\rm r}$&                    &                 &               &              &              &              &              &     &     &                                                                                      \\
1864 & 21:36:16.15 &  57:30:26.8$^{\rm r}$&      &                           &                               &                 &                 &                 &       &                 &14.524\,(36) &14.174\,(43) & 13.995\,(57)$^{\rm r}$&                    &                 &               &0.7\,(3.8)    &1.4\,(3.8)    &              &              &     &     &                                                                                      \\
1865 & 21:36:38.61 &  57:30:27.2$^{\rm r}$&      &                           &                               &                 &                 &                 &       &                 &16.818\,(181)&15.402\,(138)&14.967\,(126)$^{\rm r}$&                    &                 &               &              &              &              &              &     &     &                                                                                      \\
1866 & 21:37:11.74 &  57:30:35.1$^{\rm r}$&      &                           &                               &                 &                 &                 &       &                 &16.175\,(115)&15.520\,()   &14.928\,(148)$^{\rm r}$&                    &                 &               &              &              &              &              &     &     &                                                                                      \\
1867 & 21:36:44.72 &  57:30:37.3$^{\rm r}$&      &                           &                               &                 &                 &                 &       &                 &16.253\,(105)&15.375\,(117)&14.805\,(102)$^{\rm r}$&                    &                 &               &              &              &              &              &     &     &                                                                                      \\
1868 & 21:36:44.09 &  57:30:38.2$^{\rm r}$&      &                           &                               &                 &                 &                 &       &                 &15.996\,(90) &15.314\,(135)&14.671\,(106)$^{\rm r}$&                    &                 &               &              &              &              &              &     &     &                                                                                      \\
1869 & 21:37:03.04 &  57:30:48.7$^{\rm a}$&      &                           &                               &                 &                 &                 &       &                 &             &             &                       &                    &                 &               &              &              &              &              &     &     &                                                                                      \\
1870 & 21:36:01.65 &  57:30:49.7$^{\rm r}$&      &                           &                               &                 &                 &                 &       &                 &15.617\,(83) &14.986\,(101)&14.778\,(116)$^{\rm r}$&                    &                 &               &-55.1\,(5.7)  &-39.2\,(5.7)  &              &              &     &     &                                                                                      \\
1871 & 21:36:45.86 &  57:31:03.5$^{\rm a}$&      &                           &                               &                 &                 &                 &       &                 &             &             &                       &                    &                 &               &              &              &              &              &     &     &                                                                                      \\
1872 & 21:36:12.61 &  57:31:26.5$^{\rm r}$&      &                           &                               &                 &                 &                 &       &                 &16.504\,(132)&16.161\,()   &15.136\,(140)$^{\rm r}$&                    &                 &               &              &              &              &              &     &     &                                                                                      \\
1873 & 21:36:54.58 &  57:31:50.1$^{\rm r}$&      &                           &                               &                 &                 &                 &       &                 &16.840\,(187)&15.577\,()   &15.077\,(129)$^{\rm r}$&                    &                 &               &              &              &              &              &     &     &                                                                                      \\
1874 & 21:36:52.62 &  57:31:50.3$^{\rm r}$&      &                           &                               &                 &                 &                 &       &                 &14.915\,(46) &13.728\,(37) & 13.186\,(35)$^{\rm r}$&                    &                 &               &-81.5\,(8.6)  &56.7\,(8.6)   &              &              &     &     &                                                                                      \\
1875 & 21:36:56.53 &  57:31:51.4$^{\rm r}$&      &                           &                               &                 &                 &                 &       &                 &17.485\,()   &15.178\,(103)& 13.512\,(44)$^{\rm r}$&                    &                 &               &              &              &              &              &     &     &                                                                                      \\
1876 & 21:36:36.35 &  57:32:09.3$^{\rm a}$&      &                           &                               &                 &                 &                 &       &                 &             &             &                       &                    &                 &               &              &              &              &              &     &     &                                                                                      \\
1877 & 21:37:05.87 &  57:32:12.4$^{\rm r}$&      &                           &                               &                 &                 &                 &       &                 &15.130\,(51) &14.284\,(59) & 14.139\,(78)$^{\rm r}$&                    &                 &               &              &              &              &              &     &     &                                                                                      \\
1878 & 21:36:28.43 &  57:32:13.5$^{\rm r}$&      &                           &                               &                 &                 &                 &       &                 &14.384\,(27) &13.535\,(31) & 13.166\,(35)$^{\rm r}$&                    &                 &               &-4.3\,(5.2)   &-15.5\,(5.2)  &              &              &     &     &                                                                                      \\
1879 & 21:37:00.27 &  57:32:23.8$^{\rm a}$&      &                           &                               &                 &                 &                 &       &                 &             &             &                       &                    &                 &               &              &              &              &              &     &     &                                                                                      \\
1880 & 21:37:09.44 &  57:32:25.2$^{\rm r}$&      &                           &                               &                 &                 &                 &       &                 &16.032\,(138)&15.207\,(183)&14.635\,(112)$^{\rm r}$&                    &                 &               &1\,(9.3)      &-10.5\,(9.3)  &              &              &     &     &                                                                                      \\
1881 & 21:36:54.65 &  57:32:29.1$^{\rm r}$&      &                           &                               &                 &                 &                 &       &                 &17.506\,()   &15.976\,()   &14.963\,(150)$^{\rm r}$&                    &                 &               &              &              &              &              &     &     &                                                                                      \\
1882 & 21:36:51.54 &  57:32:53.4$^{\rm r}$&      &                           &                               &                 &                 &                 &       &                 &16.082\,(81) &15.203\,(94) &14.749\,(101)$^{\rm r}$&                    &                 &               &              &              &              &              &     &     &                                                                                      \\
1883 & 21:36:59.85 &  57:32:56.1$^{\rm r}$&      &                           &                               &                 &                 &                 &       &                 &15.700\,(76) &14.874\,(85) &14.851\,(111)$^{\rm r}$&                    &                 &               &-1.6\,(4.3)   &-1.4\,(4.3)   &              &              &     &     &                                                                                      \\
1884 & 21:36:25.97 &  57:33:10.3$^{\rm r}$&      &                           &                               &                 &                 &                 &       &                 &14.883\,(46) &14.195\,(54) & 13.958\,(58)$^{\rm r}$&                    &                 &               &2.2\,(4)      &2.9\,(4)      &              &              &     &     &                                                                                      \\
1885 & 21:36:36.40 &  57:33:14.4$^{\rm r}$&      &                           &                               &                 &                 &                 &       &                 &16.611\,(148)&15.853\,()   &   15.672\,()$^{\rm r}$&                    &                 &               &              &              &              &              &     &     &                                                                                      \\
1886 & 21:36:48.84 &  57:33:17.4$^{\rm a}$&      &                           &                               &                 &                 &                 &       &                 &17.542       &16.693       &       16.584$^{\rm a}$&                    &                 &               &              &              &              &              &     &     &                                                                                      \\
1887 & 21:36:34.84 &  57:33:57.1$^{\rm r}$&      &                           &                               &                 &                 &                 &       &                 &15.127\,(45) &14.563\,(58) & 14.316\,(66)$^{\rm r}$&                    &                 &               &11\,(3.9)     &3.6\,(3.9)    &              &              &     &     &                                                                                      \\
1888 & 21:36:12.98 &  57:34:05.5$^{\rm r}$&      &                           &                               &                 &                 &                 &       &                 &15.055\,(47) &14.118\,(41) & 13.690\,(40)$^{\rm r}$&                    &                 &               &-17.8\,(5.5)  &6.7\,(5.5)    &              &              &     &     &                                                                                      \\
1889 & 21:36:16.09 &  57:34:48.6$^{\rm r}$&      &                           &                               &                 &                 &                 &       &                 &11.836\,(21) &11.068\,(26) & 10.835\,(22)$^{\rm r}$&                    &                 &               &-2.7\,(4)     &-2.2\,(4)     &-7.5\,(7.2)   &3.7\,(7)      &     &     &                                                                                      \\
1890 & 21:36:45.97 &  57:34:55.1$^{\rm r}$&      &                           &                               &                 &                 &                 &       &                 &15.153\,(73) &14.501\,(67) & 14.303\,(72)$^{\rm r}$&                    &                 &               &-41.5\,(5.2)  &-3.8\,(5.2)   &              &              &     &     &                                                                                      \\
1891 & 21:36:25.59 &  57:35:46.4$^{\rm r}$&      &                           &                               &                 &                 &                 &       &                 &14.715\,(40) &14.417\,(48) & 14.290\,(65)$^{\rm r}$&                    &                 &               &-7.6\,(4)     &-13.1\,(4)    &              &              &     &     &                                                                                      \\
                                                                                                                                                                                                                                                                                                                                                               
 \end{mpsupertabular}

 \label{supertabfull}                                                                                                                                                                                                                                                                                                                                           
% \end{longtable}  
                                                                                                                                                                                                                                                                                                                                            
%\end{center}
\end{@twocolumnfalse}                                                                                                                                                                                                                                                                                                                                          
\end{landscape}

%\newpage

\begin{@twocolumnfalse}
  %\footnotesize
   \scriptsize
%\setlength{\tabcolsep}{0.mm}

%dot M: [c] in Header???
%                                No                  PM                  RV               &  EW(Li)           &  EW(Li)           &  EW(H$\alpha$)    &  EW(H$\alpha$)    &  $\dot M$        &   L_x,c            &  TTS              &   Li       &           H$\alpha$       &   RV         &        $\dot M$      &     X-ray        &       IR-Excess      &     Varia-              PM                  OWN aV              Mass
   
\tablefirsthead{\multicolumn{10}{l}{\bf Table~\ref{supermem} Literature data and membership predictions for stars in Trumpler 37}\\
 No. &  RV  & PM        &EW(Li)&EW(Li)&EW(H$\alpha$)&EW(H$\alpha$)& $\dot M$    &$L_{\mathrm{X,c}}$& TTS & Li    &H$\alpha$&   RV  &$\dot M$&X-ray&IR ex-&Varia- & PM  &$A_{\mathrm{V}}$ & Mass          \\
     &      & [j]       & max  & min  & max         & min         &$10^{-8}$    &$10^{30}$         &     &       &         &       &       &      &cess  &bility &     &(\textit{JHK})   & (models)      \\
     & km/s & \%        & \AA  & \AA  & \AA         & \AA         &M$_\odot$/yr & erg/s            &     &       &         &       &[c]    &[b]   &[d]   &[a,e,f]& [j] &mag              & M$_\odot$    \\
   \hline}
\tablehead{\multicolumn{10}{l}{\bf Table~\ref{supermem} Literature data and membership predictions for stars in Trumpler 37 -- continued}\\
 No. &  RV  & PM        &EW(Li)&EW(Li)&EW(H$\alpha$)&EW(H$\alpha$)& $\dot M$    &$L_{\mathrm{X,c}}$& TTS & Li    &H$\alpha$&   RV  &$\dot M$&X-ray&IR ex-&Varia- & PM  &$A_{\mathrm{V}}$ & Mass          \\
     &      & [j]       & max  & min  & max         & min         &$10^{-8}$    &$10^{30}$         &     &       &         &       &       &      &cess  &bility &     &(\textit{JHK})   & (models)      \\
     & km/s & \%        & \AA  & \AA  & \AA         & \AA         &M$_\odot$/yr & erg/s            &     &       &         &       &[c]    &[b]   &[d]   &[a,e,f]& [j] &mag              & M$_\odot$    \\
   \hline}

\bottomcaption{\small Literature data and membership probabilities for stars in Trumpler 37.      \newline
Remarks: The literature sources and numbering are the same as in Table~\ref{supertab}, empty lines were omitted.
  The proper motion (PM) membership probability as it is given in {$[$j$]$}.
  If the literature gives more than one value for Li or H$\alpha$ equivalent width, the minimal and maximal values are given, otherwise the value is written in the maximum columns. 
  The mass accretion $\dot M$ is only from {$[$c$]$}, the corrected X-ray luminosity only from {$[$b$]$}.
  Column TTS indicates a classical (c) or a weak (w) T Tauri star. If an additional T Tauri state follows in parentheses, the classification differs between low and high resolution spectra (see source literature for more details), colons indicate uncertainty.    \newline
  The next to last column gives the re-calculated extinction as decribed in the text. %An anoted value of ``neg'' indicates negative extinction.
  The last column contains the masses determinded by the models by Siess et al. (\cite{sie00}) from the infrared color-magnitude diagram (Fig.~\ref{CMD-IR_ownAV-iso}).       \newline
  \textbf{The membership prediction:} h, m and l stand for high, medium and low membership probability, as a result of the following criteria:                                               \newline
  $\cdot$ Lithium absorption: see Table~\ref{tabli}. \newline
  %for spectral type early or equal G3: EW(Li)$>$0.15\AA $\rightarrow$ h, EW(Li)$\le$0.05\AA $\rightarrow$ l;
  % $\ge$G8: EW(Li)$>$0.2\AA $\rightarrow$ h, EW(Li)$\le$ 0.1\AA $\rightarrow$ l;
  % $\ge$K3: EW(Li)$>$0.3\AA $\rightarrow$ h, EW(Li)$\le$ 0.2\AA $\rightarrow$ l;
  % $\ge$K7: EW(Li)$>$0.3\AA $\rightarrow$ h, EW(Li)$\le$ 0.2\AA $\rightarrow$ l;
  % $\ge$M4: EW(Li)$>$0.2\AA $\rightarrow$ h, EW(Li)$\le$ 0.1\AA $\rightarrow$ l;
  % $\ge$M9: EW(Li)$>$0.15\AA $\rightarrow$ h, EW(Li)$\le$ 0.1\AA $\rightarrow$ l; otherwise m. \newline
%  $\cdot$ H$\alpha$ emission: if EW(H$\alpha$)$<0 \rightarrow$ h, otherwise l. \newline
  $\cdot$ H$\alpha$ emission: {\bf if spectral type earlier than K0 and EW(H$\alpha$)$<0 \rightarrow$ h, if spectral type later than K0 we follow White \& Basri (\cite{whi03}) to distingish between h and l}. \newline
  $\cdot$ radial velocity (RV): if within $1\sigma$ (3.6\,km/s) around $-15\,\rm{km/s} \rightarrow$ h, if within $3\sigma \rightarrow$ m, otherwise l. \newline
  $\cdot$ Accretion: if $\dot M>0.05\cdot10^{-8}$M$_\odot$/yr $\rightarrow$ h, if $\dot M>0\cdot10^{-8}$M$_\odot$/yr $\rightarrow$ m, if $\dot M=0\cdot10^{-8}$M$_\odot$/yr $\rightarrow$ l. \newline
  $\cdot$ X-ray: {$[$b$]$} analyzed only bright X-ray sources with corrected luminosity $L_{x,c}>0.75\cdot10^{30}$\,erg/s, so all $\rightarrow$ h.\newline
  $\cdot$ Infrared excess: if excess visible in SEDs from Sicilia-Aguilar et al. (\cite{sic06-1}), then h, otherwise l. \newline
  $\cdot$ Variability: if marked as ``$V$'' or ``\textit{RI}'' in the source literature $\rightarrow$ h, if ``$I$'' $\rightarrow$ m, if marked as ``N'' or ``No'' $\rightarrow$ l.  \newline
  $\cdot$ Proper motion: if $p\ge75$\% $\rightarrow$ h, if $p\ge50$\% $\rightarrow$ m, otherwise l.                    %\newline
}
\begin{supertabular}{@{\hspace{0.8mm}}l @{\hspace{1.6mm}} l @{\hspace{1.6mm}} l @{\hspace{1.6mm}} l @{\hspace{1.6mm}} l @{\hspace{1.6mm}} l @{\hspace{1.6mm}} l @{\hspace{1.6mm}} l @{\hspace{1.6mm}} l @{\hspace{1.6mm}} l | @{\hspace{1.0mm}} l @{\hspace{1.6mm}} l @{\hspace{1.6mm}} l @{\hspace{1.6mm}} l @{\hspace{1.6mm}} l  @{\hspace{1.6mm}} l  @{\hspace{1.6mm}} l @{\hspace{1.6mm}} l | @{\hspace{1.0mm}} l @{\hspace{1.6mm}} l @{\hspace{0.8mm}} } 
  % No  &    RV                    & PM &EW(Li          &EW(Li          &EW(Ha          &EW(Ha)          & $\dot M$   &L_x,c  & TTS           &Li     & Halpha & RV    & Mdot  & X-ray & IR-E  & Var & PM &own Av&MassMod                                           \\
%1    &                          &    &               &               &               &                &            &       &               &       &        &       &       &       &       &     &    &      &                                                  \\                                                                                                                                                                                                                                                                                   
 2    &                          & 93 &               &               &               &                &            &       &               &       &        &       &       &       &       &     & h  &      &                                                  \\
\end{supertabular}
  \label{supermemfull}
 \end{@twocolumnfalse}

%\appendix

%\section{This is the title of the first appendix}
%Larger tables, collections of images, spectra or similar kind of data shall be 
%presented in the appendix section rather than in the main body of the 
%text. Several appendices can be separated by the \verb+\section{+{\it title
%of appendix}\verb+}+ command. They are enclosed in the 
%\verb+appendix+ environment.

\end{document}